\documentclass{aa}

\usepackage{hyperref}
\usepackage{natbib}
\usepackage[varg]{txfonts}
\usepackage{booktabs}
\usepackage{bm}
\usepackage{comment}
\usepackage{graphicx}  
\usepackage{subfig}
\usepackage{upgreek}
\usepackage{multirow}
\bibpunct{(}{)}{;}{a}{}{,} 
\usepackage{color, colortbl}

\usepackage{footmisc}

\usepackage{url}

\definecolor{LightCyan}{rgb}{0.88,1,1}

\begin{document}

\title{PENELLOPE\,VI. - Searching the PENELLOPE/UVES sample with spectro-astrometry: Two new microjets of Sz\,103 and XX\,Cha\thanks{Based on observations collected at the European Southern Observatory under ESO programmes 106.20Z8.009, 106.20Z8.010, 106.20Z8.011, and 106.20Z8.012.}} 
 
\author{T.\,Sperling\inst{1}
\and J.\,Eislöffel\inst{1} \and C.\,F.\,Manara\inst{2}\and J.\,Campbell-White\inst{2} \and C.\,Schneider\inst{3} \and A.\,Frasca\inst{4} \and K.\,Maucó\inst{2} \and M.\,Siwak\inst{5,}\inst{6} \and B.\,Fuhrmeister\inst{1} \and R.\,Garcia\,Lopez\inst{7}}
\institute{Thüringer Landessternwarte, Sternwarte 5, D-07778, Tautenburg, Germany
\and
European Southern Observatory, Karl-Schwarzschild-Straße 2, D-85748, Garching bei München, Germany
\and
Hamburger Sternwarte, Gojenbergsweg 112, D-21029, Hamburg, Germany
\and INAF -- Osservatorio Astrofisico di Catania, via S.\,Sofia 78, 95123 Catania, Italy
\and HUN-REN Research Centre for Astronomy and Earth Sciences, Konkoly Observatory, Konkoly-Thege Mikl\'os \'ut 15-17., 1121 Budapest, Hungary
\and
CSFK, MTA Centre of Excellence, Budapest, Konkoly Thege Miklós út 15-17., H-1121, Hungary
\and School of Physics, University College Dublin, Belfield, Dublin 4,
Ireland}
\date{Received: 19/03/2024  / Accepted: 16/04/2024}

\abstract {Young stars accrete matter from their surrounding protoplanetary disk and drive powerful outflows. These two processes shape the final system architecture, and studying how these processes interact is the goal of the ESO Large programme PENELLOPE. PENELLOPE complements the ULLYSES legacy programme on the Hubble Space Telescope (HST) by providing ground-based -- Very Large Telescope (VLT) -- optical and near-infrared spectroscopy of more than 80 low-mass young stars.} {The main goal of this study is to screen the PENELLOPE/UVES targets for outflow activity and find microjets. A spectro-astrometric analysis in the [OI]$\lambda$6300 line in the velocity components of the microjet can give insights into the origin of the line emission, that is, if they originate from a magneto-hydrodynamical (MHD) wind or a photoevaporative wind.} {In total, 34 T\,Tauri stars of the PENELLOPE survey have been observed with the high-resolution optical slit spectrograph UVES ($R\sim 65\,000$, $\lambda = 3300-6800\,\AA$). We formulated four criteria to rank the targets according to their outflow activity. Most of the targets have been observed in three different slit positions rotated by $120^\text{o}$ with UVES. Using spectro-astrometric techniques in the [OI]$\lambda$6300 and H$\alpha$ emission lines in each slit position of each target, we searched for outflow signatures, that is, an offset emission with respect to the continuum contribution of the associated T\,Tauri star. We checked all spectra for the presence of other wind line emission of [SII]$\lambda$4068, [SII]$\lambda$4076, [OI]$\lambda$5577, [OI]$\lambda$6300, [OI]$\lambda$6363, [SII]$\lambda$6716, [SII]$\lambda$6731, [NII]$\lambda$6548, and [NII]$\lambda$6583. Line profiles of H$\alpha$ were inspected for a P-Cygni signature. All [OI]$\lambda$6300 line profiles were decomposed into their constituent high-velocity component (HVC) and low-velocity component (LVC).} {Our spectro-astrometric analysis in the [OI]$\lambda$6300 wind line reveals two newly discovered microjets associated with Sz\,103 and XX\,Cha. Both microjets have an extent of about $0\farcs 04$, that is, $<10\,\text{au}$, and we confined their orientation by the three slit observations. We identified two other interesting targets for which all four outflow criteria are fulfilled: Sz\,98 and Sz\,99. These targets display peculiar wind lines in their spectra with multiple velocity components, however, with the lack of a spectro-astrometric outflow signature. Furthermore, we confirm the binary nature of VW\,Cha and CVSO\,109. We present (further) evidence that DK\,Tau\,B and CVSO\,104\,A are spectroscopic binaries. Sz\,115 is tentatively a spectroscopic binary. We find that the P-Cygni line profile in the H$\alpha$ line is not a robust indicator for the presence of outflows.} {The utilised observing strategy (rotating the UVES slit in three different positions) is very powerful in detecting microjets in T Tauri stars. The three slit positions can confine the spatial extend of the forbidden emission line regions. The introduced metric to rank targets according to their outflow activity is useful for follow-up observations. The origin  of the LVC, that is, MHD winds versus photoevaporative winds, of the Sz\,103 and XX\,Cha microjets remains unclear.}   
\keywords{ISM: jets and outflows, Stars: spectroscopic binaries, Stars: winds, outflows, Stars: protostars, T\,Tauri stars}

\titlerunning{short title}
\authorrunning{name(s) of author(s)}
\maketitle

\section{Introduction}

Protostellar outflows are crucial elements of star formation \citep[e.g.][]{frank_2014, bally_2016, ray_2021}. They carry away the bulk part of angular momentum from the accreting disk-star system, thereby inhibiting the spin-up of the forming protostar to excessive rotational velocities. We see protostellar outflows as extended emission regions in atomic and ionic (e.g. [OI], [FeII]) and molecular lines (e.g. H$_2$ or CO) towards young stellar objects (YSOs) of all early evolutionary stages.

In forbidden emission lines (FELs) such as the oxygen line at $6300\,\AA$, they often exhibit two distinct velocity components \citep[e.g.][]{hartigan_1995, natta_2014, rigliaco_2013, banzatti_2019}: a low-velocity component (LVC, $|v_p| < 30$\,km\,s$^{-1}$), which is attributed to the wide-angle, slow winds, and a high-velocity component (HVC, $|v_p|$  up to $300$\,km\,s$^{-1}$) tracing jets, that is, a highly collimated, fast outflow component. Both outflow components (winds and jets) are thought to originate within a few au of the star-disk system, but their  physical connection is still not fully understood \citep[see e.g. the review by][]{pascucci_2023}. It is unclear if the winds are directly collimated into the jets, or if they just represent an independent outflow component. Probing protostellar outflows in the innermost region down to $<$10\,au scales (i.e. $<0\farcs 07$ at 150\,pc), close to their driving source where the wind first interacts with the ambient medium, is therefore of major importance to understand how or if winds turn into jets. Microjets are small-scale (less than a few $100\,\text{au}$), compact jets driven by YSOs such as T\,Tauri stars and they are key elements to study the ejection process in action \citep[e.g.][]{ray_2007_review}. Unfortunately, only a few microjets are known, because they are hard to detect \citep{nisini_2018}. The discovery and analysis of new microjets is crucial for understanding protostellar outflows. 
 
 Over the past decades, much attention has been paid to surveys of YSOs in the [OI]$\lambda$6300 emission line to investigate their outflow activity. \citet{hartigan_1995} used the high-resolution echelle spectrograph at the Kitt Peak 4-m telescope (resolving power: $R\sim 25\,000$) to study 42\,T\,Tauri stars in the optically forbidden wind lines in the range of $5\,000-6800\,\AA$. That data were later re-processed and analysed together with seven VLT/UVES and one VLT/FEROS\footnote{FEROS - Fibre-fed, Extended Range, Échelle Spectrograph; HIRES - High Resolution Echelle Spectrometer; FLAMES - Fibre Large Array Multi Element Spectrograph; TNG - Telescopio Nazionale Galileo; GIARPS - GIAno and haRPS; ESPRESSO - Echelle Spectrograph for Rocky Exoplanets and Stable Spectroscopic Observations\label{first_footnote}} observations of YSOs by \citet{rigliaco_2013}. \citet{simon_2016}, \citet{fang_2018}, and \citet{banzatti_2019} used the high-resolution spectrograph Keck/HIRES\footref{first_footnote} ($R\sim 45\,000$) to study the FELs in more than 60 YSOs. \citet{natta_2014} and \citet{nisini_2018} utilised the medium-resolution long slit spectrograph VLT/X-Shooter ($R\sim 5\,000-18\,000$) to study 131 YSOs in the [OI]$\lambda$6300 line. \citet{mcginnis_2018} investigated 184 young stars of NGC 2264 using VLT/FLAMES\footref{first_footnote} ($R\sim 26\,500$). The highest spectral resolution was achieved in the surveys of \citet{giannini_2019, gangi_2020}, and \citet{2023_nisini} using the spectrograph TNG/GIARPS\footref{first_footnote} ($R\sim 115\,000$). \citet[][]{pascucci_2023} have summarised what can be learned from these studies:  1) Accretion and ejection processes in YSOs are connected as the [OI]$\lambda$6300 line has been detected towards almost all accreting T\,Tauri stars. 2) Outflows may evolve together with their driving source, that is, the star-disk-envelope system. 3) Photoevaporative or magneto-hydrodynamical disk (MHD) winds may explain the origin of the LVC. However, the precise origin of the LVC, which often again shows two separate components itself (a narrow and a broad component), is still unclear. So far, only in the case of RU\,Lup could \cite{whelan_2021} show strong evidence using spectro-astrometry that the LVC for that source originates from MHD winds. 
 
In this context, the ULLYSES/HST legacy survey \citep{roman_duval_2020} is an extensive programme using 500 Hubble Space Telescope orbits to observe young low-mass stars ($\sim 0.5-1M_\odot$) at optical and UV wavelengths, for an unprecedented dataset to study the evolution of YSOs in their accretion and outflow properties. The ESO Large programme PENELLOPE/VLT \citep{manara_2021} complements the ULLYSES project by providing optical and near-infrared spectroscopic data of the ULLYSES targets. Thus, ULLYSES and PENELLOPE are delivering a benchmark sample of young stars with well-known accretion and disk properties that will, undoubtedly, form the core group of objects to be studied in detail in the coming years. PENELLOPE provides high-resolution spectroscopy observations of about 90 low-mass YSOs using the two VLT instruments UVES and ESPRESSO\footref{first_footnote}. Medium resolution spectra were obtained for 83 targets using VLT/X-Shooter. For observational details of the PENELLOPE survey see \citet{manara_2021}. This paper focuses on investigating the outflow activity and the detection of microjets in the 34 PENELLOPE/UVES targets. As a spectro-astrometric and kinematical study we present 1D line profile decompositions of the [OI]$\lambda$6300 line towards all targets including the width at half maximum (FWHM) and peak velocities. 

The paper is structured as follows: Section\,\ref{sec:observations} provides details on the targets and the utilised $0^\text{o}$/$120^\text{o}$/$240^\text{o}$ observation strategy. Most of the UVES targets have been observed in these three different slit positions offering the unique opportunity to search for microjets via spectro-astrometry in the [OI]$\lambda$6300 line (which is not possible for ESPRESSO as a fibre-fed instrument). In Sect.\,\ref{sec:data_reduction} we briefly describe the data reduction steps involved after reducing it standardly with ESOREFLEX (e.g. telluric and photospheric correction). In Sect.\,\ref{sec:data_analysis} we outline the implemented spectro-astrometric workflow and we formulate four criteria to screen for interesting outflow sources. The results of our search, that is, a ranking of the sources according to their outflow activity and the detection of microjets, are presented and discussed in Sect.\,\ref{sec:results} and Sect.\,\ref{sec:discussion}, respectively. For the newly detected microjets we discuss their spectro-astrometric signals in the HVC and LVC of [OI]$\lambda$6300, [OI]$\lambda$5577, [SII]$\lambda$6731, and [NII]$\lambda$6584. In a follow-up paper, we will discuss the physical parameters derived from line fluxes (not included here) that can be derived for the interesting outflow sources.  

\section{Observations}\label{sec:observations}

\subsection{UVES setup}

In the course of the PENELLOPE survey 34 young stellar objects have been observed with the Ultraviolet and Visual Echelle Spectrograph \citep[UVES,][]{dekker_2000} on the European Southern Observatory's Very Large Telescope (VLT). 
UVES is a high resolution spectrograph with two arms (UV-Blue, Visual-Red) operating at $3000-5000\,\AA$ and $4200-11000\,\AA$, respectively. Since the CCD detector in the red arm of UVES consists of two chips with a physical gap in between two separate spectral datasets for the red arm (REDL, REDU) were obtained.  All targets have been observed in both arms (UVES settings: DIC1\,390+580) covering a spectral range of $3280-4560\,\AA$ (BLUE data), $4730-5750\,\AA$ (REDL data), and $5765-6835\,\AA$ (REDU data).  
 The slit width was $0\farcs 6$ providing a spectral resolution of $R\sim 65\,000$ ($\Delta v \sim 4.6$\,km\,s$^{-1}$) in both arms (UVES User manual, Doc. No. VLT-MAN-ESO-13200-1825). The pixel scale for the 2-dimensional data (short 2D) were $0\farcs 246\,\text{pixel}^{-1}$ for the BLUE data and $0\farcs 182\,\text{pixel}^{-1}$ for the REDL and REDU data.  The observations have been carried out between the end of 2020 and mid 2022 (programme IDs: 106.20Z8.009, 106.20Z8.010, 106.20Z8.011, 106.20Z8.012; PI: C.\,Manara). 
 
 \subsection{Targets}
 
The targets are part of the ULLYSES sample\footnote{\url{https://ullyses.stsci.edu/ullyses-targets-ttauri.html}}. The target list with the observational details (coordinates, exposure times, seeing, and slit position angles) and source properties (e.g. spectral type, extinction values, systemic velocities) are presented in Tables\,\ref{table:observation_details_I} and \ref{table:observation_details_II} of the Appendix. Further target parameters such as disk inclination, stellar masses, distances, and accretion rates are compiled in Table\,1 in \citet{france_2023}. The targets are located in well studied star forming regions (number in brackets) of Lupus (14), Chameleon (8), Orion\,OB1 (7), $\sigma$ Ori (2), Taurus (1), and the TW\,Hydrae Association (1). These star forming regions are about 1-10\,Myr old \citep[see Fig.\,5 in][]{manara_2023}.
We note that 19 targets of this study are also included in the X-Shooter survey of 131 class II sources carried out by \citet{nisini_2018}.

\subsection{Multiplicity of the targets}\label{sec:multiplicity}

 Ten targets are known (wide/spectroscopic) binaries \citep[][]{brandner_1996, correia_2006, guenther_2007, kraus_2007, lafreniere_2008, manara_2019, tokovinin_2020, manara_2021}: CSVO\,17 ($ 8\farcs 1$), CSVO\,36 ($ 3\farcs 4$), CSVO\,104 ($2\farcs 4$), CSVO\,109 ($ 0\farcs 7$), DK\,Tau ($2\farcs 4$), VW\,Cha ($0\farcs 7$), CS\,Cha ($\sim 4\,\text{au}$ or $\sim 0\farcs 023$ at 176\,pc), CV\,Cha ($11\farcs 4$), XX\,Cha ($24\farcs 4$), and Sz\,68 ($ 2\farcs 8$). Their projected separations are indicated in brackets after the source name. Given the slit lengths (BLUE: $8''$, RED: $12''$) and the seeing conditions of the observations the binaries Sz\,68, CVSO\,36, CVSO\,104, and DK\,Tau are spatially resolved in the 2D data. For Sz\,68 and CVSO\,36 the secondary components are very faint, so that we focus on the bright primary towards these targets. The primary and secondary visual components of CVSO\,104 (A, B) are comparably bright in our UVES spectra. \citet{frasca_2021} recently found CVSO\,104\,B to be a background Sun-like star - we therefore focus our analysis on CVSO\,104\,A. In their study, \citet{frasca_2021} also discovered that CVSO\,104\,A is a double-lined spectroscopic binary with an orbital period of about five days.
DK\,Tau\,A \citep{manara_2019} is a young accreting classical T Tauri star with a disk \citep[e.g.][]{nelissen_2023_dktau_I, nelissen_2023_dktau_II} featuring multiple wind line detections \citep{fang_2018, banzatti_2019}. DK\,Tau\,B is the less studied companion. We decided to study DK\,Tau\,A and DK\,Tau\,B separately in our search for outflows. 
  The binaries CS\,Cha, VW\,Cha, and CVSO\,109 are not spatially resolved and in the cases of CVSO\,17, CV\,Cha, and XX\,Cha the UVES slits do not cover the secondaries, which are too far separated. For the binaries CVSO\,36, CVSO\,104, CVSO\,109, and DK\,Tau the slit was orientated along the line connecting the binaries and not using the $0^\text{o}$/$120^\text{o}$/$240^\text{o}$ strategy.

\subsection{Spectro-astrometry in three slit positions}

\begin{figure}  
\resizebox{\hsize}{!}{\includegraphics[trim=0 0 0 0, clip, width=0.9\textwidth]{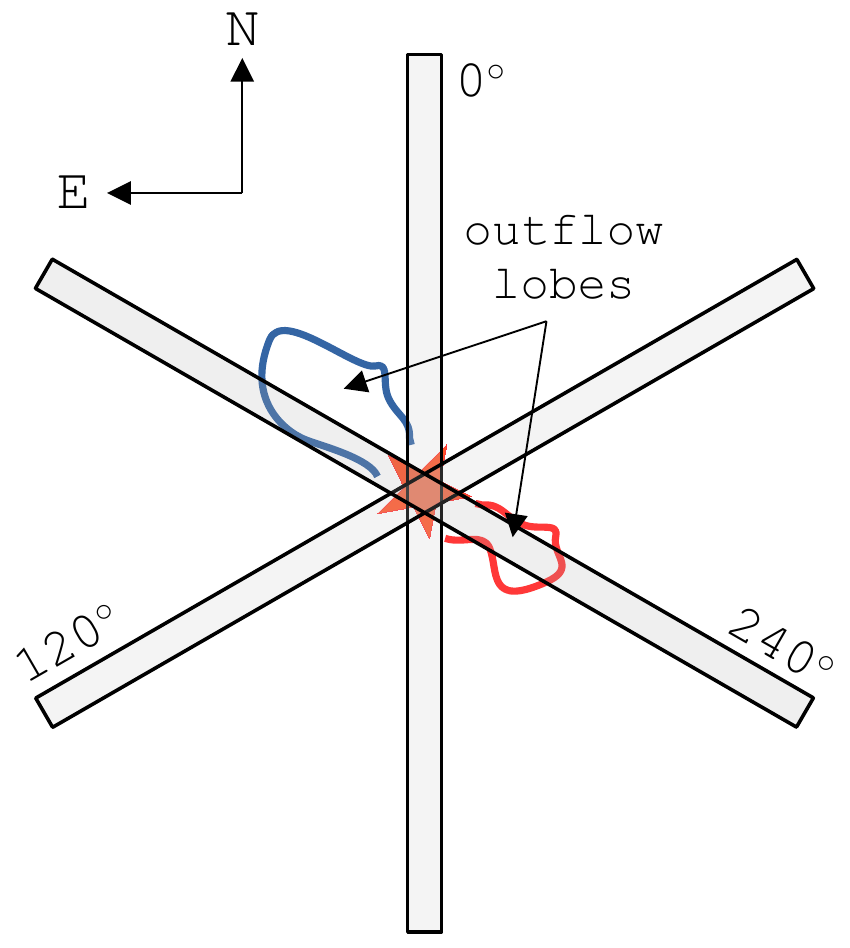}}
\caption{\small{UVES slit positions to detect microjets. Marked are arbitrary positioned outflow lobes with the one directed towards us indicated by the blue line and the one directed in the other direction indicated by the red line. The star is always located at the centre of the slit.}}\label{fig:slit_positions} 
\end{figure}

Since neither the presence nor the orientation of a microjet towards the targets of this study were known initially, we observed 
24 targets in three different slit positions rotated by $120^\text{o}$ (Fig.\,\ref{fig:slit_positions}). For the remaining nine targets only one or two slit positions were obtained. The three slit observations were taken a few days apart from each other to study also the aspect of short term variability \citep[see overall strategy described in][]{manara_2021}. Only in the case of failed observations the observations were repeated a few days later. The few exceptions with only one or two slit positions comprise the monitoring target TW\,Hya, the binaries Sz\,68, CVSO\,17, CVSO\,36, CVSO\,104, CVSO\,109, and DK\,Tau and two other targets (Sz\,40, 2MASS\,J16000060-4221567), for which only two slit position were taken due to failed observations. 

The adopted observing strategy offers the unique chance to discover microjets via spectro-astrometry even if they are only partially aligned with the spectrograph slit \citep[e.g.][]{hirth_1997, whelan_2021}. \citet{bailey_binaries_1998} utilised this technique to detect binaries.

 A spectro-astrometric analysis in the [OI]$\lambda$6300 line in only one slit position of X-Shooter was also undertaken in the  \citet{nisini_2018} study. Among the 131 targets \citet{nisini_2018} found five targets (Sz\,22, Sz\,73, Sz\,83 = Ru\,Lup, Eso\,Ha\,562, SSTc2dJ160708.6) with a spectro-astrometric shift with respect to the continuum, that is, a (micro-)jet. However, no spectro-astrometric offset was detected in any of the overlapping 19 targets of this study. 

\begin{figure}  
\resizebox{\hsize}{!}{\includegraphics[trim=0 0 0 0, clip, width=0.9\textwidth]{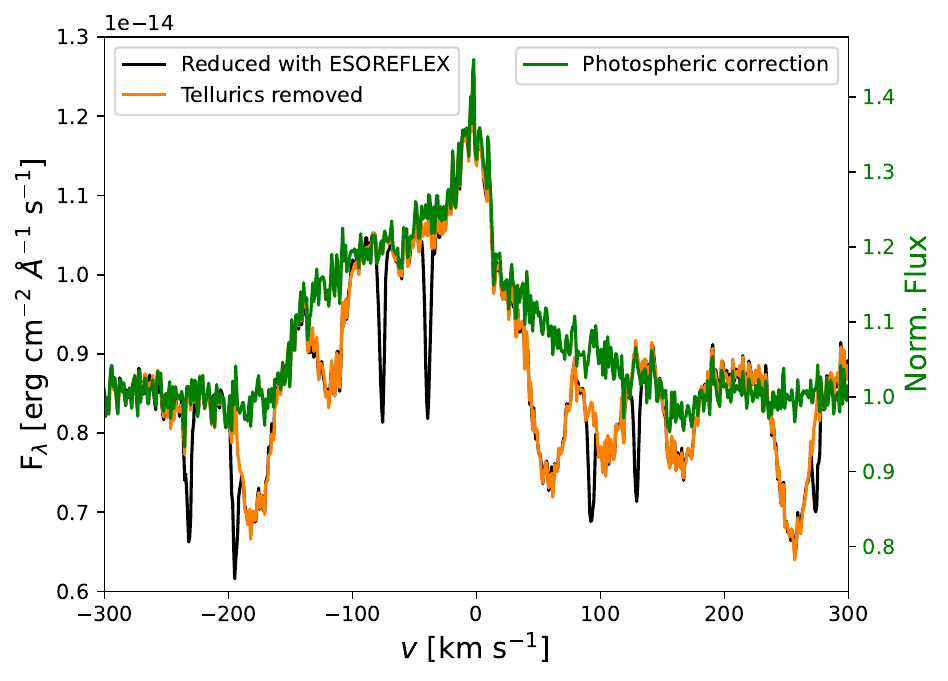}}
\caption{\small{Correction of the [OI]$\lambda$6300 spectra shown for Sz\,98 in slit position 3 as an example for all targets. \textit{Black:} The 1D spectrum after the reduction with the ESOREFLEX/UVES workflow. \textit{Orange:} The 1D spectrum after the telluric correction. The narrow telluric absorption features have been mitigated. The remaining absorption features are due to the stellar photosphere. \textit{Green:} The 1D spectrum after the photospheric correction. The broad photospheric absorption features have been removed and the flux is normalised to the continuum.}}\label{fig:correction} 
\end{figure}

\section{Data reduction}\label{sec:data_reduction}

The obtained data were standardly reduced with the ESOREFLEX environment  \citep[v.\,2.11.5,][]{esoreflex_2013} using the UVES pipeline \citep[v.\,6.1.8,][]{ballester_2000} to extract the 1D and 2D spectra for further analysis, that is, line detection, line profile inspection, and position-velocity (PV) diagrams. The data reduction done by ESOREFLEX includes bias subtraction, flat fielding, wavelength calibration, and optimal spectrum extraction. Two consecutive exposures for each observation were coadded. Cosmics in the 2D data were removed with the lacosmic routine implemented in PYTHON based on  \citet{van_dokkum_2001}. The bright terrestrial night-skylines at $6300\,\AA$ and $6363\,\AA$ were removed manually by fitting and subtracting a 2D Gaussian or 2D Voigt profile to the skyline. In a few cases the skyline subtraction left artefacts such as spikes in this spectral region, which we masked in our line analysis. We used the \ion{Li}{i} absorption line feature centred at $6707.856\,\AA$ \citep{champbell_white_lithium} to correct the spectra for systemic velocity shifts, listed in Tables\,\ref{table:observation_details_I} and \ref{table:observation_details_II} of the Appendix. 

We corrected the 1D data of the REDU arm for telluric (O$_2$, H$_2$O bands) and photospheric absorption features as described in \citep[][]{manara_2021}. This correction step is displayed in Fig.\,\ref{fig:correction}. 
In the REDL data only the photospheric correction was applied, since the telluric features are only marginally present. No photospheric correction nor telluric correction was applied in the BLUE data, due to the low signal-to-noise and the greater presence of (other) emission lines. 

In order to apply the spectro-astrometry method in interesting emission line regions we also corrected the 2D REDU data for telluric and photospheric absorption by applying the 1D correction row-by-row to the 2D data of the stellar continuum (see Fig.\,\ref{fig:2d_correction_steps_Sz98}). This correction is indeed necessary before analysing the data via spectro-astrometry. Our experiments on the obtained UVES data show that false spectro-astrometric signals can be induced by steep gradients in the line profiles (see Appendix\,A). Strong tellurics or imprints of the stellar photosphere itself can have such effects and therefore must be mitigated.

\section{Data analysis}\label{sec:data_analysis}

\subsection{The spectro-astrometric workflow and line fitting}\label{sec:sa}
 
\begin{figure*} 
\centering
\subfloat{\includegraphics[trim=0 0 0 0, clip, width=0.99 \textwidth]{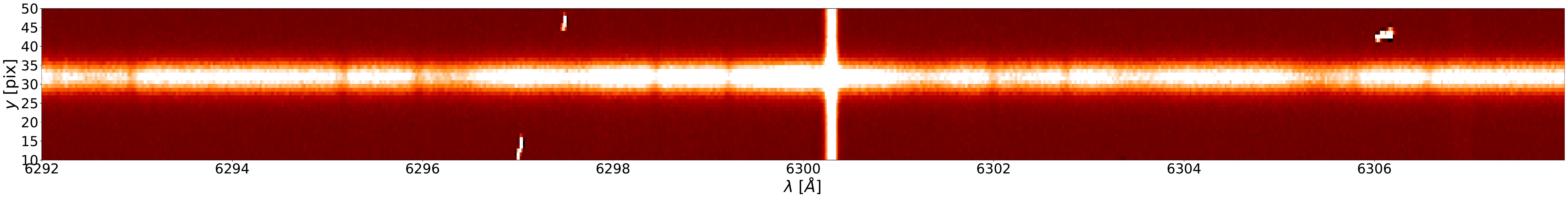}}
\hfill
\subfloat{\includegraphics[trim=0 0 0 0, clip, width=0.99 \textwidth]{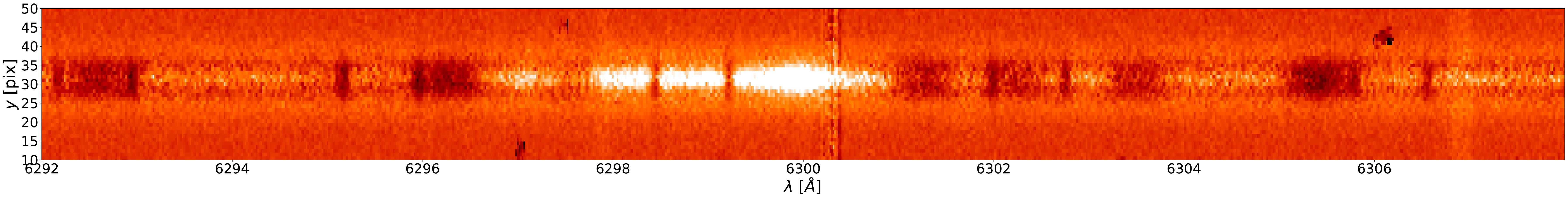}} 
\hfill 
\subfloat{\includegraphics[trim=0 0 0 0, clip, width=0.99 \textwidth]{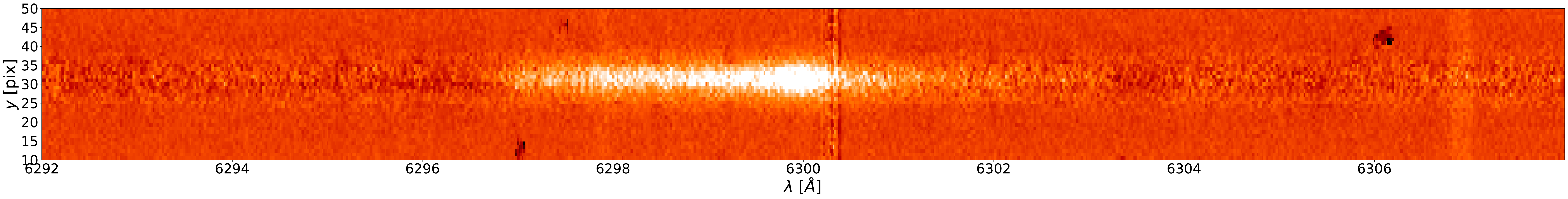}}
\hfill  
\caption{\small{Illustration of the correction steps involved for the spectro-astrometric anylysis of the [OI]$\lambda$6300 emission line. All rows are showing the same 2D data of Sz\,98 in slit position 3 at consecutive data reduction steps. \textit{Top:} The combined 2D data after the reduction with the ESOREFLEX/UVES workflow is shown. The continuum contribution of the star plus the line emission is seen as a bright horizontal stripe between the pixel rows $y=25-40\,\text{pix}$. The continuum of Sz\,98 displays several absorption features. The narrow absorption lines are tellurics and the broader lines are originating from the photosphere of Sz\,98.
The bright oxygen night-skyline at about 6300$\AA$ is prominently present as a vertical bright stripe in the spectrum. The impact of three cosmics is seen in the spectral region between $6296\,\AA$ and $6298\,\AA$ as well as at about $6306\,\AA$. \textit{Middle:} The night-skyline and the cosmics have been removed. The continuum contribution (2D Gaussian) of Sz\,98 has been removed without correcting for telluric and photospheric absorption lines.  \textit{Bottom:} The continuum contribution of Sz\,98 has been removed (2D Voigt profile) including the correction for tellurics and photospherics row-by-row (1D correction in Fig.\,\ref{fig:correction}). A spectro-astrometric analysis is done by fitting 1D Gaussians in spatial direction to the remaining [OI]$\lambda$6300 emission, which is displayed in Figs.\,\ref{fig:all_minispectra_CSCha}-\ref{fig:all_minispectra_CVSO104A}.}}\label{fig:2d_correction_steps_Sz98}
\end{figure*}

The spectro-astrometric PV-diagrams were extracted following \citet[e.g.][]{whelan_2005, whelan_2006_brown_dwarf, whelan_2007}, however, without smoothing the data. The continuum contribution of the star in the 2D data was removed by fitting a single Gaussian or Voigt profile in a narrow region around the emission line while excluding the emission line region itself. For a Gaussian fit the 2D fit profile function is given by
\begin{equation}\label{eq:2d_gaussian}
\phi(y,\lambda) = \frac{A}{\sigma\sqrt{2\pi}}\text{exp}\left[-\frac{1}{2}\left( \frac{y-y_c}{\sigma}\right)^2\right] + C,
\end{equation}
with $A$, $\sigma$, $y_c$, and $C$ being the Gaussian fit parameters ($y$ spatial direction along the slit and $\lambda$ spectral direction). The right side of Eq.\,\ref{eq:2d_gaussian} is independent of $\lambda$, since we assume that the 2D Gaussian around the emission line is a series of the same 1D Gaussian in dispersion direction. In some cases the Gaussian profile function left residuals after the continuum contributions (Fig.\,\ref{fig:2d_correction_steps_Sz98}, second row). In these cases the Voigt profile better described the continuum (Fig.\,\ref{fig:2d_correction_steps_Sz98}, third row). Since there is no closed form for the Voigt profile, we used the standard relation to the complex Faddeeva function, $w(z)$, that is, given by
\begin{equation}
\phi(y,\lambda) = A\cdot \text{Re}\left[w(z)\right]/(\sigma\sqrt{2\pi}) + C
\end{equation}
with
\begin{equation}
z = \left(y-y_c+i\gamma\right)/\left(\sigma\sqrt{2}\right)  
\end{equation}
to remove the continuum. Here, $A$, $\sigma$, $\gamma$, $y_c$, and $C$ are the Voigt fit parameters. 
We then fitted 1D Gaussians to the remaining line emission in spatial direction and determined their centroids with respect to the centroid of the removed continuum for each spectral slice.  We used the Non-Linear Least-Square Minimisation tool implemented in Python \citep{newville_2014} for the fitting. In order to find the minimum amount of Gaussian components in an emission line, we fitted up to four Gaussians to the [OI]$\lambda$6300 line profiles and compared the quality of the fits \citep{banzatti_2019}. Our analysis indicates that at most three Gaussian components were sufficient to fit all [OI]$\lambda$6300 line profiles.

 \subsection{Outflow signatures}\label{sec:outflow_criteria}
 
Protostellar outflows are traced by a variety of methods \citep[e.g.][]{ray_2007_review, frank_2014, bally_2016} In order to find the targets which are potentially interesting outflow sources we included the following four criteria in our search for microjets in the PENELLOPE/UVES sample:

 \begin{itemize}
 \item[1.] \textbf{P-Cygni signature:}  The geometry (inclination of the accretion disk and viewing angle), the accretion activity, and the outflow activity have a substantial influence on the observed line profiles and therefore can be very useful confirming microjets.  The presence of a P-Cygni-like line profile in for example H$\alpha$, H$\beta$, or He\,I\,10830$\AA$ can indicate strong outflow activity \citep[e.g.][]{muzerolle_2001, edwards_2006, kursosawa_2006, wilson_2022}. In this context classification schemes for observed line profiles have been put forward, which order line profiles according to the depth and spectral position of absorption features \citep[e.g.][]{reipurth_1996, erkal_2022}.
 
  We classified the H$\alpha$ line profiles of the observed targets in all three slit positions by eye following \citet{reipurth_1996}.  The classical P-Cygni line profile is an outflow signature seen as a dip in the blue-shifted wing of the emission line below the continuum, that is, classified as IVB in \citet{reipurth_1996}. In many cases of YSOs, however, the described dip does not fall below the continuum (case IIIB therein) but still are labelled as a P-Cygni-like dip \citep[e.g.][]{whelan_2005}. We follow that notion by looking for IVB and IIIB profiles as outflow signature. An inverse P-Cygni line profile, that is, a  red-shifted sub-continuum absorption feature, can be attributed to magnetospheric accretion \citep[e.g][]{edwards_1994, fohla_2001}. We denote the appearance of multiple secondary peaks in the 1D line profiles with the letter \textit{m} behind the profile class.
\item[2.] \textbf{Detection of the [OI] lines}: The [OI]$\lambda$6300 line is usually the brightest optical FEL detected towards protostellar outflows \citep[e.g.][]{nisini_2018, giannini_2019, banzatti_2019, pascucci_2023} and thus has proven to be a powerful outflow tracer. The total observed [OI]$\lambda$6300 emission can, however, originate from very distinct physical processes in action (e.g. MHD disk winds, photoevaporative winds, shocks, dissociation of OH molecules) and potentially very distinct regions around the YSO. The two complementary FELs of oxygen are [OI]$\lambda$6363 and [OI]$\lambda$5577. Coming from the same upper atomic level ($^1$D$_2$ at $E/k_B = 22\,830\,K$), the [OI]$\lambda$6363 line is expected to be fainter than the [OI]$\lambda$6300 line by a factor of about $\sim 0.32$.
The [OI]$\lambda$5577 line on the other hand emerges from a higher excited level of the oxygen atom ($^1\text{S}_0$ at $E/k_B = 48\,620\,K$). As a result its critical density differs from the [OI]$\lambda$6300 line and thus they both can trace different gas components \citep[see discussions in e.g.][]{natta_2014, simon_2016}. If both lines are thermally excited their line ratio [OI]$\lambda$5577/[OI]$\lambda$6300 depends on the physical conditions prevailing (e.g. electron density, electron temperature, ionisation fraction). Non-thermal processes such as the dissociation of OH molecules due to the presence of UV photons can, however, produce similar line ratios \citep[e.g.][]{gorti_2011}. Thus, the detection of [OI] emission alone is not enough to unambiguously indicate the presence of an outflow but nevertheless a strong indicator.  
 \item[3.] \textbf{Detection of additional wind lines:} Protostellar outflows are often prominently traced via other ionic FELs of, for example, [NII], [OII], and [SII], which can be collisionally excited in shocks \citep[e.g.][]{bally_2016, liu_2016, fang_2018, pascucci_2023}. In this scenario, their line ratios with [OI] are connected with each other by ionisation balance and thus provide an invaluable diagnostic tool to derive gas parameters of the outflow region \citep{be_1999}. We constrained our search for outflow activity by looking at strong wind emission lines falling in the spectral range of UVES. Explicitly, the FELs of nitrogen ([NII]$\lambda$6548, [NII]$\lambda$6583)\footnote{The nitrogen line [NII]$\lambda$5754 was not covered in the UVES observations.}, oxygen ([OII]$\lambda$3726, [OII]$\lambda$3729), and sulphur ([SII]$\lambda$6716, [SII]$\lambda$6731, [SII]$\lambda$4068, [SII]$\lambda$4076) are very useful outflow tracers.  We note that we did not check for the presence of optical [FeII] or [FeIII] lines in our spectra since even the brightest iron lines falling in the UVES spectral range are expected to be much fainter than the already hardly detected wind lines \citep[e.g.][]{whelan_2014}. 
 \item[4.] \textbf{Presence of a high velocity component:} A high-resolution line profile analysis in FELs towards YSOs often reveals the presence of multiple velocity components \citep[e.g.][]{hartigan_1995, hirth_1997, banzatti_2019}, that is, the low velocity  component (LVC, peak velocities: $|v_p|\lesssim 30\,\text{km}\,\text{s}^{-1}$) and the high velocity component (HVC, $|v_p| \geq 50\,\text{km}\,\text{s}^{-1}$). It is widely accepted that the HVC is attributed to the collimated jet, which is more often detected via the blue-shifted HVC (HVCB) than the red-shifted HVC (HVCR).  The LVC can often be decomposed in two  Gaussians \citep{rigliaco_2013, simon_2016, banzatti_2019}, that is, a narrow component (NLVC, full width half maximum of the line  $\Delta v\sim 10-70\,\text{km}\,\text{s}^{-1}$) and a broad component (BLVC, $\Delta v \sim 40-300\,\text{km}\,\text{s}^{-1}$). The origin of the LVC is still under debate and its relation to the HVC needs to be clarified \citep[see review of][]{pascucci_2023}. For the analysis of outflow activity it is interesting to see in which FELs we see the HVC and for which we see only the LVC \citep{giannini_2019}.
 \end{itemize}
 
 The main motivation behind these four criteria is that it introduces a scientifically motivated metric to rank targets according to their outflow activity from less complex sources to very complex outflow sources. Sources fulfilling more criteria are physically more interesting for outflow studies, since they potentially allow us to study their system geometry, outflow kinematics, origin of line emission, and gas parameters in more detail. All four criteria should in theory be physically connected to each other, but observationally one or two criteria may fail (e.g. S/N issues). 
The four criteria give no information about the spatial extent of the outflow. 

The introduced metric can be applied to a population of YSOs. The resulting distribution within this metric, that is, the number of YSO fulfilling a certain number of these criteria, can give insights on evolutionary trends within the population. Older sources tend to fulfill less criteria as their outflows die out with time.  On the contrary, younger sources are expected to fulfill almost all criteria.   
 
  \subsection{Outflow detection via spectro-astrometry}\label{sec:microjet_criteria}
  
  Outflows and microjets are detected as extended emission line regions. As such they most likely fulfill most of the four outflow criteria in Sect.\,\ref{sec:outflow_criteria}. Integral field spectroscopy or slit spectroscopy in  FELs can reveal information about the orientation of the outflow on sub-arcsecond scale \citep[e.g.][]{solf_boehm_1993, bailey_1998, whelan_2005, murphy_2021, kirwan_2023}. In our case three different slit positions rotated $120^\text{o}$ potentially cover part of the emission region, which thus could be detected via spectro-astrometry. In order to rule out spectro-astrometric artefacts in our data we checked also the bright lines of helium (He\,I\,5876, He\,I\,6678) for spectro-astrometric features. These lines should not show any spectro-astrometric signature as they are connected to accretion \citep{alcala_2014, alcala_2017}. 
  We therefore analysed the H$\alpha$, H$\beta$, [OI]$\lambda$6300, and helium lines (He\,I\,5876, He\,I\,6678) for the presence of a spectro-astrometric offset in the PV-diagrams.  
  In principle, all detected FELs could be used for a spectro-astrometric analysis. If they are associated with an extended outflow region they should show similar but not identical spectro-astrometric signals. 

As for the study performed by \citet{whelan_2012}, our observations were not customised for detecting FELs that trace outflow activity in for example optical [OI], [SII], [NII]. Thus, exposure times geared at investigating the central source make it challenging to detect these wind lines. In turn outflow activity in form of a microjet cannot be ruled out for all targets. Reasons for this could be: 1. The microjets may be too faint to be detected, 2. The inclination of the star-disk system is unfortunate, 3. The microjet is yet too small, that is, smaller than about $0\farcs01$ \citep{whelan_2007}, or 4. The outflow has already died out.

\section{Results}\label{sec:results}

 We present the result of our search for outflow activity with regard to the four criteria in Tables\,\ref{table:outflow_criteria_I}-\ref{table:outflow_criteria_IV} of the Appendix. Figs. \ref{fig:all_minispectra_CSCha}-\ref{fig:all_minispectra_CVSO104A} of the Appendix are showing the spectro-astrometric analysis in H$\alpha$ and [OI]$\lambda$6300 for all targets and slit positions together with their 1D line profiles. The centroid positions of the 1D Gaussians with respect to the position of the continuum centroid are shown in crosses therein. 
   We present the decomposition of the [OI]$\lambda$6300 line in all targets and slit positions in Figs. \ref{fig:all_minispectra_CSCha}-\ref{fig:all_minispectra_CVSO104A} in the Appendix. The Tables\,\ref{table:OI6300components_part_I}-\ref{table:OI6300components_part_IV} in the Appendix summarise the measured [OI]$\lambda$6300 line peak velocities and the full width half maximum in each component for all targets. We do not present absolute flux values in the [OI]$\lambda$6300 or other emission lines, because the chosen UVES slit width is too narrow to rule out substantial slit-losses.

 \subsection{Outflow signatures}\label{lines}

\textbf{P-Cygni-like line profiles of H$\alpha$:} 
The observed H$\alpha$ line profiles of our targets show a variety of features such as symmetries/asymmetries, the presence of (multiple) blue-shifted and red-shifted peaks and dips. The classification scheme of \citet{reipurth_1996} does not capture the complexity of the observed H$\alpha$ profiles. We also see substantial changes in the line profiles for the same target at different slit positions/day of observation in about half of the targets. These changes comprise a change in the relative height of the peaks, or number of peaks, or the depth of dips (see e.g. line profiles of Sz\,68, Sz\,100, Sz\,103).  

A true P-Cygni signature (IVB type) is only detected in VW\,Cha in slit position 3 at about $-200\,\text{km}\,\text{s}^{-1}$. 
A relaxed P-Cygni-like outflow signature whereby the secondary blue-shifted dip does not fall below the continuum but has less than half the strength of the primary peak (IIIB type) is seen in ten targets. Inverse P-Cygni signatures are present in four targets: DK\,Tau\,A, CVSO\,107, CVSO\,109, and SO583.

\textbf{Detection of [OI]:} The highest detection rates of FELs are coming from [OI]. We detect [OI] emission towards most sources:  [OI]$\lambda$6300 in about $79\,\%$ (27/34), [OI]$\lambda$6363 in about $65\,\%$  (22/34), and [OI]$\lambda$5577 in about $56\,\%$  (19/34).  We did not detect any [OI] emission towards seven sources. 

\textbf{Detection of additional wind lines:}
Eighteen of our targets are showing the presence of wind lines of nitrogen or sulphur. In all these targets the [OI]$\lambda$6300 has been detected. Among the nitrogen and sulphur lines the [SII]$\lambda$4068 line has the highest detection rate (15/34). The complementary [SII]$\lambda$4076 line, however, is only detected in four targets.   
The  nitrogen lines ([NII]$\lambda$6548, [NII]$\lambda$6583]) are detected in eight targets. The [SII]$\lambda$6731 and [SII]$\lambda$6716 lines are detected in ten and five targets, respectively. We also checked for lines coming from ionised oxygen ([OII]$\lambda$3726, [OII]$\lambda$3729, [OIII]$\lambda$5007) but none was detected.  We note that we also detect compact [NII] low-velocity emission towards VW\,Cha.  We present the 1D wind line detections of XX\,Cha, Sz\,103, Sz\,100, Sz\,98, Sz\,99, CVSO\,58, CVSO\,107, and SO518 in the REDL and REDU data in Fig.\,\ref{fig:XXCha}-\ref{fig:SO518} in the Appendix.

\textbf{Presence of HVCs:}
We inspected all FELs for the presence of multiple velocity components. We detected a HVC in [OI]$\lambda$6300 in 11 targets. Almost all targets have a HVCB (10/11). Two targets have a HVCR. SO518 is the only target with both a HVCR and a HVCB. All targets with a HVC in [OI]$\lambda$6300 also display a LVC in [OI]$\lambda$6300. Highest peak velocities of $|v_p|> 100\,\text{km}\,\text{s}^{-1}$ are detected in the Orion\,OB\,1 targets (CVSO\,58, CVSO\,107, CVSO\,176), Sz\,98, and DK\,Tau\,A. All other targets with a HVC are showing peak velocities of $|v_p|\sim 40-80\,\text{km}\,\text{s}^{-1}$ (see Tables\,\ref{table:OI6300components_part_I}--\ref{table:OI6300components_part_IV} in the Appendix).\\
With the exception of VW\,Cha, we only see the HVC in [NII]$\lambda$6548 and [NII]$\lambda$6583. The [NII]$\lambda$6583 line features slightly higher peak velocities as compared to [NII]$\lambda$6548. The [SII]$\lambda$6716 and [SII]$\lambda$6731 can show a HVC and a LVC. A HVC in [OI]$\lambda$5577 is tentatively detected towards XX\,Cha, Sz\,103, Sz\,98, and Sz\,99.  

\textbf{Spectro-astrometric signals:} 
Nine targets display peculiar features in their spectro-astrometric analysis. For the spatially unresolved binary CVSO\,109 we detect the spectro-astrometric offset of about $0\farcs 2$ in H$\alpha$, [OI]$\lambda$6300, and He\,I. The triple system VW\,Cha displays different spectro-astrometric offsets in the three slit positions in H$\alpha$, [OI]$\lambda$6300, and He\,I. The spectro-astrometric offsets are: $\rho\sim 0\farcs 2$ for slit 1, $\rho\sim 0\farcs 05$ for slit 2, and $\rho\sim 0\farcs 1$ for slit 3.  The binary CS\,Cha shows some interesting features (Fig.\,\ref{fig:all_minispectra_CSCha} in the Appendix, slit 1: at $v \sim -50\,\text{km}\,\text{s}^{-1}$, slit 2: at $v \sim +50\,\text{km}\,\text{s}^{-1}$, slit 3: at $v \sim -100\,\text{km}\,\text{s}^{-1}$) only in H$\alpha$. CVSO\,104\,A shows an astrometric-offset only in H$\alpha$ in the first two slit positions. The binary Sz\,68 also shows a peculiar spectro-astrometric result in H$\alpha$ in slit positions 1 and 2. 
We detect a peculiar spectro-astrometric feature in Sz\,115 in slit positions 2 and 3 in H$\alpha$, which seems similar to the feature in CVSO\,109. The spectra of Sz\,115 show very weak helium lines (He\,I\,5875, He\,I\,6678) so that these spectro-astrometric signals could not be checked for being artefacts. DK\,Tau\,B displays a spectro-astrometric offset in H$\alpha$ and [OI] only in slit position 1. Interestingly, the lithium absorption feature at $6707\,\AA$ was only detected in the other two slit positions in DK\,Tau\,B.  
We found two targets with strong spectro-astrometric signatures in H$\alpha$ and [OI]$\lambda$6300 but not in He\,I: Sz\,103 and XX\,Cha.  

 \subsection{Distribution of outflow targets}\label{distribution}
 
\begin{figure}  
\resizebox{\hsize}{!}{\includegraphics[trim=0 0 0 0, clip, width=0.9\textwidth]{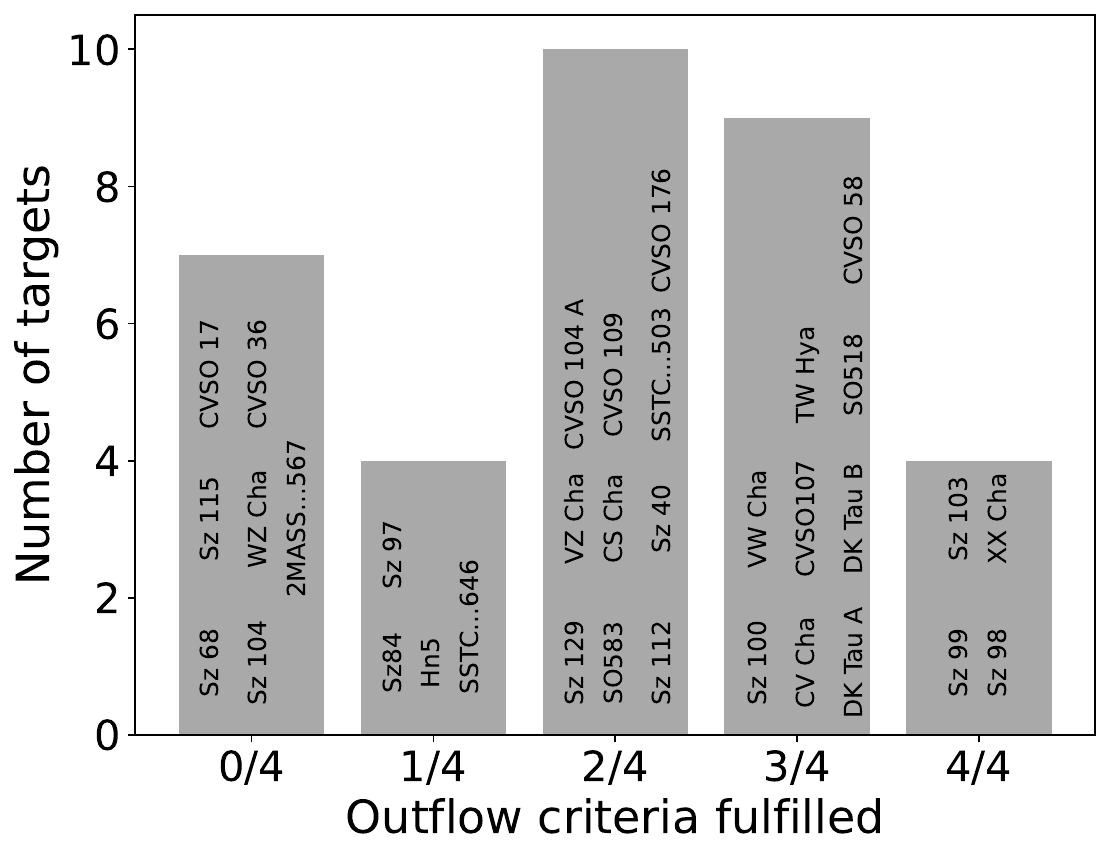}}
\caption{\small{Targets ranked by our outflow criteria.}}\label{fig:bar_plot} 
\end{figure}

The number of targets fulfilling none, one, two, three, or all four outflow criteria is displayed in Fig.\,\ref{fig:bar_plot}. Figure\,\ref{fig:bar_plot} ranks the targets according to their outflow activity. We found four targets, for which all four outflow criteria are fulfilled: Sz\,98, Sz\,99, Sz\,103, XX\,Cha. These targets are not from the same star forming region -- they are associated with Lupus\,III (Sz\,98, Sz\,99, Sz\,103) and Chameleon\,I (XX\,Cha). In general, the star forming region of the target does not provide a strong correlation with the number of criteria fulfilled.   Evidently, the metric filters out the interesting outflow targets.  

\section{Discussion}\label{sec:discussion}

\subsection{Outflow signatures and detections}

\textbf{P-Cygni-like line profiles of H$\alpha$:}  The zoo of different line profiles indicates that multiple phenomena such as accretion absorption, jet or wind absorption, jet or wind emission as well as viewing geometry are simultaneously in action \citep{erkal_2022}. 
The changes in the line profiles are most likely related to short-term variability on time scales of days rather than to the slit orientation.

The detected P-Cygni line profile towards VW\,Cha can be interpreted as originating from strong winds. VW\,Cha is a young binary with strong accretion variability. Its accretion properties and the detection of the [OI] line was recently discussed in \citet{zsidi_2022}. We found no evidence that VW\,Cha drives a microjet and the compact [OI] and  [NII] emission may be connected to disk winds as suggested by \citet{zsidi_2022}.

 The relaxed P-Cygni signal (IIIB) is seen in ten targets -- all of these show [OI]$\lambda$6300 line emission. However, sources with a HVC in [OI]$\lambda$6300 do not necessarily show a IIIB or IVB type line profile in H$\alpha$ (e.g. SO518). Among the ten sources with a P-Cygni signature in H$\alpha$ three are showing only a low velocity [OI]6300 component and no other wind lines, that is, they ranked very low in the metric. We therefore conclude that the P-Cygni line profile in H$\alpha$ is not a strong indicator for outflow activity in form of microjets.
 
 The inverse P-Cygni profile in CVSO\,107, CVSO\,109, DK\,Tau\,A, and SO583 is likely caused by strong accretion. These four targets indeed display high accretion rates and accretion variability \citep{fischer_2011, manara_2021, campbell_white_2021, froebrich_2022, mauco_2023}. 
 
\textbf{Detection of the [OI] lines}: Our detection rates of [OI] are consistent with (but slightly smaller than) the studies of \citet{nisini_2018} ([OI]$\lambda$6300 with $77\,\%$) or \citet{fang_2018} ([OI]$\lambda$6300 with $94\,\%$, [OI]$\lambda$5577 with $54\,\%$), or \citet{banzatti_2019} ([OI]$\lambda$6300 with $94\,\%$, [OI]$\lambda$5577 with $58\,\%$). The small differences in detection rates can be explained by the different sample sizes, S/N, and spectral resolution of the studies. Overall we see an agreement in the detection rates.

\textbf{Detection of additional wind lines:} As in the study of \cite{nisini_2018} the wind lines of sulphur and nitrogen are much fainter than the [OI] lines and therefore much rarer to detect. These lines are shock excited in the outflow material and therefore trace different outflow components depending on the physical conditions prevailing \citep[e.g.][]{giannini_2019}. In the targets of this study they most likely trace low-ionisation material, since we do not detect lines of [OII] or [OIII]. 

\textbf{Presence of HVCs:} \citet{nisini_2018} detected the HVC in [OI]$\lambda$6300 in 30\,\% of their targets. This is in agreement with our detection rate of (11/34). The presence of a HVC is connected to outflow activity \citep{hartigan_1995}. Most targets show only a HVCB. Standardly, one would assume bipolar outflows, that is, the HVCR should be present in these sources as well. Observationally, the non-detection of the HVCR in these cases is interpreted as obscuration effect by the disk. Only in the case of SO518 we see both a blue-shifted and red-shifted HVC. The disk inclination for this target is not known \citep{france_2023}, but the detection of the HVCR could indicate that the inclination is almost edge-on. The high peak velocities of the Orion OB1 targets (CVSO\,58, CVSO\,107, CVSO\,176) could indicate that the Orion OB1 targets are younger than the other targets -- contrary to the expectation as Orion OB\,1  is older than Cha\,I, Lupus, $\sigma$\,Ori. 

\textbf{Spectro-astrometric signals:} From the spectro-astrometric analysis we identified nine targets of interest: VW\,Cha, CVSO\,109, CS\,Cha, CVSO\,104A, Sz\,68, Sz\,115, DK\,Tau\,B, Sz\,103, and XX\,Cha. We think that most of the spectro-astrometric signals can be explained by multiplicity, that is, they are for example binaries (see Sect.\,\ref{sec:multiplicity}). Only in Sz\,103 and XX\,Cha the signals are consistent with the presence of a microjet. In the following we discuss the spectro-astrometric features for each of the nine targets individually. 

For the already known spectroscopic binaries VW\,Cha and CVSO\,109 we detect an offset emission, which is consistent with the binary separation of $0\farcs 7$.  The spectro-astrometric offsets in the case of VW\,Cha can be explained by the individual slit position with respect to the three constituents A+(B+C) of VW\,Cha (see Fig.\,\ref{fig:vwcha_slits} in the Appendix).  CVSO\,109 is a binary \citep{tokovinin_2020} with the secondary being about 0.6\,mag fainter than the primary \citep{proffitt_2021}. Due to the different brightnesses of the two constituents of CVSO\,109 the spectro-astrometric signal is measured at about $0\farcs 2$ and not at $0\farcs 7$. Furthermore, we see the offset emission not only in H$\alpha$ and [OI] but also in He\,I. VW\,Cha and CVSO\,109 do not display a HVC in [OI]. These facts confirm the binary nature of VW\,Cha and CVSO\,109. 

The CS\,Cha feature may also be induced by the binary nature of the system \citep{guenther_2007}. The separation of the binary system is very small ($\sim 0\farcs 02$) implying that the spectro-astrometric signal induced by that should also only be marginally present (which we see in the UVES data). We do not see a HVC towards CS\,Cha -- only a very narrow LVC in [OI]. Therefore we conclude that these spectro-astrometric features are not connected to a micro-jet.

In the case of Sz\,68 the second (very faint) continuum source in slit position 1 and 2 is potentially causing the false spectro-astrometric signals in H$\alpha$. No [OI] emission has been detected in Sz\,68 pointing to the conclusion that winds or jets are only very weakly present, if at all.\\
Sz\,115 displays a spectro-astrometric signal in H$\alpha$ which is very similar to the one of the spectroscopic binary CVSO\,109. Since no wind lines have been detected towards Sz\,115 and the H$\alpha$ line shows no P-Cygni-like line profile, we think that this signal is not connected to outflow activity. Sz\,115 could be a spectroscopic binary.

DK\,Tau\,B shows a spectro-astrometric feature in H$\alpha$ and [OI] in slit position 1. No HVC in [OI] is present in DK\,Tau\,B. This could suggest that DK\,Tau\,B is also a spectroscopic binary, implying that DK\,Tau\,A and DK\,Tau\,B potentially form a triple system.  The same interpretation holds for CVSO\,104\,A, which is in agreement with \citet{frasca_2021}. The non-detection of the lithium line in DK\,Tau\,B is puzzling as one would expect to always detect the lithium line towards a YSO. A possible explanation is that DK\,Tau\,B is a binary with one young companion and one other companion without lithium absorption. Alternatively, it is the effect of very low S/N ratios of DK\,Tau\,B in two spectra.

Sz\,103 and XX\,Cha are the only sources that fulfill all four outflow criteria and display FELs detected via spectro-astrometry. The spectro-astrometric signals are absent in He\,I and clearly connected to the blue-shifted HVC of the [OI]$\lambda$6300 line. Therefore we have strong evidence that Sz\,103 and XX\,Cha host a microjet. 

\begin{figure}  
\resizebox{\hsize}{!}{\includegraphics[trim=0 0 0 0, clip, width=0.9\textwidth]{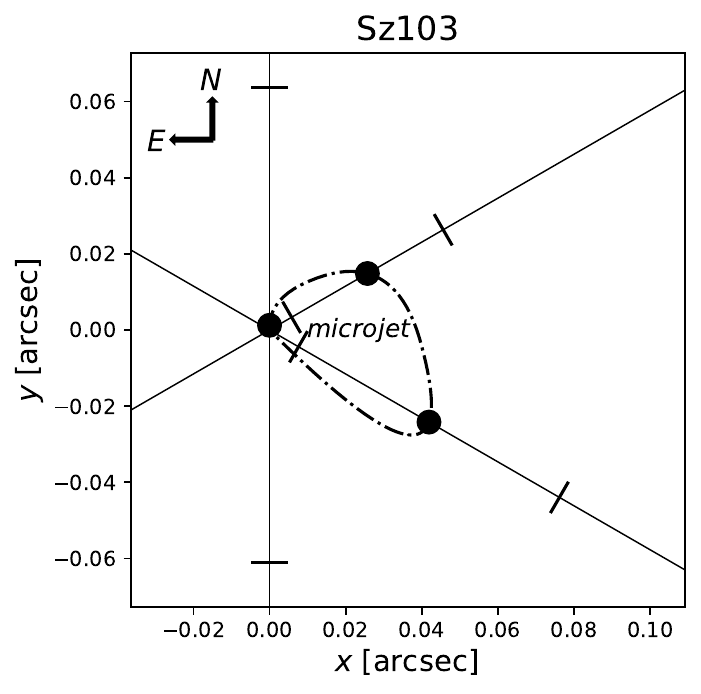}}
\caption{\small{Microjet of Sz\,103. The three UVES slit position angles are indicated as solid lines. For each slit position the spectro-astrometric offset in the [OI]$\lambda$6300 emission line is indicated as filled circle together with their $1\sigma$ error bars (thick lines). We determined its position by calculating the mean of all offsets (1D Gaussians) with respect to the continuum between $v= -80\dots 10\,\text{km}\,\text{s}^{-1}$ (see Fig.\ref{fig:all_minispectra_Sz103}). We interpolated a quadratic curve that connects the mean detected positions of the spectro-astrometric offsets. It thus represents a lower limit of the extension of the microjet.}}\label{fig:microjet_Sz103} 
\end{figure}

\begin{figure}  
\resizebox{\hsize}{!}{\includegraphics[trim=0 0 0 0, clip, width=0.9\textwidth]{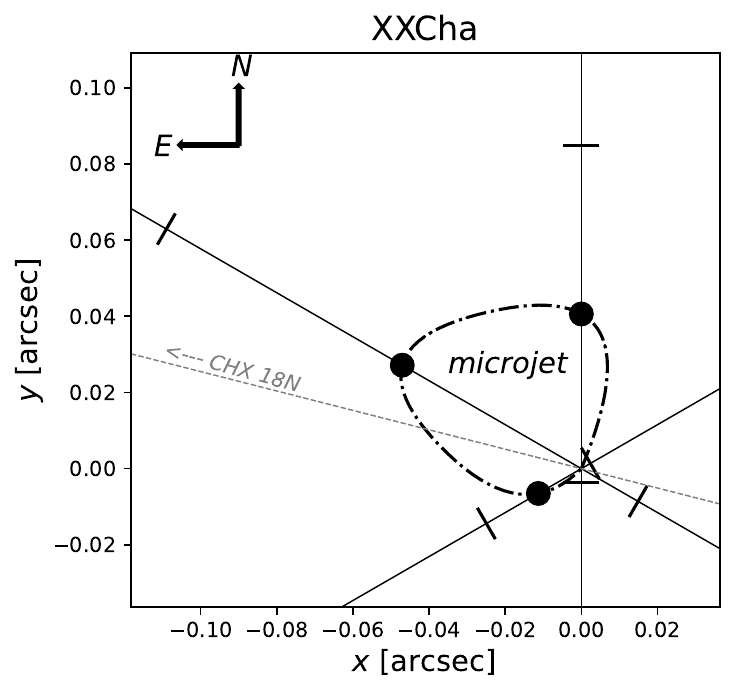}}
\caption{\small{Same as for Fig.\,\ref{fig:microjet_Sz103} but for XX\,Cha. The spectro-astrometric offset is determined in the velocity range of  $v= -110\dots -40\,\text{km}\,\text{s}^{-1}$ (see Fig.\,\ref{fig:all_minispectra_XXCha}).}}\label{fig:microjet_XXCha} 
\end{figure}

In conclusion, the targets Sz\,98, Sz\,99, Sz\,100, SO518, and CVSO\,58 show strong indication of outflow activity, but no evidence for a microjet. \citet{nisini_2018} already identified Sz\,98, and Sz\,99 as such sources since they feature a HVC in [OI]$\lambda$6300 or multiple FEL (see Table\,C.1 therein). For Sz\,103 and XX\,Cha we detected extended FELs in the [OI]$\lambda$6300 line, that is, the first detection of their microjets. We analysed their spectro-astrometric offsets in the three slit positions to spatially confine their outflow extent (Figs.\,\ref{fig:microjet_Sz103} and \ref{fig:microjet_XXCha}). They will be discussed separately in Sects.\,\ref{source_information_Sz103} and \ref{source_information_XXCha}.

It is instructive to compare the results of this study with the findings of \citet{nisini_2018}. We focus here on the detection of the [OI]$\lambda$6300 line in its velocity components, that is, Table\,B.2\,therein. Table\,\ref{table:comparison_nisini2018} compares  the [OI]$\lambda$6300 line detections and velocity constituents for the 19 overlapping targets of this study. Table\,\ref{table:comparison_nisini2018} displays an overall agreement with the results of \citet{nisini_2018}, that is, for most targets the [OI]$\lambda$6300 line shows qualitatively the same velocity components. Clearly, due to the much higher spectral resolution of UVES the low velocity components could be decomposed in the narrow and broad components in this study. However, there are a few disagreements. Among the 19 overlapping targets, \citet{nisini_2018} detected a HVC in 5/19 targets, whereas we found 7/19. In the case of Sz\,103 \citet{nisini_2018} found only a LVC, whereas we identified three components: NLVC, BLVC, and a HVCB. The reason for that is, that \cite{nisini_2018} set the limit of the HVC to $|v_p| > 40\,\text{km}\,\text{s}^{-1}$. Our spectro-astrometric analysis in Sz\,103 clearly connects the velocity component of Sz\,103 at $|v_p|\lesssim 40\,\text{km}\,\text{s}^{-1}$ to the jet. In the case of CV\,Cha we found a previously undetected HVCB at $|v_p|\sim 60\,\text{km}\,\text{s}^{-1}$. One explanation could be that too short exposure times in the X-Shooter spectra prevented the detection of the comparably weak HVC in CV\,Cha. Alternatively, CV\,Cha displays outflow variability and the HVCB was really not present at the time the X-Shooter data were taken. The discrepancy in the HVC identification of SSTc2dJ161344.1-373646 can be explained by the way \cite{nisini_2018} classified the velocity components.  Whenever two components were detected, \citet{nisini_2018} classified one component as a HVC. Following that reasoning SSTc2dJ161344.1-373646 would also feature a HVCB in the UVES data even though its peak velocity is $|v_p|\lesssim 30\,\text{km}\,\text{s}^{-1}$. In two sources (Sz\,104, 2MASS,J16000060-4221567) we do not detect the LVC in [OI]$\lambda$6300 albeit \citet{nisini_2018} did detect it. This is likely due to the higher  spectral resolution (and thus lower S/N) of UVES as compared to X-Shooter.


\begin{table}
\centering
{\def\arraystretch{2}\tabcolsep=1pt
\tiny
\caption{\small{Comparison of the detections of the [OI]$\lambda$6300 velocity components for  the 19 overlapping targets of the study of \citet{nisini_2018}.}}\label{table:comparison_nisini2018}
\centering
\begin{tabular}{|c||c|c||}
\hline
 \multirow{1}{*}{\textbf{Target}} &
      \multicolumn{2}{c||}{\textbf{[OI]$\lambda$6300 components}}  \\  
       &  Nisini+18  &    This work    \\ 
 \hline 
 \hline  
 \textbf{Sz98} &  LVC, HVCB   &     NLVC+BLVC, HVCB   \\ 
  \textbf{Sz100} &  LVC, HVCB  &    NLVC+BLVC, HVCB    \\
 \textbf{XXCha}  &  LVC, HVCB  &    NLVC+BLVC, HVCB     \\ 
  \textbf{Sz99}  & LVC, HVCR  &      BLVC, HVCR  \\ 
 \textbf{SSTc...646}  &  LVC, HVCB  &      LVC  \\
 \hline
  \textbf{Sz103} & LVC  &     NLVC+BLVC, HVCB    \\  
  \textbf{CVCha} & LVC  &      NLVC, HVCB   \\
 \textbf{Sz112}  & LVC   &    NLVC, BLVC    \\  
 \textbf{Sz129}  & LVC   &    NLVC, BLVC    \\  
 \textbf{VWCha} &  LVC  &     NLVC, BLVC   \\
\textbf{VZCha} &  LVC  &    NLVC, BLVC     \\
 \textbf{Hn5} & LVC  &     NLVC, BLVC   \\
  \textbf{Sz84}  & LVC   &    LVC    \\ 
   \textbf{Sz97}  & LVC  &      LVC   \\  
  \textbf{SSTC...503}  & LVC  & LVC  \\
   \textbf{Sz104} & LVC   &    not detected    \\
 \textbf{2MASS...567} & LVC  &   not detected     \\  
 \hline
   \textbf{Sz68} & not detected  &  not detected      \\
     \textbf{Sz115} &  not detected &    not detected    \\
\hline\hline
 \end{tabular}
 }
 \end{table}

\subsection{The microjet of Sz\,103}\label{source_information_Sz103}

 Sz\,103 (THA\,15-29) is a young accreting T\,Tauri star ($L_\star = 0.188\,L_\odot$) of spectral type M4 (age about $1-3$\,Myr) located in the Lupus Cloud III \citep{alcala_2014,frasca_2017}. Based on the second data release of {\it Gaia}, \citet{luhman_2020} refined the distance for YSOs associated with the Lupus Clouds to be $D\sim 157\,\text{pc}$, which is agrees with its individual distance stated in \citet{manara_2023}. Sz\,103 is associated with a protoplanetary disk with an inclination angle of $i_\text{disk}\sim 50^\text{o}$ \citep{yen_2018}. \citet{erkal_2022} found an inverse P-Cygni signature in the He\,I 10830$\AA$ line which traces accreting material. 
Sz\,103 is among the X-Shooter targets of \citet{natta_2014, nisini_2018}. Therein, a LVC centred at $v_p = -32.4\,\text{km}\,\text{s}^{-1}$ in [OI]$\lambda$6300 was detected, which is consistent with the position of the HVC detected with UVES in this study (HVC ep2: $v_p = -36.3\,\text{km}\,\text{s}^{-1}$). The microjet of Sz\,103 (Fig.\,\ref{fig:microjet_Sz103}) was not detected by \citet{nisini_2018} in their spectro-astrometric analysis, which could be explained by several reasons: 1. The X-Shooter slit in the spectra of \citet{nisini_2018} was disadvantageously aligned with the outflow. Indeed, the X-Shooter observations were undertaken at parallactic angle, that is, at about $84.5^\text{o}$ for Sz\,103. Our observations suggest that the Sz\,103 microjet is directed towards slit position 3, that is, at $240^\text{o}$ P.A, which is nearly orthogonal to the slit orientation in the X-Shooter spectrum of \citet{nisini_2018}.  2. The spectral and spatial resolution of X-Shooter were too low. In the VIS arm X-Shooter observations featured a medium velocity resolution of $34\,\text{km}\,\text{s}^{-1}$. In their study, \citet{nisini_2018} were able to detect spectro-astrometric offsets in the [OI]$\lambda$6300 line down to $0\farcs 2$. Our analysis revealed the extent of the microjet to be of the order of $0\farcs04$,  which is by a factor of five smaller than the smallest detected X-Shooter microjet. The projected distance of the microjet, that is, $0\farcs04$, at 160\,pc translates to about 6.4\,au.

 \subsection{The microjet of XX\,Cha}\label{source_information_XXCha}
  
XX\,Cha (2MASS\,J11113965-7620152, Ass\,Cha\,T\,2-49, Sz\,39, T49) is a classical T\,Tauri star ($L_\star = 0.29\,L_\odot$) of spectral type M3.5 located in Chameleon I ($D\sim 191\,\text{pc}$), for example \citet{, manara_2017, nisini_2018}. A disk around XX\,Cha has been detected \citep{pascucci_2016}, however, its inclination is unknown \citep[see Table\,1 in][]{france_2023}. XX\,Cha and CHX18N form a wide binary system with a separation of about 24$\arcsec$ and a position angle of $255.7^\text{o}$ \citep{kraus_2007}. Our spectro-astrometric analysis suggests that the XX\,Cha microjet is most likely aligned in slit position 3, that is, $240^\text{o}$. The direction of the jet seems therefore aligned with the line joining XX Cha and its wide companion CHX18 (Fig.\,\ref{fig:microjet_XXCha}). \citet{claes_2022} reports an extreme accretion variability for XX\,Cha and concludes that XX\,Cha shows similarities with EXor-variables. As discussed therein, XX\,Cha may have been in an outburst phase between 2010 and 2021 lasting for about several months. In this regard, the dynamical time scale of the XX\,Cha microjet may give further hints for a potential connection of the outbursts and the appearance of the jet. The projected distance of the XX\,Cha microjet is about $0\farcs 05$ which is 9.6\,au at 192\,pc. For a jet of $100\,\text{km}\,\text{s}^{-1}$ only 166 days are needed to reach that distance. That implies that the XX\,Cha microjet may have launched about 5 months before the UVES observations in 2021 and it is linked to the outbursts. We checked the ASAS-SN light curve of XX Cha, whether there was a signature of substantially enhanced accretion about 5 months before UVES observations, but found no substantial evidence for that. XX\,Cha could still be a burster \citep{burster_cody_2014}, that is, XX\,Cha varies on much shorter time scales than months.
On the other hand \citet{nisini_2018} detected the LVC and the HVC towards XX\,Cha in the X-Shooter data of 2010. This points to the conclusion that already in 2010 a jet was present towards XX\,Cha and due to the reasons (1-2) mentioned in Sect.\,\ref{source_information_Sz103} the jet was not detected in spectro-astrometry. We note that due to the unknown inclination of the system this line of reasoning is highly speculative. The accretion variability of XX\,Cha, however, may be linked to an outflow variability. 

\begin{figure*} 
\centering
\subfloat{\includegraphics[trim=0 0 0 0, clip, width=0.3 \textwidth]{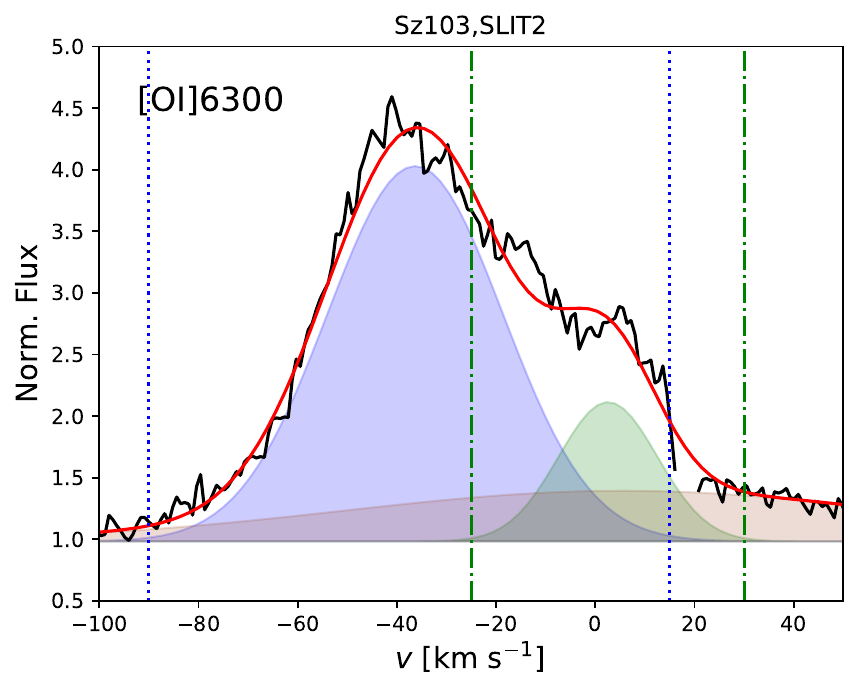}}
\hfill
\subfloat{\includegraphics[trim=0 0 0 0, clip, width=0.3 \textwidth]{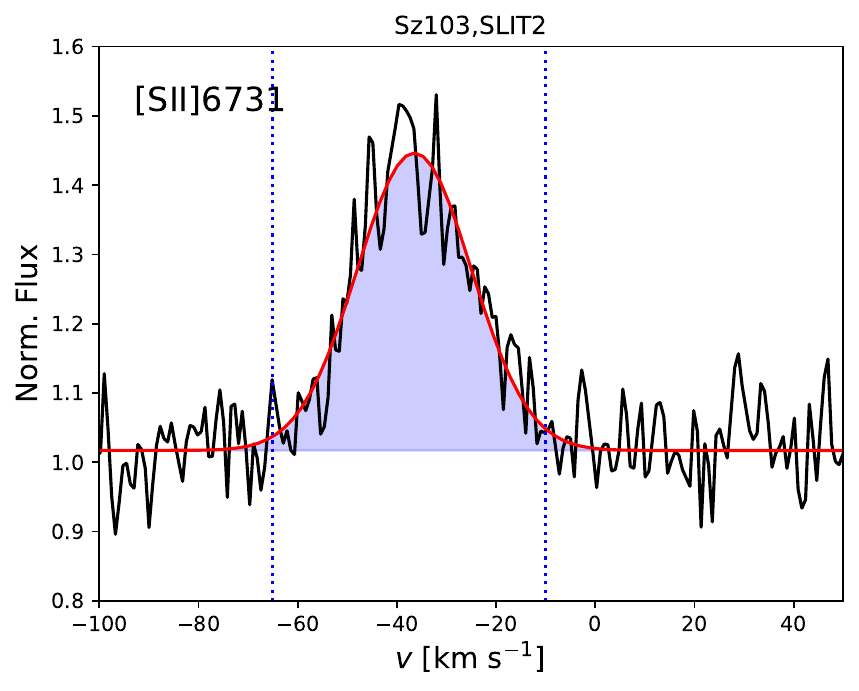}}
\hfill
\subfloat{\includegraphics[trim=0 0 0 0, clip, width=0.3 \textwidth]{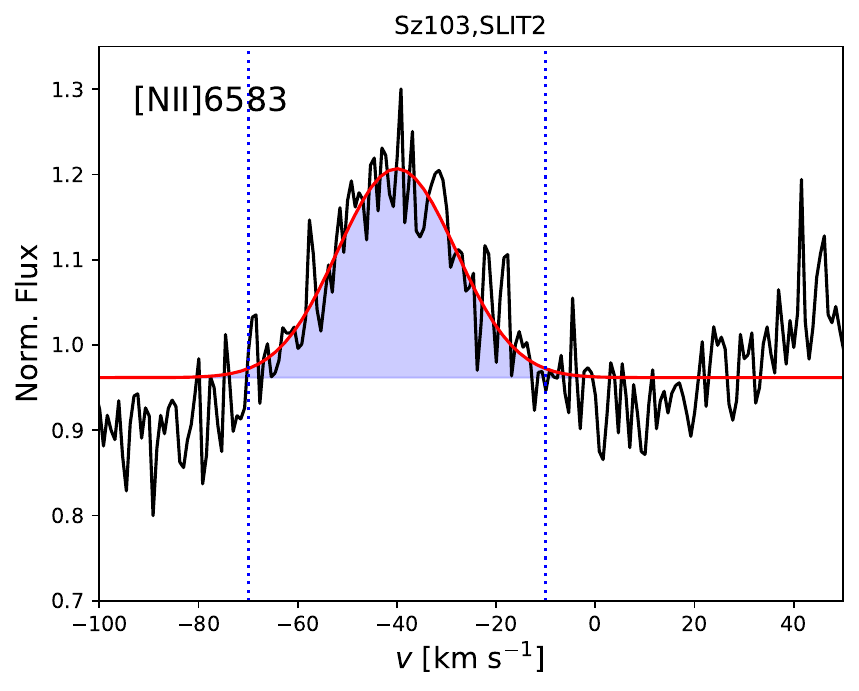}}
\hfill 
\subfloat{\includegraphics[trim=0 0 0 0, clip, width=0.3 \textwidth]{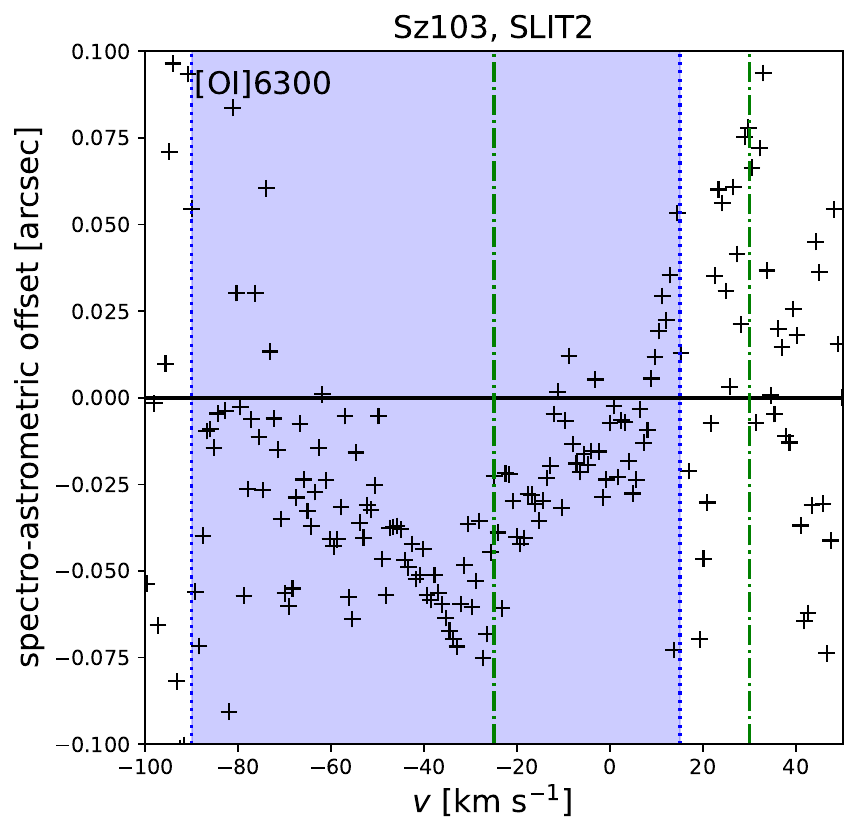}}
\hfill
\subfloat{\includegraphics[trim=0 0 0 0, clip, width=0.3 \textwidth]{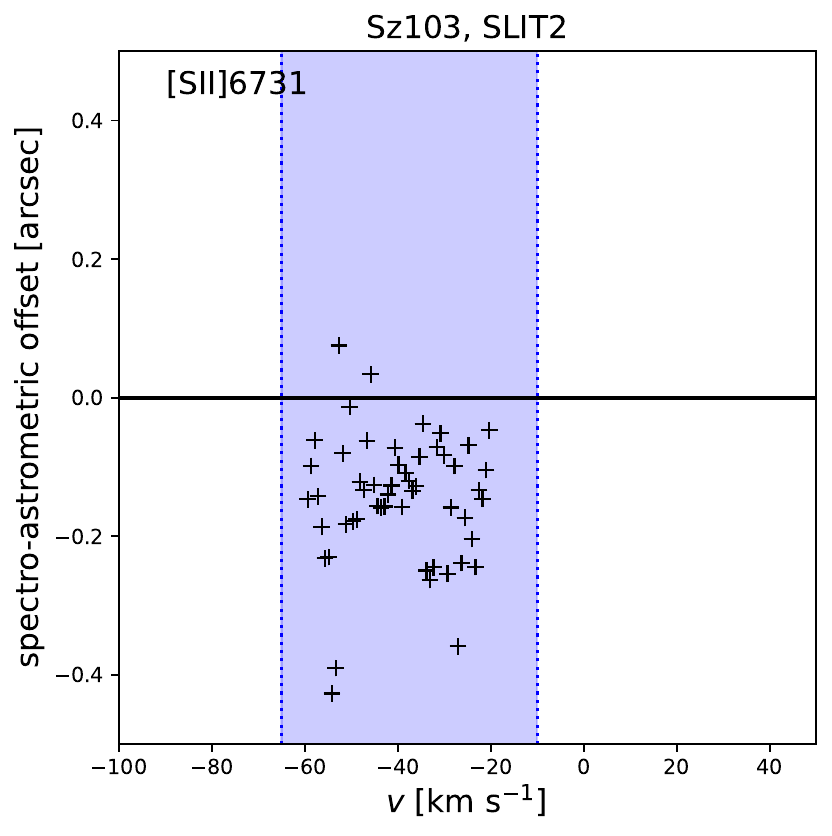}}
\hfill
\subfloat{\includegraphics[trim=0 0 0 0, clip, width=0.3 \textwidth]{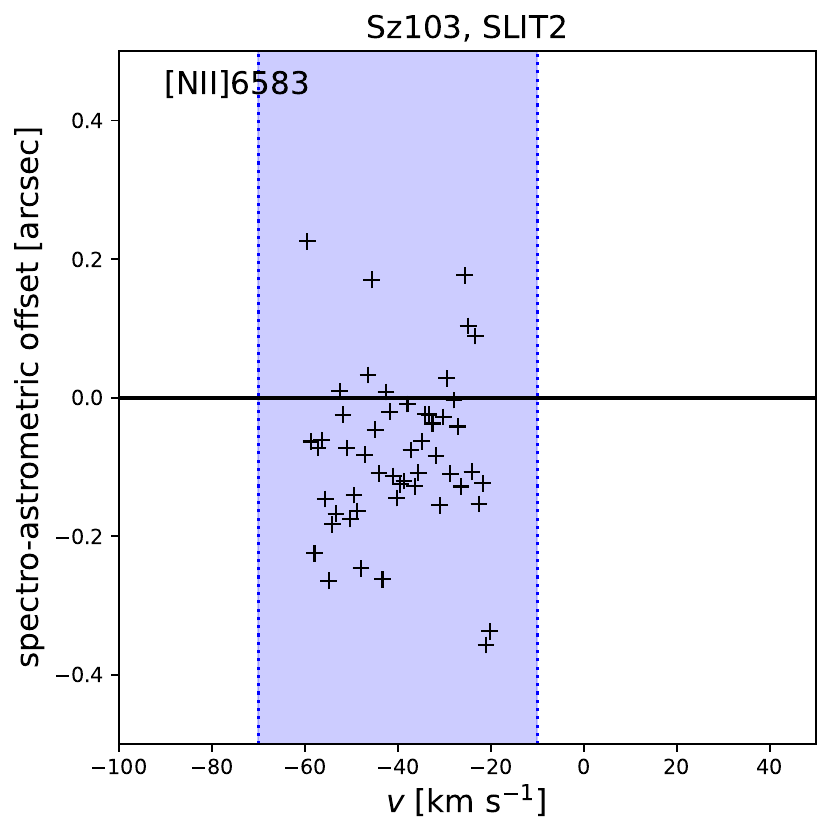}} 
\hfill  
\caption{\small{Spectro-astrometry in slit position 2 for three wind lines of Sz103. \textit{Top row:} Decomposition of the FELs of [OI]$\lambda$6300, [SII]$\lambda$6731, and [NII]$\lambda$6583 in its HVC (blue), NLVC (green), and BLVC (sienna). \textit{Bottom row:} Spectro-astrometric offsets with respect to the continuum. The blue shaded area indicates the extend of the HVC. Green vertical lines are showing the extend of the NLVC.}}\label{fig:sa_sz103}
\end{figure*} 

\begin{figure*} 
\centering
\subfloat{\includegraphics[trim=0 0 0 0, clip, width=0.24 \textwidth]{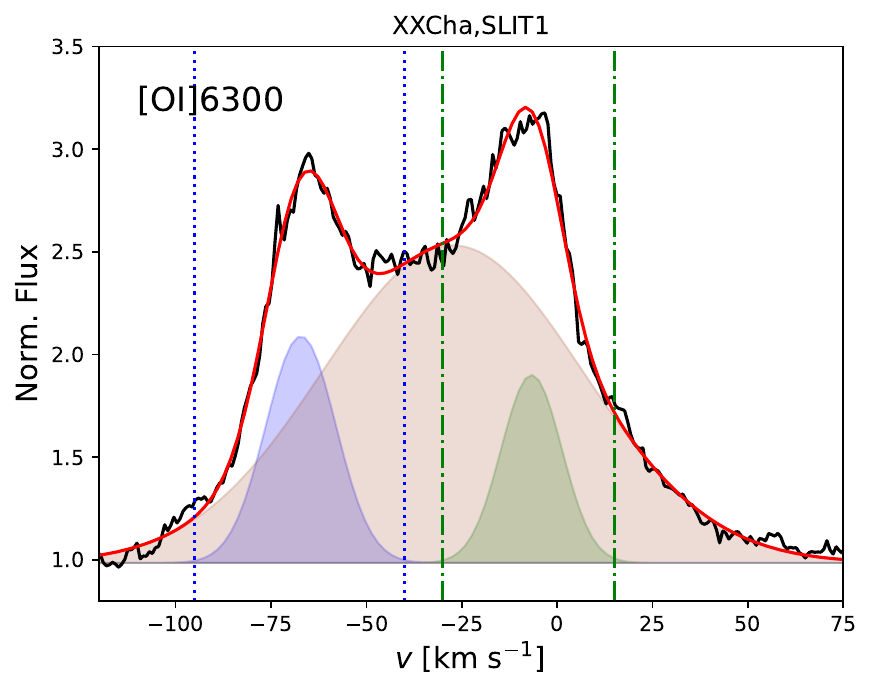}}
\hfill
\subfloat{\includegraphics[trim=0 0 0 0, clip, width=0.24 \textwidth]{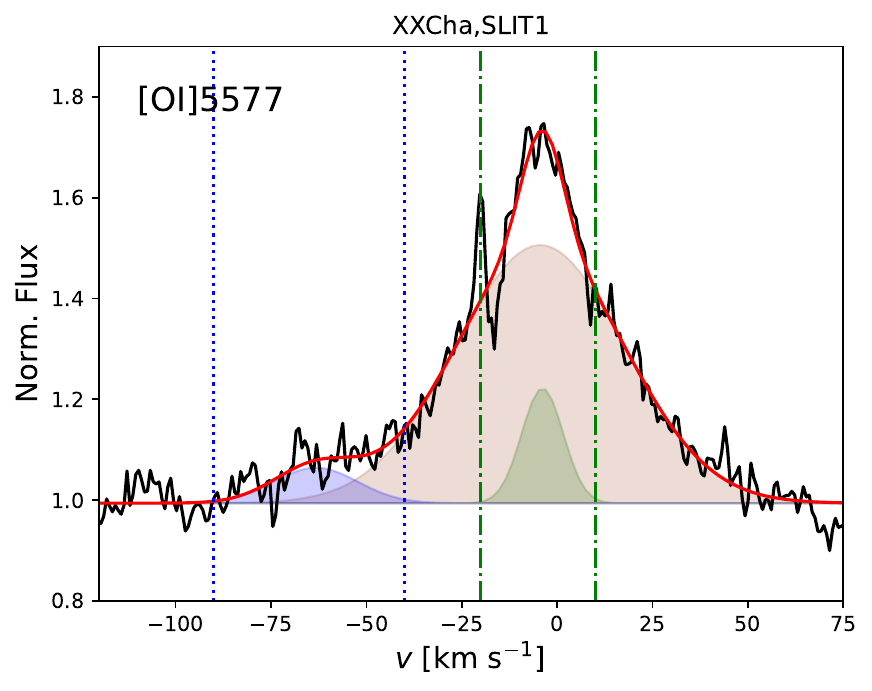}}
\hfill
\subfloat{\includegraphics[trim=0 0 0 0, clip, width=0.24 \textwidth]{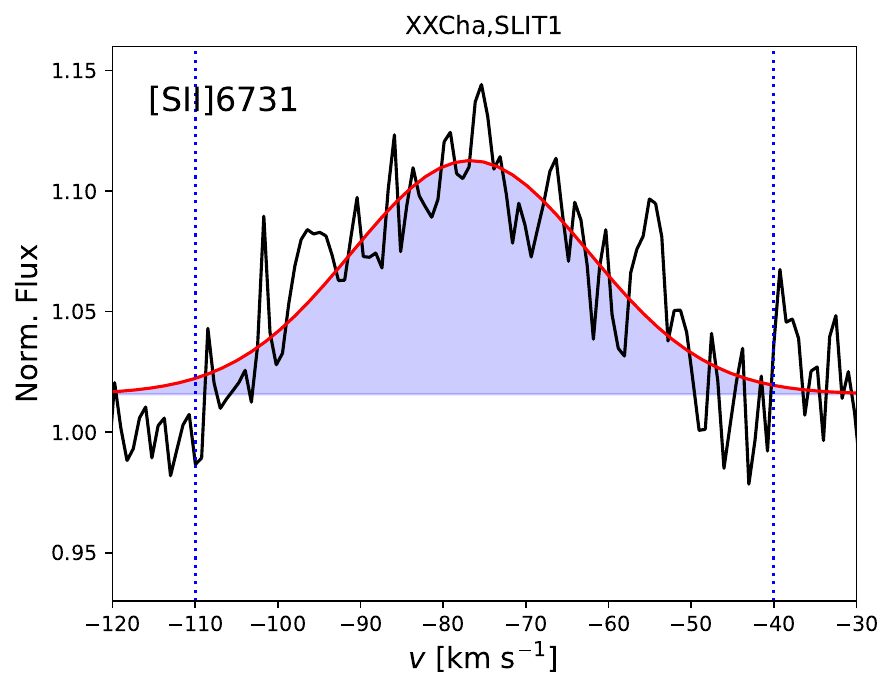}}
\hfill
\subfloat{\includegraphics[trim=0 0 0 0, clip, width=0.24 \textwidth]{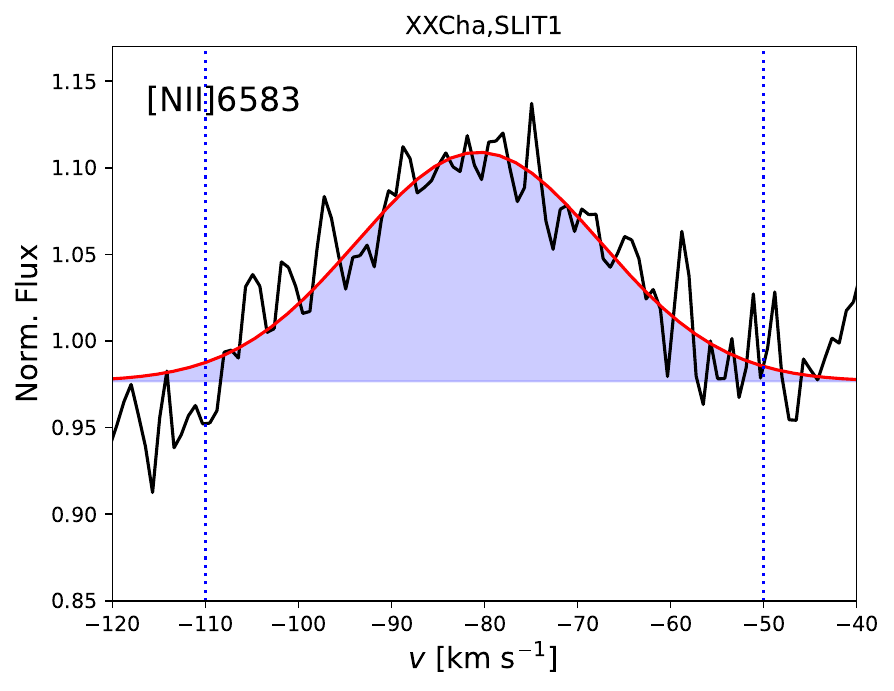}}
\hfill \\
\subfloat{\includegraphics[trim=0 0 0 0, clip, width=0.24 \textwidth]{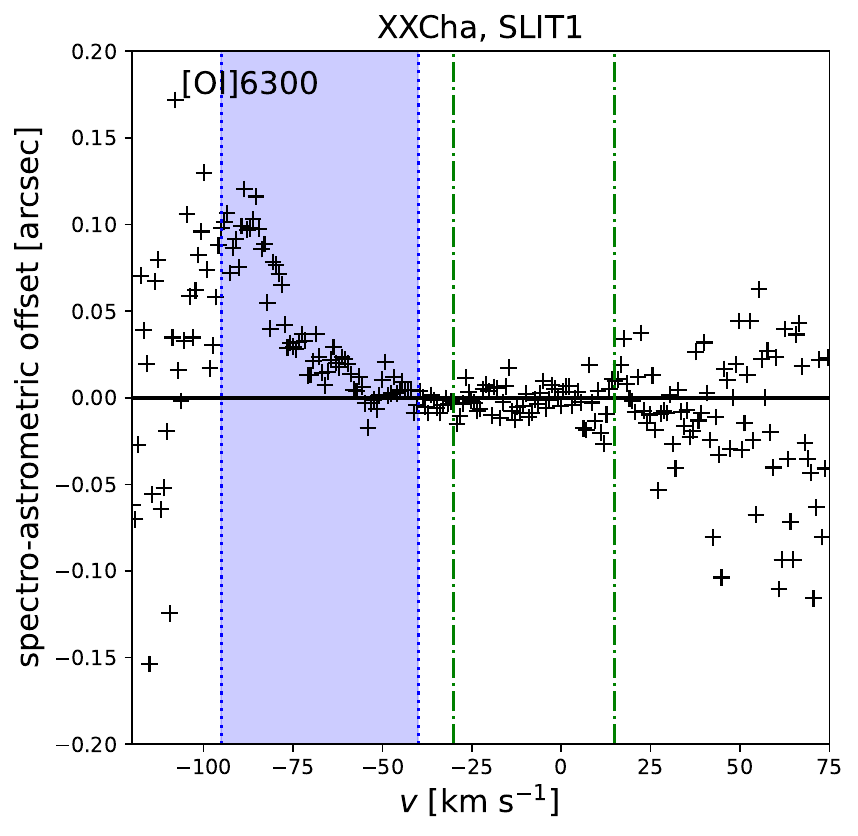}}
\hfill
\subfloat{\includegraphics[trim=0 0 0 0, clip, width=0.24 \textwidth]{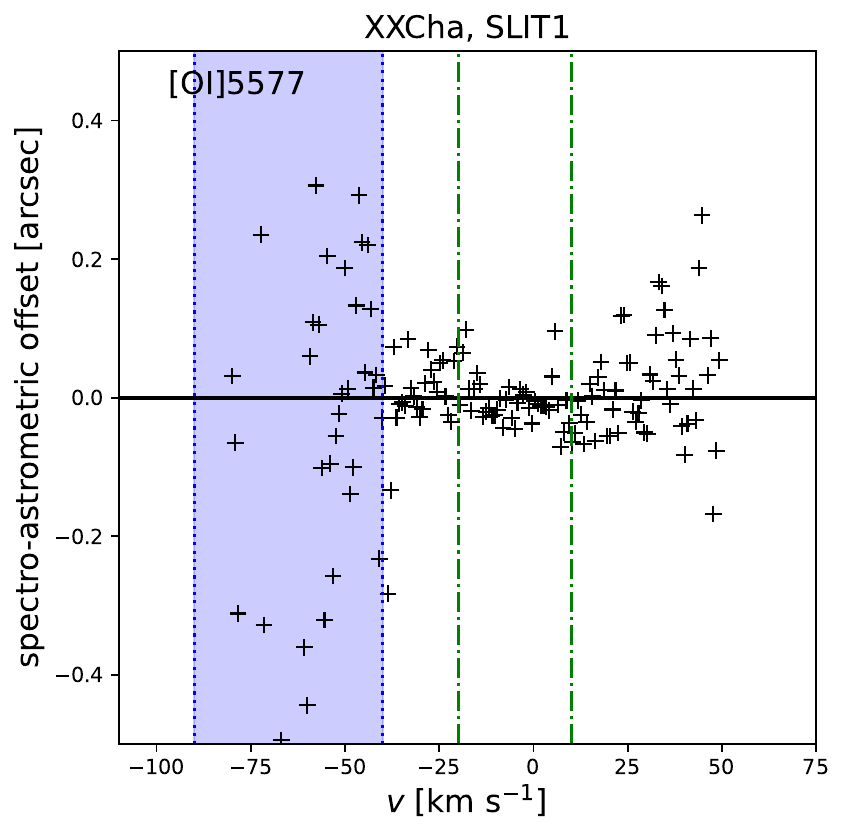}}
\hfill
\subfloat{\includegraphics[trim=0 0 0 0, clip, width=0.24 \textwidth]{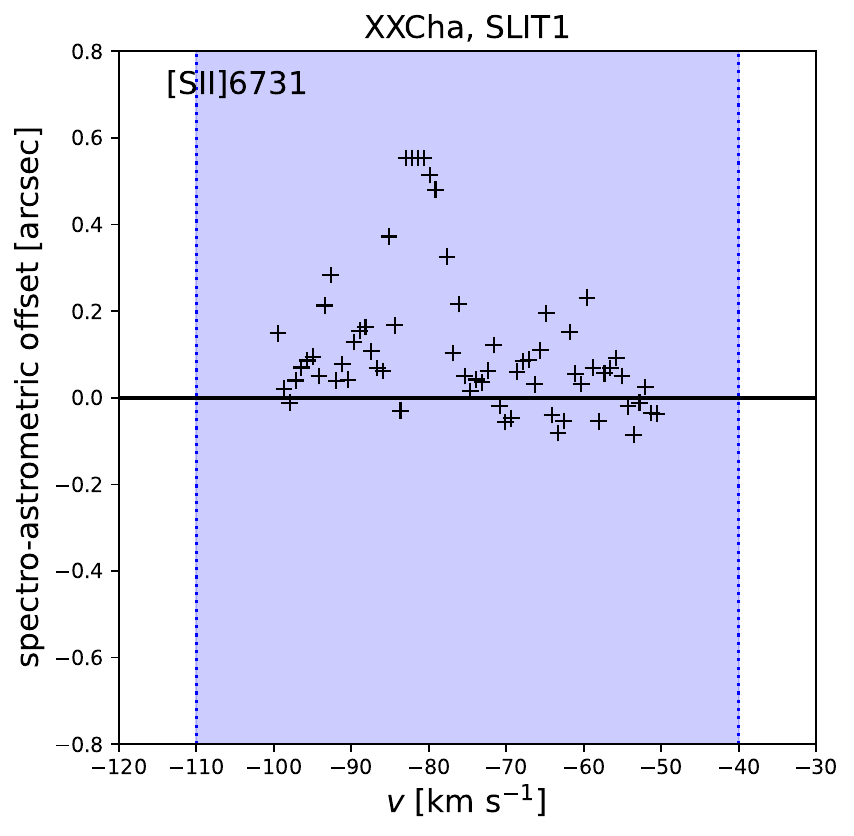}}
\hfill
\subfloat{\includegraphics[trim=0 0 0 0, clip, width=0.24 \textwidth]{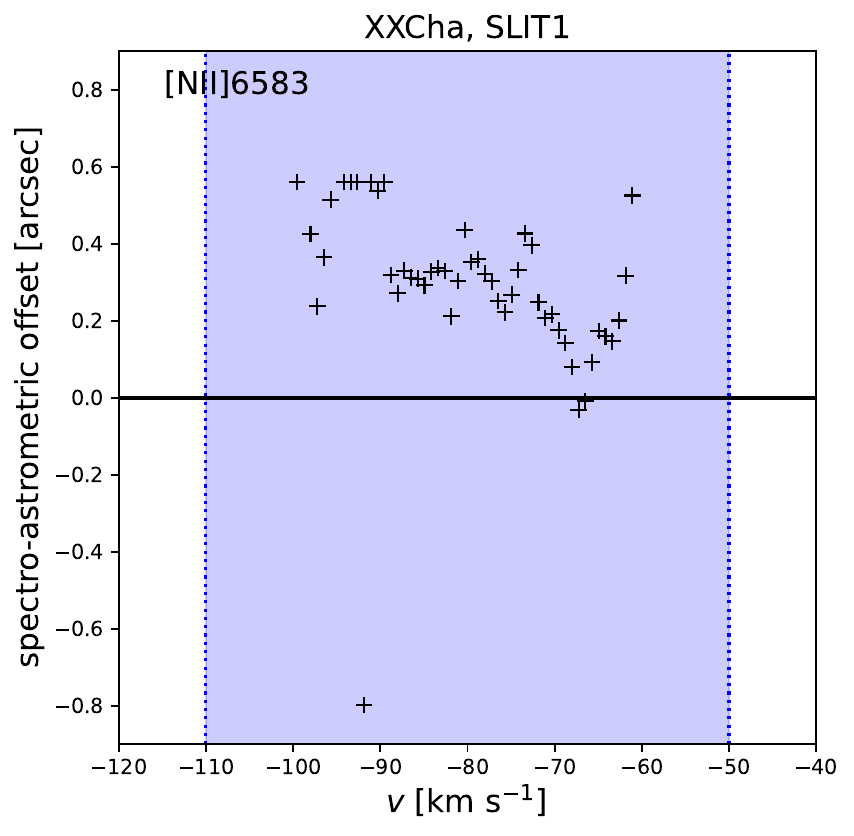}} 
\hfill  
\caption{\small{Same as for Fig.\,\ref{fig:sa_sz103} but for for XX\,Cha in slit position.}}\label{fig:sa_xxcha}
\end{figure*} 

\subsection{Spectro-astrometry in the HVC and the LVC of Sz\,103 and XX\,Cha}\label{sa_sz103_xxcha}

Following \citet{whelan_2021, whelan_2023} we can apply the spectro-astrometric technique separately in the HVC and the LVC of Sz\,103 and XX\,Cha. The main motivation for that approach is to get more insights on the origin of the line emission. In the case of RU\,Lup \citet{whelan_2021} could show evidence that the narrow LVC on the [OI]$\lambda$6300 line emerges from MHD disk winds and not photoevaporative winds. We display the results for the bright wind lines of oxygen, nitrogen, and sulphur in Figs.\,\ref{fig:sa_sz103} and \ref{fig:sa_xxcha}. For XX\,Cha we investigate four lines: [OI]$\lambda$6300, [OI]$\lambda$5577, [SII]$\lambda$6731, and [NII]$\lambda$6583. In the case of Sz\,103 the [OI]$\lambda$5577 line was too noisy for spectro-astrometry so that we could use only the remaining three lines of that list.  

\textbf{LVC:} For both targets a LVC is only detected in [OI]. This is different to the two targets RU\,Lup and AS\,205 of the study of \cite{whelan_2021}, where they see the LVC prominently also in [SII] and [NII]. This may indicate that the LVC of the microjets of this study display different physical gas parameters or are even of different origins. For XX\,Cha the NLVC is largely blended by the BLVC, whereas for Sz\,103 the NLVC is only partially blended by the much broader HVC. In the case of XX\,Cha the LVC in [OI]$\lambda$6300 and [OI]$\lambda$5577 clearly display no spectro-astrometric offset, that is, it is not extended emission and we do not see a velocity gradient. Unfortunately, we cannot deproject distances to the disk or velocities since the inclination of XX\,Cha is unknown. This showcases the limitations of the analysis method of \cite{whelan_2021}. In order to investigate if the NLVC originates from a MHD disk wind or a photoevaporative wind one needs very close systems where the disk inclination is known and the velocity components are spectrally separated. For XX\,Cha we interpret it to originate close to the star-disk system but cannot make a decisive statement on its origin.
 
Similarly, for Sz\,103 the LVC and the HVC of the [OI]$\lambda$6300 line overlap in a region where a spectro-astrometric offset is detected. In addition, the NLVC in Sz\,103 is not the dominant component, that is, it is less bright than the HVCB. It would be a speculation to connect the observed spectro-astrometric offset with one or the other component alone. Potentially it is connected to both. However, since the position of the peak of the spectro-astrometric offset coincides with the peak of the HVC and the overall trend matches with the HVC we think that the spectro-astrometric offset detected in the overlap region of the HVC and the LVC is connected to the HVC. In turn, the narrow LVC of Sz\,103 is not extended and as for XX\,Cha connected to the close in region of the star-disk system.

\textbf{HVC:} For both targets, the HVC of [OI]$\lambda$6300 is associated with the spectro-astrometric offset, that is, its emission is extended and connected to their jets. We also see the extended emission in the HVC of [SII]$\lambda$6731 and [NII]$\lambda$6583 in both sources. In these lines the spatial extend of the detected FEL is larger, but less pronounced as compared to the [OI]$\lambda$6300 emission. The HVC in [OI]$\lambda$6300 of XX\,Cha is comparably narrow (FWHM$\sim 20\,\text{km}\,\text{s}^{-1}$) and can therefore be associated with a highly collimated, fast jet. In [SII] and [NII] the FWHM of the HVC is much larger indicating that it not only traces the collimated fast flow but also outer layers of the onion-like outflow of XX\,Cha. For Sz\,103 its FWHM of the HVC is much larger ($\sim 40\,\text{km}\,\text{s}^{-1}$) and comparable in [SII] and [NII].

\section{Conclusions}

We have screened all 34 UVES targets in the PENELLOPE ESO VLT Large programme for outflow signatures. In doing so, we formulated four criteria to find the interesting targets that display strong protostellar outflows. We ranked the 34 targets according to their outflow activity and found four promising targets: Sz\,98, Sz\,99, Sz\,103, and XX\,Cha. 
All these targets display outflow activity in form of a HVC in [OI]$\lambda$6300 and detections of other wind lines of [SII] or [NII]. 
We identified two sources (Sz\,103, XX\,Cha)
that host small-scale jets (projected extent $\sim 0\farcs 04$), which were detected in the HVCB of the [OI]$\lambda$6300 line and consistently as spectro-astrometric offset with respect to the continuum of the associated T\,Tauri star. With the three UVES slit positions we could constrain the direction of the jet for both targets. However, only observations with for example the integral-field unit such as MUSE/NFM operating in the optical could reveal the true extent of their microjets. The detection rate of microjets of this study ($2/34 \approx 6\,\%$)  is comparable with the detection rate in the \citet{nisini_2018} study ($5/131\approx 4\,\%$).

None of the promising targets display a true P-Cygni line profile in H$\alpha$. Only four of the six interesting sources show a relaxed P-Cygni line profile in H$\alpha$. We conclude that the P-Cygni line profile in H$\alpha$ is not a strong indicator for outflow activity. 

Our spectro-astrometric analysis confirms the binary nature of VW\,Cha and CVSO109.  In addition we found evidence that DK\,Tau\,B and CVSO\,104\,A are also spectroscopic binaries. Sz\,115 tentatively is a spectroscopic binary.

For the microjets Sz\,103 and XX\,Cha we investigated the spectro-astrometric signals separately in their HVC and LVC \citep{whelan_2021}. Our analysis does not show clear evidence of the origin of the LVC, that is, if it comes from MHD winds or photoevaporative winds.

In principle, however, our methods provide offsets separately for different line components. Thus, higher S/N observations utilising the information on the outflow P.A. from this study would provide the desired information.

\begin{acknowledgements}  
We are grateful to  Ignacio Mendigutia, and to an anonymous
referee for their suggestions and constructive discussions. This work has been supported by the Deutsche Forschungsgemeinschaft (DFG) in the framework of the YTTHACA Project (Young stars at Tübingen, Tautenburg,
Hamburg \& ESO – A Coordinated Analysis) under the programme IDs: EI 409/20-1, MA 8447/1-1, and SCHN 1382/4-1. This work benefited from discussions with the ODYSSEUS
team (\url{https://sites.bu.edu/odysseus/}). This project has received funding by the European Union (ERC, WANDA, 101039452). Views and opinions expressed are however those of the author(s) only and do not necessarily reflect those of the European Union or the European Research Council Executive Agency. Neither the European Union nor the granting authority can be held responsible for them.   
\end{acknowledgements}

\bibliographystyle{aa} 
\bibliography{papers}  
    
\appendix

\setcounter{table}{0}
\renewcommand{\thetable}{A.\arabic{table}}
  
 \setcounter{figure}{0}
\renewcommand\thefigure{\thesection A.\arabic{figure}}  

\setcounter{equation}{0}
\renewcommand{\theequation}{A.\arabic{equation}}  

\section*{Appendix A - Spectro-astrometric concepts}\label{appendix:s_a_concepts}

The main motivation for the subsequent remarks are twofold. First of all, we wish to investigate the question of whether the continuum contribution of the star has to be removed for the spectro-astrometric analysis or not. So far the spectro-astrometric method carried out in the literature standardly includes this data reduction step. Without a doubt this explicit removal best showcases to the reader extended emission line regions such as outflows. However, as can be seen in the following Section this is mathematically not necessary. Secondly and more importantly, we study the influence of artefacts in our UVES spectra on the spectro-astrometric analysis.

\subsection*{The spectro-astrometric centroid}

\begin{figure}
    \includegraphics[width=0.49\textwidth]{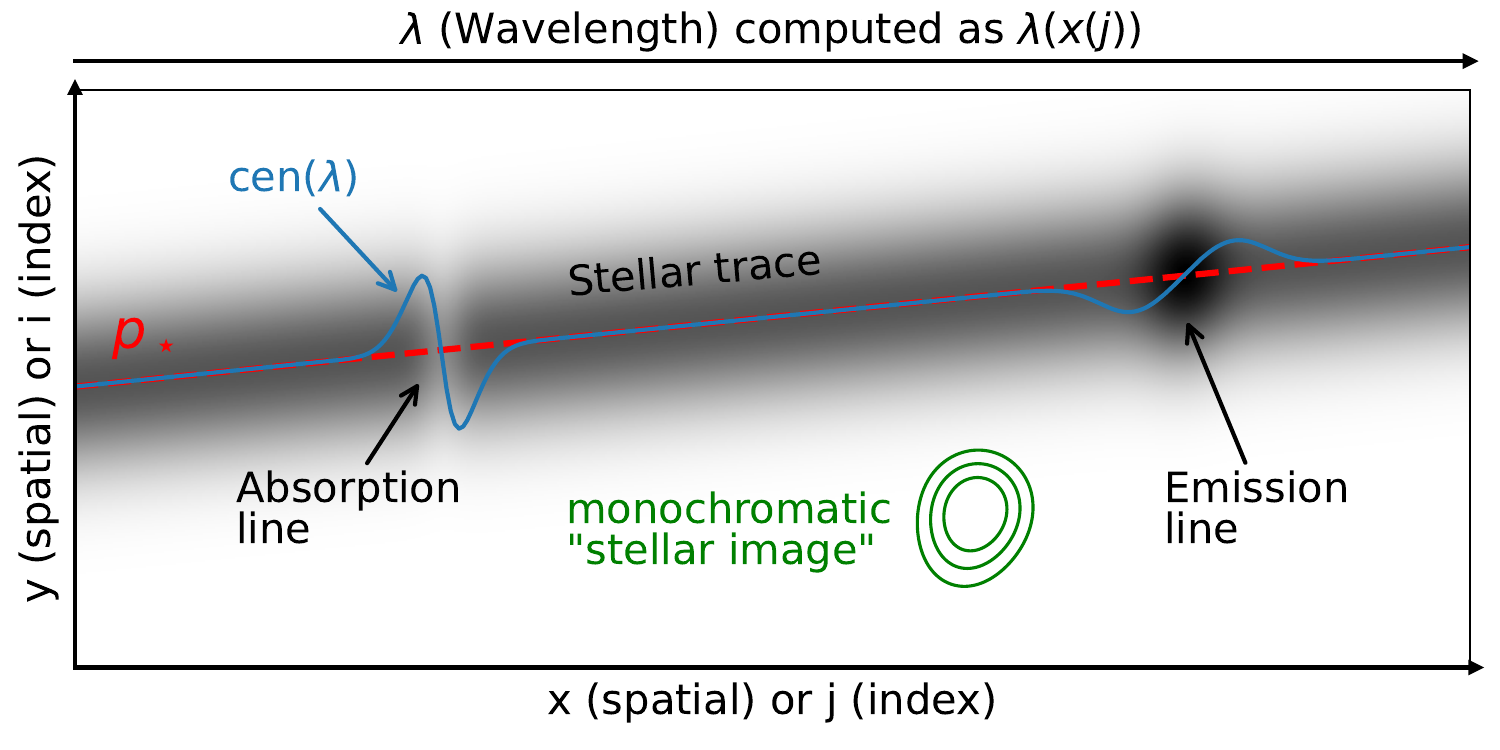}
    \caption{Sketch of the coordinate systems with respect to the spectral trace including artefacts introduced 
    due to an absorption and an emission line. In addition, the contours show how a monochromatic emission line emitted by a point-source
    would be recorded on the detector. Artefacts and asymmetry are exaggerated for displaying purposes. \label{fig:sketch}}
\end{figure}

To discuss the measurement of the spectro-astrometric offset we first introduce the coordinate system as shown in Fig.~\ref{fig:sketch}. A position on 
the detector is described by its row and column indices $(i, j)$. The spectral trace is recorded mainly
along a particular row so that the wavelength changes with index $j$ and the nominal wavelength $\lambda_j$ is some function
of the column index. The row index $i$ describes the cross-dispersion direction. The projected sky distance is $y_i$.  
The wavelength-dependent weighted centroid $cen(\lambda)$
of the observed emission as recorded on the detector with the spatial extend $Y$ is given by
\begin{equation}
cen(\lambda)  = \frac{\int_Y y F(y, \lambda) \text{d}y}{\int_Y  F(y, \lambda) \text{d}y}  \,, \label{eq:cen_01}
\end{equation}
where the recorded flux is $F(\lambda,\,y)$, that is, the 2D-image recorded on the detector (see also Fig.~\ref{fig:sketch}). 
We note that Eq.\,\ref{eq:cen_01} represents a general way of finding a centroid for an arbitrary function $F(y, \lambda)$. If, for example, the function $F(y, \lambda)$ is well described by a 1D Gaussian, that is,
\begin{equation}
F(y, \lambda) = \frac{1}{\sigma\sqrt{2\pi}}\text{exp}\left(-\frac{1}{2}\frac{(y-y_c)^2}{\sigma^2}\right)\, ,
\end{equation} 
then a fit to that Gaussian - as done in this paper - will give the same result for the centroid, $cen(\lambda)=y_c$.  \\
We can approximate $cen(\lambda)$ with a pixelised version where the slit is confined to $Y=\{y_0,y_1, \dots, y_N\}$ spatial pixels. Thus, Eq.\,\ref{eq:cen_01} reduces to
\begin{equation}
cen(\lambda_j) = \frac{\sum_{i=0}^{N-1} y_i F(y_i, \lambda_j) \Delta y_i}{\sum_{i=0}^{N-1}  F(y_i, \lambda_j)\Delta y_i} = \frac{\sum_{i=0}^{N-1} y_i F(y_i, \lambda_j) }{\sum_{i=0}^{N-1}  F(y_i, \lambda_j) } \,, \label{eq:cen_02}
\end{equation}
since $\Delta y_i = y_{i+1}-y_i = (y_N-y_0)/N$. Measuring $cen(\lambda_j$) is quite straightforward once properly rectified 2D-spectra (wavelength, cross-dispersion)
are available, see Sect.\,\ref{sec:sa}. \\
The measured $cen(\lambda_j)$ will have a numeric value depending on the somewhat arbitrarily defined zero position\footnote{One typically assigns
a zero offset to the on-axis position. However, this 'zero' may already depend on 
the wavelength where the on-axis position is measured. Furthermore, 
the telescope data is often insufficient to define the exact on-axis position with the accuracy needed for spectro-astrometry.} of the coordinates $y_i$. 
Because we are interested in the offset of the jet emission with respect to the star, we define the spectro-astrometric offset with respect to the star as
\begin{equation}
off(\lambda_j) = cen(\lambda_j) - p_\star(\lambda_j) , \label{eq:off}
\end{equation}
where $p_\star(\lambda_j)$ denotes the (ab initio only coarsely known) wavelength-dependent cross-dispersion position of the star. Combining  Eq.~\ref{eq:cen_02} with Eq.~\ref{eq:off} shows that in general
\begin{equation}
off(\lambda_j) =  \frac{\sum_{i=0}^{N-1} \big(y_i - p_\star(\lambda_j)\big)F(y_i, \lambda_j) }{\sum_{i=0}^{N-1}  F(y_i, \lambda_j)  } \,.\label{eq:off_01}
\end{equation}
We introduce a new coordinate $y_i'= y_i - p_\star(\lambda_j)$ in the above Eq.\,\ref{eq:off_01}, which measures the spatial distance to the centre of the stellar trace at the spectral position $\lambda_j$ and which needs to be extrapolated from nearby wavelength regions without extended emission.\\
For the case of (point-like) stellar and (extended) jet emission,
$F(y_i,\,\lambda_j)$ becomes
\begin{equation}
F(y_i, \lambda_j) = F^{jet}(y_i, \lambda_j) + F^\star(y_i, \lambda_j) \,.
\end{equation}
In our scenario, the jet (or any other extended emission) is assumed to contribute only in certain 
wavelength intervals around (strong) emission lines, for example around [OI]$\lambda6300$,  while no significant jet emission 
is present in continuum regions. Hence, $p_\star(\lambda_j)$ around the emission line is simply the interpolation of  
$cen(\lambda_j)$ between the neighbouring continuum regions so that $off(\lambda_j)=0$ in the continuum.\\
We are interested in the spatial offset of the jet (extended) emission $off^{jet}(\lambda_j)$, defined as
the mean position of the jet emission with respect to the star
\begin{eqnarray}
off^{jet}(\lambda_j) & = & \frac{\sum_{i=0}^{N-1} y_i' F^{jet}(y_i, \,\lambda_j)}{\sum_{i=0}^{N-1}  F^{jet}(y_i,\,\lambda_j)} \,.
\end{eqnarray}
Rewriting Eq.~\ref{eq:off} as
\begin{eqnarray}
off(\lambda_j) & = & \frac{\sum_{i=0}^{N-1} y_i' \left(F^{jet}(y_i,\,\lambda_j) + F^\star(y_i,\,\lambda_j)\right)}{\sum_{i=0}^{N-1} F(y_i,\,\lambda_j)} \\
&=& \frac{\sum_{i=0}^{N-1} y_i' F^{jet}(y_i,\,\lambda_j)}{\sum_{i=0}^{N-1} F(y_i,\,\lambda_j)}    + \overbrace{\frac{\sum_{i=0}^{N-1} y_i'    F^\star(y_i,\,\lambda_j)}{\sum_{i=0}^{N-1} F(y_i,\,\lambda_j)}}^{=0} \nonumber \\
&=& \frac{\sum_{i=0}^{N-1} y_i' F^{jet}(y_i,\,\lambda_j)}{\sum_{i=0}^{N-1} F^\star(y_i,\,\lambda_j) + F^{jet}(y_i,\,\lambda_j)} \label{eq:off2}\,
\end{eqnarray}
shows that $off^{jet}$ and $off$ measure the very same property so that $off$ and $off^{jet}$ are 
mathematically equivalent. They can be converted into each other according to the following equation
\begin{equation}
off^{jet}(\lambda_j) =  \frac{\sum_{i=0}^{N-1} F^{jet}(y_i,\,\lambda_j) + F^\star(y_i,\,\lambda_j)}{\sum_{i=0}^{N-1} F^{jet}(y_i,\,\lambda_j)} off(\lambda_j)\label{eq:conv}\,,
\end{equation}
that is, the $off$ is smaller than $off^{jet}$ by the fraction of jet to total emission. \\

\subsection*{Spectro-astrometric artefacts}

Real observations suffer from a distorted stellar image in the slit, uneven illumination, tracking inaccuracies,
and inaccurate centring to some degree, which we call PSF imperfections in the following. Therefore, any spectro-astrometric signal will contain artefacts to some degree, that is, offsets that are not caused by
a true change in the spatial distribution of the emission. \\
Such artefacts are caused by gradients in the spectrum together with PSF imperfections in cross-dispersion direction.
The grating geometry implies 
that a spatial displacement $\delta x$ with respect to the nominal aim point is directly proportional to 
a shift in wavelength $\delta \lambda \propto \delta x$, that is, PSF imperfections cause flux from
different wavelength to be recorded at the same location in dispersion direction. Thus, 
PSF imperfections in cross-dispersion direction together with spectral gradients cause
a spatial shift in the recorded emission. The artefacts scale with the flux gradient so 
that absorption and emission lines show inverse pattern (see Fig.~\ref{fig:sketch}).

To illustrate how a distorted stellar image affects the spectro-astrometric signal, 
we follow \cite{Brannigan_2006} and \citet{whelan_2015} and develop the stellar picture seen by the detector as the super-position of two 2D-Gaussians. The two 2D-Gaussians resemble the first two elements of a Taylor series decomposition. Most of the flux is assumed to be perfectly centred and circular. Some fraction of the flux is registered offset in spatial as well as in dispersion direction. 

We denote the (intrinsic) normalised stellar spectrum by $f(\lambda)$ and the spatial distribution 
recorded by the detector as
\begin{eqnarray}
S^\star(x,\,y) & = & \frac{1}{2\pi \sigma^2}\left(A_1 e^{-\frac{(x - x_1)^2 + (y-y_1)^2}{2\sigma^2} }\right) \nonumber \\
&& + \frac{1}{2\pi \sigma^2}\left(A_2 e^{-\frac{(x - x_2)^2 + (y-y_2)^2}{2\sigma^2} }\right)\,,
\end{eqnarray}
where $(x_i, y_i)$ are the Gaussians centres, $\sigma$ is the Gaussian width (assumed to be equal for
both Gaussians and both spatial directions), and $A_0$, $A_1$ are the Gaussian amplitudes. 
At a specific nominal wavelength, the detector sees the following 'image'
\begin{eqnarray}
F(\lambda,\, y) & = & \int_X f(\lambda + \Delta\lambda(\text{d}x)) \, S^\star(x,\,y) \text{d}x
\end{eqnarray}
and the flux weighted centroid at a specific wavelength will be
\begin{eqnarray}
cen^\star(\lambda) = \frac{\int_Y   \int_X y f(\lambda + \Delta\lambda(\text{d}x)) \, S^\star(x,\,y) \text{d}x\,\text{d}y}{\int_Y \int_X f(\lambda + \Delta\lambda(\text{d}x)) \, S^\star(x,\,y) \text{d}x\,\text{d}y}\, \label{eq:arti},
\end{eqnarray}
which is essentially Eq.~5 of \citet{Brannigan_2006}.\\
To estimate the magnitude of spectro-astrometric artefacts in our data, we evaluate
Eq.~\ref{eq:arti} numerically for different $(x_1, y_1)$, $\sigma$, and $A_1/A_2$-ratios.
Since absorption as well emission lines cause spectro-astrometric artefacts, but do
not involve spatially extended emission, they effectively limit the magnitude of potential
spectro-astrometric artefacts. Similarly, true spatial offsets do not necessarily fall into 
regions of strong spectral gradients, that is, we checked that observed signals cannot be explained
by spectro-astrometric artefacts. \\
These experiments showed that some minor artefacts may be present in the UVES data, they 
are generally smaller than the observed signals and would also be located at different
wavelengths, mostly in the line wings rather than close to the centre where our strongest
signals are.

 
 \section*{Appendix B - Target list}\label{appendix:target_list}  
 
  \section*{Appendix C - Outflow criteria}\label{appendix:outflow_criteria} 
 
\section*{Appendix D - H$\alpha$ and [OI]$\lambda$6300 line profiles}\label{appendix:line_profiles}

\section*{Appendix E - Line detections towards Sz\,103, XX\,Cha, Sz\,98, Sz\,99, Sz\,100, CVSO58, CVSO107, and SO518}\label{appendix:line_detections}    

\section*{Appendix F - [OI]$\lambda$6300 kinematical information}\label{appendix:OI6300_information}

\section*{Appendix G - Explanation of the spectro-astrometric signal in VW\,Cha}\label{appendix:vw_cha_sa}

\setcounter{table}{0}
\renewcommand{\thetable}{B.\arabic{table}}
  
 \setcounter{figure}{0}
\renewcommand\thefigure{\thesection B.\arabic{figure}}  

\setcounter{equation}{0}
\renewcommand{\theequation}{B.\arabic{equation}}  

 {\renewcommand{\arraystretch}{1.2}
\begin{table*}    \tiny
\caption{\small{List of the 18 targets of programme ID 106.20Z8.012.}}\label{table:observation_details_I}
\centering
\begin{tabular}{  |c||c|c|c|c|c|c|c|c|c|   }
\hline\hline
 \textbf{Target} (other names)   & RA (J2000) &  DEC (J2000) & SpT   & $A_V$ & Date of Obs. & Exp. Time  & Seeing   & P.A.$^1$  & $v_\text{sys}$ \\  
      & [hh:mm:ss]  &  [dd:mm:ss]   &  & [mag] & slit pos. 1/2/3  &  (NDIT x DIT) [s] &   [arcsec]    &  [deg] &  [km s$^{-1}$] \\ \hline 
 \hline
 \textbf{Sz\,68}  (HT\,Lup) & 15:45:12.87   & $-$34:17:30.8    & K2  & 1.0 &  2022-06-30  &  2x160 &  $\sim 0.8$   & 118 & $+15.5$ \\
Lupus\,1 &    &     &  &   &  2022-07-01    &   2x160 & $\sim 0.6$  &  298 & $+15.3$ \\
  (binary)  &    &     &   &  & 2022-07-04 & 2x160  &  $\sim 0.6$ & 178 & $+17.4$ \\
    \hline
 \textbf{Sz\,84} & 15:58:02.53   &  $-$37:36:02.7  & M5.0  & 0.0 &  2022-05-10  &   2x1800 & $\sim 1.1$  &  0 & $-9.0$\\
Lupus\,2 &    &     &    & &  2022-05-12 &  2x1800 &  $\sim 0.9$ & 120 & $-7.8$ \\
    &    &     &   &   & 2022-05-15 & 2x1800 & $\sim 0.5$  & 240  & $-6.8$\\
    \hline  
 \textbf{Sz\,97} & 16:08:21.79   &   $-$39:04:21.5  & M4.0    & 0.0 &  2022-05-11  &  2x1400 & $\sim 1.0$  &  0 & $-11.8$ \\  
 Lupus\,3 &    &     &  &   & 2022-05-12 &  2x1400 &  $\sim 0.7$ & 120   & $-10.0$\\
   &    &   &   &  &  2022-05-14  & 2x1400  &  $\sim 0.7$ &  240 & $-9.8$ \\
    \hline
 \textbf{Sz\,98}  (V1279 Sco) & 16:08:22.50   & $-$39:04:46.0  &  K7  & 1.0 & 2022-05-03   & 2x750  & $\sim 1.0$  & 0 &  $-15.2$\\ 
Lupus\,3 &    &     &  &   & 2022-05-06 &   2x750 &  $\sim 1.1$   &  120 & $-14.1$\\
   &    &  &    &  &   2022-05-10 &   2x750 & $\sim 1.3$  & 240  & $-12.3$\\
    \hline
 \textbf{Sz\,99} & 16:08:24.04   & $-$39:05:49.4   &  M4.0  & 0.0 &   2022-06-03 &  2x1750 & $1.3-1.7$  & 0  & $-1.2$  \\ 
Lupus\,3 &    &     &  &  &  2022-07-03 &  2x1750 &  $\sim 1.2$ &  120 & $+12.8$ \\
    &    &  &    &  &  2022-06-29  & 2x1750  & $\sim 0.8$  & 240 & $+12.2$   \\
    \hline
 \textbf{Sz\,100} & 16:08:25.76    & $-$39:06:01.1  & M5.5  & 0.0  &  2022-06-17  &  2x1700 &  $1.5-1.7$ & 8 &  $+6.8$ \\  
 Lupus\,3 &    &     &  &  &  2022-06-30 &   2x1700  & $0.6-0.9$  & 128 &  $+11.9$  \\
    &    &   &   &  &   2022-07-04 &  2x1700  &  $0.7-1.1$ & 248 & $+14.6$  \\
    \hline
 \textbf{Sz\,103} & 16:08:30.26   & $-$39:06:11.1   & M4.0   & 0.7 &  2022-04-28  & 2x1500 & $\sim 0.7$  &  0 & $-14.8$ \\  
Lupus\,3 &    &     &  &   & 2022-05-01 & 2x1500  & $\sim 0.4$  & 120 &  $-15.8$ \\
    &    &   &    & &   2022-05-04 &  2x1500 & $\sim 0.7$  & 240  & $-14.7$ \\
    \hline
 \textbf{Sz\,104} & 16:08:30.81   & $-$39:05:48.8    & M5.0   & 0.0 &  2022-06-24  &  2x1650 &  $1.0-1.3$ & 0 & $+9.9$ \\
Lupus\,3 &    &     & &  &  2022-07-05  &  2x1650 &  $\sim 0.6$ & 120 & $+15.6$ \\
   &   &  &     &  &   2022-06-30 & 2x1650  &  $\sim 0.7$ & 240 & $+13.0$ \\
    \hline
 \textbf{Sz\,112} &  16:08:55.52  & $-$39:02:33.9    & M5.0   & 0.0 &   2022-07-23 &  2x1700 &  $\sim 1.0$ & 0 &  $+21.0$ \\ 
 Lupus\,3 &    &    &   &  &  2022-07-24 &  2x1700 &  $0.9-1.8$ & 120 & $+21.2$ \\
   &    &    &   &  &  2022-07-25  & 2x1700  &  $\sim 0.7$ & 240  & $+21.4$\\
    \hline
 \textbf{Sz\,115} &  16:09:06.21  & $-$39:08:51.8    & M4.5  & 0.5 &  2022-06-03  &  2x1700  & $\sim 1.7 $  & 0  & $-1.1$ \\   
 Lupus\,3 &    &      & &   &  2022-06-30 &  2x1700  &  $\sim 0.5$ & 120 & $+12.3$ \\
     &    &   &   &  &  2022-06-09  & 2x1700   & $\sim 0.8$  & 240 &  $+2.6$ \\
    \hline 
\textbf{SSTC2DJ161243.8-381503} & 16:12:43.75  &     $-$38:15:03.3   & M1  & 0.8 &   2022-04-27 &  2x750 &  $0.9-1.5$  & 0  & $-18.1$\\   
 Lupus\,3 &    &     &  &   & 2022-05-04 & 2x750  & $0.7-1.0$  &  120 & $-14.9$\\
     &    &  &    &  & 2022-05-02   & 2x750  &  $0.6-0.9$ & 240  & $-15.3$ \\
    \hline
\textbf{SSTc2dJ161344.1-373646} &  16:13:44.11   &  $-$37:36:46.4  &   M5  & 0.6 & 2022-05-02   &  2x1800 &  $0.6-0.9$ &  55 &  $-12.5$\\ 
   Lupus\,3 &    &      & &   & 2022-05-04 & 2x1800   & $\sim 0.8$  & 175  & $-11.3$ \\
   &    &   & &    &   2022-05-07 & 2x1800  &  $\sim 1.4$  & 295 & $-11.1$ \\
   \hline
\textbf{Sz\,129} & 15:59:16.48   &  $-$41:57:10.3  & K7  & 0.9 &    2022-05-01 &  2x400 &  $\sim 0.5$ &  0 &  $-12.8$\\  
 Lupus\,4 &    &    &  &  &   2022-05-03 &   2x400 &  $\sim 0.9$ & 120 & $-12.2$ \\
      &   &  &  &     &  2022-05-07  & 2x400   & $\sim 1.4$  & 240  & $-9.5$ \\
    \hline 
\textbf{CS\,Cha} (Ass\,Cha\,T\,2-11) &  11:02:24.88  & $-$77:33:35.7  & K2 & 0.8 &   2022-05-11 &  2x350 &  $\sim 1.1-1.4$ &  0 & $+13.2$ \\  
Chameleon\,I &    &     &  &  & 2022-05-12  &  2x350 &  $\sim 0.8$ &  120 & $+14.8$ \\
  (binary)     &   &  &    &   & 2022-05-16   & 2x350  &  $\sim 1.1$ & 240  & $+15.0$ \\
    \hline
\textbf{CV\,Cha} (Ass\,Cha\,T\,2-52) &  11:12:27.72  & $-$76:44:22.30   &  K0  & 1.0 & 2022-05-11  &  2x200 &  $\sim 0.5$ &  0 & $+14.0$ \\  
   Chameleon I &    &  &     &   & 2022-05-13 &  2x200 &  $\sim 0.6$ & 120  & $+15.0$ \\
 (binary)   &   &  &    &   &  2022-05-16  &  2x200 & $\sim 1.0$  & 240 & $+15.4$ \\
    \hline
\textbf{VW\,Cha} (Ass\,Cha\,T\,2-31) &  11:08:01.49  & $-$77:42:28.8   & K7  & 1.9 & 2022-05-11   &  2x500 & $1.1-1.5$  & 0 &  $+21.6$\\  
Chameleon\,I &    & &     &  &  2022-05-12  & 2x500  &  $\sim 0.9$ & 120  & $+14.9$ \\
 (binary)     & &    &     &  &  2022-05-16  &  2x500 &  $\sim 1.0$ & 240 & $+11.8$ \\
    \hline 
\textbf{VZ\,Cha} (Ass\,Cha\,T\,2-40) & 11:09:23.79   & $-$79:23:20.8   & M0.5  & 1.2 &  2022-05-04 &  2x600 &  $1.0-1.2$ & 0 & $+14.5$ \\ 
 Chameleon\,I &    &    &   &  &  2022-05-07 &  2x600 &  $\sim 1.3$ & 120 & $+13.8$ \\
    &   &  &    &   & 2022-05-11   & 2x600  &  $\sim 0.5$ & 240 & $+15.6$ \\
    \hline
\textbf{WZ\,Cha} (Ass\,Cha\,T\,2-48) & 11:10:53.34   &  $-$76:34:32.0   & M3   & 1.2 & 2022-06-24   &  2x1700 &  $\sim 0.7-1.4$  & 0 & $+22.8$\\  
 Chameleon\,I   &    &      & &  & 2022-05-07  & 2x1700  &  $\sim 1.4$ & 120  & $+13.9$ \\
    &   &  &    &   &   2022-05-11 &  2x1700 & $\sim 0.6-1.2$  & 240 &  $+14.4$ \\
 \hline\hline
 \end{tabular}
 \tablefoot{$^1$ The position angle (P.A.) is measured positive from north to east, that is, the slit is aligned in north-south direction for P.A.$=0^{o}$.
}
\end{table*}

 \pagebreak

 {\renewcommand{\arraystretch}{1.2}
\begin{table*}    \tiny
\caption{\small{Same as for Table\,A.1 but for the remaining 16 targets associated with the PIDs 106.20Z8.009, 106.20Z8.010, and 106.20Z8.011.}}\label{table:observation_details_II}
\centering
\begin{tabular}{  |c||c|c|c|c|c|c|c|c|c|   }
\hline\hline
 \textbf{Target} (other names)  & RA (J2000) &  DEC (J2000) & SpT  & $A_V$ & Date of Obs. & Exp. Time  & Seeing   & P.A.$^1$  & $v_\text{sys}$ \\  
      & [hh:mm:ss]  &  [dd:mm:ss] &  & [mag] & slit pos. 1/2/3  & (NDIT x DIT) [s] &   [arcsec]    &  [deg] &  [km s$^{-1}$] \\ \hline 
 \hline
 \textbf{Hn\,5} &   11:06:41.64  &  -76:35:49.9   &  M5 &  0.0 & 2021-06-03 &  2x1700 & $\sim 0.7$    &  0 & +20.9  \\
 Chameleon\,I     &                 &                &  & & 2021-06-04  &  2x1700    & $\sim 0.6$  &  120  & +21.5 \\
      &                 &                &  & & 2021-06-06  &  2x1700    &  $\sim 0.8$ &  240  & +20.5 \\
    \hline  
\textbf{XX\,Cha} (Sz\,39, Ass\,Cha\,T\,2-49)&   11:11:39.51  &   -76:20:16.0  &  M3.5  & 1.0 &  2021-06-03  & 2x1700  &  $\sim 0.5$   & 0 &  +20.9 \\
Chameleon\,I     &                 &               &  &  & 2021-06-04  &  2x1700    & $\sim 0.5$    &  120    & +20.5 \\
 (binary)     &                 &               &  &  & 2021-06-06  &  2x1700    &  $\sim 1.1$ &  240  &  +20.4 \\
    \hline  
\textbf{2MASS\,J16000060-4221567}   & 16:00:00.49    &  -42:21:56.2   & M4.5   & 0.0 &      2021-07-21 &  2x1800 &  $\sim 0.6$   & 0 &  +22.3 \\
 Lupus\,4  &            &                &         & &  2021-07-22    & 2x1800    & $\sim 0.4$  &  120  &  +23.2 \\
    \hline   
\textbf{Sz\,40}  (In\,Cha, Ass\,Cha\,T\,2-50) & 11:12:09.76    & -76:34:37.1    &   M5  & 0.1 & 2021-06-06    & 2x1800  & $\sim 0.9$   &  0 &  +20.8 \\
 Chameleon\,I     &                &       &  &   &   2021-06-07      &  2x1800   & $\sim 0.5$  &   120  & +20.2 \\
 \hline\hline
 \textbf{CVSO\,17} &    05:23:04.77 & +01:37:14.8    &  M3  & 0.0  &  2020-12-04  & 2x1800   & $\sim 0.7$   & 350 &  +21.0 \\
  Orion\,OB1   &                &                &  & &  2020-12-05       & 2x1800    & $\sim 0.9$  & 110   & +20.9 \\
    (binary)     &                &                &  &  & 2020-12-06   &  2x1800  &  $\sim 0.9$  & 290 & +21.2  \\
    \hline
\textbf{CVSO\,36} & 05:25:50.35    &  +01:49:37.1   &  M3  & 0.0 &    2020-12-02  &   2x1800  &    $\sim 0.9$    & 109 &  +13.5 \\
   Orion\,OB1  &    &     &  &      &    2020-12-03 &   2x1800    &  $\sim 0.8$ &   289 &  +13.5 \\
 (binary)   &    &     &    &   &  2020-12-04 &    2x1800  &  0.64  &   109 & +14.1  \\
    \hline
\textbf{CVSO\,58} &  05:29:23.36   & -01:25:15.1    &   K7  & 0.12 &  2020-11-30  &  2x1300 &  $\sim 0.9$   & 0 & +15.2  \\
  Orion\,OB1 &    &     &  & &  2020-12-01  &   2x1300  & $\sim 0.8$  &   120 &  +13.9 \\
    &    &     &   &  &  2020-12-02 &  2x1300  &  $\sim 0.9$  & 240  & +18.1  \\
    \hline
\textbf{CVSO\,104A} &  	05:32:06.48   &   -01:11:00.7  &   K7  & 0.32 &   2020-11-25 &  2x1100 &  $\sim 1.0$   & 109 &  -22.9 \\
  Orion\,OB1 &    &     &  &    &    2020-11-26 &    2x1100  & $\sim 0.8$ &   109 &  +13.2 \\
 (binary)   &    &     &     & &  2020-11-27 &    2x1100 &   $\sim 0.8$ & 109 &  -26.0 \\
        &    &     &  &   &  2021-01-29 &    2x1100 &  $\sim 0.6$  & 109 & +5.6  \\
    \hline
\textbf{CVSO\,107} &  05:32:25.82   &    -00:36:52.5  &  K7   & 0.77 &  2020-12-03   & 2x1250  & $\sim 1.0$  & 0 &  +11.2 \\
      Orion\,OB1           &                &               &   &   &  2020-12-04  &  2x1250   & $\sim 0.5$  & 120   & +10.0 \\
                &                &              &    &   & 2020-12-05   &  2x1250  & $\sim 1.1$  &  240 & +10.3  \\
    \hline
\textbf{CVSO\,109} &  05:32:32.66    &  -01:13:45.3   &  M0  & 0.0 &    2020-11-26 &   2x1050 &  $\sim 0.7$   &  42 & +6.8  \\
 Orion\,OB1  &    &     &  &     &   2020-11-27 &     2x1050 &  $\sim 0.8$ &   42 &  +7.4\\
  (binary)  &    &     &  &   &   2020-11-28 &     2x1050 &  $\sim 0.8$   &  42 &  +8.0 \\
    \hline
\textbf{CVSO\,176} &  05:40:24.23   & -00:31:20.0    & M3   & 0.0 &   2020-11-28  &  2x1800  &   $\sim 0.8$  & 0 &  +1.0 \\
  Orion\,OB1 &    &     &  &    &    2020-11-29 &    2x1800  & $\sim 0.6$  &   120 & +3.3  \\
    &    &     &    & &  2020-11-30 &  2x1800  & $\sim 0.6$  & 240 & +4.4  \\
    \hline
\textbf{SO518} (V505\,Ori)  &  05:38:27.27   &  -02:45:08.7   & K7   & 0.0 &   2020-11-29 &  2x1100 & $\sim 0.7$    & 0 &  +22.9 \\
  $\sigma$\,Ori  &    &     &  &     &  2020-11-30  &    2x1100  &  $\sim 1.1$ &  120  &  +22.5\\
   &    &     &    &   & 2020-12-01 &   2x1100  & $\sim 0.6$   & 240 & +25.4  \\
    \hline
\textbf{SO583}  (TX\,Ori)&   05:38:33.72  &  -02:44:13.2   &  K4.5   & 0.0 &  2020-11-29  & 2x420  &  $\sim 0.5$   & 0 &  +23.0 \\
  $\sigma$\,Ori  &    &     &  &    &    2020-11-30 &     2x420 & $\sim 0.9$ &    120 & +23.0 \\
   &    &     &     & &  2020-12-01  &    2x420 &   $\sim 0.7$ &  240 &  +22.7 \\
    \hline\hline
\textbf{TW\,Hya} &  11:01:51.91   & $-$34:42:17.0  & K7  & 0.0 &  2022-03-29  & 2x170  & $\sim 1.0$  &  0 & $+12.6$ \\ 
  &    &      & &  &  2022-03-31 &  2x170 &  $\sim 0.8$ & 0 & $+14.6$ \\
  \hline
\textbf{DK\,Tau\,A+B} &   04:30:44.27  & +26:01:25.5     &  K8.5, M1.5 & 1.3 &    2021-11-25   &  2x800 &  $\sim 0.5$    & 117.6 &  +13.4$^2$ \\
  Taurus-Auriga       &     &      &     &  &   2021-12-01   &  2x800     &  $\sim 0.6$  &  117.6   & +14.0$^2$ \\
   (binary)     &     &      &     &   &   2021-12-02 &    2x800   &   $\sim 0.6$ &   117.6  & +16.9$^2$ \\
  \hline\hline
 \end{tabular}
 \tablefoot{$^1$ The position angle (P.A.) is measured positive from north to east, that is, the slit is aligned in north-south direction for P.A.$=0^{o}$. $^2$ The stated values for $v_{\text{sys}}$  refer to DK\,Tau\,A. For DK\,Tau\,B a $v_{\text{sys}}$  could only be derived in slit positions 2 and 3, that is, $+13.9\,\text{km}\,\text{s}^{-1}$ and $+15.7\,\text{km}\text{s}^{-1}$, respectively. In slit position 1 no Lithium line nor the alternative \ion{Ca}{i} (6572.78$\AA$) absorption line is present in the spectrum of DK\,Tau\,B.
}
\end{table*}

\pagebreak

\setcounter{table}{0}
\renewcommand{\thetable}{C.\arabic{table}}
  
 \setcounter{figure}{0}
\renewcommand\thefigure{\thesection C.\arabic{figure}}  

\setcounter{equation}{0}
\renewcommand{\theequation}{C.\arabic{equation}}  

\begin{sidewaystable*}
\caption{\small{Outflow activity in the targets. }}\label{table:outflow_criteria_I}
\centering \tiny
\begin{tabular}{  |c||c|c|c|c|c|c|c|c|c|c|c|c|c |c||   }
\hline\hline
 Target  &  Slit  & [SII] &  [SII]& [OI] &  [OI] &   [OI]  &  [NII] & [NII] &  [SII] &   [SII] & H$\alpha$  & Spectro-astrometric   &  [OI]$\lambda$6300 &   Outflow       \\   
      &  & 4068  & 4076  & 5577  &   6300  &  6363  & 6548  &   6583  &  6716  & 6731 & Profile  &  Signal    &  Components     &      Criteria           \\ \hline 
 \hline
 \textbf{Sz\,68} & 1   & -  &- &  -& - & - & - &  -&-  &  -   &   IIB & H$\alpha^1$, not in He\,I     &  - &   0/4    \\
        & 2   &  -& - &  -&  -&  -& - & - & - &  -   &    IIIR     & H$\alpha^1$, not in He\,I     & -  &       \\
        & 3   &  - &  -& - &-  &-  &  -&-  &-  & -   &    IIB    &  -    &   - &       \\
\hline
 \textbf{Sz\,84} & 1   & -  &- &  -& yes & yes & - &  -&-  &  - &   IIIRm  &  -     &  LVC  &     1/4 \\
        & 2   & -  &- &  -& yes &  yes & - &  -&-  &  -  &     IIIRm  & -    &     LVC  &          \\
        & 3   & -  &- &  -& yes&  yes & - &  -&-  &  -  &     IIIR  &  -   &  LVC  &         \\
 \hline  
 \textbf{Sz\,97}  & 1  & -  &- &  -& yes & - & - &  -&-  &    -  &   IIR  & -    &  NLVC, BLVC  &   1/4     \\
        & 2   & - &- &  -& yes & - & - &  -&-  &   -    &     IIIR   &  -      &   NLVC, BLVC  &         \\
        & 3   & -  &- &  -& yes & - & - &  -&-  &  -   &    IIR/IIIR &  -    &   NLVC, BLVC &       \\
\hline
\textbf{Sz\,98}  & 1  & -  &- & yes & yes & yes & yes &  yes & -     &   yes &  IIIB     & -    &   NLVC, BLVC, HVCB    &  4/4    \\
           & 2   & yes  &- &  yes & yes& yes & yes &  yes & -  & yes &     IIB/IIIB   & -     &  NLVC, BLVC, HVCB   &           \\
          & 3   & yes &- &  yes & yes & yes & yes &  yes & -   & yes  &    IIB/IIIB  &  -   &   NLVC, BLVC, HVCB  &        \\
    \hline
\textbf{Sz\,99}  & 1   & -  &- &  -& yes & - & - &  -&-  &    -  &   I  & -      &   BLVC, HVCR &    4/4   \\
        & 2   & yes  &- &  yes & yes & yes & - &  -& yes  &   yes  &   IIB   &  -     &  BLVC, HVCR   &      \\
        & 3   & yes  &- &  yes & yes & yes & - &  -& yes  &   yes  &   IIIB  &  -    &   BLVC, HVCR   &     \\
    \hline
 \textbf{Sz\,100}  & 1    & yes  &- &  yes & yes & yes & yes &  yes & yes   &  yes &    IIRm  &  -   & NLVC, BLVC, HVCB   &     3/4 \\
        & 2    & yes & yes &  yes & yes & yes & yes &  yes & yes &  yes &    I    &  -     &  NLVC, BLVC, HVCB  &     \\
        & 3    & yes  & - &  yes & yes & yes & yes &  yes & yes  &   yes   &   I   & -     &   NLVC, BLVC, HVCB &     \\
    \hline
\textbf{Sz\,103}  & 1  & yes  &- &   yes & yes & yes & yes &  yes & yes &    yes  & IIBm   & -    &   NLVC, BLVC, HVCB &    4/4  \\
        & 2   & yes  & yes &  yes & yes & yes & yes &  yes & yes  &    yes  &    IIIBm  & H$\alpha$, [OI], not in He\,I     &  NLVC, BLVC, HVCB    &    \\
        & 3 & yes & yes &  yes & yes & yes & yes &  yes & yes  &    yes  &   IIB  & H$\alpha$, [OI], not in He\,I      &  NLVC, BLVC, HVCB    &      \\
    \hline
 \textbf{Sz\,104}  & 1   & -  &- &  -& - & - & - &  -&-  &  -   & I    & -       & -  &   0/4    \\
        & 2   & -  &- &  -& - & - & - &  -&-  &  -   &  I    &  -      & -  &         \\
        & 3   & -  &- &  -& - & - & - &  -&-  &  -   &   I   &  -     &  - &         \\
    \hline
\textbf{Sz\,112}  & 1   & yes &- & yes & yes & yes & - &  -&-  &  -   & I    &   -     &  NLVC, BLVC    &   2/4      \\
        & 2   & -  & - &  yes & yes & yes & - &  -&-  &  -   &  I  & -      &   NLVC, BLVC  &         \\
        & 3   & yes  &- &  yes & yes & yes & - &  -&-  &  -   &   I  &  -     &  NLVC, BLVC   &          \\
    \hline
 \textbf{Sz\,115}  & 1   & -  &- &  -& - & - & - &  -&-  &  -   & I   & -  &    - &   0/4    \\
        & 2  & -  &- &  -& - & - & - &  -&-  &  -   &     I &  H$\alpha$ feature     & -  &      \\
        & 3   & -  &- &  -& - & - & - &  -&-  &  -   &    I & H$\alpha$ feature     & -  &      \\  
 \hline\hline
 \end{tabular}
\end{sidewaystable*}
 

 \pagebreak

\begin{sidewaystable*}
\caption{\small{Outflow activity in the targets.  }}\label{table:outflow_criteria_II}
\centering \tiny
\begin{tabular}{  |c||c|c|c|c|c|c|c|c|c|c|c|c|c| c||   }
\hline\hline
 Target  & Slit  & [SII] &  [SII]& [OI] &  [OI] &   [OI]  &  [NII] & [NII] &  [SII] &   [SII] & H$\alpha$  & Spectro-astrometric   &  [OI]$\lambda$6300 &  Outflow       \\  
      &  & 4068  & 4076  & 5577  &   6300  &  6363  & 6548  &   6583  &  6716  & 6731 & Profile  &  Signal  &  Components &     Criteria            \\ \hline 
 \hline 
\textbf{Sz\,129}  & 1    &yes &- &  yes& yes & yes & - &  -&-  &  -   &  I    &  weak offset in [OI]       & NLVC, BLVC & 2/4  \\
        & 2    & yes  &- &  yes& yes & yes & - &  -&-  &  -   &    I    & -        & NLVC, BLVC &     \\
        & 3    & yes  &- & yes& yes & yes & - &  -&-  &  -   &    I   & offset in H$\alpha$     & NLVC, BLVC &    \\
    \hline 
\textbf{CS\,Cha}  & 1    & -  &- &  -& yes & yes & - &  -&-  &  -   & I  & weak in H$\alpha$ at $-50\,\text{km}\,\text{s}^{-1}$     & LVC  &   2/4 \\
        & 2    & -  &- &  -& yes &  yes & - &  -&-  &  -   &   I   &    weak in H$\alpha$ at $+50\,\text{km}\,\text{s}^{-1}$      & LVC &     \\
        & 3    & -  &- &  -& yes & yes & - &  -&-  &  -   &   IIIB  &  weak in H$\alpha$ at $-100\,\text{km}\,\text{s}^{-1}$       & LVC &    \\
    \hline
\textbf{CV\,Cha}  & 1   & -  &- &  yes & yes & yes & - &  -&-  &  -   & IIB/IIIB    &   -      & NLVC, HVCB  &  3/4  \\
        & 2   & -  &- &  yes & yes & yes & - &  -&-  &  -   &      IIB &   -       & NLVC, HVCB &      \\
        & 3   & -  &- &  yes & yes & yes & - &  -&-  &  -   &   IIB/IIIB     & -       &   NLVC, HVCB  &      \\
    \hline
\textbf{VW\,Cha} & 1    &- &- & yes & yes & yes& - &  yes & -  &  -   &   IIIB & offset in H$\alpha$, [OI]$\lambda$6300, He\,I        & NLVC, BLVC  &     3/4  \\
        & 2    &- &- & yes & yes & yes & yes &  yes & -  &  -   &    IIIB  & offset in H$\alpha$, [OI]$\lambda$6300, He\,I       &  NLVC, BLVC &    \\
        & 3    &-  &- & yes & yes & yes& - &  yes &-  &  -   & IVB   & offset in  H$\alpha$, [OI]$\lambda$6300, He\,I        &  NLVC, BLVC   &     \\
    \hline 
\textbf{VZ\,Cha}  & 1 &- &- & yes & yes & yes & - &  -&-  &  -   & I  &  -       & NLVC, BLVC   &  2/4   \\
        & 2   &- &- & yes & yes & yes & - &  -&-  &  -   &  IIR  &  -       & NLVC, BLVC   &        \\
        & 3   &- &- & yes & yes & yes & - &  -&-  &  -   &   IIIB  & -      &  NLVC, BLVC   &        \\
    \hline
\textbf{WZ\,Cha}  & 1   & -  &- &  -& - & - & - &  -&-  &  -   &   IIR & weak offset in H$\alpha$, not in He\,I        & -  &  0/4   \\
        & 2 & -  &- &  -& - & - & - &  -&-  &  -   &   I  &   weak offset in H$\alpha$, not in He\,I    & - &      \\
        & 3  & -  &- &  -& - & - & - &  -&-  &  -   &  I   & weak offset in H$\alpha$        , not in He\,I &  -&     \\
    \hline
\textbf{SSTC2D...503}  & 1  & -  &- &  -& yes & yes & - &  -&-  &  -   &  IIIB  &  -       &  LVC &    2/4   \\
        & 2   & -  &- &  -& yes & yes & - &  -&-  &  -   &     IIIB    &    -       & LVC  &       \\
        & 3 & -  &- &  -& yes & yes & - &  -&-  &  -   &    IIB   &  -      &  LVC &      \\
    \hline
\textbf{SSTc2d...646} & 1    & -  &- &  yes & yes & - & - &  -&-  &  -   &  I  &  -      & LVC  &    1/4    \\
        & 2    & -  &- &  -&yes & - & - &  -&-  &  -   &    I &  -      &   LVC &        \\
        & 3    & -  &- &  -&yes & - & - &  -&-  &  -   &   I  & -     &   LVC &        \\
 \hline\hline
 \end{tabular}
\end{sidewaystable*}
 

\pagebreak

\begin{sidewaystable*}
\caption{\small{Outflow activity in the targets.  }}\label{table:outflow_criteria_III}
\centering \tiny
\begin{tabular}{  |c||c|c|c|c|c|c|c|c|c|c|c|c|c| c||   }
\hline\hline
 Target  &  Slit  & [SII] &  [SII]& [OI] &  [OI] &   [OI]  &  [NII] & [NII] &  [SII] &   [SII] & H$\alpha$  & Spectro-astrometric   &  [OI]$\lambda$6300 &  Outflow       \\  
      &  & 4068  & 4076  & 5577  &   6300  &  6363  & 6548  &   6583  &  6716  & 6731 & Profile  & Signal    &  Components &     Criteria            \\ \hline 
 \hline 
\textbf{CVSO\,17}  & 1    & - & - & -  & - & -& - &  - & - & -   & I     &     -   & -  & 0/4   \\
        & 2    & - &-  &  - & - & - & - & - & - & -   &  I    &     -  &  - &    \\
        & 3    & -& -&  - & - & -& - &  -& - &  -  &  I    &      -  & - &    \\
    \hline 
\textbf{CVSO\,36}   & 1    & - & - & -  & - & -& - &  - & - & -   & I     &     -   & -  &  0/4  \\
        & 2    & - &-  &  - & - & - & - & - & - & -   &  I    &   -  & - &    \\
        & 3    & -& -&  - & - & -& - &  -& - &  -  &   I   &     -  & - &    \\
    \hline
\textbf{CVSO\,58}  & 1 & yes  & -&  yes  & yes  & yes  &  yes  &  yes & - &  yes  &  IIRm   &   weak in H$\alpha$     & NLVC, BLVC, HVCB  &  3/4 \\
        & 2 &  yes & -&  yes  & yes  & yes &  - &  yes & - &   yes &   IIB  &   weak  in H$\alpha$   &  NLVC, BLVC, HVCB &   \\
        & 3   &  yes & -&  yes  & yes  & yes  &  -  & yes & - &  yes  &  IIR   &   weak in H$\alpha$   &  NLVC, BLVC, HVCB &   \\
    \hline
\textbf{CVSO\,104\,A}  & 1    & yes & -&  yes &  yes & yes & - &  -& - &-    &  IIB  &  offset in H$\alpha$       & NLV, BLV  &  2/4 \\
        & 2  & yes  &- &    yes & yes & yes &  - & - & - & -   &  I  &      offset in H$\alpha$    &   NLV, BLV &   \\
          & 3 &  yes & -& yes  & yes & yes &  - & - & - &  -  & IIR    &     -    & NLV, BLV  &   \\
            & 4 &  yes & -& yes  & yes & yes &  - & - & - &  -  & I    &     -  & NLV, BLV  &   \\
    \hline
\textbf{CVSO\,107} & 1  & yes&- &  yes &  yes & yes & - &-  & -  & yes   &  IIB   & -      & NLVC, BLVC, HVCB  &  3/4 \\
        & 2  &yes & -&  yes &  yes & yes &  -&  -&  - & yes   &  IIB   &    -   &  NLVC, BLVC, HVCB &   \\
        & 3  &yes & -&  yes &  yes & yes & - & - & -  &  yes  &  IIR   &   -     &  NLVC, BLVC, HVCB &   \\
    \hline 
\textbf{CVSO\,109} & 1  & yes   & -   & yes  & yes & yes& - & - & - &-    &  I   &  offset in H$\alpha$, [OI]$\lambda$6300, HeI      & NLVC, BLVC  &  2/4 \\
        & 2 & yes & -&  yes & yes  &yes  &   -  &  -  & - &  -  &  I  &    offset in H$\alpha$, [OI]6300, He\,I      &  NLVC, BLVC &   \\
        & 3 & yes & -&  yes & yes & yes & - & - &  -&   - &   I  &     offset in H$\alpha$, [OI]$\lambda$6300, He\,I     &  NLVC, BLVC &   \\
    \hline
\textbf{CVSO\,176} & 1 & -& -&  yes & yes & yes & - & - & - &  -  &  IIBm   &  -      & BLVC, HVCB  & 2/4  \\
        & 2 & -& -&  yes & yes & yes & - & - & - & -   &   IIR    &    -   &  BLVC, HVCB  &   \\
        & 3 & -&- &  yes & yes & yes & - & - & - &   - &  IIR    &   -    &  BLVC, HVCB  &   \\
    \hline
\textbf{SO518}  & 1 & yes& yes &  yes & yes & yes&  yes &  yes&  yes& yes   &  IIR   &    -    &  BLVC, HVCB, HVCR &  3/4 \\
       & 2 &yes & yes&   yes & yes & yes & yes & yes & yes &  yes  &   IIR  &    -   & BLVC, HVCB, HVCR  &   \\
       & 3 & yes& yes&   yes & yes & yes & yes & yes & yes &  yes  &  IIR   &    H$\alpha$ at $-90\,\text{km}\,\text{s}^{-1}$ (cosmics)   & BLVC, HVCB, HVCR  &   \\
    \hline
\textbf{SO583} & 1 & -&- & -  & yes & - & yes  & yes &-  & -   &    IIRm/IVR &  feature in H$\alpha$     & BLVC &  2/4 \\
        & 2   & -& -& -  & yes  & -& yes & yes & - & -   & IIRm/IVR    &   -    &  BLVC &   \\
        & 3   &- & -& -  & yes & -&  yes & yes & - &  -  &   IIRm/IVR  &  offset in H$\alpha$     &  BLVC &   \\
 \hline\hline
 \end{tabular}
\end{sidewaystable*}
 

\pagebreak

\begin{sidewaystable*}
\caption{\small{Outflow activity in the targets. }}\label{table:outflow_criteria_IV}
\centering \tiny
\begin{tabular}{  |c||c|c|c|c|c|c|c|c|c|c|c|c|c| c||   }
\hline\hline
 Target  &  Slit  & [SII] &  [SII]& [OI] &  [OI] &   [OI]  &  [NII] & [NII] &  [SII] &   [SII] & H$\alpha$  & Spectro-astrometric   &  [OI]$\lambda$6300 &  Outflow       \\  
      &  & 4068  & 4076  & 5577  &   6300  &  6363  & 6548  &   6583  &  6716  & 6731 & Profile  & Signal    &  Components &     Criteria           \\ \hline 
 \hline 
\textbf{Hn\,5}      & 1 & - & - &  -  & yes  & -  & -   &  - &  - &  -   &   I  &   -     &  NLVC, BLVC & 1/4  \\
        & 2 & - &  -&  -  & yes  & -  &  -  &  - &  - &  -   &   I  &   -   & NLVC, BLVC   &   \\
        & 3 & - & - &  -  & yes  &  - &  -  &  - &  - &  -   &   I  &   -    & NLVC, BLVC   &   \\
 \hline
\textbf{Sz\,40}    & 1 & - &  -&  -  & yes  & -  & -   &  - &  feature & yes    &  I   &  -    & LVC  & 2/4  \\
        & 2 & yes & - &  -  & yes  & -  &  -  &  - &  feature & yes   &   I  &    -   &  LVC &   \\
 \hline
\textbf{XX\,Cha}   & 1 & yes &yes  &  yes  & yes  & yes  & yes  &  yes &  - &  yes   &  IIIB  &   H$\alpha$, [OI]$\lambda$6300 at $-80\,\text{km}\,\text{s}^{-1}$, not in HeI   &  NLV, BLV, HVB &  4/4 \\
        & 2 &yes  &yes  &  yes  &  yes & yes  &  yes  & yes  & -  &  yes   &   IIIB  &     H$\alpha$ at $0\,\text{km}\,\text{s}^{-1}$, not in He\,I    &  NLV, BLV, HVB &   \\
        & 3 &yes  & yes &   yes &  yes & yes  &  yes  &  yes &  - &  yes   &   IIIB  &  H$\alpha$, [OI]$\lambda$6300 at $-90\,\text{km}\,\text{s}^{-1}$, not in He\,I       &  NLV, BLV, HVB &   \\
 \hline
\textbf{2MASS...567} & 1 & - & - &  -  & -  & -  & -   & feature &   -&  -   &   I  &  -       & -  & 0/4  \\
         & 2 & - & - &  -  &  - &  - & -   &  feature & -  &  -   & I    &    -   &  - &   \\
 \hline\hline
\textbf{TW\,Hya}  & 1   & yes  &- & yes & yes & yes& - &  -&-  &  -   &  IIIB   &  -      &  LVC & 3/4  \\
        & 2   & yes  &- &  yes & yes & yes & - &  -&-  &  -   &    IIIB   & -        & LVC &       \\
 \hline
\textbf{DK\,Tau\,A}  & 1 & -   & - & yes & yes & yes &  -  & -  &  - &  ?   &    IIR  &    -    &  NLV, BLV, HVB &  3/4 \\
                     & 2 & yes & - & yes & yes & yes &  -  & -  & -  &  ?   &      IIB &   -   & NLV, BLV, HVB &   \\
                     & 3 & yes & - & yes & yes & yes &  -  & -  & -  &  ?   &     IIB  &    -   & NLV, BLV, HVB  &   \\
        \hline
\textbf{DK\,Tau \,B}  & 1  & - & - & -   &  yes &  yes &   - &  - & yes  &  yes   &    IIR  &    offset in H$\alpha$, [OI]$\lambda$6300   &  NLV, BLV, HVB & 3/4  \\
        & 2 &  - &  -&  -  & yes  &  yes &   - & -  & yes  &  -   &    IIR  &   s-shaped feature in [OI]$\lambda$6300     & NLV, BLV, HVB   &   \\
        & 3 & - & - &  -  & yes  & yes  &  -  &  - &  - & -    &     IIR  &   s-shaped feature in [OI]$\lambda$6300     &  NLV, BLV, HVB  &   \\
 \hline\hline
 \end{tabular}
\end{sidewaystable*}
 

\pagebreak

\setcounter{table}{0}
\renewcommand{\thetable}{D.\arabic{table}}
  
 \setcounter{figure}{0}
\renewcommand\thefigure{\thesection D.\arabic{figure}}  

\setcounter{equation}{0}
\renewcommand{\theequation}{D.\arabic{equation}}    
 

\begin{figure*} 
\centering
\subfloat{\includegraphics[trim=0 0 0 0, clip, width=0.3 \textwidth]{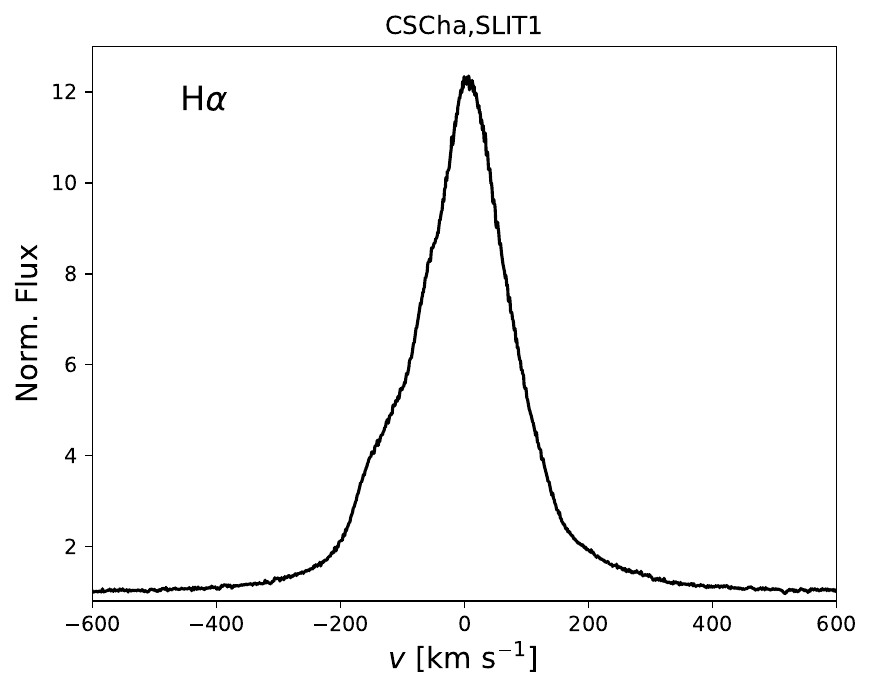}}
\hfill
\subfloat{\includegraphics[trim=0 0 0 0, clip, width=0.3 \textwidth]{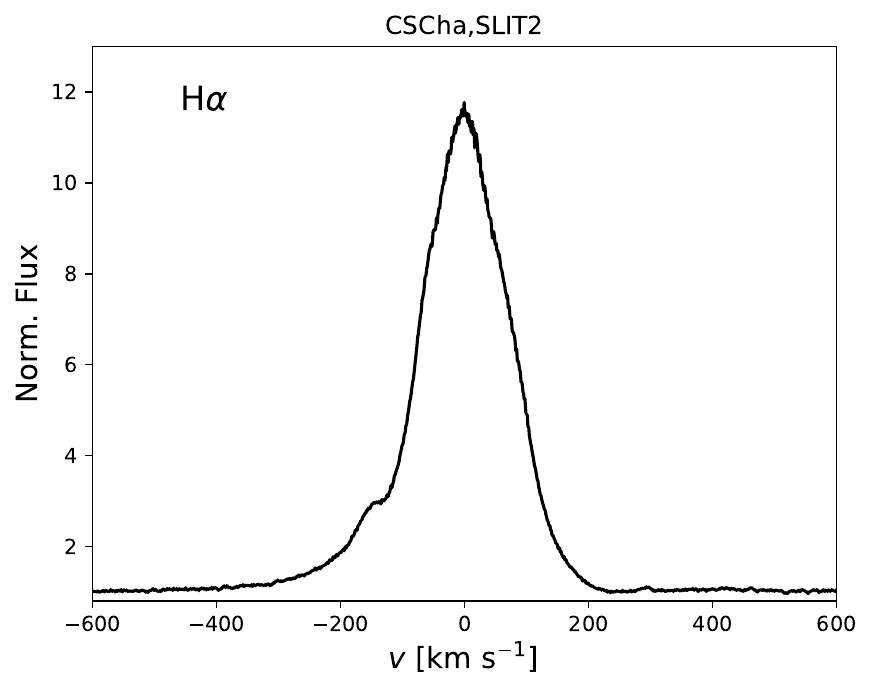}}
\hfill
\subfloat{\includegraphics[trim=0 0 0 0, clip, width=0.3 \textwidth]{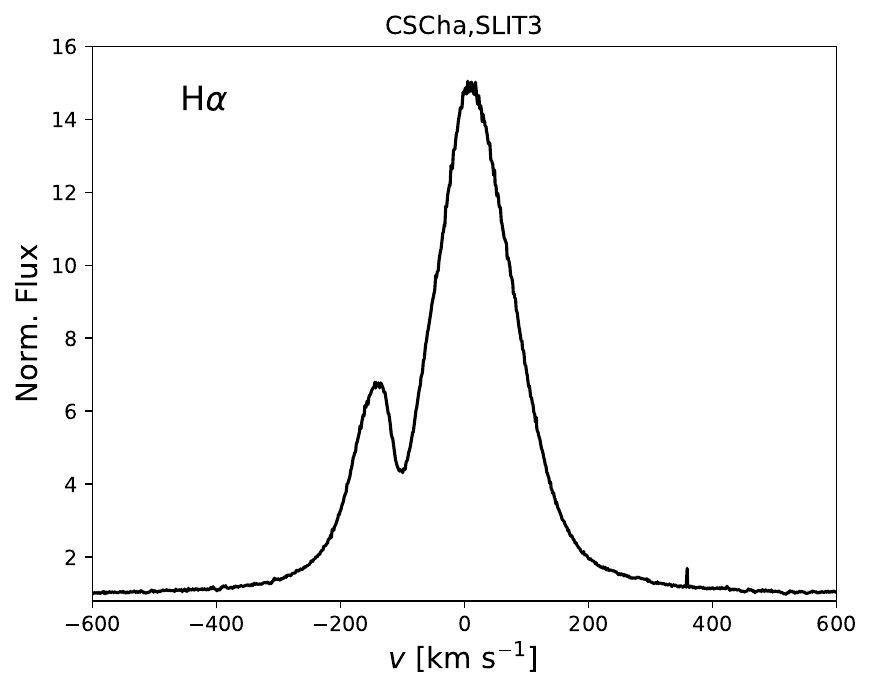}}
\hfill 
\subfloat{\includegraphics[trim=0 0 0 0, clip, width=0.3 \textwidth]{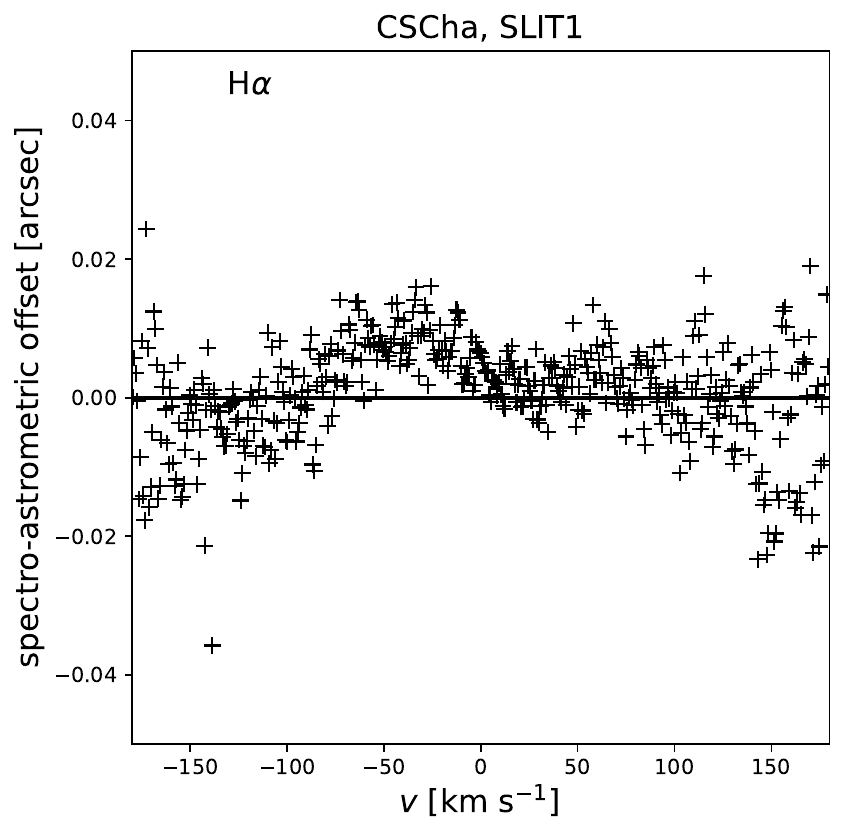}}
\hfill
\subfloat{\includegraphics[trim=0 0 0 0, clip, width=0.3 \textwidth]{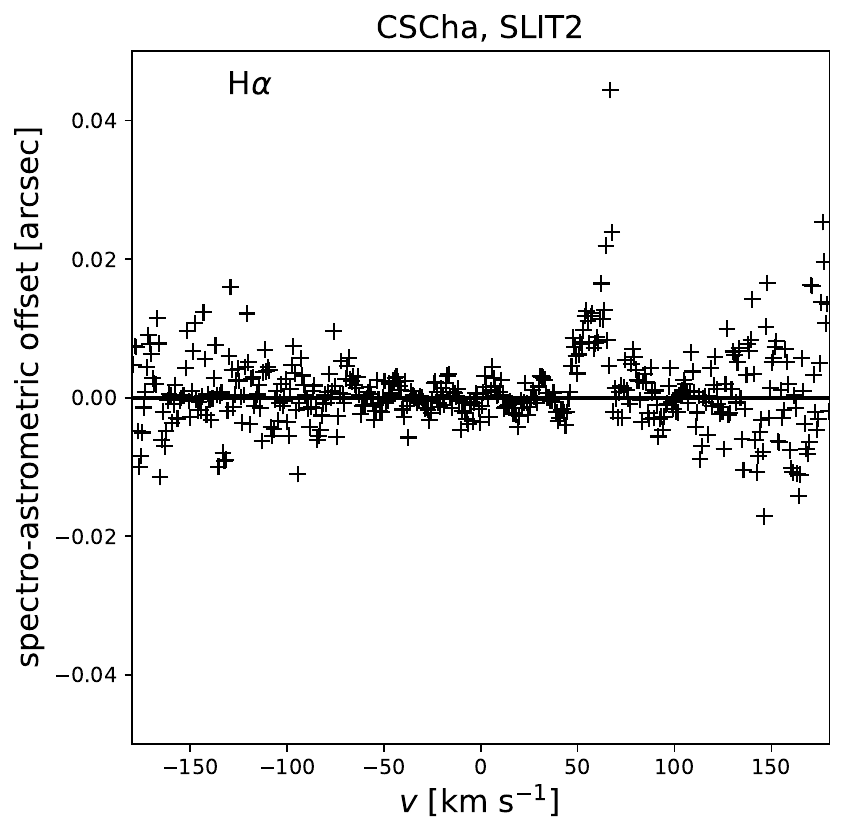}}
\hfill
\subfloat{\includegraphics[trim=0 0 0 0, clip, width=0.3 \textwidth]{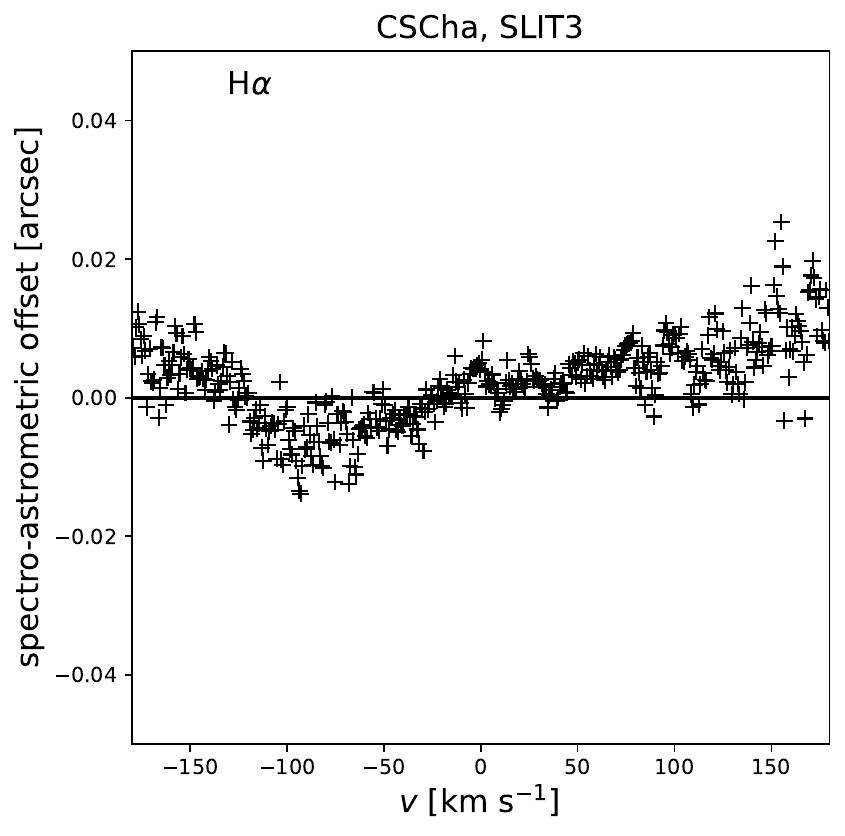}} 
\hfill
\subfloat{\includegraphics[trim=0 0 0 0, clip, width=0.3 \textwidth]{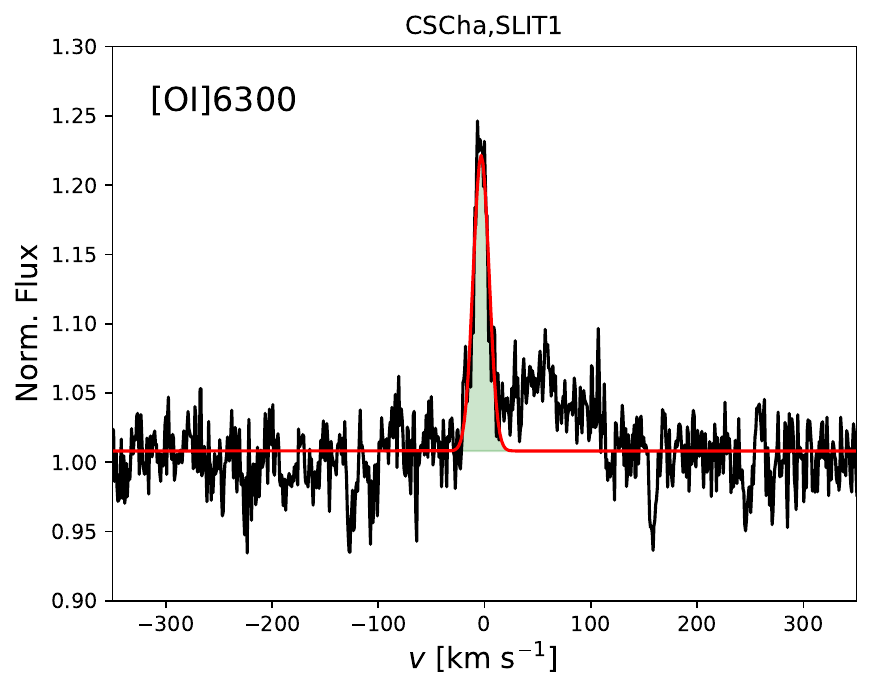}}
\hfill
\subfloat{\includegraphics[trim=0 0 0 0, clip, width=0.3 \textwidth]{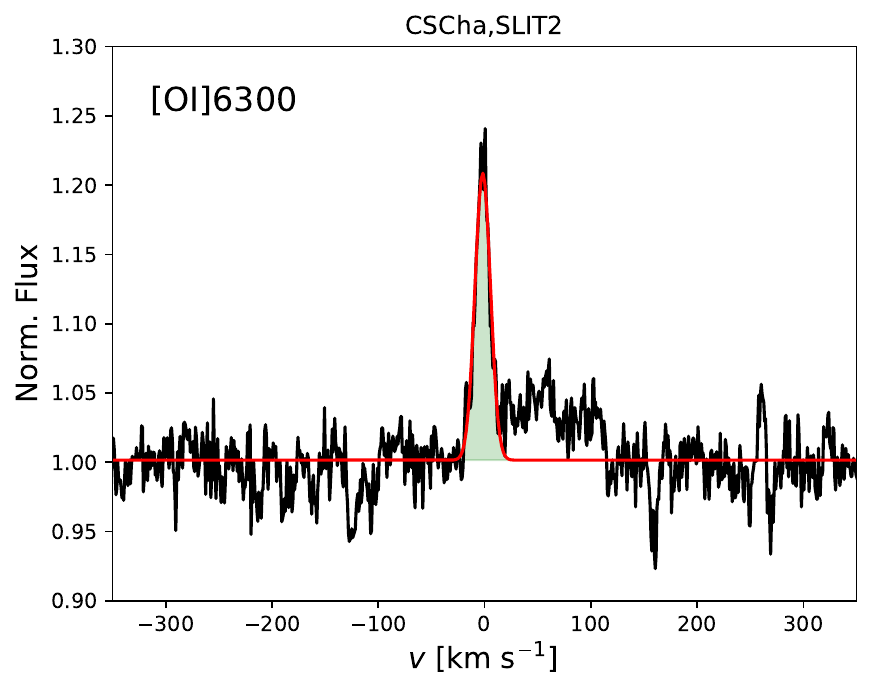}}
\hfill
\subfloat{\includegraphics[trim=0 0 0 0, clip, width=0.3 \textwidth]{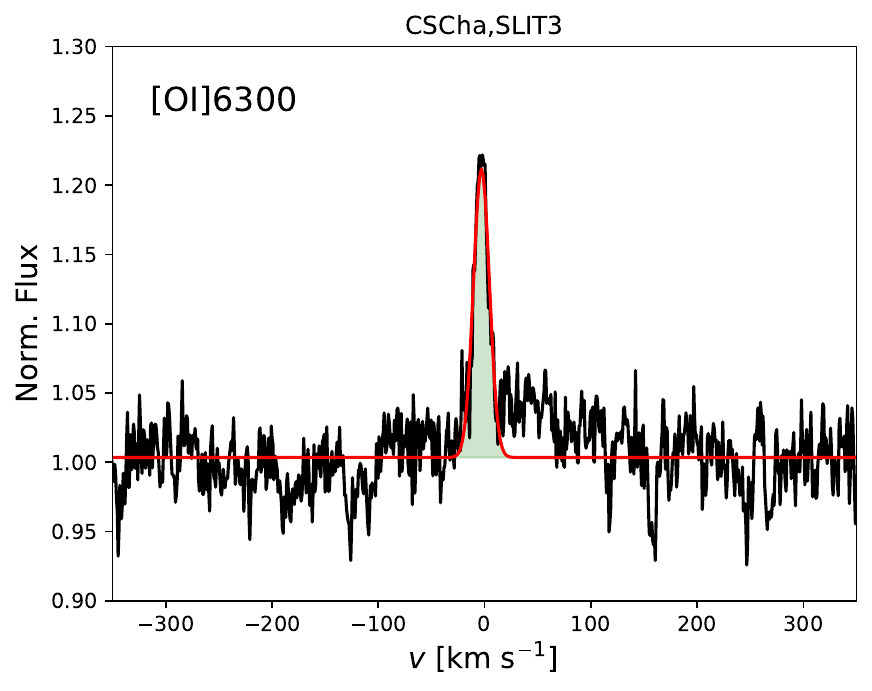}} 
\hfill 
\subfloat{\includegraphics[trim=0 0 0 0, clip, width=0.3 \textwidth]{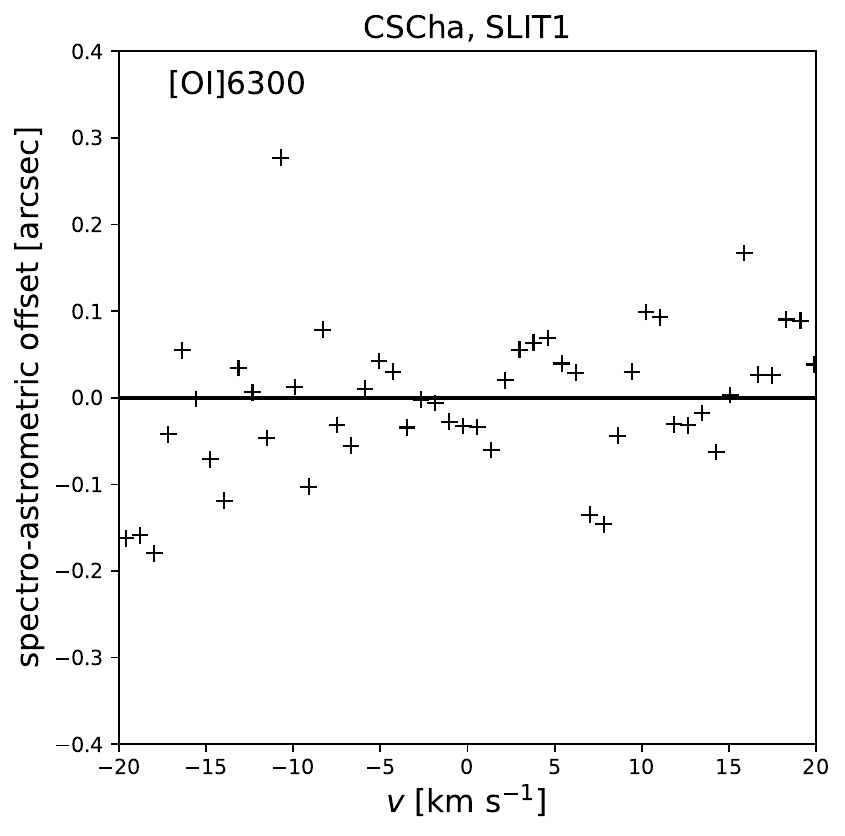}}
\hfill
\subfloat{\includegraphics[trim=0 0 0 0, clip, width=0.3 \textwidth]{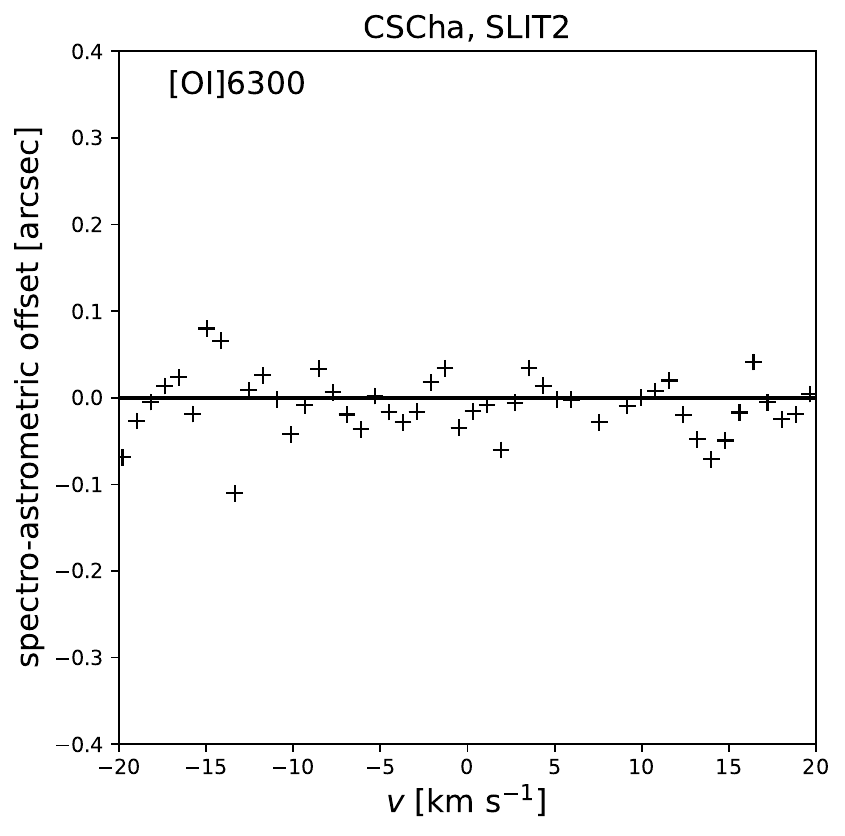}}
\hfill
\subfloat{\includegraphics[trim=0 0 0 0, clip, width=0.3 \textwidth]{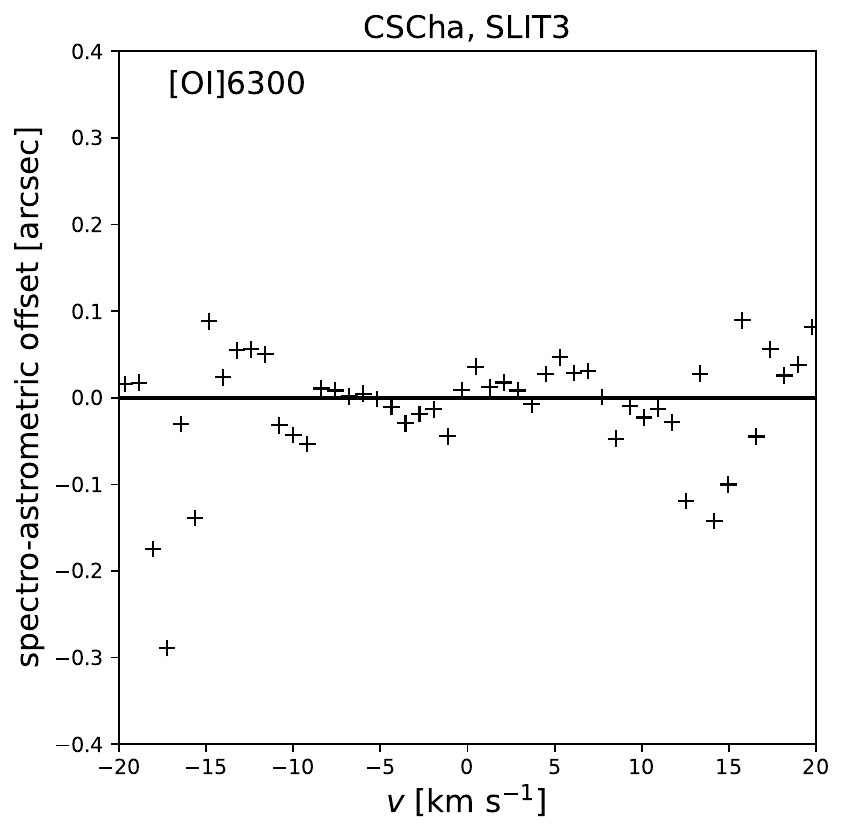}} 
\hfill
\caption{\small{Line profiles of H$\alpha$ and [OI]$\lambda$6300 for all slit positions of CS\,Cha.}}\label{fig:all_minispectra_CSCha}
\end{figure*} 

\begin{figure*} 
\centering
\subfloat{\includegraphics[trim=0 0 0 0, clip, width=0.3 \textwidth]{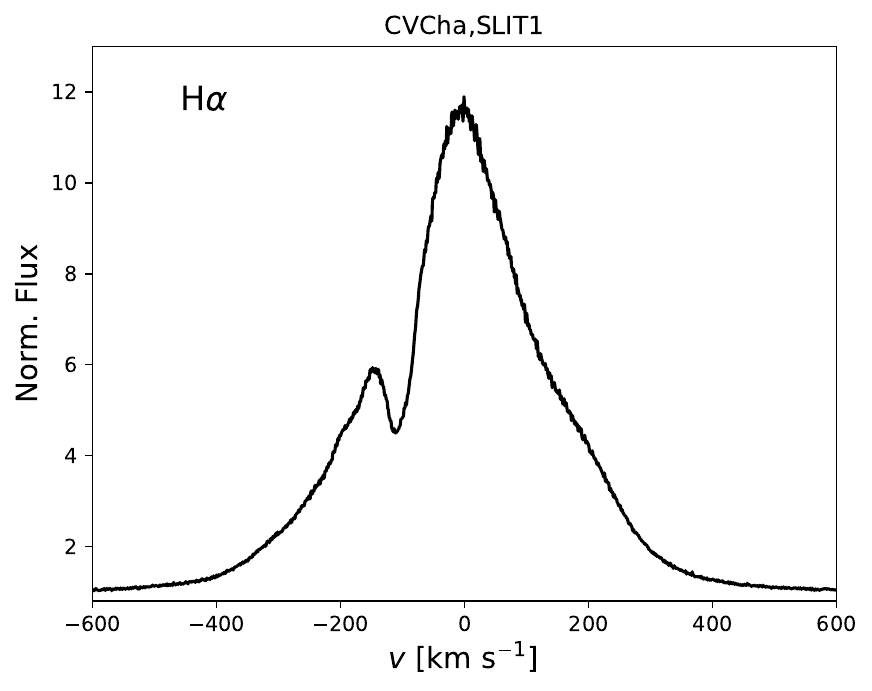}}
\hfill
\subfloat{\includegraphics[trim=0 0 0 0, clip, width=0.3 \textwidth]{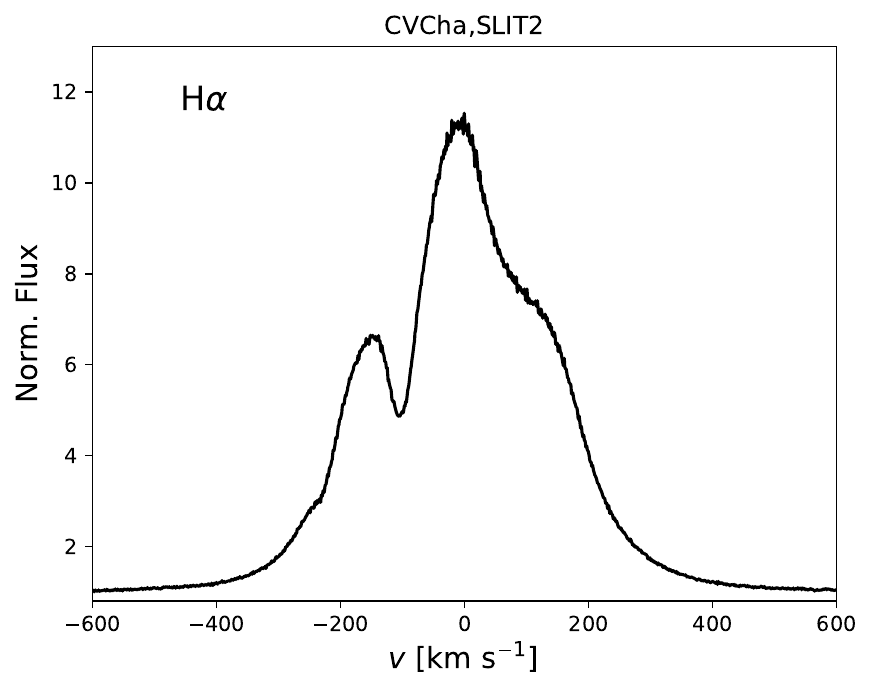}}
\hfill
\subfloat{\includegraphics[trim=0 0 0 0, clip, width=0.3 \textwidth]{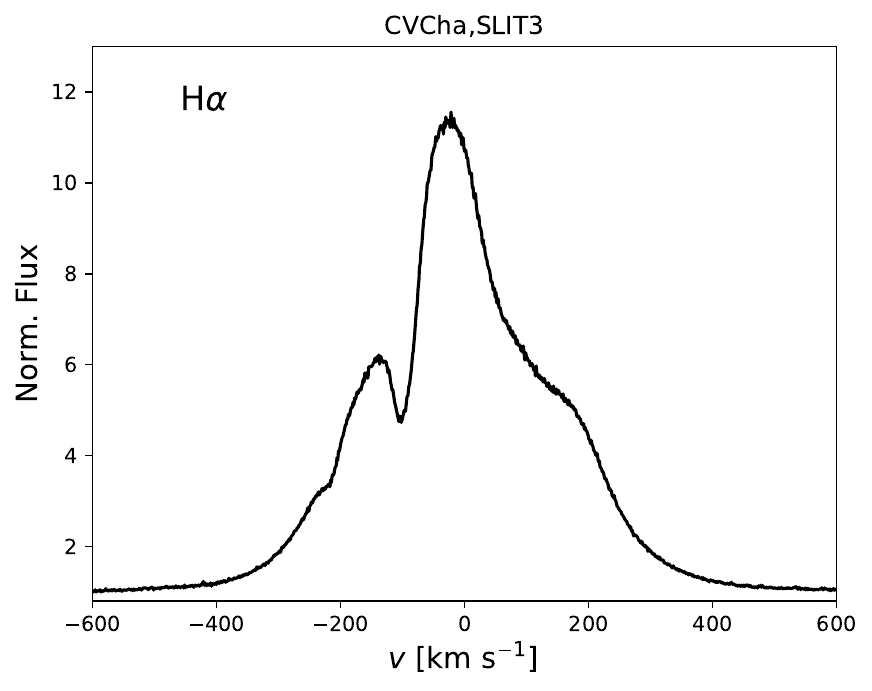}}
\hfill 
\subfloat{\includegraphics[trim=0 0 0 0, clip, width=0.3 \textwidth]{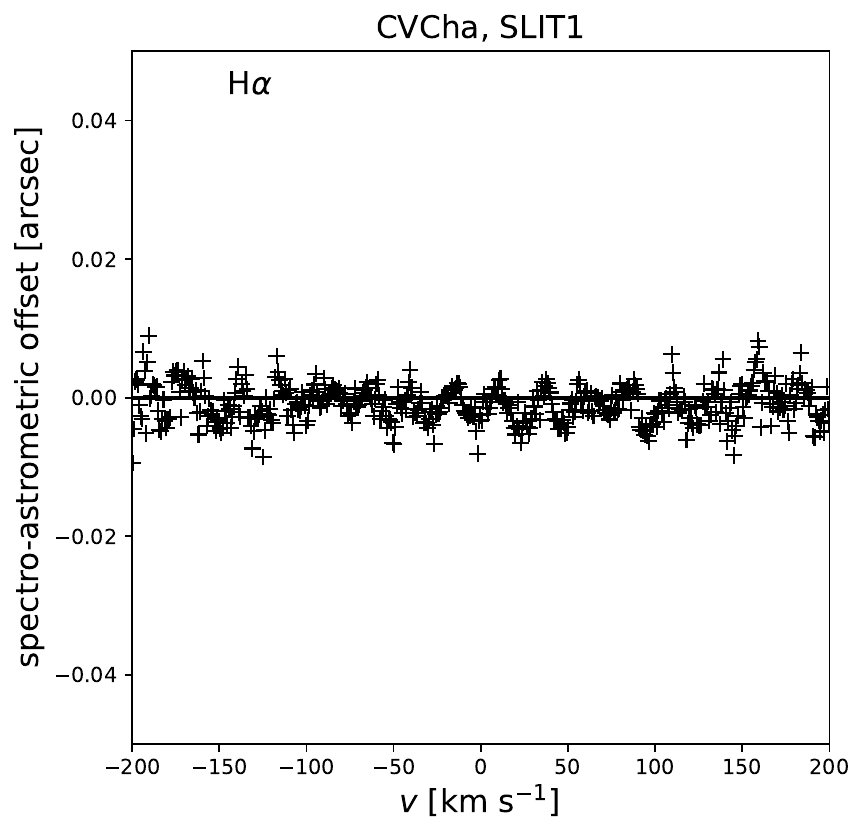}}
\hfill
\subfloat{\includegraphics[trim=0 0 0 0, clip, width=0.3 \textwidth]{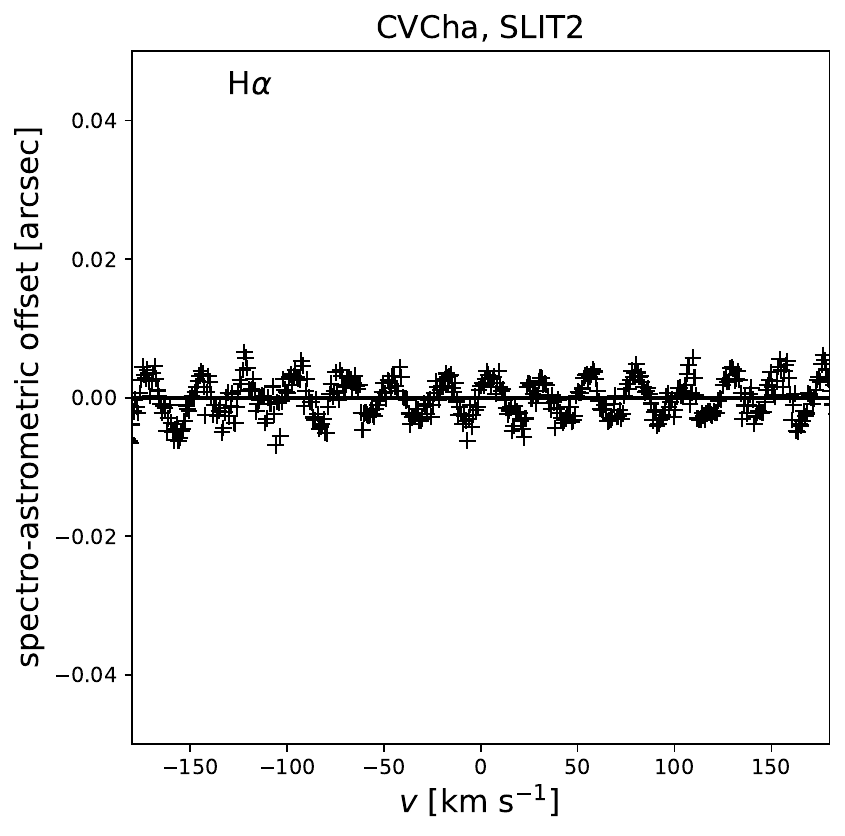}}
\hfill
\subfloat{\includegraphics[trim=0 0 0 0, clip, width=0.3 \textwidth]{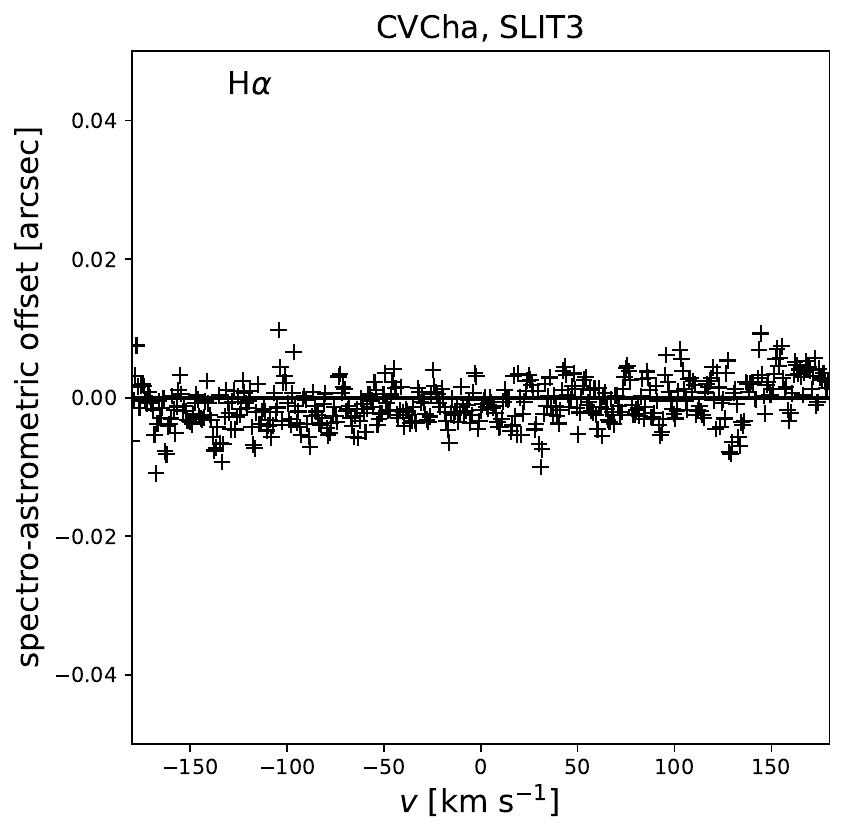}} 
\hfill
\subfloat{\includegraphics[trim=0 0 0 0, clip, width=0.3 \textwidth]{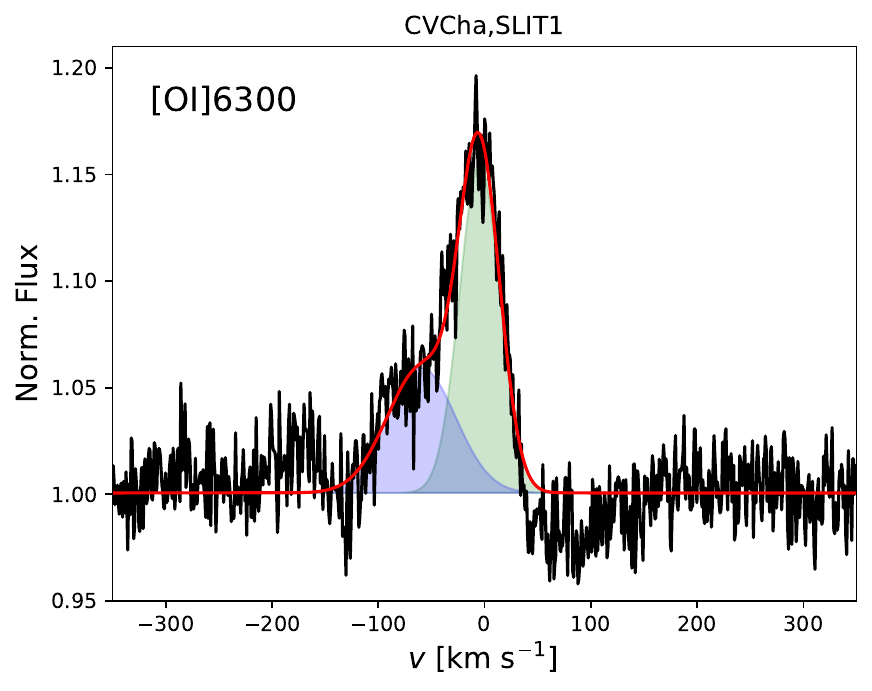}}
\hfill
\subfloat{\includegraphics[trim=0 0 0 0, clip, width=0.3 \textwidth]{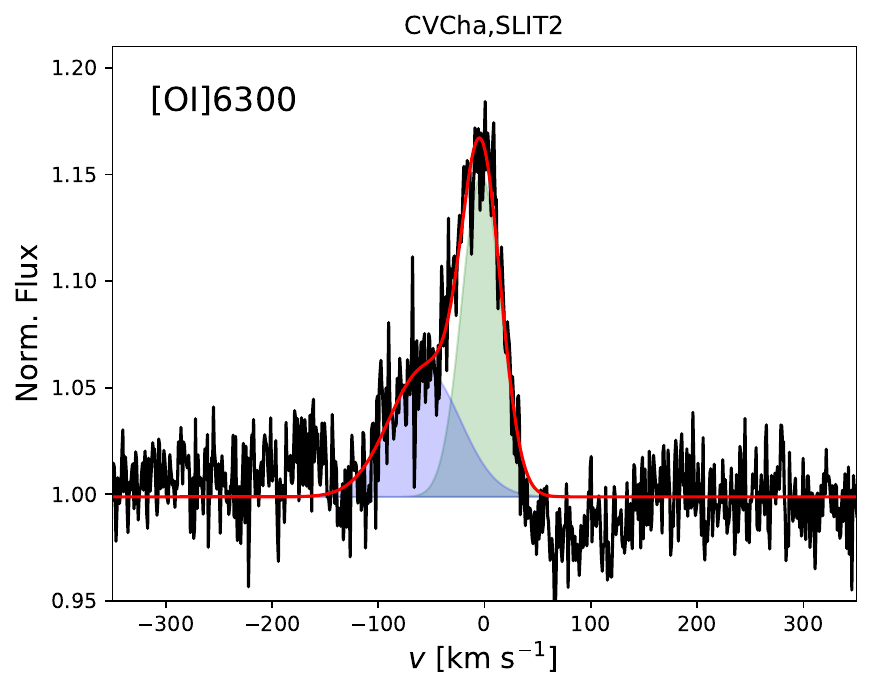}}
\hfill
\subfloat{\includegraphics[trim=0 0 0 0, clip, width=0.3 \textwidth]{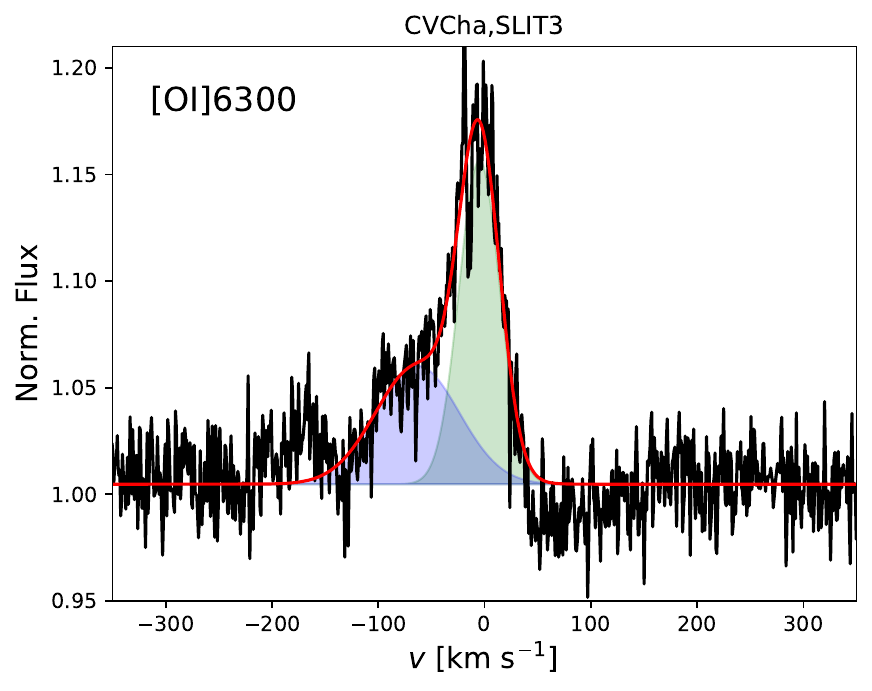}} 
\hfill  
\subfloat{\includegraphics[trim=0 0 0 0, clip, width=0.3 \textwidth]{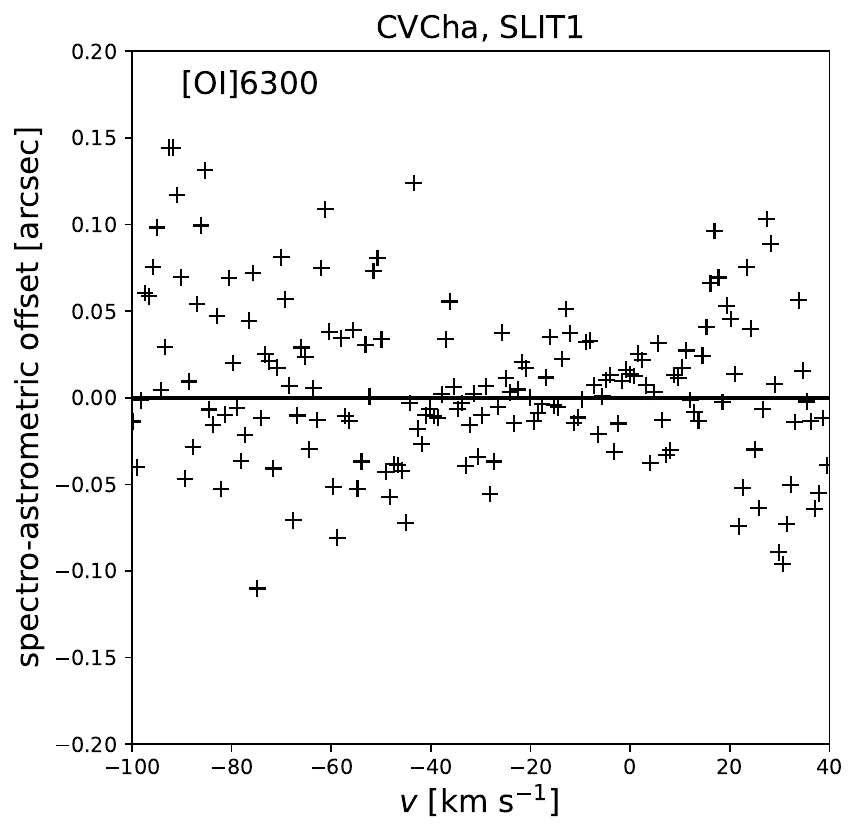}}
\hfill
\subfloat{\includegraphics[trim=0 0 0 0, clip, width=0.3 \textwidth]{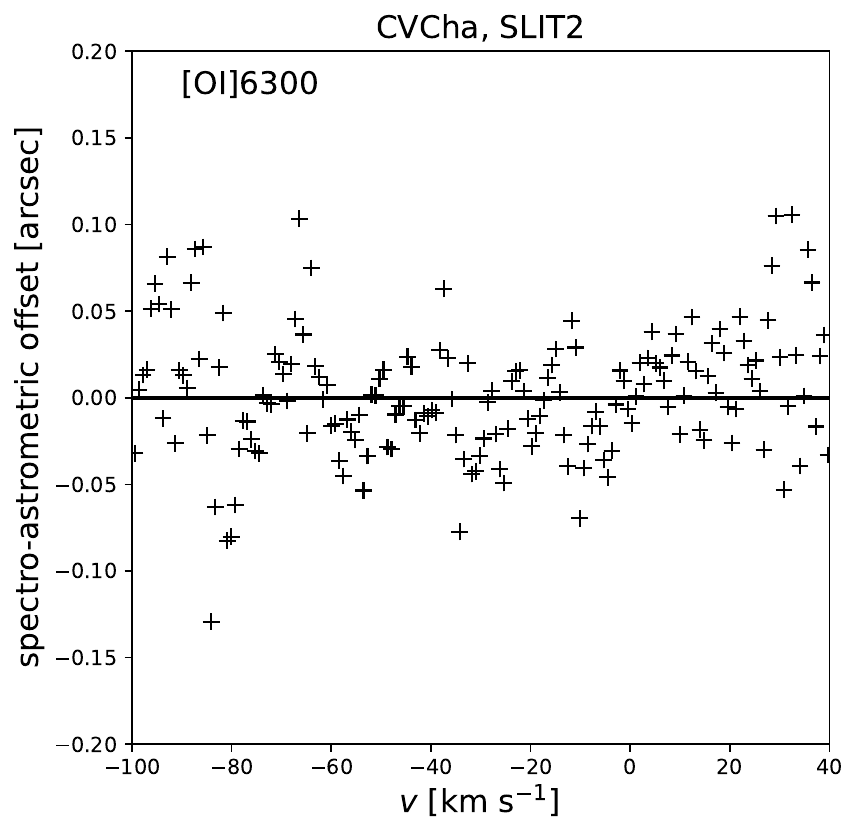}}
\hfill
\subfloat{\includegraphics[trim=0 0 0 0, clip, width=0.3 \textwidth]{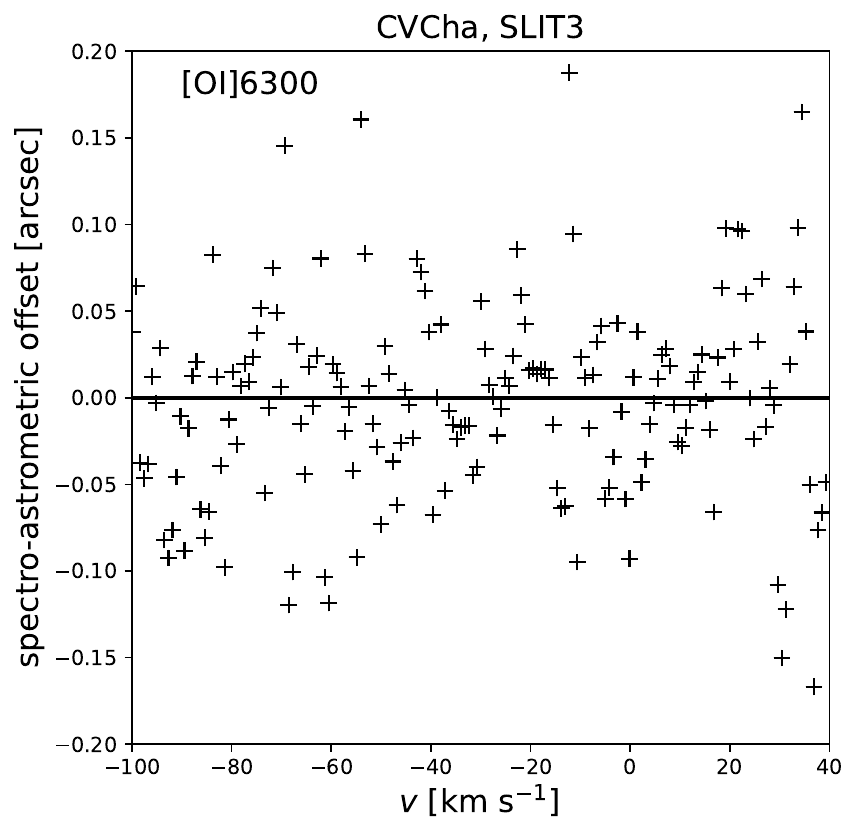}} 
\hfill
\caption{\small{Line profiles of H$\alpha$ and [OI]$\lambda$6300 for all slit positions of CV\,Cha.}}\label{fig:all_minispectra_CVCha}
\end{figure*} 

\begin{figure*} 
\centering
\subfloat{\includegraphics[trim=0 0 0 0, clip, width=0.3 \textwidth]{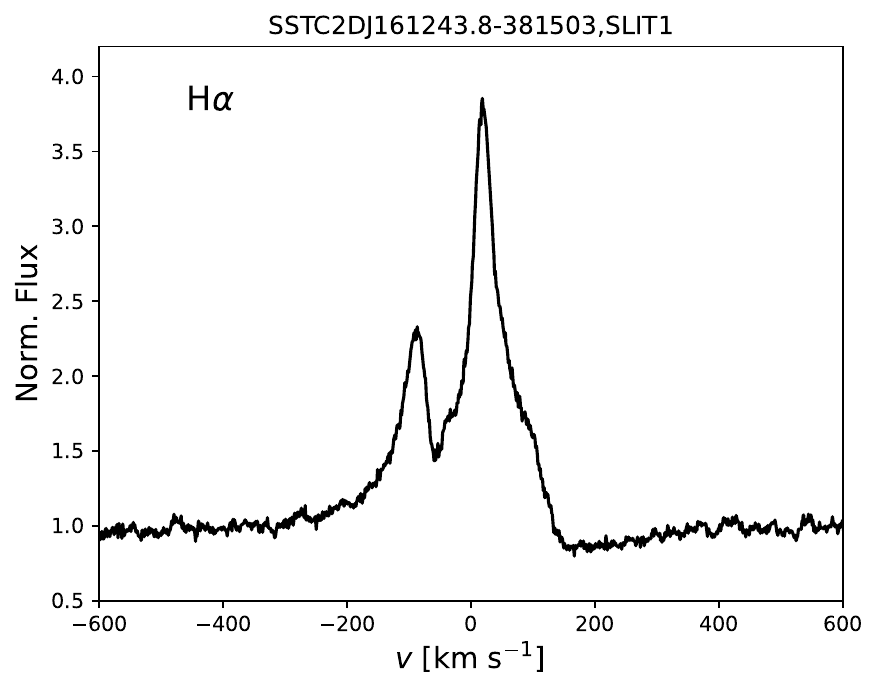}}
\hfill
\subfloat{\includegraphics[trim=0 0 0 0, clip, width=0.3 \textwidth]{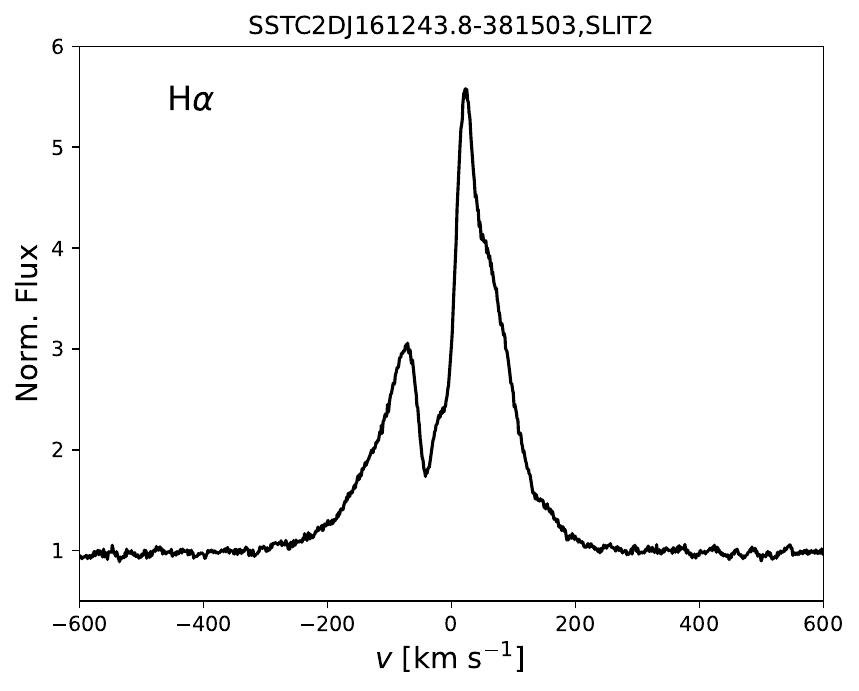}}
\hfill
\subfloat{\includegraphics[trim=0 0 0 0, clip, width=0.3 \textwidth]{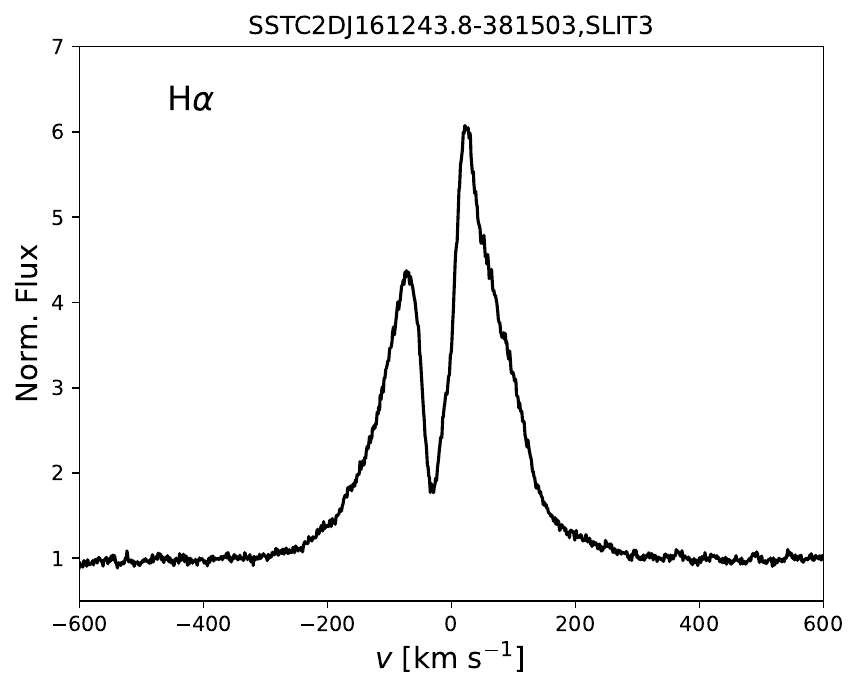}}
\hfill 
\subfloat{\includegraphics[trim=0 0 0 0, clip, width=0.3 \textwidth]{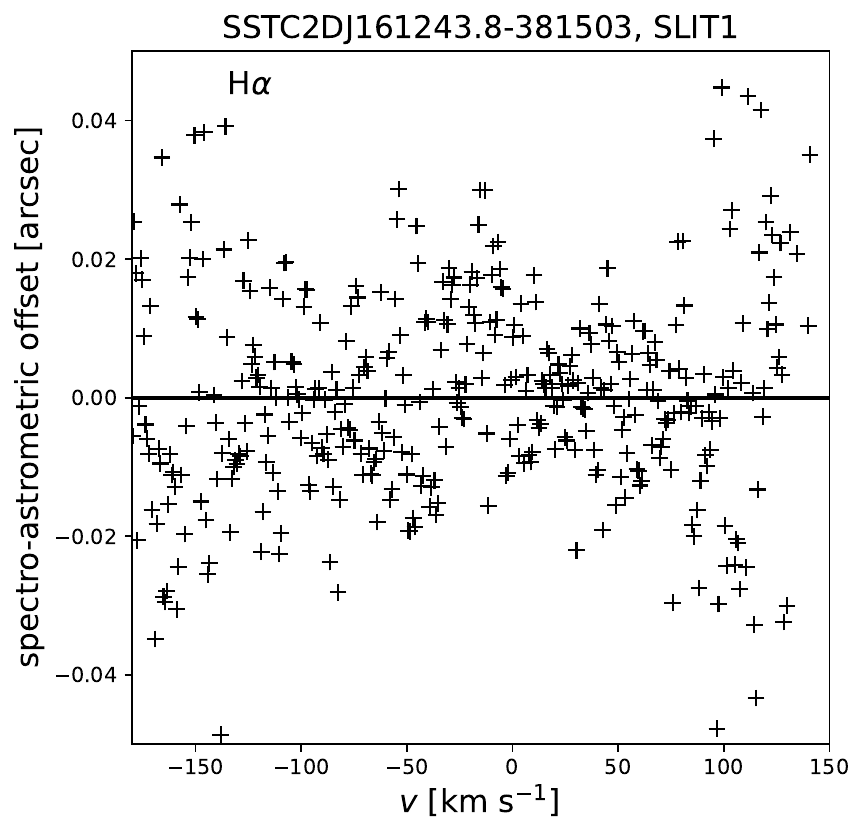}}
\hfill
\subfloat{\includegraphics[trim=0 0 0 0, clip, width=0.3 \textwidth]{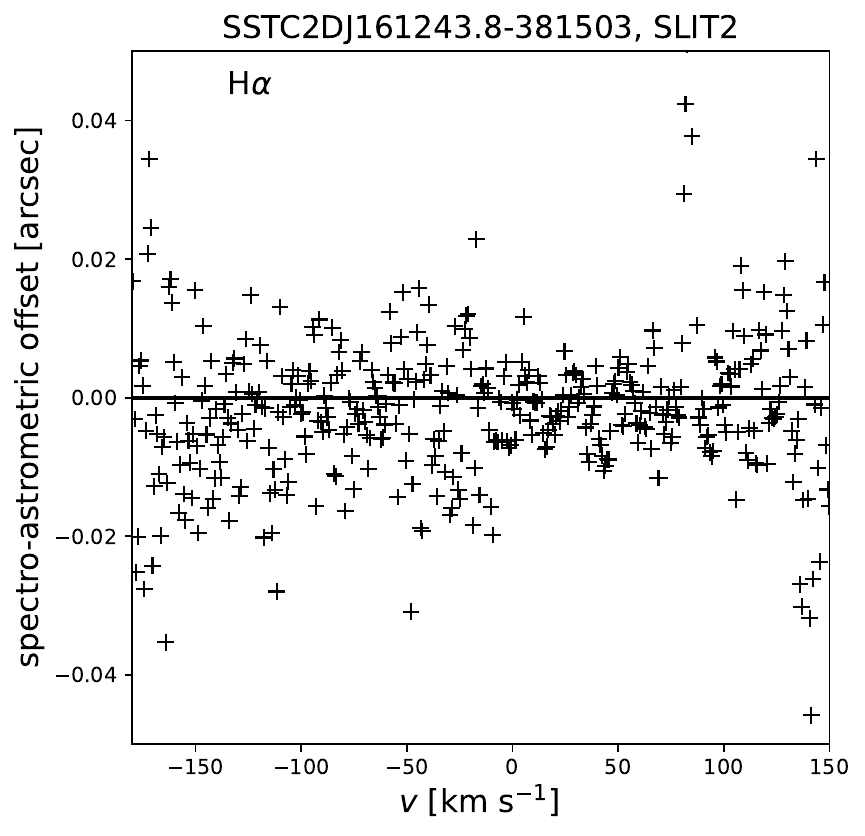}}
\hfill
\subfloat{\includegraphics[trim=0 0 0 0, clip, width=0.3 \textwidth]{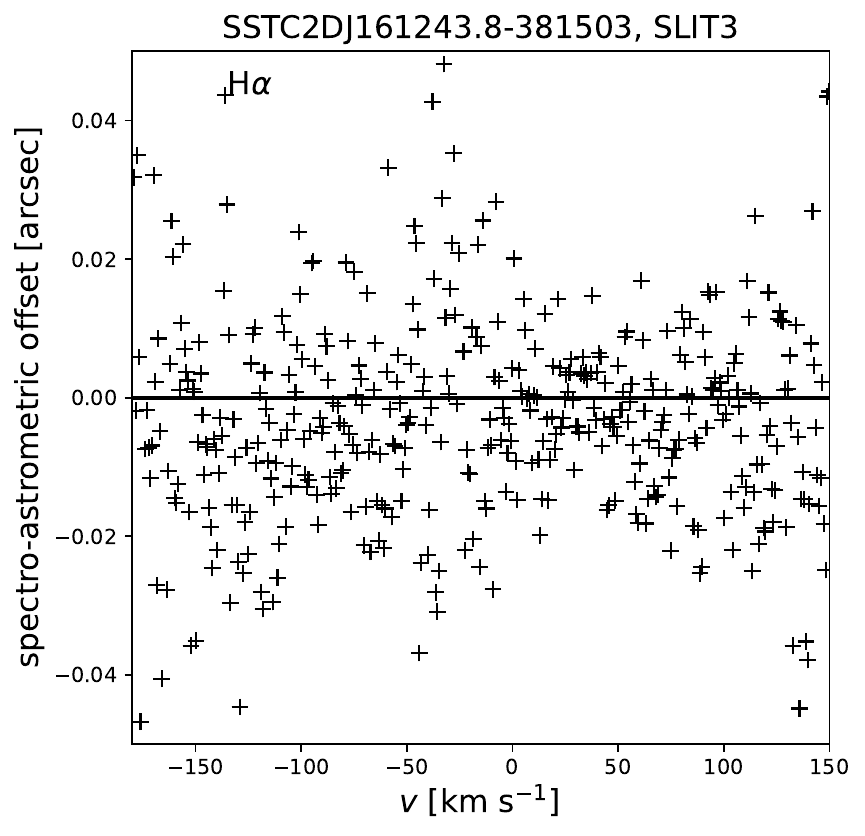}} 
\hfill
\subfloat{\includegraphics[trim=0 0 0 0, clip, width=0.3 \textwidth]{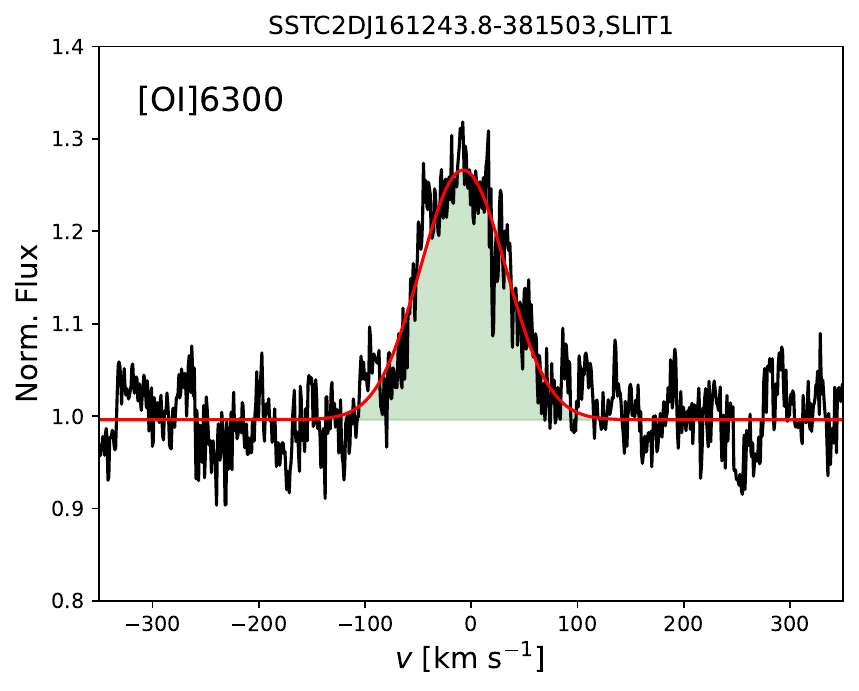}}
\hfill
\subfloat{\includegraphics[trim=0 0 0 0, clip, width=0.3 \textwidth]{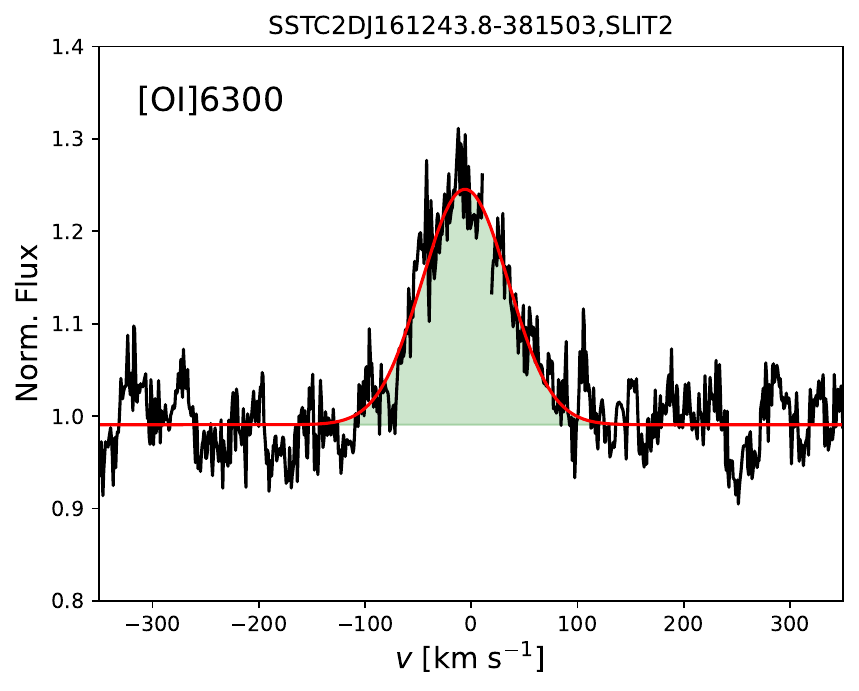}}
\hfill
\subfloat{\includegraphics[trim=0 0 0 0, clip, width=0.3 \textwidth]{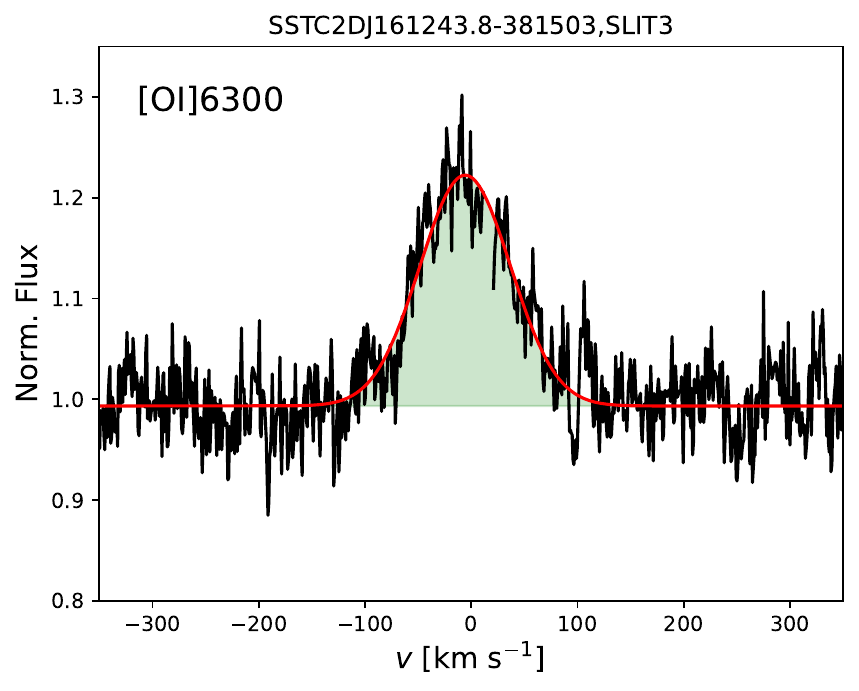}} 
\hfill  
\subfloat{\includegraphics[trim=0 0 0 0, clip, width=0.3 \textwidth]{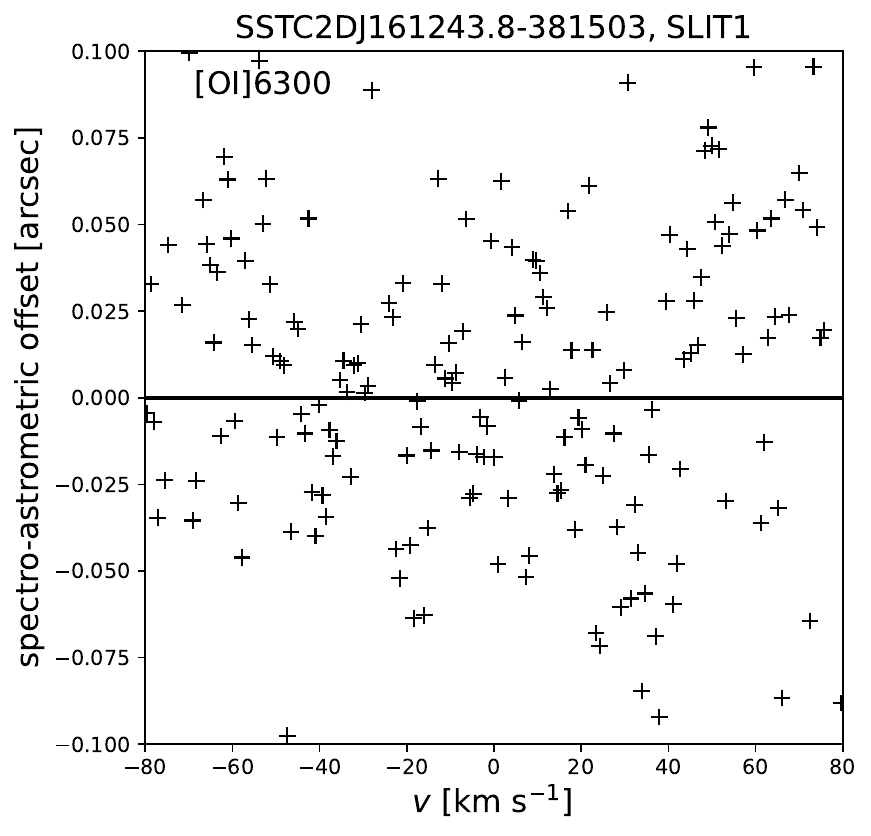}}
\hfill
\subfloat{\includegraphics[trim=0 0 0 0, clip, width=0.3 \textwidth]{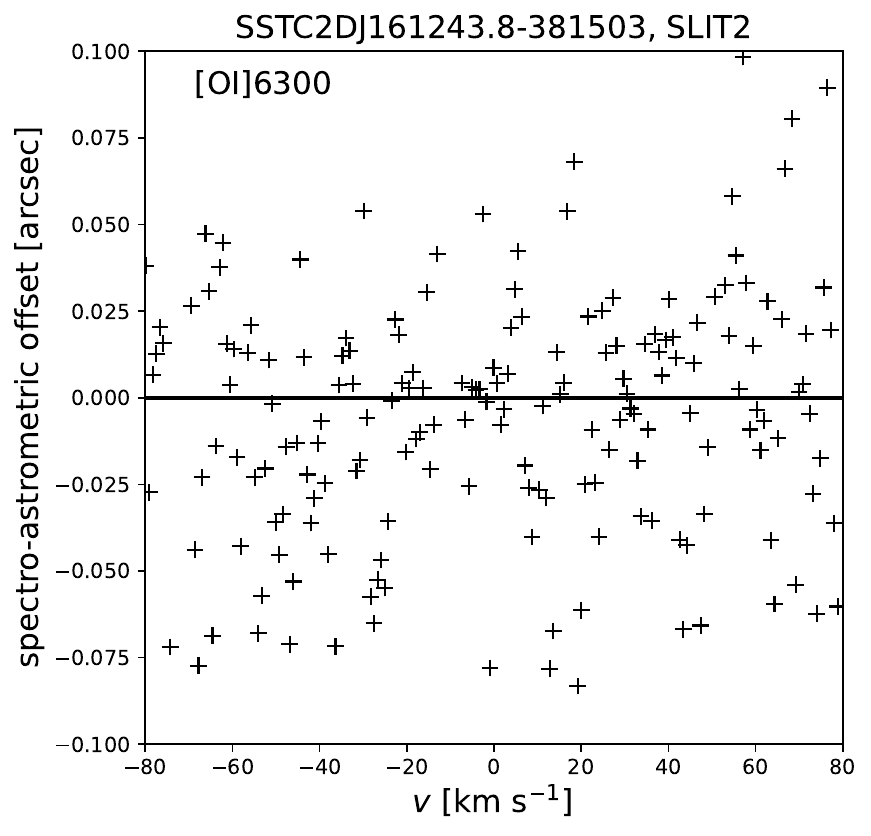}}
\hfill
\subfloat{\includegraphics[trim=0 0 0 0, clip, width=0.3 \textwidth]{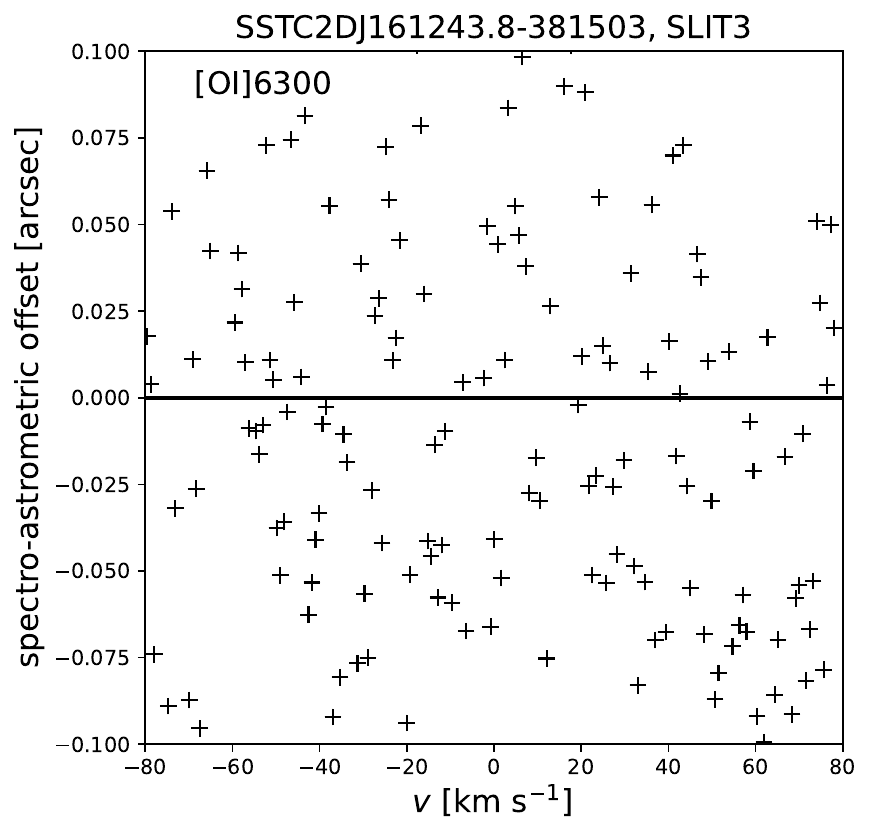}} 
\hfill
\caption{\small{Line profiles of H$\alpha$ and [OI]$\lambda$6300 for all slit positions of SSTC2DJ161243.8-381503.}}\label{fig:all_minispectra_SS503}
\end{figure*} 

\begin{figure*} 
\centering
\subfloat{\includegraphics[trim=0 0 0 0, clip, width=0.3 \textwidth]{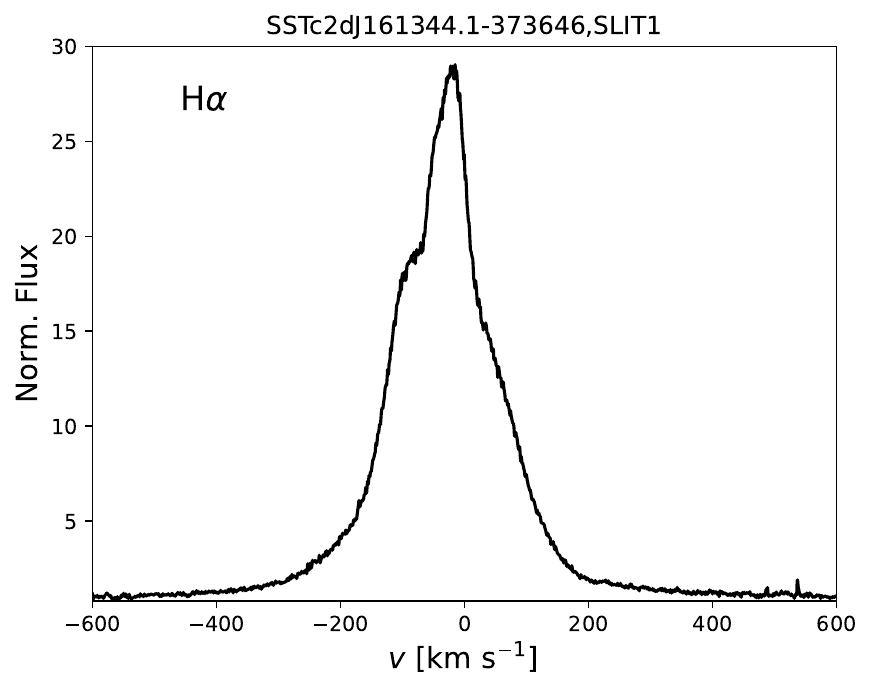}}
\hfill
\subfloat{\includegraphics[trim=0 0 0 0, clip, width=0.3 \textwidth]{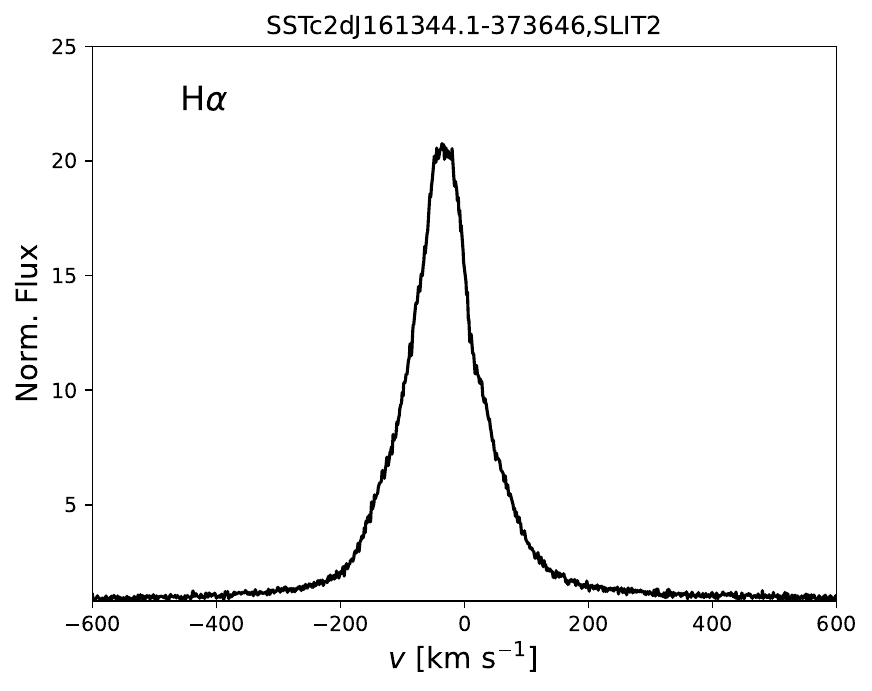}}
\hfill
\subfloat{\includegraphics[trim=0 0 0 0, clip, width=0.3 \textwidth]{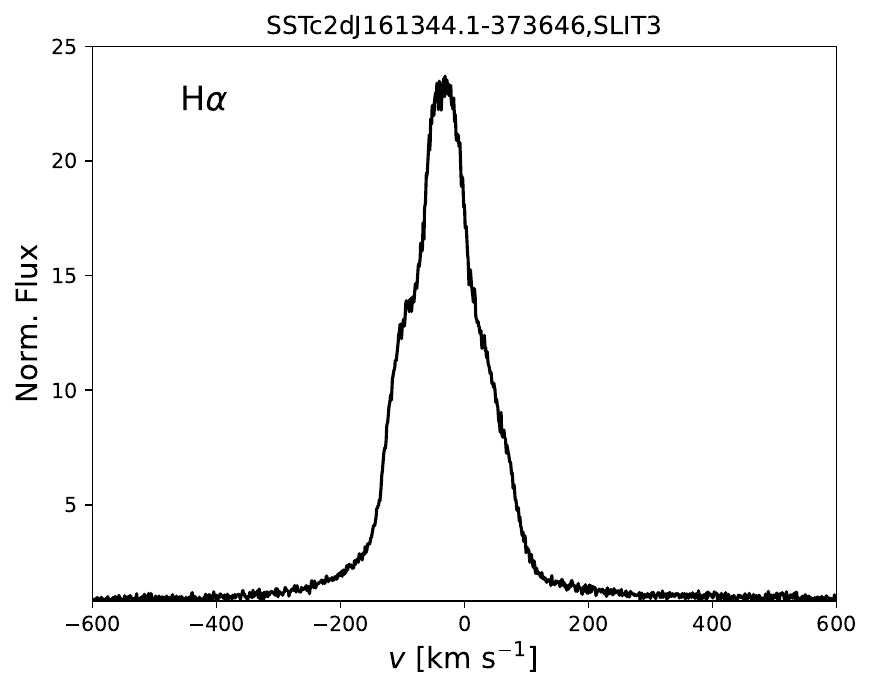}}
\hfill 
\subfloat{\includegraphics[trim=0 0 0 0, clip, width=0.3 \textwidth]{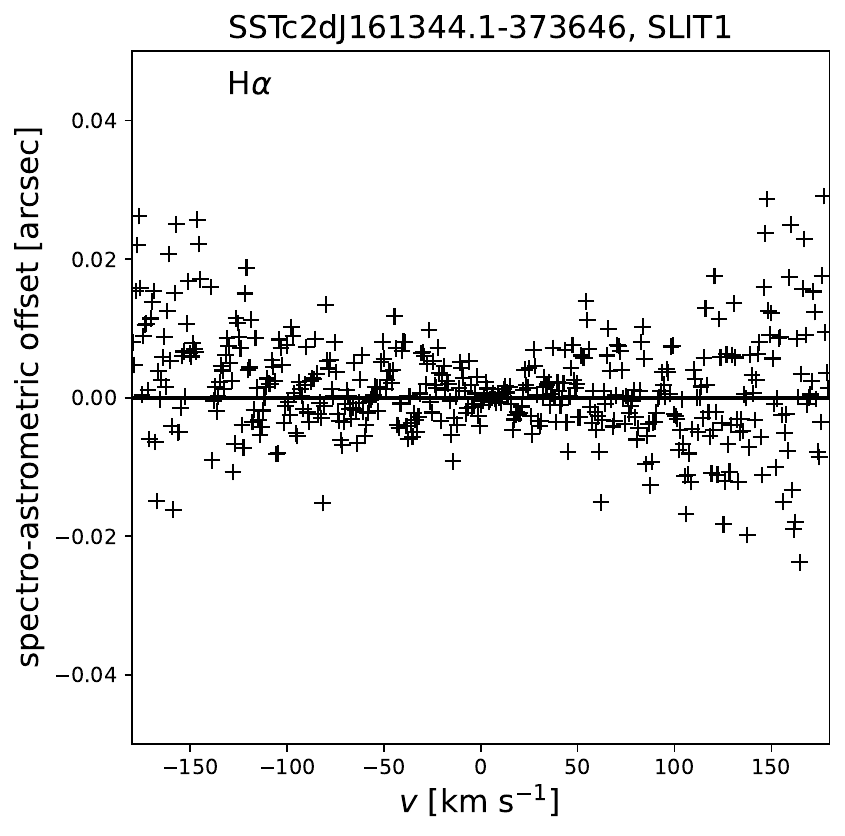}}
\hfill
\subfloat{\includegraphics[trim=0 0 0 0, clip, width=0.3 \textwidth]{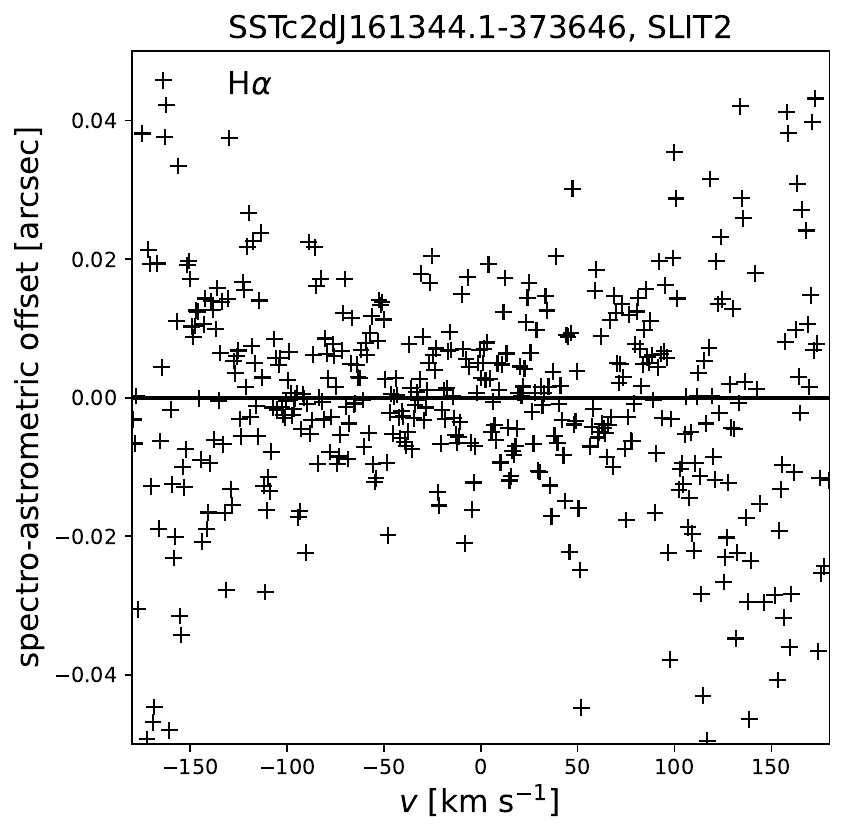}}
\hfill
\subfloat{\includegraphics[trim=0 0 0 0, clip, width=0.3 \textwidth]{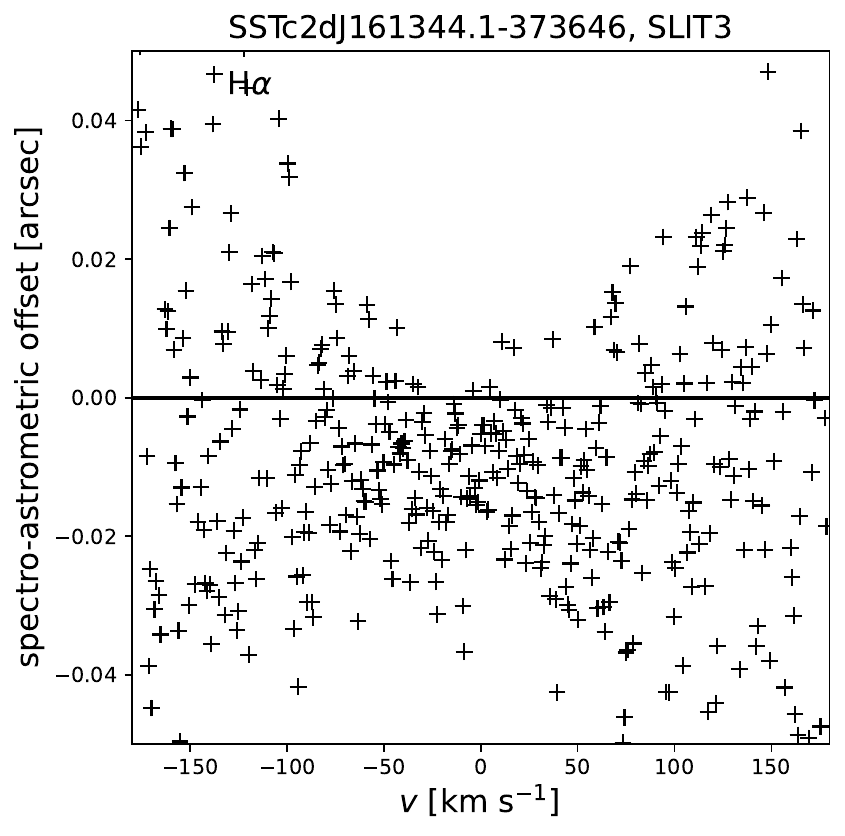}} 
\hfill
\subfloat{\includegraphics[trim=0 0 0 0, clip, width=0.3 \textwidth]{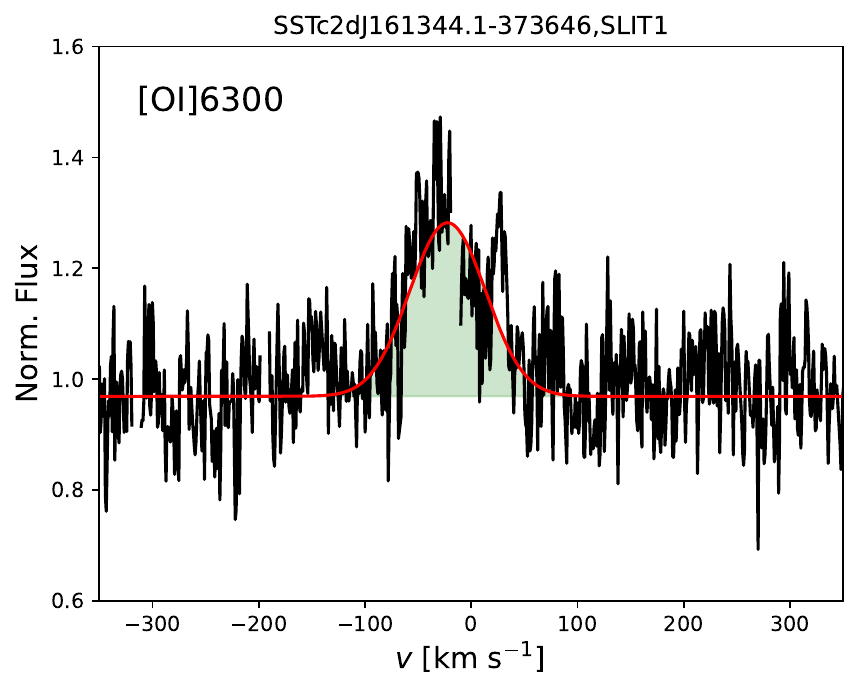}}
\hfill
\subfloat{\includegraphics[trim=0 0 0 0, clip, width=0.3 \textwidth]{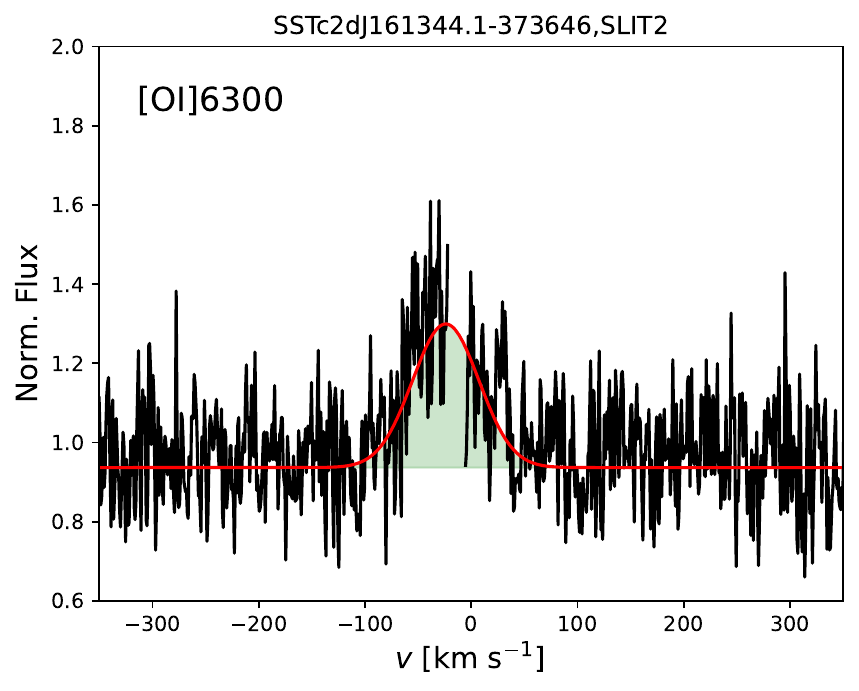}}
\hfill
\subfloat{\includegraphics[trim=0 0 0 0, clip, width=0.3 \textwidth]{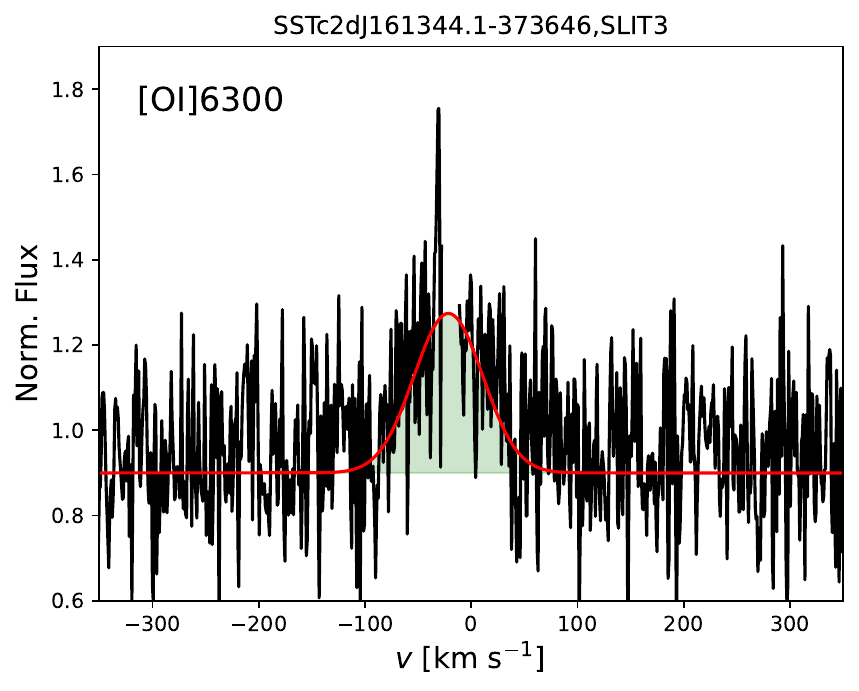}} 
\hfill  
\subfloat{\includegraphics[trim=0 0 0 0, clip, width=0.3 \textwidth]{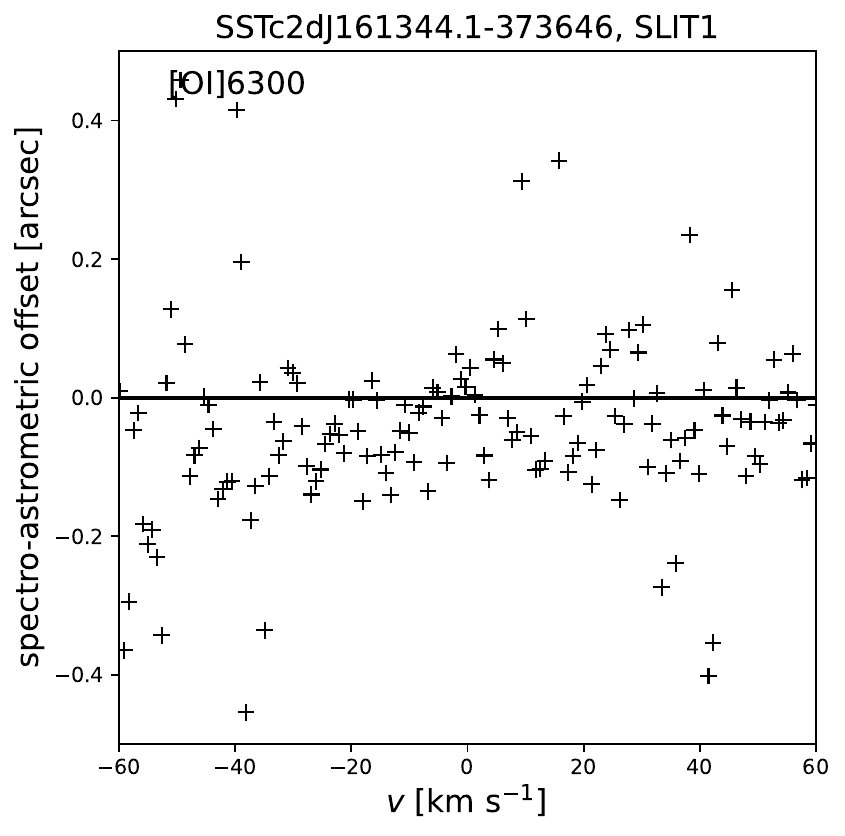}}
\hfill
\subfloat{\includegraphics[trim=0 0 0 0, clip, width=0.3 \textwidth]{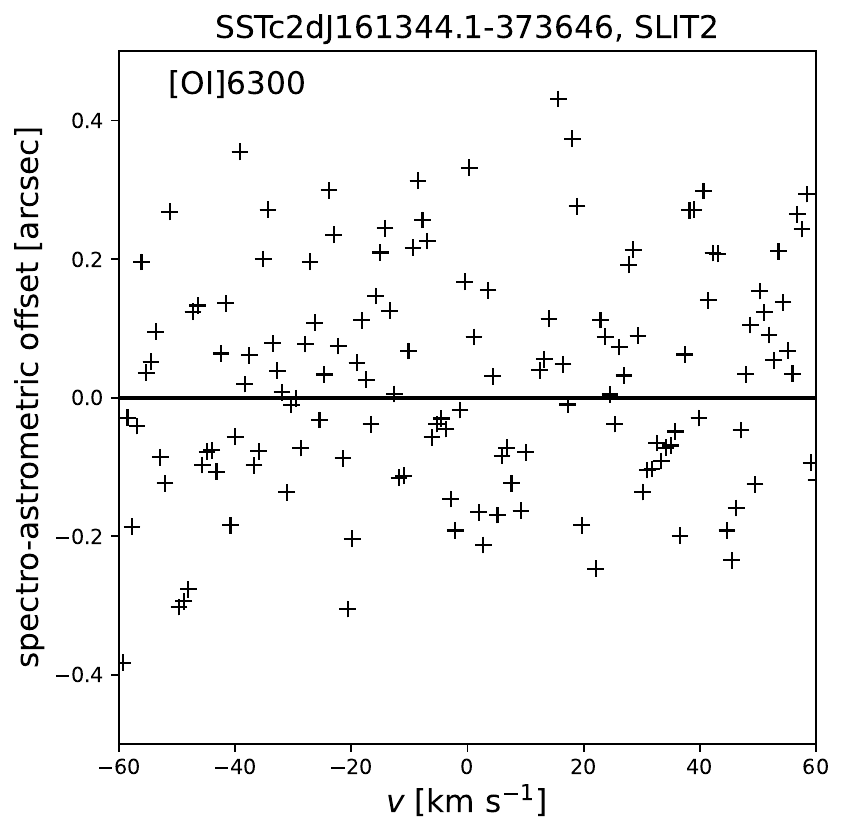}}
\hfill
\subfloat{\includegraphics[trim=0 0 0 0, clip, width=0.3 \textwidth]{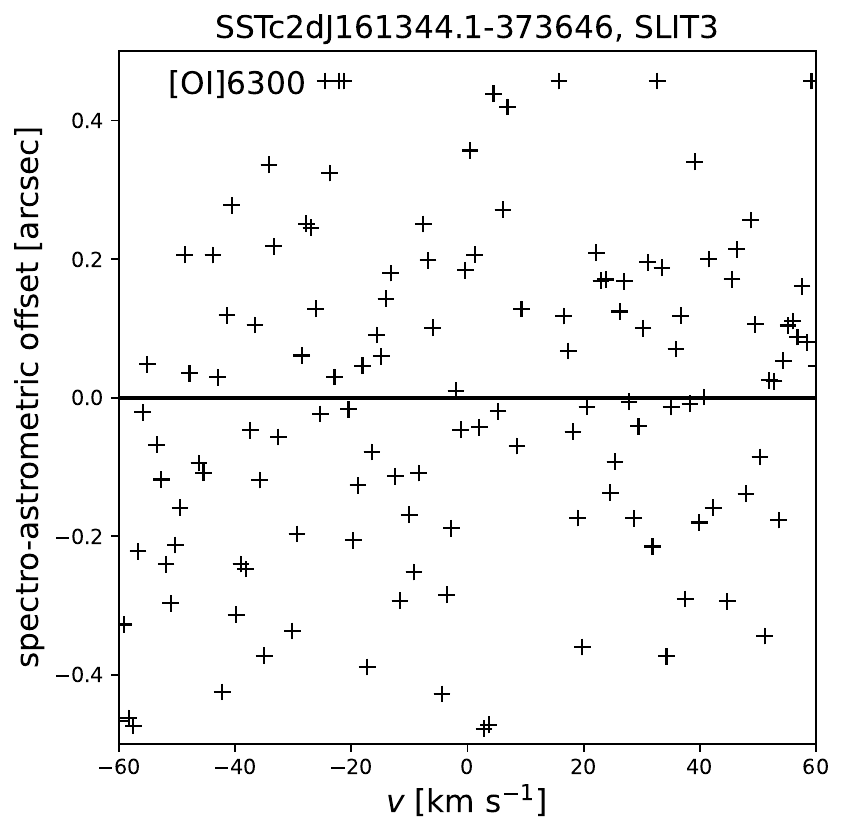}} 
\hfill
\caption{\small{Line profiles of H$\alpha$ and [OI]$\lambda$6300 for all slit positions of SSTc2dJ161344.1-373646.}}\label{fig:all_minispectra_ss344}
\end{figure*} 

\begin{figure*} 
\centering
\subfloat{\includegraphics[trim=0 0 0 0, clip, width=0.3 \textwidth]{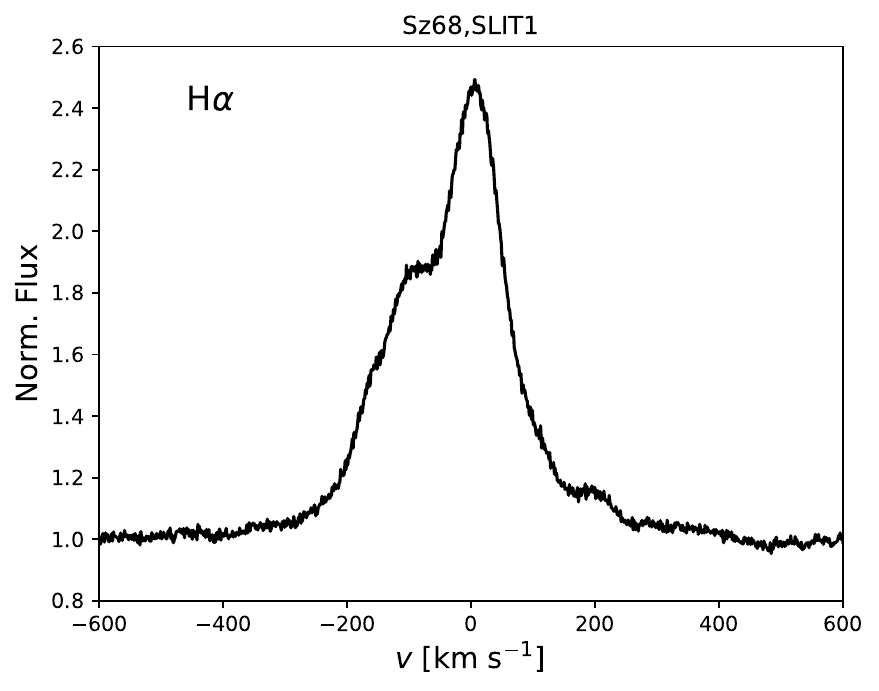}}
\hfill
\subfloat{\includegraphics[trim=0 0 0 0, clip, width=0.3 \textwidth]{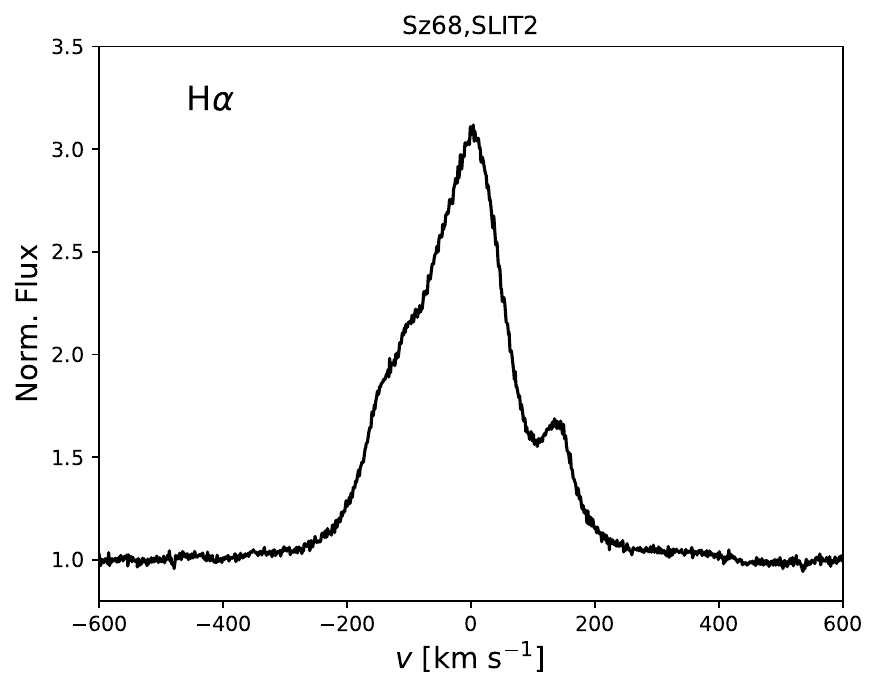}}
\hfill
\subfloat{\includegraphics[trim=0 0 0 0, clip, width=0.3 \textwidth]{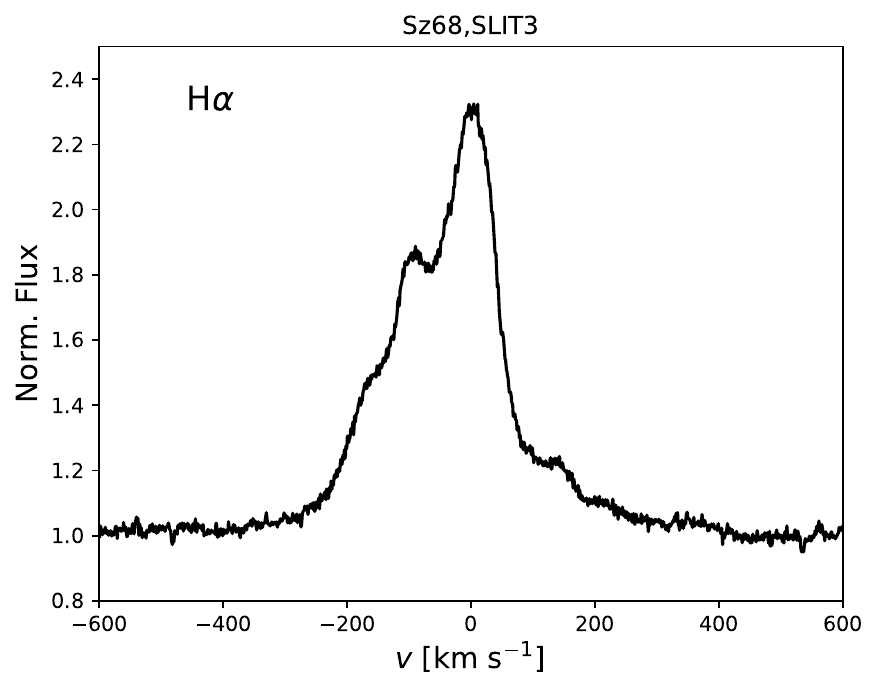}}
\hfill
\subfloat{\includegraphics[trim=0 0 0 0, clip, width=0.3 \textwidth]{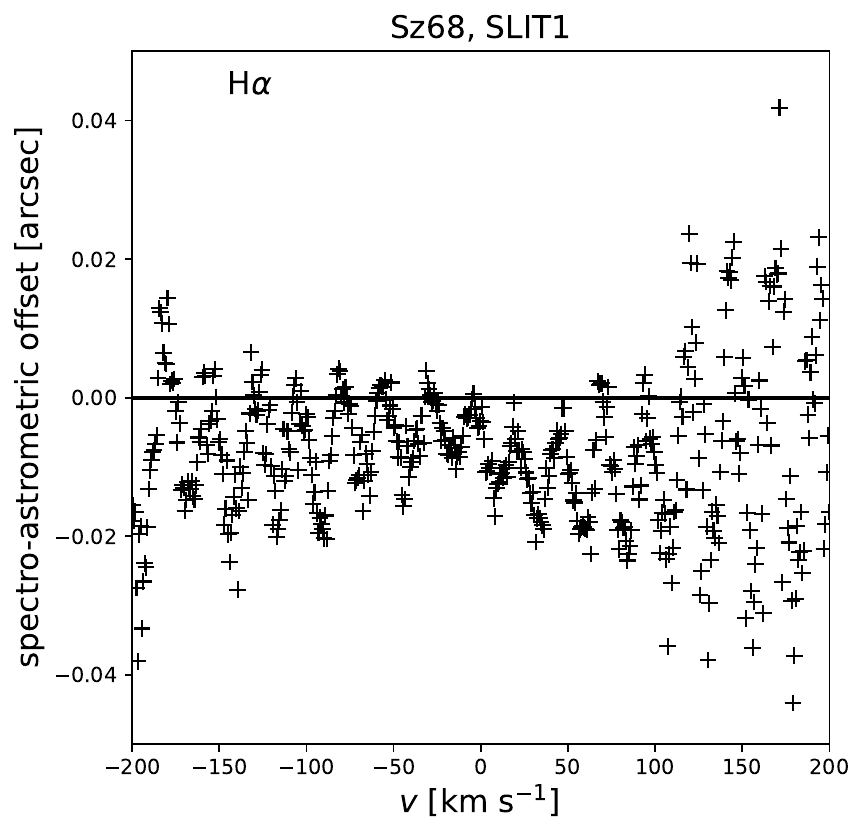}}
\hfill
\subfloat{\includegraphics[trim=0 0 0 0, clip, width=0.3 \textwidth]{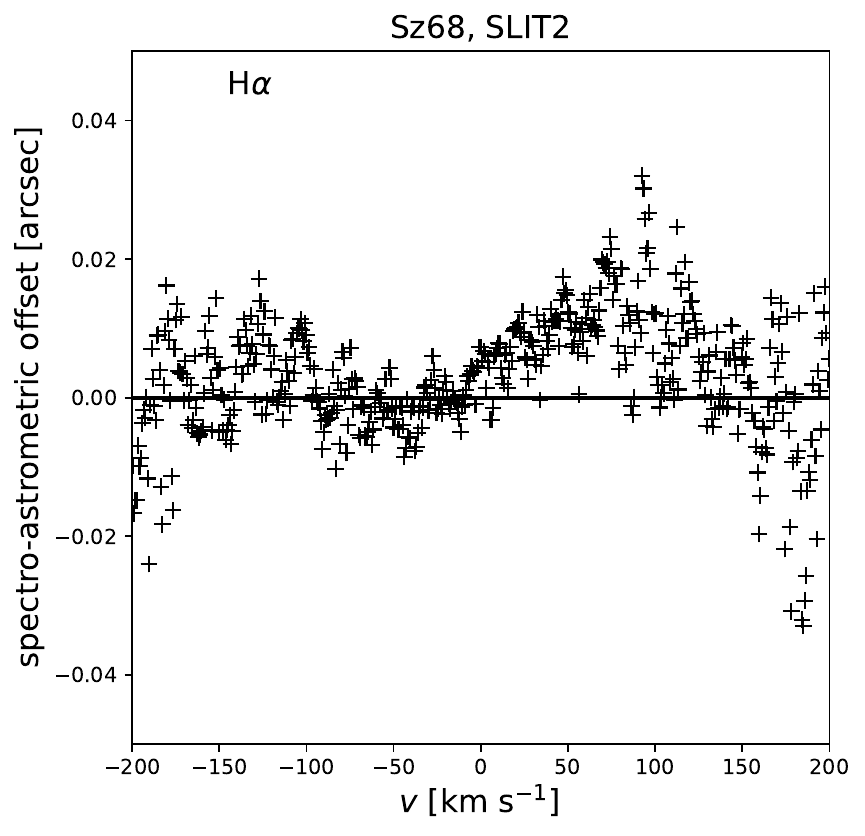}}
\hfill
\subfloat{\includegraphics[trim=0 0 0 0, clip, width=0.3 \textwidth]{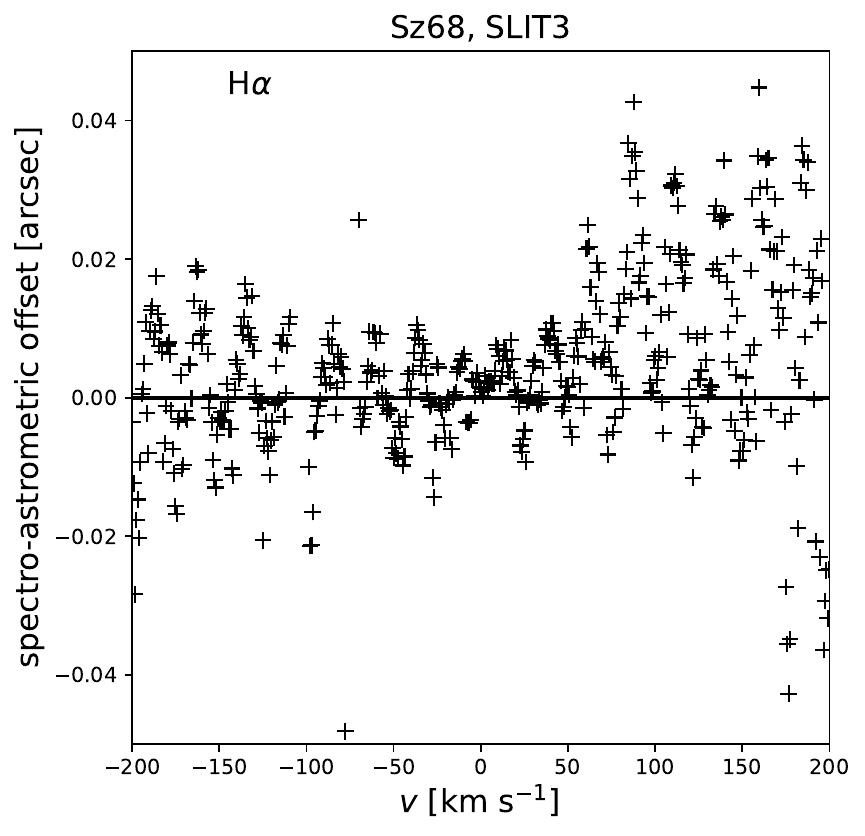}} 
\hfill
\subfloat{\includegraphics[trim=0 0 0 0, clip, width=0.3 \textwidth]{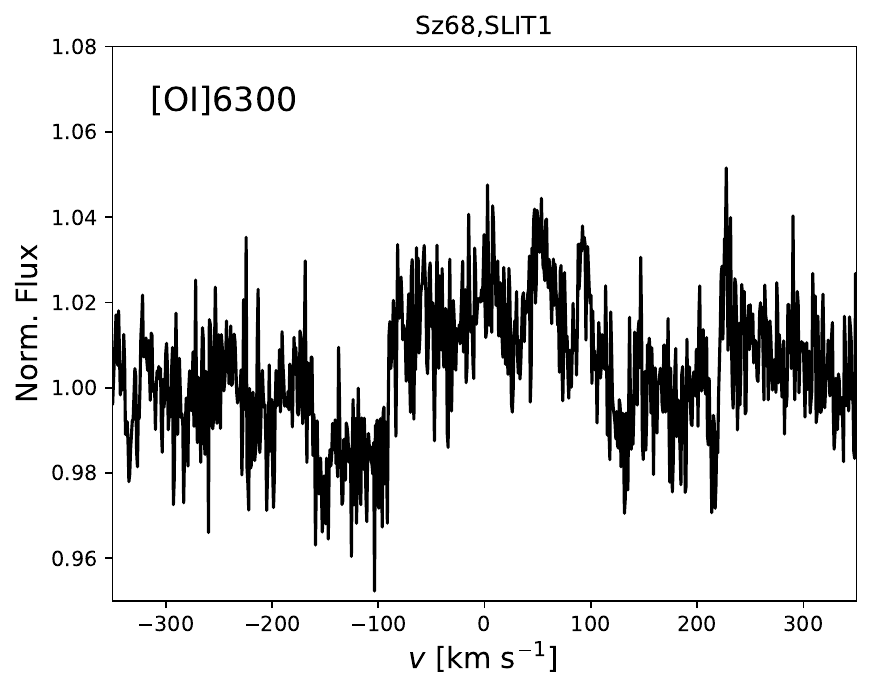}}
\hfill
\subfloat{\includegraphics[trim=0 0 0 0, clip, width=0.3 \textwidth]{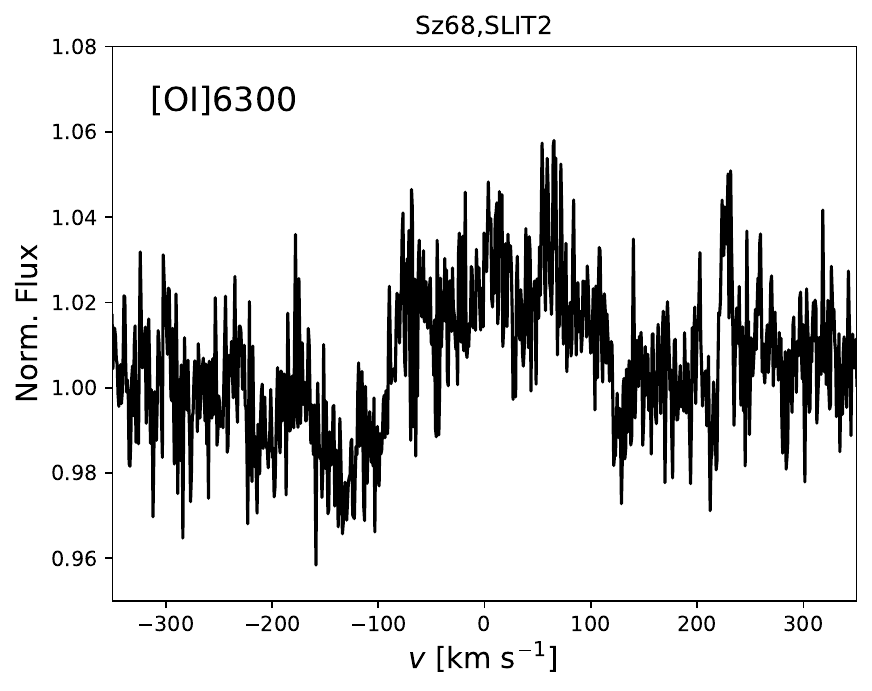}}
\hfill
\subfloat{\includegraphics[trim=0 0 0 0, clip, width=0.3 \textwidth]{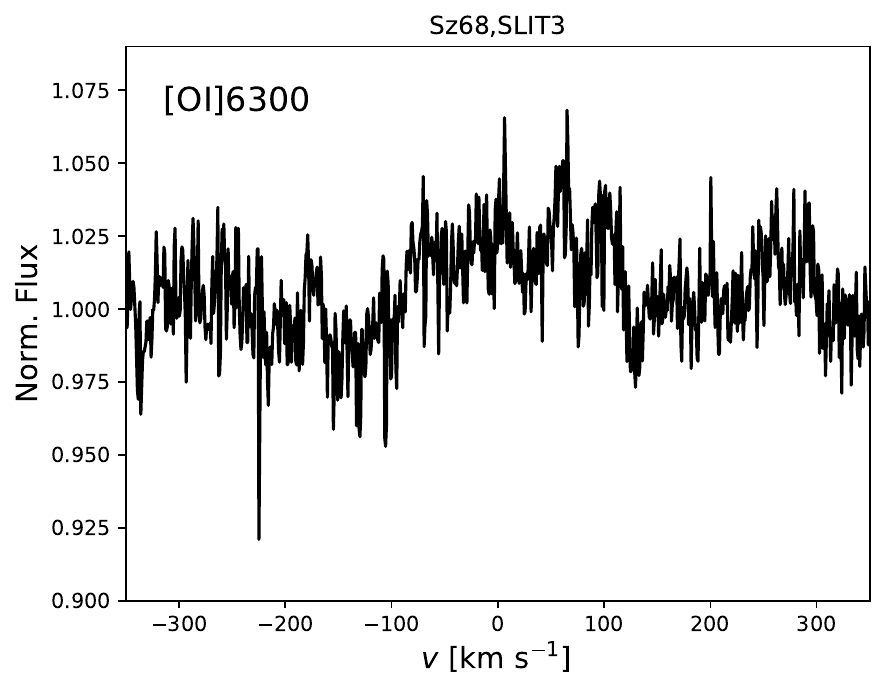}} 
\hfill  
\caption{\small{Line profiles of H$\alpha$ and [OI]$\lambda$6300 for all slit positions of Sz\,68.}}\label{fig:all_minispectra_Sz68}
\end{figure*} 

\begin{figure*} 
\centering
\subfloat{\includegraphics[trim=0 0 0 0, clip, width=0.3 \textwidth]{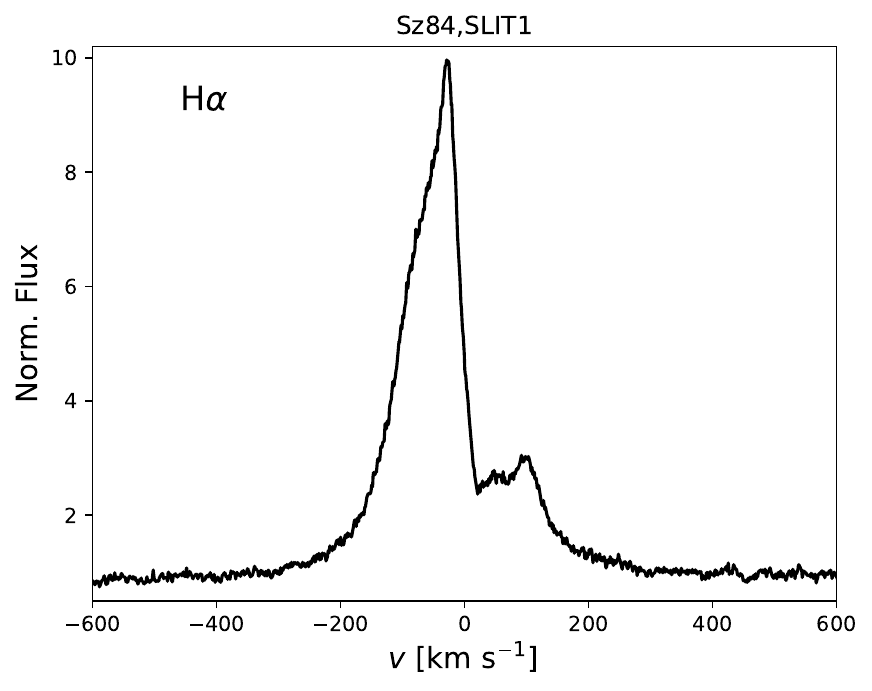}}
\hfill
\subfloat{\includegraphics[trim=0 0 0 0, clip, width=0.3 \textwidth]{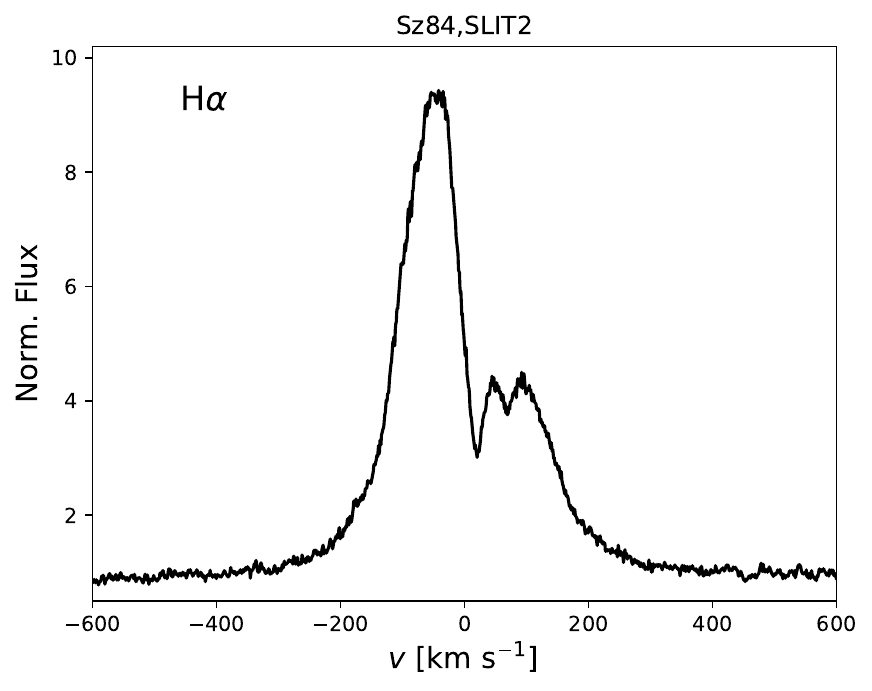}}
\hfill
\subfloat{\includegraphics[trim=0 0 0 0, clip, width=0.3 \textwidth]{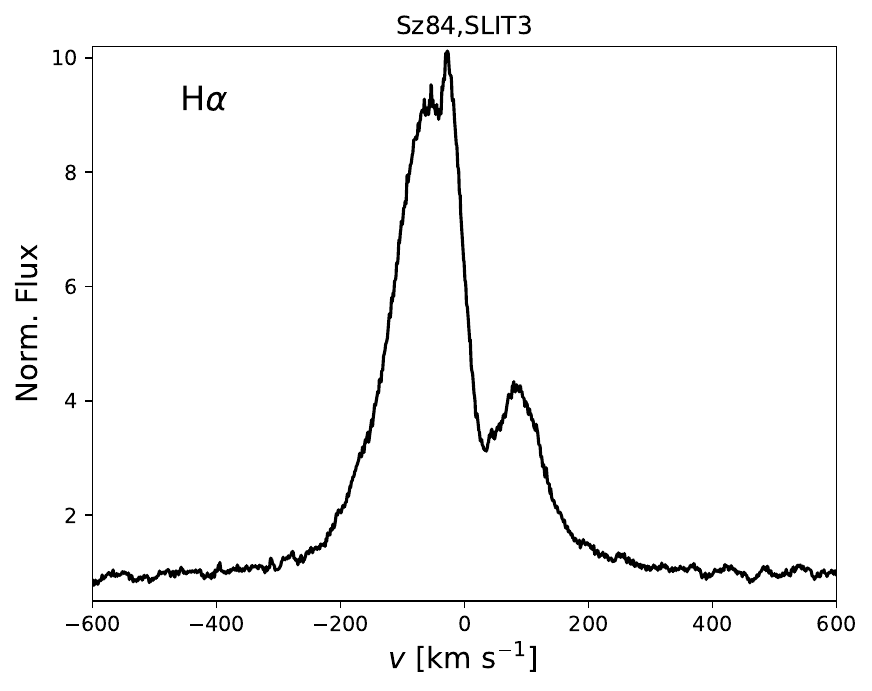}} 
\hfill
\subfloat{\includegraphics[trim=0 0 0 0, clip, width=0.3 \textwidth]{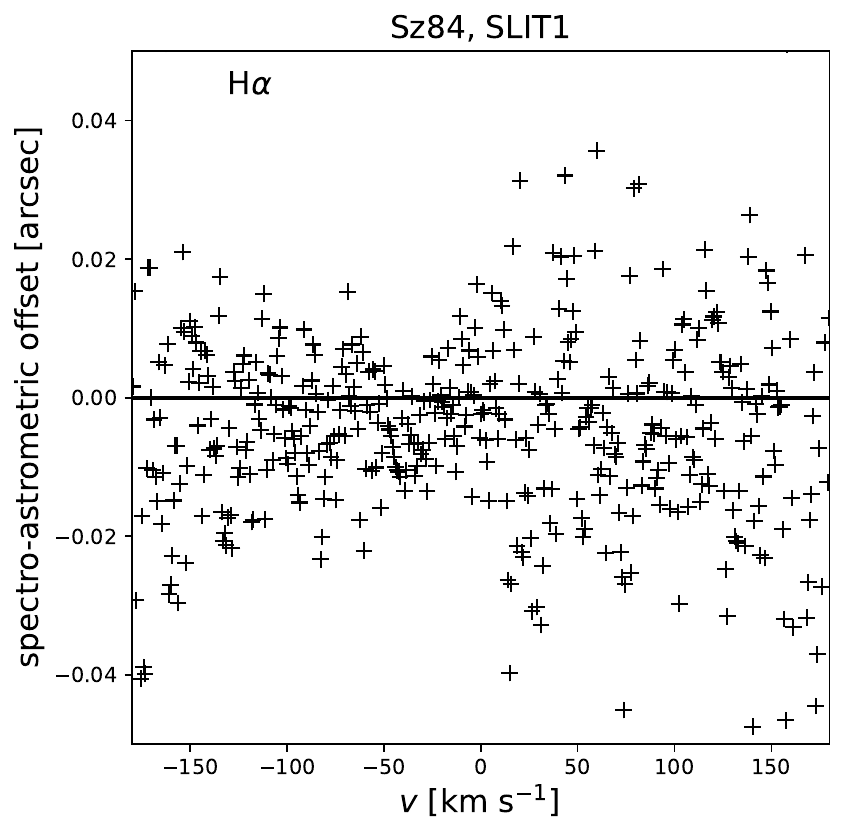}}
\hfill
\subfloat{\includegraphics[trim=0 0 0 0, clip, width=0.3 \textwidth]{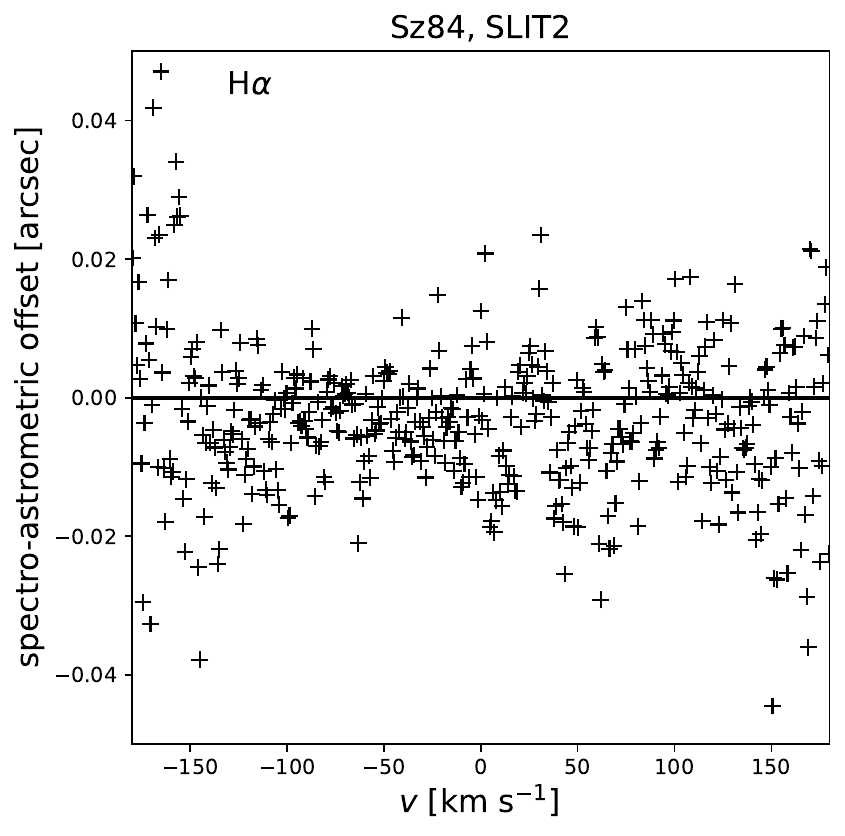}}
\hfill
\subfloat{\includegraphics[trim=0 0 0 0, clip, width=0.3 \textwidth]{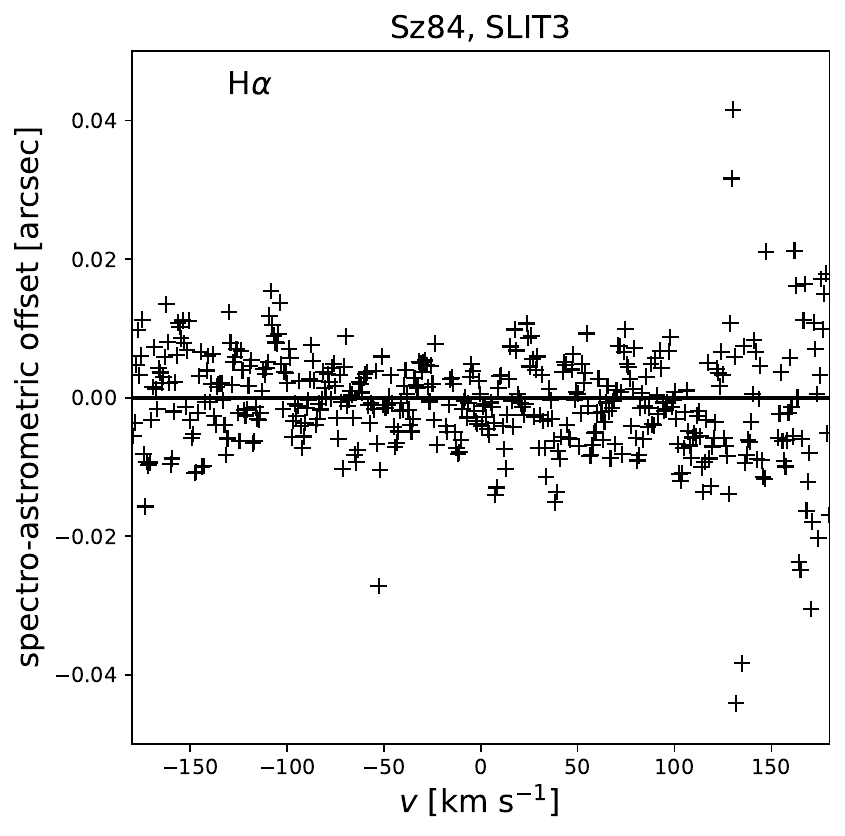}} 
\hfill
 \subfloat{\includegraphics[trim=0 0 0 0, clip, width=0.3 \textwidth]{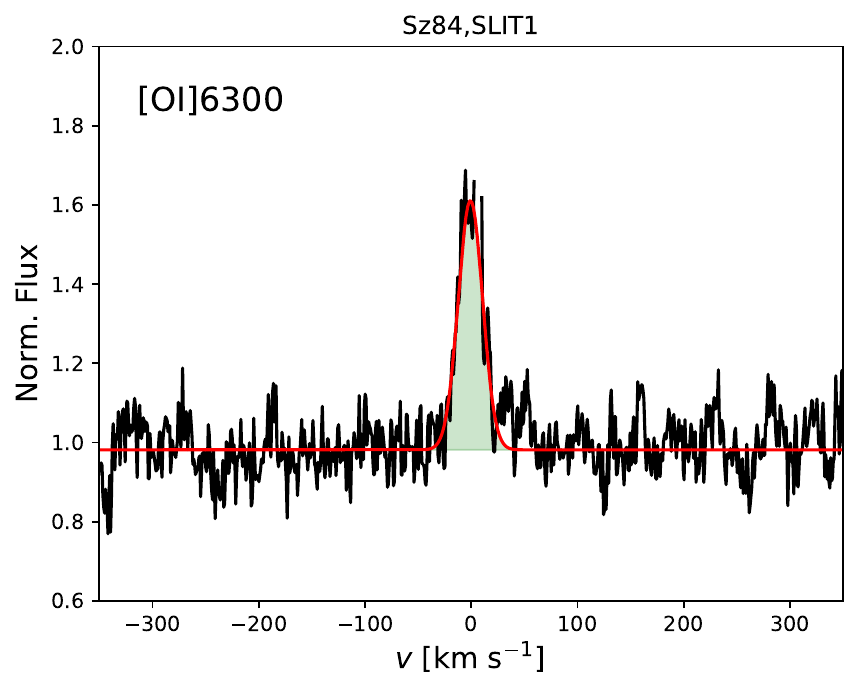}}
\hfill
\subfloat{\includegraphics[trim=0 0 0 0, clip, width=0.3 \textwidth]{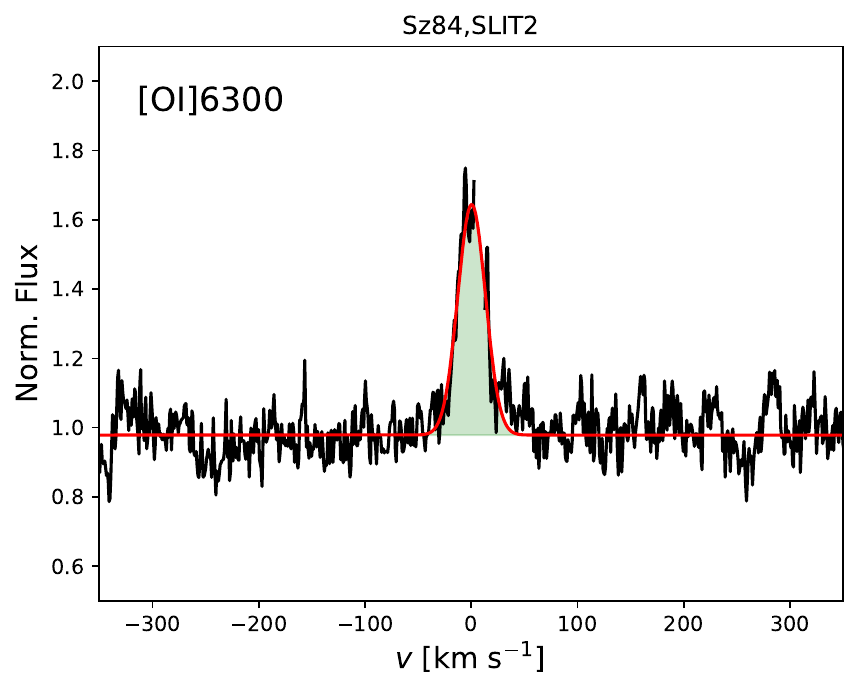}}
\hfill
\subfloat{\includegraphics[trim=0 0 0 0, clip, width=0.3 \textwidth]{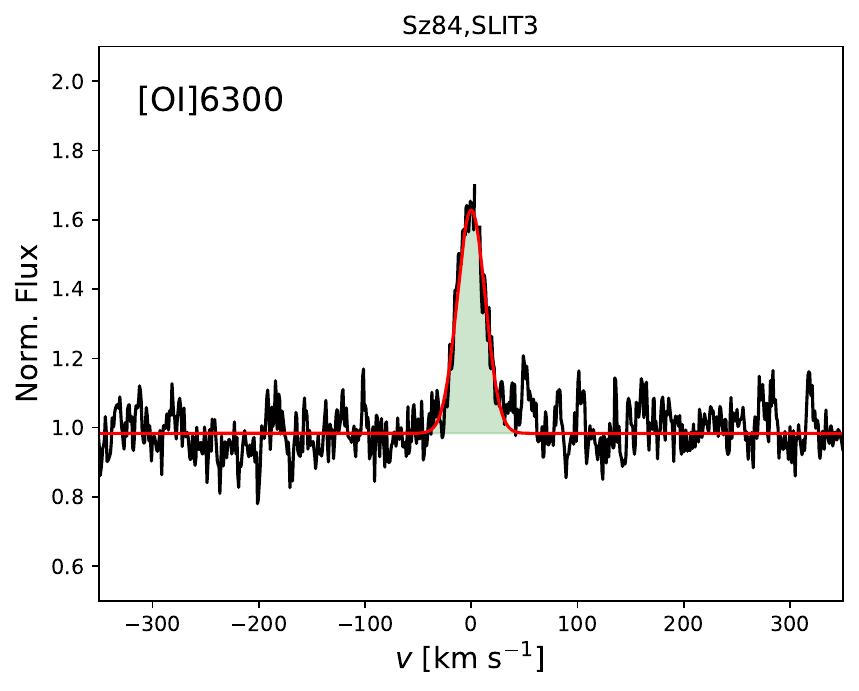}} 
\hfill  
\subfloat{\includegraphics[trim=0 0 0 0, clip, width=0.3 \textwidth]{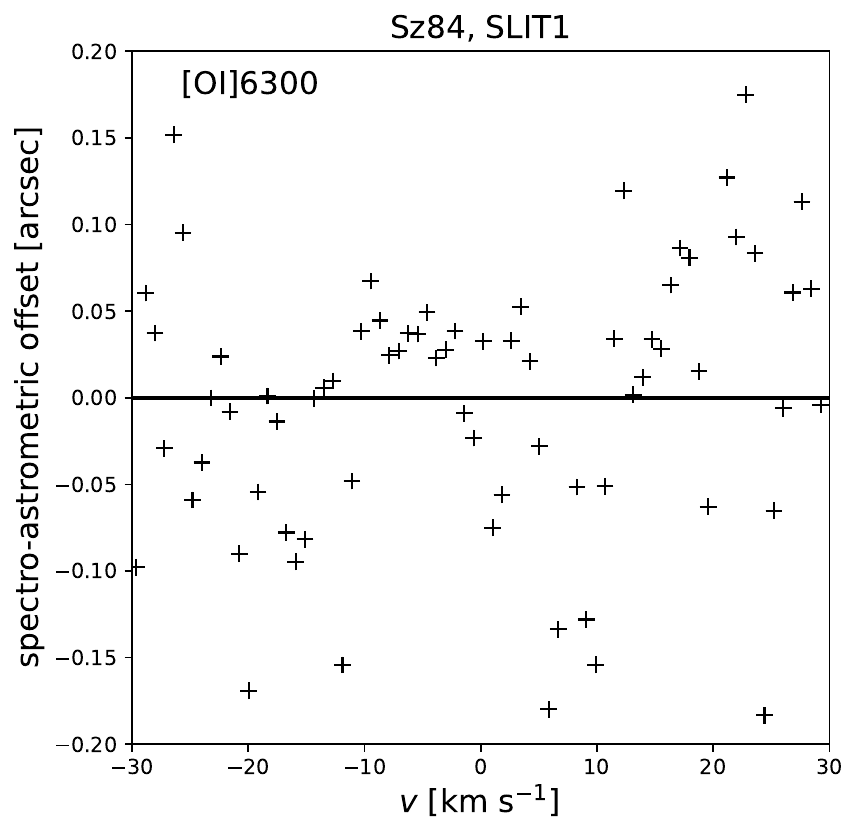}}
\hfill
\subfloat{\includegraphics[trim=0 0 0 0, clip, width=0.3 \textwidth]{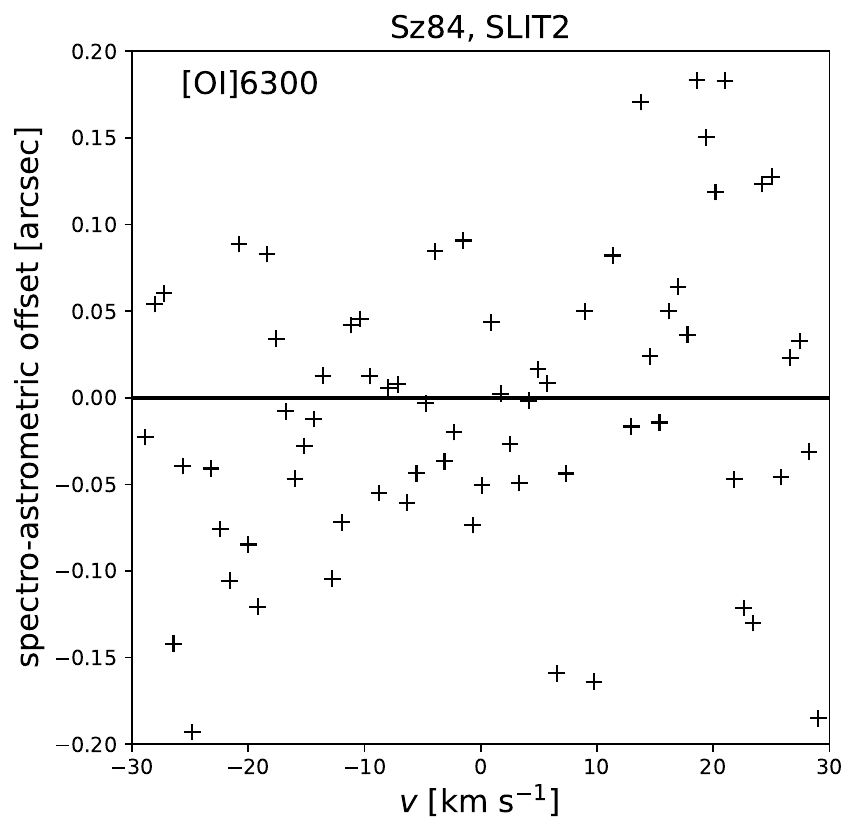}}
\hfill
\subfloat{\includegraphics[trim=0 0 0 0, clip, width=0.3 \textwidth]{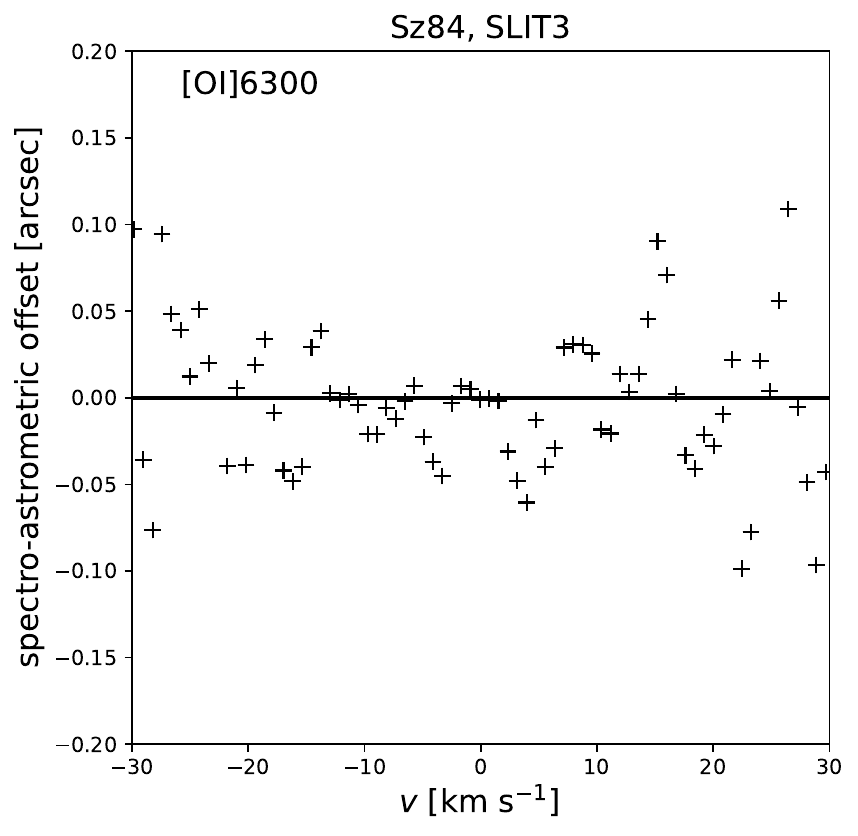}} 
\hfill
\caption{\small{Line profiles of H$\alpha$ and [OI]$\lambda$6300 for all slit positions of Sz\,84.}}\label{fig:all_minispectra_Sz84}
\end{figure*}

\begin{figure*} 
\centering
\subfloat{\includegraphics[trim=0 0 0 0, clip, width=0.3 \textwidth]{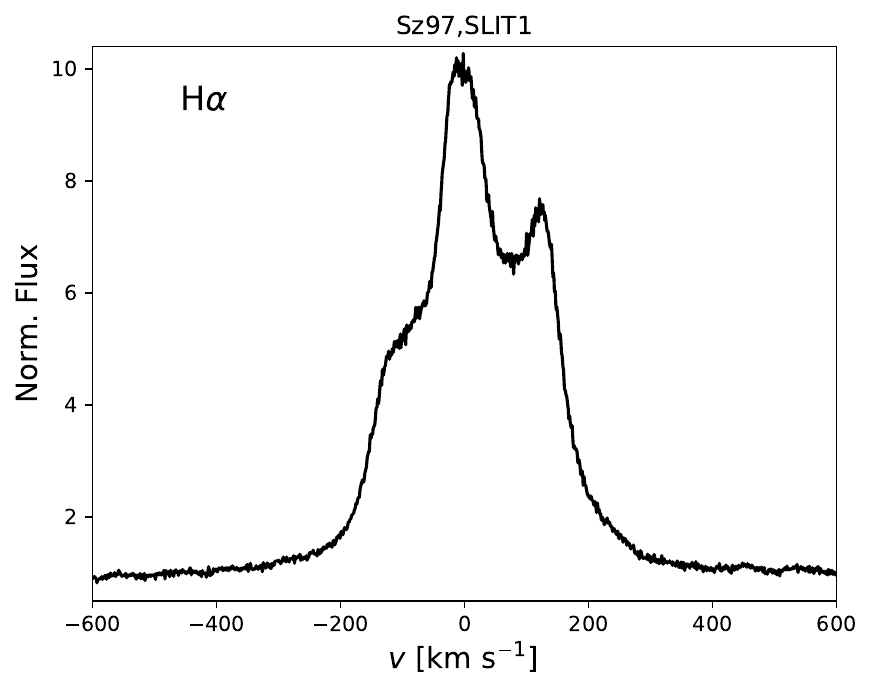}}
\hfill
\subfloat{\includegraphics[trim=0 0 0 0, clip, width=0.3 \textwidth]{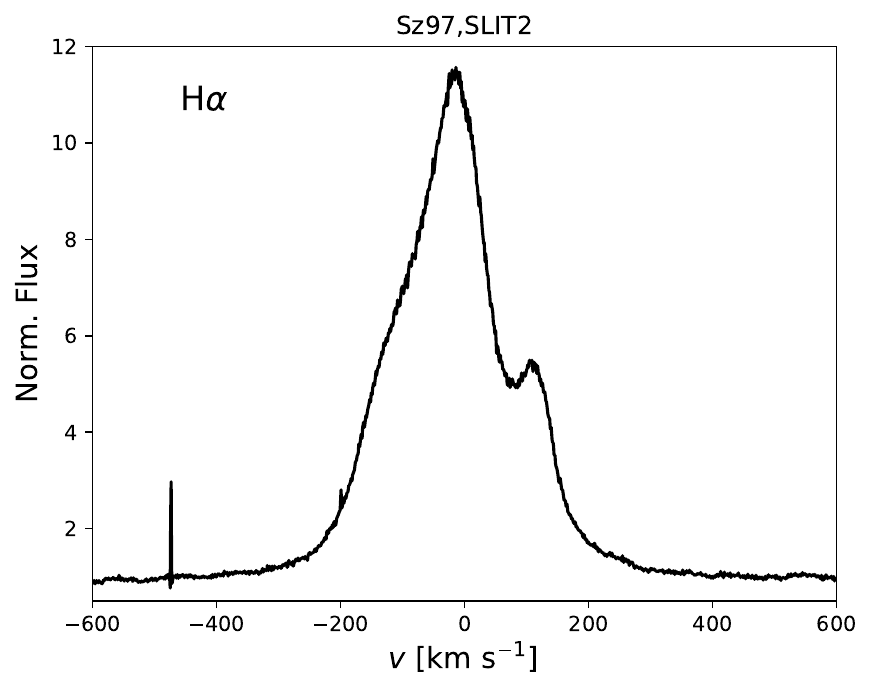}}
\hfill
\subfloat{\includegraphics[trim=0 0 0 0, clip, width=0.3 \textwidth]{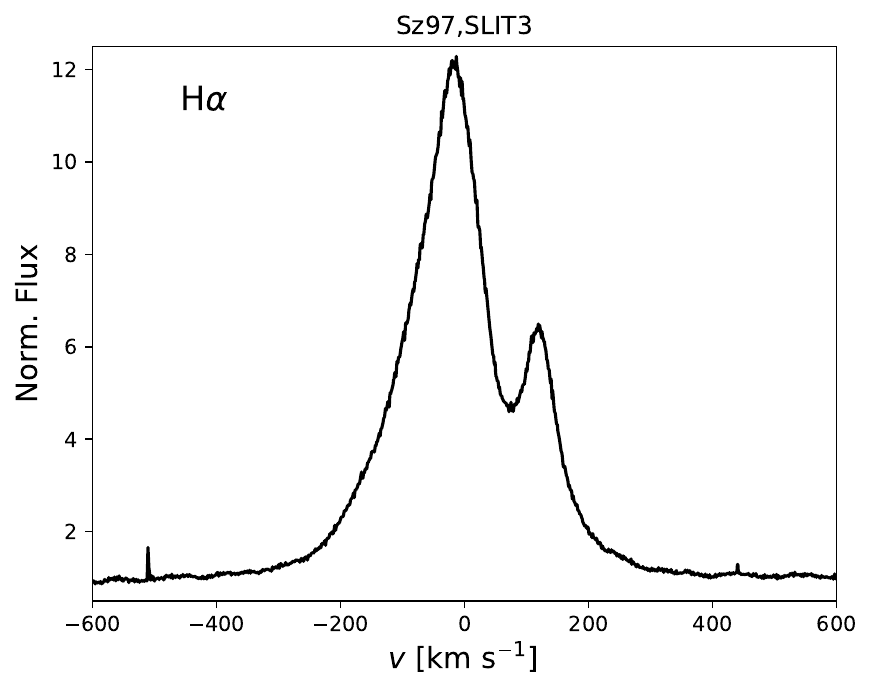}} 
\hfill
\subfloat{\includegraphics[trim=0 0 0 0, clip, width=0.3 \textwidth]{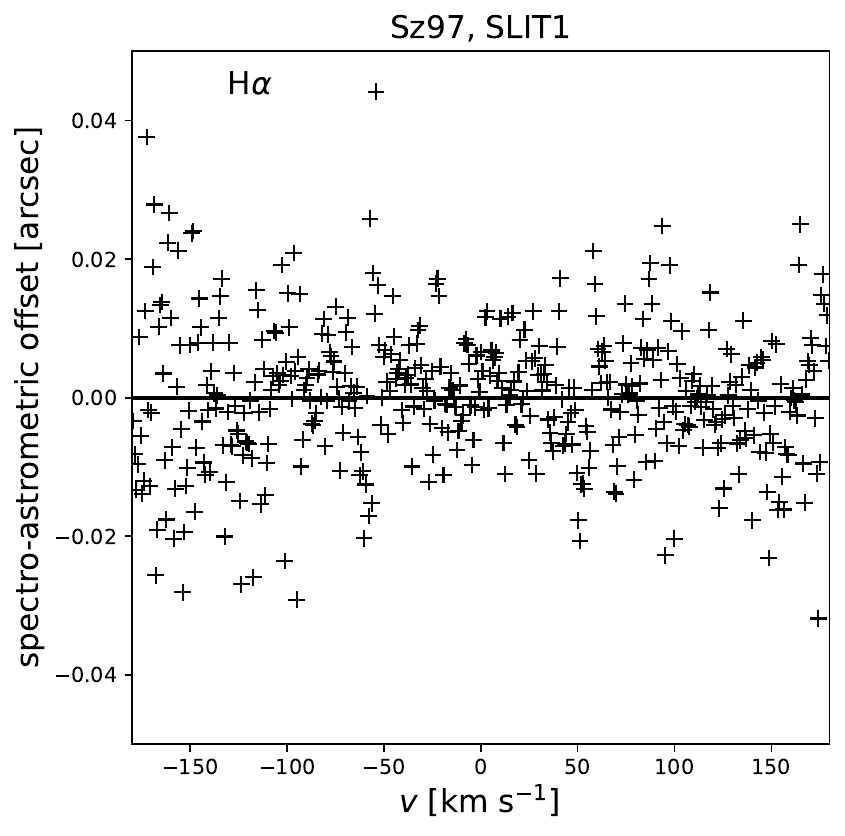}}
\hfill
\subfloat{\includegraphics[trim=0 0 0 0, clip, width=0.3 \textwidth]{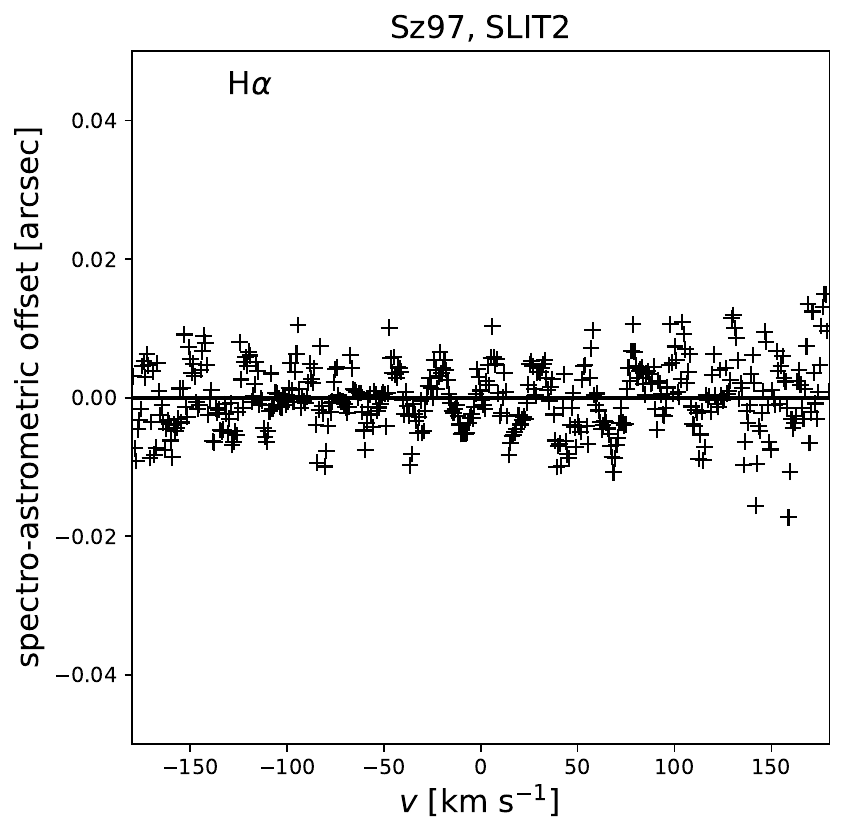}}
\hfill
\subfloat{\includegraphics[trim=0 0 0 0, clip, width=0.3 \textwidth]{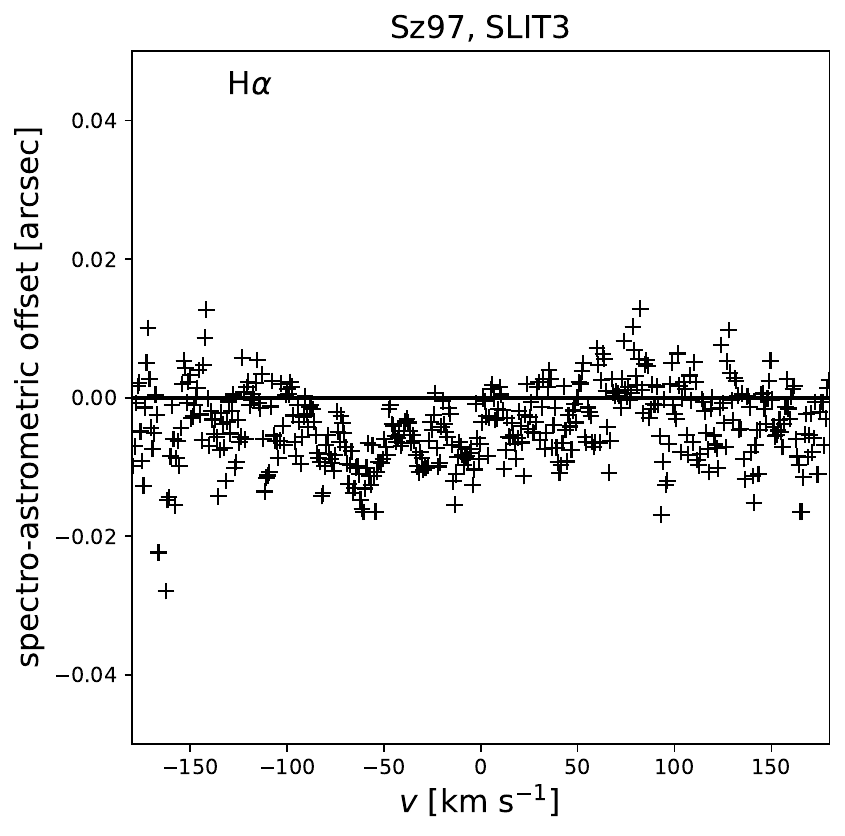}} 
\hfill
\subfloat{\includegraphics[trim=0 0 0 0, clip, width=0.3 \textwidth]{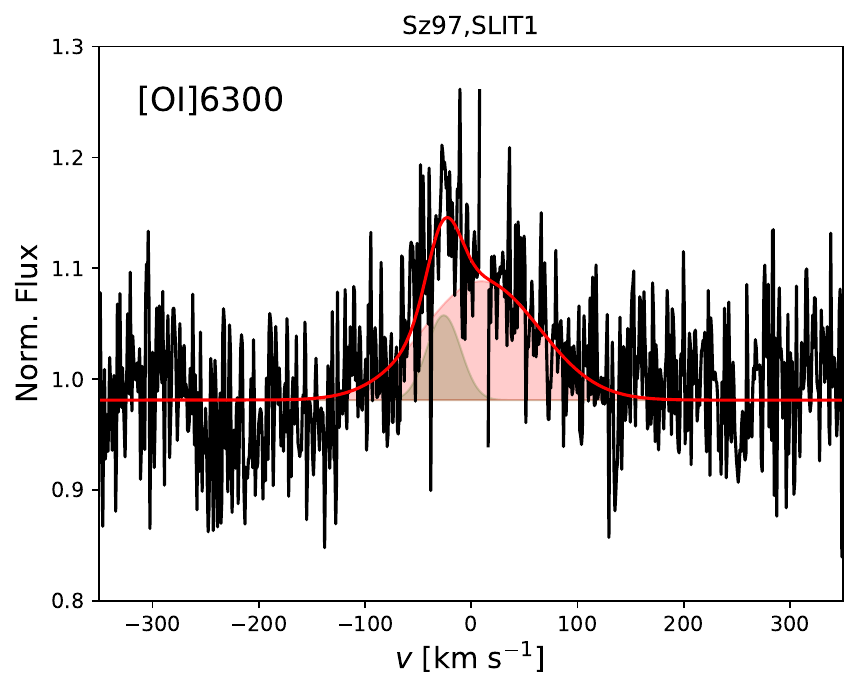}}
\hfill
\subfloat{\includegraphics[trim=0 0 0 0, clip, width=0.3 \textwidth]{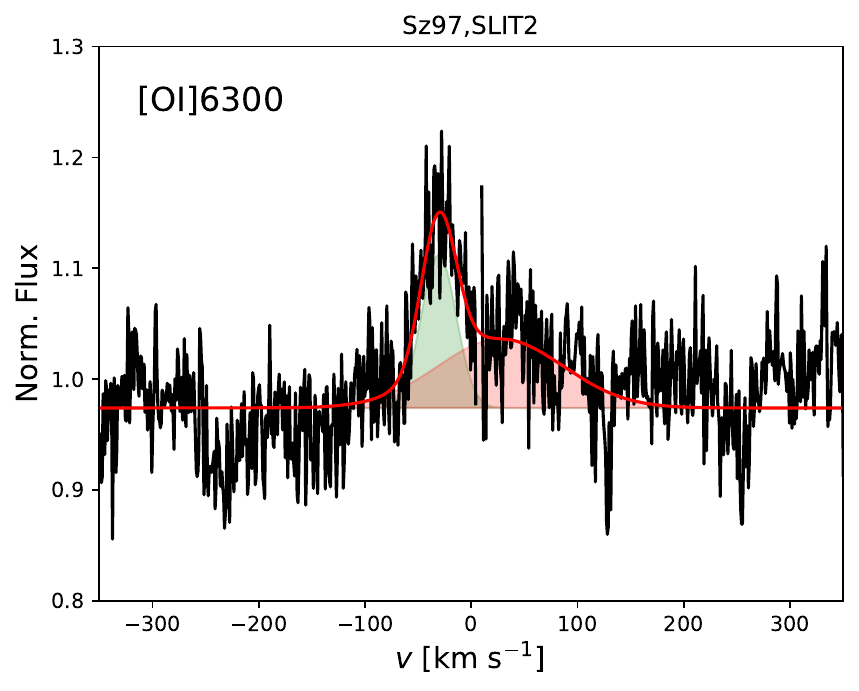}}
\hfill
\subfloat{\includegraphics[trim=0 0 0 0, clip, width=0.3 \textwidth]{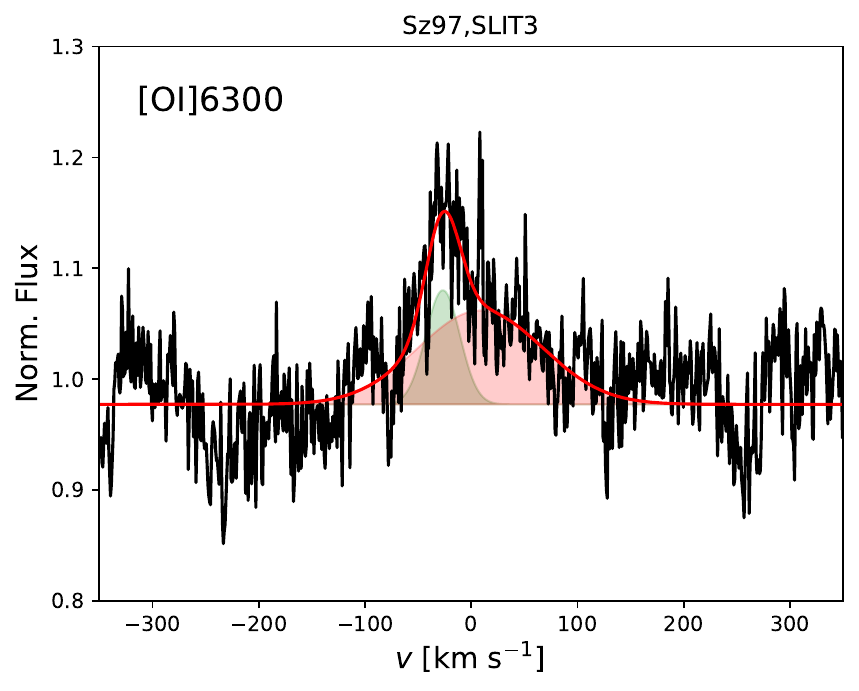}} 
\hfill 
\subfloat{\includegraphics[trim=0 0 0 0, clip, width=0.3 \textwidth]{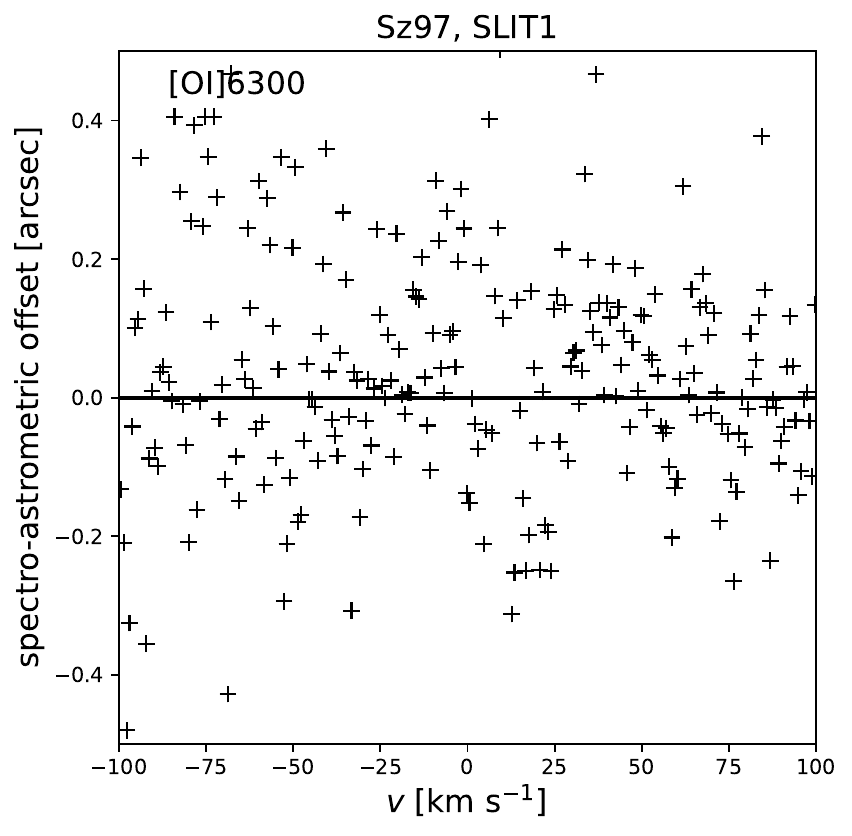}}
\hfill
\subfloat{\includegraphics[trim=0 0 0 0, clip, width=0.3 \textwidth]{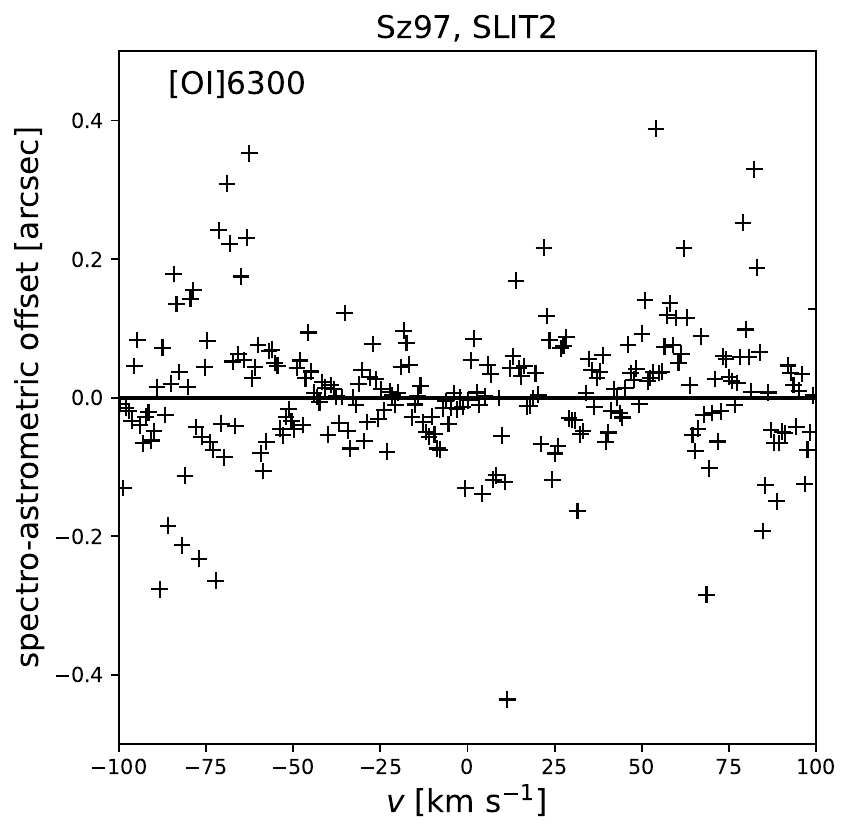}}
\hfill
\subfloat{\includegraphics[trim=0 0 0 0, clip, width=0.3 \textwidth]{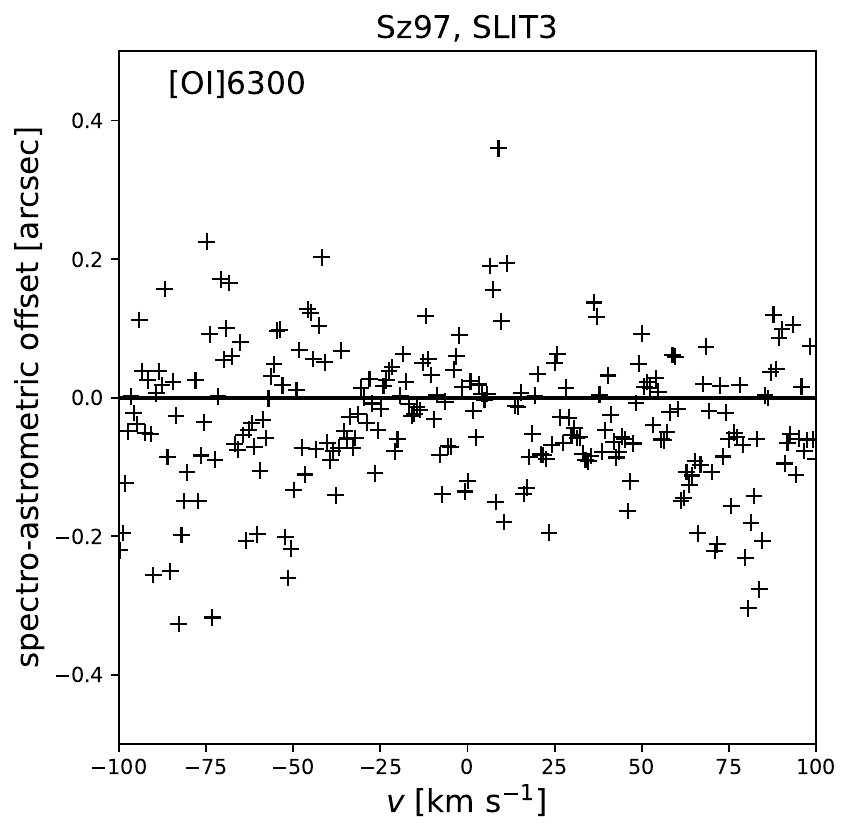}} 
\hfill
\caption{\small{Line profiles of H$\alpha$ and [OI]$\lambda$6300 for all slit positions of Sz\,97.}}\label{fig:all_minispectra_Sz97}
\end{figure*}

\begin{figure*} 
\centering
\subfloat{\includegraphics[trim=0 0 0 0, clip, width=0.3 \textwidth]{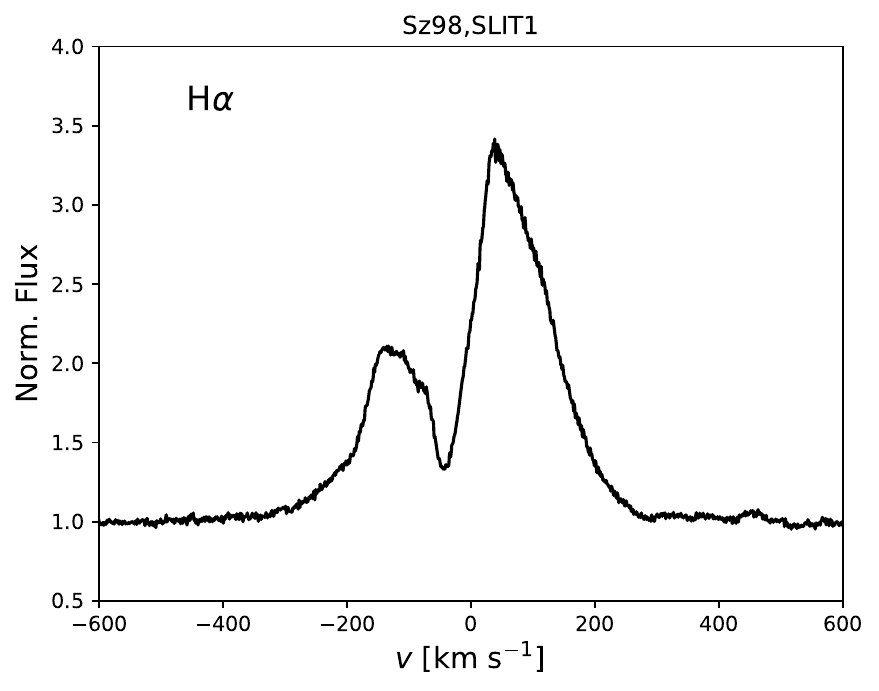}}
\hfill
\subfloat{\includegraphics[trim=0 0 0 0, clip, width=0.3 \textwidth]{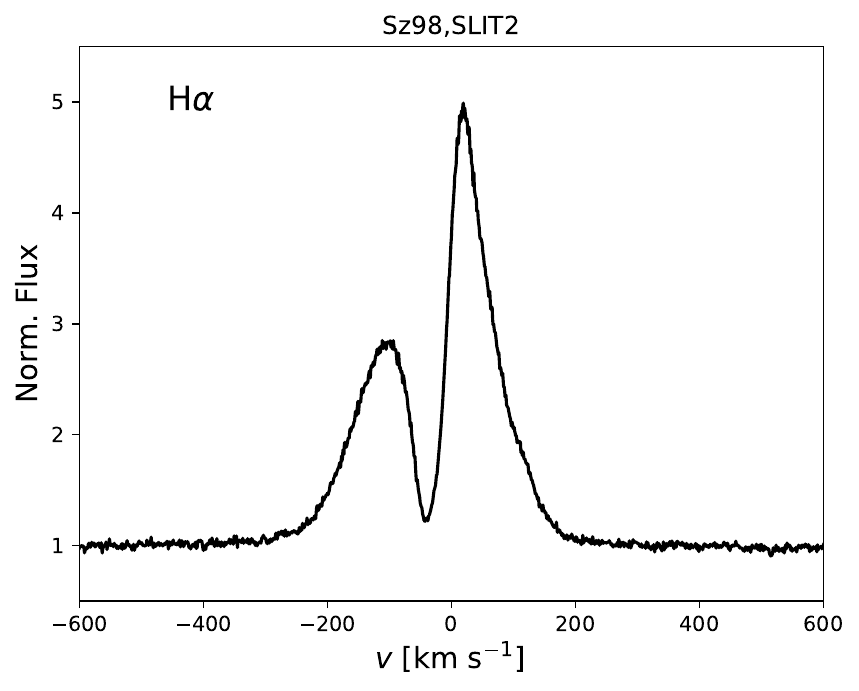}}
\hfill
\subfloat{\includegraphics[trim=0 0 0 0, clip, width=0.3 \textwidth]{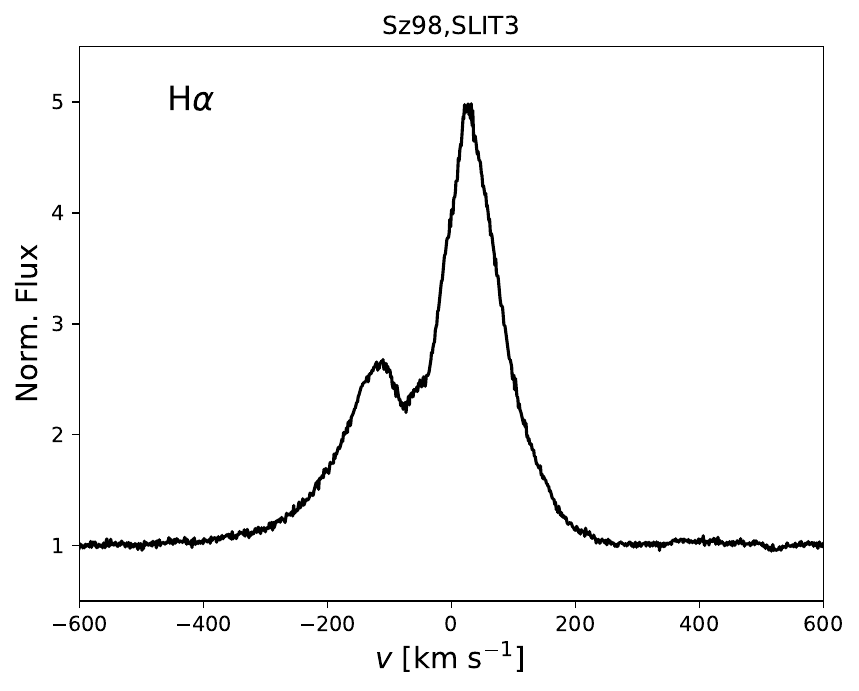}} 
\hfill
\subfloat{\includegraphics[trim=0 0 0 0, clip, width=0.3 \textwidth]{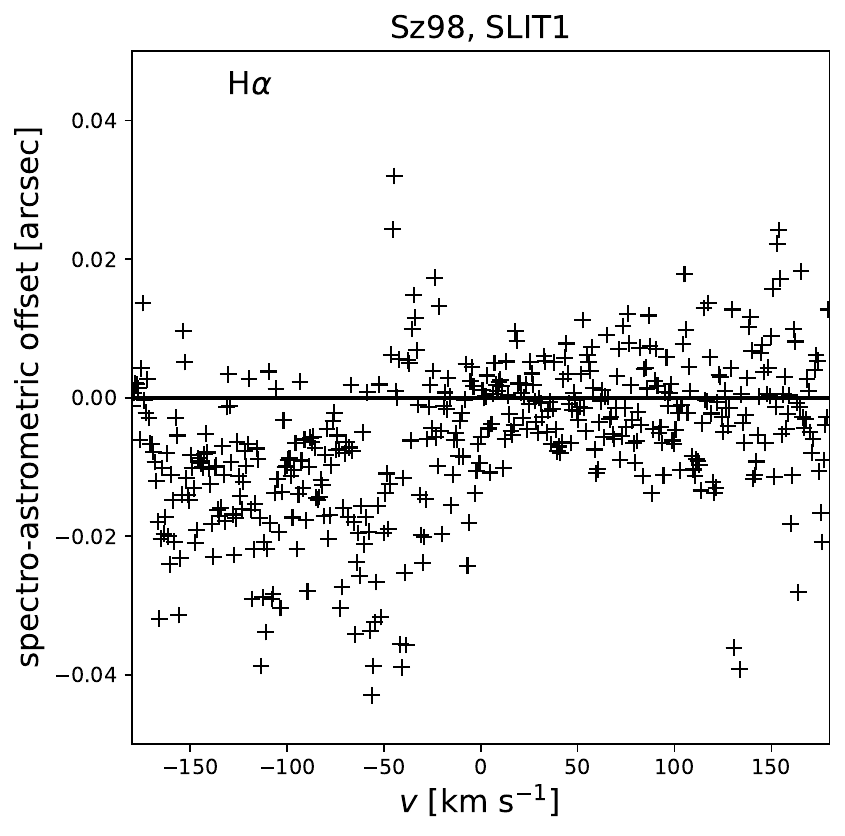}}
\hfill
\subfloat{\includegraphics[trim=0 0 0 0, clip, width=0.3 \textwidth]{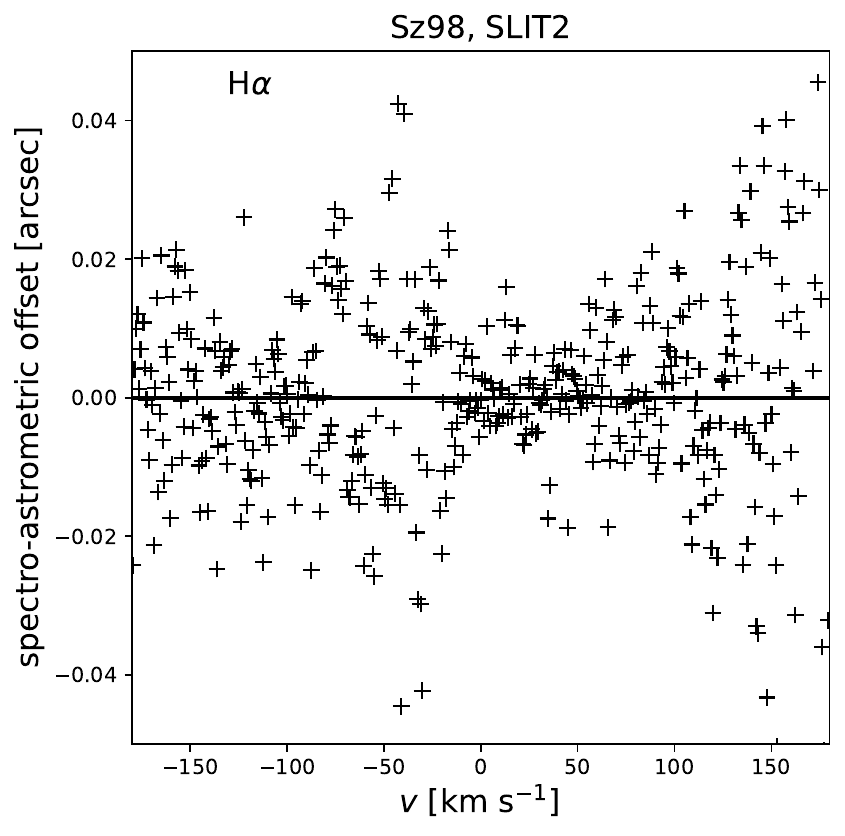}}
\hfill
\subfloat{\includegraphics[trim=0 0 0 0, clip, width=0.3 \textwidth]{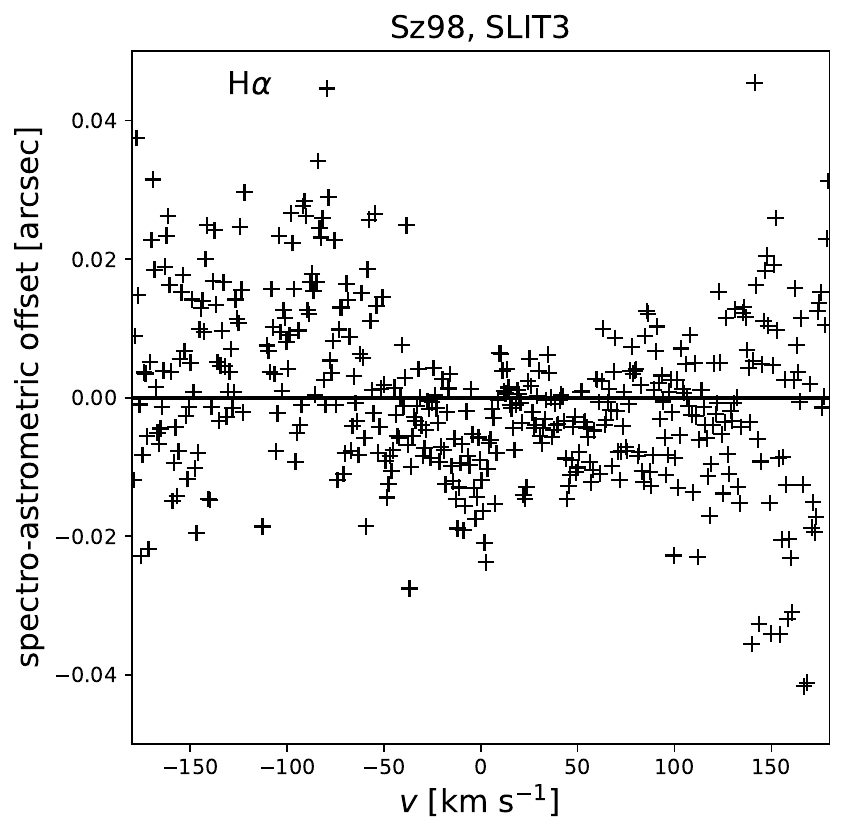}} 
\hfill
\subfloat{\includegraphics[trim=0 0 0 0, clip, width=0.3 \textwidth]{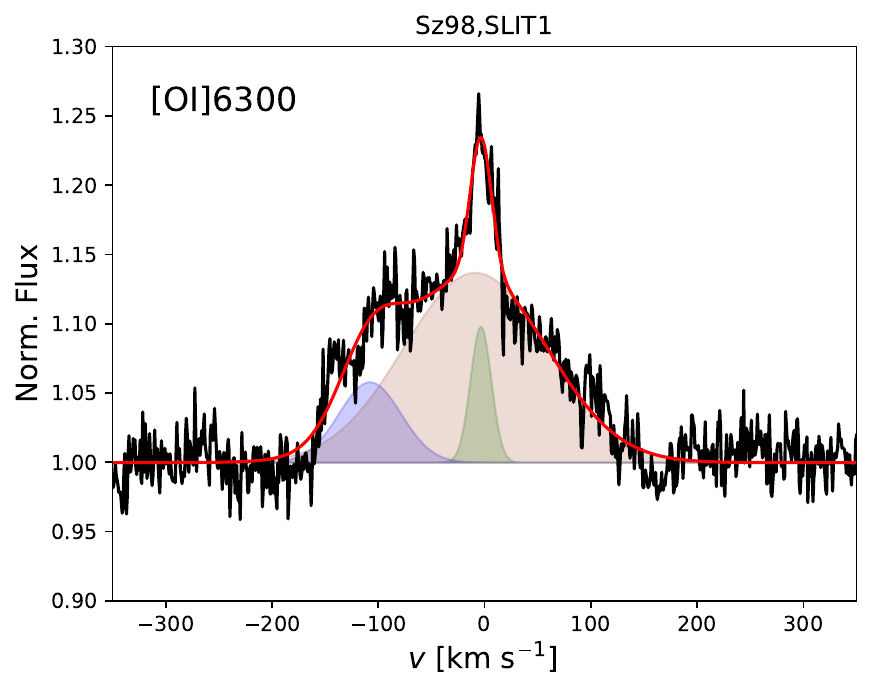}}
\hfill
\subfloat{\includegraphics[trim=0 0 0 0, clip, width=0.3 \textwidth]{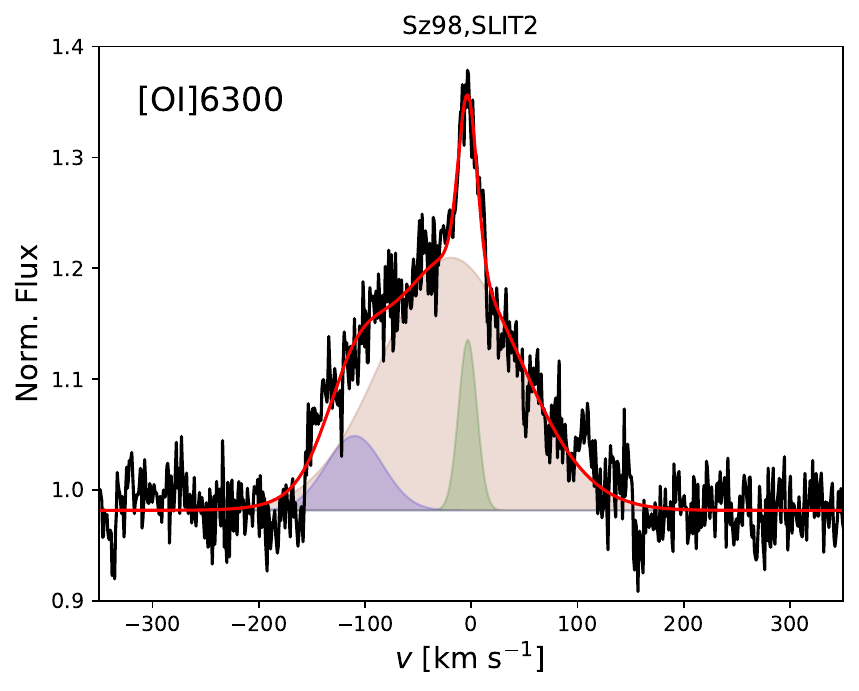}}
\hfill
\subfloat{\includegraphics[trim=0 0 0 0, clip, width=0.3 \textwidth]{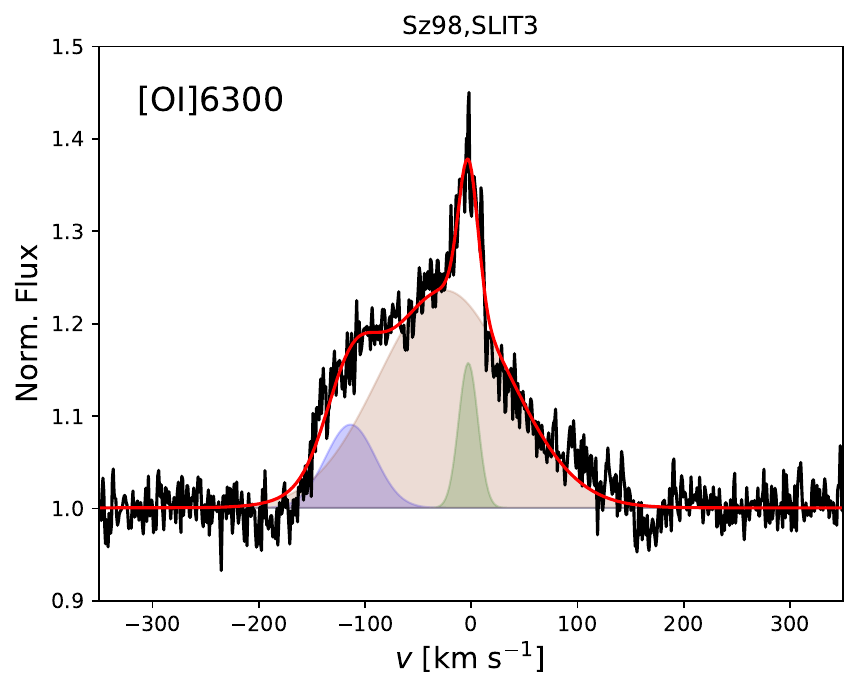}} 
\hfill 
\subfloat{\includegraphics[trim=0 0 0 0, clip, width=0.3 \textwidth]{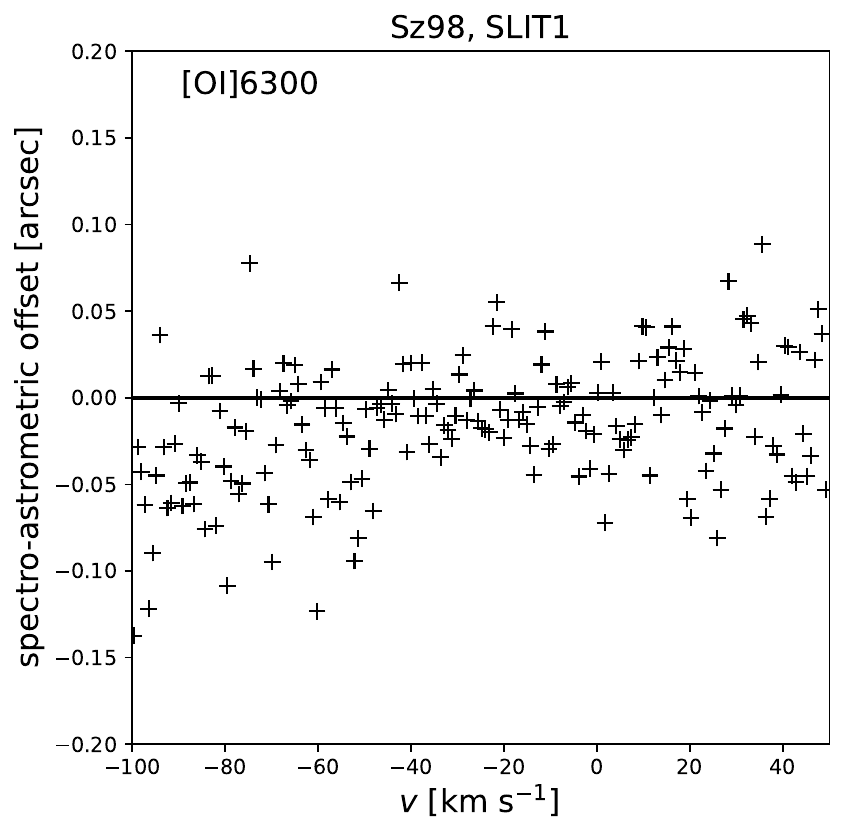}}
\hfill
\subfloat{\includegraphics[trim=0 0 0 0, clip, width=0.3 \textwidth]{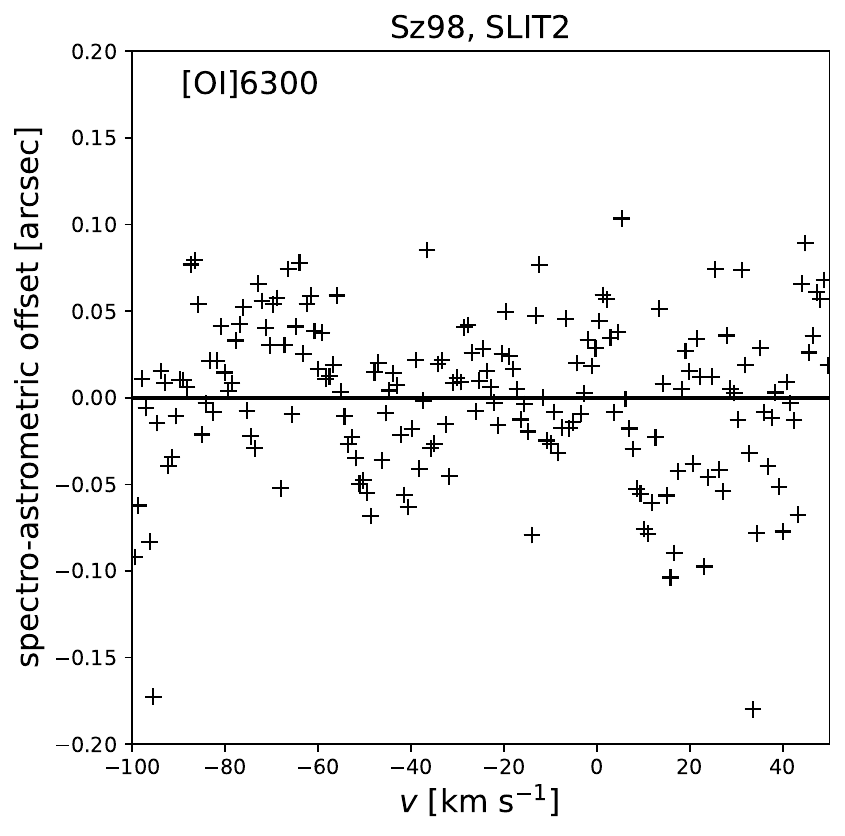}}
\hfill
\subfloat{\includegraphics[trim=0 0 0 0, clip, width=0.3 \textwidth]{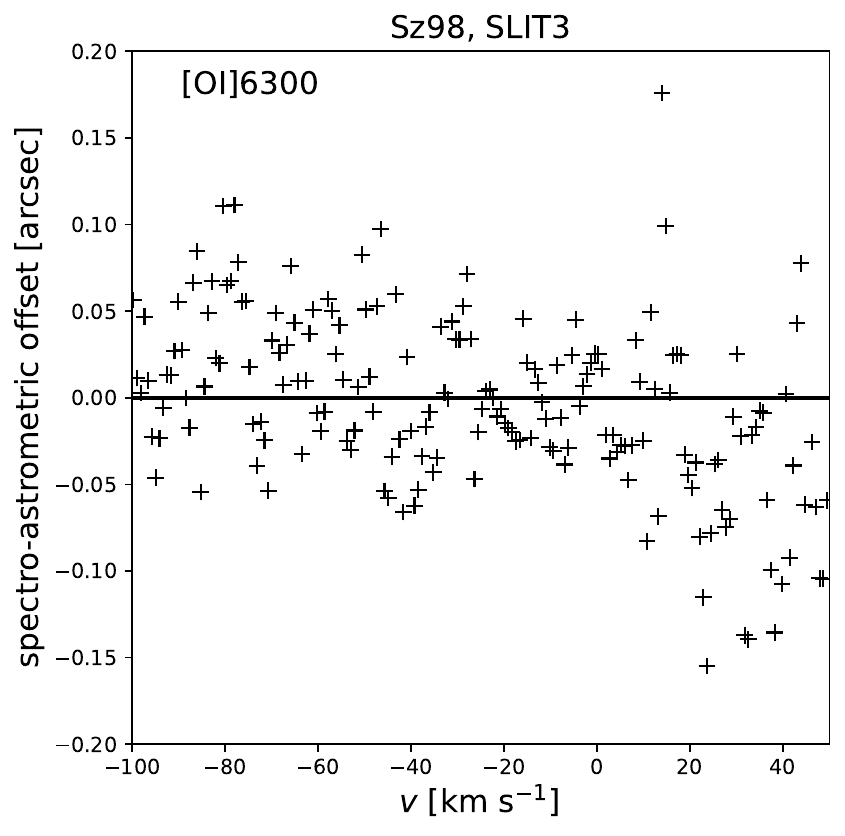}} 
\hfill
\caption{\small{Line profiles of H$\alpha$ and [OI]$\lambda$6300 for all slit positions of Sz\,98.}}\label{fig:all_minispectra_Sz98}
\end{figure*}

\begin{figure*} 
\centering
\subfloat{\includegraphics[trim=0 0 0 0, clip, width=0.3 \textwidth]{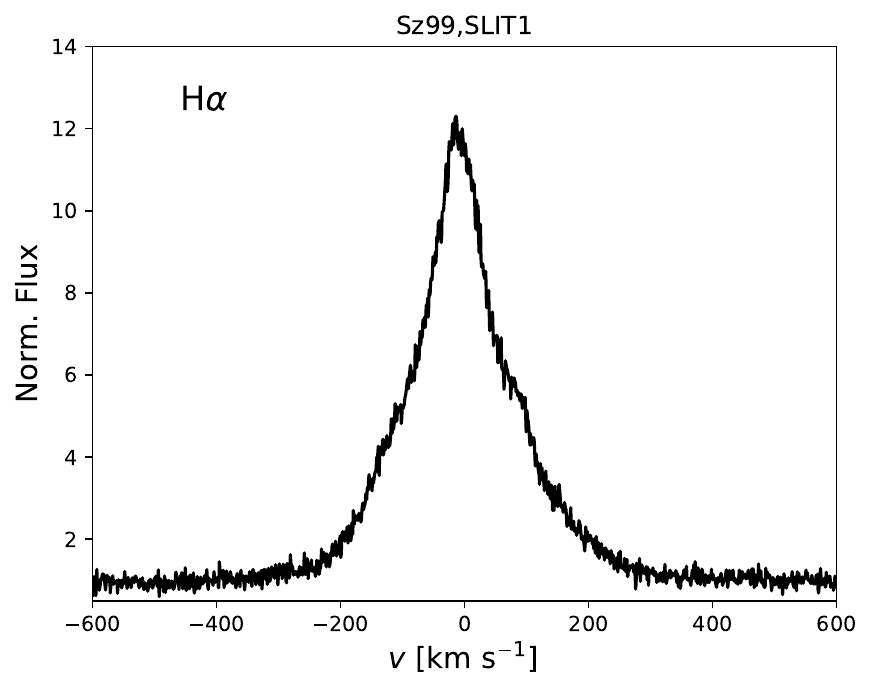}}
\hfill
\subfloat{\includegraphics[trim=0 0 0 0, clip, width=0.3 \textwidth]{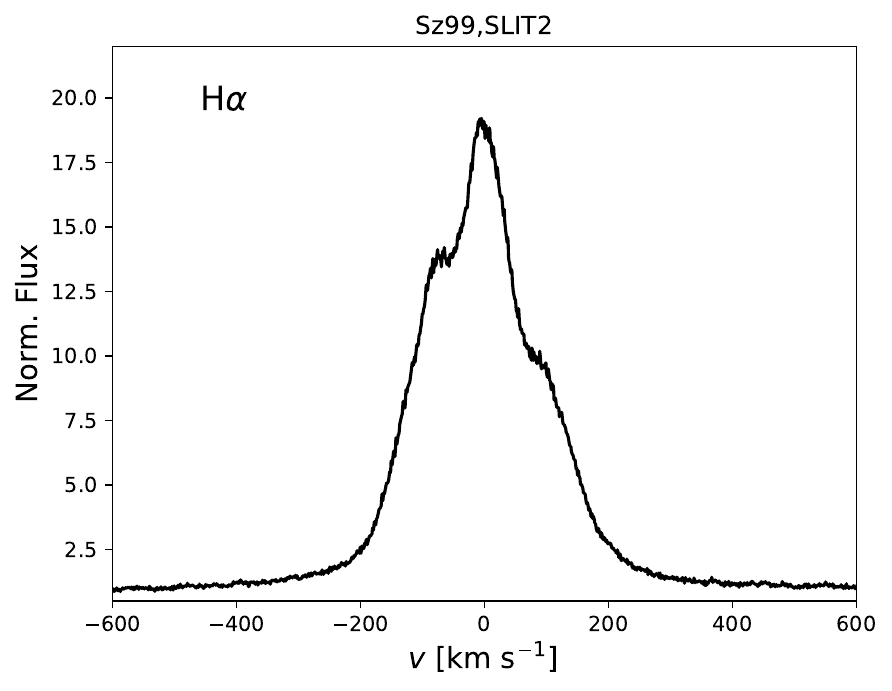}}
\hfill
\subfloat{\includegraphics[trim=0 0 0 0, clip, width=0.3 \textwidth]{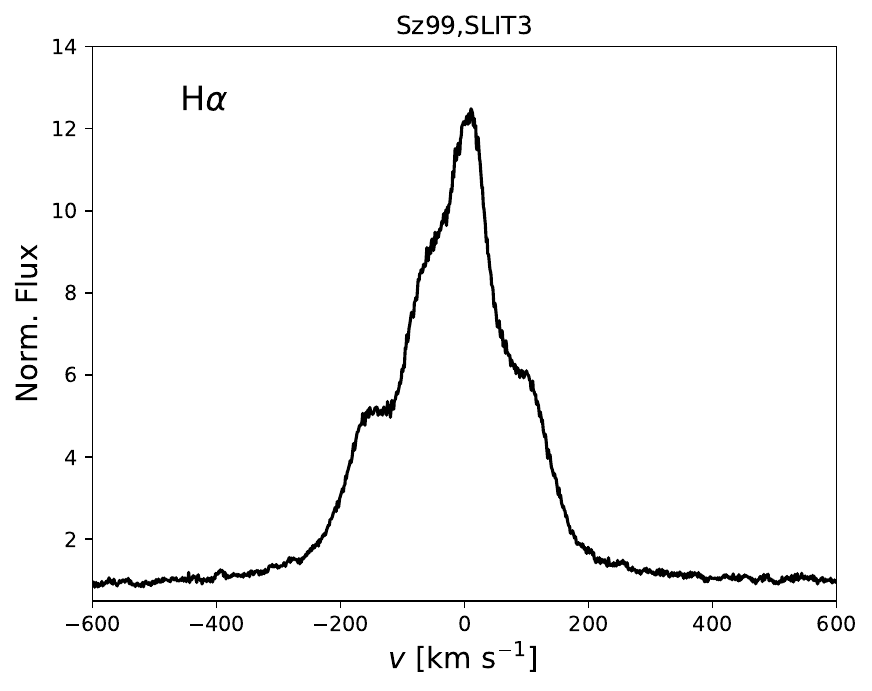}} 
\hfill
\subfloat{\includegraphics[trim=0 0 0 0, clip, width=0.3 \textwidth]{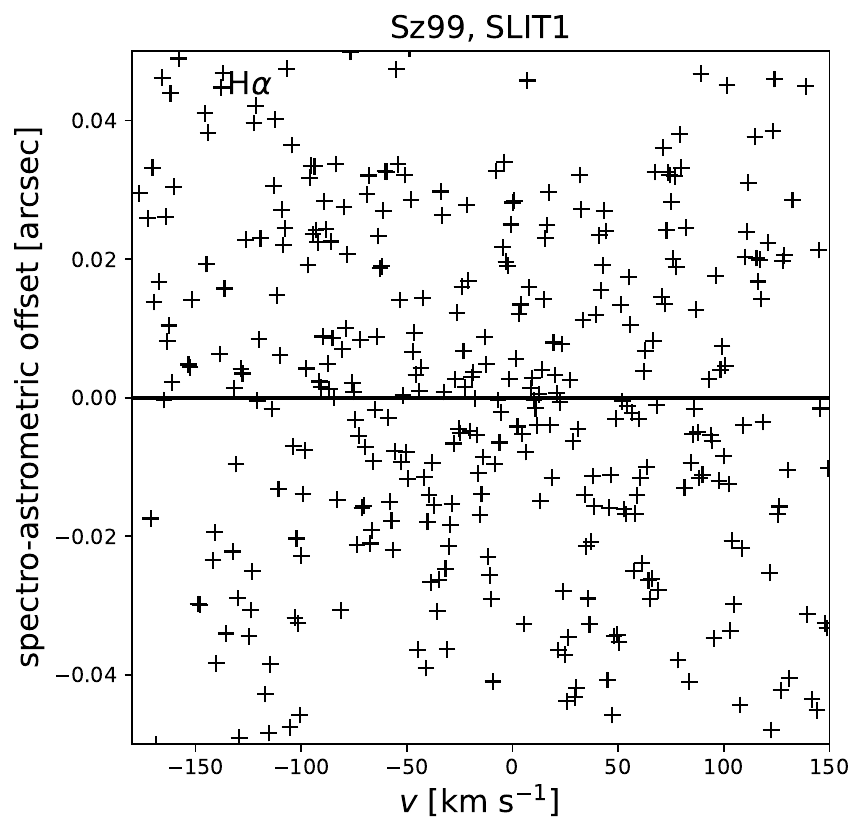}}
\hfill
\subfloat{\includegraphics[trim=0 0 0 0, clip, width=0.3 \textwidth]{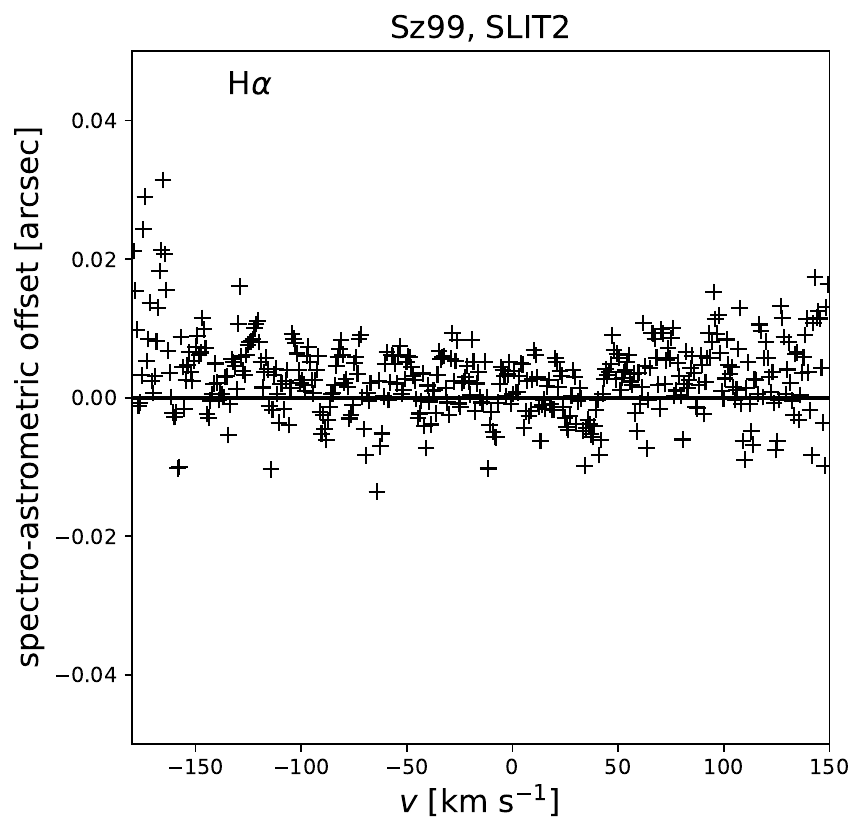}}
\hfill
\subfloat{\includegraphics[trim=0 0 0 0, clip, width=0.3 \textwidth]{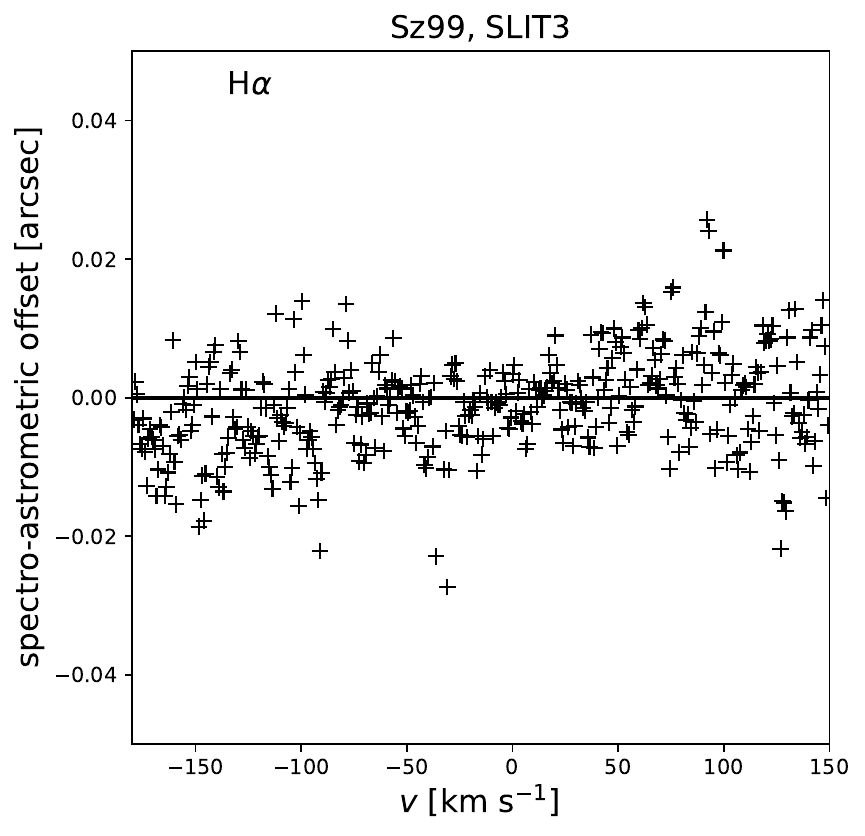}} 
\hfill
\subfloat{\includegraphics[trim=0 0 0 0, clip, width=0.3 \textwidth]{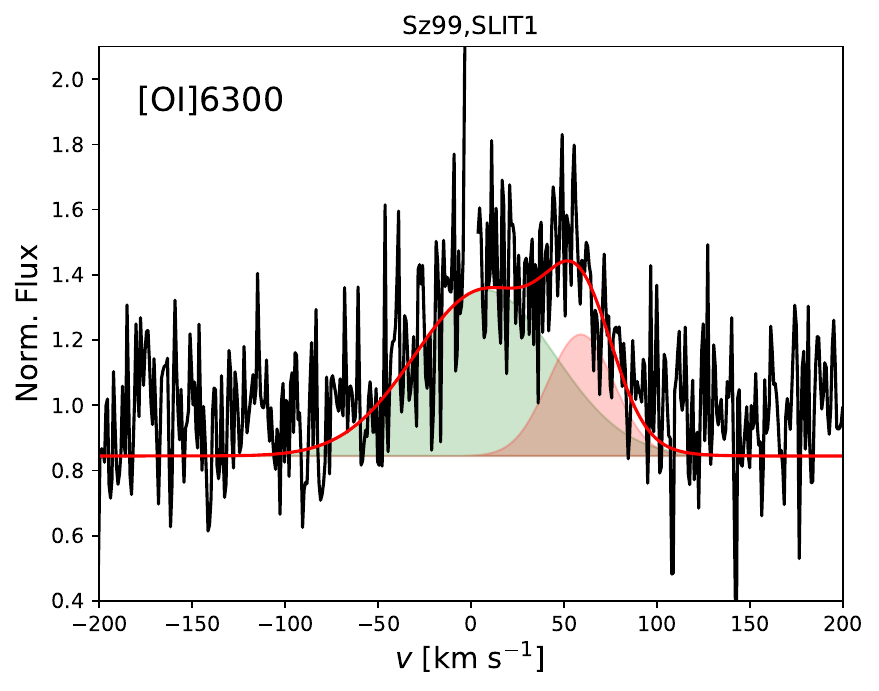}}
\hfill
\subfloat{\includegraphics[trim=0 0 0 0, clip, width=0.3 \textwidth]{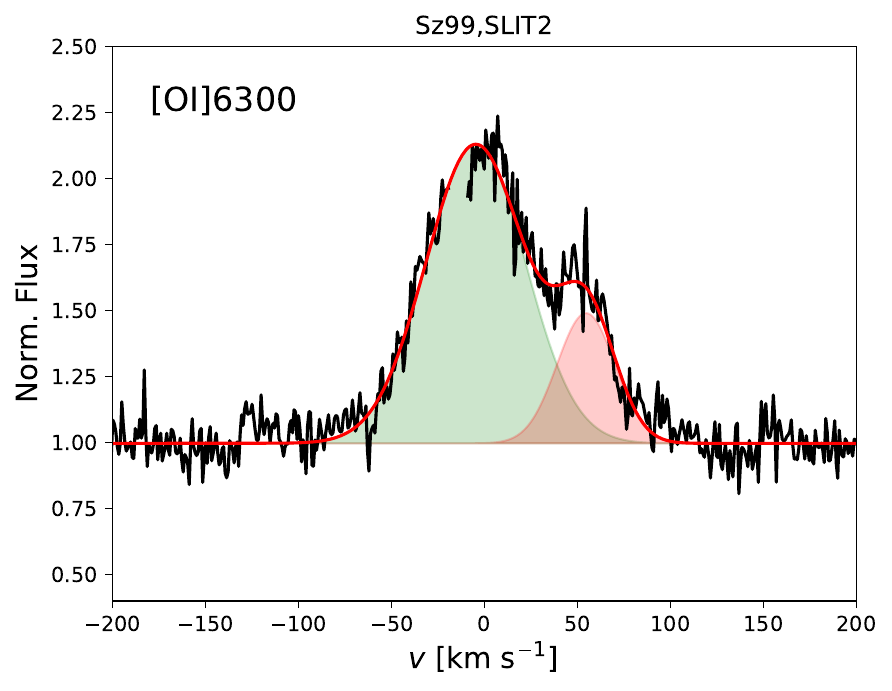}}
\hfill
\subfloat{\includegraphics[trim=0 0 0 0, clip, width=0.3 \textwidth]{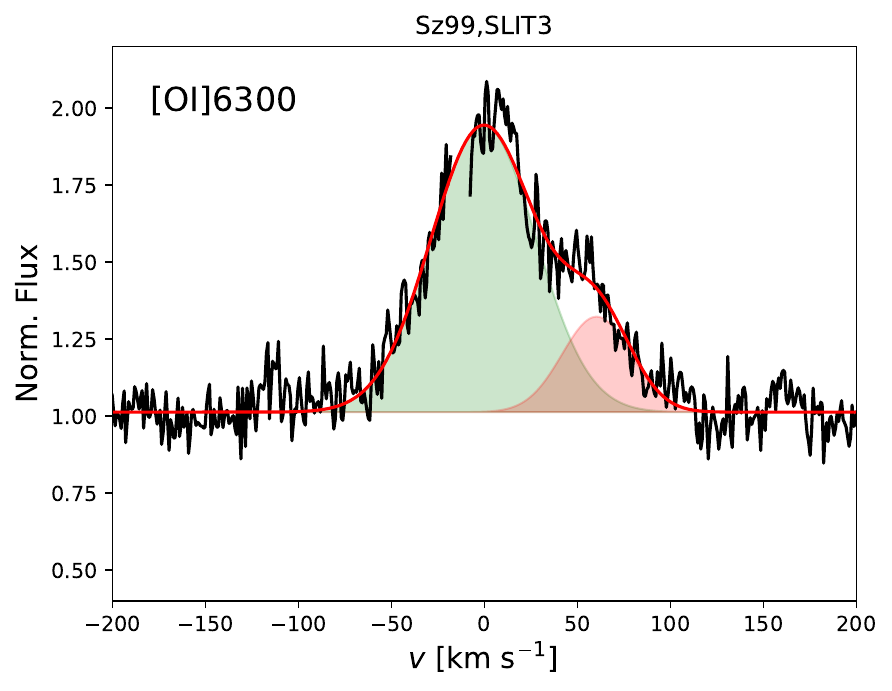}} 
\hfill 
\subfloat{\includegraphics[trim=0 0 0 0, clip, width=0.3 \textwidth]{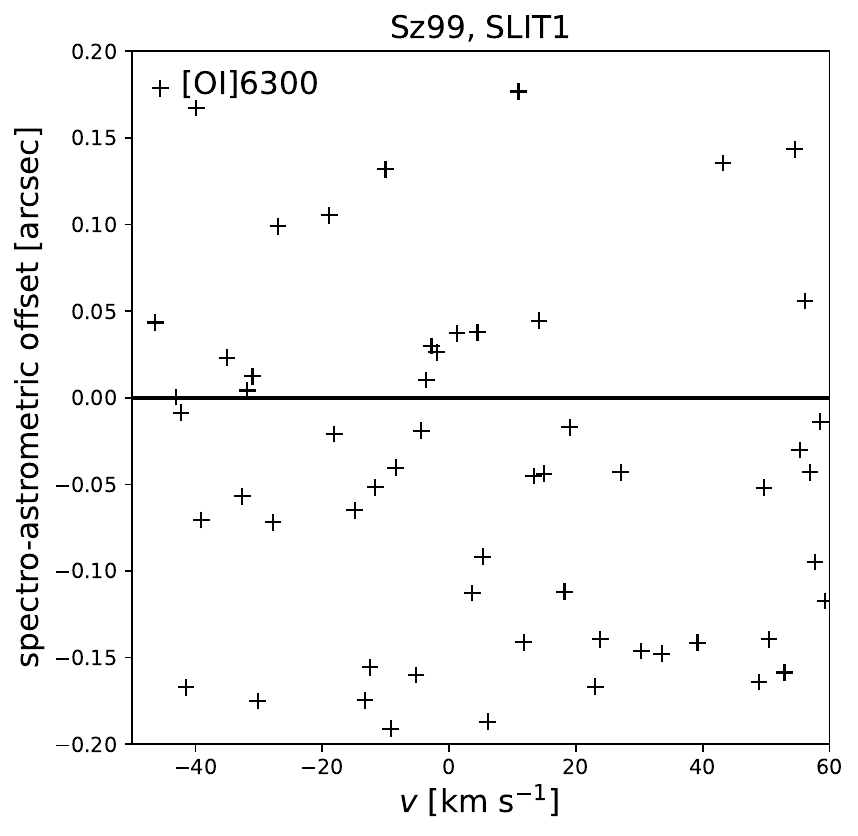}}
\hfill
\subfloat{\includegraphics[trim=0 0 0 0, clip, width=0.3 \textwidth]{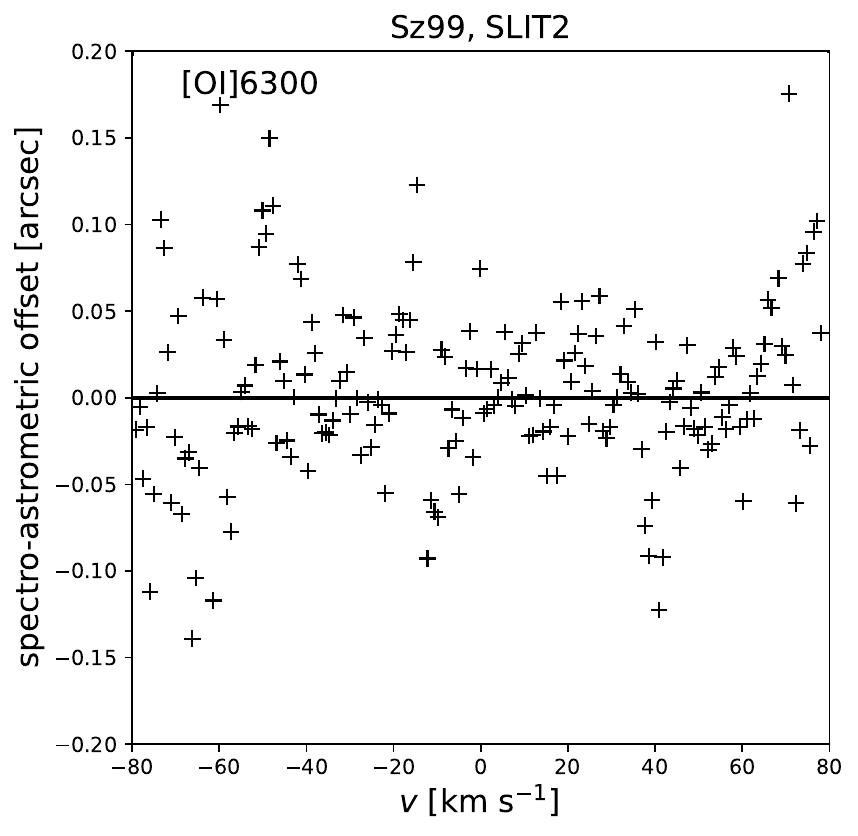}}
\hfill
\subfloat{\includegraphics[trim=0 0 0 0, clip, width=0.3 \textwidth]{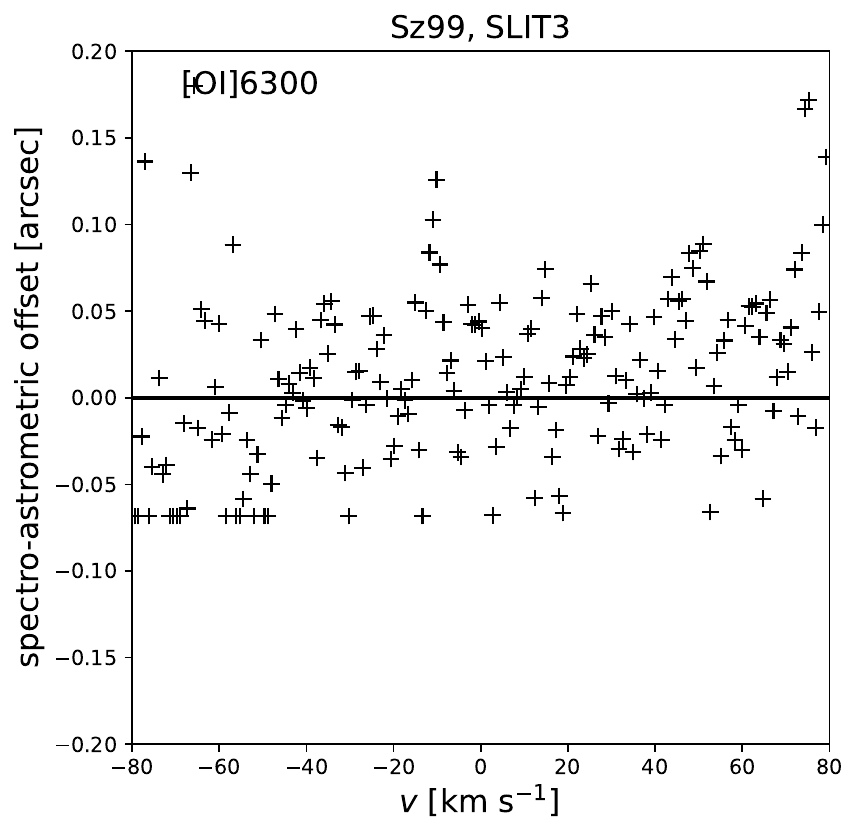}} 
\hfill
\caption{\small{Line profiles of H$\alpha$ and [OI]$\lambda$6300 for all slit positions of Sz\,99.}}\label{fig:all_minispectra_Sz99}
\end{figure*}

\begin{figure*} 
\centering
\subfloat{\includegraphics[trim=0 0 0 0, clip, width=0.3 \textwidth]{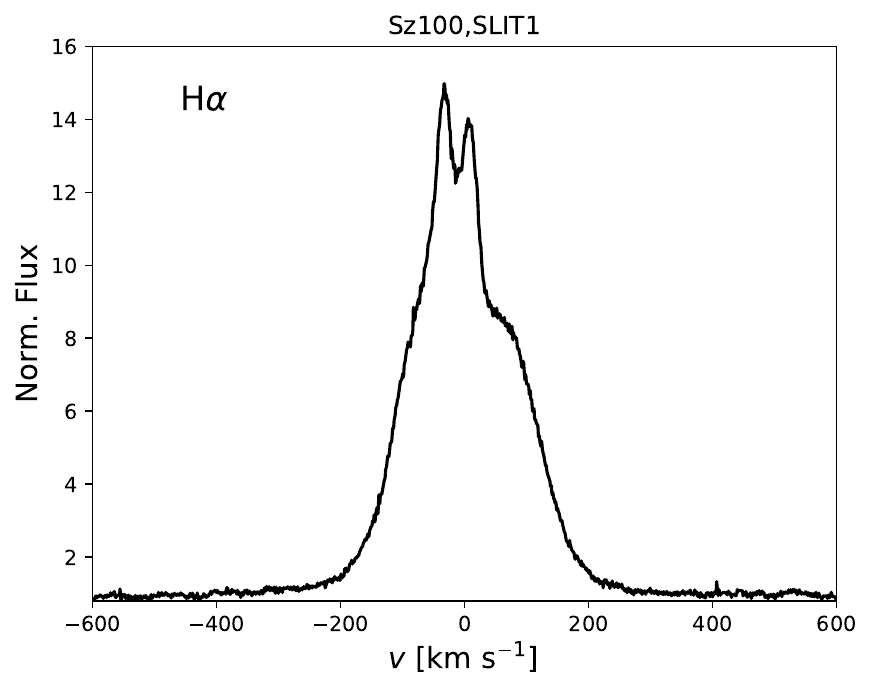}}
\hfill
\subfloat{\includegraphics[trim=0 0 0 0, clip, width=0.3 \textwidth]{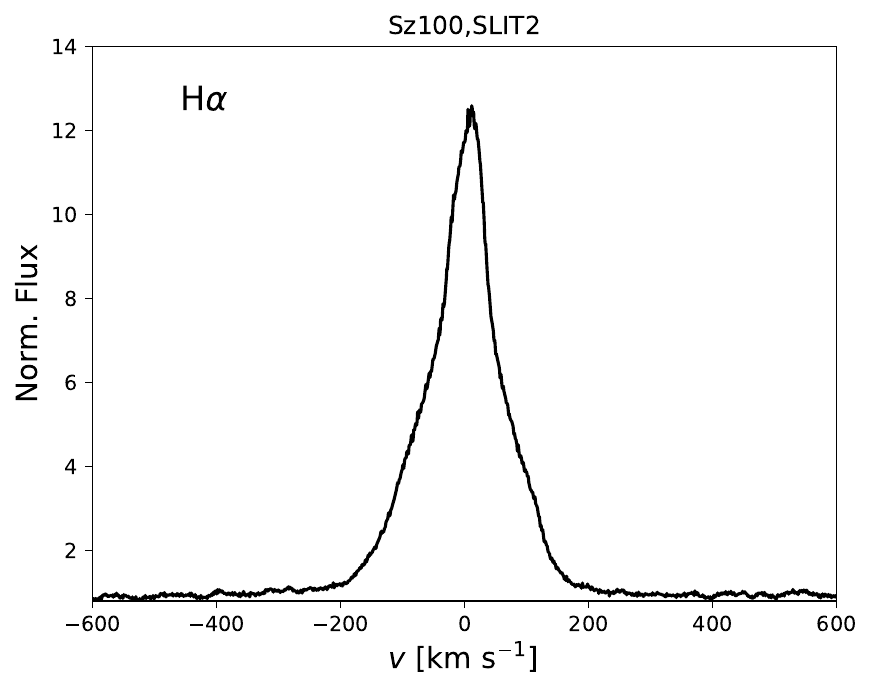}}
\hfill
\subfloat{\includegraphics[trim=0 0 0 0, clip, width=0.3 \textwidth]{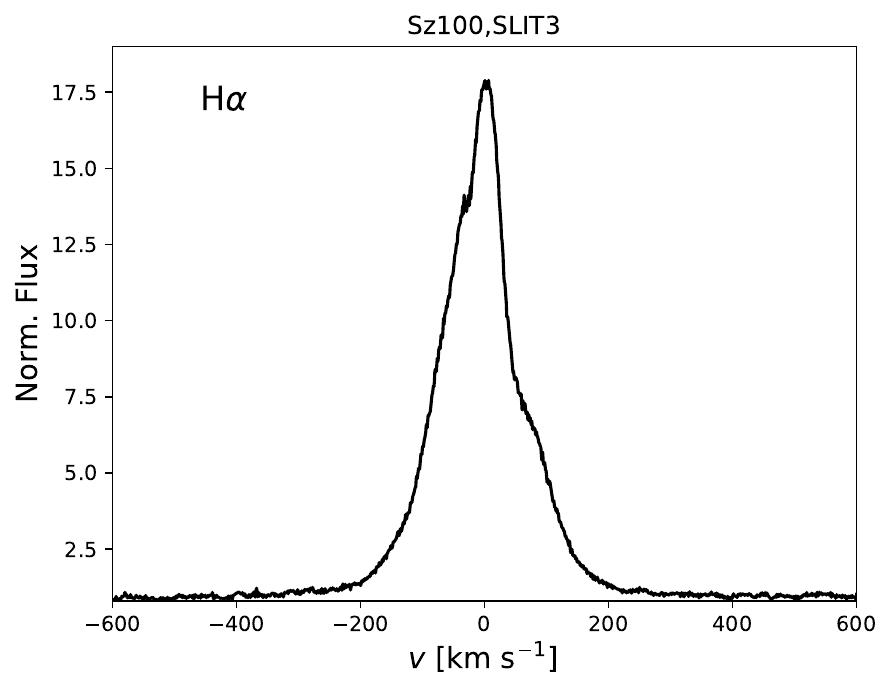}} 
\hfill
\subfloat{\includegraphics[trim=0 0 0 0, clip, width=0.3 \textwidth]{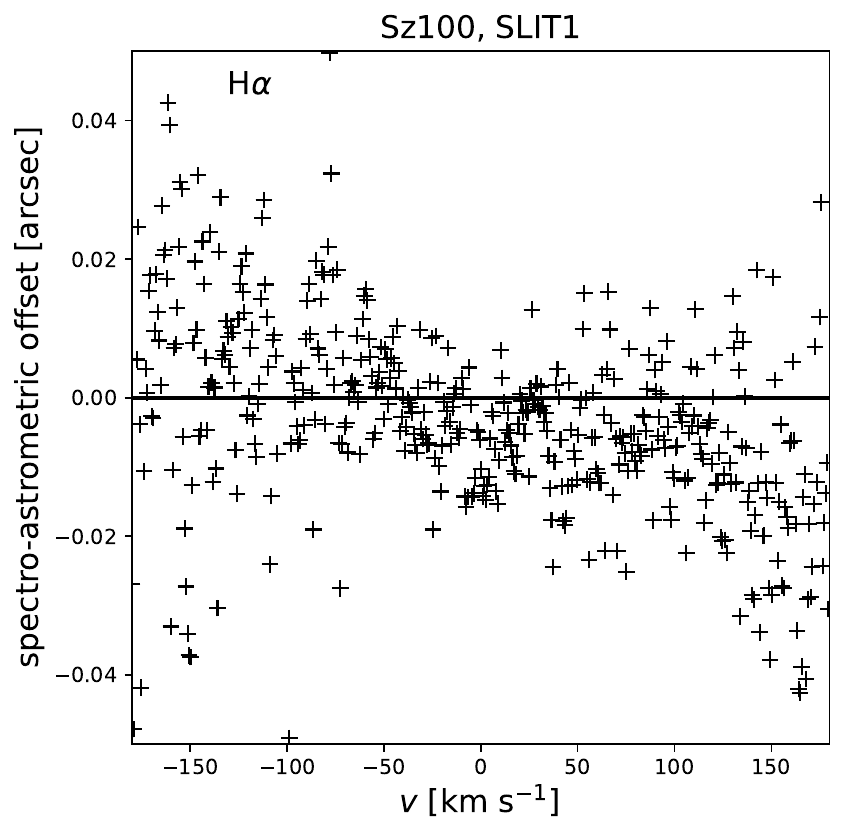}}
\hfill
\subfloat{\includegraphics[trim=0 0 0 0, clip, width=0.3 \textwidth]{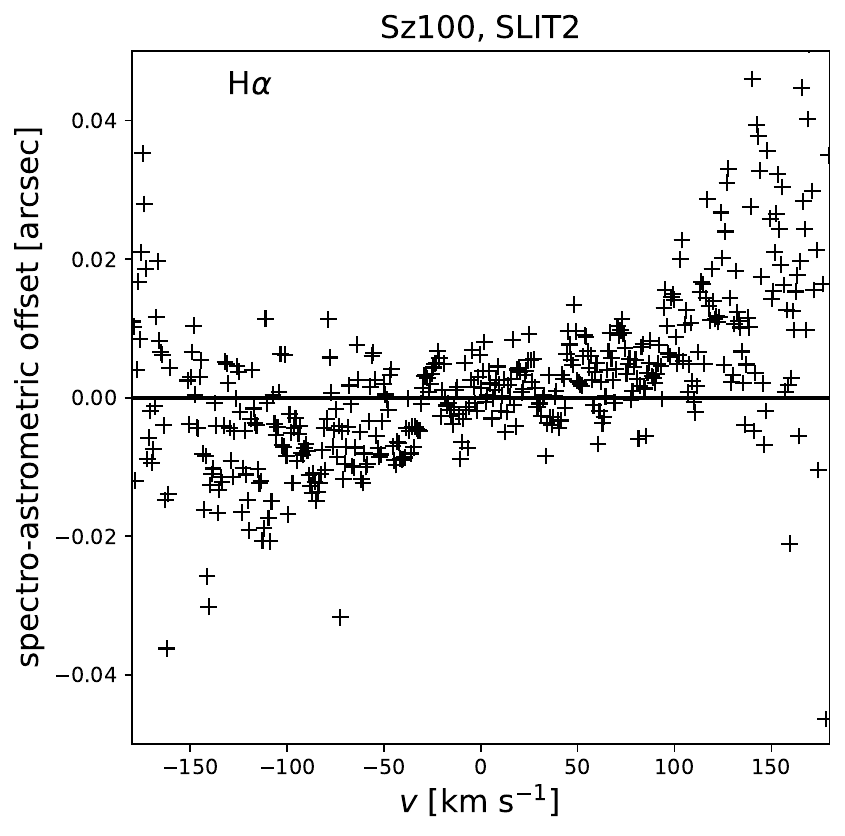}}
\hfill
\subfloat{\includegraphics[trim=0 0 0 0, clip, width=0.3 \textwidth]{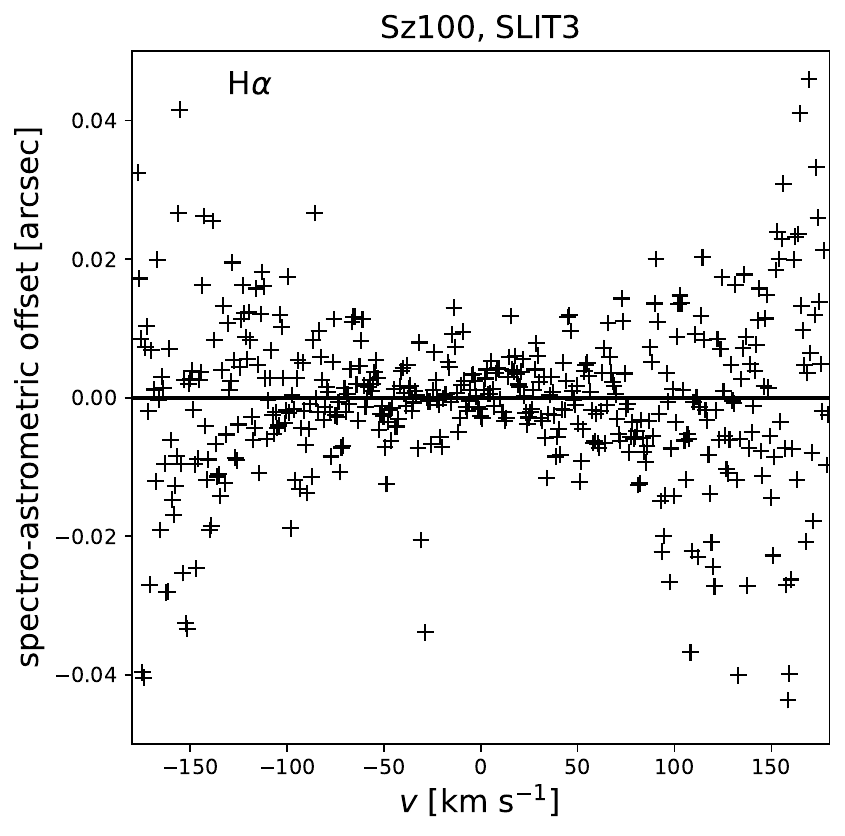}} 
\hfill
\subfloat{\includegraphics[trim=0 0 0 0, clip, width=0.3 \textwidth]{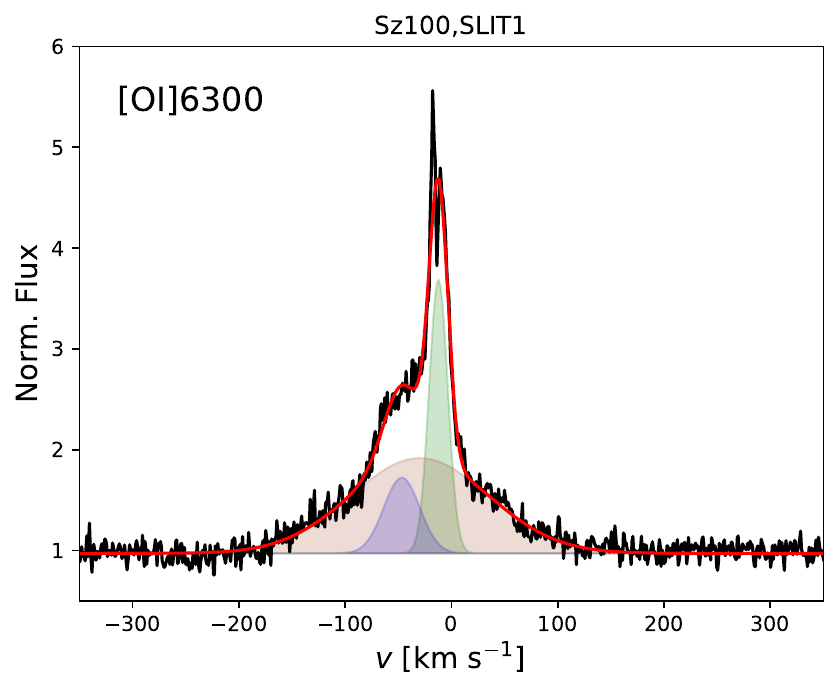}}
\hfill
\subfloat{\includegraphics[trim=0 0 0 0, clip, width=0.3 \textwidth]{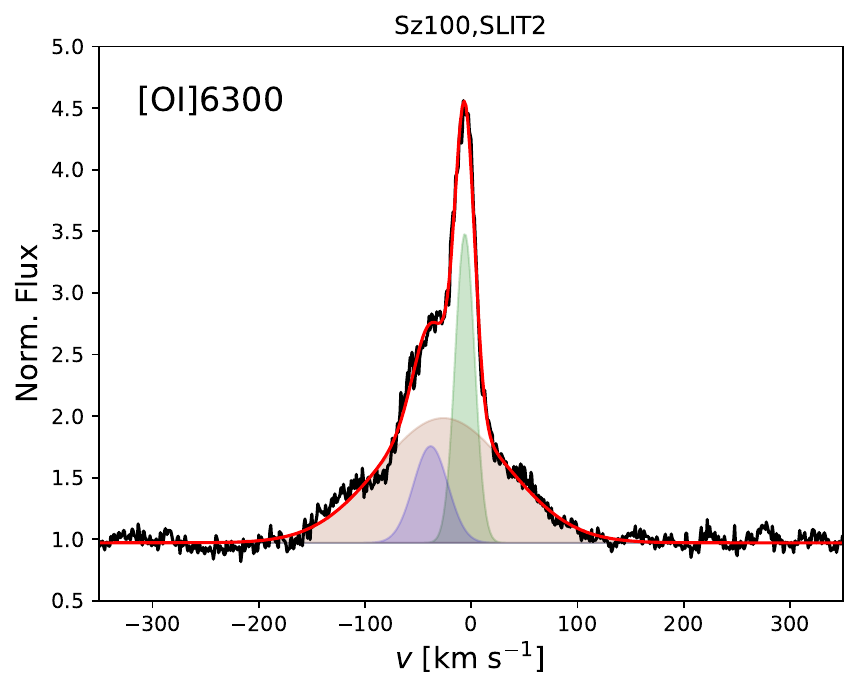}}
\hfill
\subfloat{\includegraphics[trim=0 0 0 0, clip, width=0.3 \textwidth]{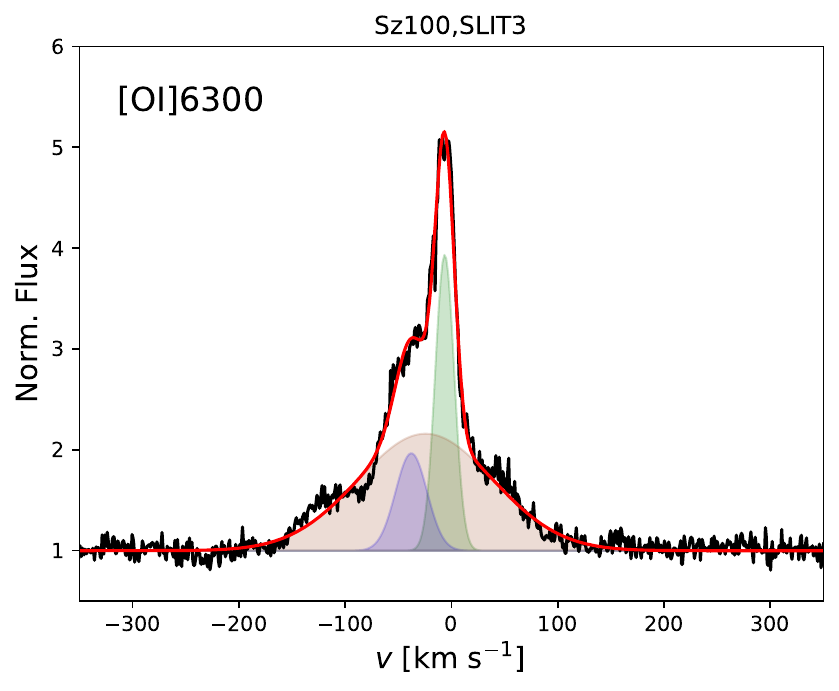}} 
\hfill 
\subfloat{\includegraphics[trim=0 0 0 0, clip, width=0.3 \textwidth]{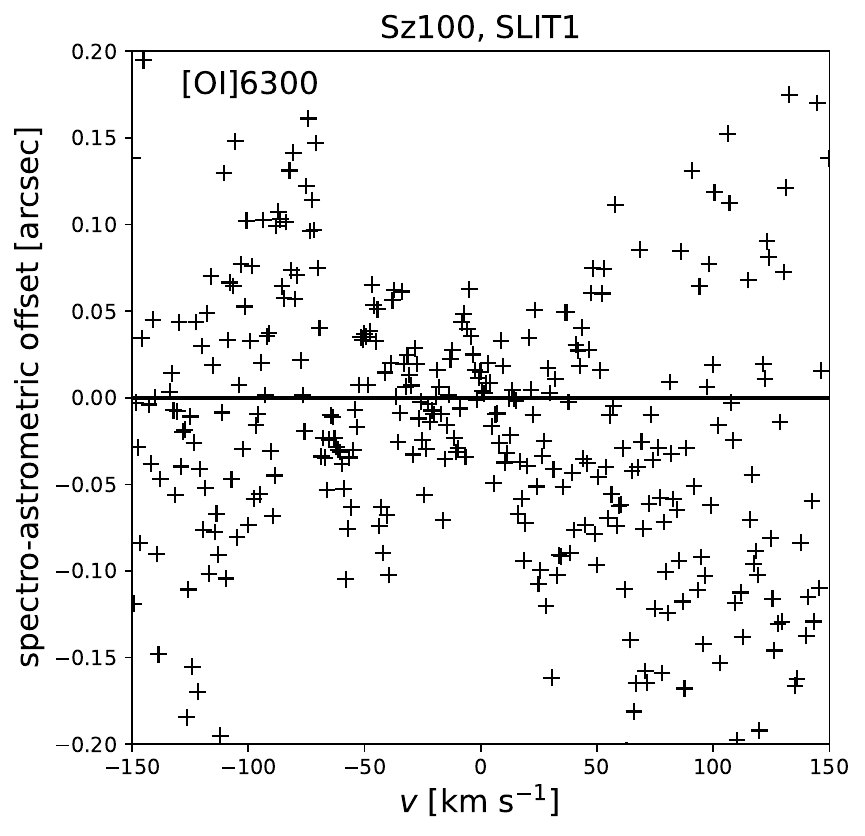}}
\hfill
\subfloat{\includegraphics[trim=0 0 0 0, clip, width=0.3 \textwidth]{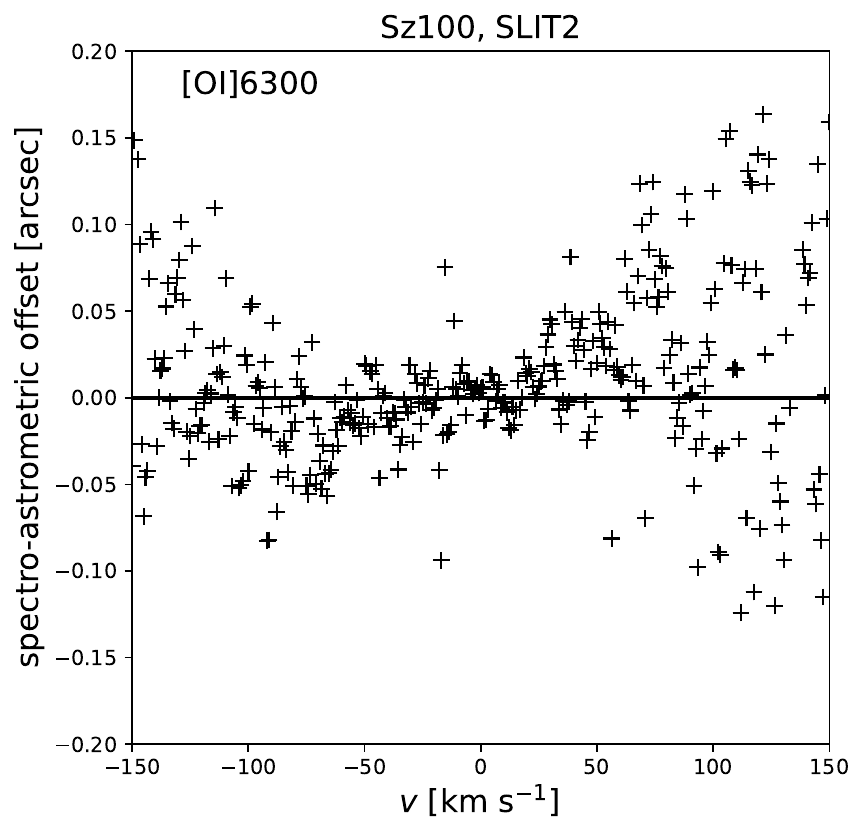}}
\hfill
\subfloat{\includegraphics[trim=0 0 0 0, clip, width=0.3 \textwidth]{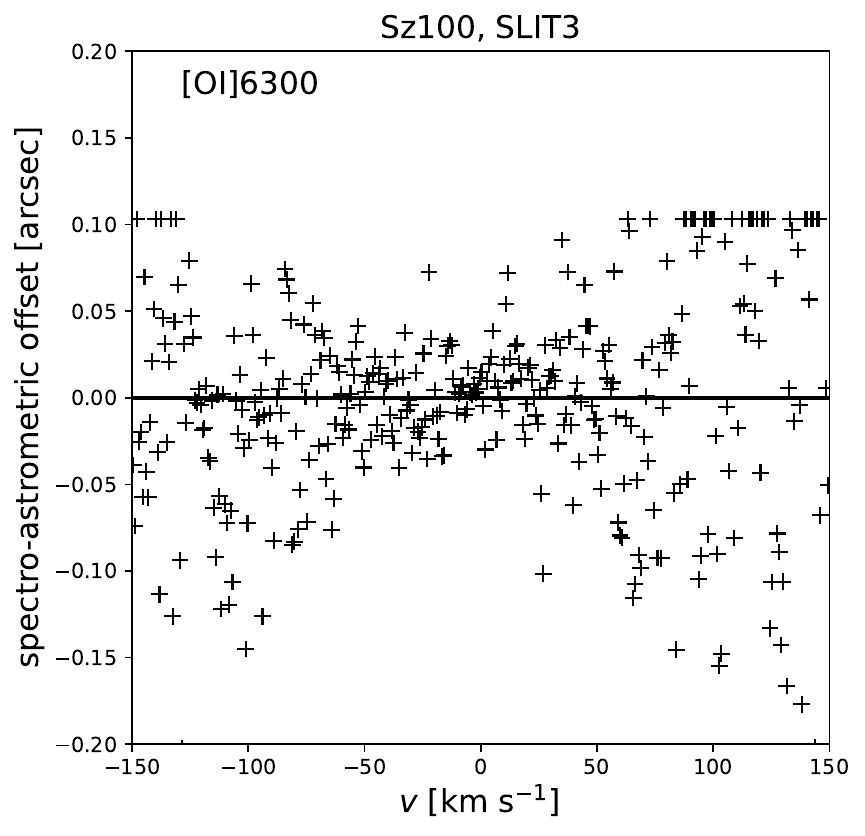}} 
\hfill
\caption{\small{Line profiles of H$\alpha$ and [OI]$\lambda$6300 for all slit positions of Sz\,100.}}\label{fig:all_minispectra_Sz100}
\end{figure*} 

\begin{figure*} 
\centering 
\subfloat{\includegraphics[trim=0 0 0 0, clip, width=0.3 \textwidth]{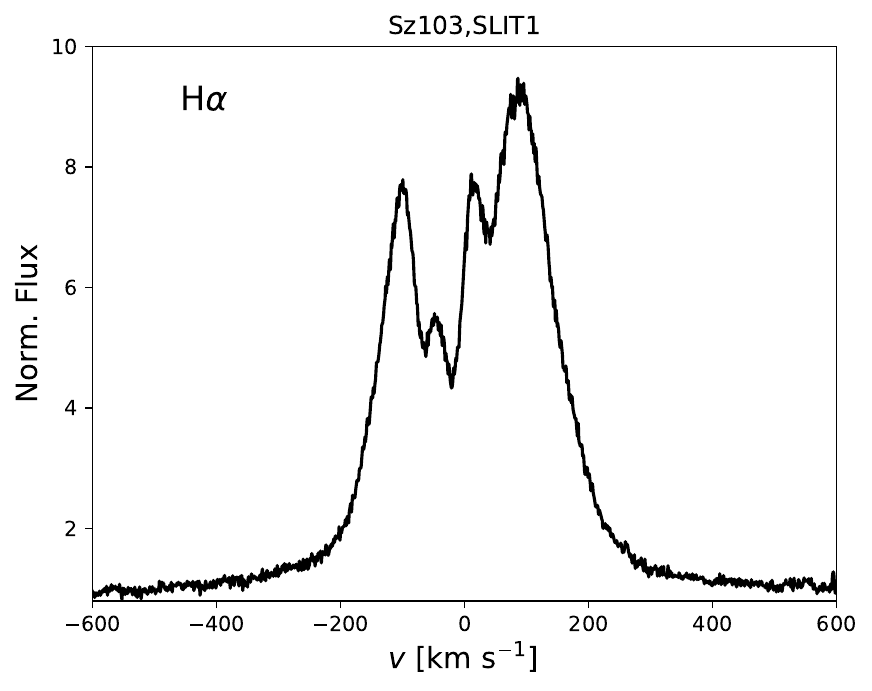}}
\hfill
\subfloat{\includegraphics[trim=0 0 0 0, clip, width=0.3 \textwidth]{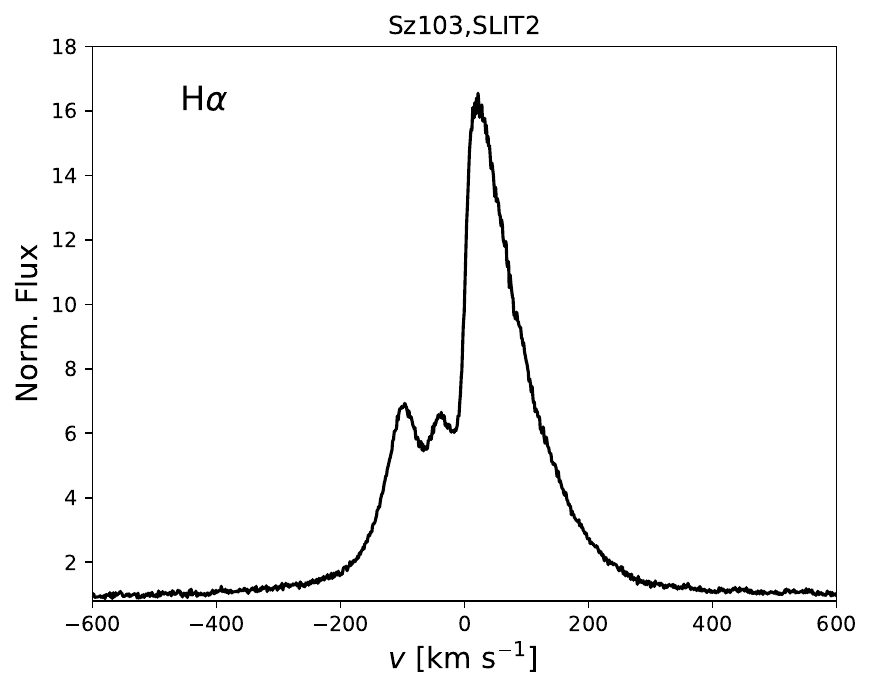}}
\hfill
\subfloat{\includegraphics[trim=0 0 0 0, clip, width=0.3 \textwidth]{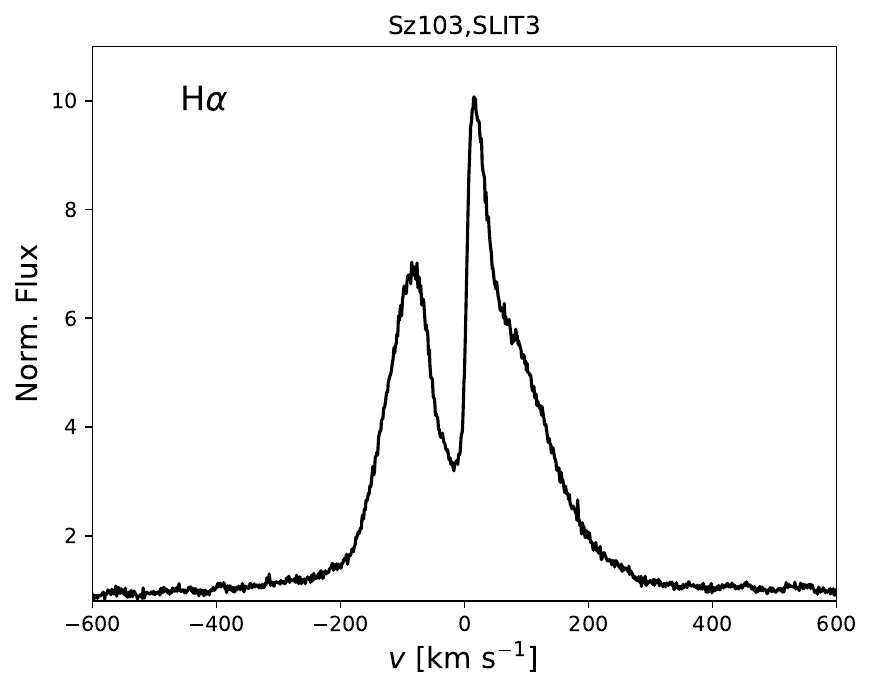}} 
\hfill
\subfloat{\includegraphics[trim=0 0 0 0, clip, width=0.3 \textwidth]{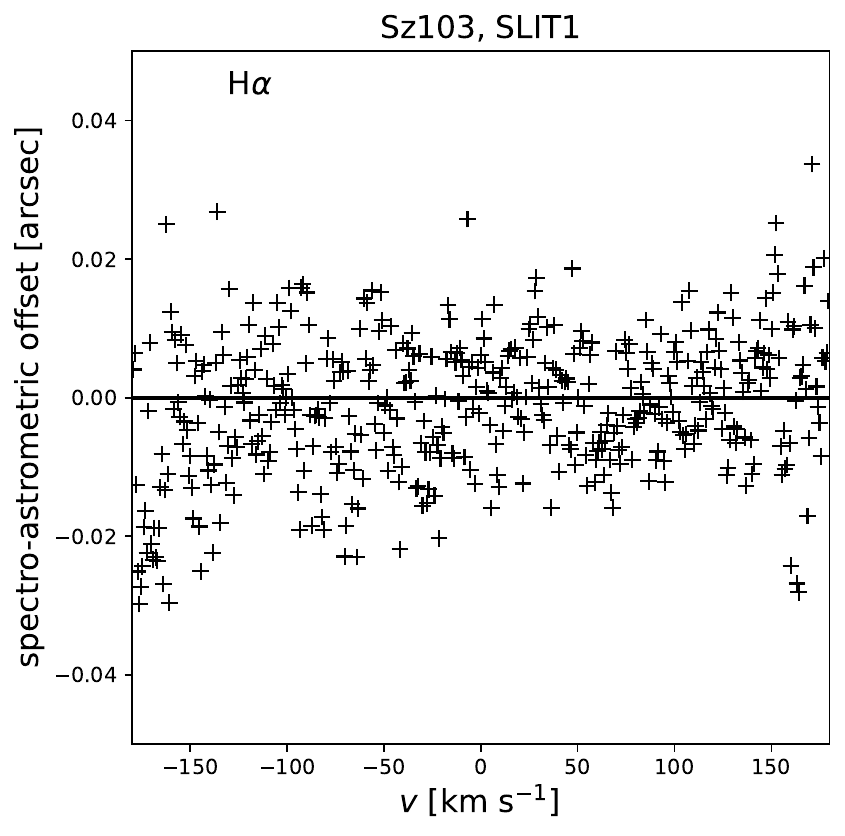}}
\hfill
\subfloat{\includegraphics[trim=0 0 0 0, clip, width=0.3 \textwidth]{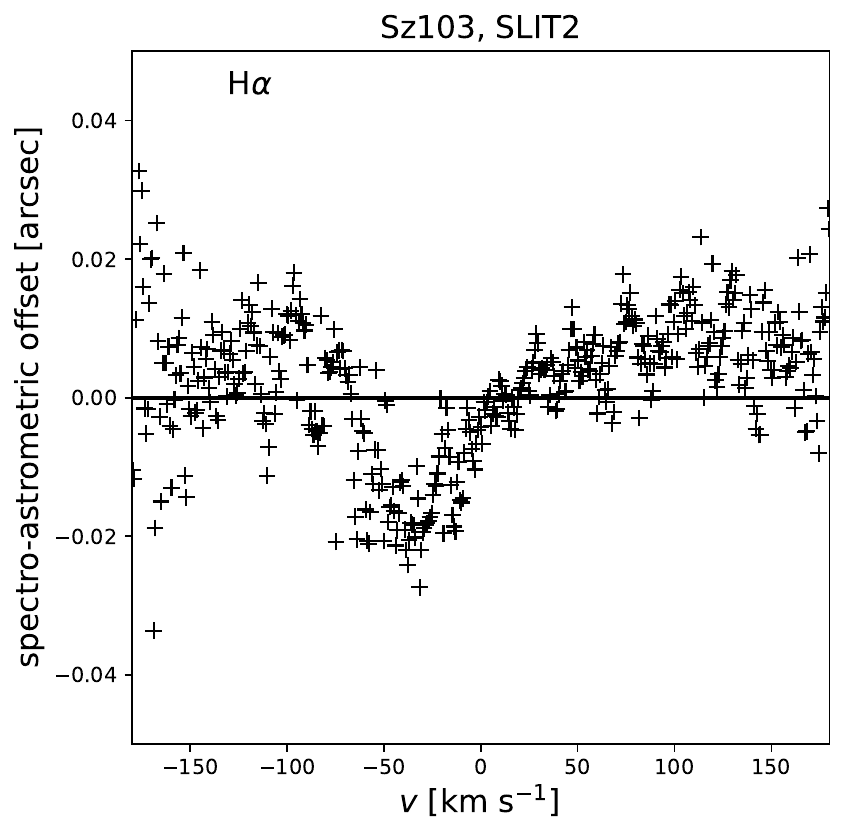}}
\hfill
\subfloat{\includegraphics[trim=0 0 0 0, clip, width=0.3 \textwidth]{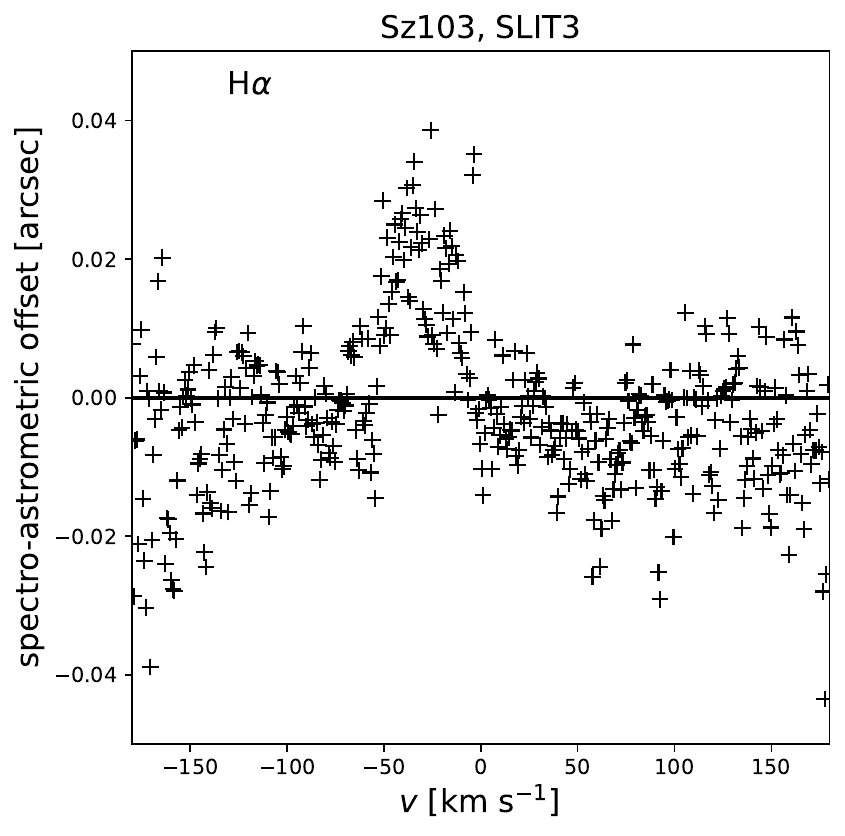}} 
\hfill
\subfloat{\includegraphics[trim=0 0 0 0, clip, width=0.3 \textwidth]{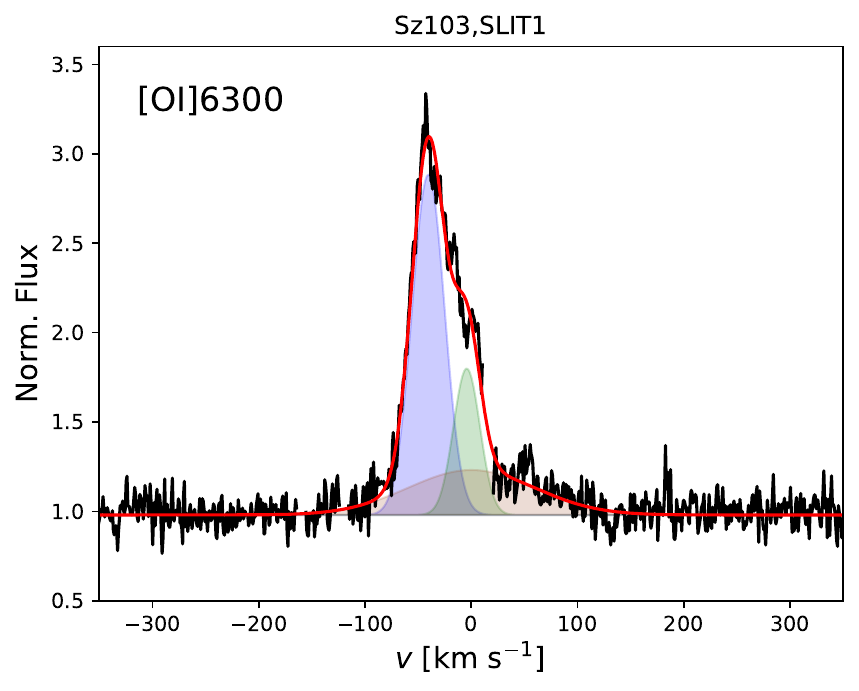}}
\hfill
\subfloat{\includegraphics[trim=0 0 0 0, clip, width=0.3 \textwidth]{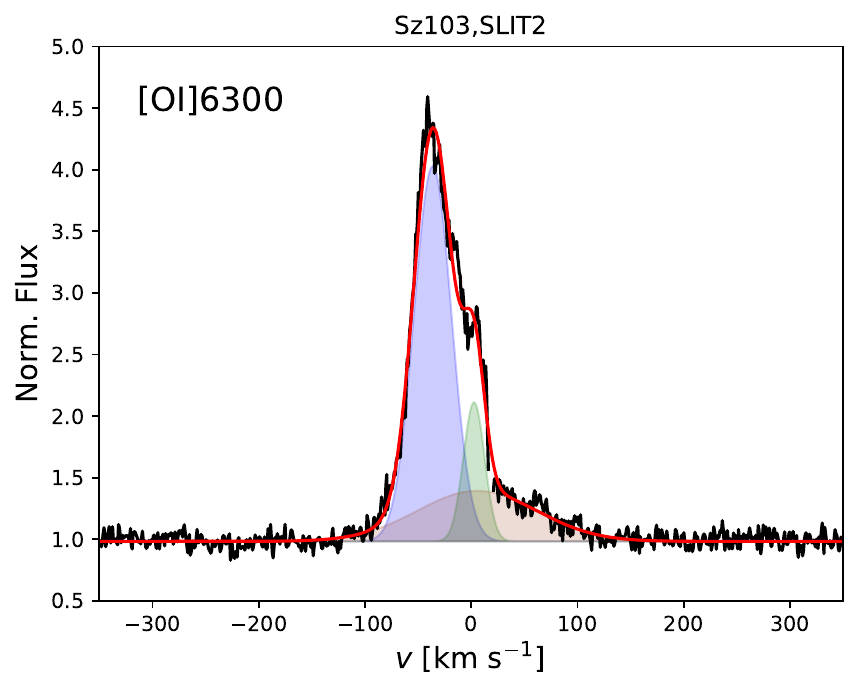}}
\hfill
\subfloat{\includegraphics[trim=0 0 0 0, clip, width=0.3 \textwidth]{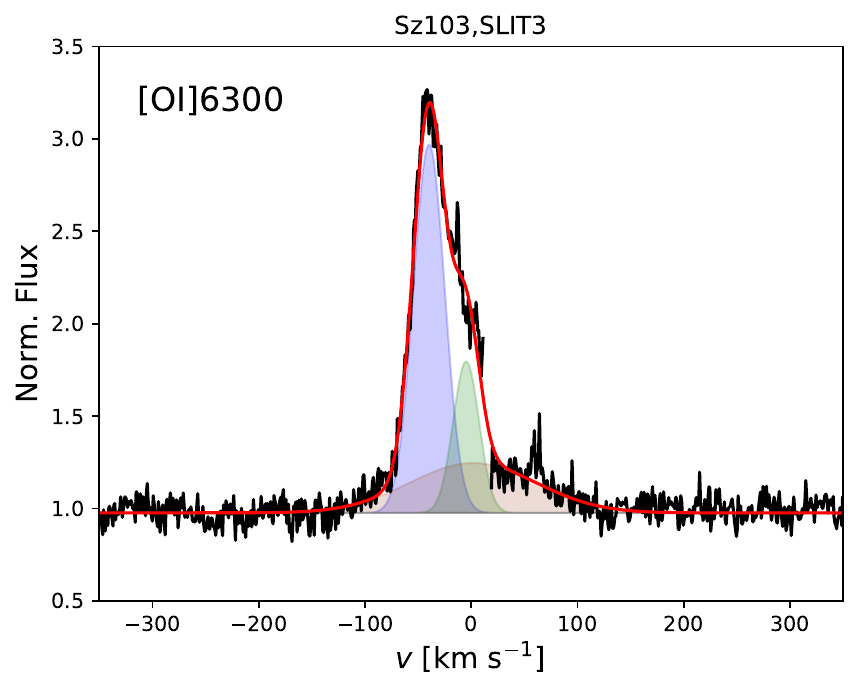}} 
\hfill 
\subfloat{\includegraphics[trim=0 0 0 0, clip, width=0.3 \textwidth]{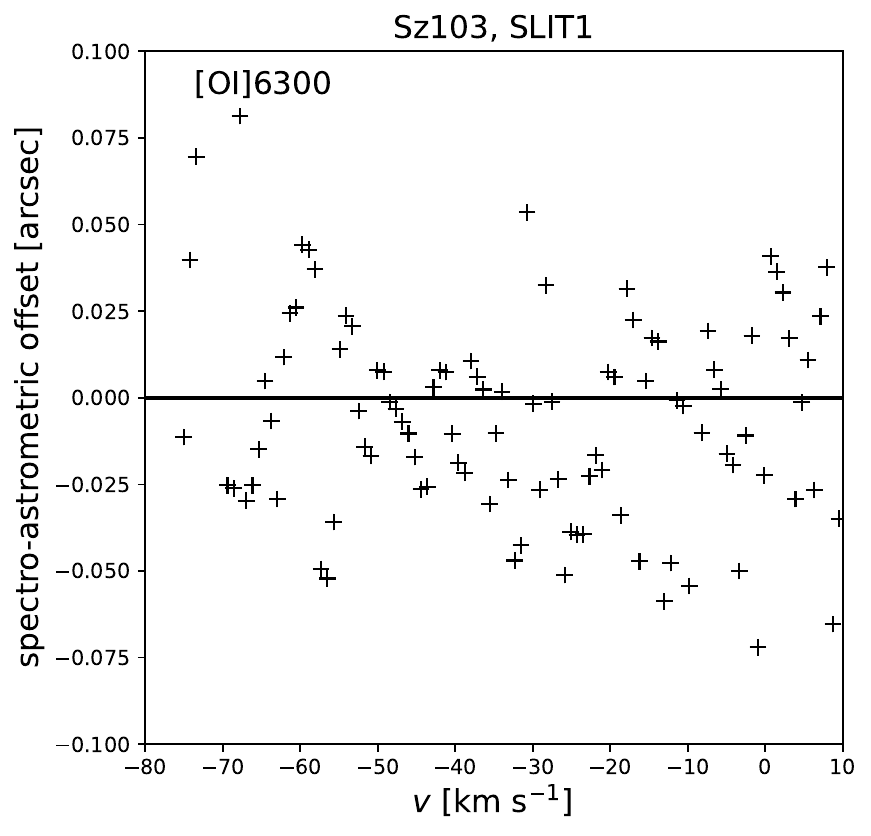}}
\hfill
\subfloat{\includegraphics[trim=0 0 0 0, clip, width=0.3 \textwidth]{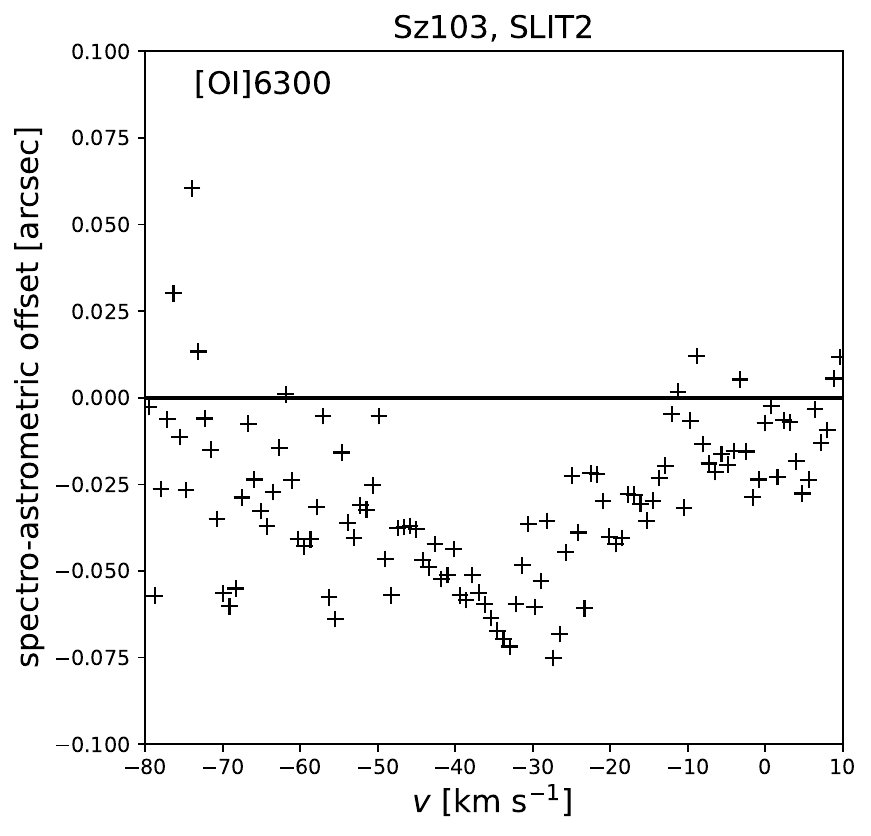}}
\hfill
\subfloat{\includegraphics[trim=0 0 0 0, clip, width=0.3 \textwidth]{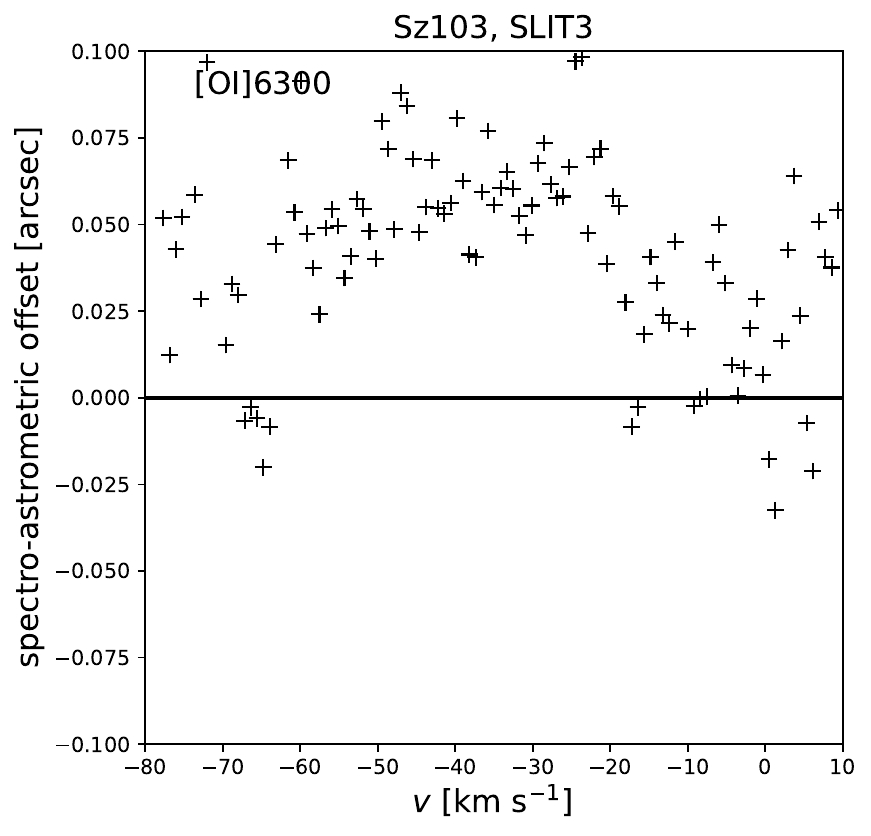}} 
\hfill
\caption{\small{Line profiles of H$\alpha$ and [OI]$\lambda$6300 for all slit positions of Sz\,103.}}\label{fig:all_minispectra_Sz103}
\end{figure*}

\begin{figure*} 
\centering 
\subfloat{\includegraphics[trim=0 0 0 0, clip, width=0.3 \textwidth]{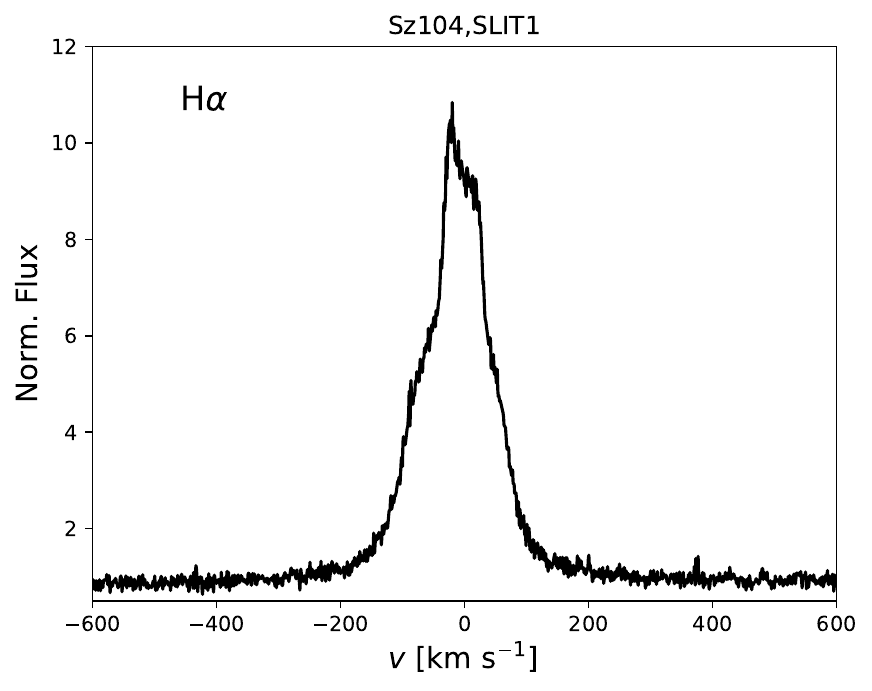}}
\hfill
\subfloat{\includegraphics[trim=0 0 0 0, clip, width=0.3 \textwidth]{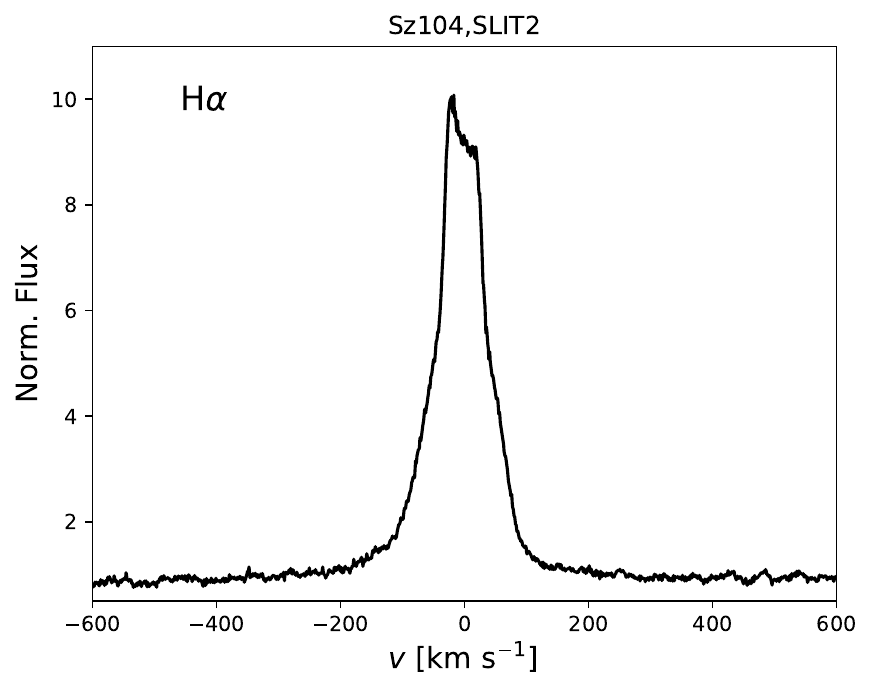}}
\hfill
\subfloat{\includegraphics[trim=0 0 0 0, clip, width=0.3 \textwidth]{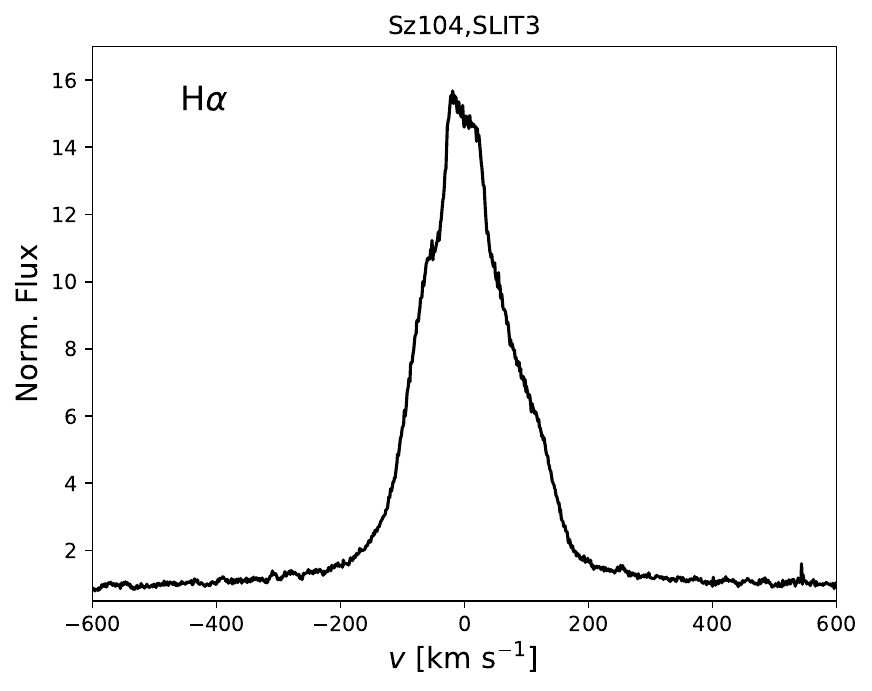}} 
\hfill 
\subfloat{\includegraphics[trim=0 0 0 0, clip, width=0.3 \textwidth]{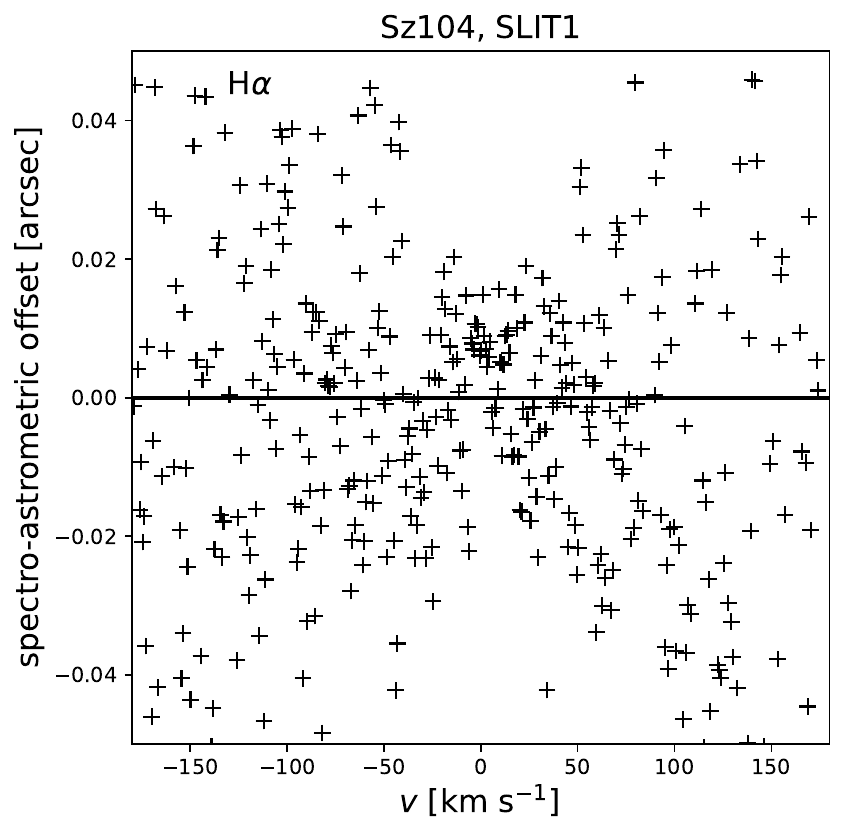}}
\hfill
\subfloat{\includegraphics[trim=0 0 0 0, clip, width=0.3 \textwidth]{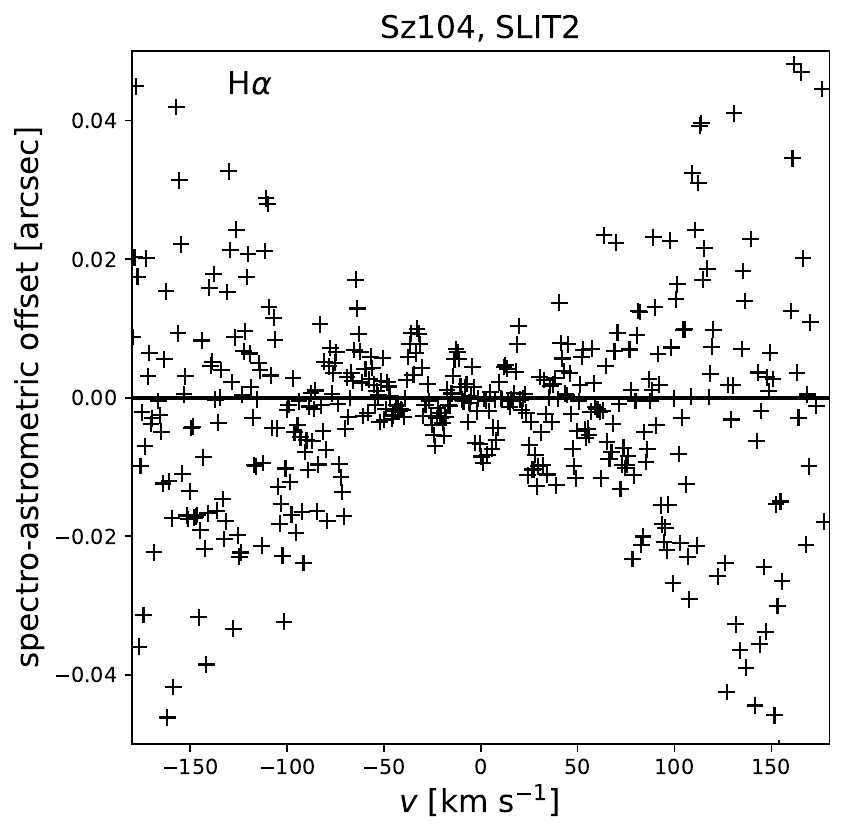}}
\hfill
\subfloat{\includegraphics[trim=0 0 0 0, clip, width=0.3 \textwidth]{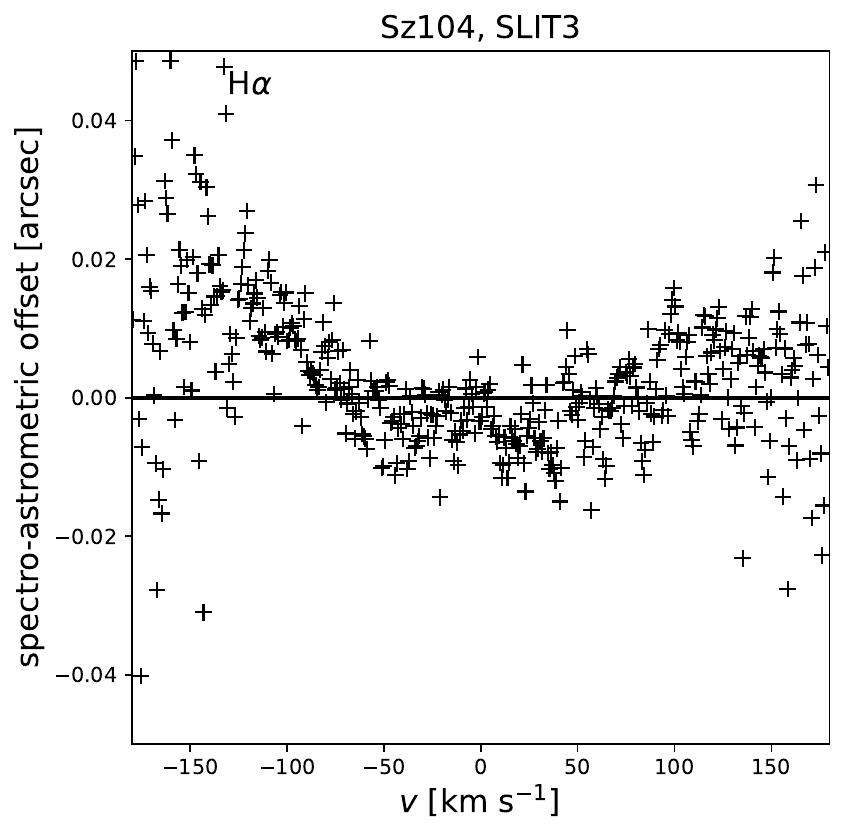}} 
\hfill
\subfloat{\includegraphics[trim=0 0 0 0, clip, width=0.3 \textwidth]{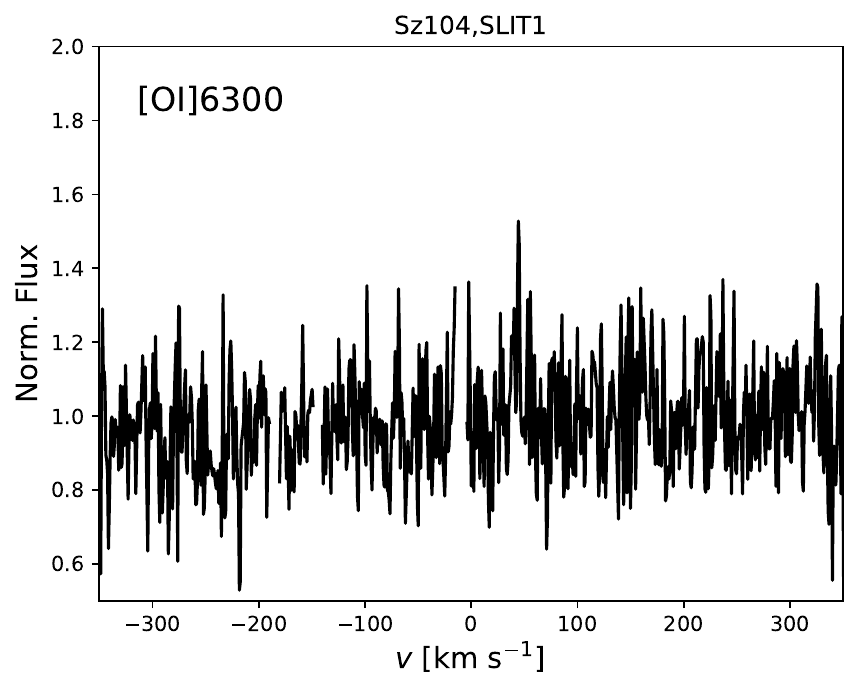}}
\hfill
\subfloat{\includegraphics[trim=0 0 0 0, clip, width=0.3 \textwidth]{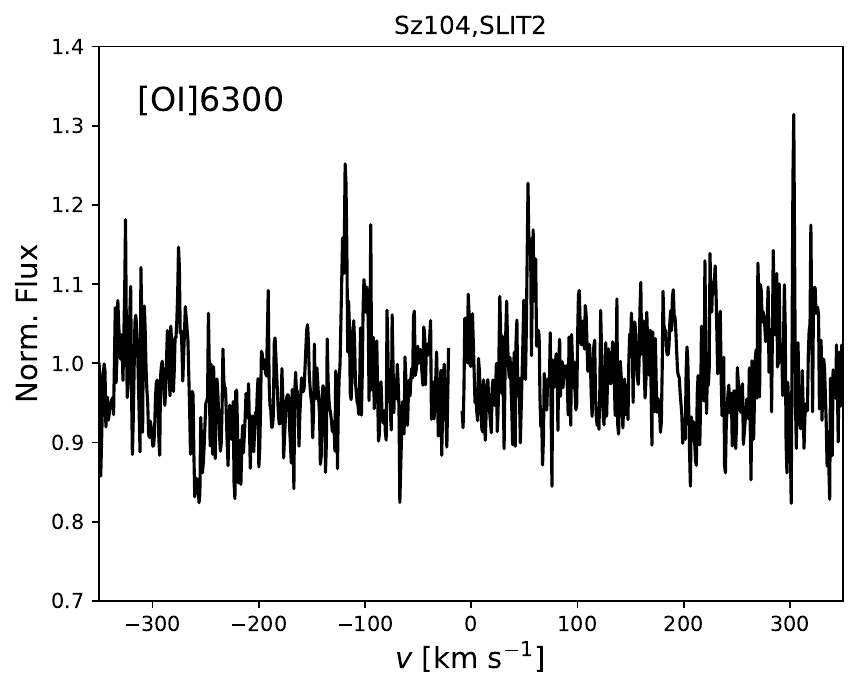}}
\hfill
\subfloat{\includegraphics[trim=0 0 0 0, clip, width=0.3 \textwidth]{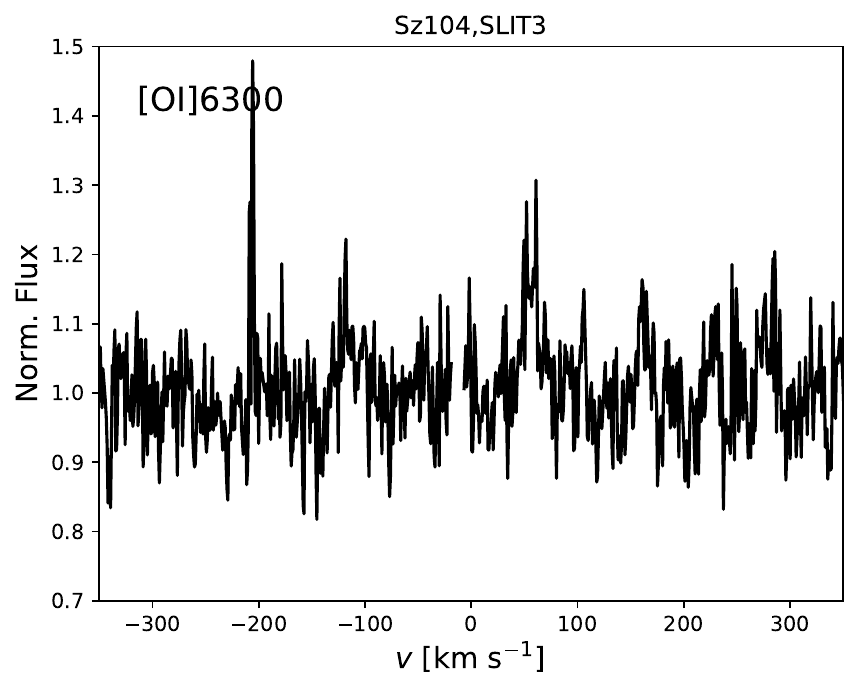}} 
\hfill 
\caption{\small{Line profiles of H$\alpha$ and [OI]$\lambda$6300 for all slit positions of Sz\,104.}}\label{fig:all_minispectra_Sz104}
\end{figure*}

\begin{figure*} 
\centering 
\subfloat{\includegraphics[trim=0 0 0 0, clip, width=0.3 \textwidth]{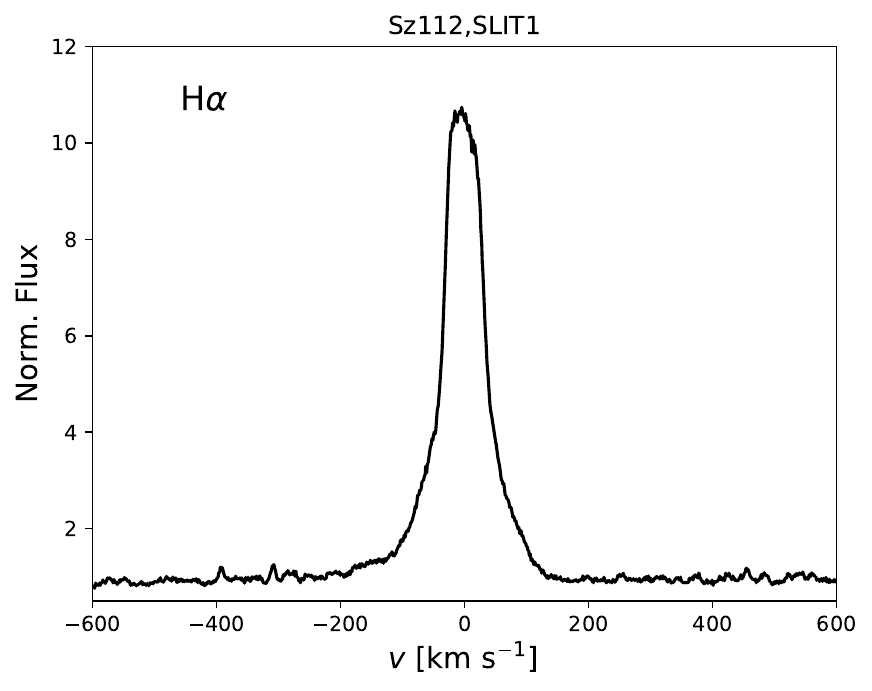}}
\hfill
\subfloat{\includegraphics[trim=0 0 0 0, clip, width=0.3 \textwidth]{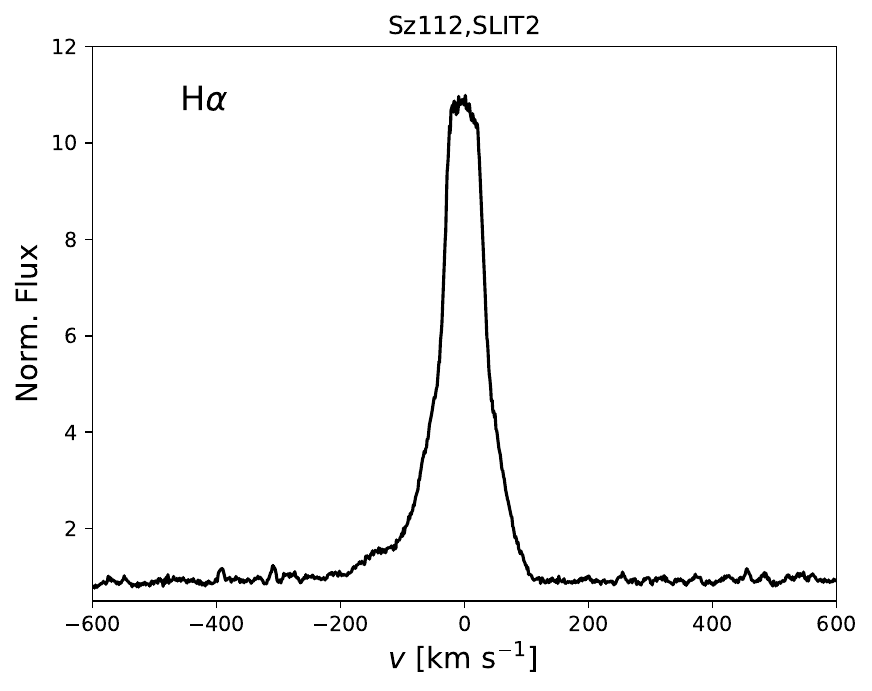}}
\hfill
\subfloat{\includegraphics[trim=0 0 0 0, clip, width=0.3 \textwidth]{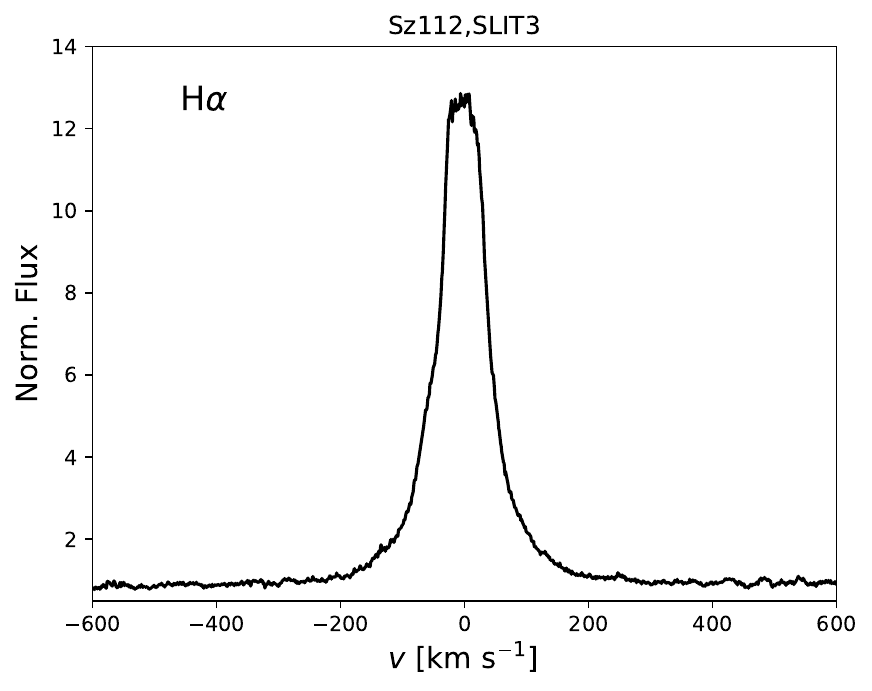}} 
\hfill
\subfloat{\includegraphics[trim=0 0 0 0, clip, width=0.3 \textwidth]{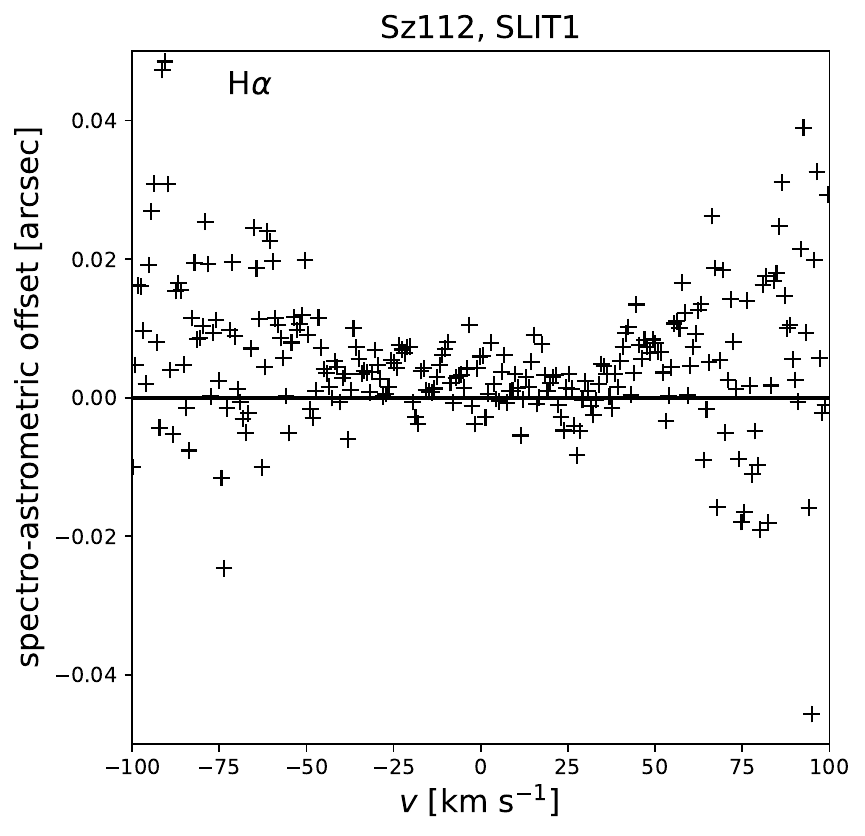}}
\hfill
\subfloat{\includegraphics[trim=0 0 0 0, clip, width=0.3 \textwidth]{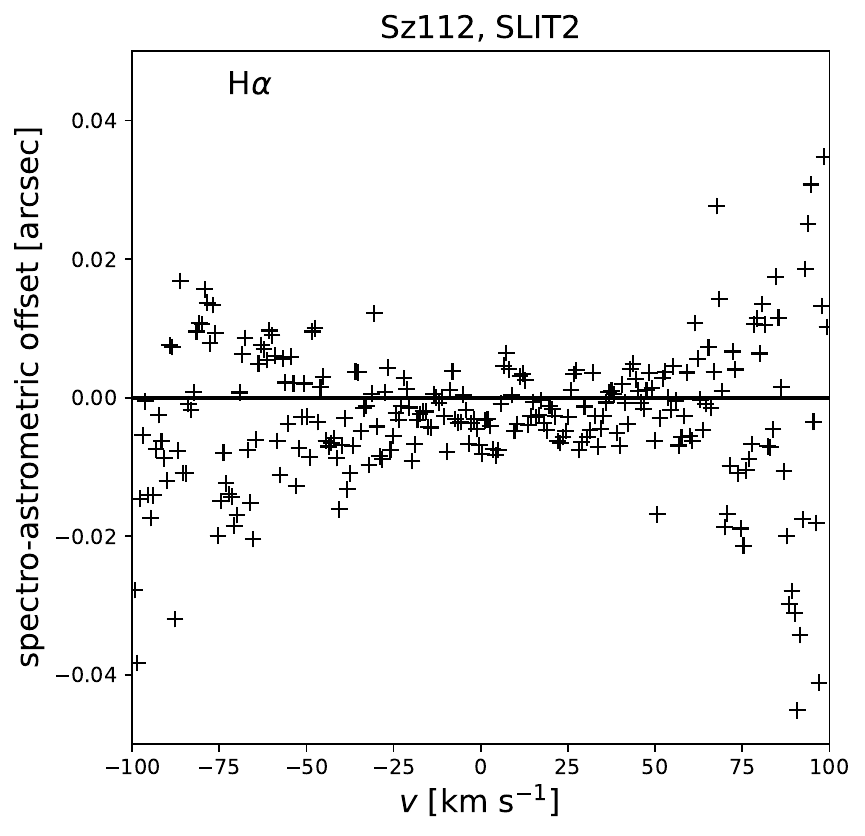}}
\hfill
\subfloat{\includegraphics[trim=0 0 0 0, clip, width=0.3 \textwidth]{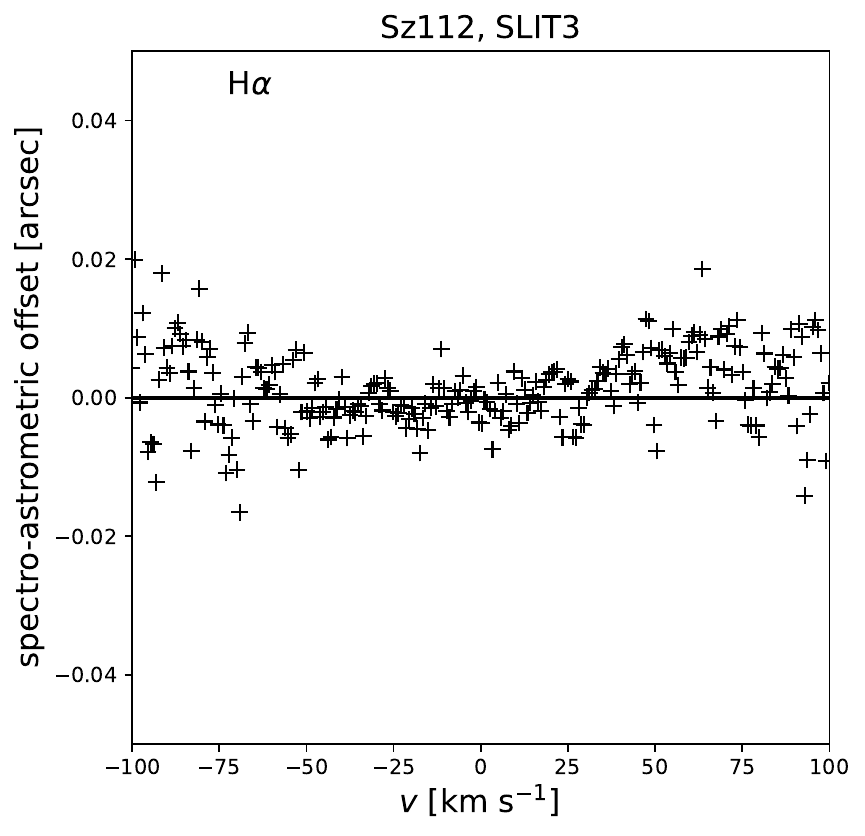}} 
\hfill
\subfloat{\includegraphics[trim=0 0 0 0, clip, width=0.3 \textwidth]{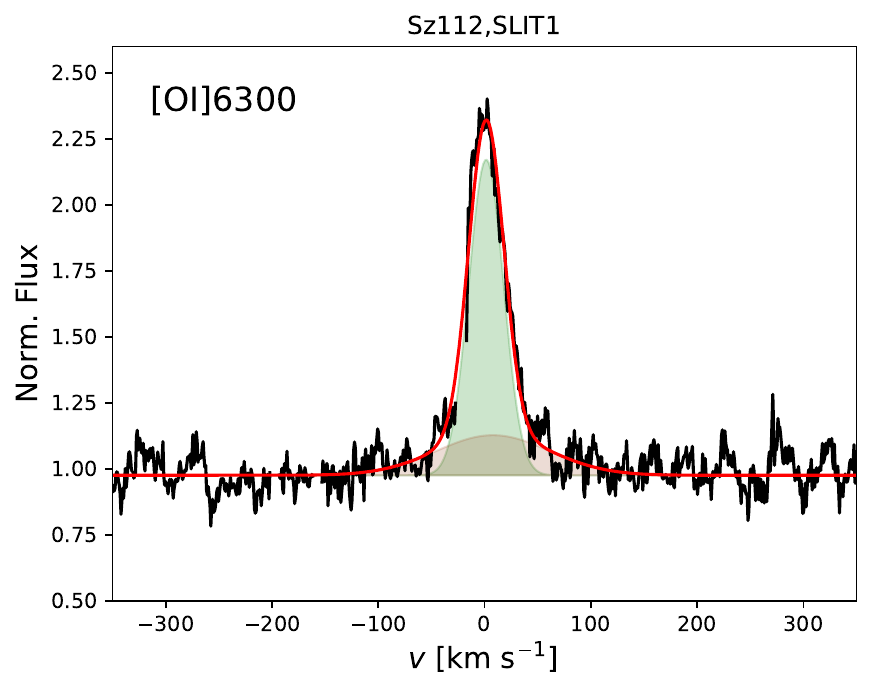}}
\hfill
\subfloat{\includegraphics[trim=0 0 0 0, clip, width=0.3 \textwidth]{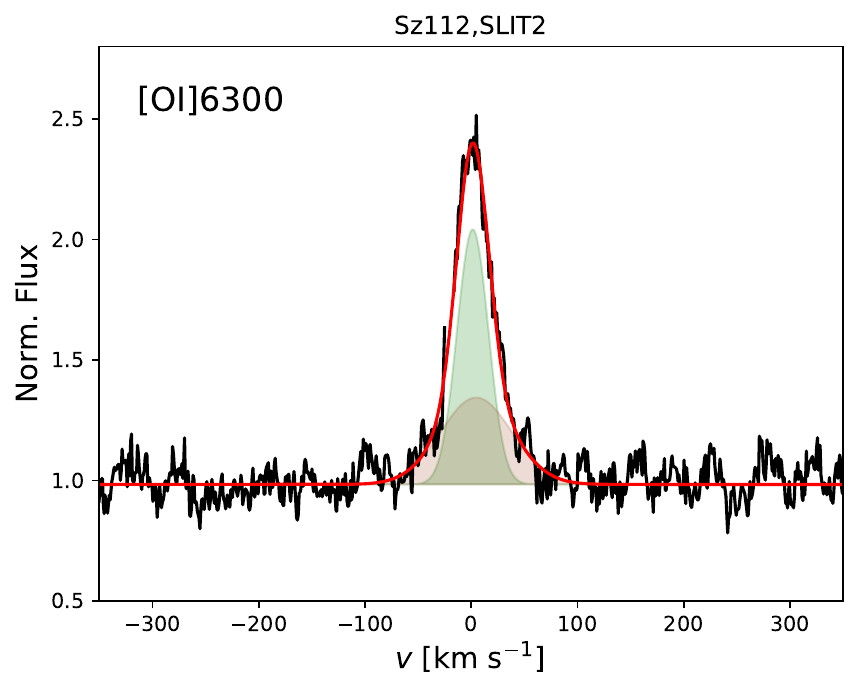}}
\hfill
\subfloat{\includegraphics[trim=0 0 0 0, clip, width=0.3 \textwidth]{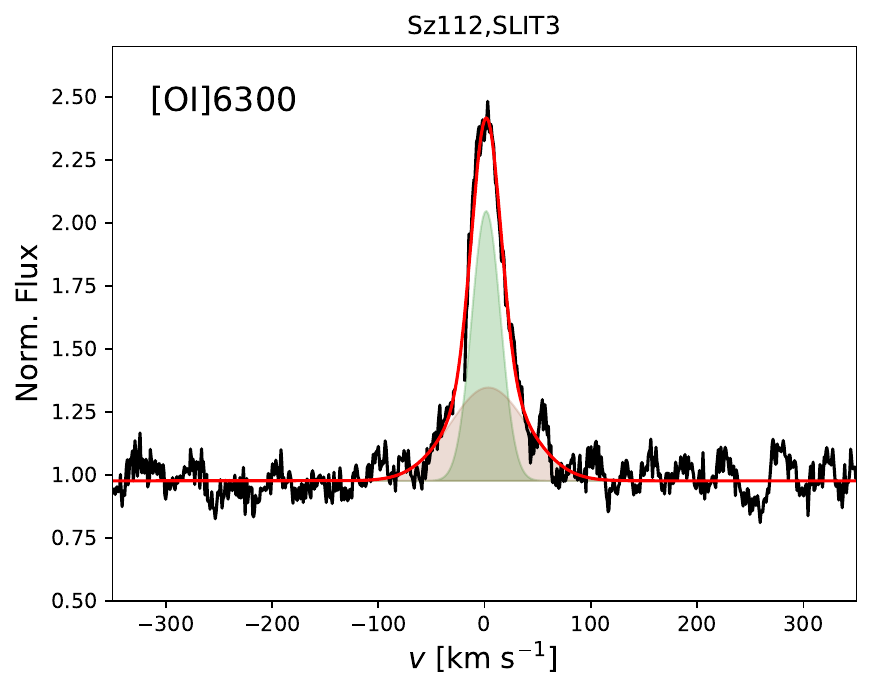}} 
\hfill 
\subfloat{\includegraphics[trim=0 0 0 0, clip, width=0.3 \textwidth]{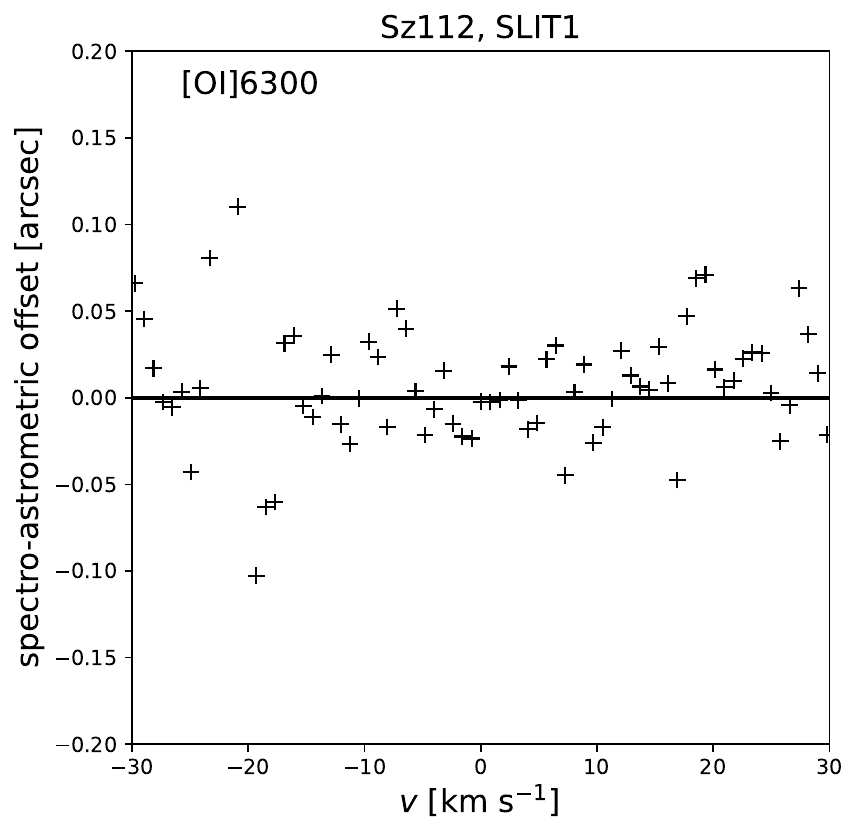}}
\hfill
\subfloat{\includegraphics[trim=0 0 0 0, clip, width=0.3 \textwidth]{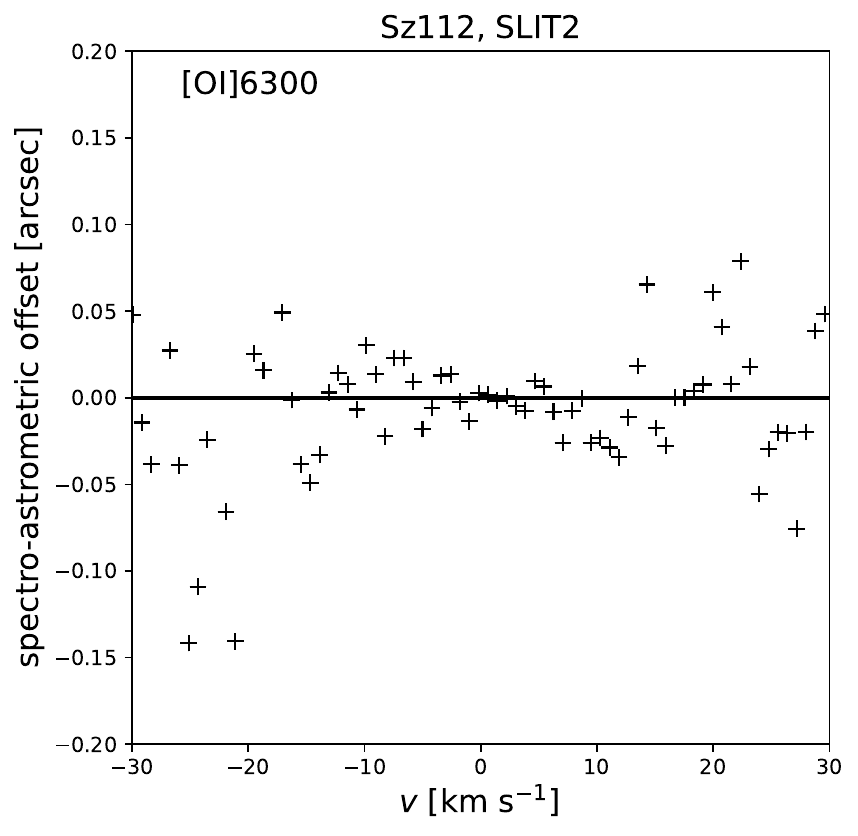}}
\hfill
\subfloat{\includegraphics[trim=0 0 0 0, clip, width=0.3 \textwidth]{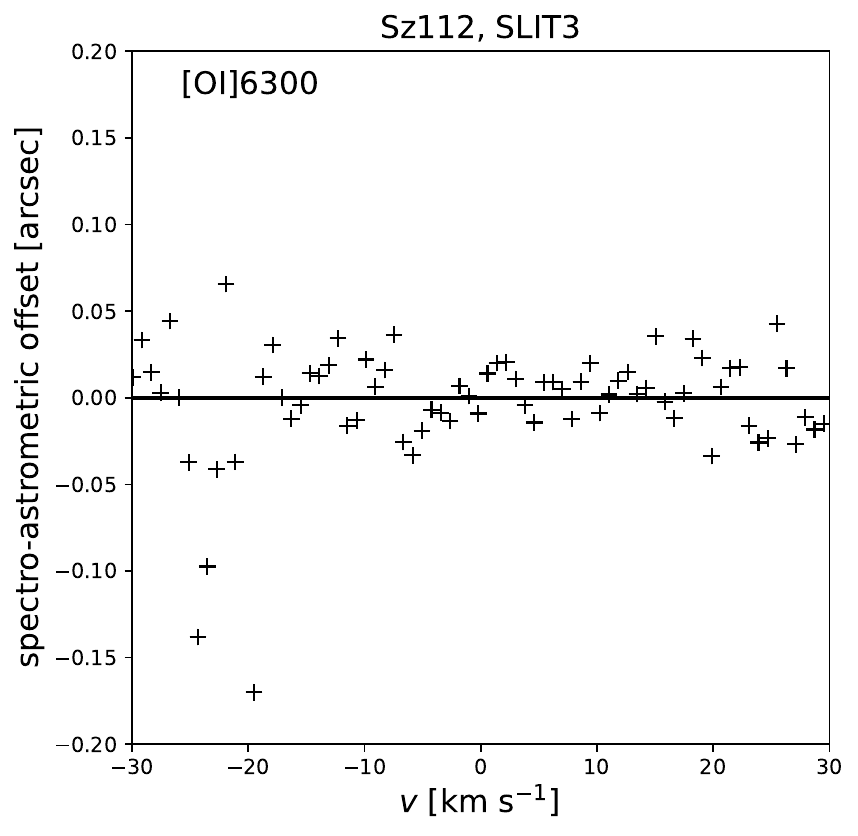}} 
\hfill
\caption{\small{Line profiles of H$\alpha$ and [OI]$\lambda$6300 for all slit positions of Sz\,112.}}\label{fig:all_minispectra_Sz112}
\end{figure*}

\begin{figure*} 
\centering 
\subfloat{\includegraphics[trim=0 0 0 0, clip, width=0.3 \textwidth]{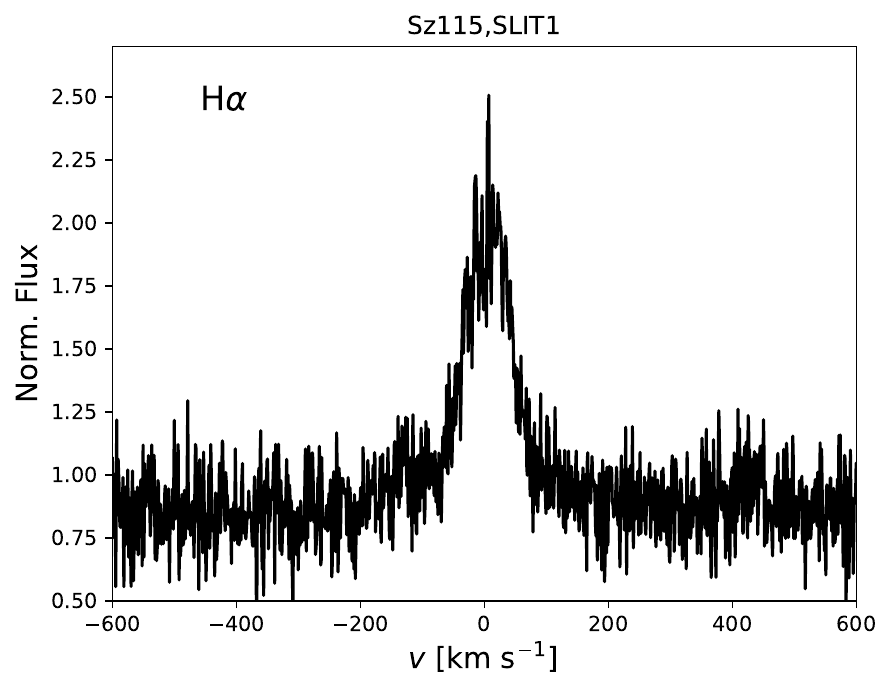}}
\hfill
\subfloat{\includegraphics[trim=0 0 0 0, clip, width=0.3 \textwidth]{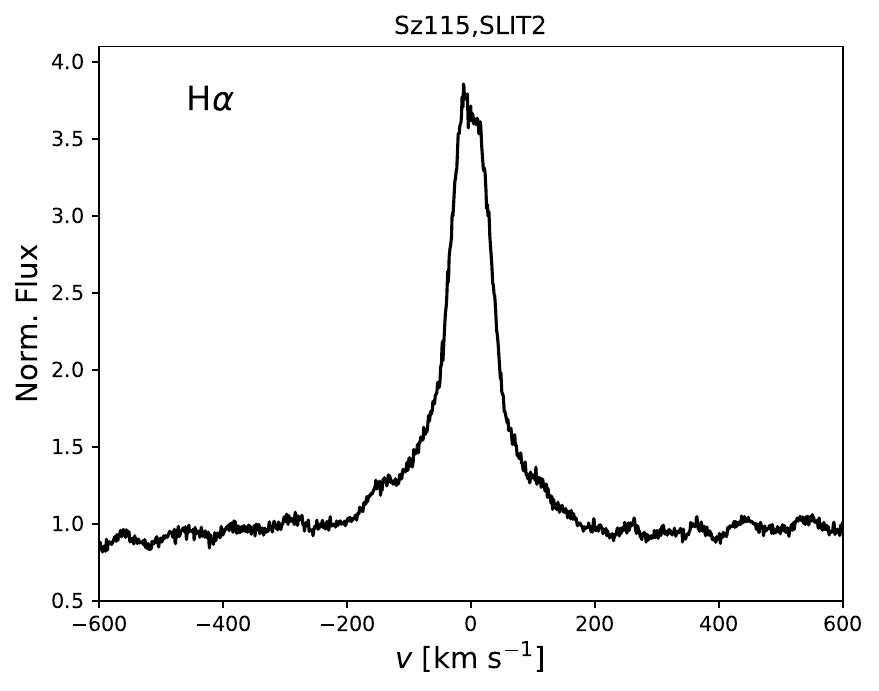}}
\hfill
\subfloat{\includegraphics[trim=0 0 0 0, clip, width=0.3 \textwidth]{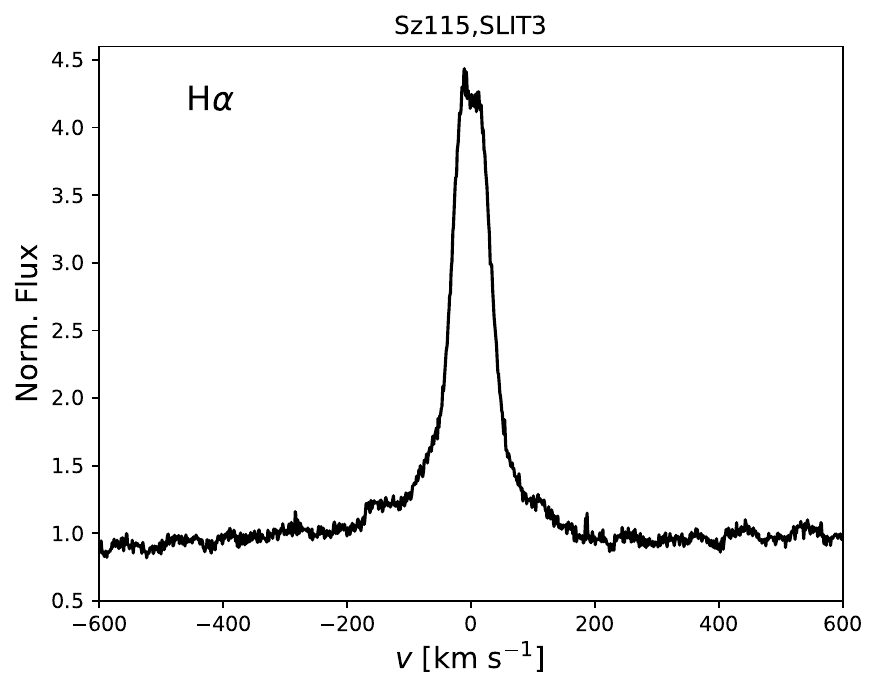}} 
\hfill
\subfloat{\includegraphics[trim=0 0 0 0, clip, width=0.3 \textwidth]{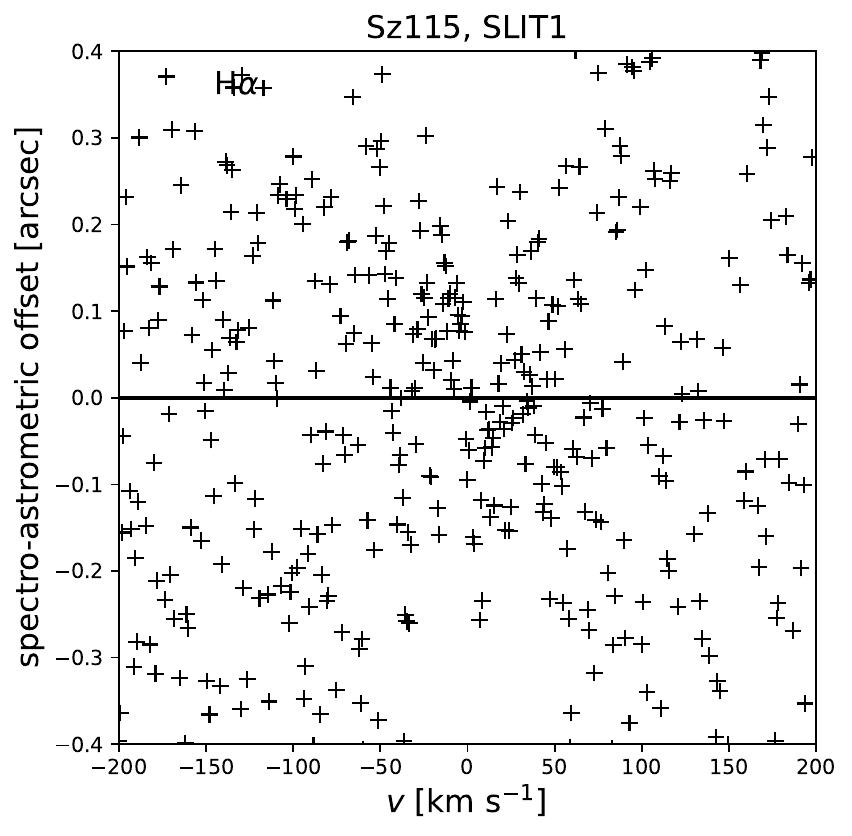}}
\hfill
\subfloat{\includegraphics[trim=0 0 0 0, clip, width=0.3 \textwidth]{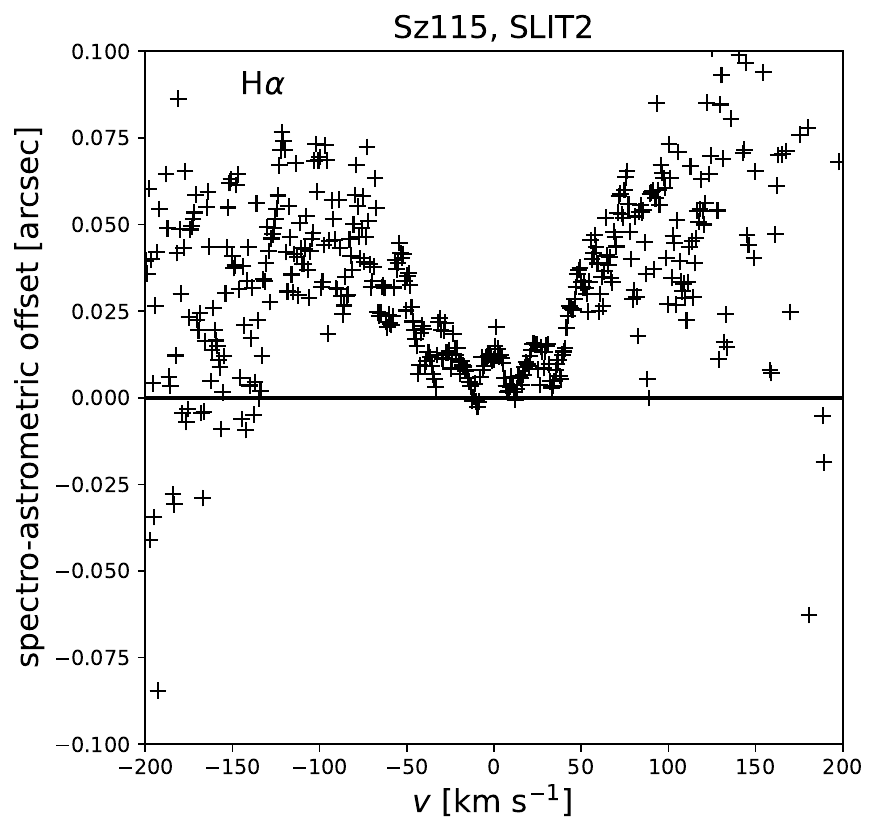}}
\hfill
\subfloat{\includegraphics[trim=0 0 0 0, clip, width=0.3 \textwidth]{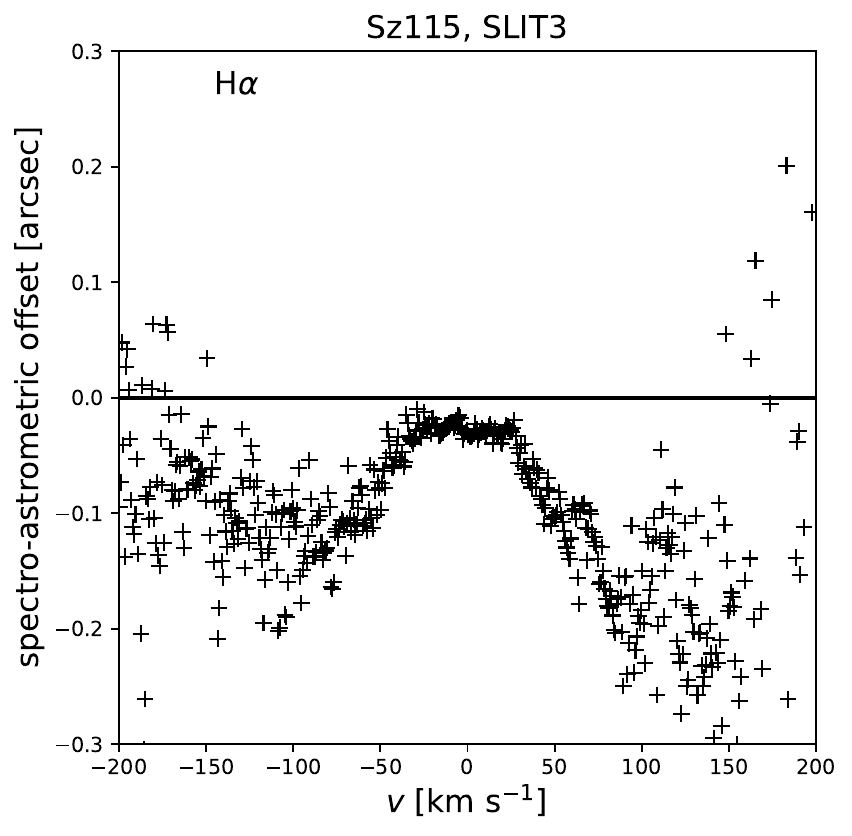}} 
\hfill
\subfloat{\includegraphics[trim=0 0 0 0, clip, width=0.3 \textwidth]{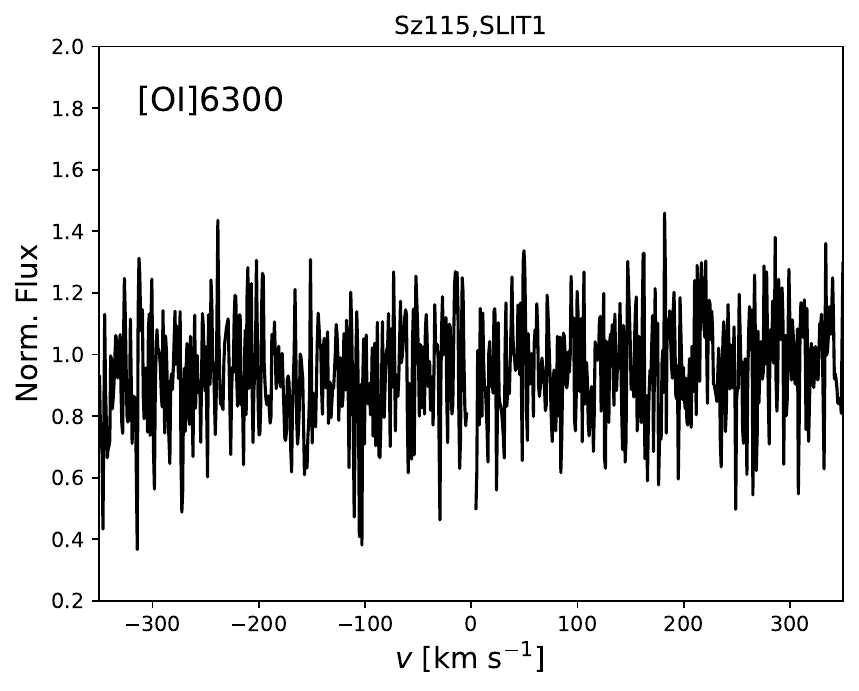}}
\hfill
\subfloat{\includegraphics[trim=0 0 0 0, clip, width=0.3 \textwidth]{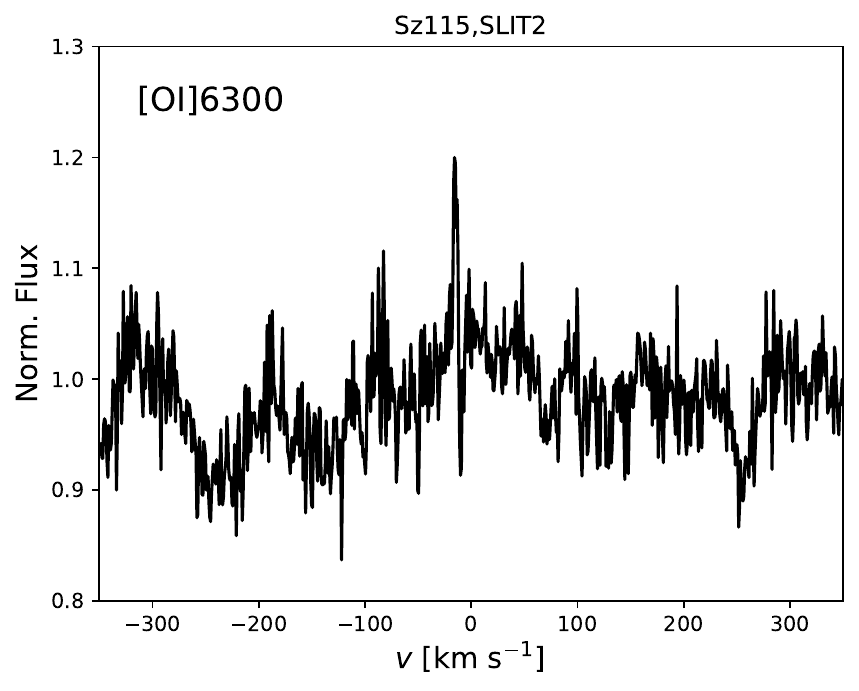}}
\hfill
\subfloat{\includegraphics[trim=0 0 0 0, clip, width=0.3 \textwidth]{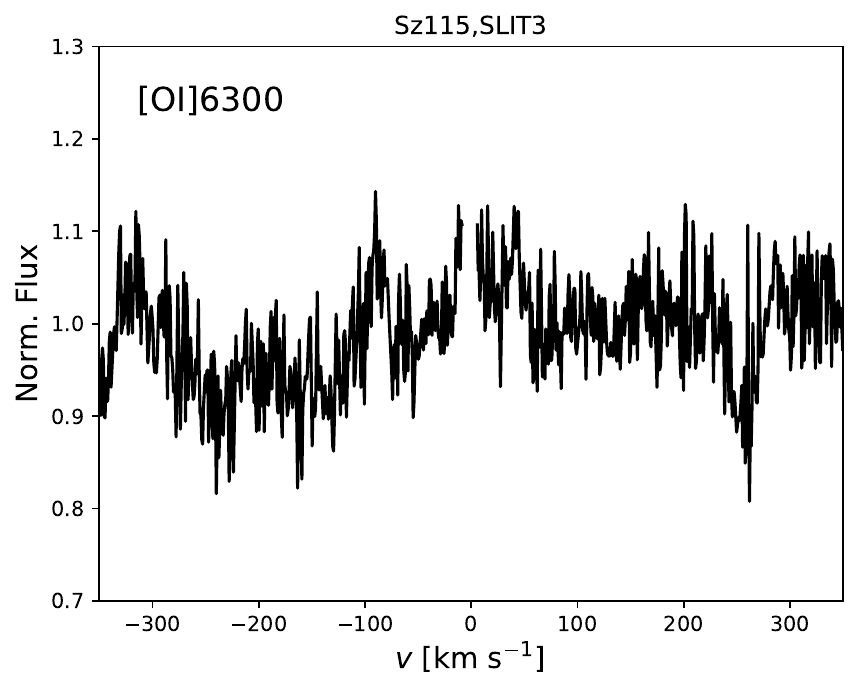}} 
\hfill 
\caption{\small{Line profiles of H$\alpha$ and [OI]$\lambda$6300 for all slit positions of Sz\,115.}}\label{fig:all_minispectra_Sz115}
\end{figure*}

\begin{figure*} 
\centering 
\subfloat{\includegraphics[trim=0 0 0 0, clip, width=0.3 \textwidth]{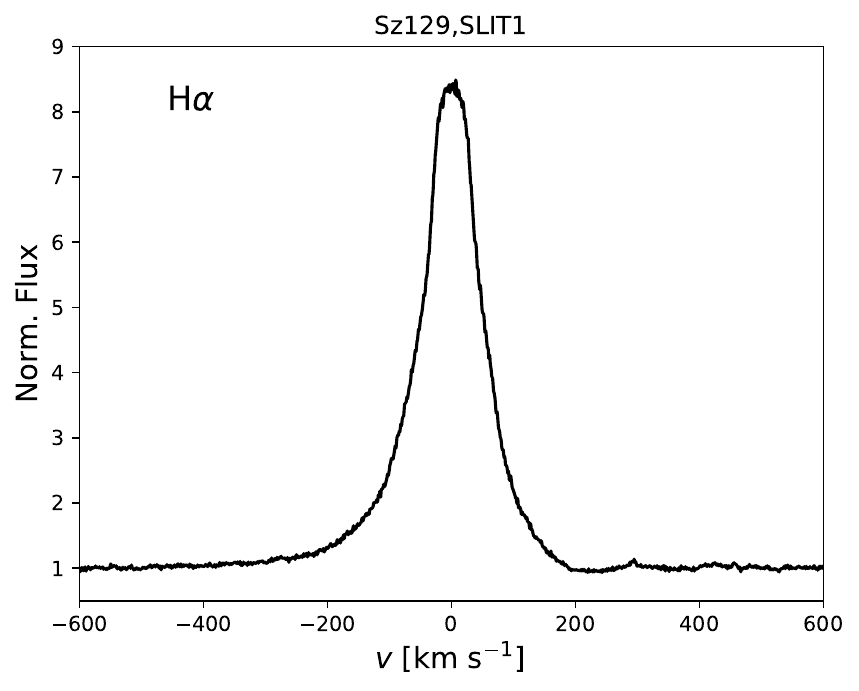}}
\hfill
\subfloat{\includegraphics[trim=0 0 0 0, clip, width=0.3 \textwidth]{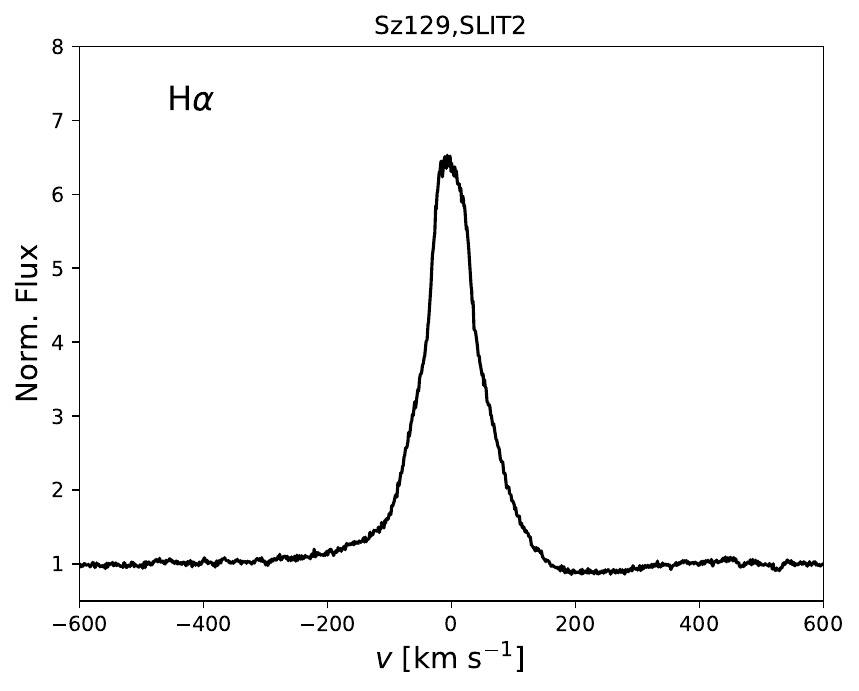}}
\hfill
\subfloat{\includegraphics[trim=0 0 0 0, clip, width=0.3 \textwidth]{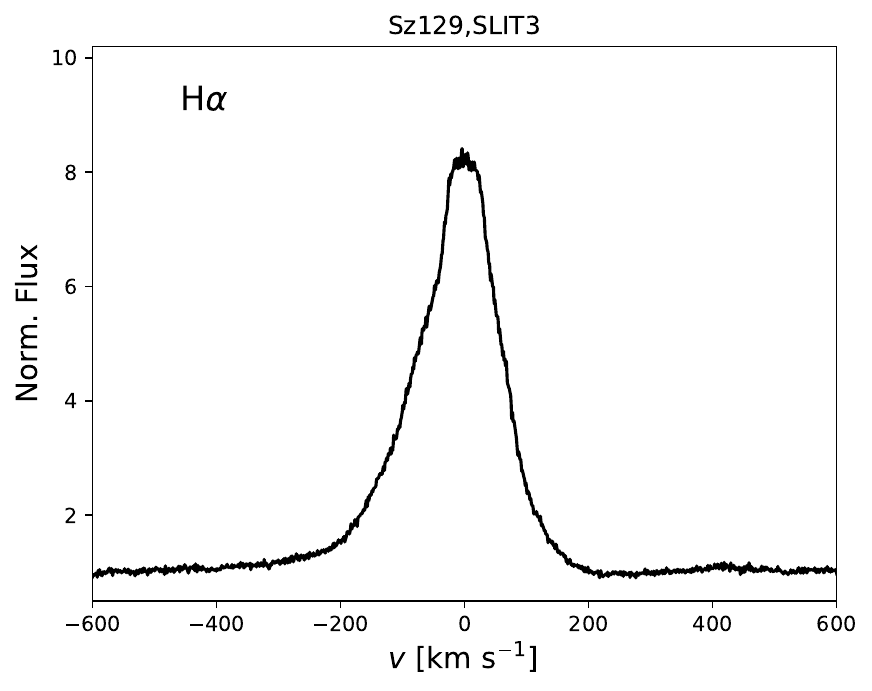}} 
\hfill
\subfloat{\includegraphics[trim=0 0 0 0, clip, width=0.3 \textwidth]{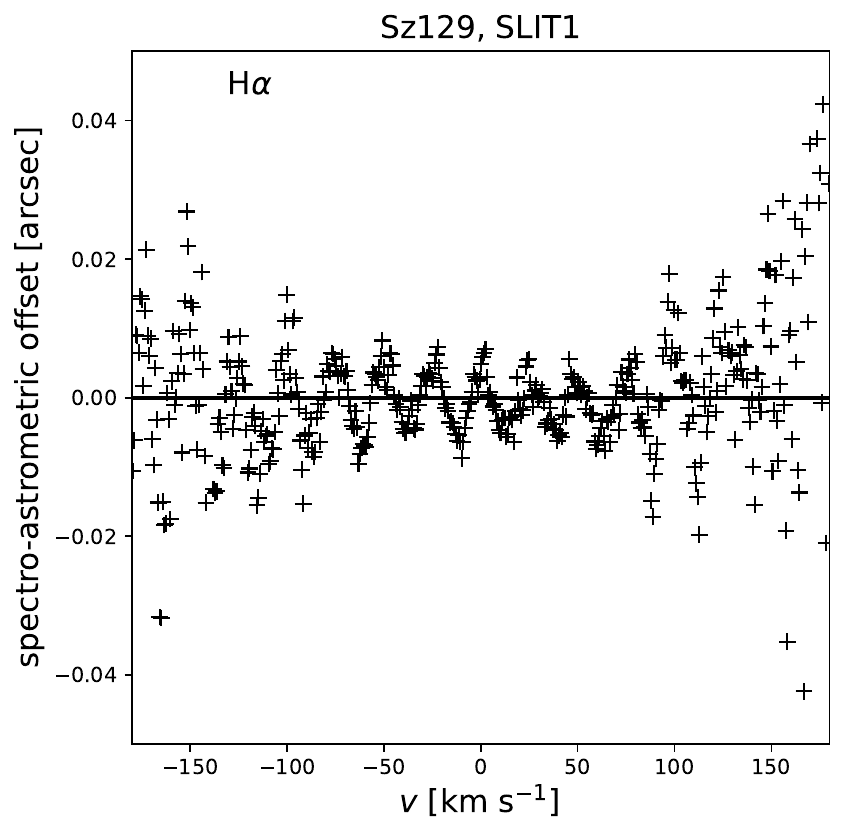}}
\hfill
\subfloat{\includegraphics[trim=0 0 0 0, clip, width=0.3 \textwidth]{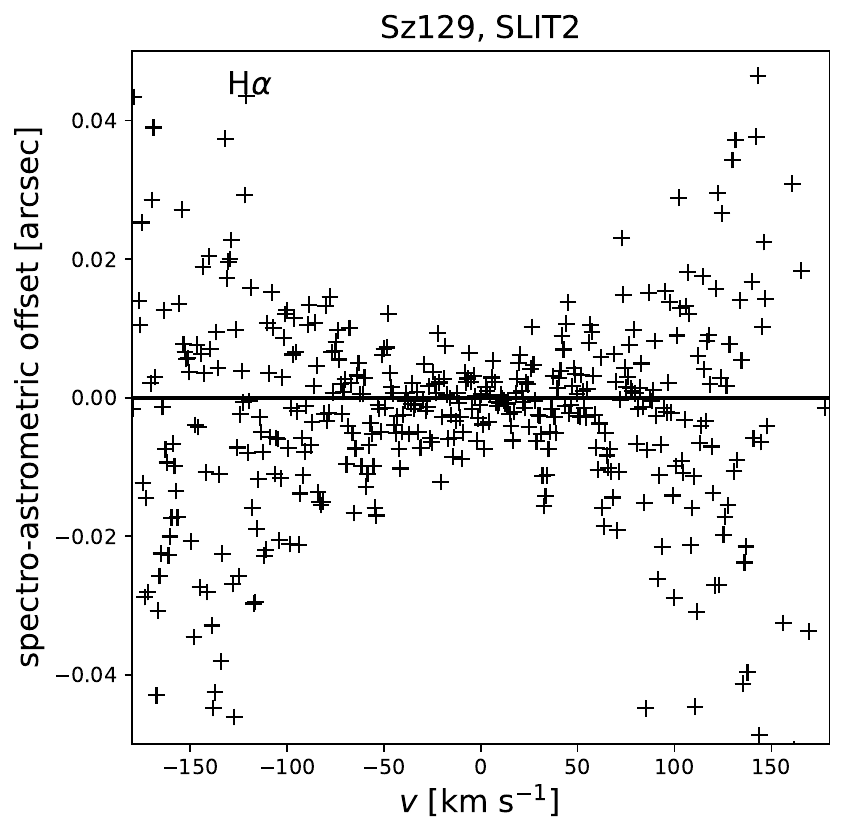}}
\hfill
\subfloat{\includegraphics[trim=0 0 0 0, clip, width=0.3 \textwidth]{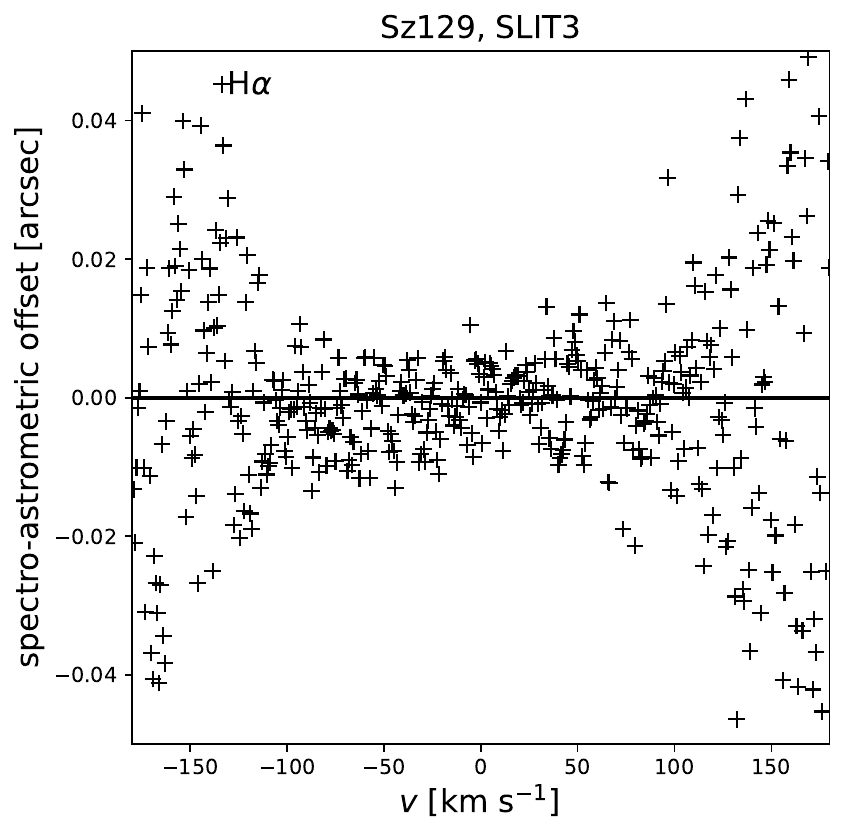}} 
\hfill
\subfloat{\includegraphics[trim=0 0 0 0, clip, width=0.3 \textwidth]{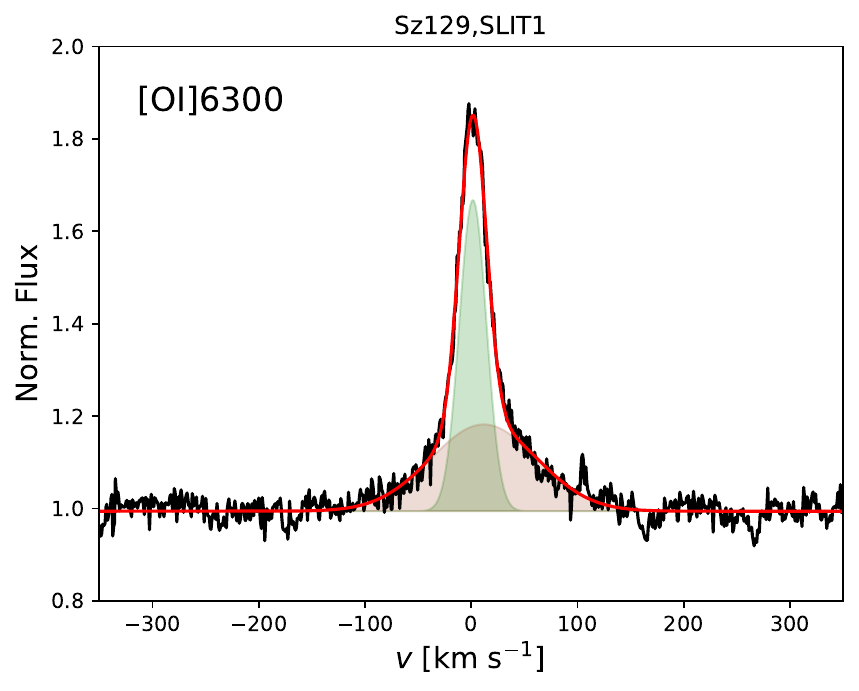}}
\hfill
\subfloat{\includegraphics[trim=0 0 0 0, clip, width=0.3 \textwidth]{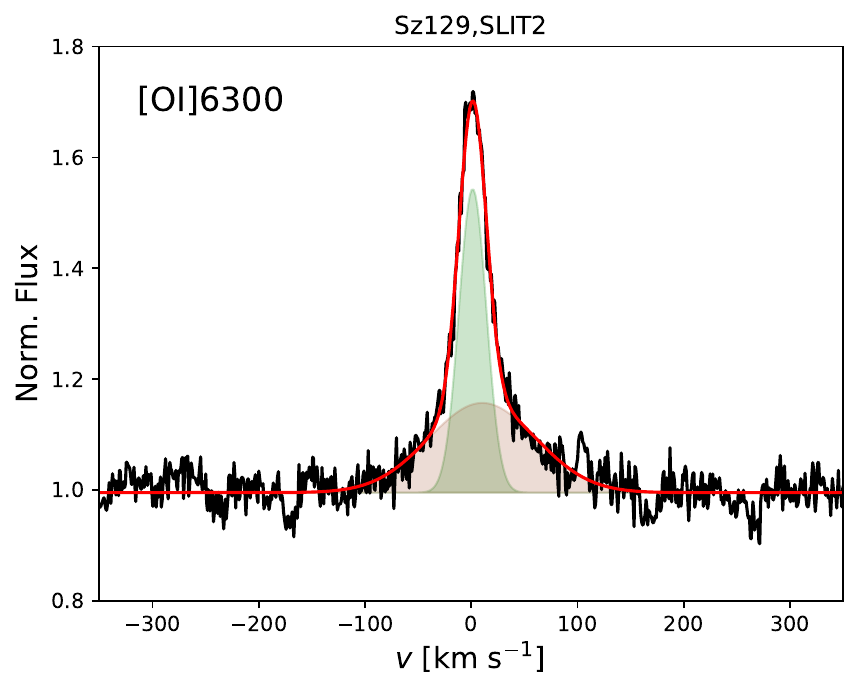}}
\hfill
\subfloat{\includegraphics[trim=0 0 0 0, clip, width=0.3 \textwidth]{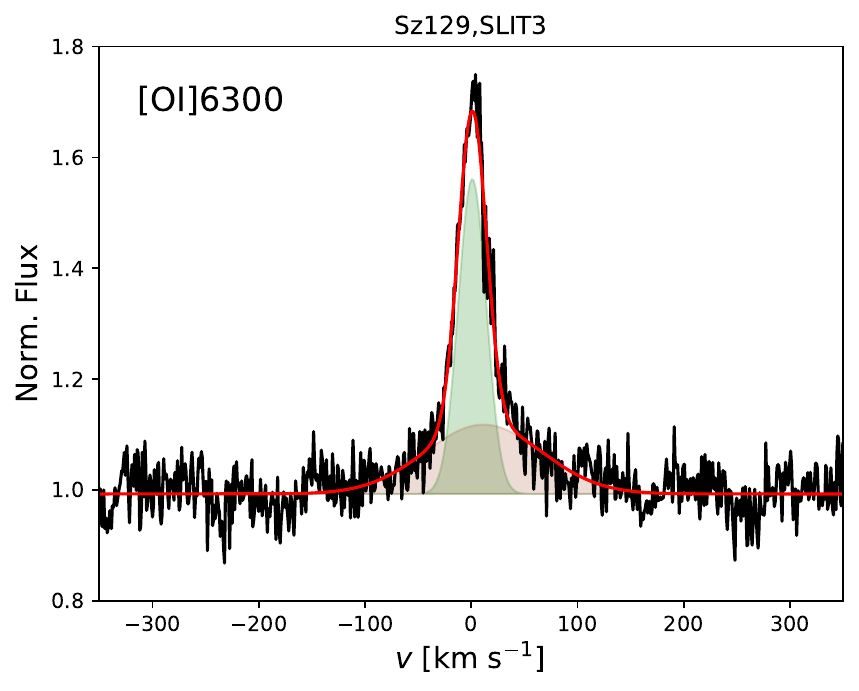}} 
\hfill 
\subfloat{\includegraphics[trim=0 0 0 0, clip, width=0.3 \textwidth]{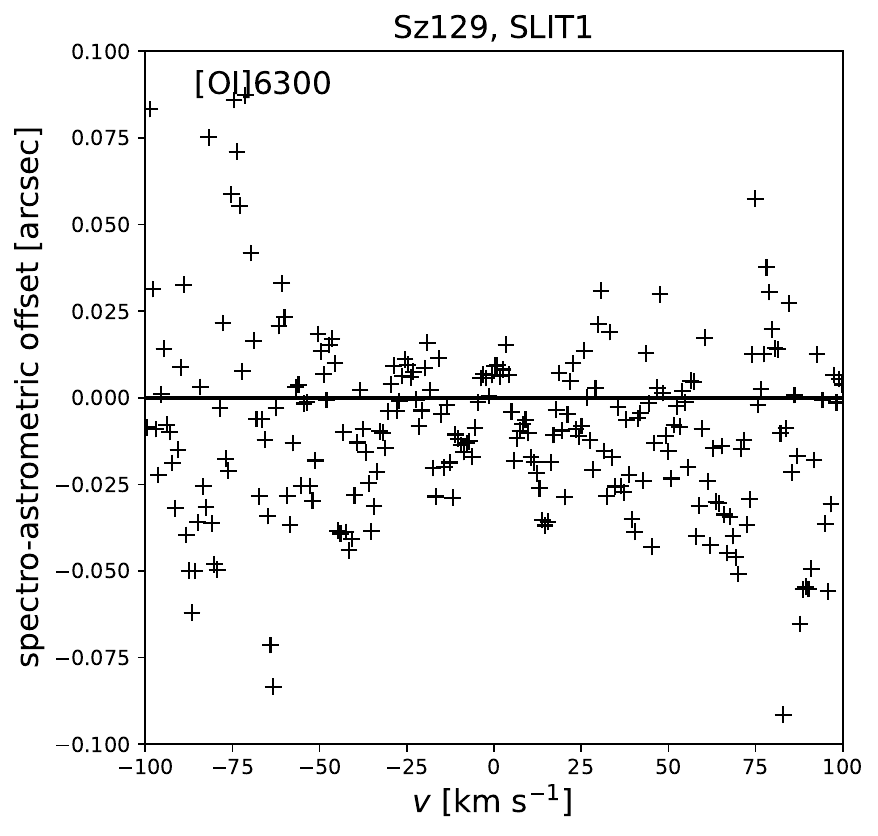}}
\hfill
\subfloat{\includegraphics[trim=0 0 0 0, clip, width=0.3 \textwidth]{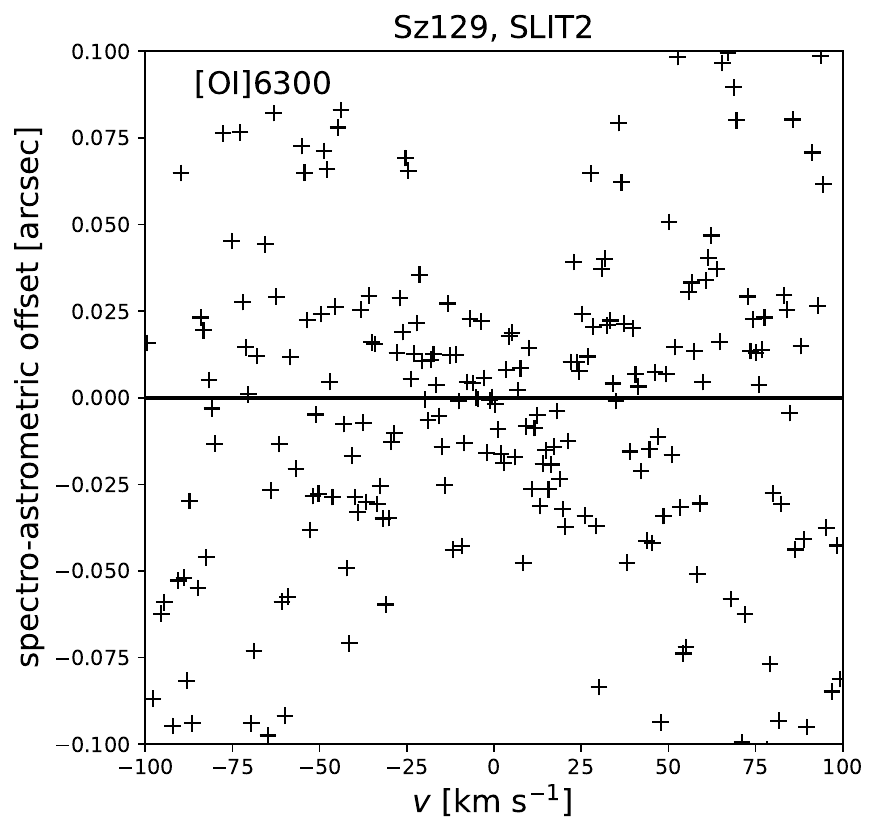}}
\hfill
\subfloat{\includegraphics[trim=0 0 0 0, clip, width=0.3 \textwidth]{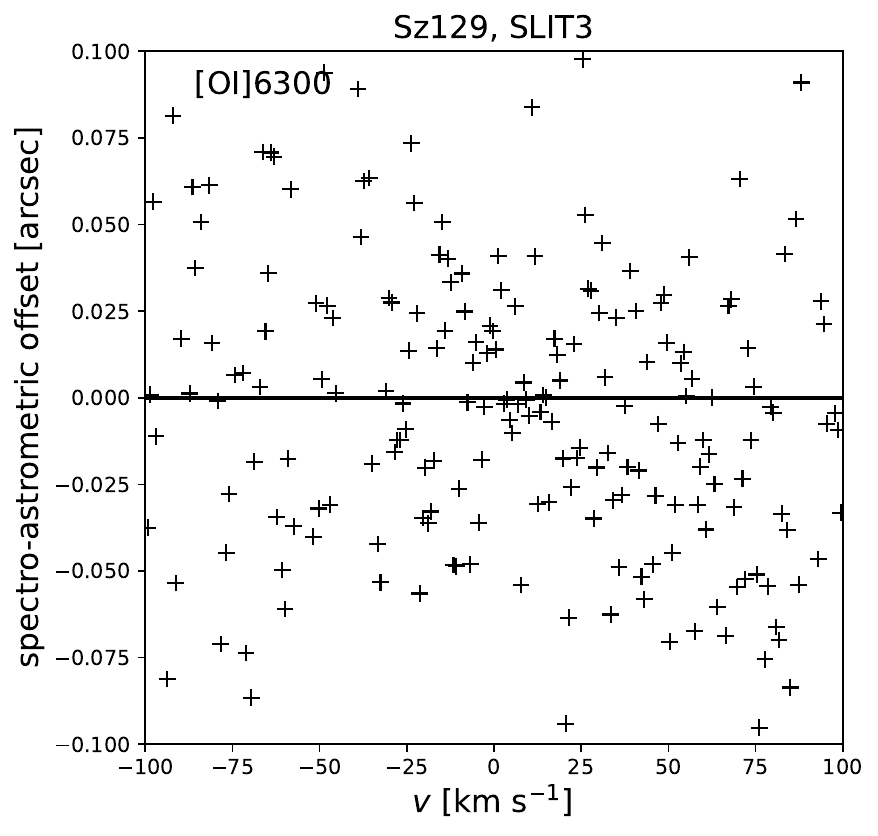}} 
\hfill
\caption{\small{Line profiles of H$\alpha$ and [OI]$\lambda$6300 for all slit positions of Sz\,129.}}\label{fig:all_minispectra_Sz129}
\end{figure*} 

\begin{figure*} 
\centering 
\subfloat{\includegraphics[trim=0 0 0 0, clip, width=0.3 \textwidth]{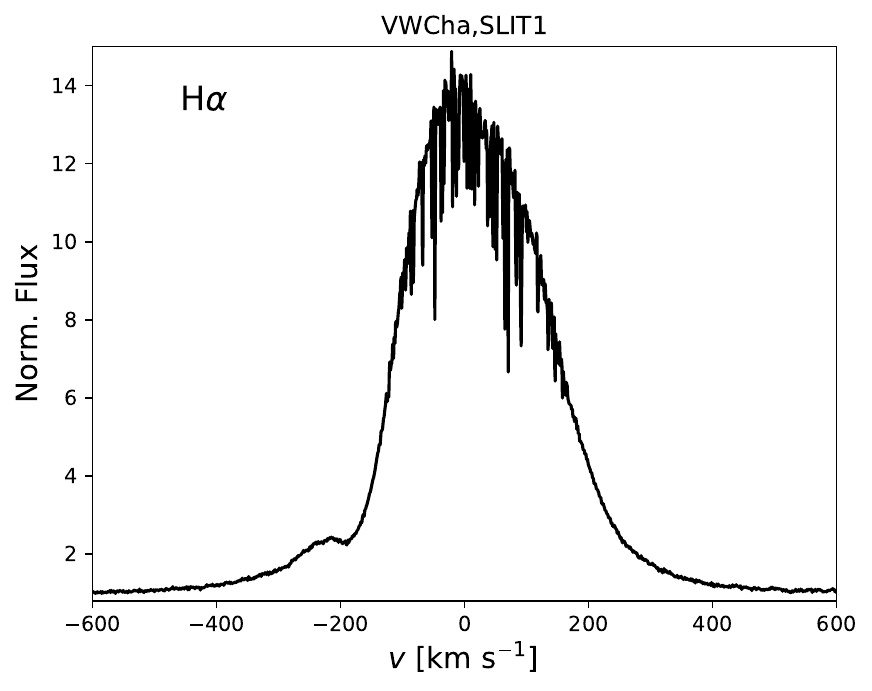}}
\hfill
\subfloat{\includegraphics[trim=0 0 0 0, clip, width=0.3 \textwidth]{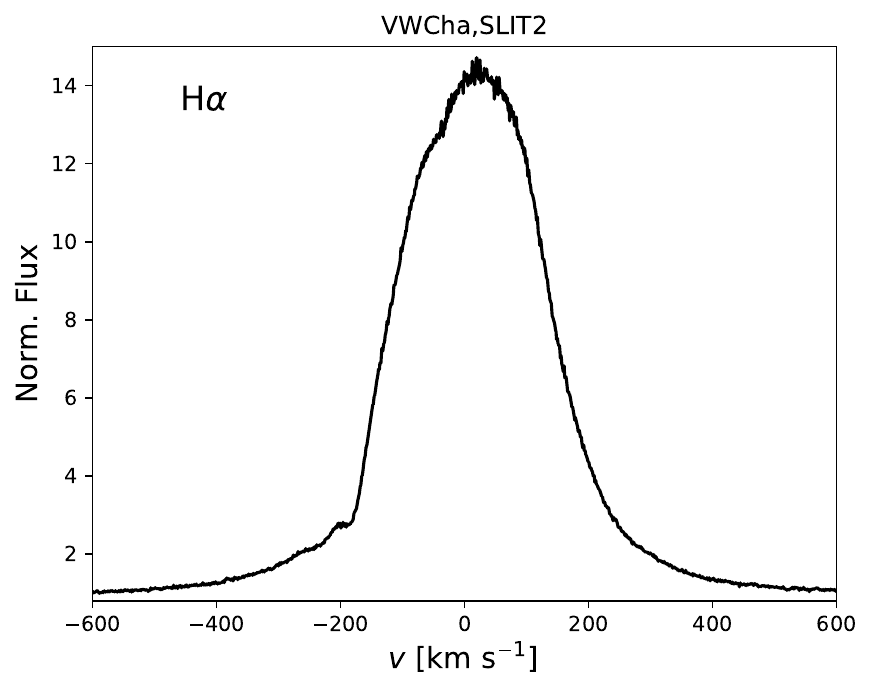}}
\hfill
\subfloat{\includegraphics[trim=0 0 0 0, clip, width=0.3 \textwidth]{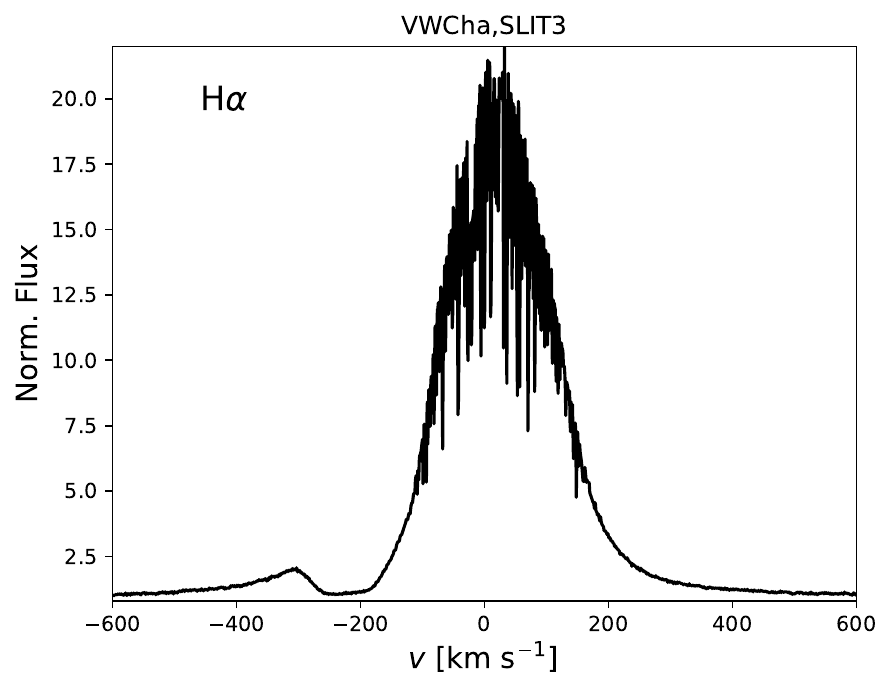}} 
\hfill 
\subfloat{\includegraphics[trim=0 0 0 0, clip, width=0.3 \textwidth]{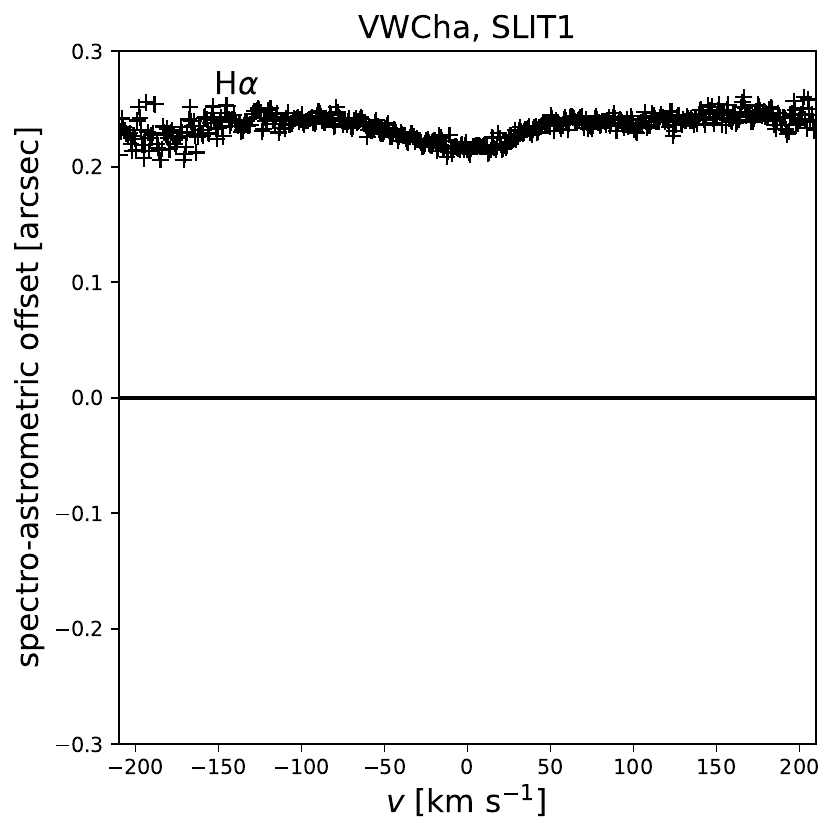}}
\hfill
\subfloat{\includegraphics[trim=0 0 0 0, clip, width=0.3 \textwidth]{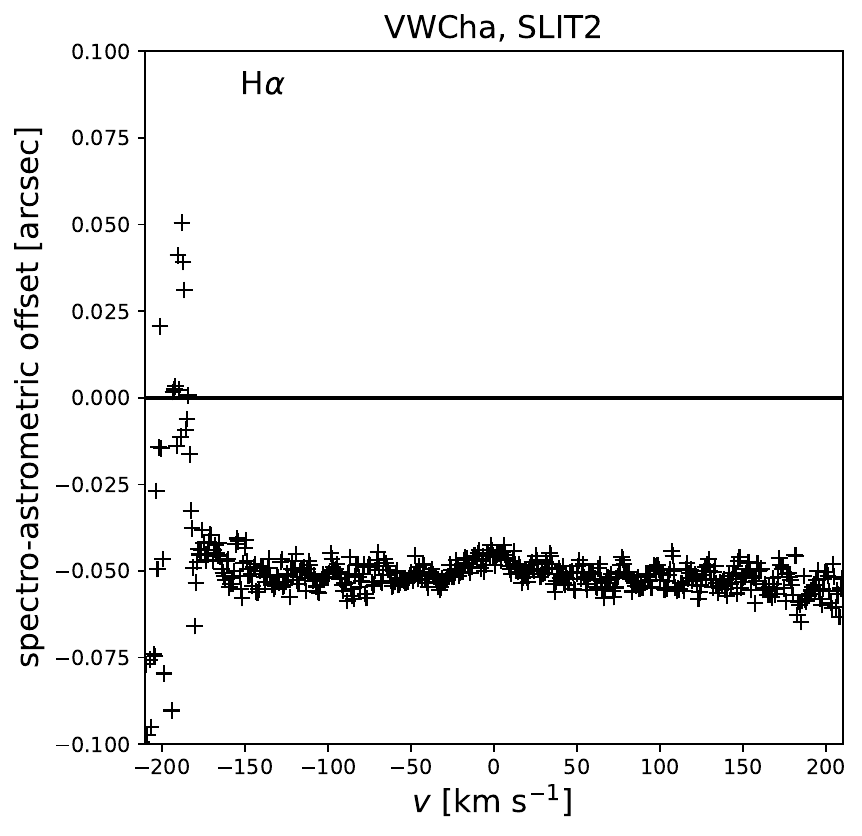}}
\hfill
\subfloat{\includegraphics[trim=0 0 0 0, clip, width=0.3 \textwidth]{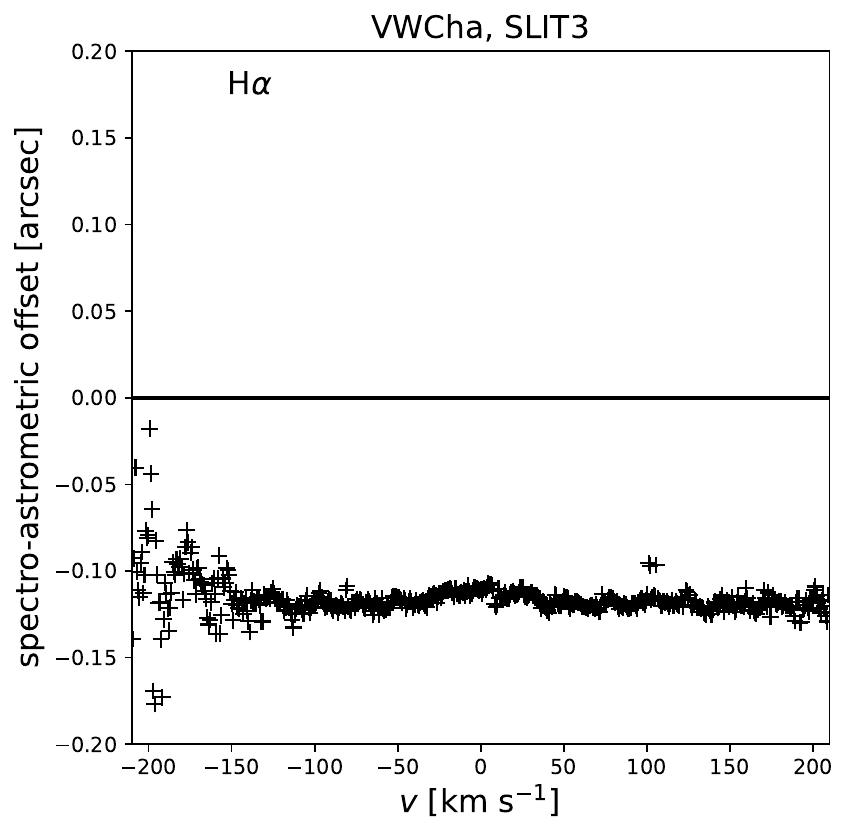}} 
\hfill
\subfloat{\includegraphics[trim=0 0 0 0, clip, width=0.3 \textwidth]{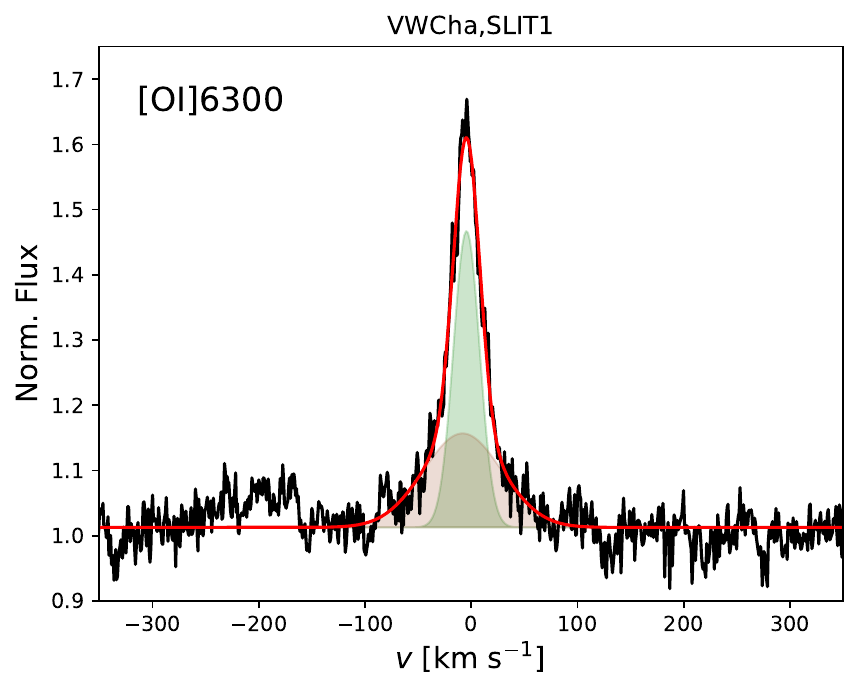}}
\hfill
\subfloat{\includegraphics[trim=0 0 0 0, clip, width=0.3 \textwidth]{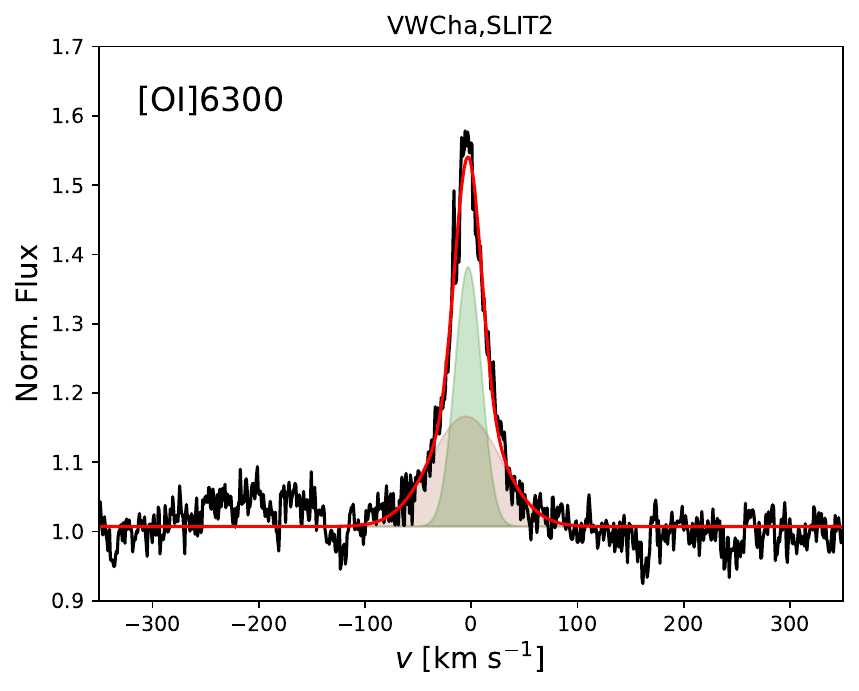}}
\hfill
\subfloat{\includegraphics[trim=0 0 0 0, clip, width=0.3 \textwidth]{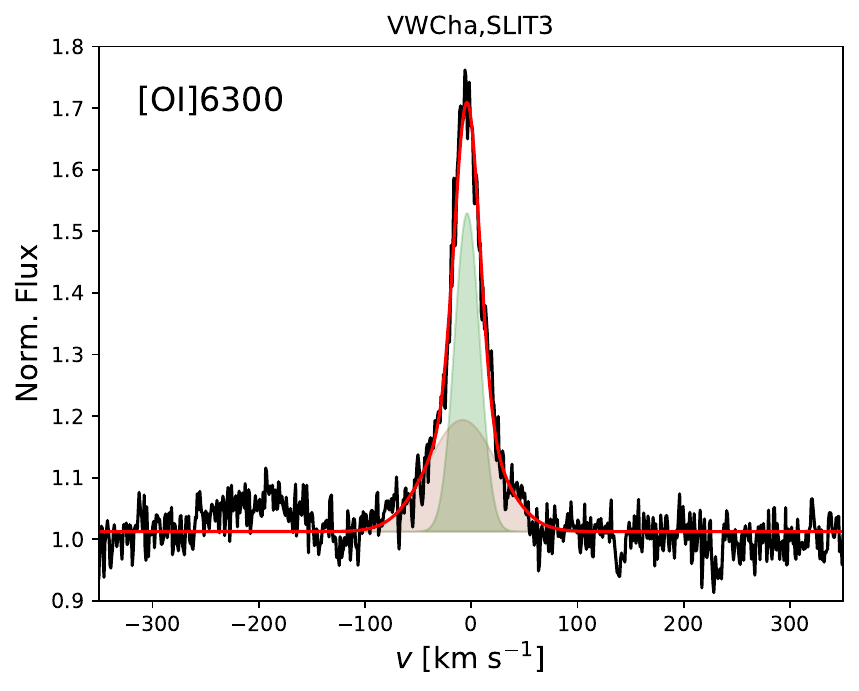}} 
\hfill 
\subfloat{\includegraphics[trim=0 0 0 0, clip, width=0.3 \textwidth]{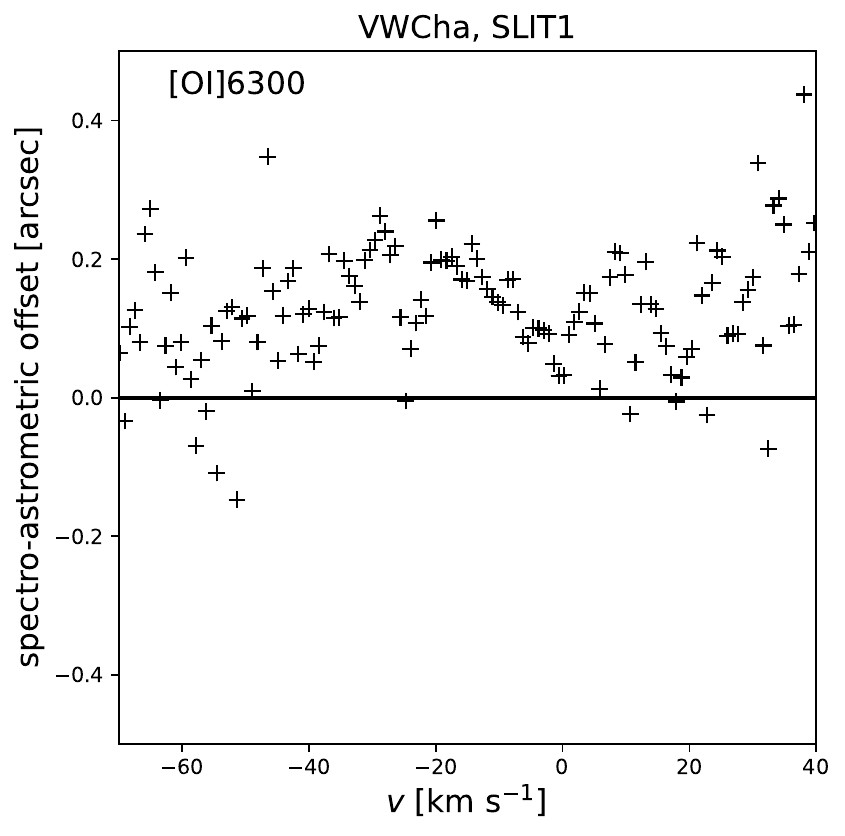}}
\hfill
\subfloat{\includegraphics[trim=0 0 0 0, clip, width=0.3 \textwidth]{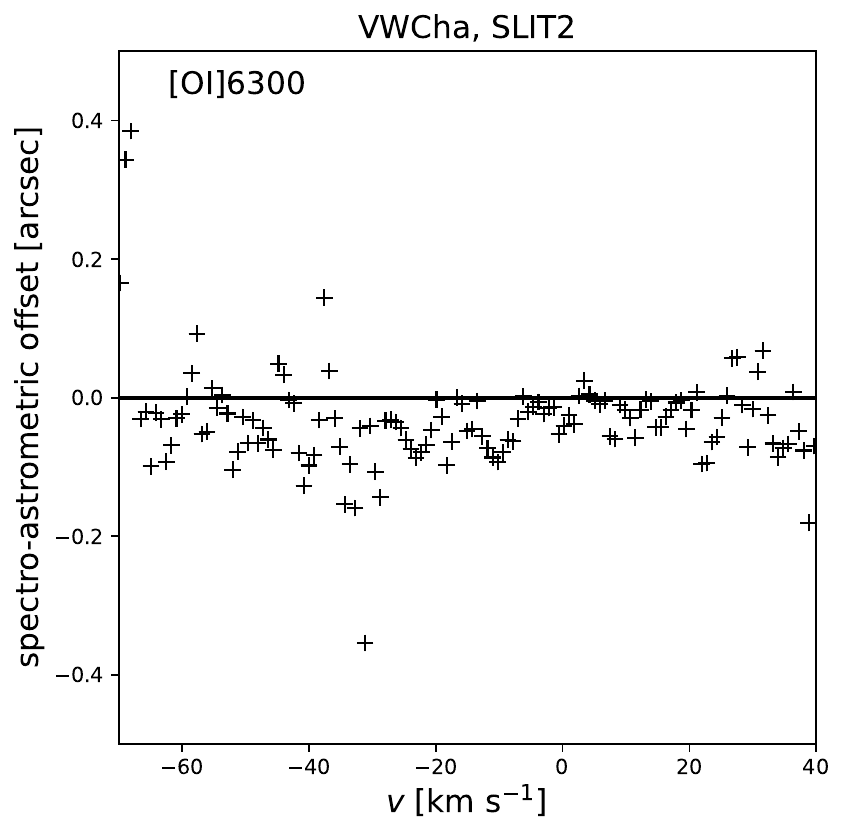}}
\hfill
\subfloat{\includegraphics[trim=0 0 0 0, clip, width=0.3 \textwidth]{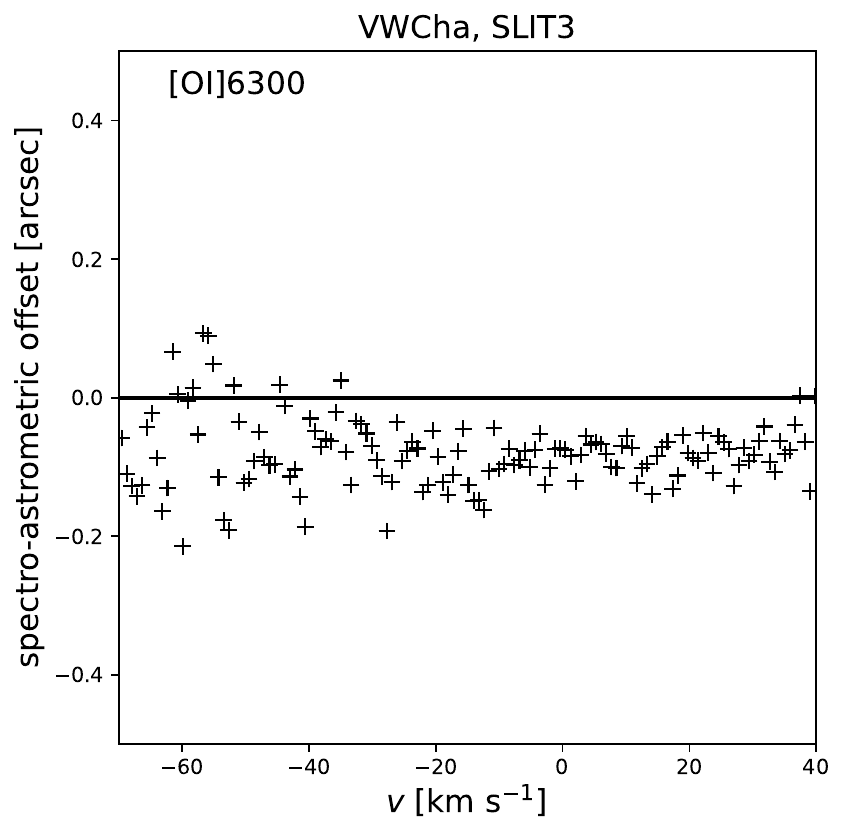}} 
\hfill
\caption{\small{Line profiles of H$\alpha$ and [OI]$\lambda$6300 for all slit positions of VW\,Cha.}}\label{fig:all_minispectra_VWCha}
\end{figure*}

\begin{figure*} 
\centering
\subfloat{\includegraphics[trim=0 0 0 0, clip, width=0.3 \textwidth]{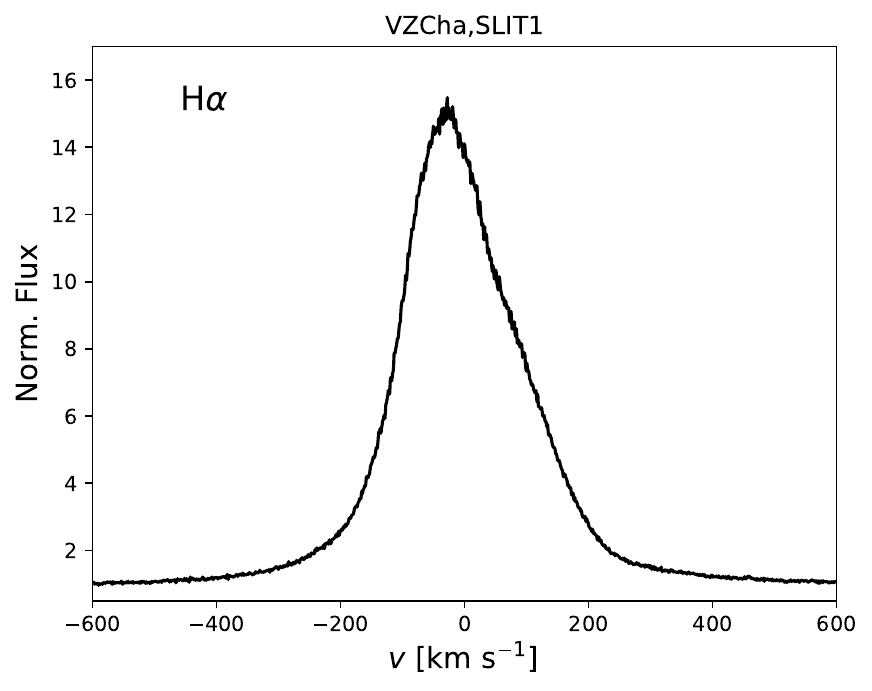}}
\hfill
\subfloat{\includegraphics[trim=0 0 0 0, clip, width=0.3 \textwidth]{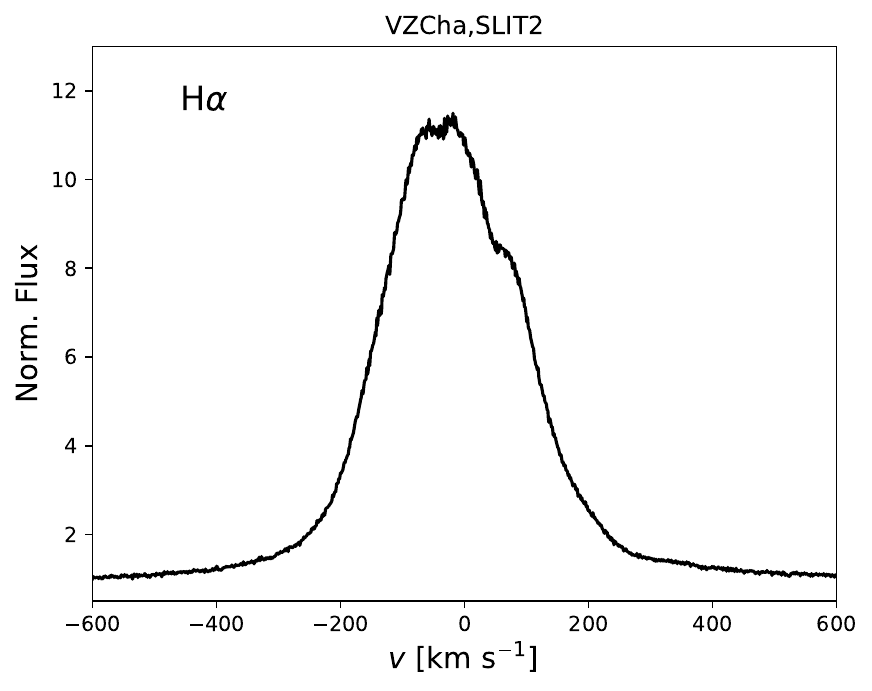}}
\hfill
\subfloat{\includegraphics[trim=0 0 0 0, clip, width=0.3 \textwidth]{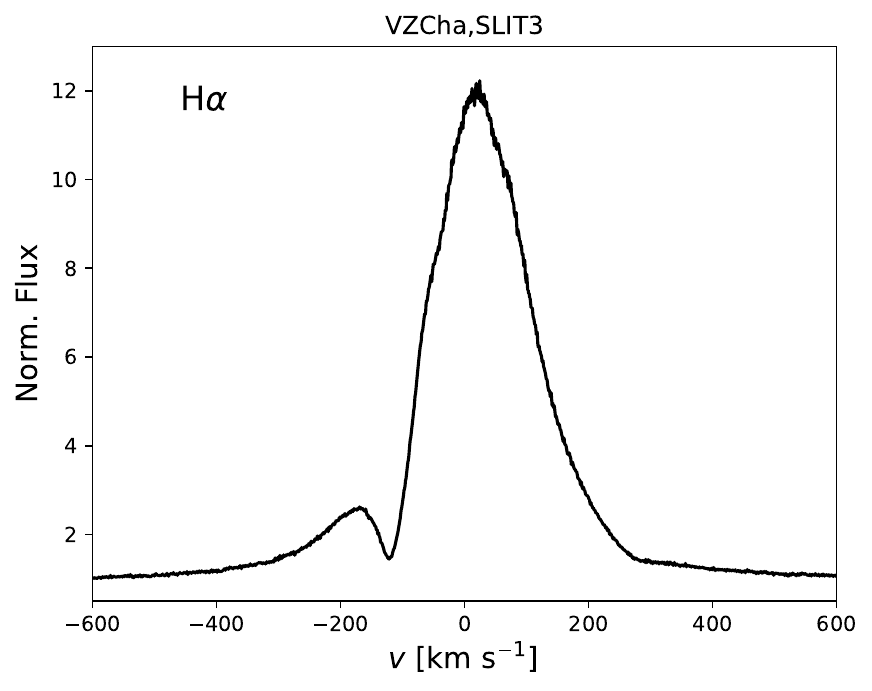}} 
\hfill 
\subfloat{\includegraphics[trim=0 0 0 0, clip, width=0.3 \textwidth]{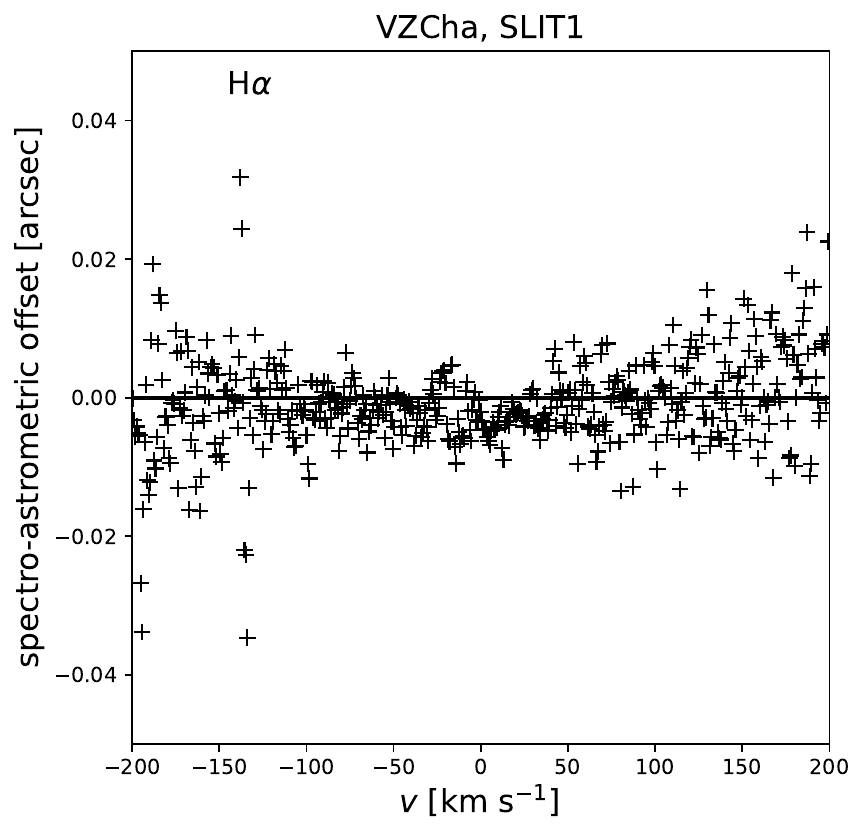}}
\hfill
\subfloat{\includegraphics[trim=0 0 0 0, clip, width=0.3 \textwidth]{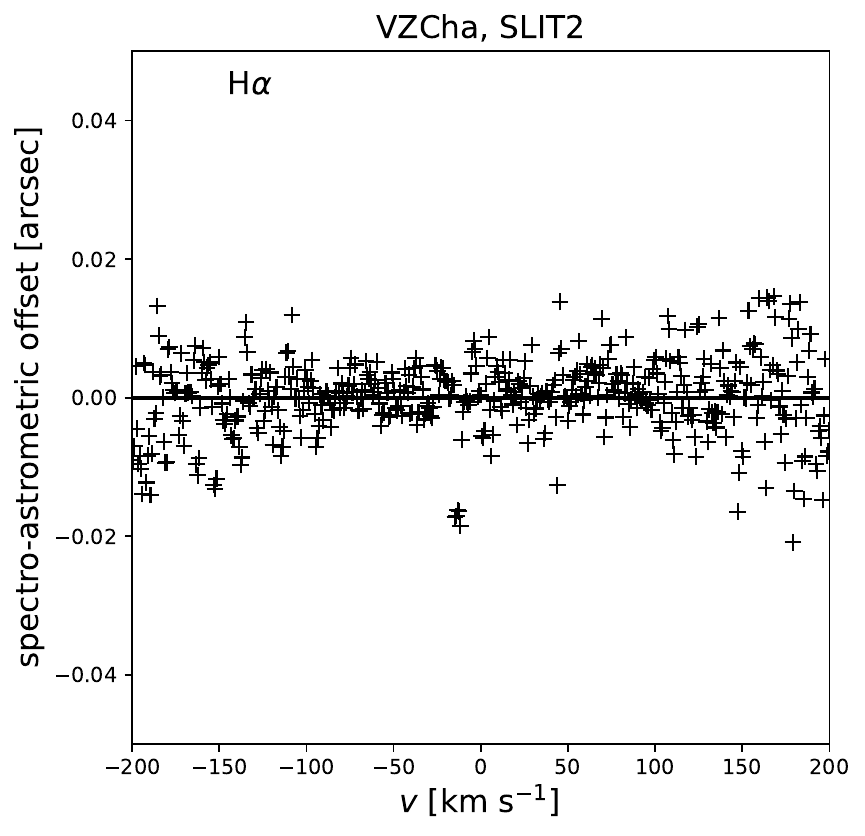}}
\hfill
\subfloat{\includegraphics[trim=0 0 0 0, clip, width=0.3 \textwidth]{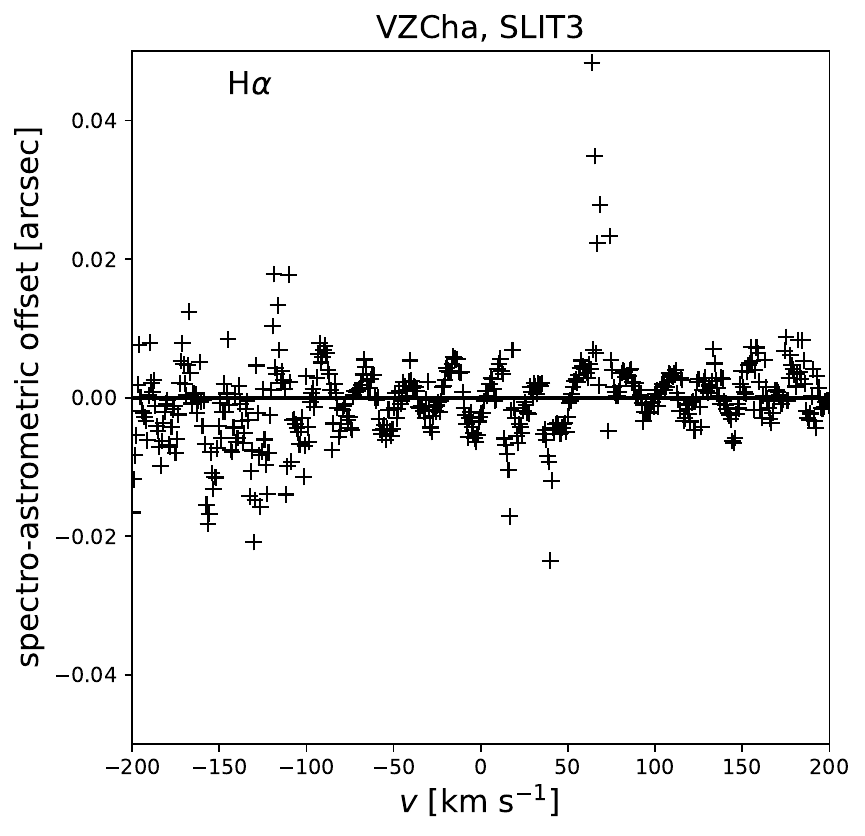}} 
\hfill
\subfloat{\includegraphics[trim=0 0 0 0, clip, width=0.3 \textwidth]{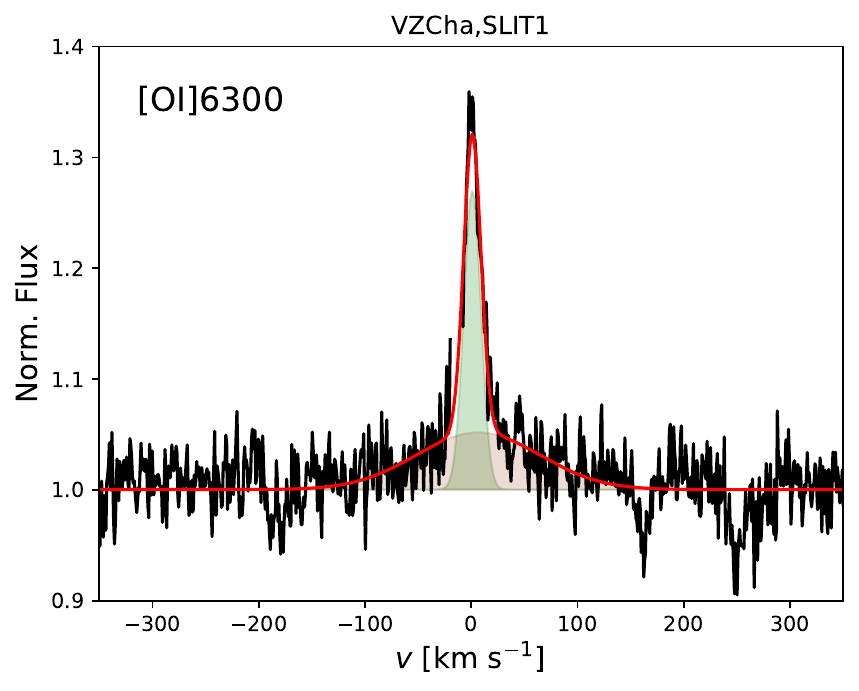}}
\hfill
\subfloat{\includegraphics[trim=0 0 0 0, clip, width=0.3 \textwidth]{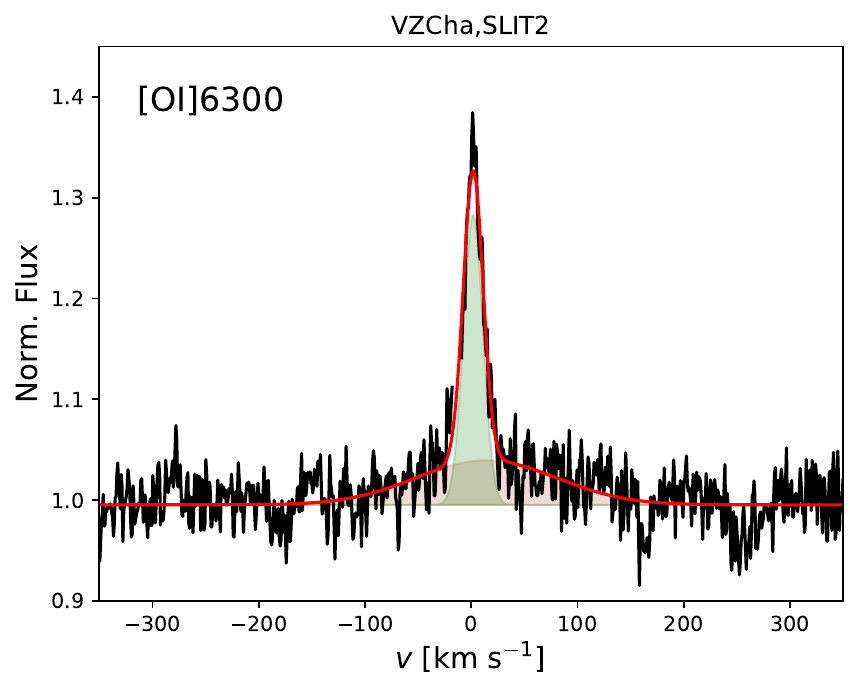}}
\hfill
\subfloat{\includegraphics[trim=0 0 0 0, clip, width=0.3 \textwidth]{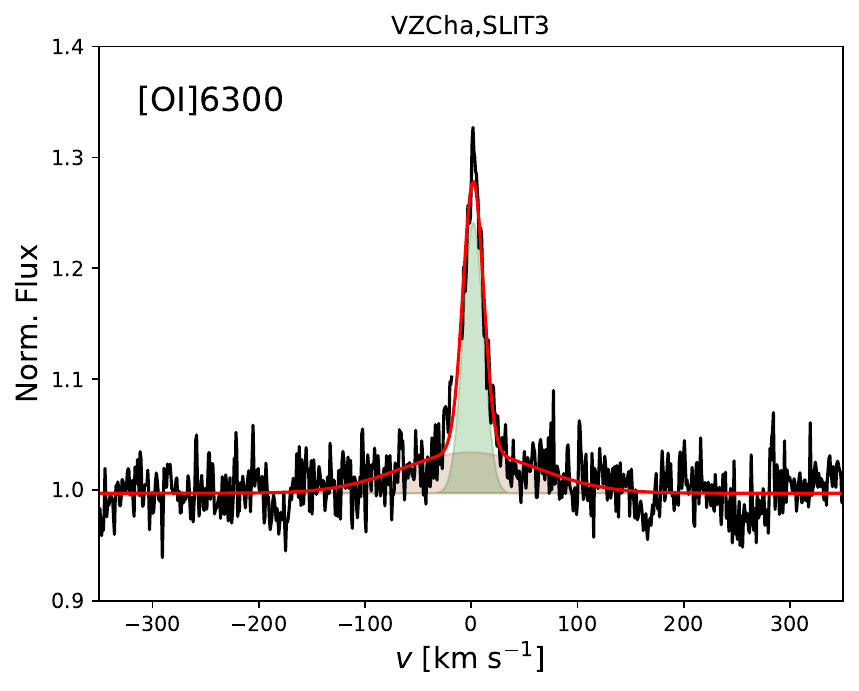}} 
\hfill  
\subfloat{\includegraphics[trim=0 0 0 0, clip, width=0.3 \textwidth]{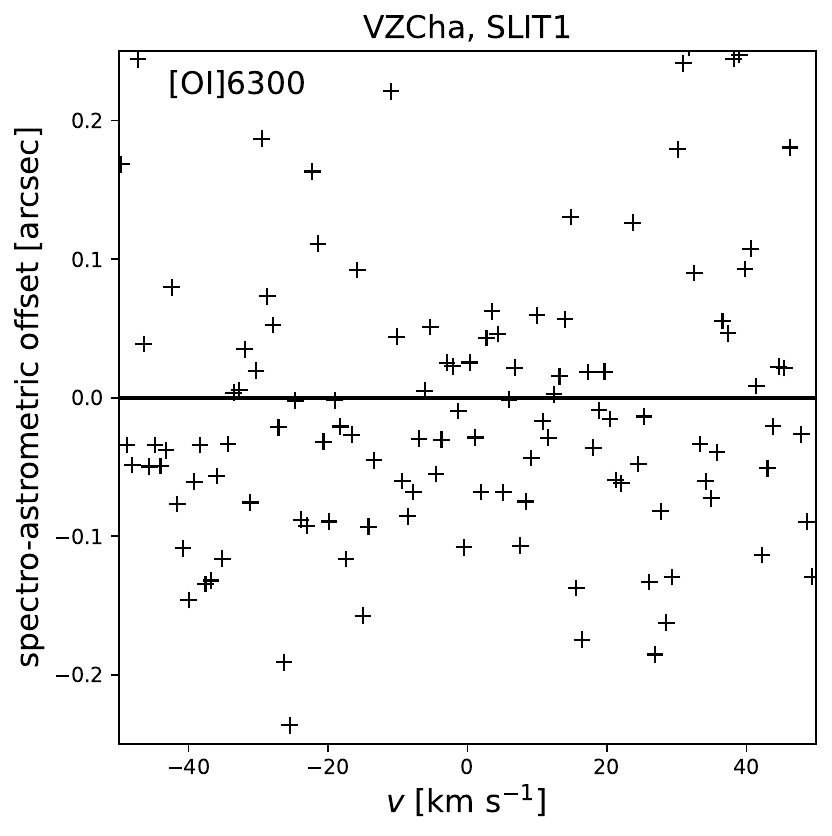}}
\hfill
\subfloat{\includegraphics[trim=0 0 0 0, clip, width=0.3 \textwidth]{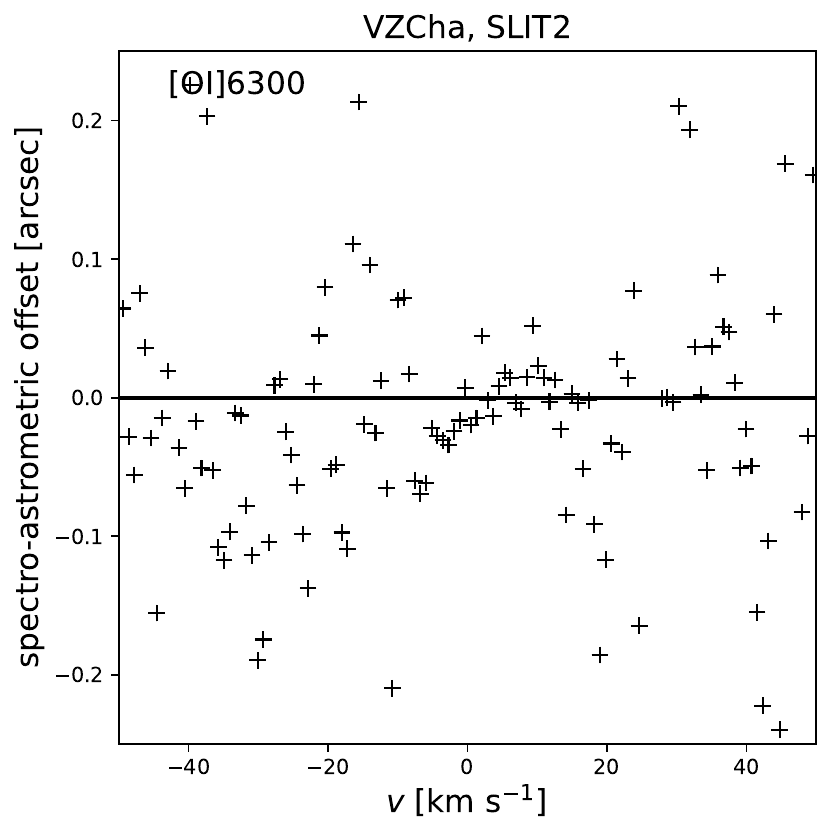}}
\hfill
\subfloat{\includegraphics[trim=0 0 0 0, clip, width=0.3 \textwidth]{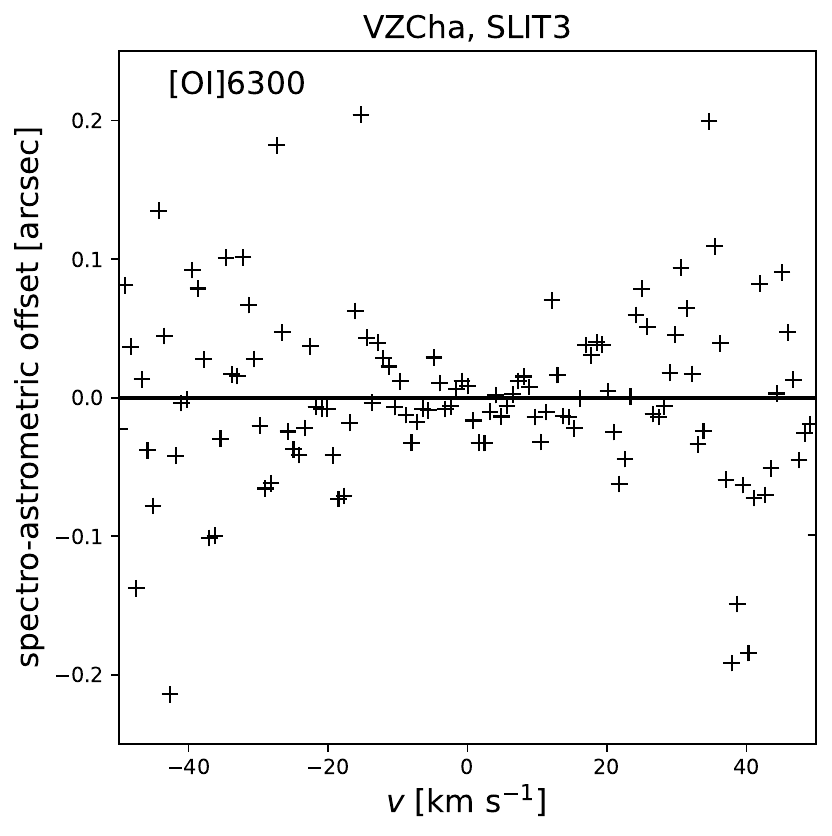}} 
\hfill
\caption{\small{Line profiles of H$\alpha$ and [OI]$\lambda$6300 for all slit positions of VZ\,Cha.}}\label{fig:all_minispectra_VZCha}
\end{figure*}

\begin{figure*} 
\centering
\subfloat{\includegraphics[trim=0 0 0 0, clip, width=0.3 \textwidth]{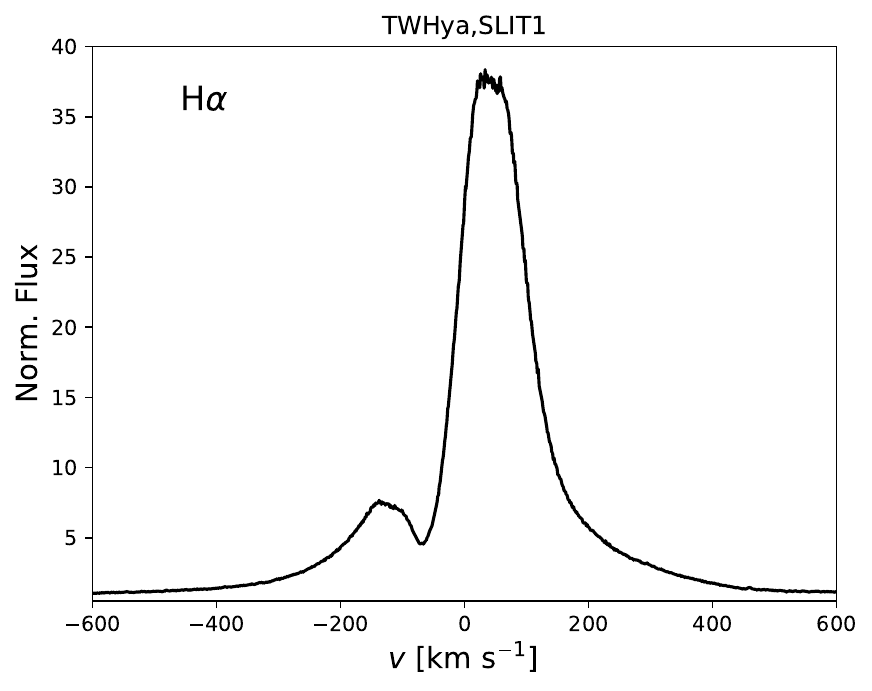}}
\hfill
\subfloat{\includegraphics[trim=0 0 0 0, clip, width=0.3 \textwidth]{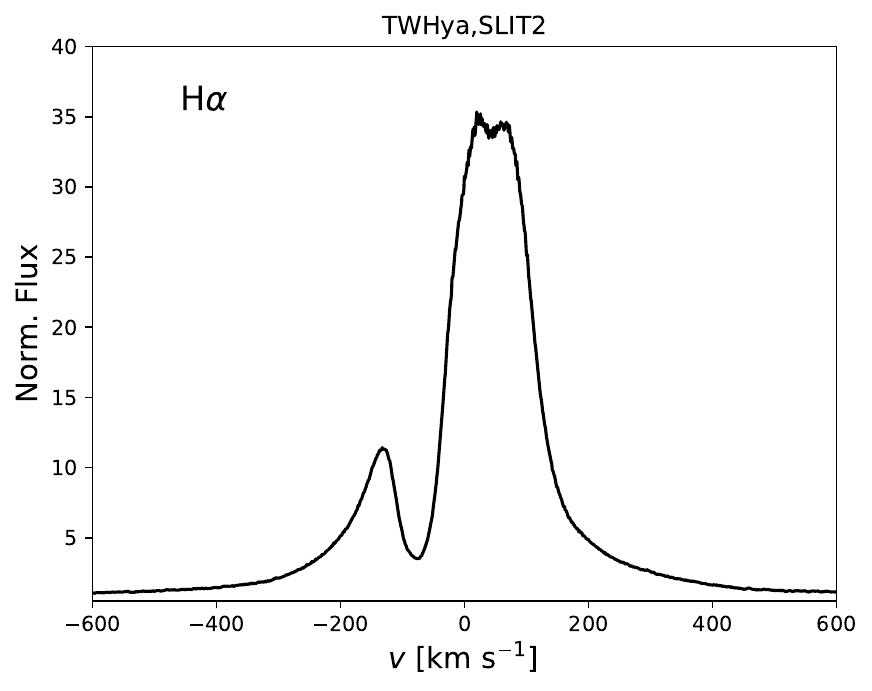}} 
\hfill \\
\subfloat{\includegraphics[trim=0 0 0 0, clip, width=0.3 \textwidth]{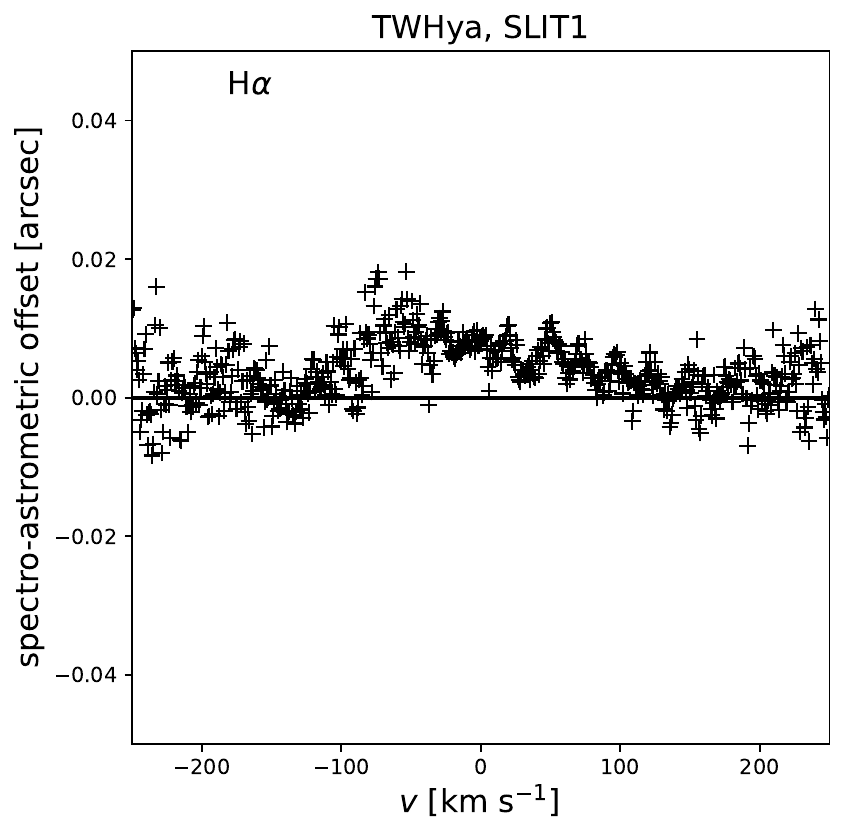}}
\hfill
\subfloat{\includegraphics[trim=0 0 0 0, clip, width=0.3 \textwidth]{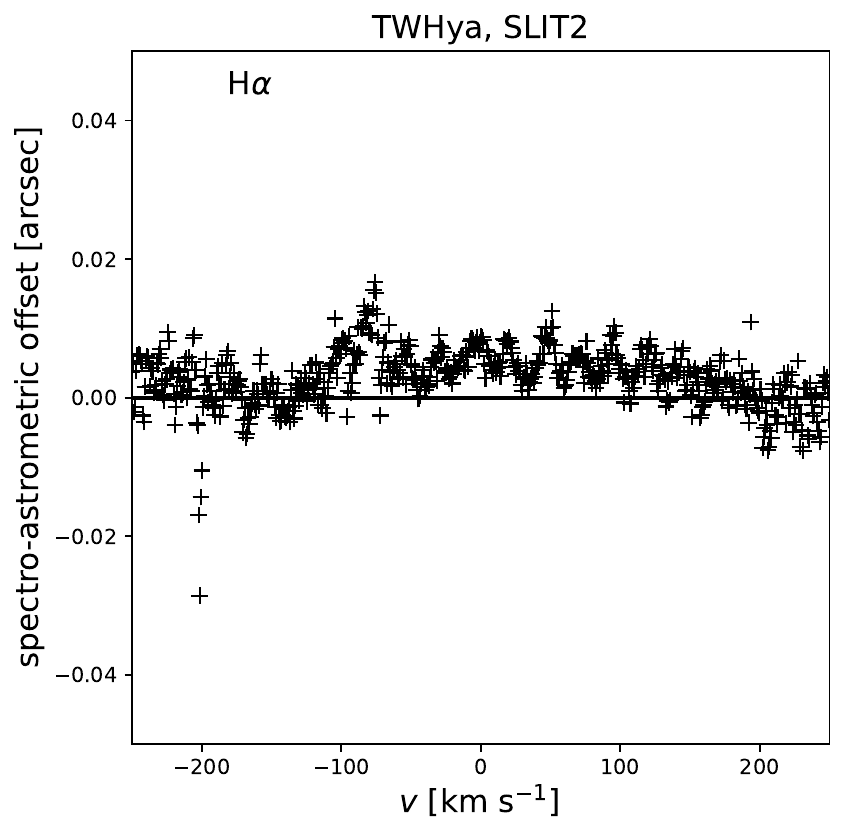}}
\hfill \\
\subfloat{\includegraphics[trim=0 0 0 0, clip, width=0.3 \textwidth]{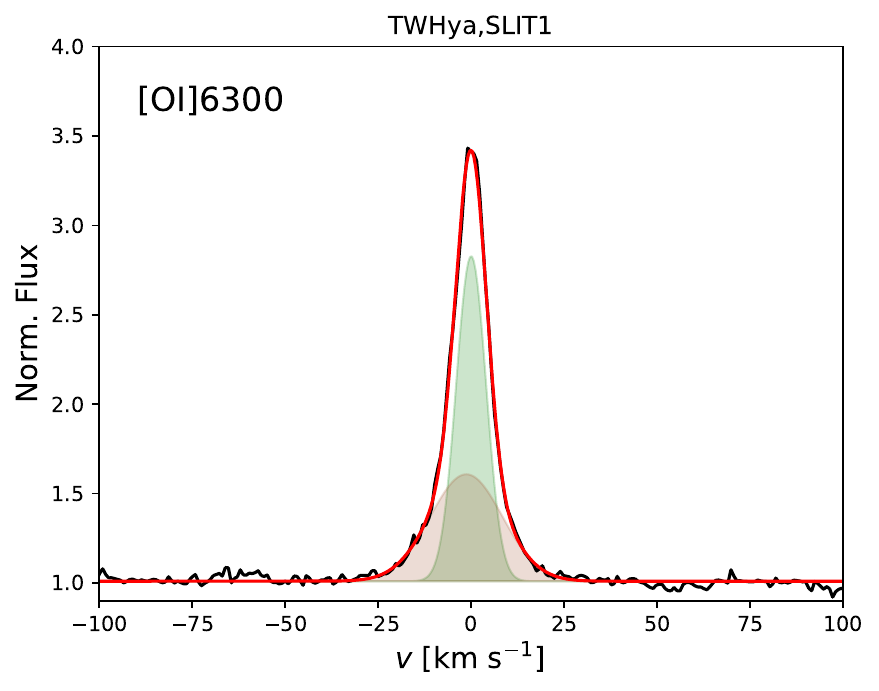}}
\hfill
\subfloat{\includegraphics[trim=0 0 0 0, clip, width=0.3 \textwidth]{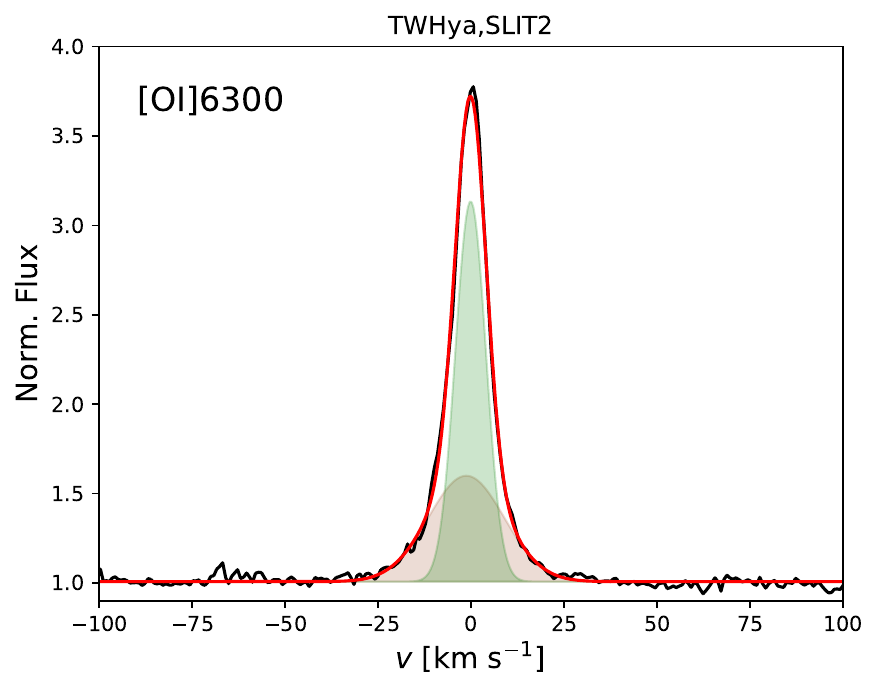}}
\hfill\\
\subfloat{\includegraphics[trim=0 0 0 0, clip, width=0.3 \textwidth]{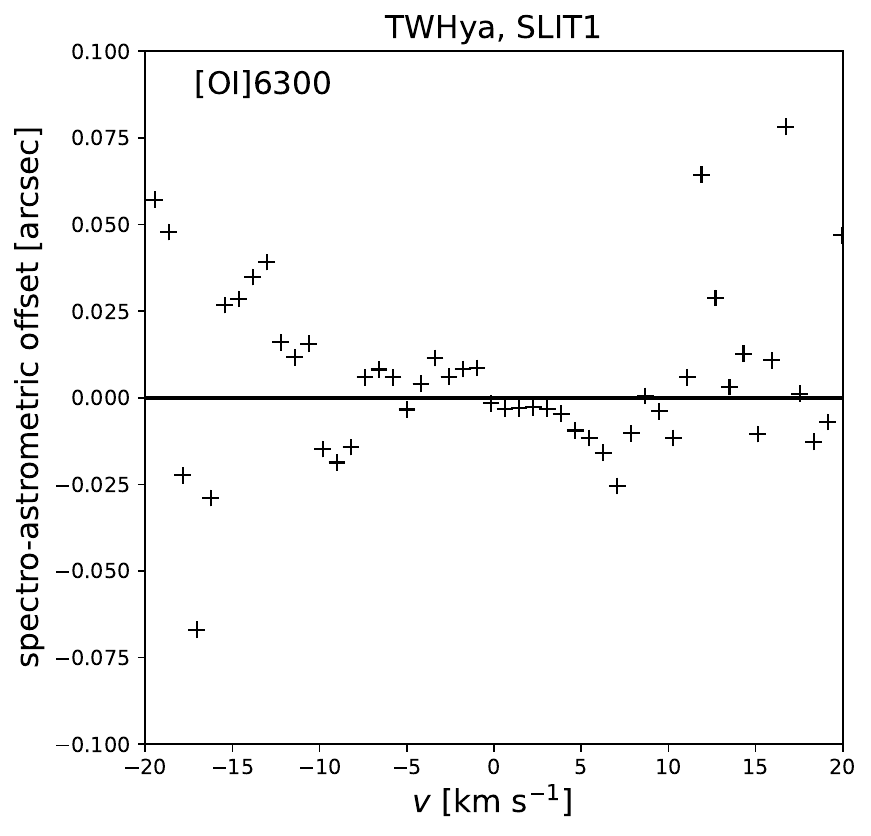}}
\hfill
\subfloat{\includegraphics[trim=0 0 0 0, clip, width=0.3 \textwidth]{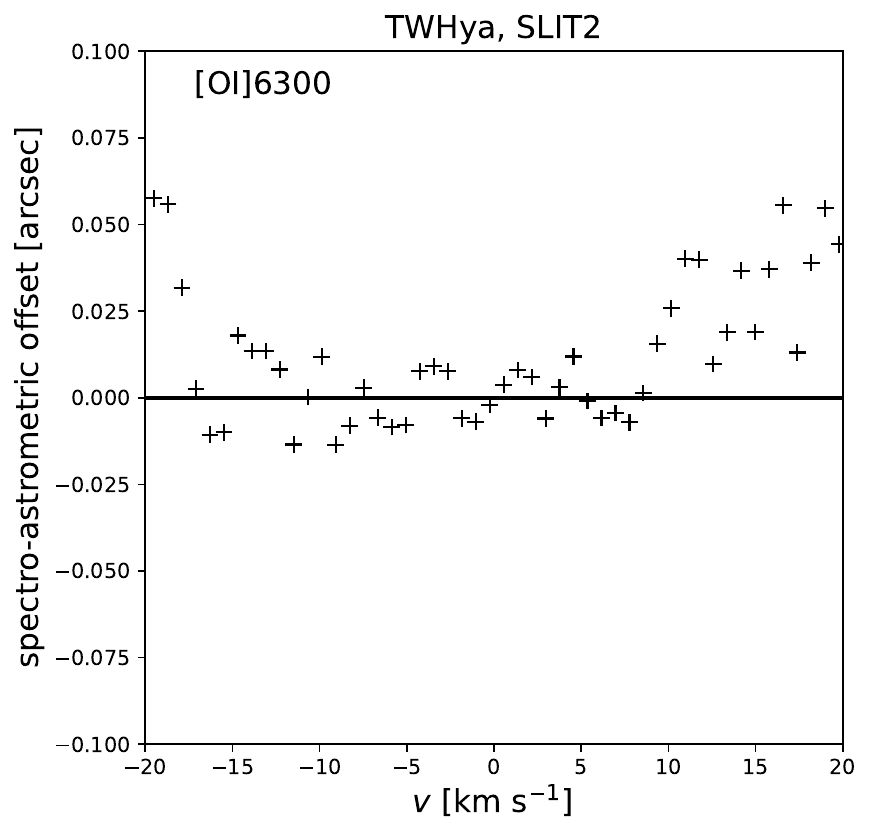}}
\hfill
\caption{\small{Line profiles of H$\alpha$ and [OI]$\lambda$6300 for all slit positions of TW\,Hya.}}\label{fig:all_minispectra_TWHya}
\end{figure*} 

\begin{figure*} 
\centering 
\subfloat{\includegraphics[trim=0 0 0 0, clip, width=0.3 \textwidth]{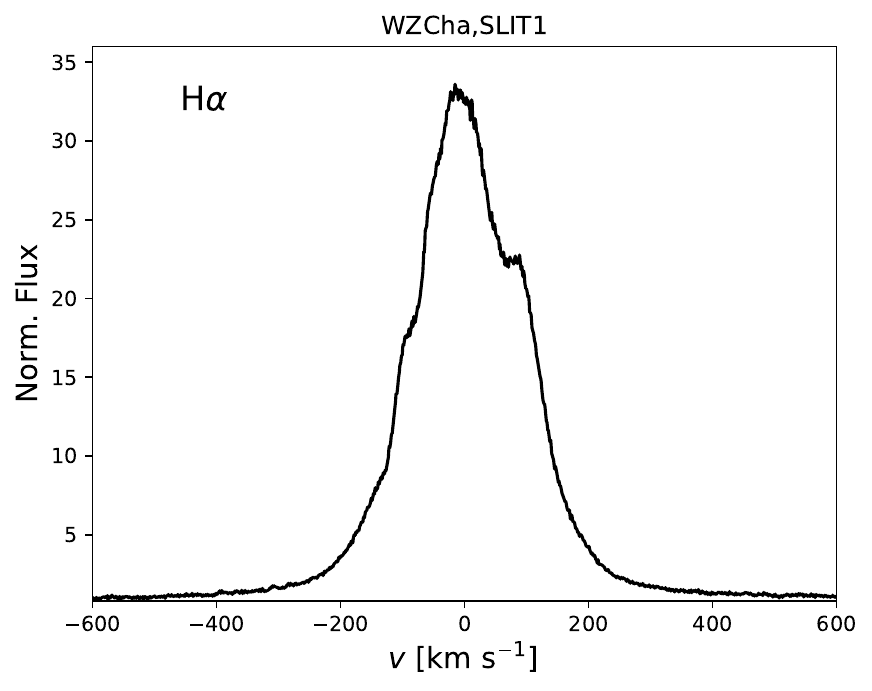}}
\hfill
\subfloat{\includegraphics[trim=0 0 0 0, clip, width=0.3 \textwidth]{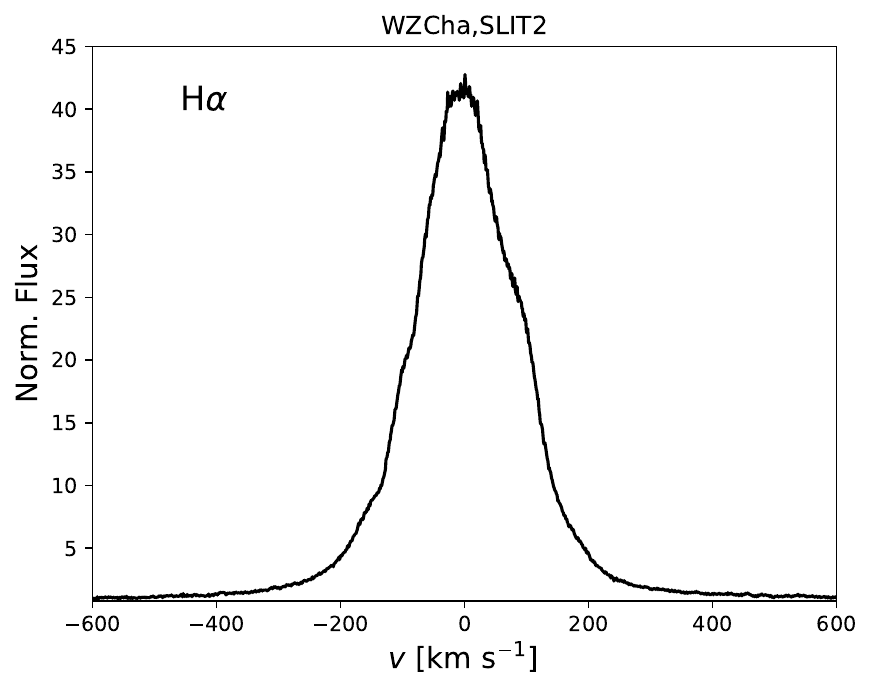}}
\hfill
\subfloat{\includegraphics[trim=0 0 0 0, clip, width=0.3 \textwidth]{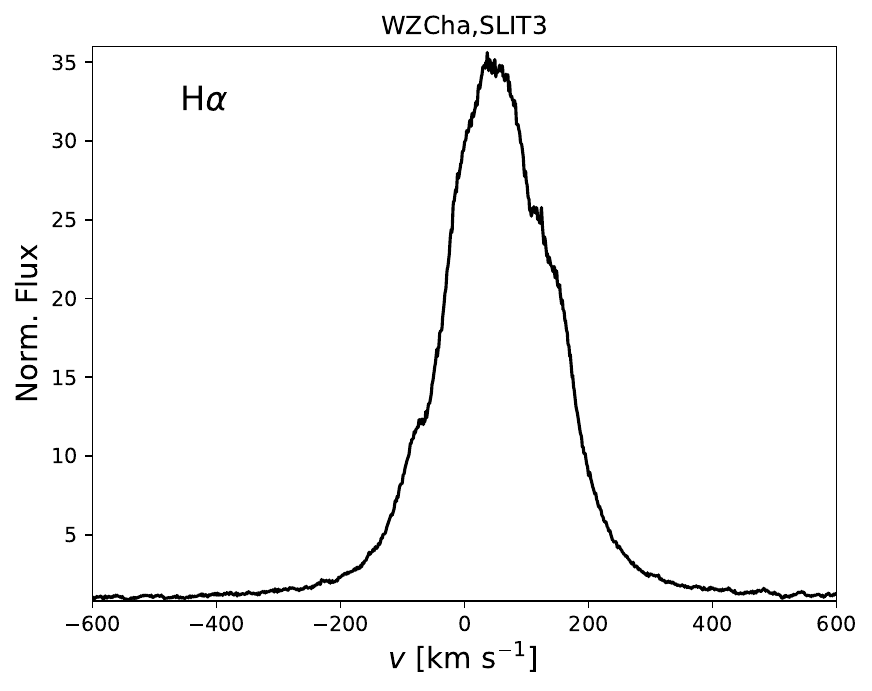}} 
\hfill  
\subfloat{\includegraphics[trim=0 0 0 0, clip, width=0.3 \textwidth]{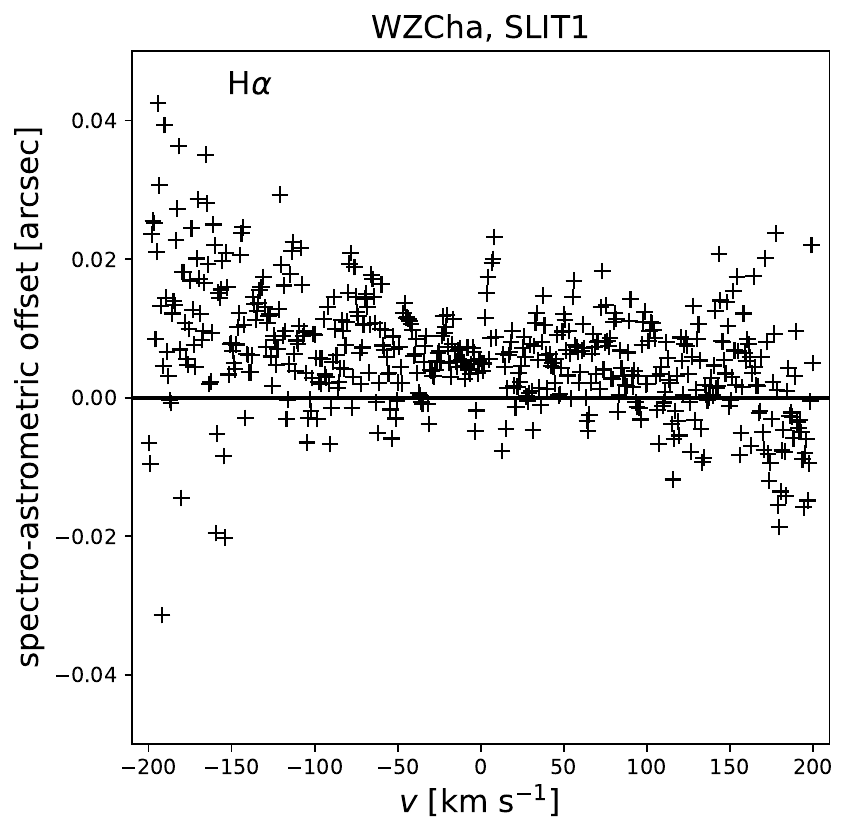}}
\hfill
\subfloat{\includegraphics[trim=0 0 0 0, clip, width=0.3 \textwidth]{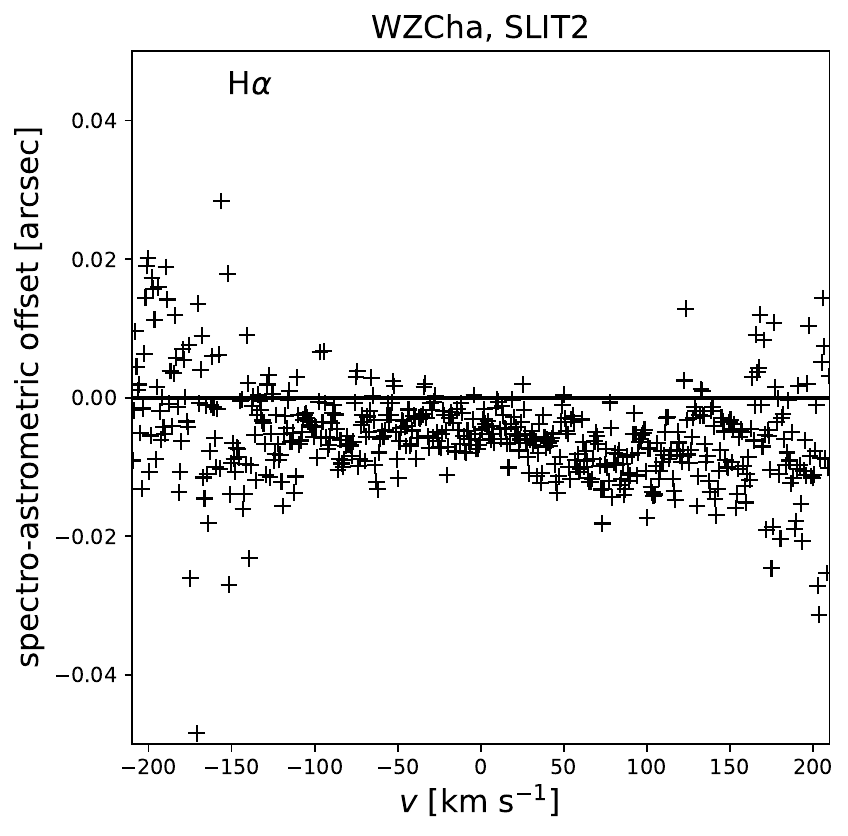}}
\hfill
\subfloat{\includegraphics[trim=0 0 0 0, clip, width=0.3 \textwidth]{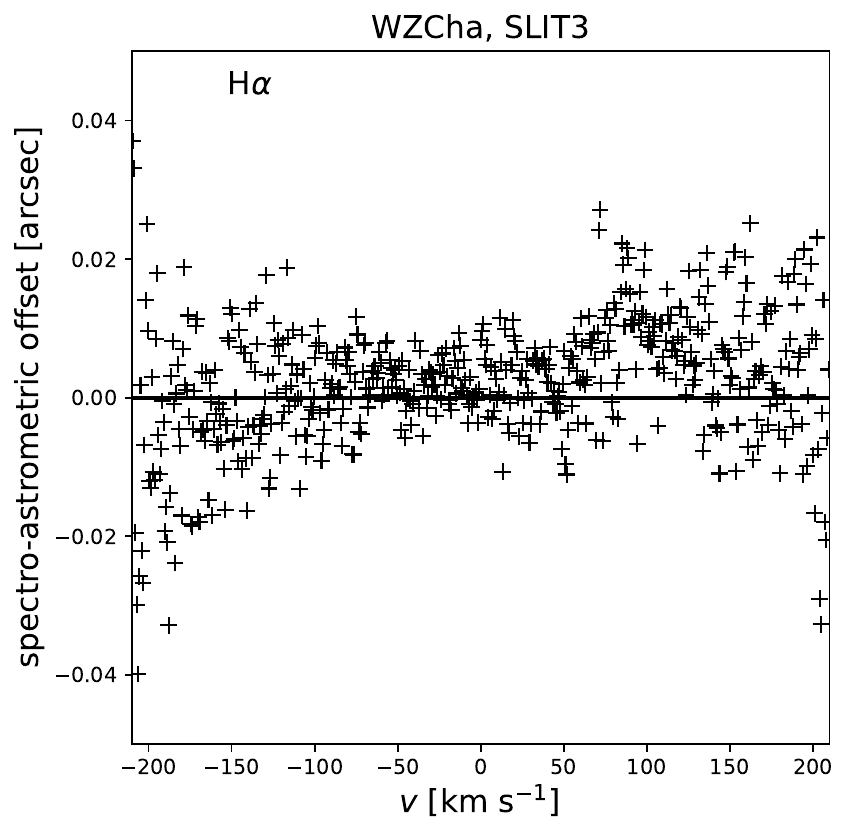}} 
\hfill
\subfloat{\includegraphics[trim=0 0 0 0, clip, width=0.3 \textwidth]{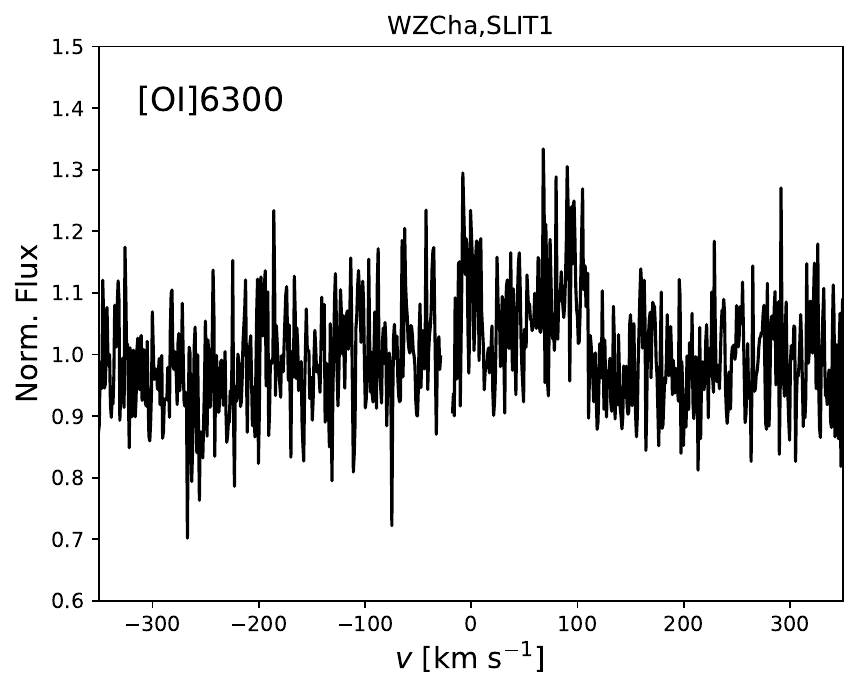}}
\hfill
\subfloat{\includegraphics[trim=0 0 0 0, clip, width=0.3 \textwidth]{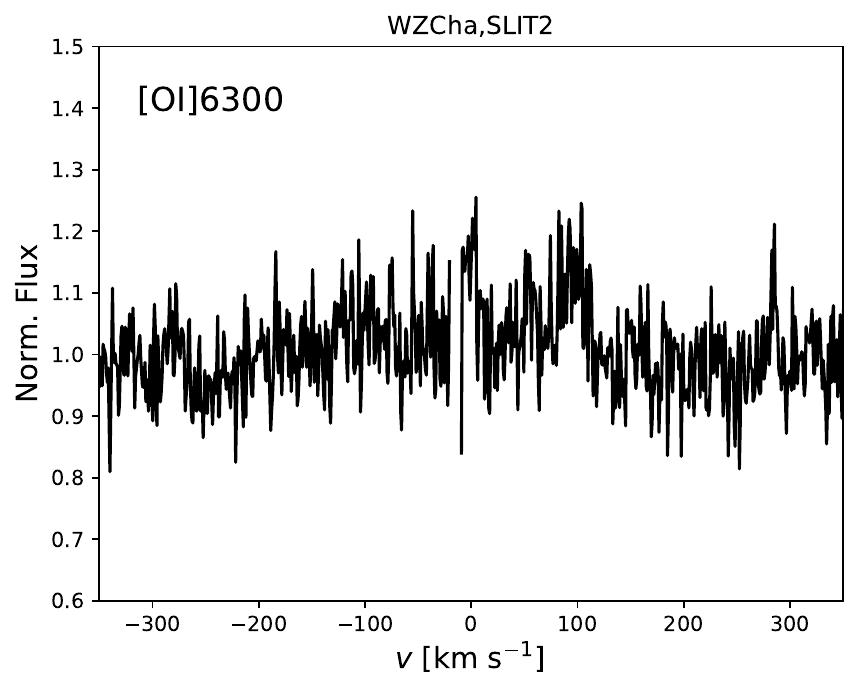}}
\hfill
\subfloat{\includegraphics[trim=0 0 0 0, clip, width=0.3 \textwidth]{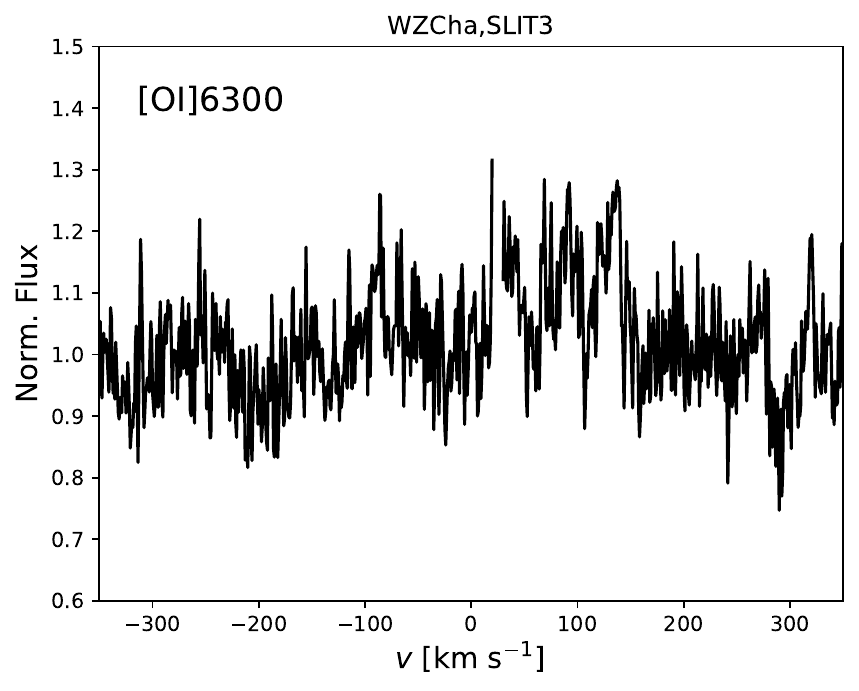}} 
\hfill  
\caption{\small{Line profiles of H$\alpha$ and [OI]$\lambda$6300 for all slit positions of WZ\,Cha.}}\label{fig:all_minispectra_WZCha}
\end{figure*}

\begin{figure*} 
\centering
\subfloat{\includegraphics[trim=0 0 0 0, clip, width=0.3 \textwidth]{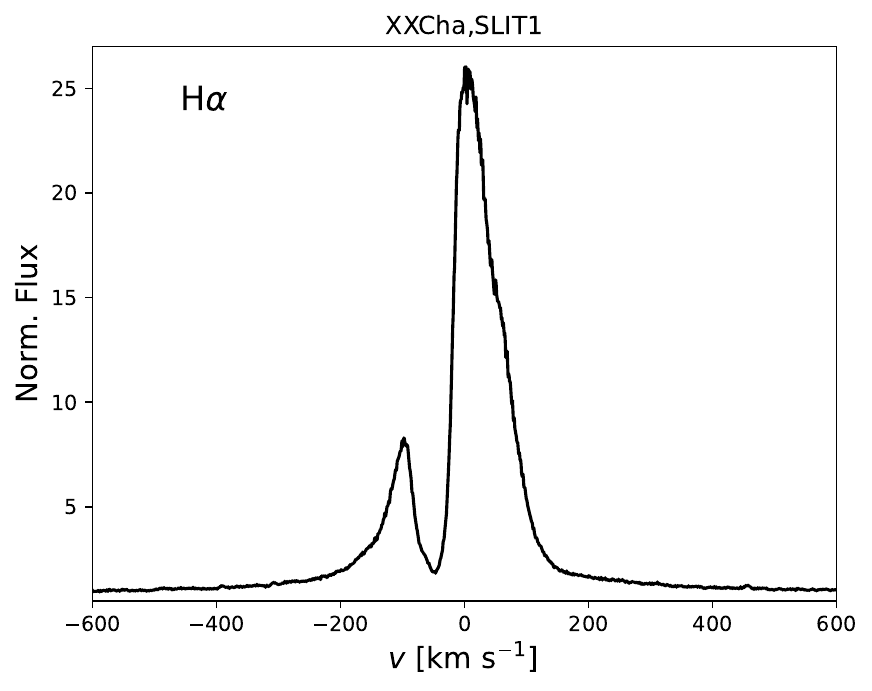}}
\hfill
\subfloat{\includegraphics[trim=0 0 0 0, clip, width=0.3 \textwidth]{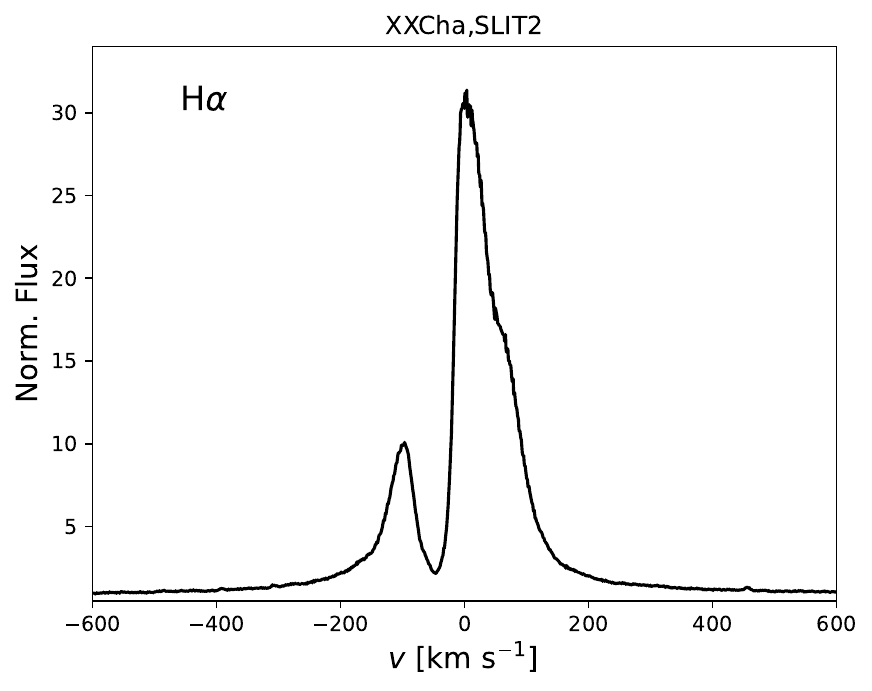}}
\hfill
\subfloat{\includegraphics[trim=0 0 0 0, clip, width=0.3 \textwidth]{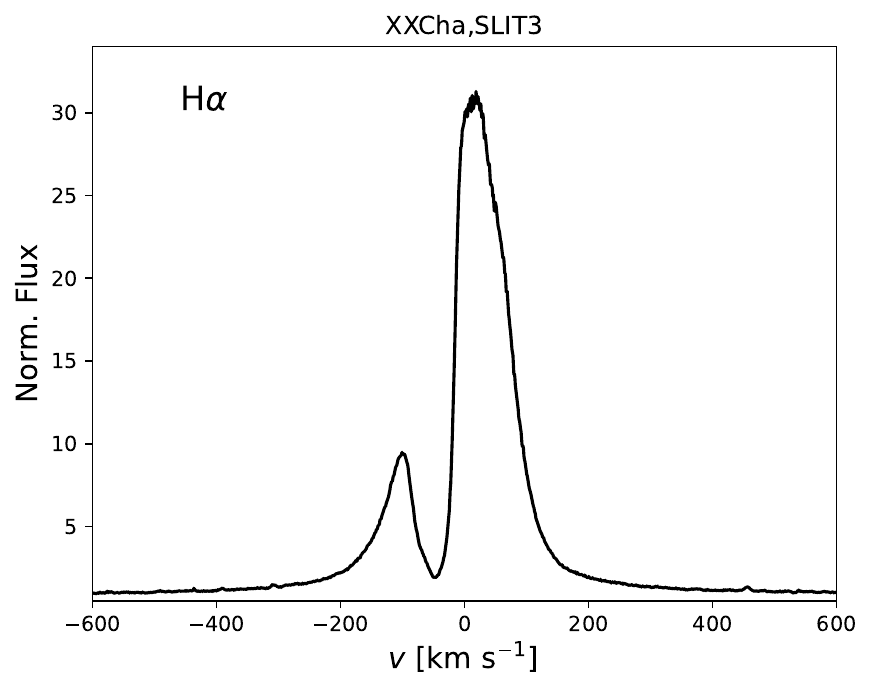}}
\hfill  
\subfloat{\includegraphics[trim=0 0 0 0, clip, width=0.3 \textwidth]{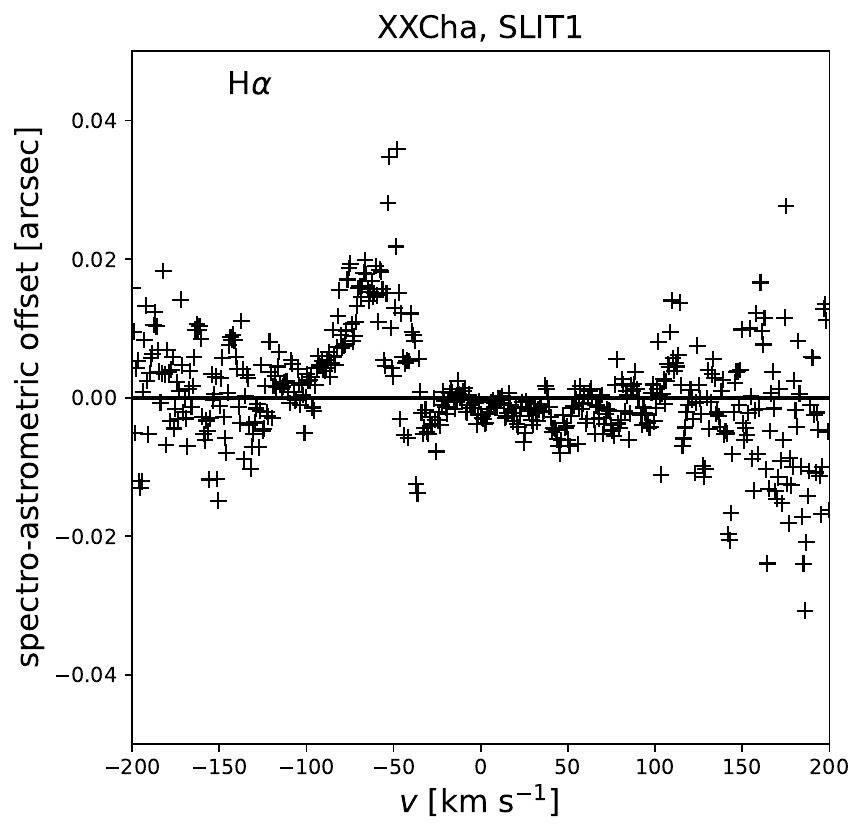}}
\hfill
\subfloat{\includegraphics[trim=0 0 0 0, clip, width=0.3 \textwidth]{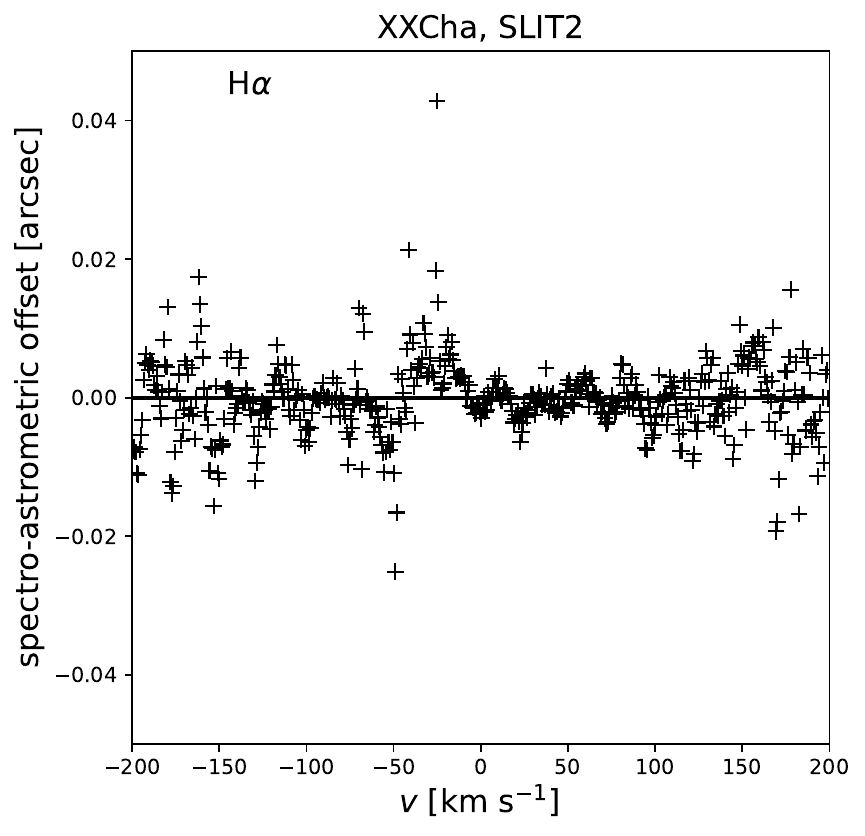}}
\hfill
\subfloat{\includegraphics[trim=0 0 0 0, clip, width=0.3 \textwidth]{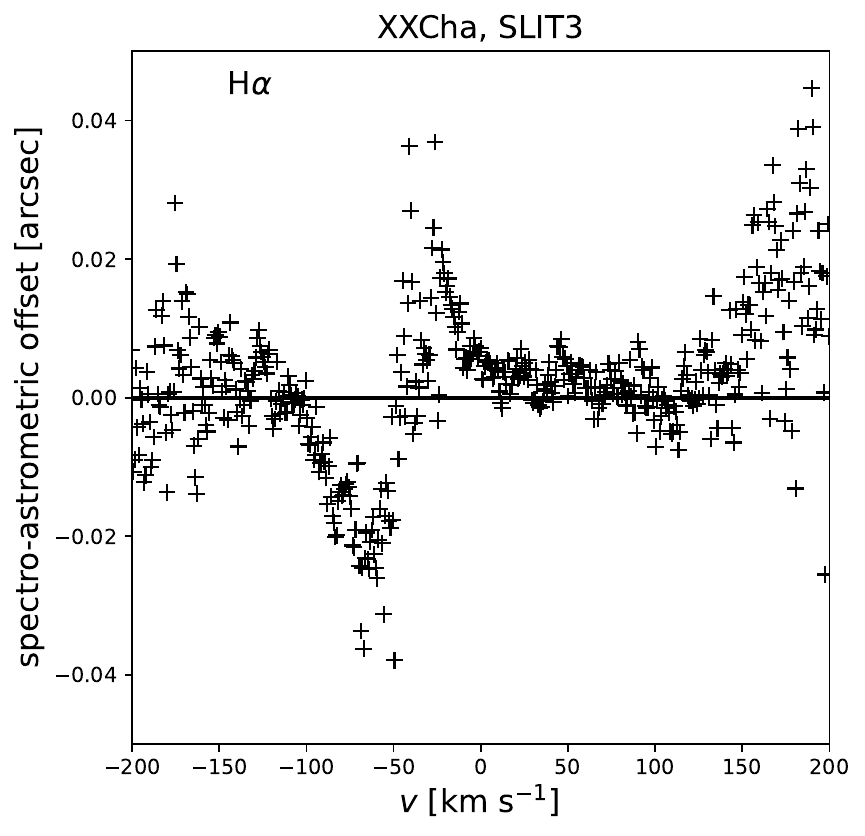}} 
\hfill
\subfloat{\includegraphics[trim=0 0 0 0, clip, width=0.3 \textwidth]{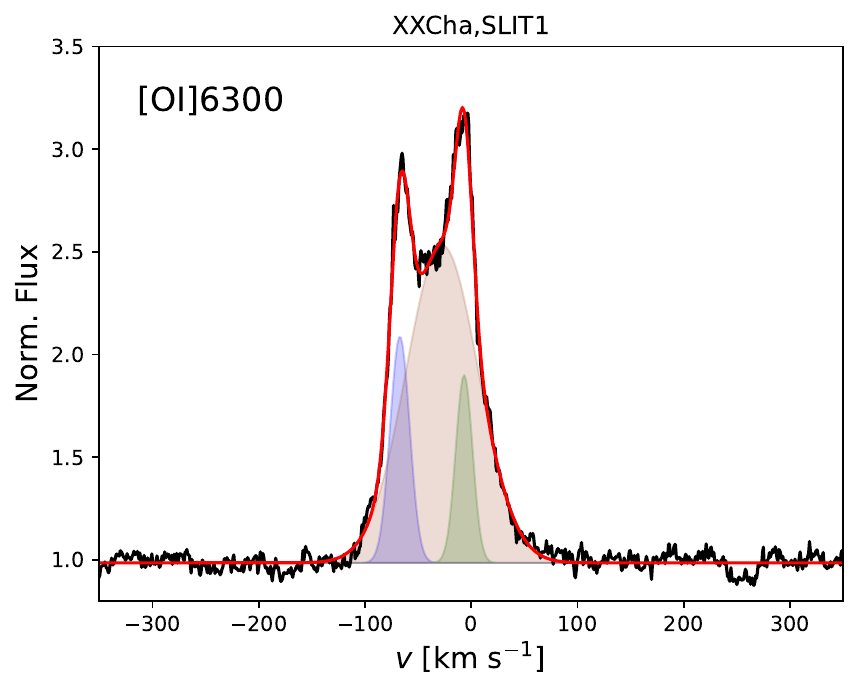}}
\hfill
\subfloat{\includegraphics[trim=0 0 0 0, clip, width=0.3 \textwidth]{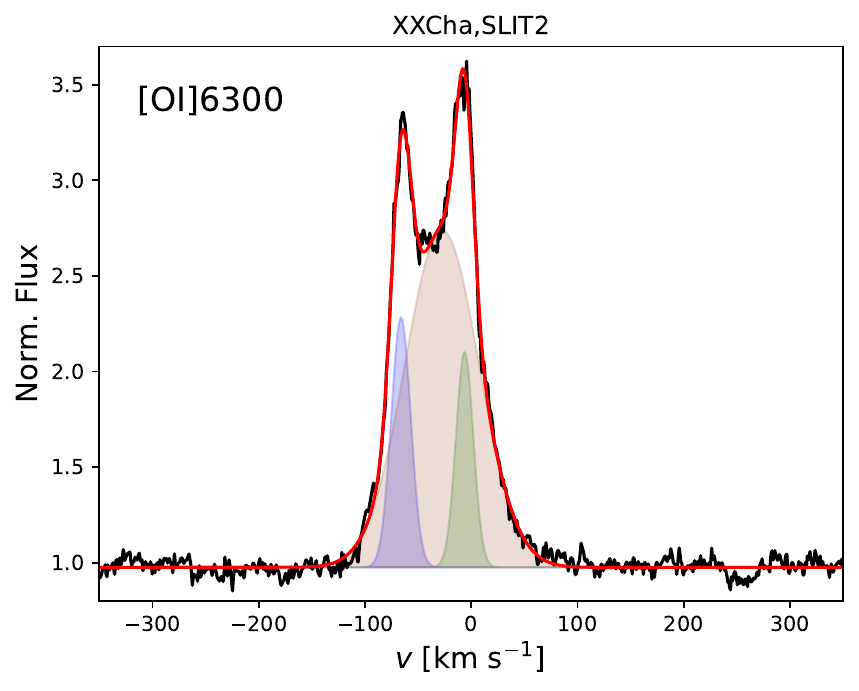}}
\hfill
\subfloat{\includegraphics[trim=0 0 0 0, clip, width=0.3 \textwidth]{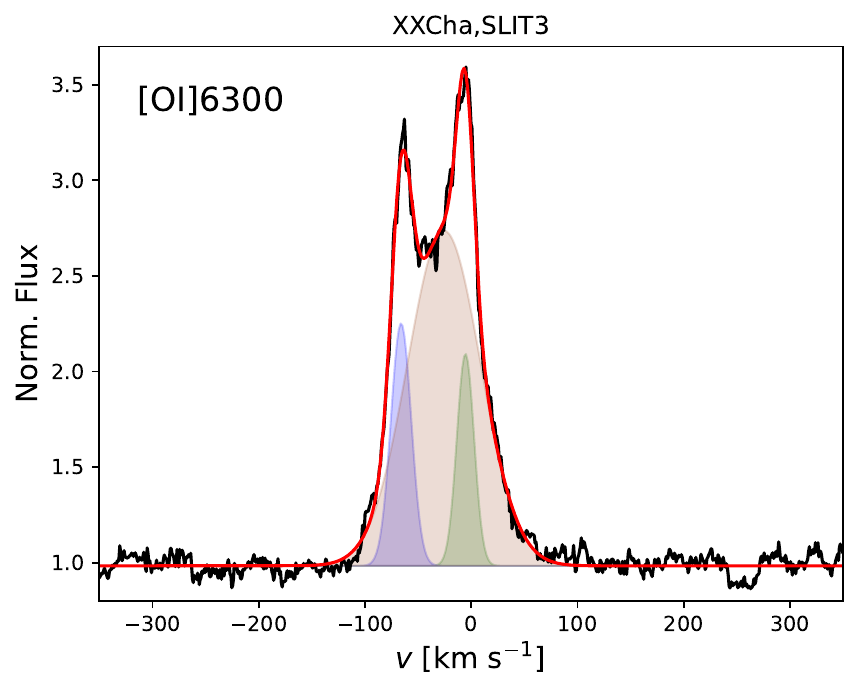}} 
\hfill   
\subfloat{\includegraphics[trim=0 0 0 0, clip, width=0.3 \textwidth]{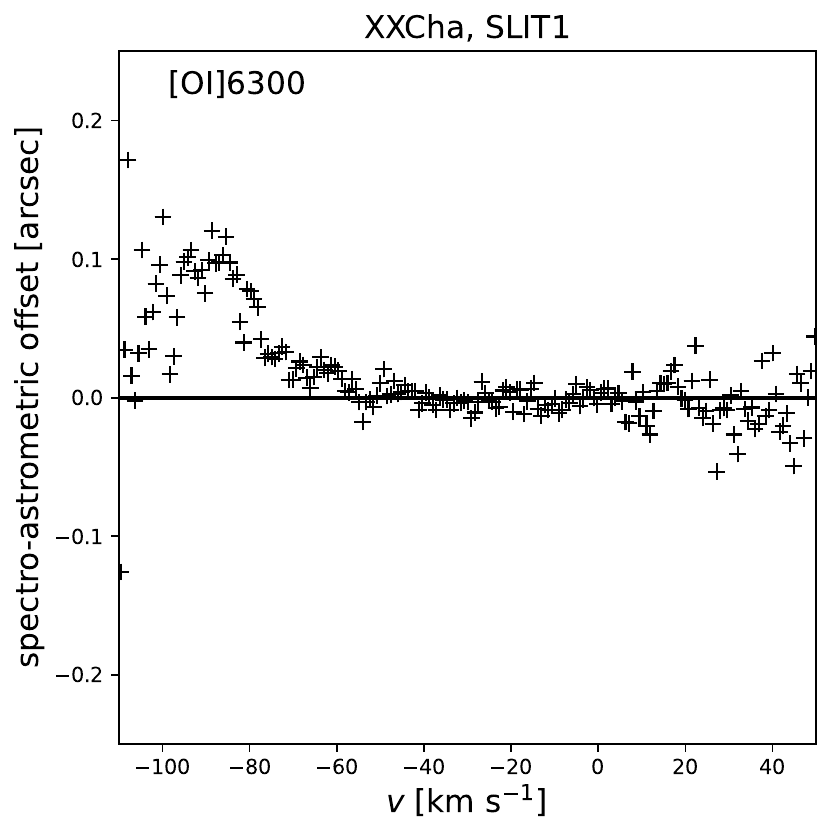}}
\hfill
\subfloat{\includegraphics[trim=0 0 0 0, clip, width=0.3 \textwidth]{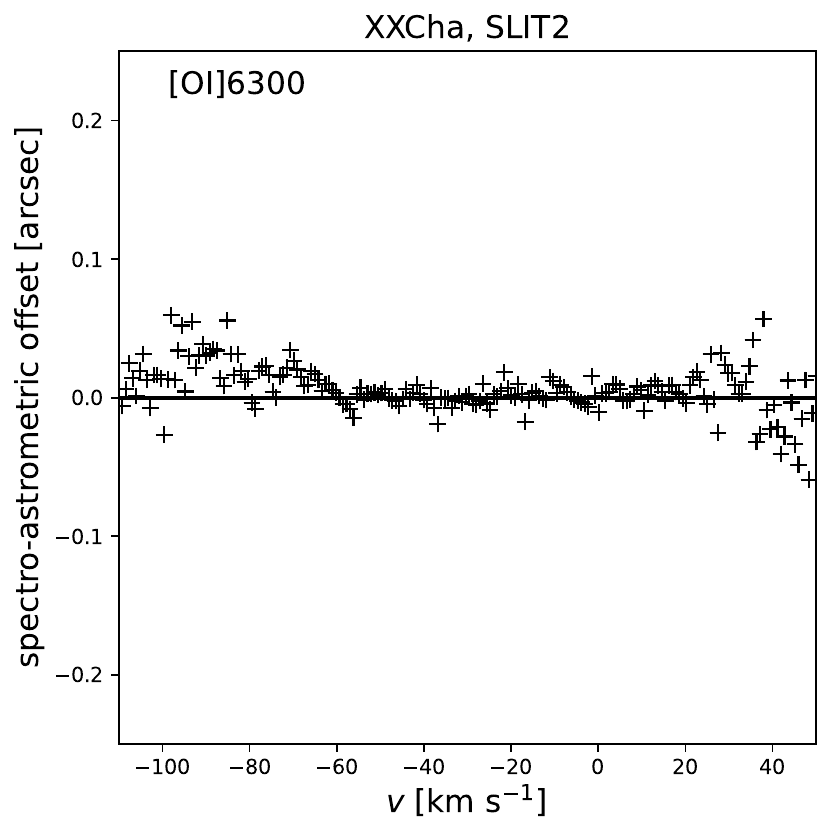}}
\hfill
\subfloat{\includegraphics[trim=0 0 0 0, clip, width=0.3 \textwidth]{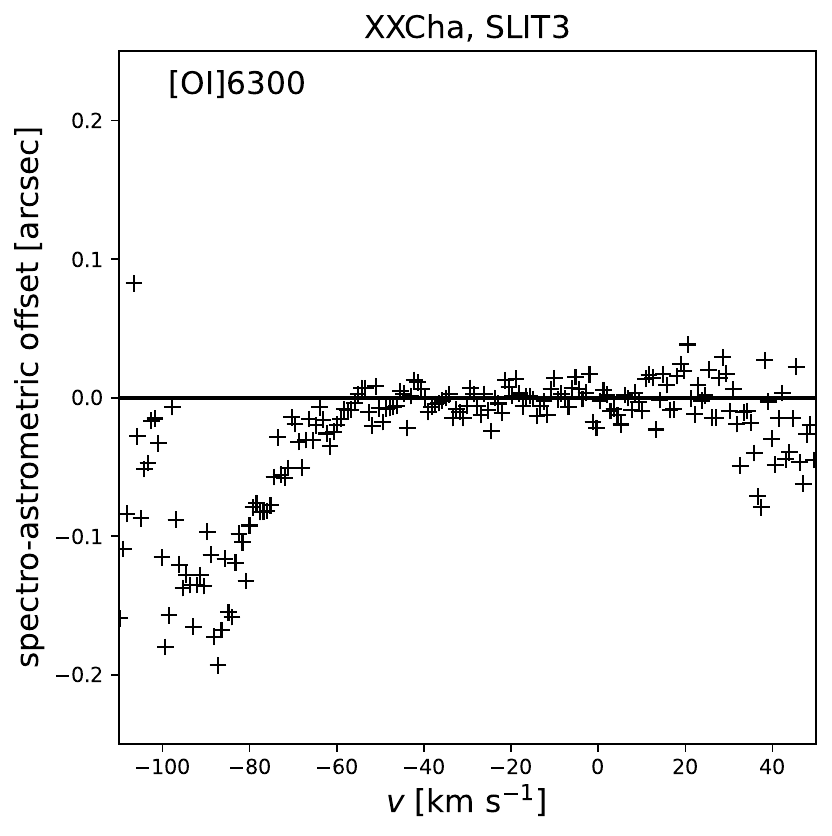}} 
\hfill
\caption{\small{Line profiles of H$\alpha$ and [OI]$\lambda$6300 for all slit positions of XX\,Cha.}}\label{fig:all_minispectra_XXCha}
\end{figure*} 

\begin{figure*} 
\centering
\subfloat{\includegraphics[trim=0 0 0 0, clip, width=0.3 \textwidth]{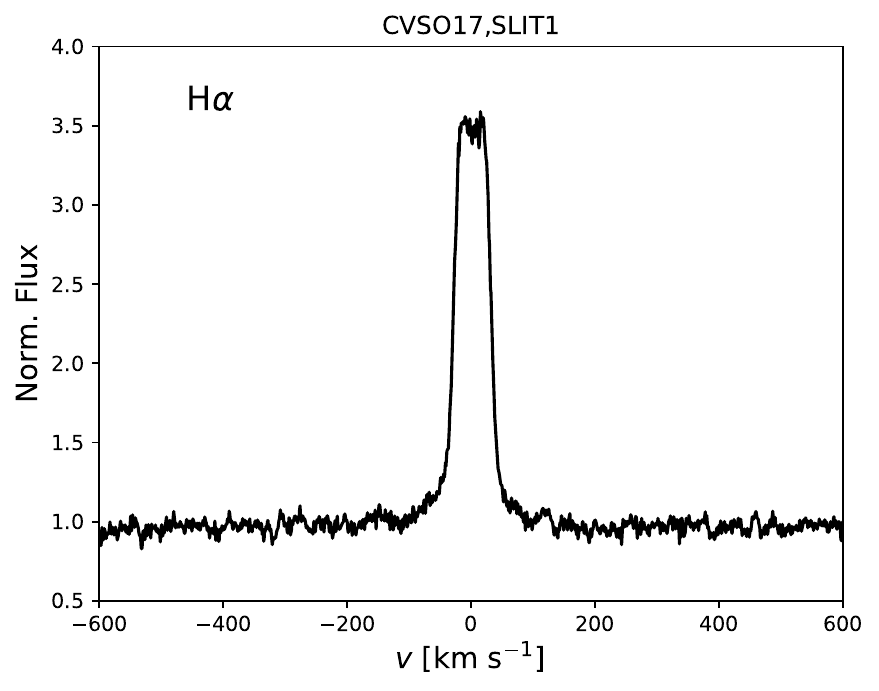}}
\hfill
\subfloat{\includegraphics[trim=0 0 0 0, clip, width=0.3 \textwidth]{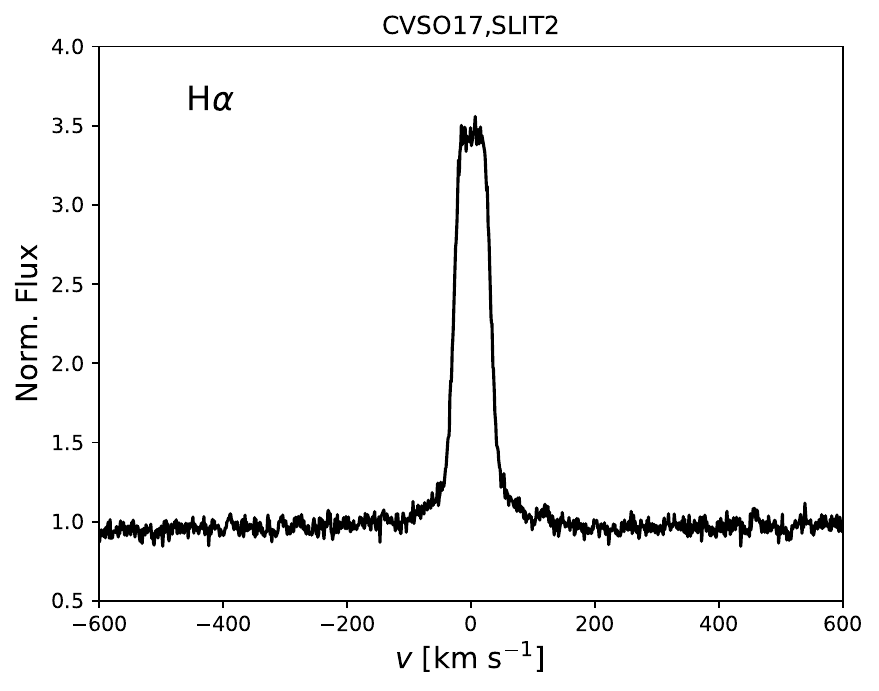}}
\hfill
\subfloat{\includegraphics[trim=0 0 0 0, clip, width=0.3 \textwidth]{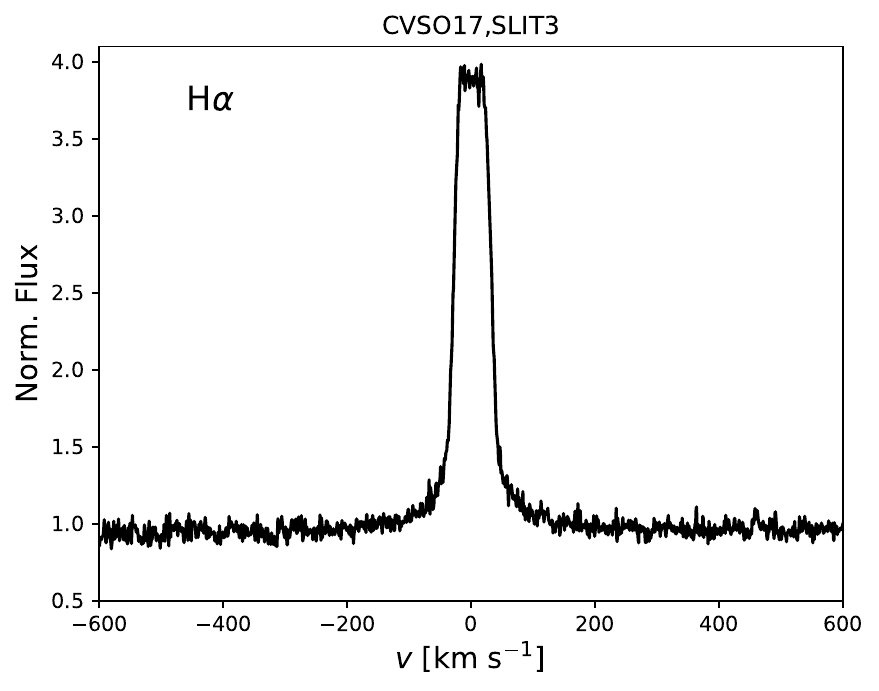}}
\hfill  
\subfloat{\includegraphics[trim=0 0 0 0, clip, width=0.3 \textwidth]{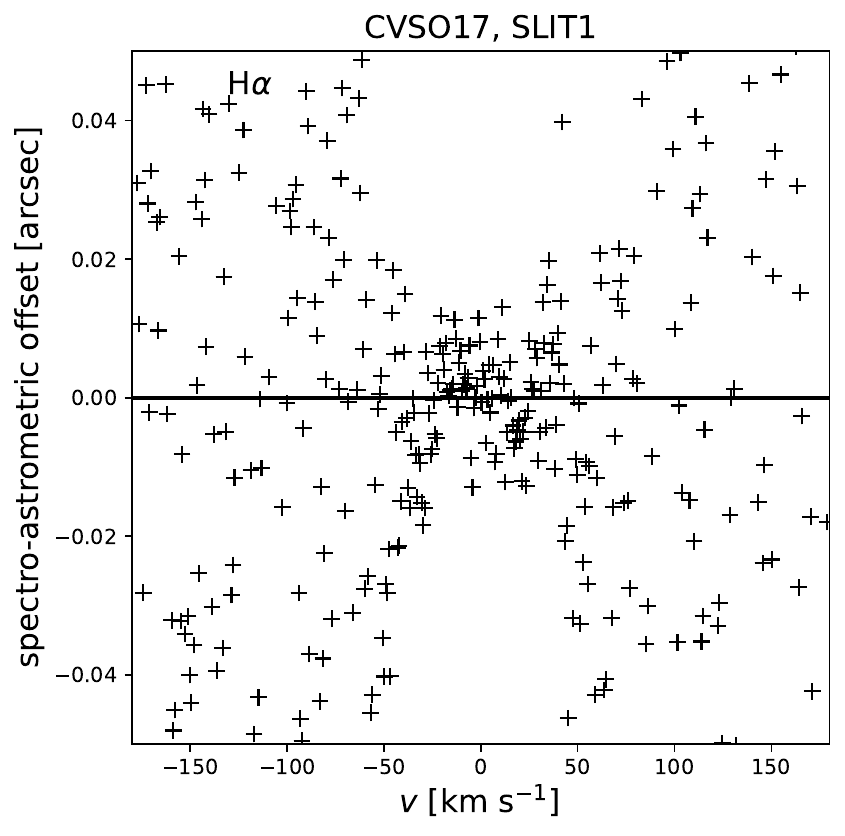}}
\hfill
\subfloat{\includegraphics[trim=0 0 0 0, clip, width=0.3 \textwidth]{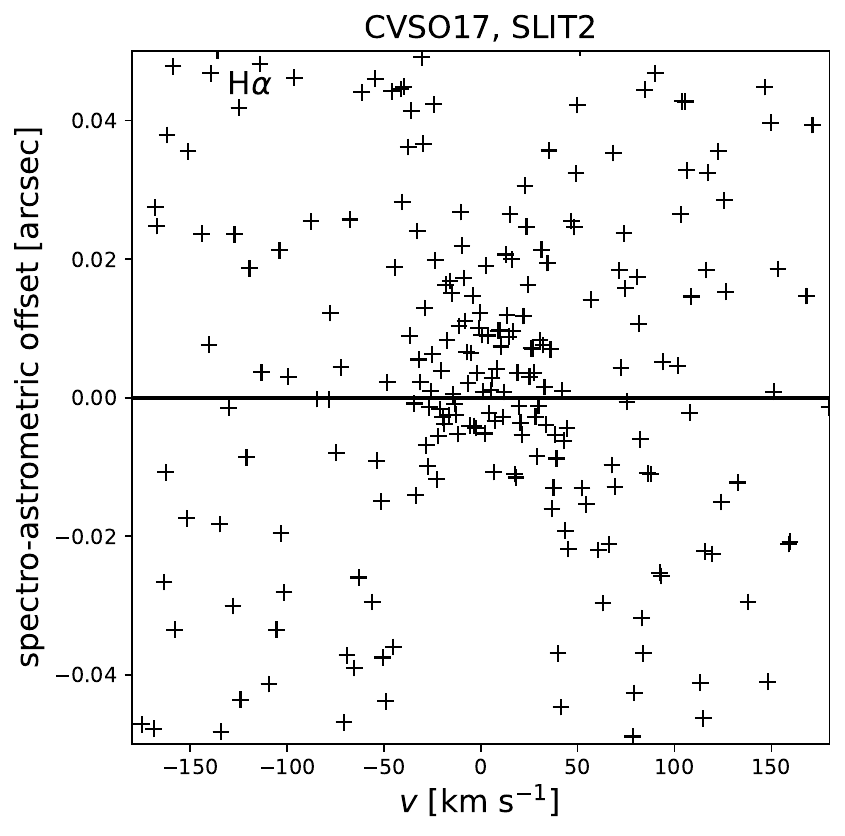}}
\hfill
\subfloat{\includegraphics[trim=0 0 0 0, clip, width=0.3 \textwidth]{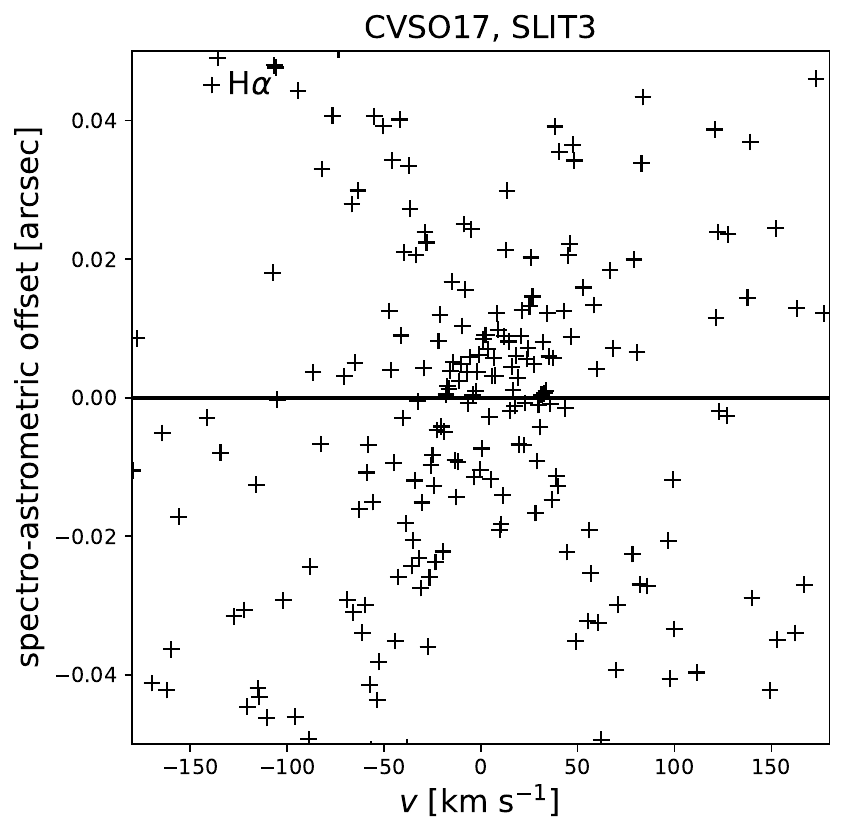}} 
\hfill
\subfloat{\includegraphics[trim=0 0 0 0, clip, width=0.3 \textwidth]{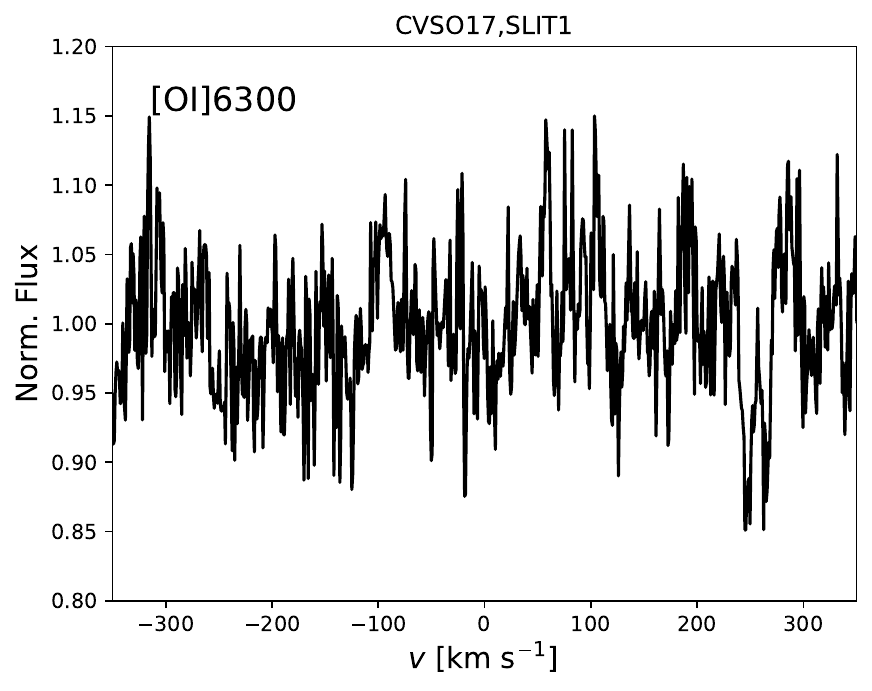}}
\hfill
\subfloat{\includegraphics[trim=0 0 0 0, clip, width=0.3 \textwidth]{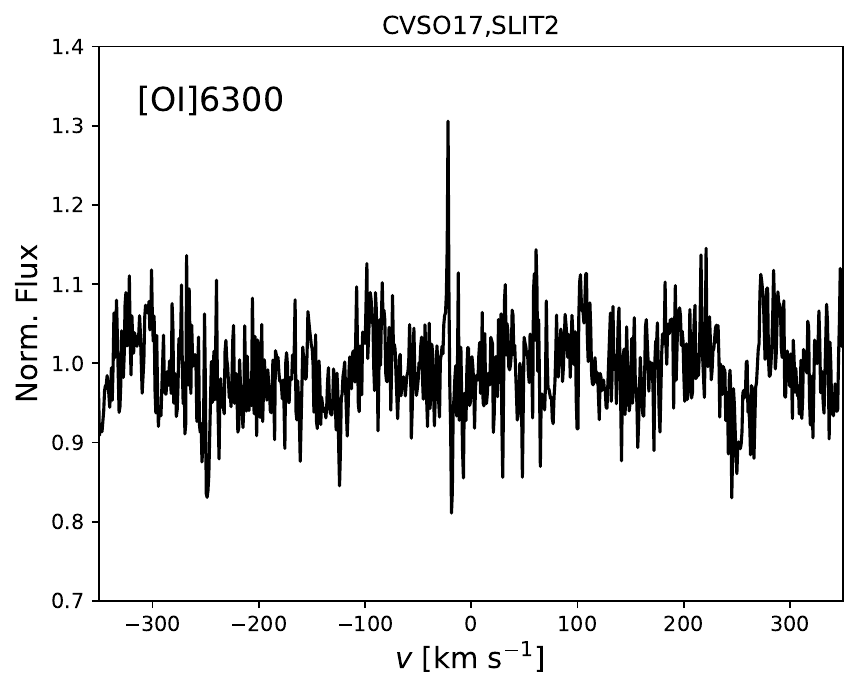}}
\hfill
\subfloat{\includegraphics[trim=0 0 0 0, clip, width=0.3 \textwidth]{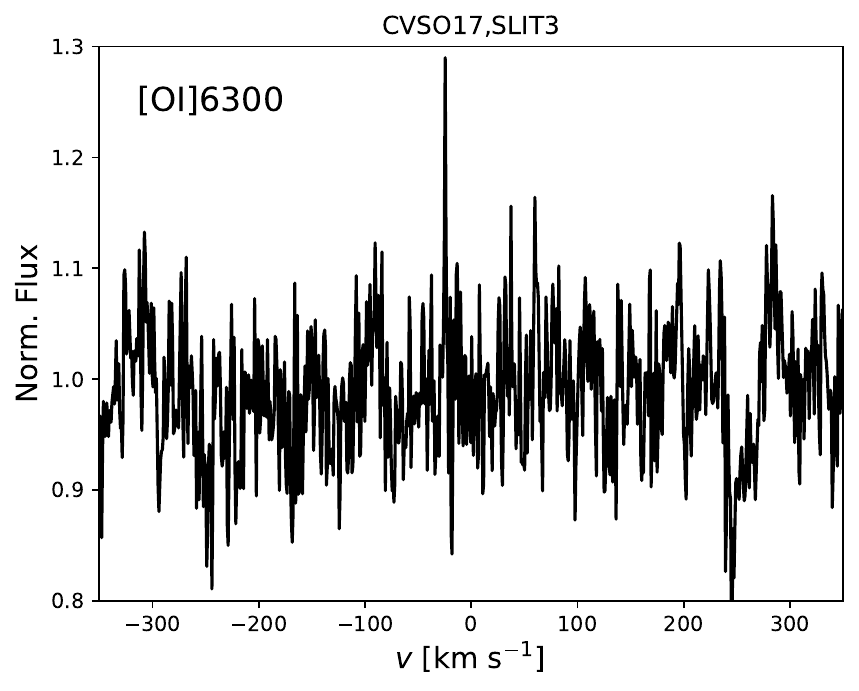}} 
\hfill   
\caption{\small{Line profiles of H$\alpha$ and [OI]$\lambda$6300 for all slit positions of CVSO\,17.}}\label{fig:all_minispectra_CVSO17}
\end{figure*} 

\begin{figure*} 
\centering
\subfloat{\includegraphics[trim=0 0 0 0, clip, width=0.3 \textwidth]{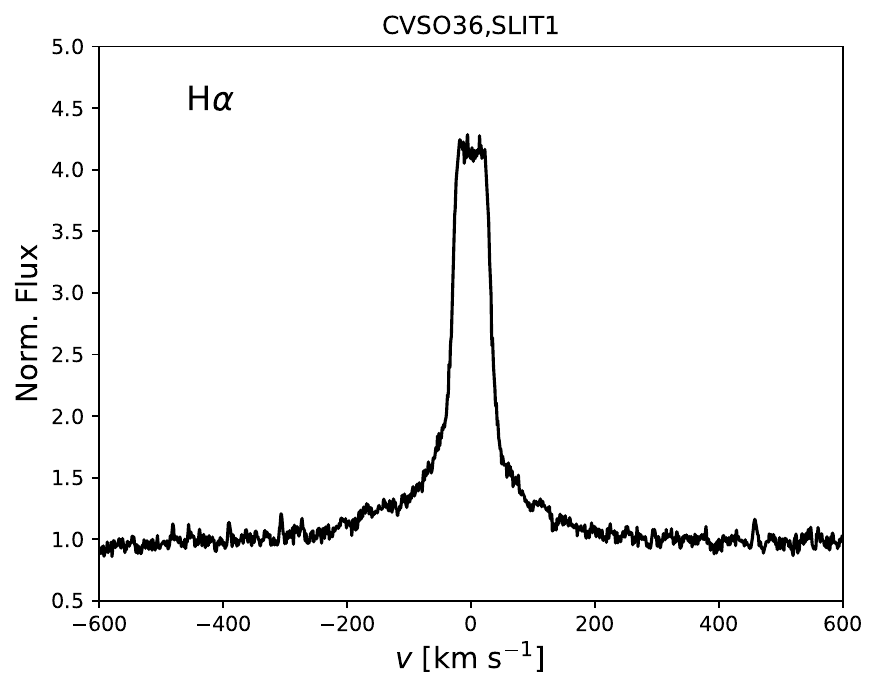}}
\hfill
\subfloat{\includegraphics[trim=0 0 0 0, clip, width=0.3 \textwidth]{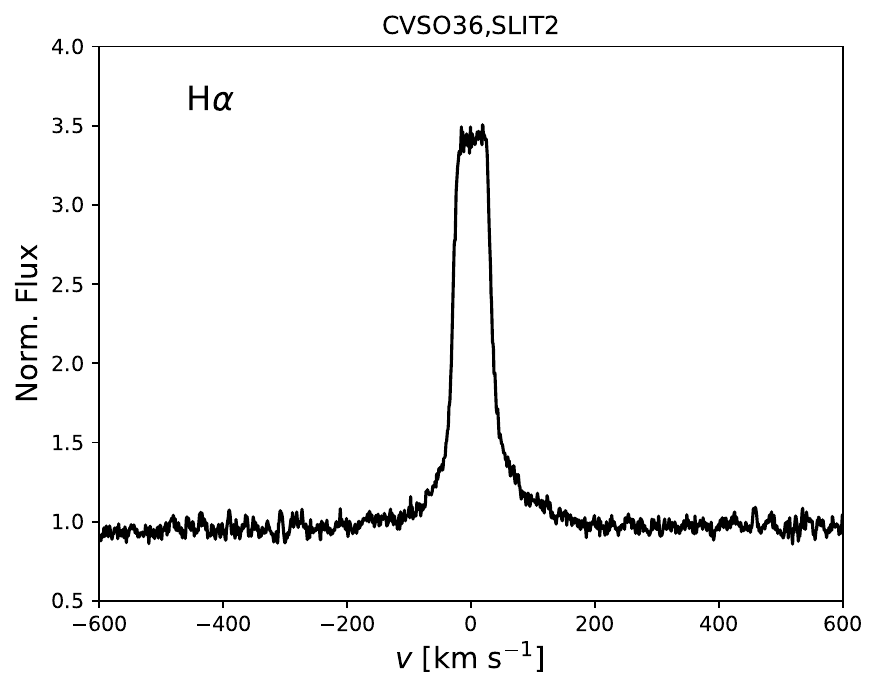}}
\hfill
\subfloat{\includegraphics[trim=0 0 0 0, clip, width=0.3 \textwidth]{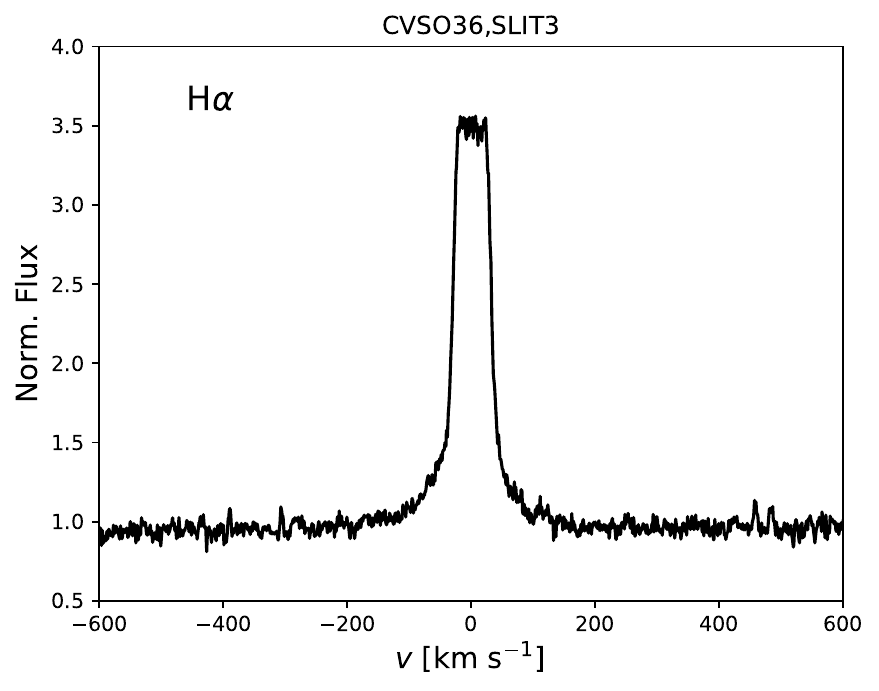}}
\hfill  
\subfloat{\includegraphics[trim=0 0 0 0, clip, width=0.3 \textwidth]{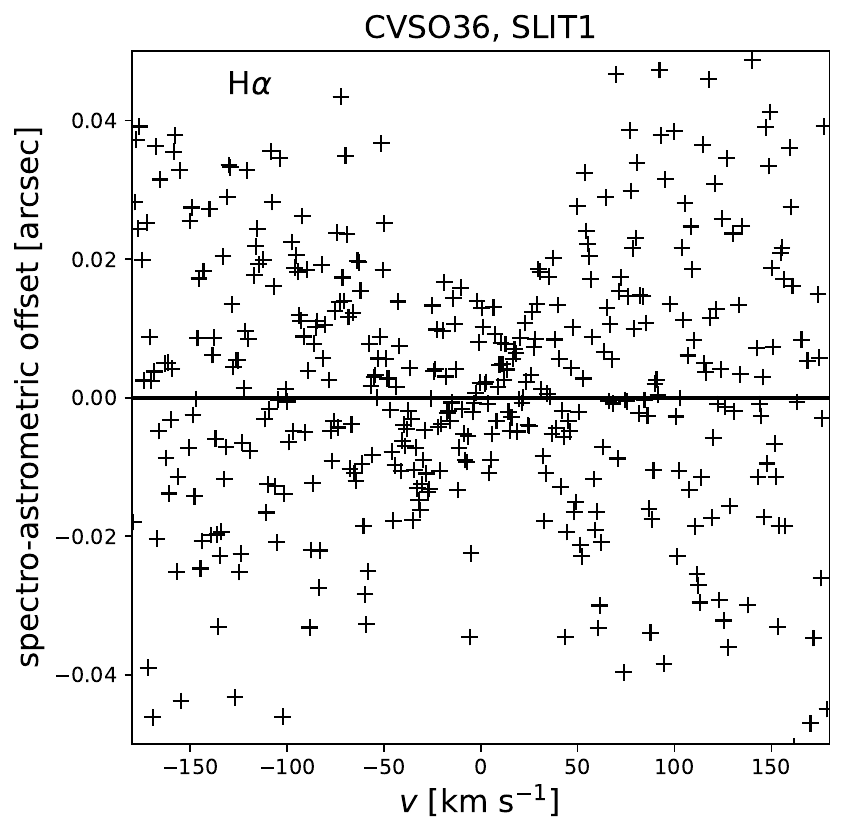}}
\hfill
\subfloat{\includegraphics[trim=0 0 0 0, clip, width=0.3 \textwidth]{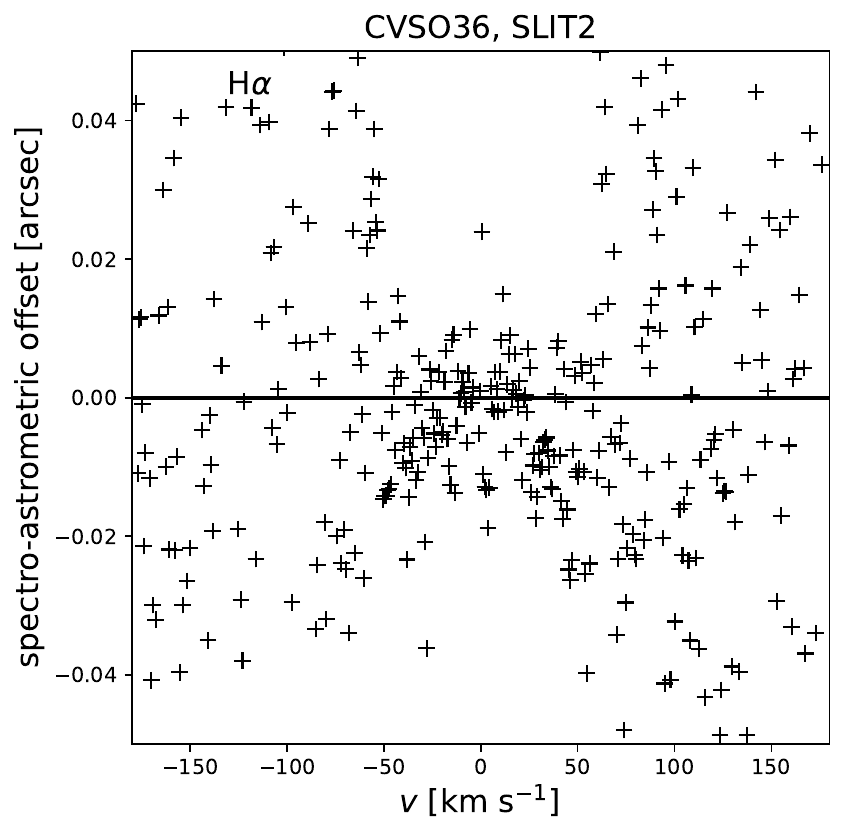}}
\hfill
\subfloat{\includegraphics[trim=0 0 0 0, clip, width=0.3 \textwidth]{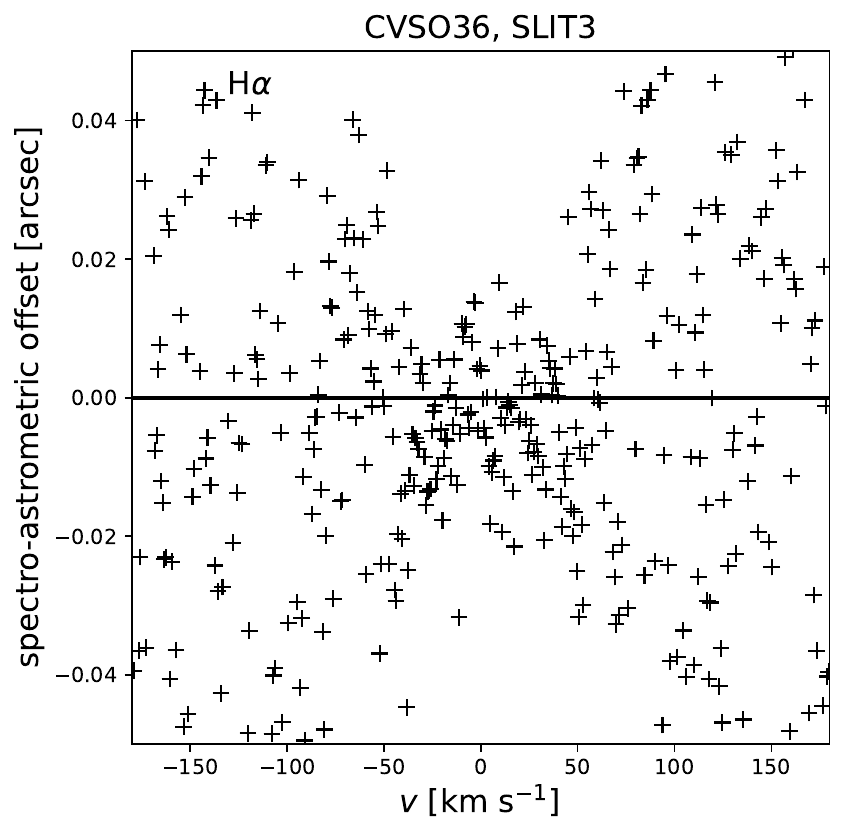}} 
\hfill
\subfloat{\includegraphics[trim=0 0 0 0, clip, width=0.3 \textwidth]{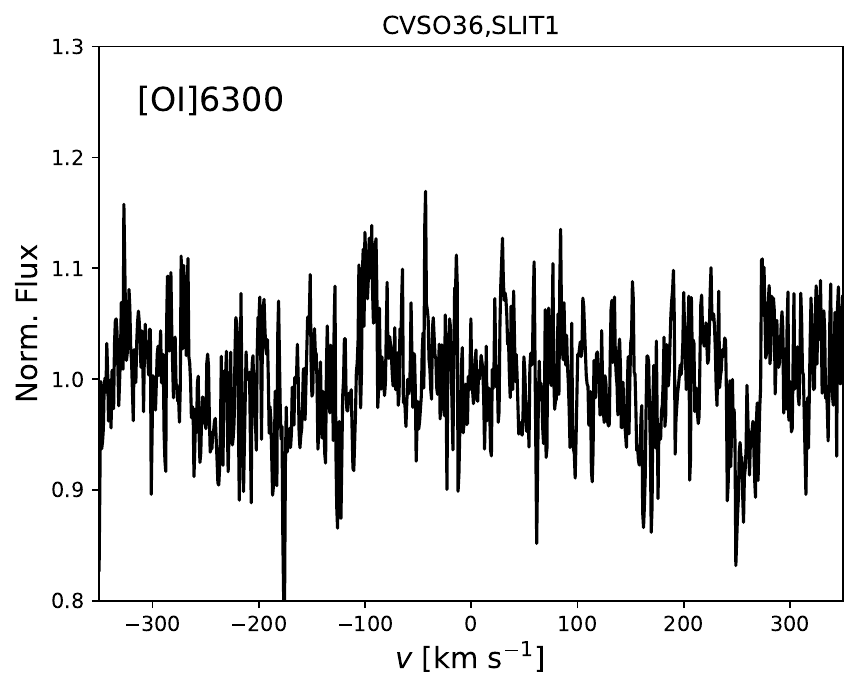}}
\hfill
\subfloat{\includegraphics[trim=0 0 0 0, clip, width=0.3 \textwidth]{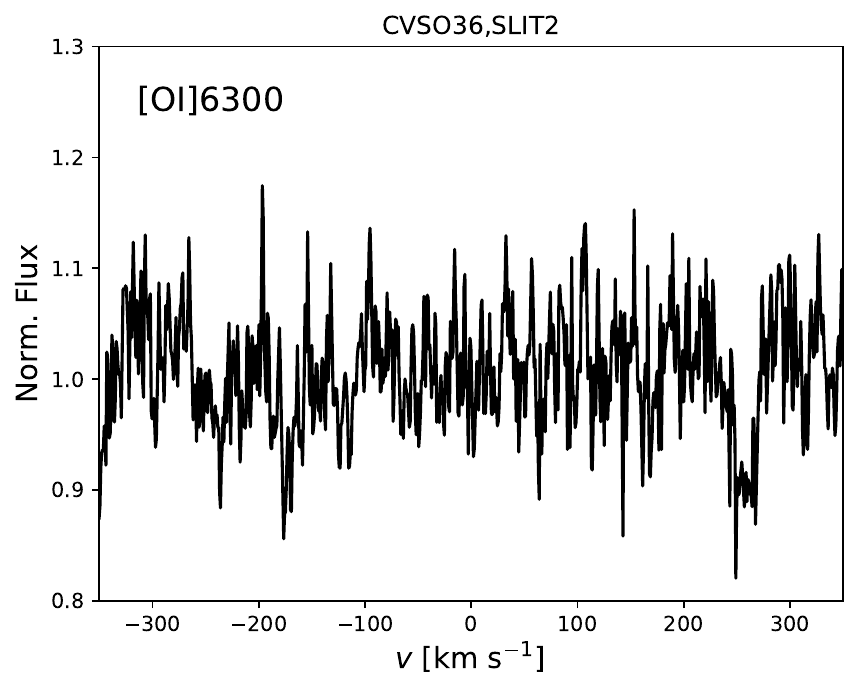}}
\hfill
\subfloat{\includegraphics[trim=0 0 0 0, clip, width=0.3 \textwidth]{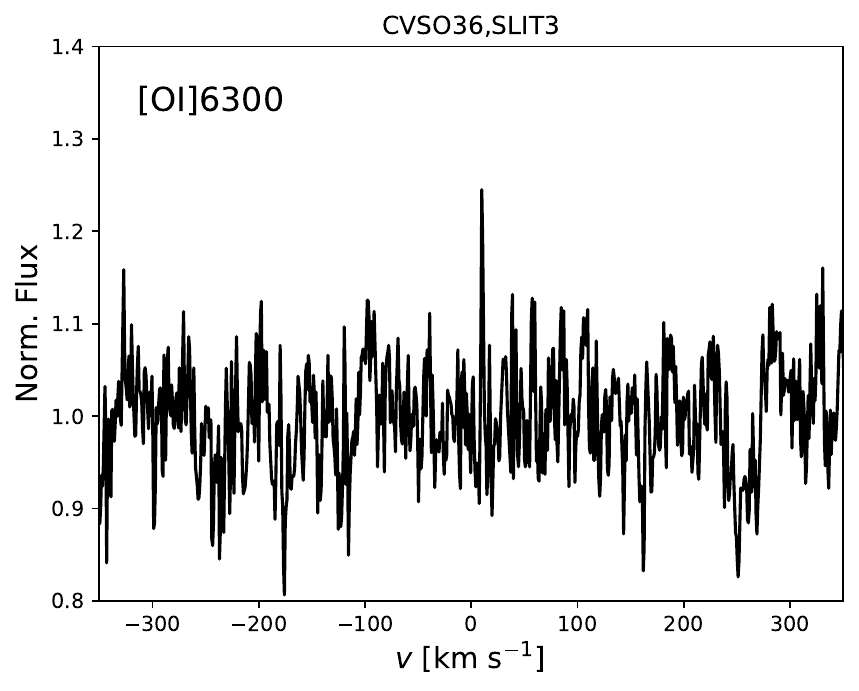}} 
\hfill   
\caption{\small{Line profiles of H$\alpha$ and [OI]$\lambda$6300 for all slit positions of CVSO\,36.}}\label{fig:all_minispectra_CVSO36}
\end{figure*} 

\begin{figure*} 
\centering
\subfloat{\includegraphics[trim=0 0 0 0, clip, width=0.3 \textwidth]{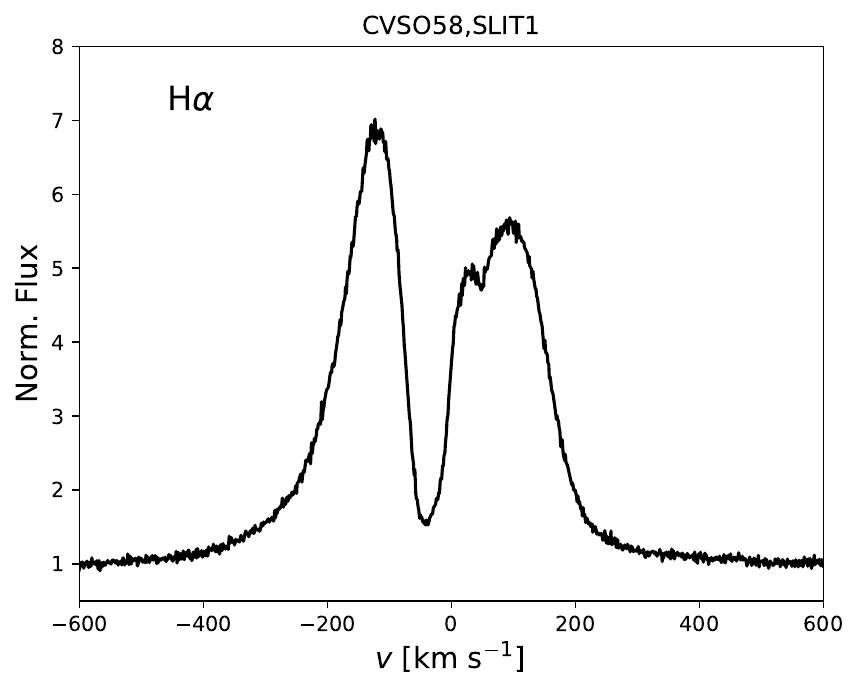}}
\hfill
\subfloat{\includegraphics[trim=0 0 0 0, clip, width=0.3 \textwidth]{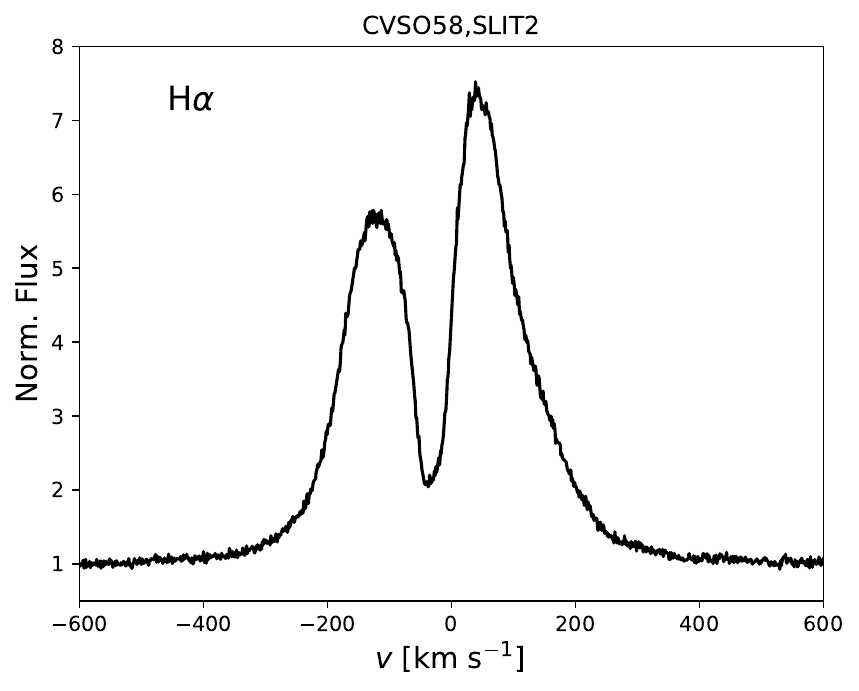}}
\hfill
\subfloat{\includegraphics[trim=0 0 0 0, clip, width=0.3 \textwidth]{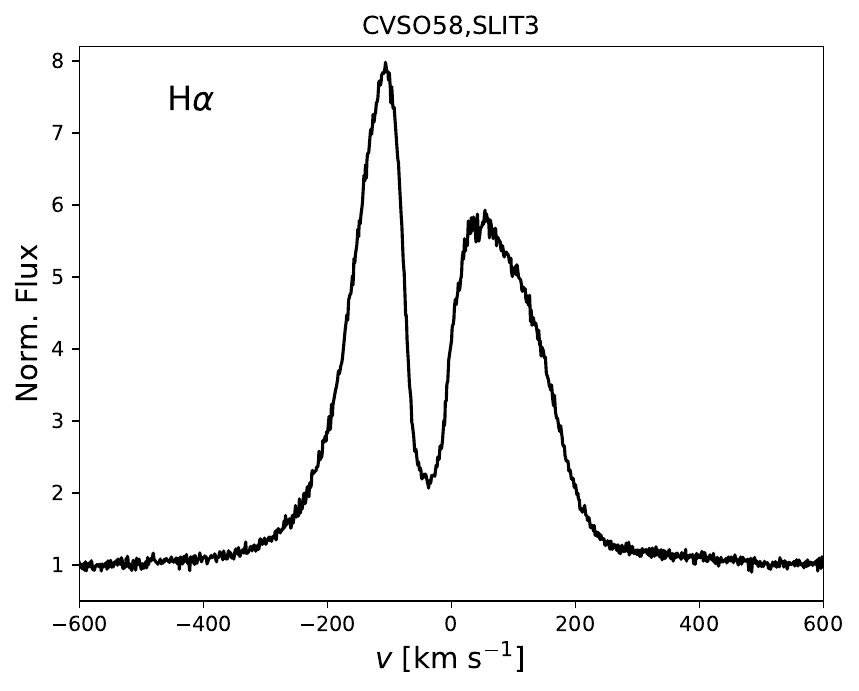}}
\hfill  
\subfloat{\includegraphics[trim=0 0 0 0, clip, width=0.3 \textwidth]{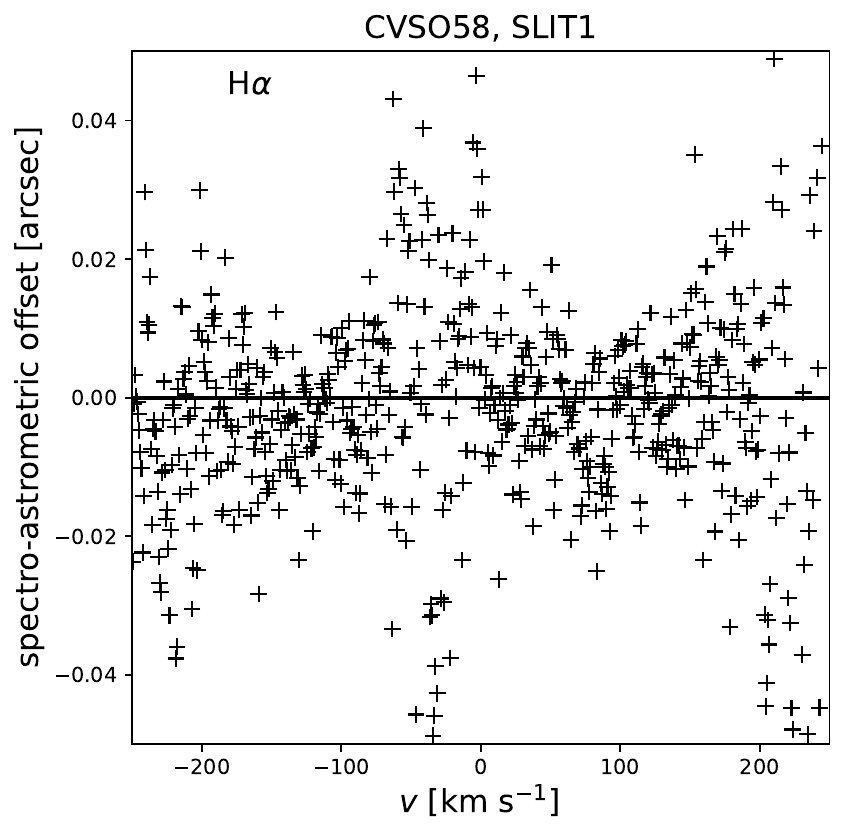}}
\hfill
\subfloat{\includegraphics[trim=0 0 0 0, clip, width=0.3 \textwidth]{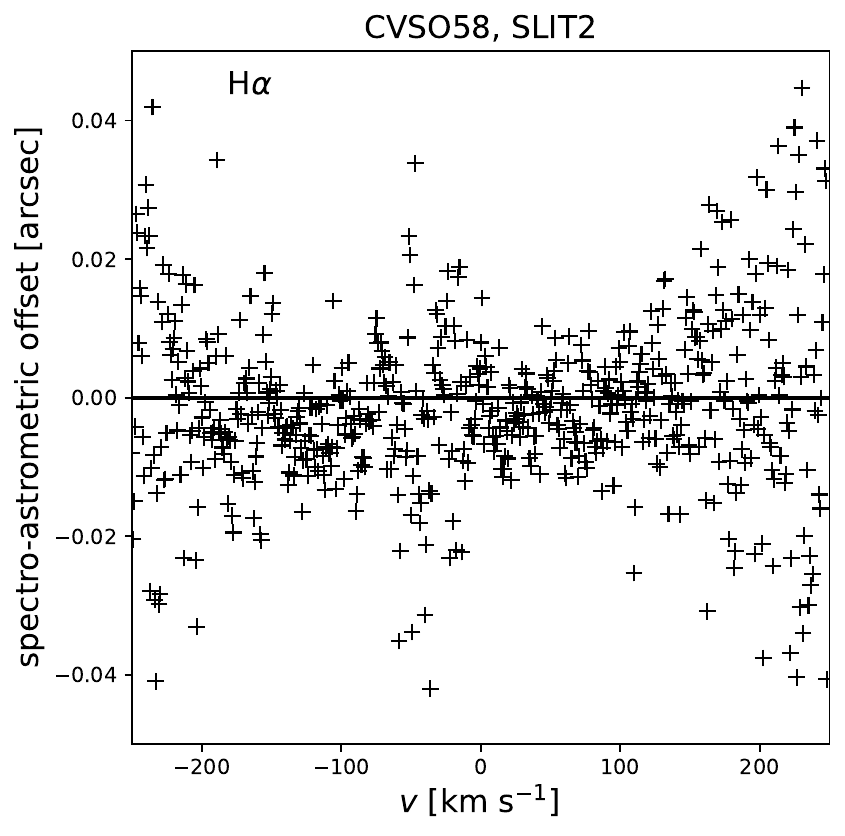}}
\hfill
\subfloat{\includegraphics[trim=0 0 0 0, clip, width=0.3 \textwidth]{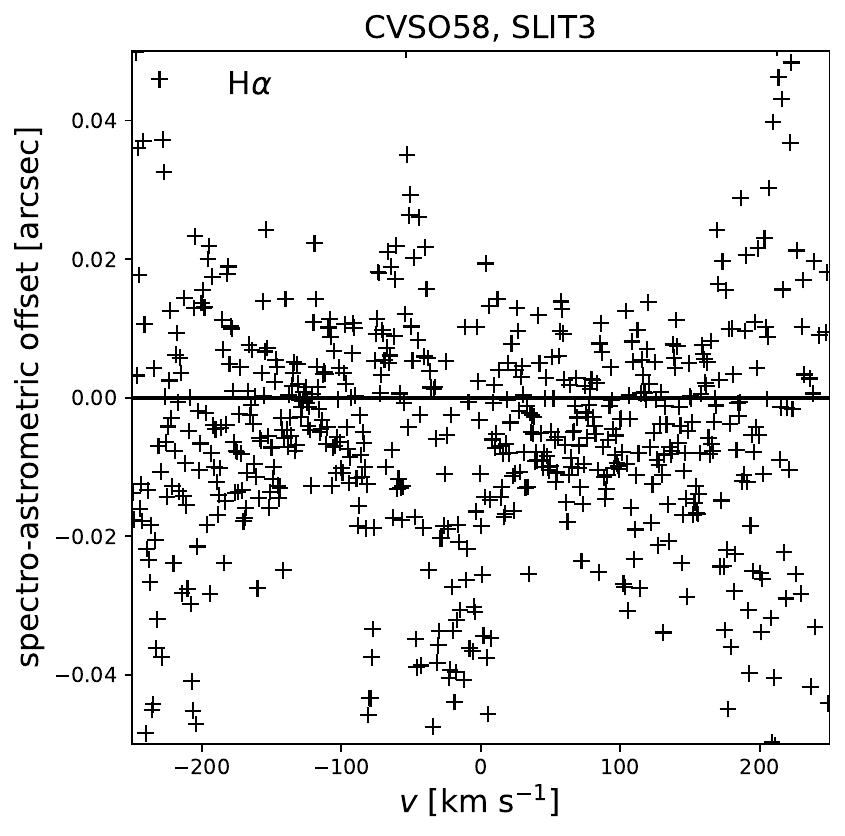}} 
\hfill
\subfloat{\includegraphics[trim=0 0 0 0, clip, width=0.3 \textwidth]{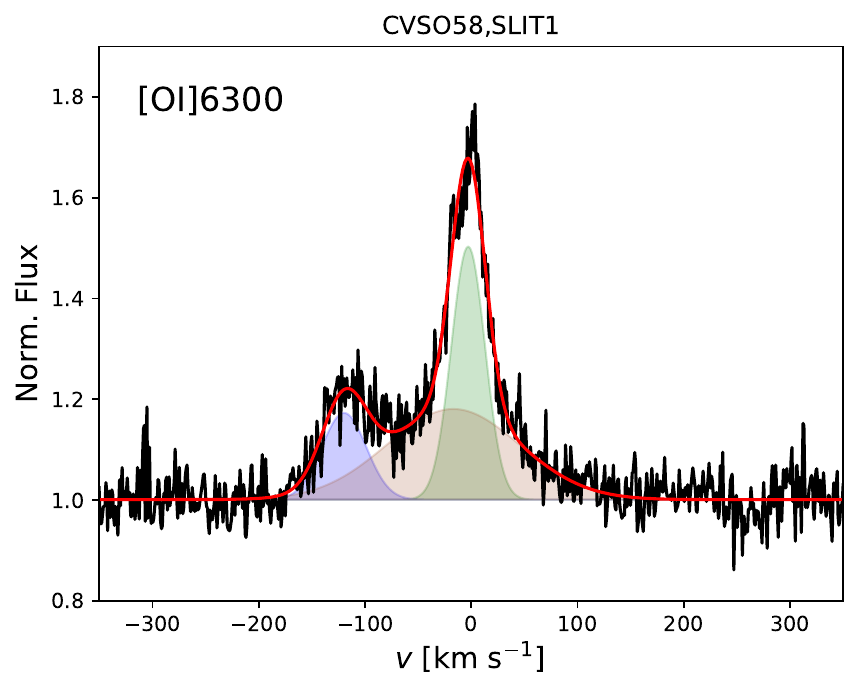}}
\hfill
\subfloat{\includegraphics[trim=0 0 0 0, clip, width=0.3 \textwidth]{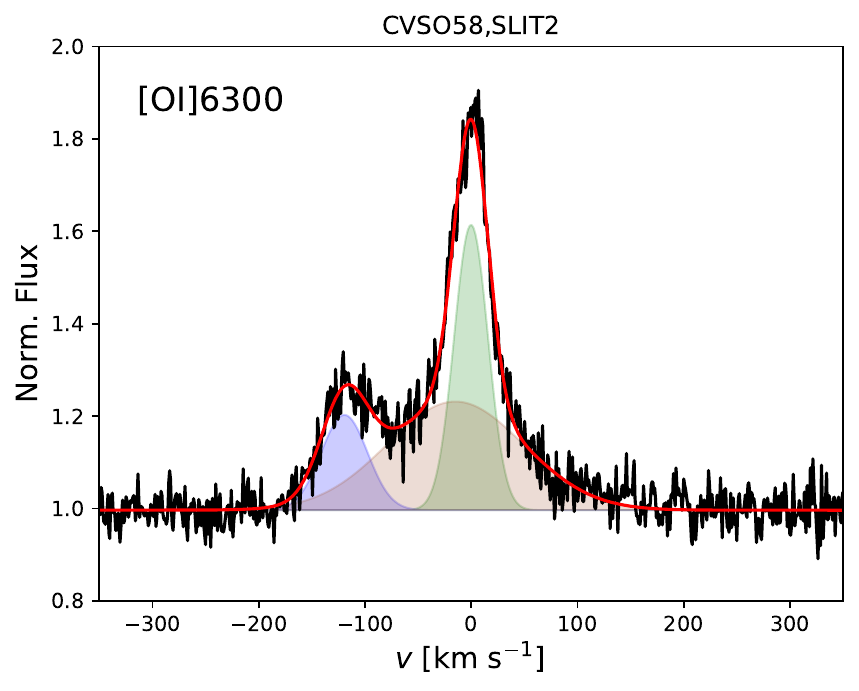}}
\hfill
\subfloat{\includegraphics[trim=0 0 0 0, clip, width=0.3 \textwidth]{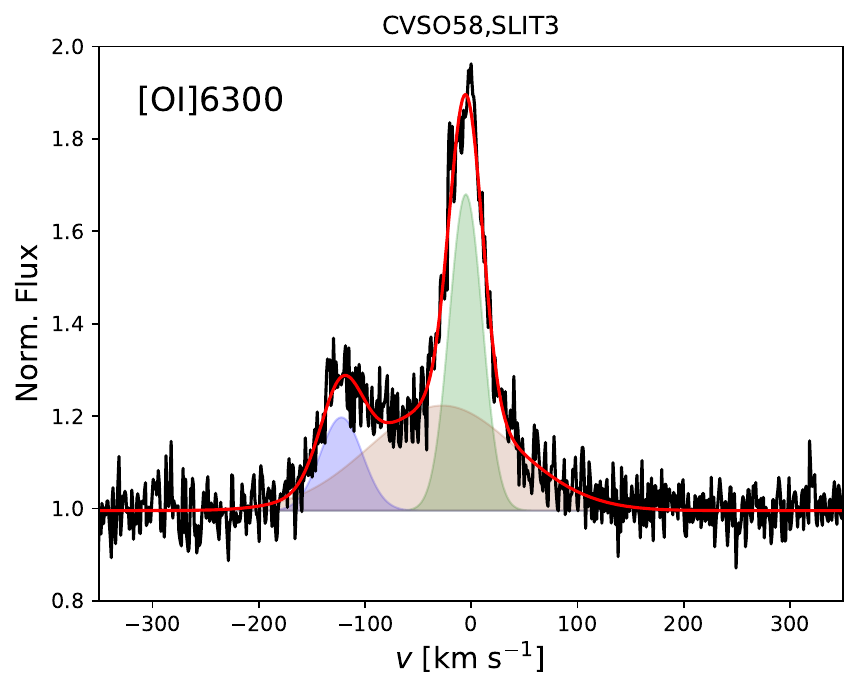}} 
\hfill   
\subfloat{\includegraphics[trim=0 0 0 0, clip, width=0.3 \textwidth]{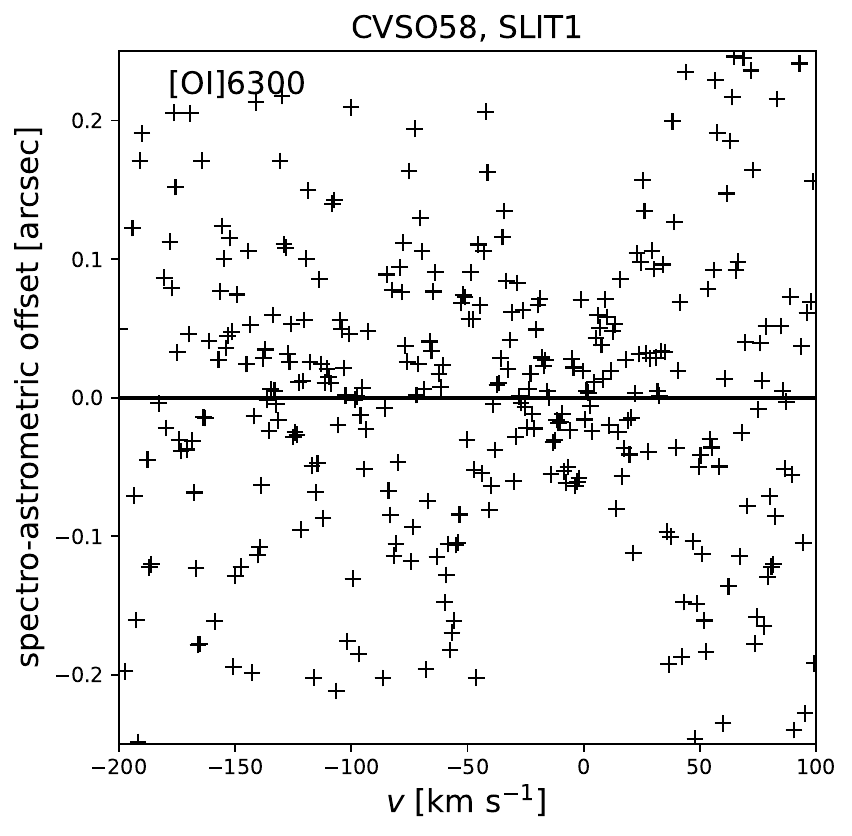}}
\hfill
\subfloat{\includegraphics[trim=0 0 0 0, clip, width=0.3 \textwidth]{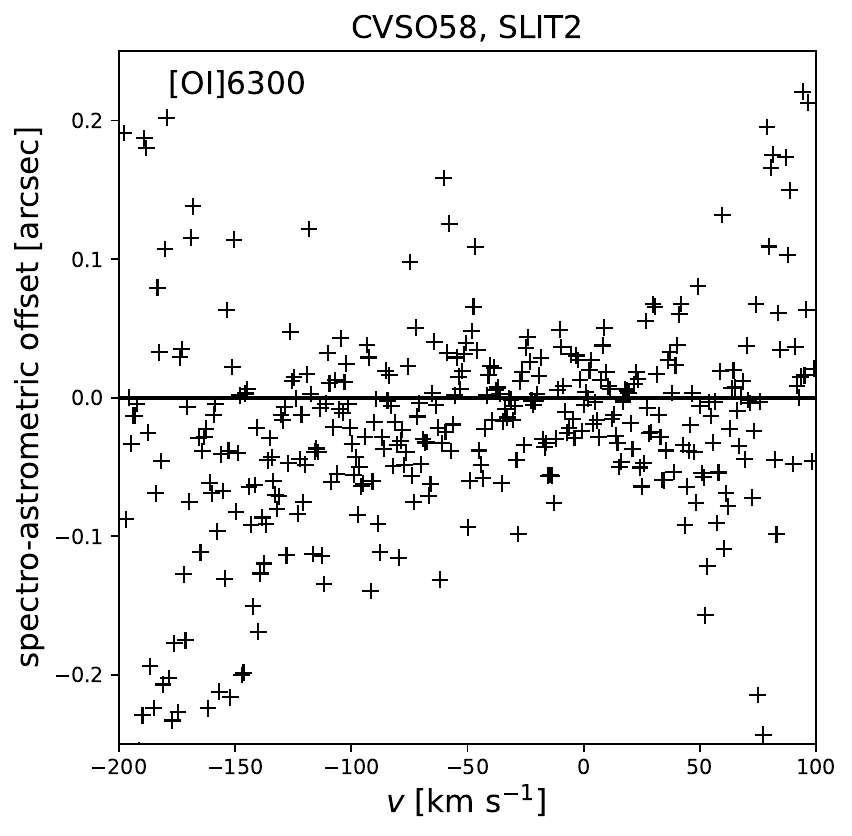}}
\hfill
\subfloat{\includegraphics[trim=0 0 0 0, clip, width=0.3 \textwidth]{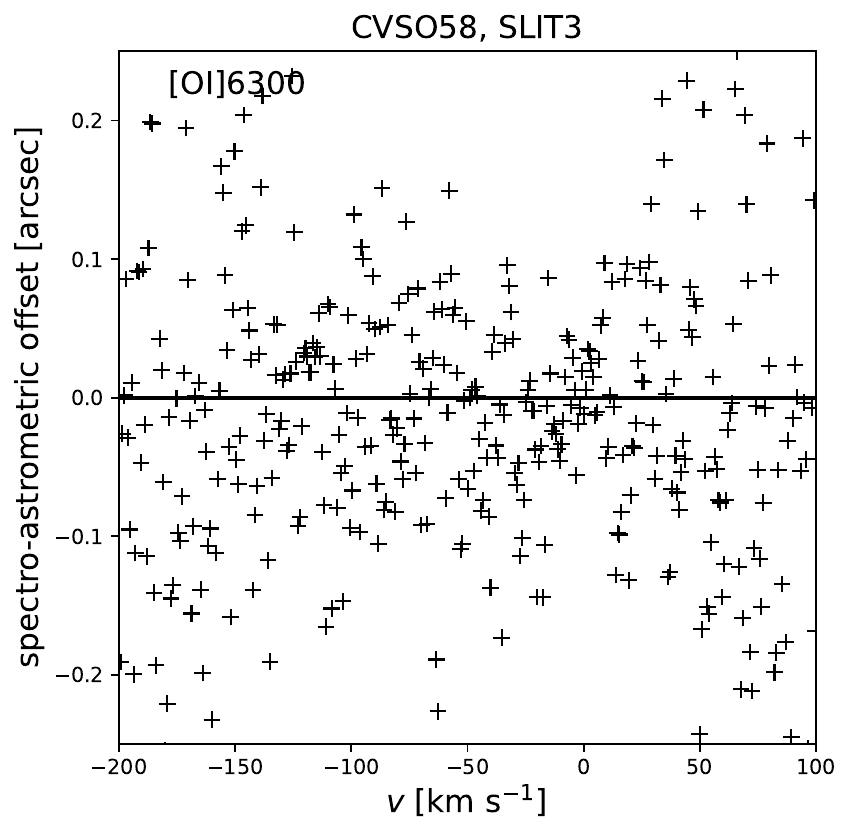}} 
\hfill
\caption{\small{Line profiles of H$\alpha$ and [OI]$\lambda$6300 for all slit positions of CVSO\,58.}}\label{fig:all_minispectra_CVSO58}
\end{figure*} 

\begin{figure*} 
\centering
\subfloat{\includegraphics[trim=0 0 0 0, clip, width=0.3 \textwidth]{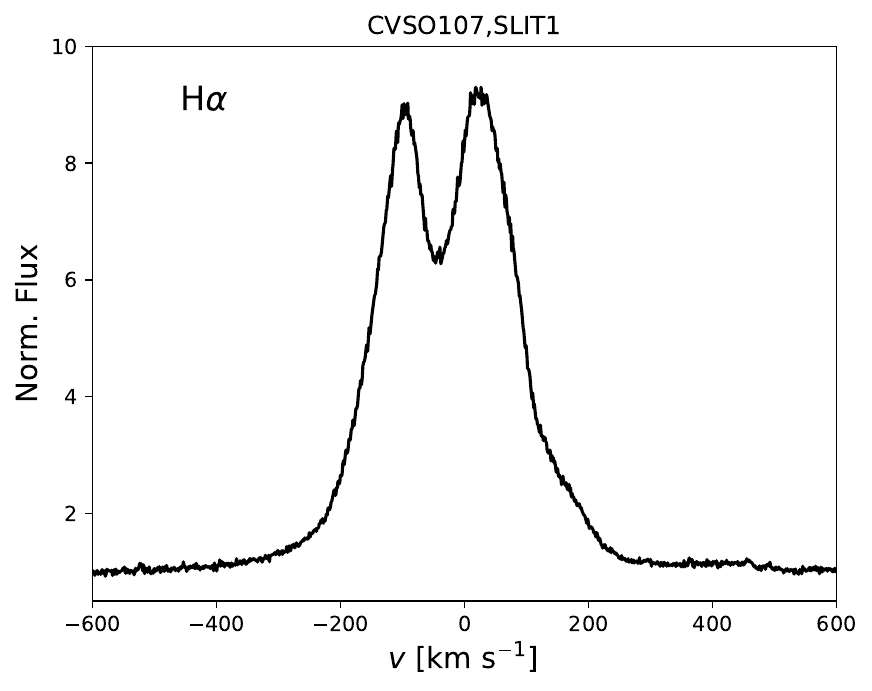}}
\hfill
\subfloat{\includegraphics[trim=0 0 0 0, clip, width=0.3 \textwidth]{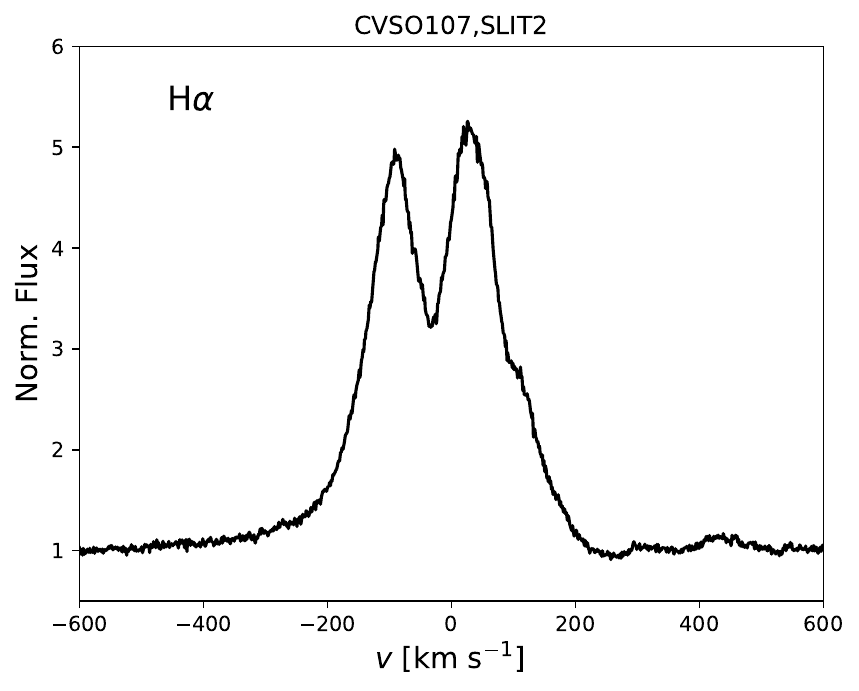}}
\hfill
\subfloat{\includegraphics[trim=0 0 0 0, clip, width=0.3 \textwidth]{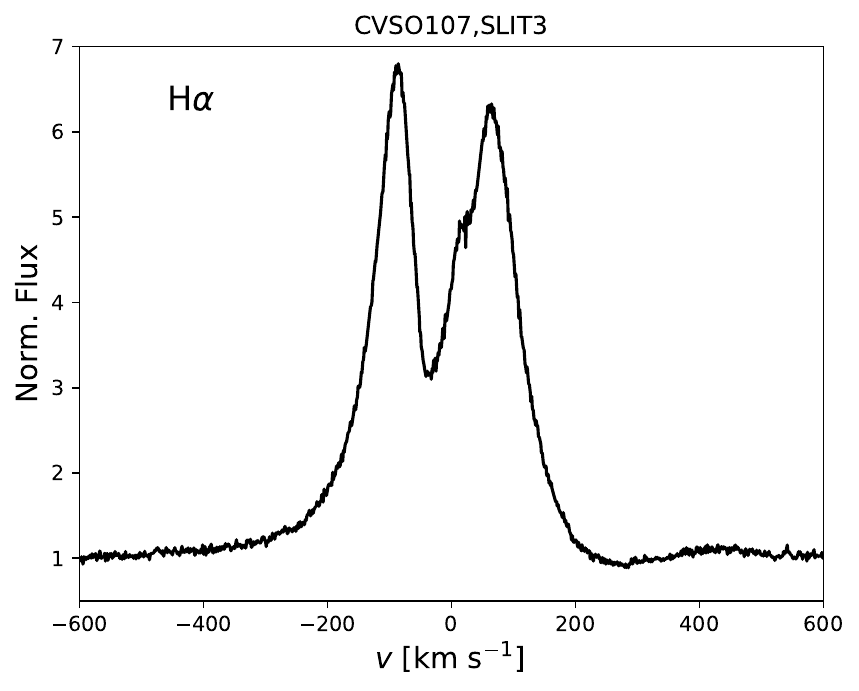}}
\hfill  
\subfloat{\includegraphics[trim=0 0 0 0, clip, width=0.3 \textwidth]{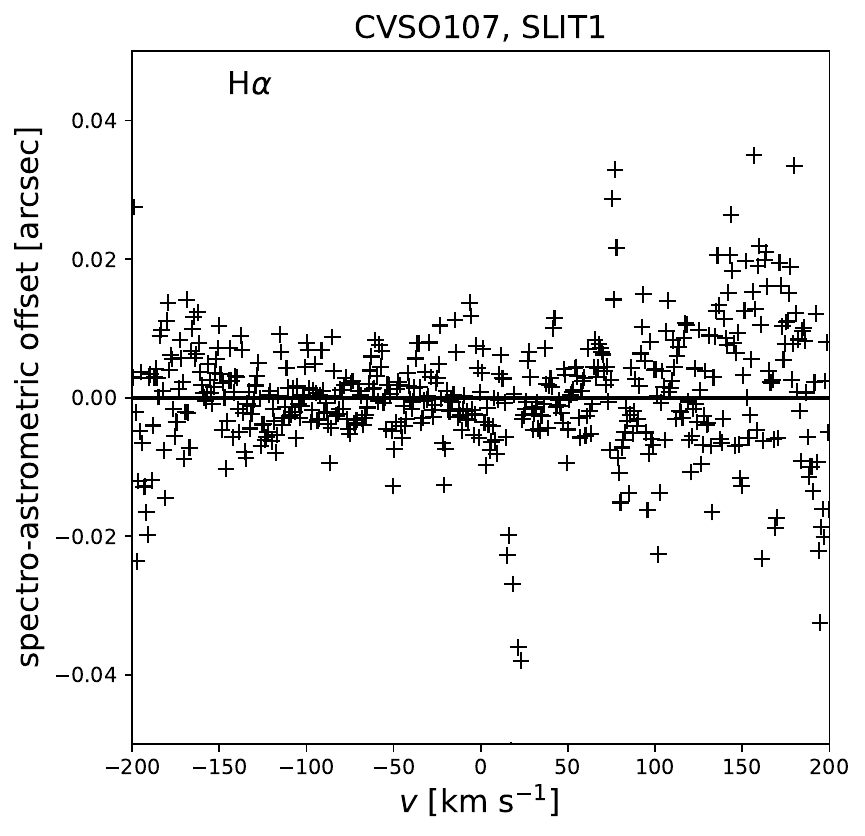}}
\hfill
\subfloat{\includegraphics[trim=0 0 0 0, clip, width=0.3 \textwidth]{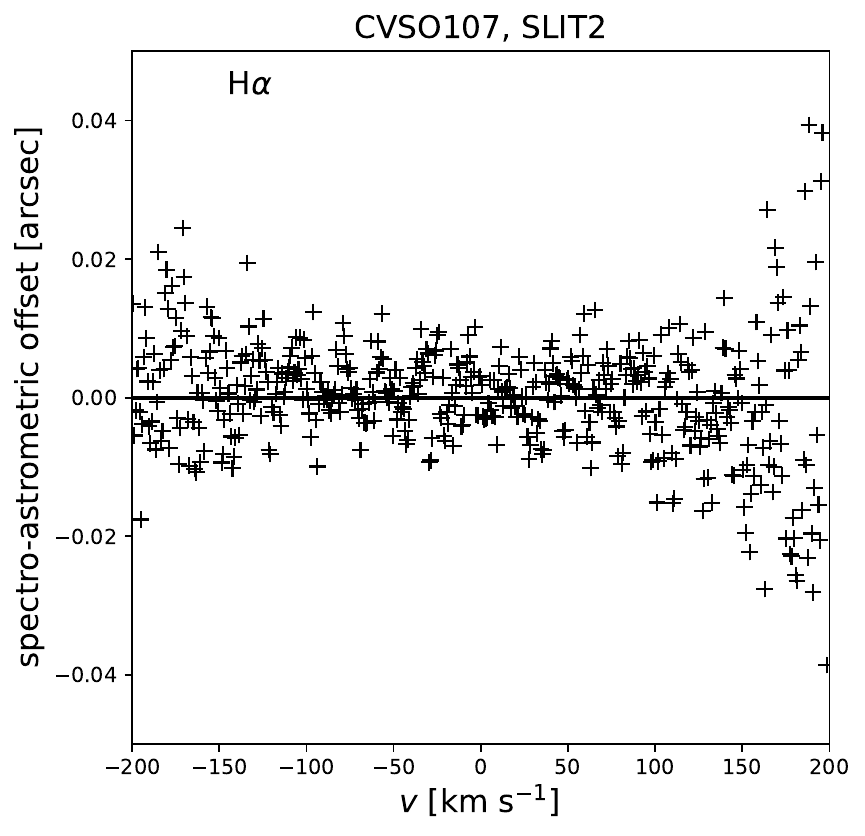}}
\hfill
\subfloat{\includegraphics[trim=0 0 0 0, clip, width=0.3 \textwidth]{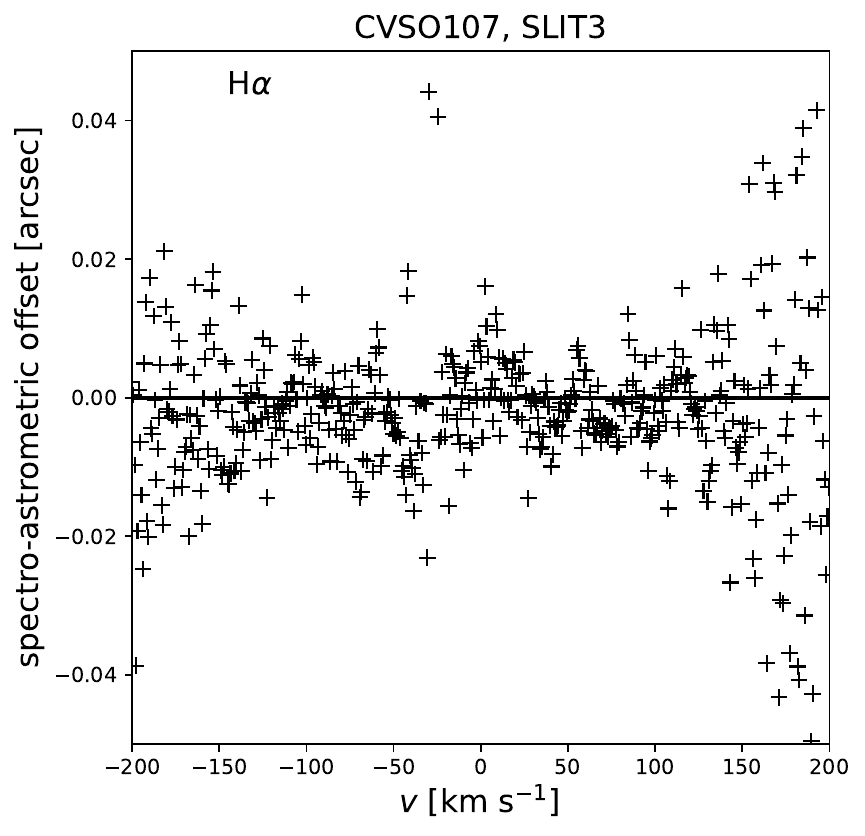}} 
\hfill
\subfloat{\includegraphics[trim=0 0 0 0, clip, width=0.3 \textwidth]{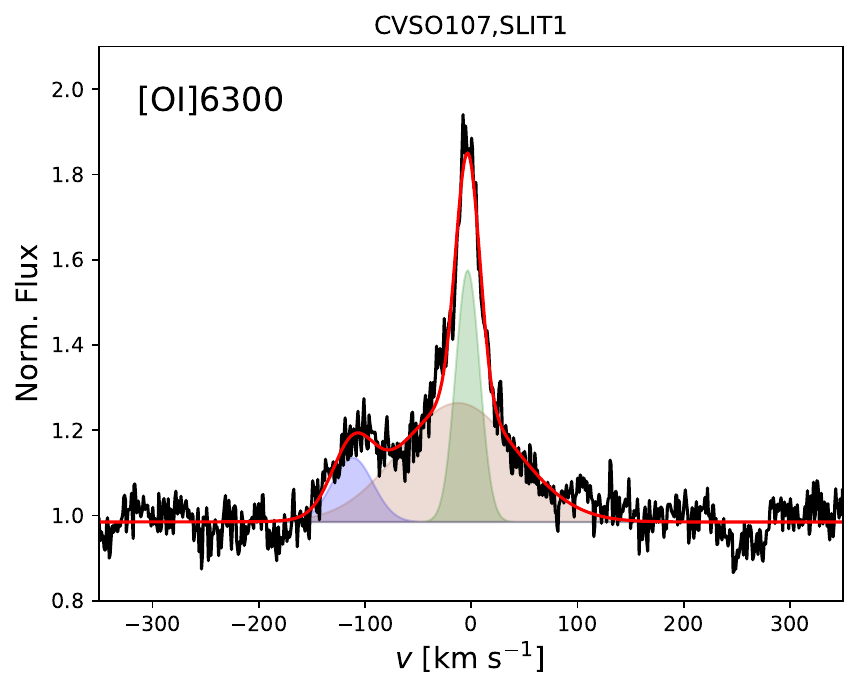}}
\hfill
\subfloat{\includegraphics[trim=0 0 0 0, clip, width=0.3 \textwidth]{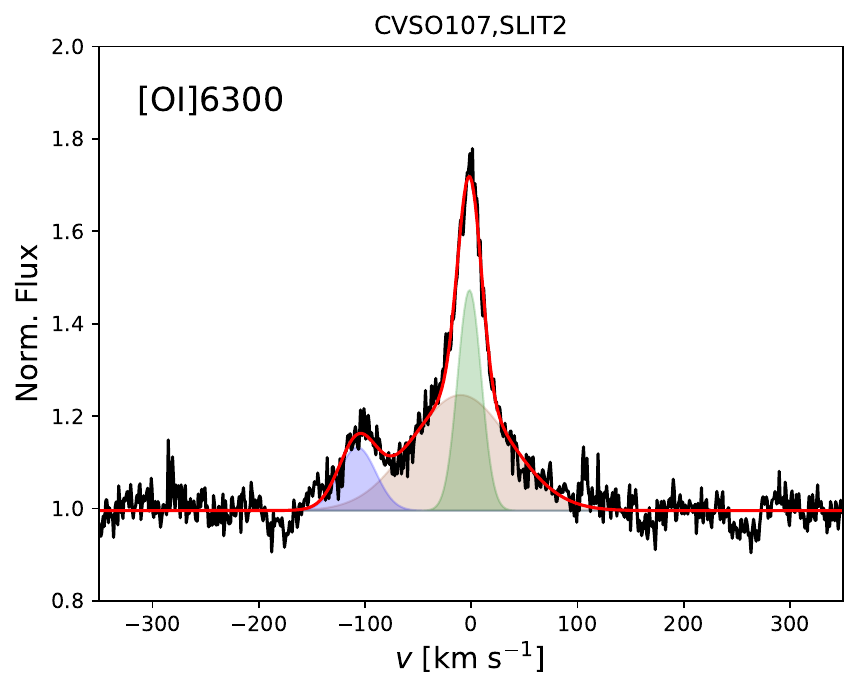}}
\hfill
\subfloat{\includegraphics[trim=0 0 0 0, clip, width=0.3 \textwidth]{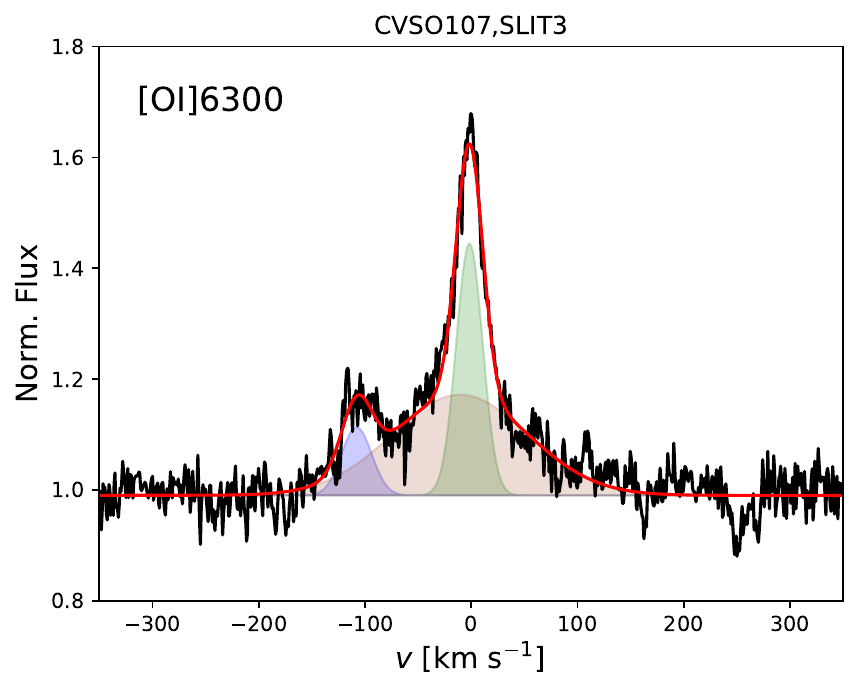}} 
\hfill   
\subfloat{\includegraphics[trim=0 0 0 0, clip, width=0.3 \textwidth]{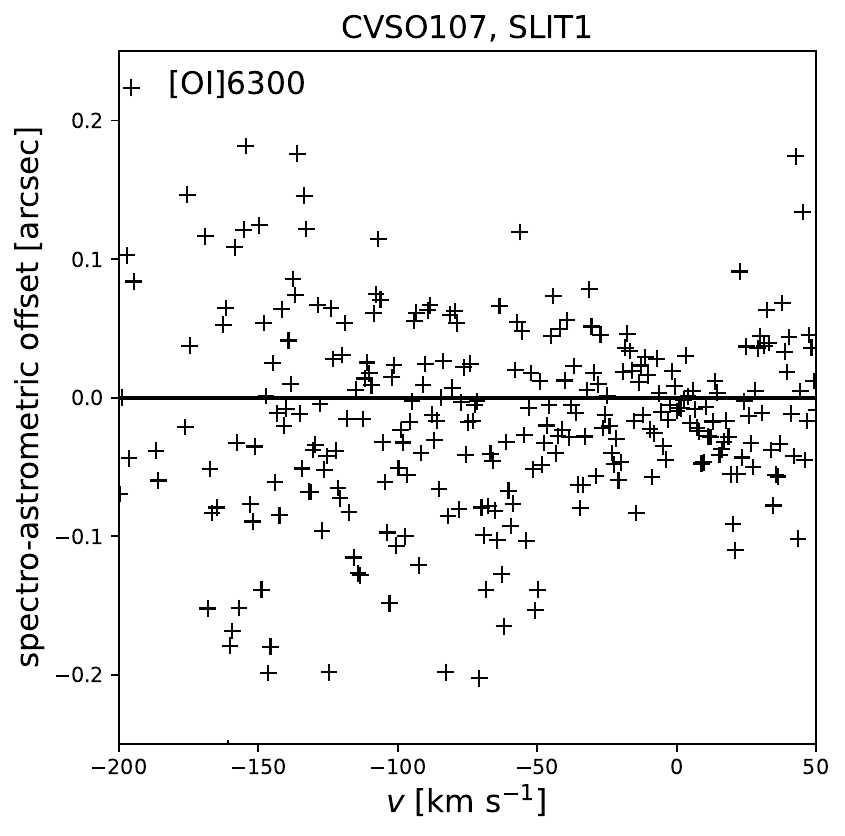}}
\hfill
\subfloat{\includegraphics[trim=0 0 0 0, clip, width=0.3 \textwidth]{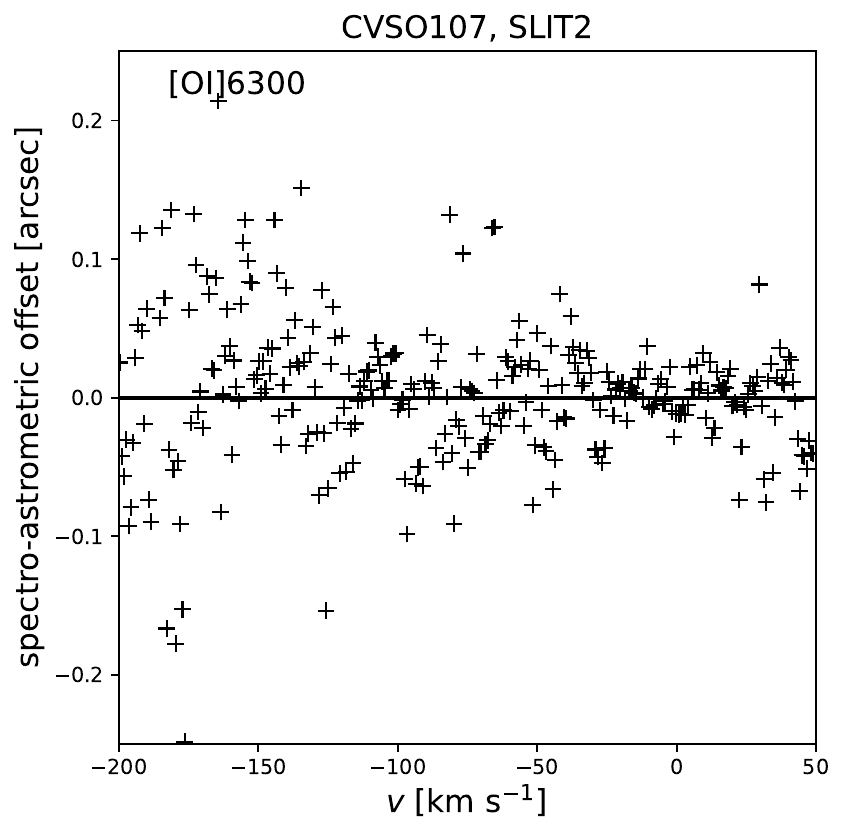}}
\hfill
\subfloat{\includegraphics[trim=0 0 0 0, clip, width=0.3 \textwidth]{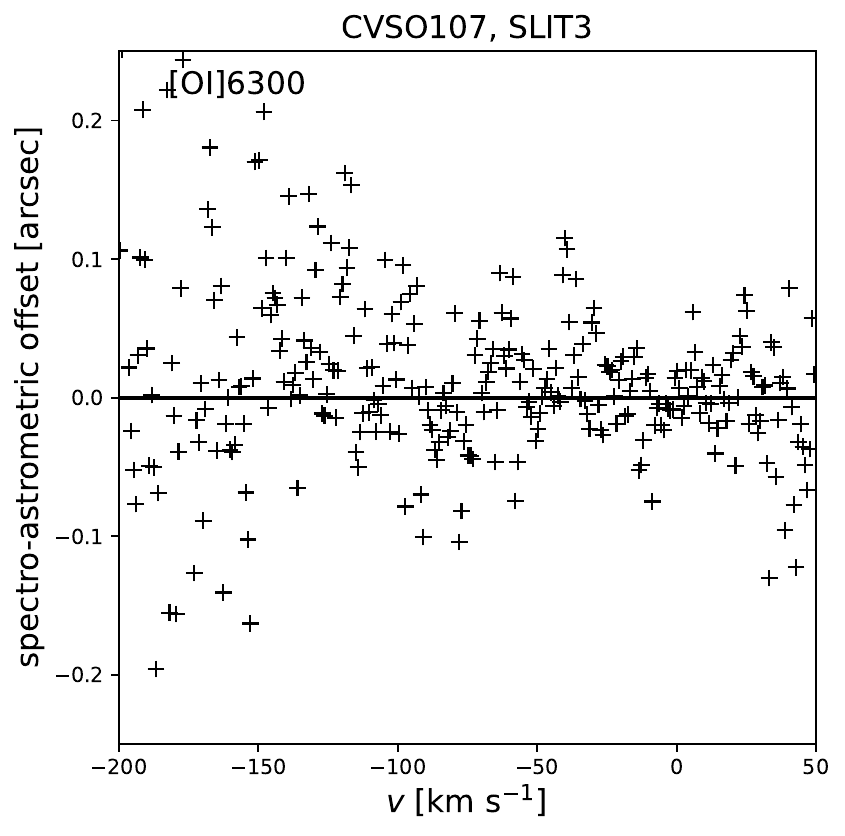}} 
\hfill
\caption{\small{Line profiles of H$\alpha$ and [OI]$\lambda$6300 for all slit positions of CVSO\,107.}}\label{fig:all_minispectra_CVSO107}
\end{figure*} 

\begin{figure*} 
\centering
\subfloat{\includegraphics[trim=0 0 0 0, clip, width=0.3 \textwidth]{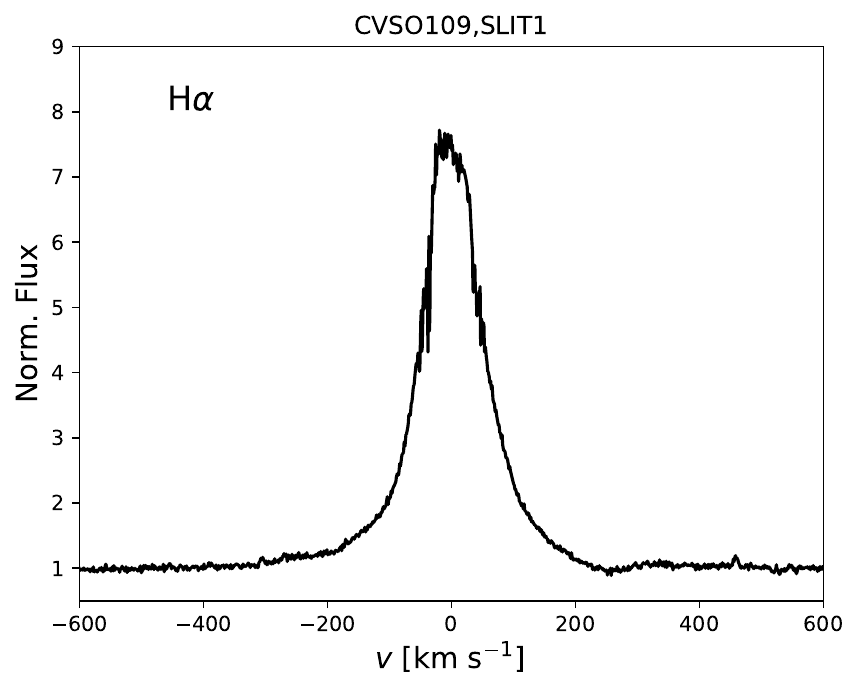}}
\hfill
\subfloat{\includegraphics[trim=0 0 0 0, clip, width=0.3 \textwidth]{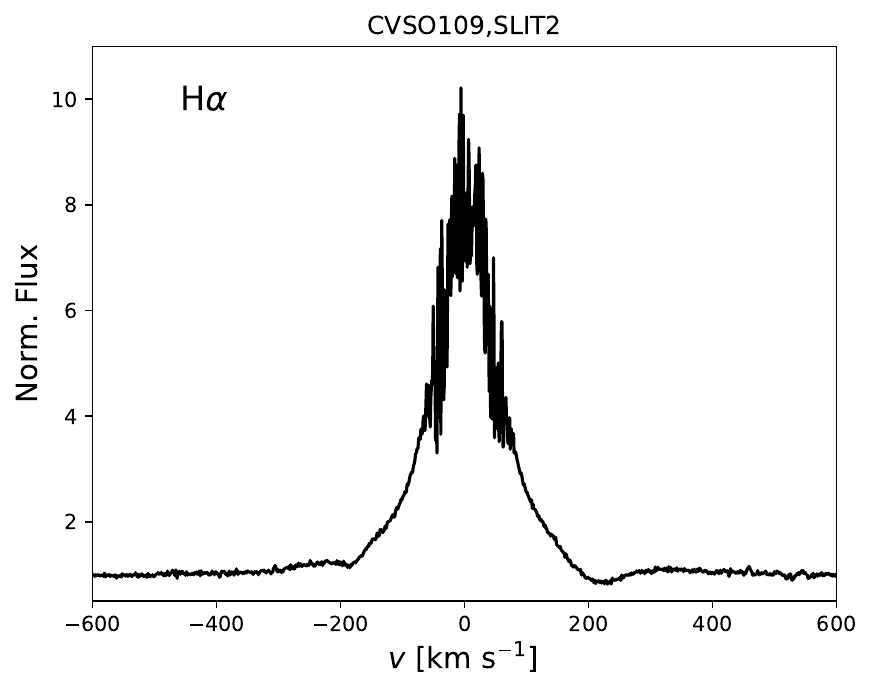}}
\hfill
\subfloat{\includegraphics[trim=0 0 0 0, clip, width=0.3 \textwidth]{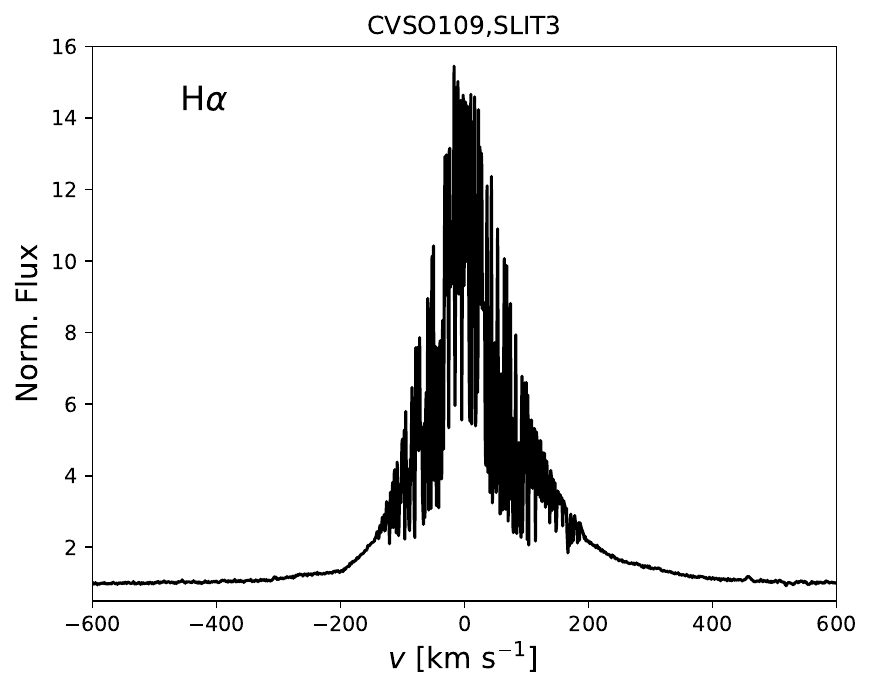}}
\hfill  
\subfloat{\includegraphics[trim=0 0 0 0, clip, width=0.3 \textwidth]{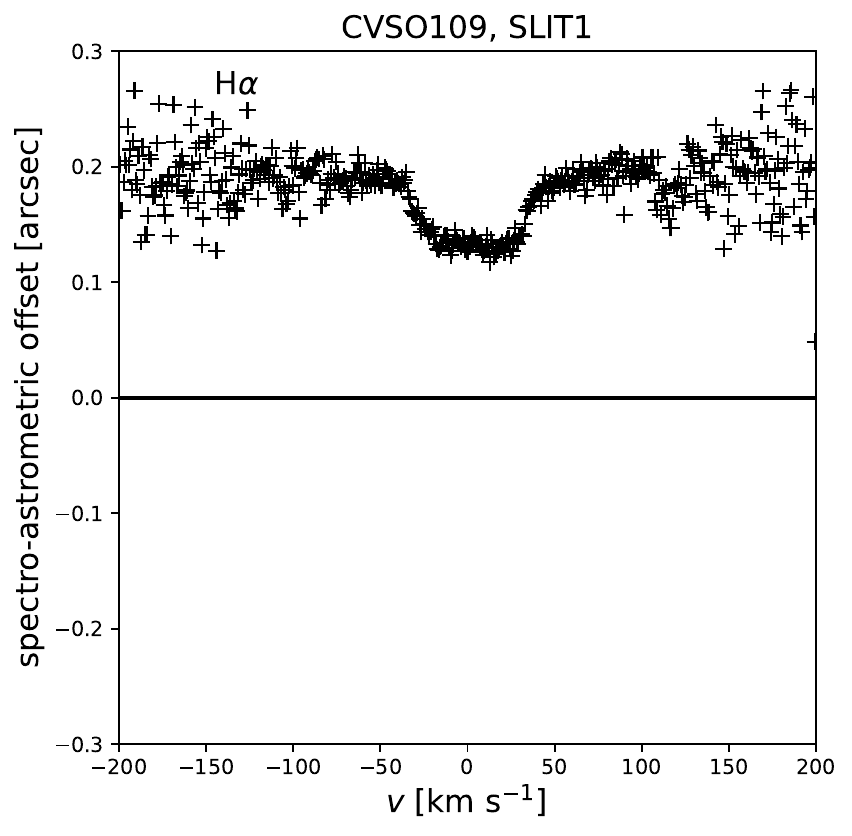}}
\hfill
\subfloat{\includegraphics[trim=0 0 0 0, clip, width=0.3 \textwidth]{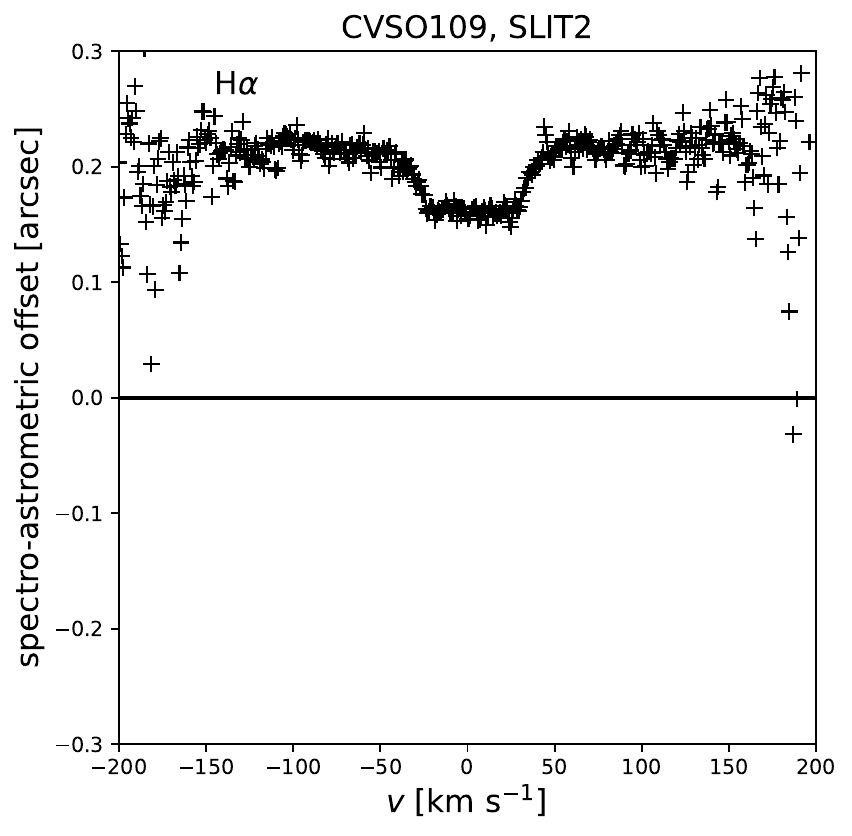}}
\hfill
\subfloat{\includegraphics[trim=0 0 0 0, clip, width=0.3 \textwidth]{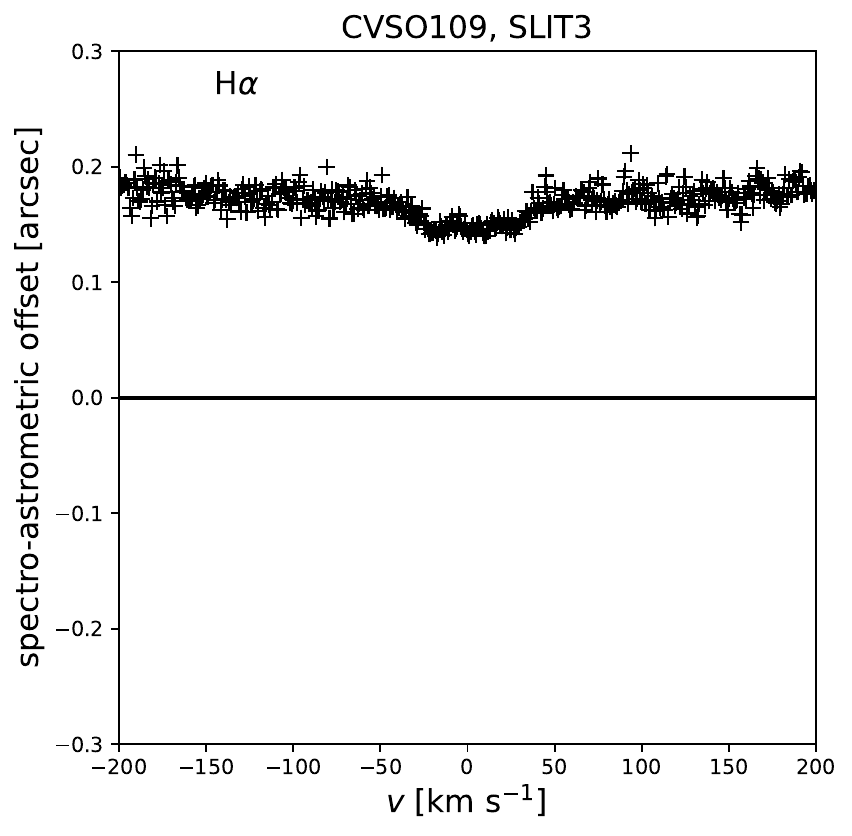}} 
\hfill
\subfloat{\includegraphics[trim=0 0 0 0, clip, width=0.3 \textwidth]{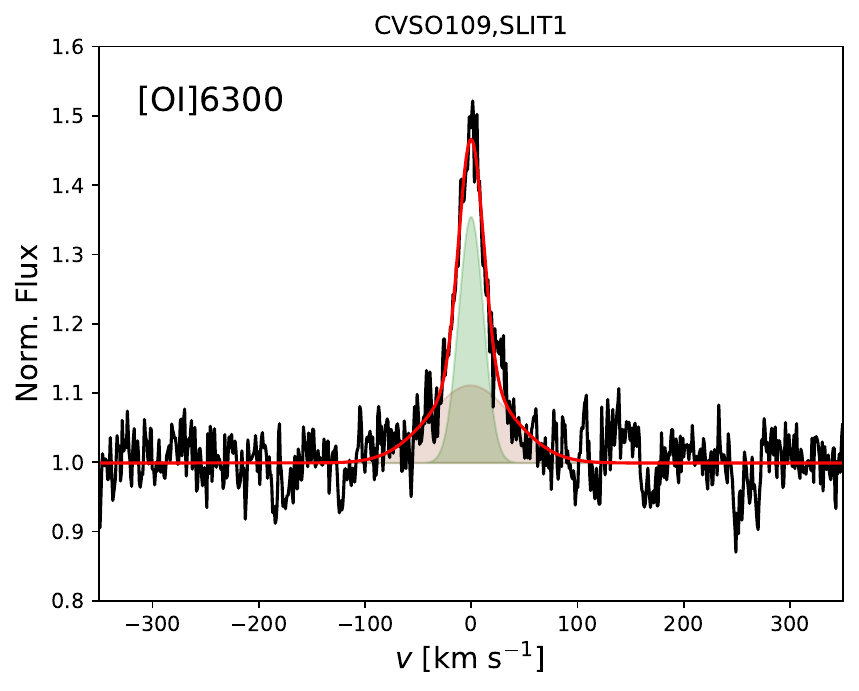}}
\hfill
\subfloat{\includegraphics[trim=0 0 0 0, clip, width=0.3 \textwidth]{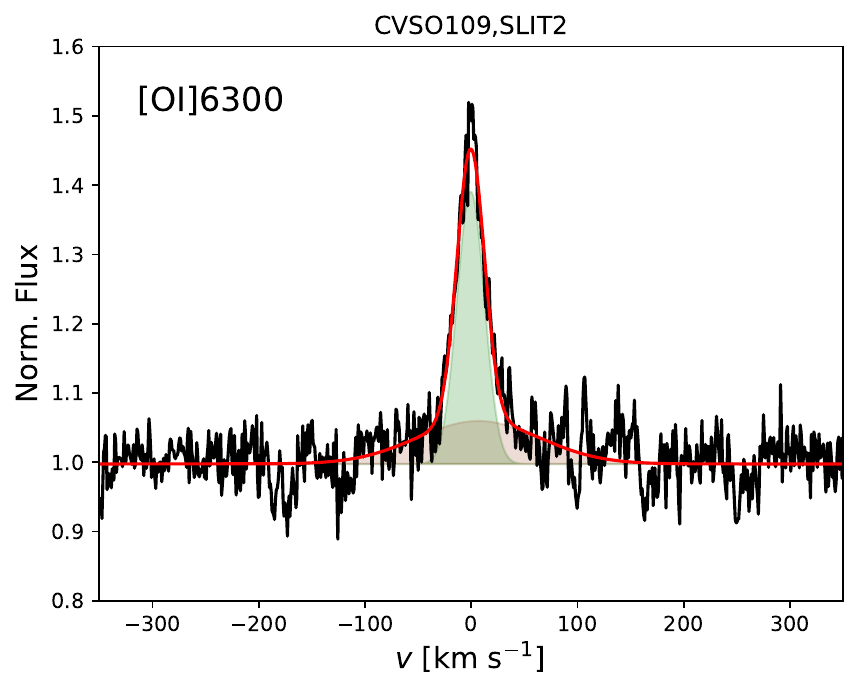}}
\hfill
\subfloat{\includegraphics[trim=0 0 0 0, clip, width=0.3 \textwidth]{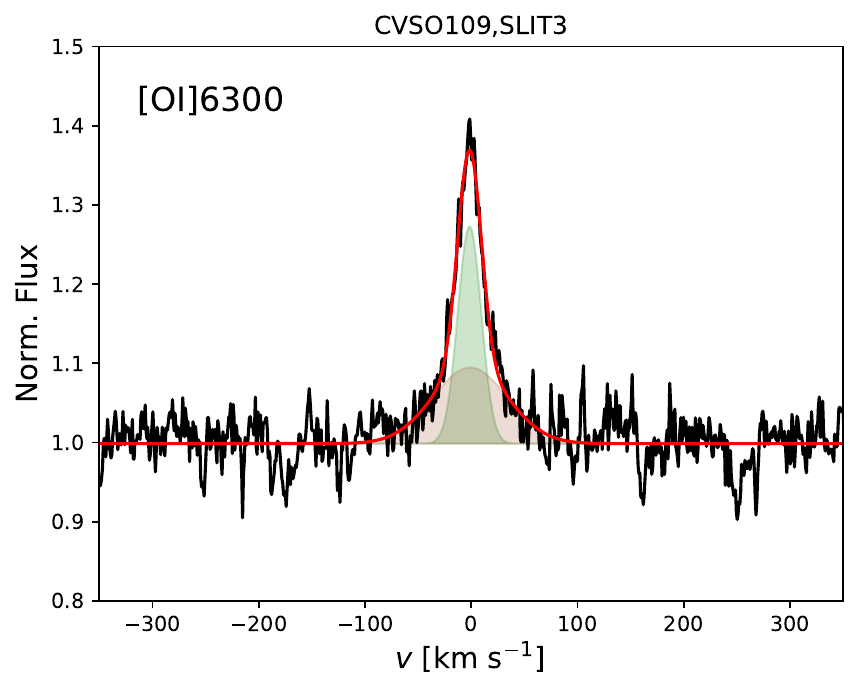}} 
\hfill 
\subfloat{\includegraphics[trim=0 0 0 0, clip, width=0.3 \textwidth]{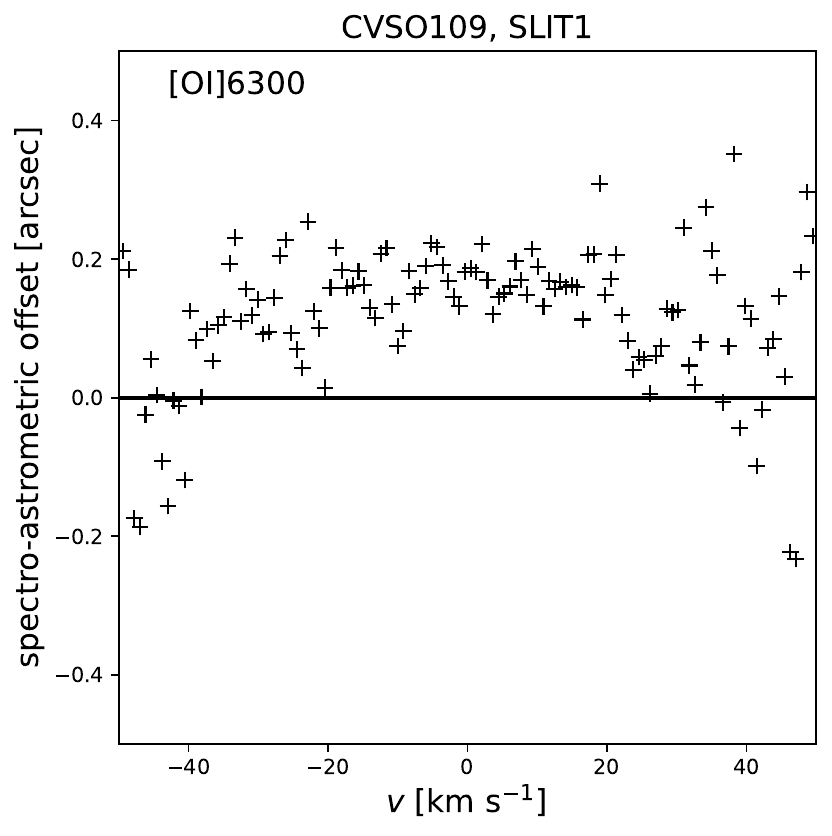}}
\hfill
\subfloat{\includegraphics[trim=0 0 0 0, clip, width=0.3 \textwidth]{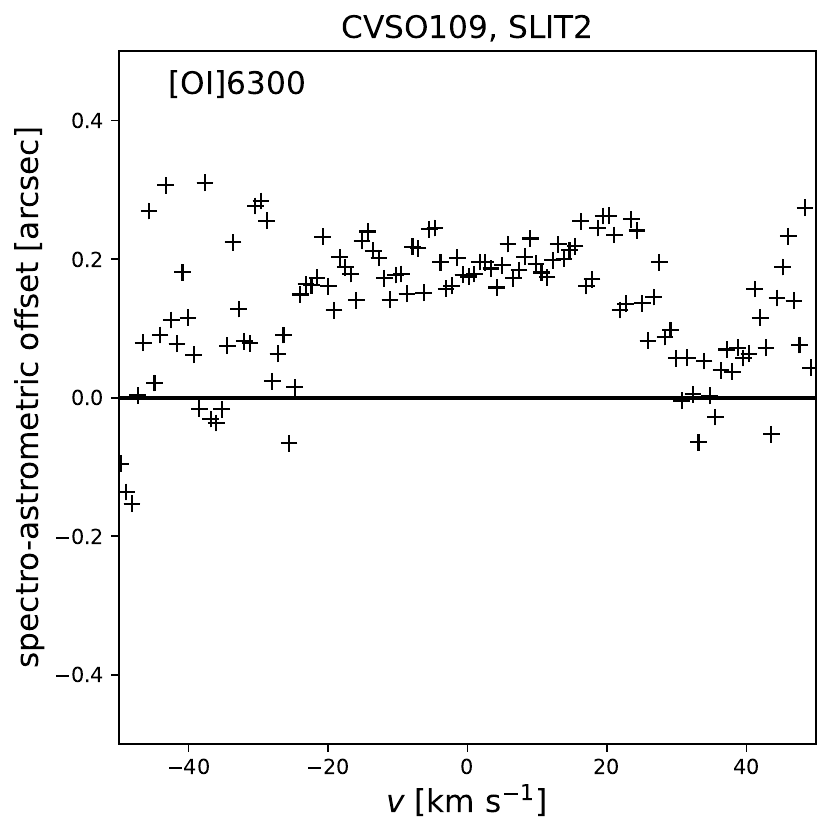}}
\hfill
\subfloat{\includegraphics[trim=0 0 0 0, clip, width=0.3 \textwidth]{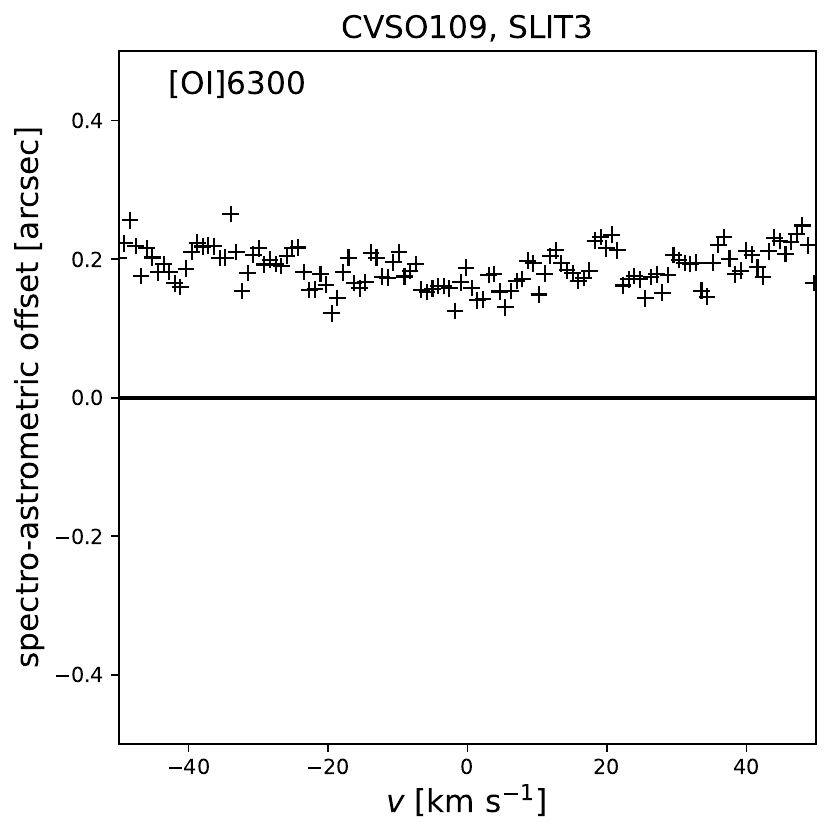}} 
\hfill  
\caption{\small{Line profiles of H$\alpha$ and [OI]$\lambda$6300 for all slit positions of CVSO\,109.}}\label{fig:all_minispectra_CVSO109}
\end{figure*} 

\begin{figure*} 
\centering
\subfloat{\includegraphics[trim=0 0 0 0, clip, width=0.3 \textwidth]{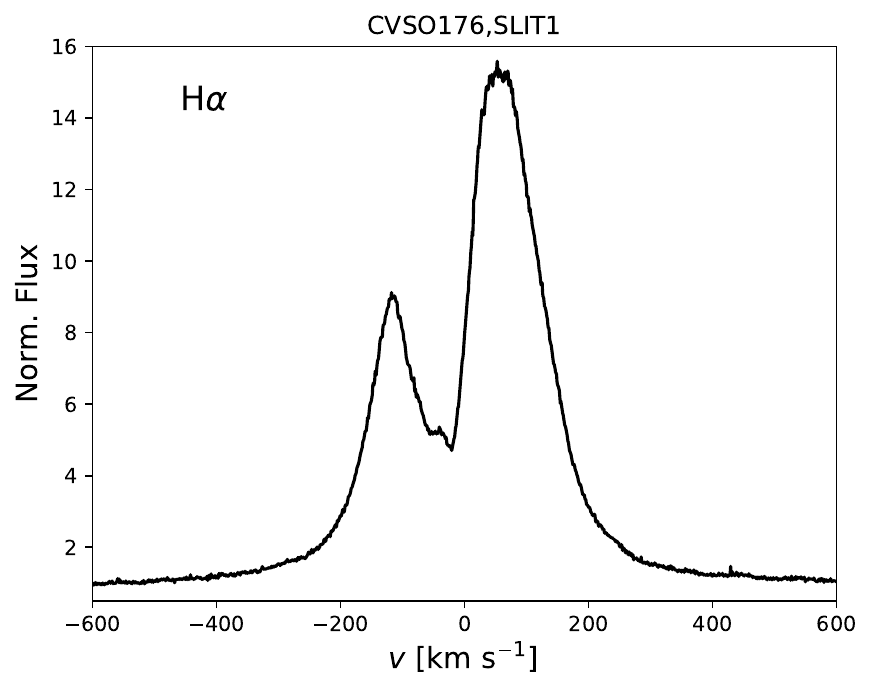}}
\hfill
\subfloat{\includegraphics[trim=0 0 0 0, clip, width=0.3 \textwidth]{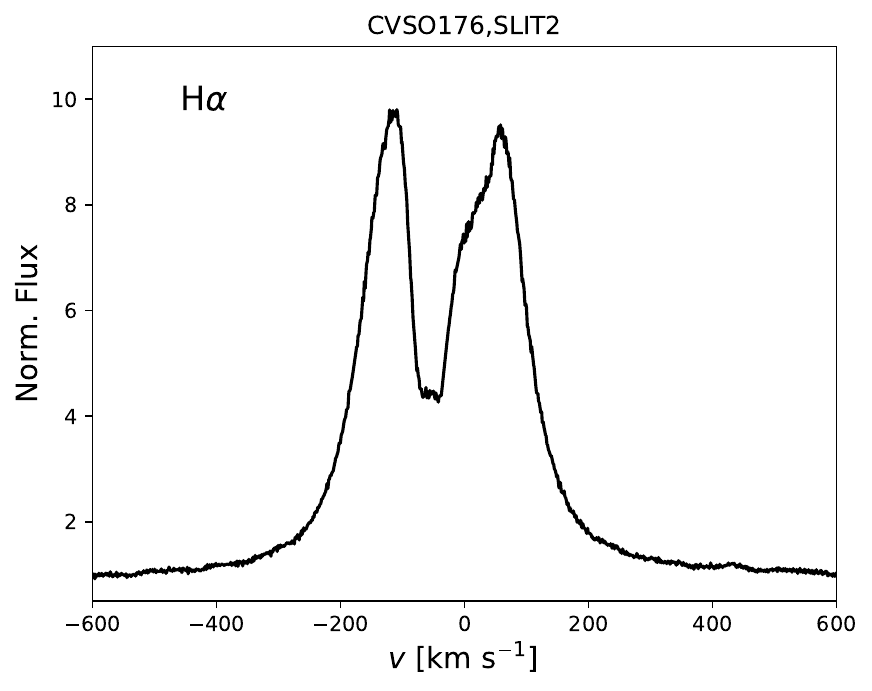}}
\hfill
\subfloat{\includegraphics[trim=0 0 0 0, clip, width=0.3 \textwidth]{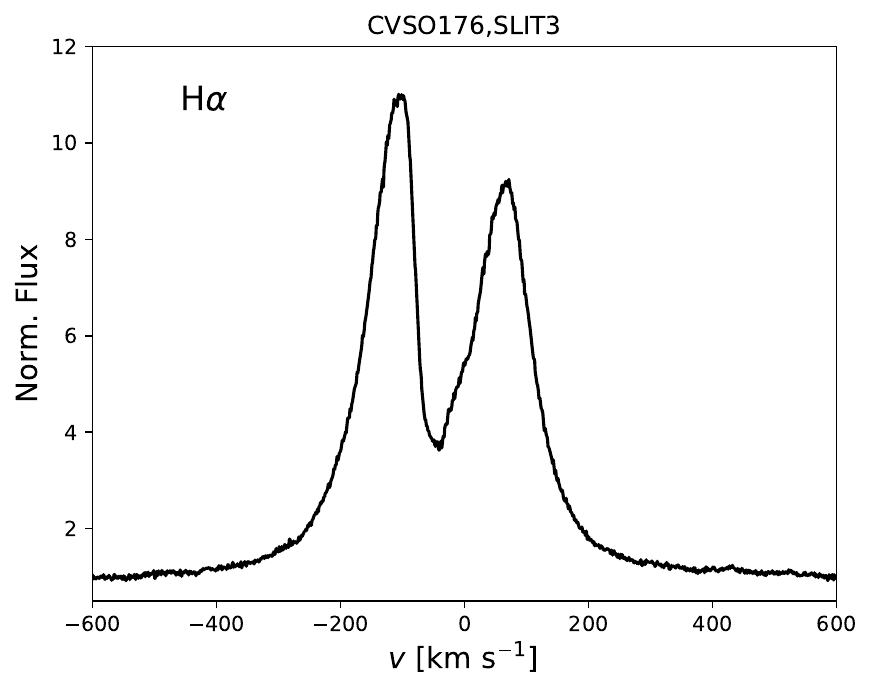}}
\hfill  
\subfloat{\includegraphics[trim=0 0 0 0, clip, width=0.3 \textwidth]{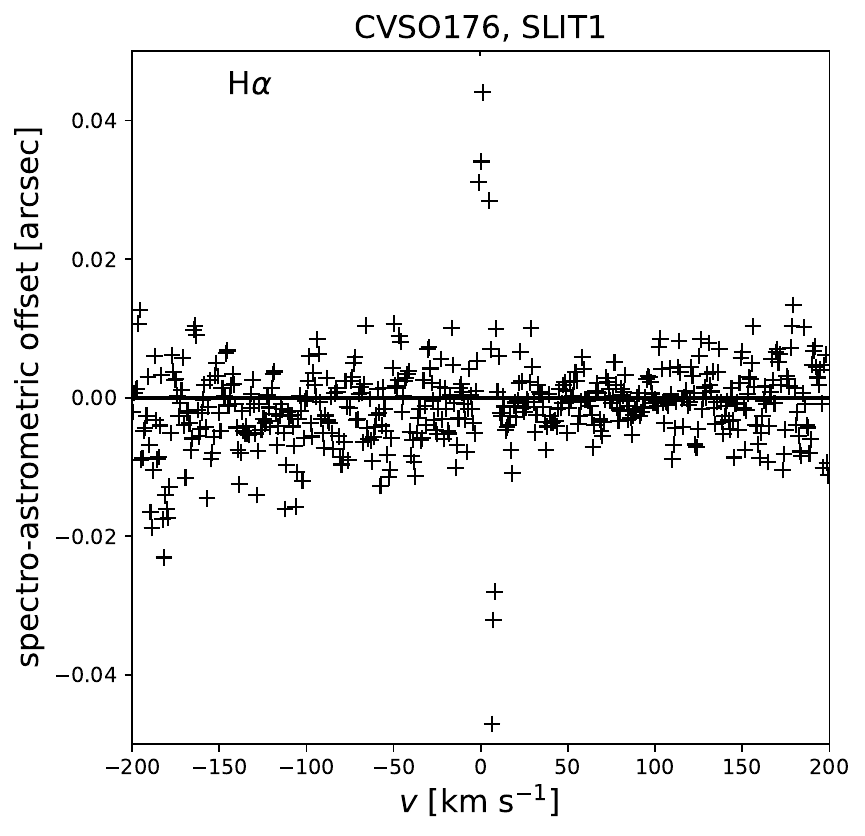}}
\hfill
\subfloat{\includegraphics[trim=0 0 0 0, clip, width=0.3 \textwidth]{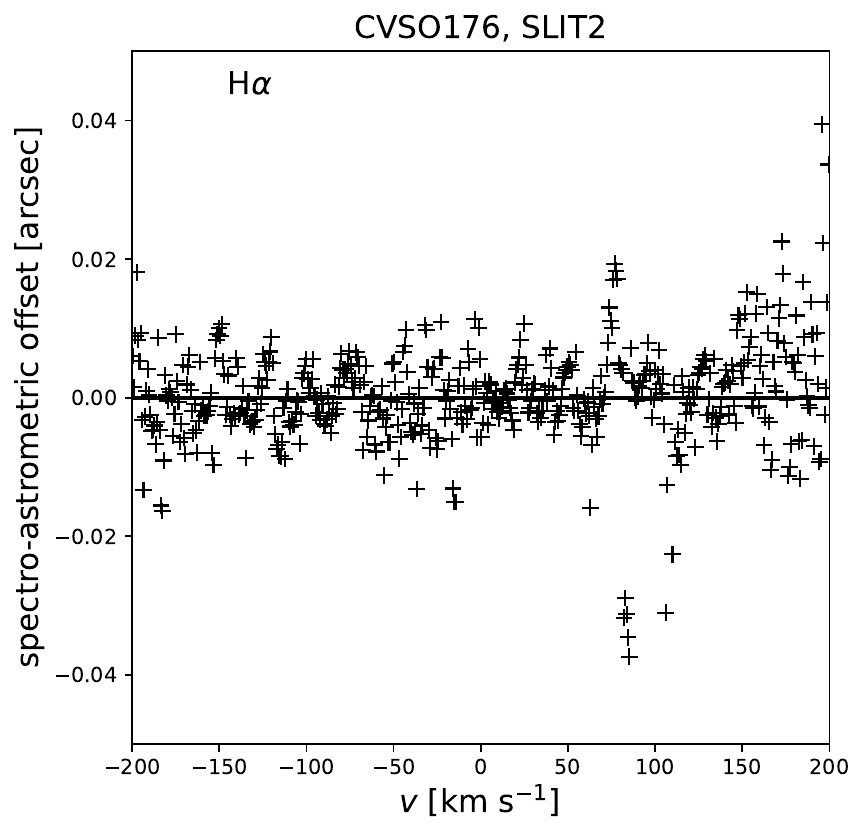}}
\hfill
\subfloat{\includegraphics[trim=0 0 0 0, clip, width=0.3 \textwidth]{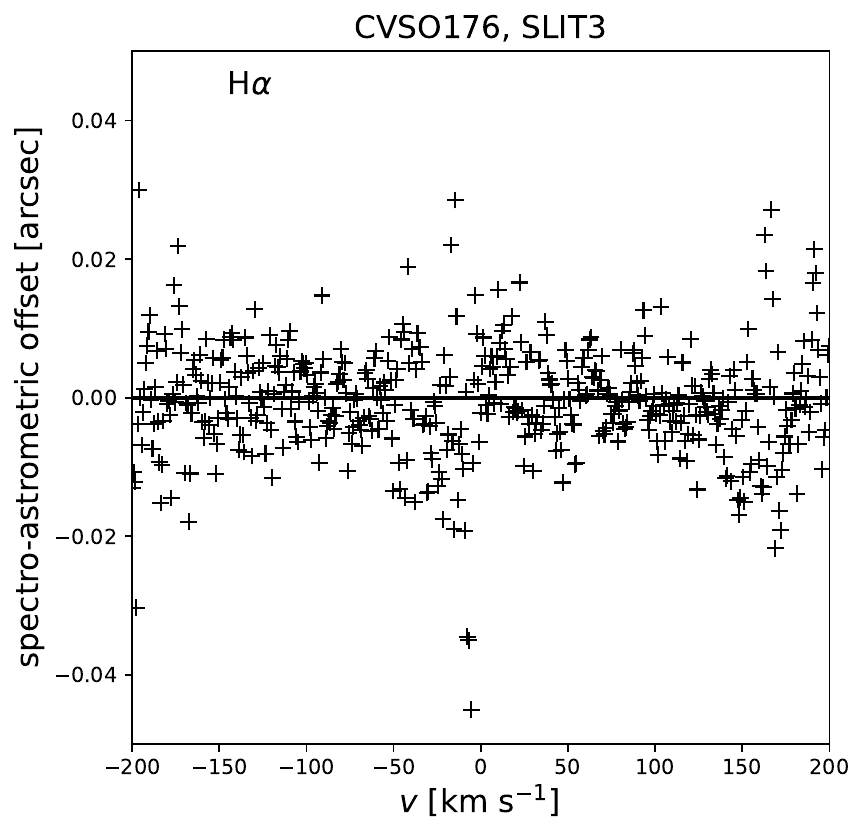}} 
\hfill
\subfloat{\includegraphics[trim=0 0 0 0, clip, width=0.3 \textwidth]{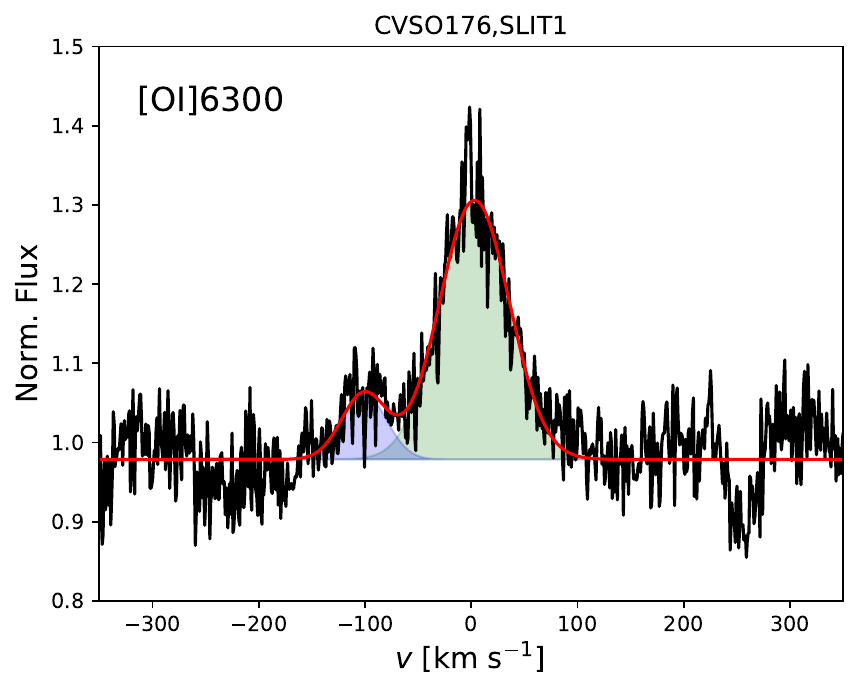}}
\hfill
\subfloat{\includegraphics[trim=0 0 0 0, clip, width=0.3 \textwidth]{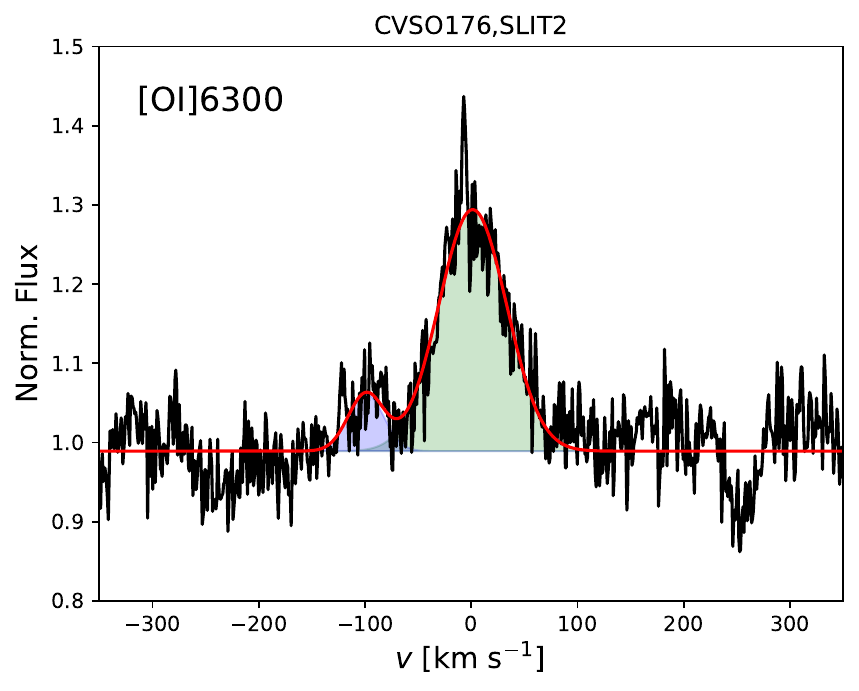}}
\hfill
\subfloat{\includegraphics[trim=0 0 0 0, clip, width=0.3 \textwidth]{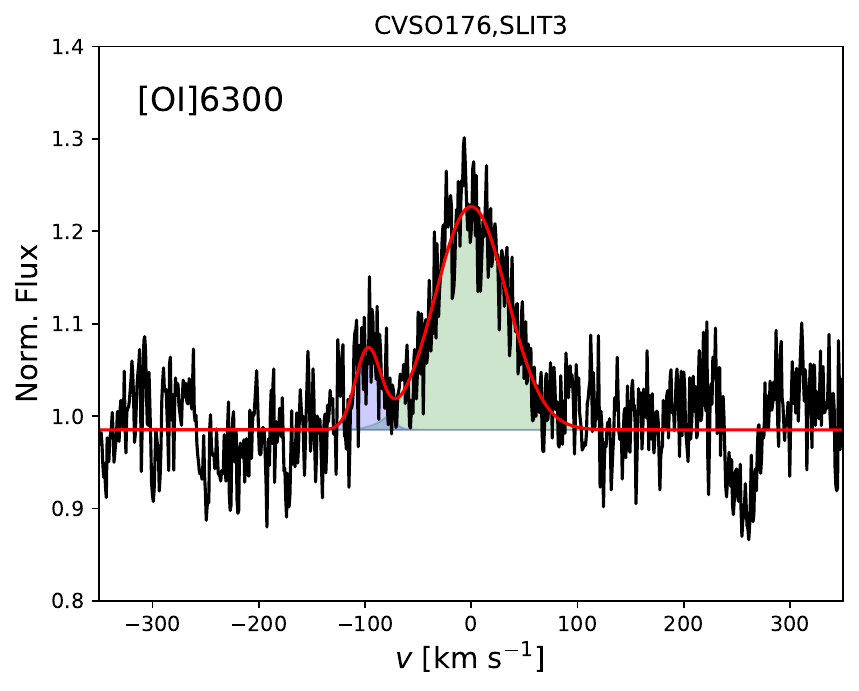}} 
\hfill   
\subfloat{\includegraphics[trim=0 0 0 0, clip, width=0.3 \textwidth]{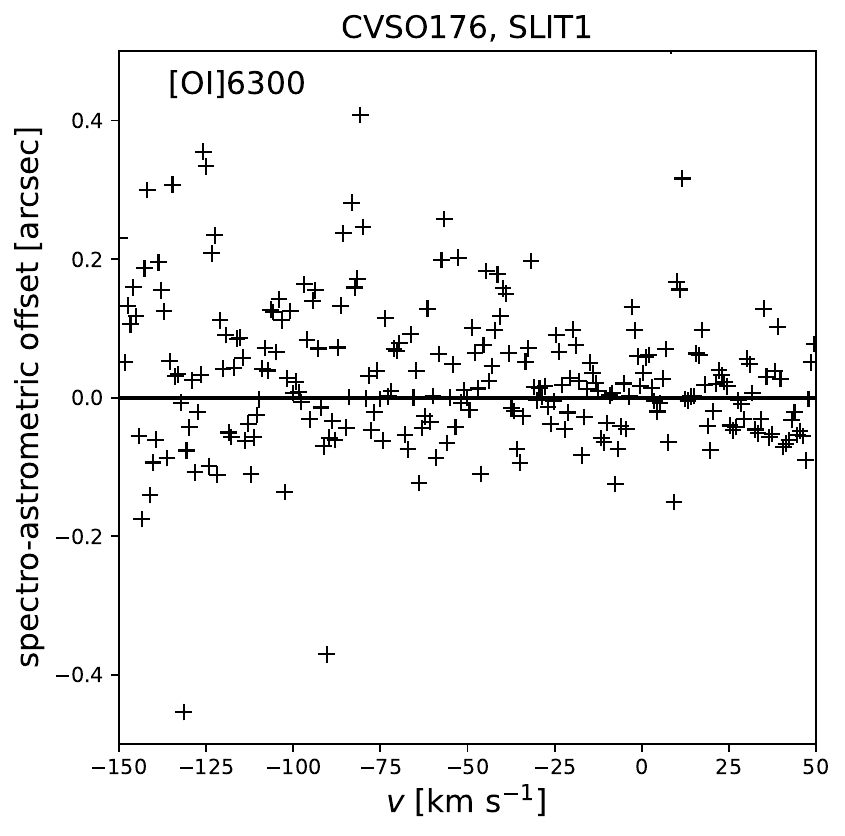}}
\hfill
\subfloat{\includegraphics[trim=0 0 0 0, clip, width=0.3 \textwidth]{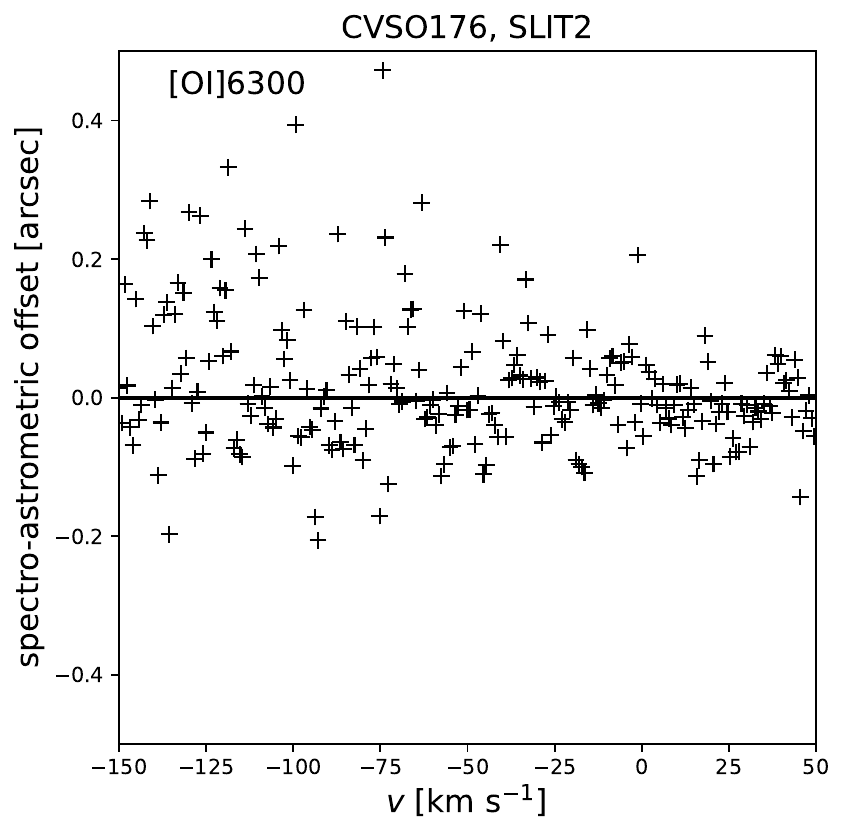}}
\hfill
\subfloat{\includegraphics[trim=0 0 0 0, clip, width=0.3 \textwidth]{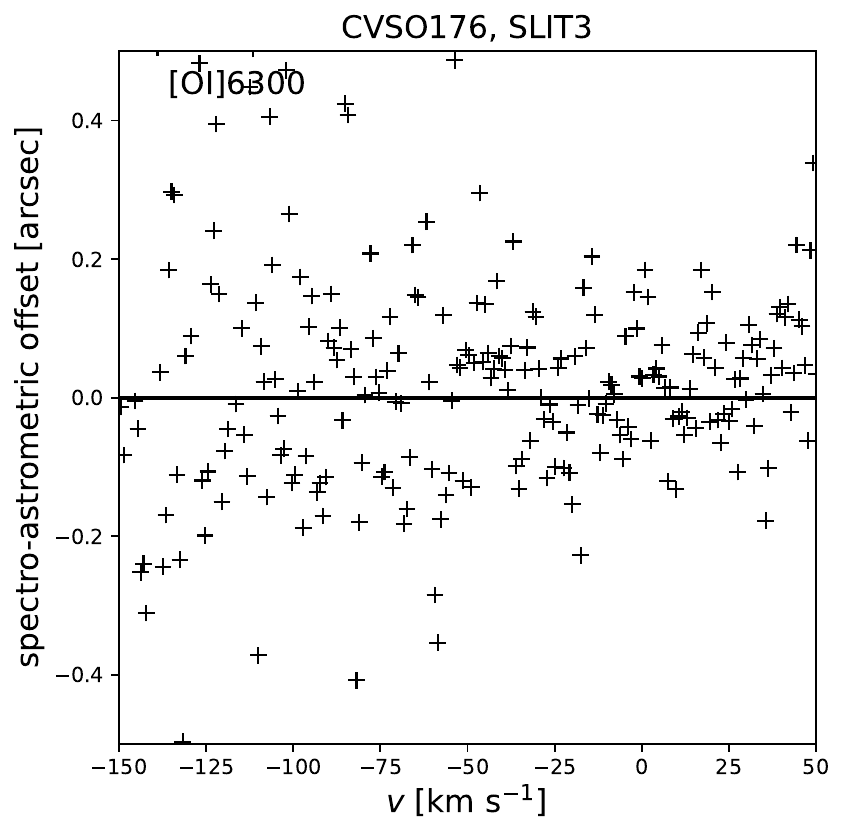}} 
\hfill  
\caption{\small{Line profiles of H$\alpha$ and [OI]$\lambda$6300 for all slit positions of CVSO\,176.}}\label{fig:all_minispectra_CVSO176}
\end{figure*} 

\begin{figure*} 
\centering
\subfloat{\includegraphics[trim=0 0 0 0, clip, width=0.3 \textwidth]{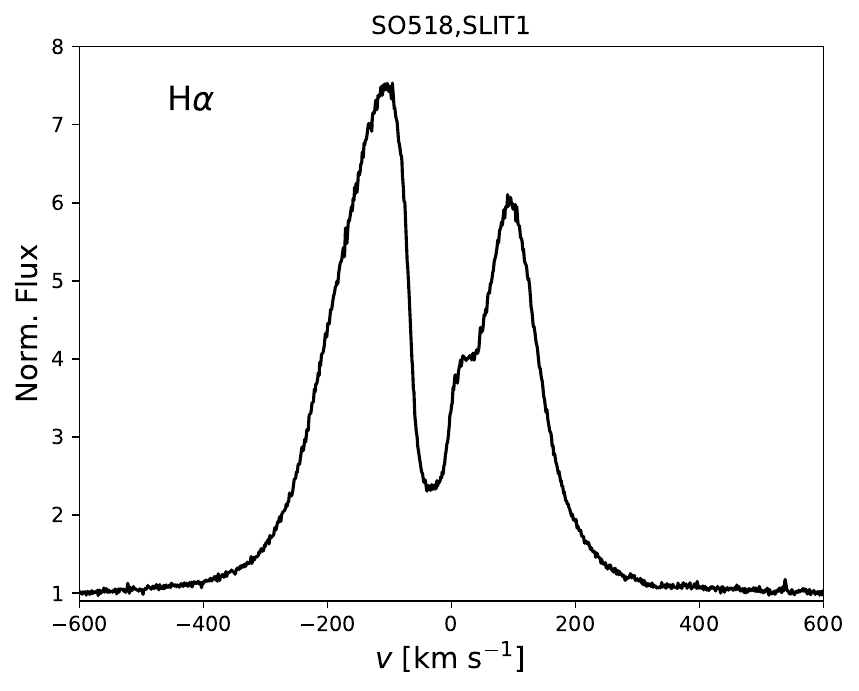}}
\hfill
\subfloat{\includegraphics[trim=0 0 0 0, clip, width=0.3 \textwidth]{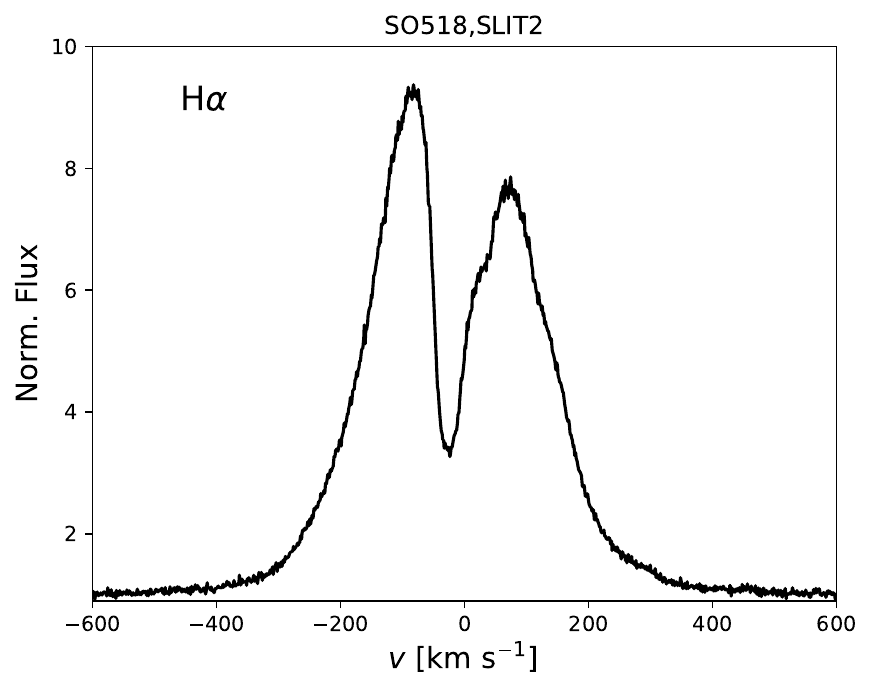}}
\hfill
\subfloat{\includegraphics[trim=0 0 0 0, clip, width=0.3 \textwidth]{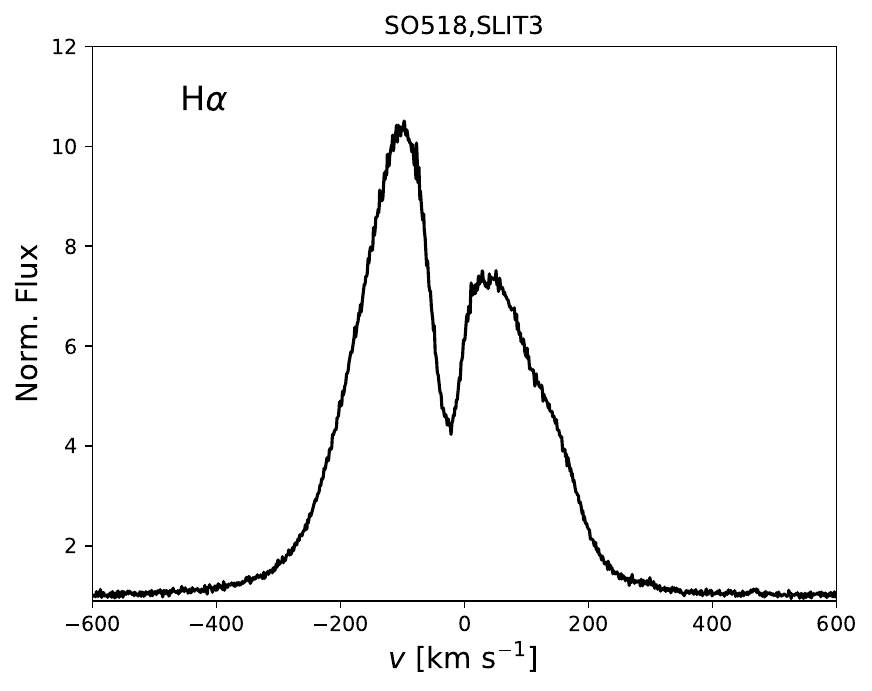}}
\hfill  
\subfloat{\includegraphics[trim=0 0 0 0, clip, width=0.3 \textwidth]{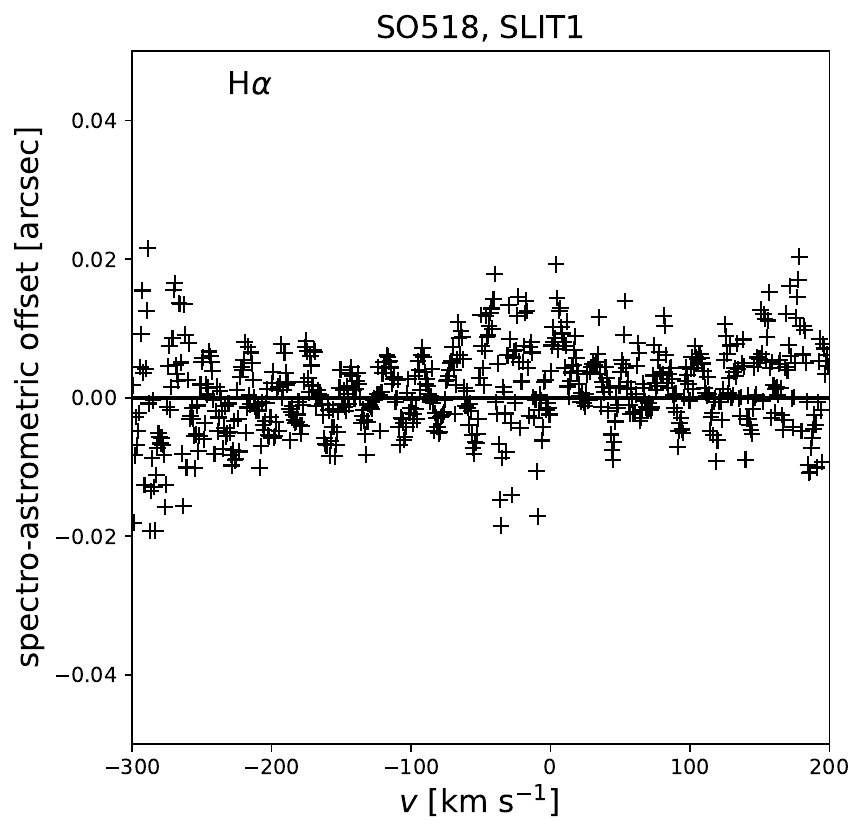}}
\hfill
\subfloat{\includegraphics[trim=0 0 0 0, clip, width=0.3 \textwidth]{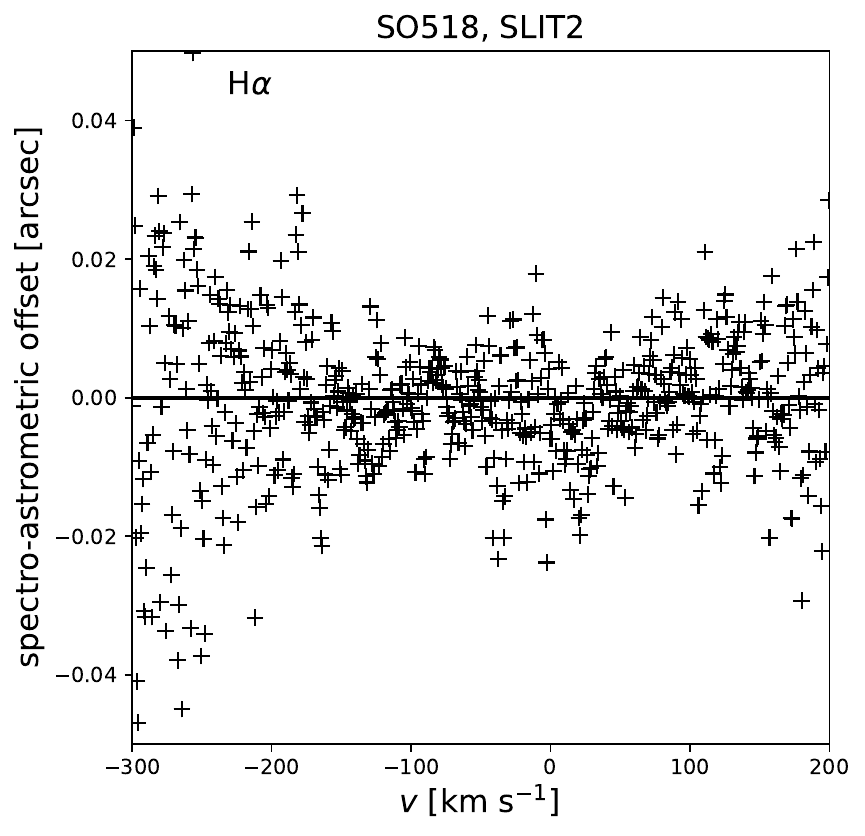}}
\hfill
\subfloat{\includegraphics[trim=0 0 0 0, clip, width=0.3 \textwidth]{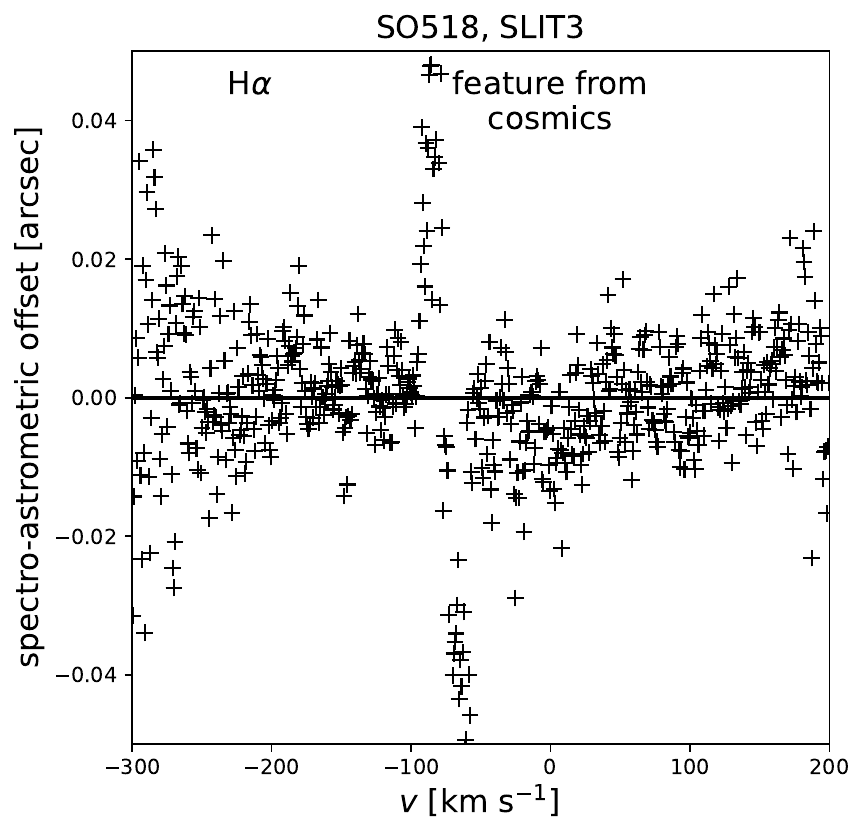}} 
\hfill
\subfloat{\includegraphics[trim=0 0 0 0, clip, width=0.3 \textwidth]{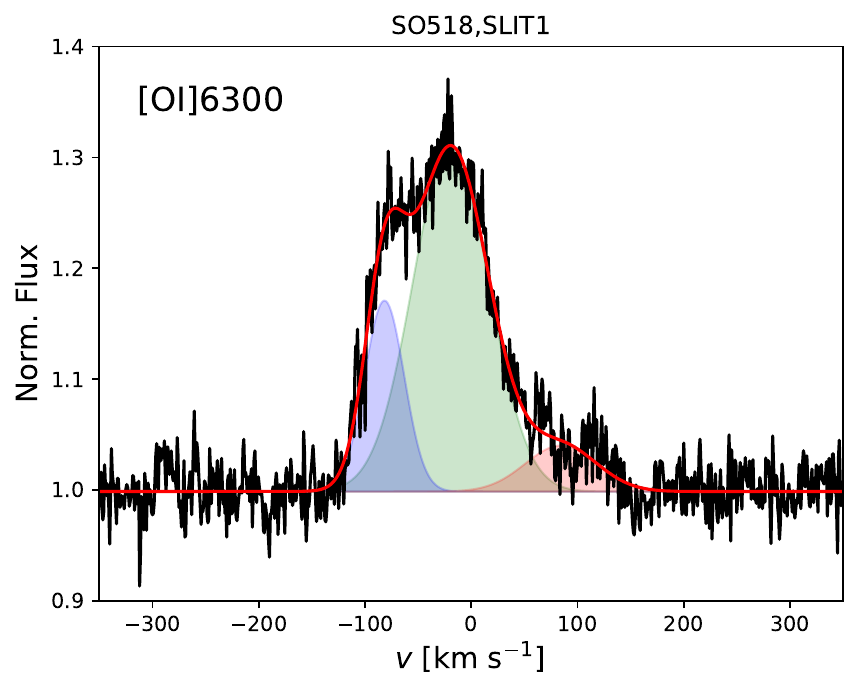}}
\hfill
\subfloat{\includegraphics[trim=0 0 0 0, clip, width=0.3 \textwidth]{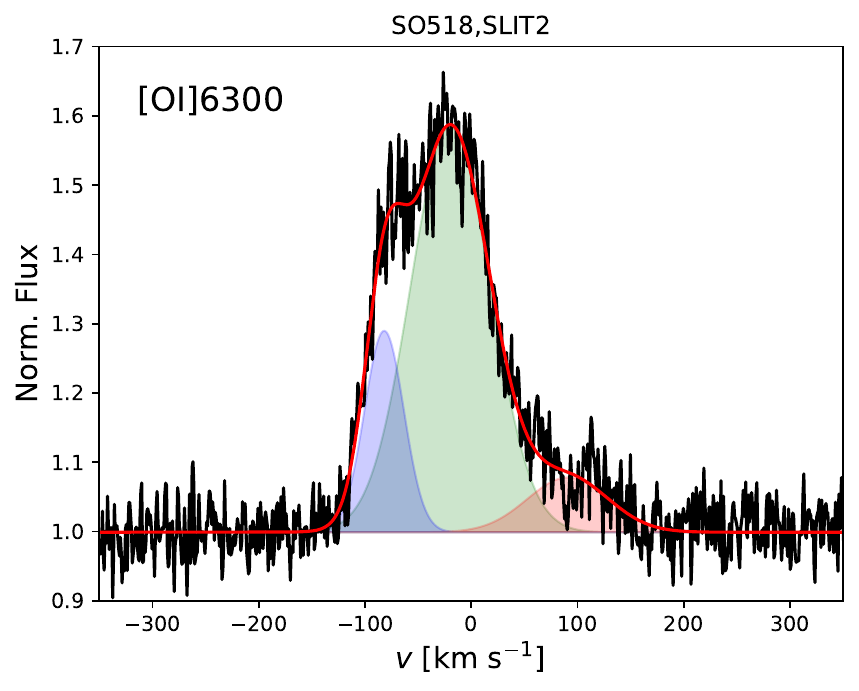}}
\hfill
\subfloat{\includegraphics[trim=0 0 0 0, clip, width=0.3 \textwidth]{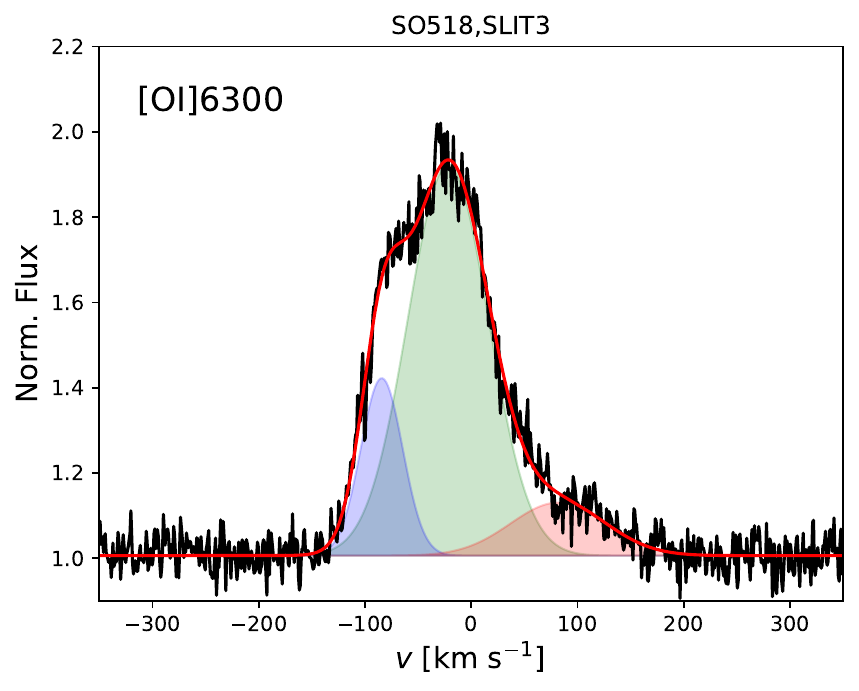}} 
\hfill  
\subfloat{\includegraphics[trim=0 0 0 0, clip, width=0.3 \textwidth]{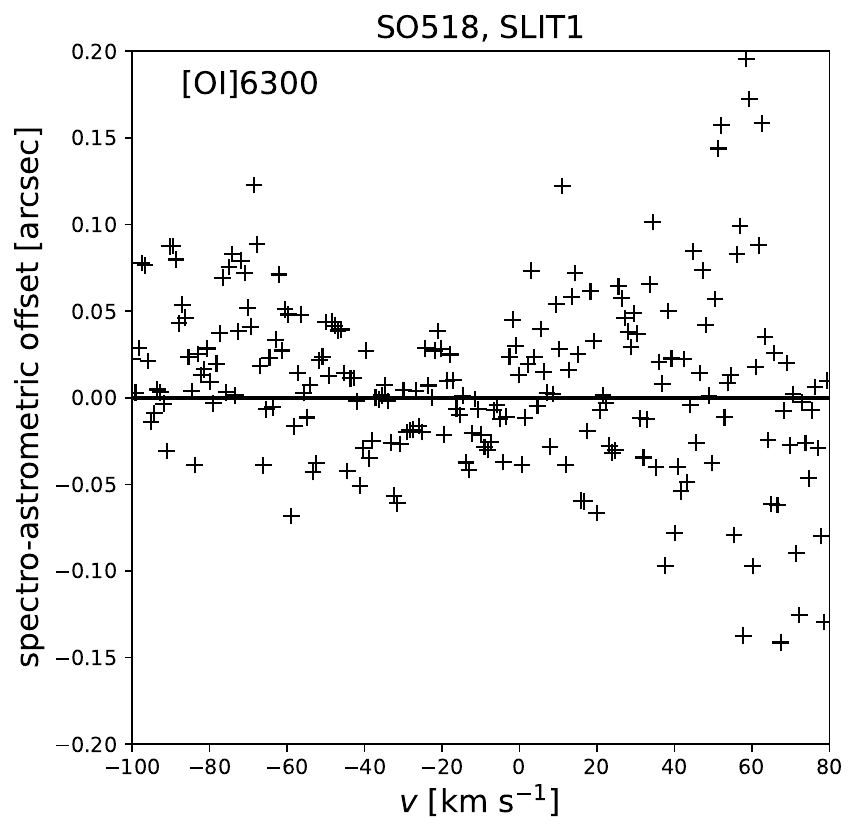}}
\hfill
\subfloat{\includegraphics[trim=0 0 0 0, clip, width=0.3 \textwidth]{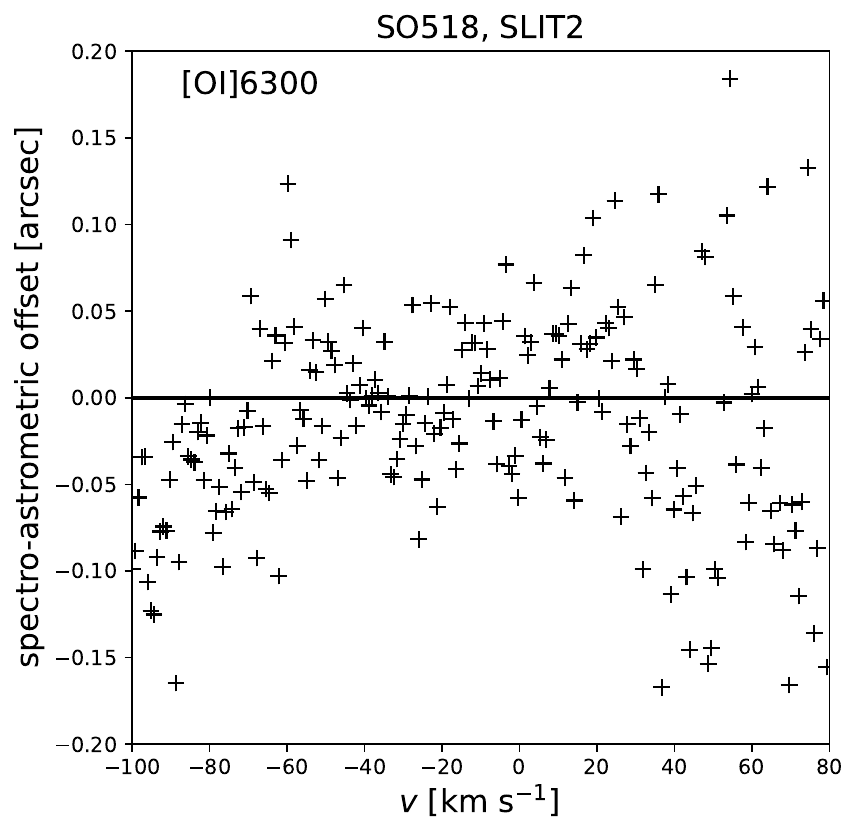}}
\hfill
\subfloat{\includegraphics[trim=0 0 0 0, clip, width=0.3 \textwidth]{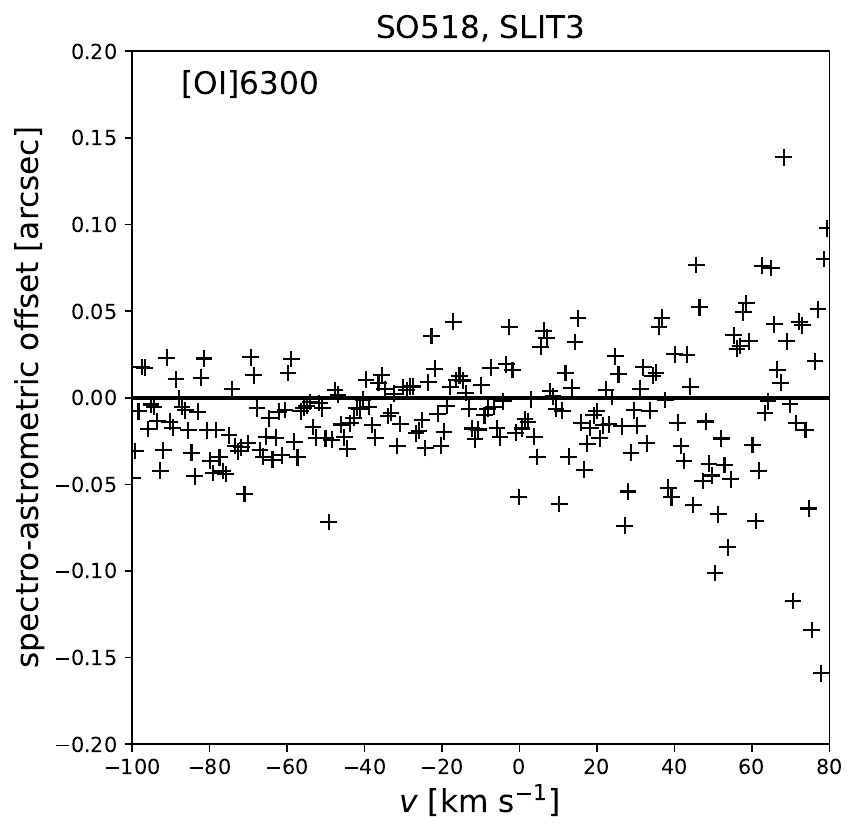}} 
\hfill 
\caption{\small{Line profiles of H$\alpha$ and [OI]$\lambda$6300 for all slit positions of SO518.}}\label{fig:all_minispectra_SO518}
\end{figure*} 

\begin{figure*} 
\centering
\subfloat{\includegraphics[trim=0 0 0 0, clip, width=0.3 \textwidth]{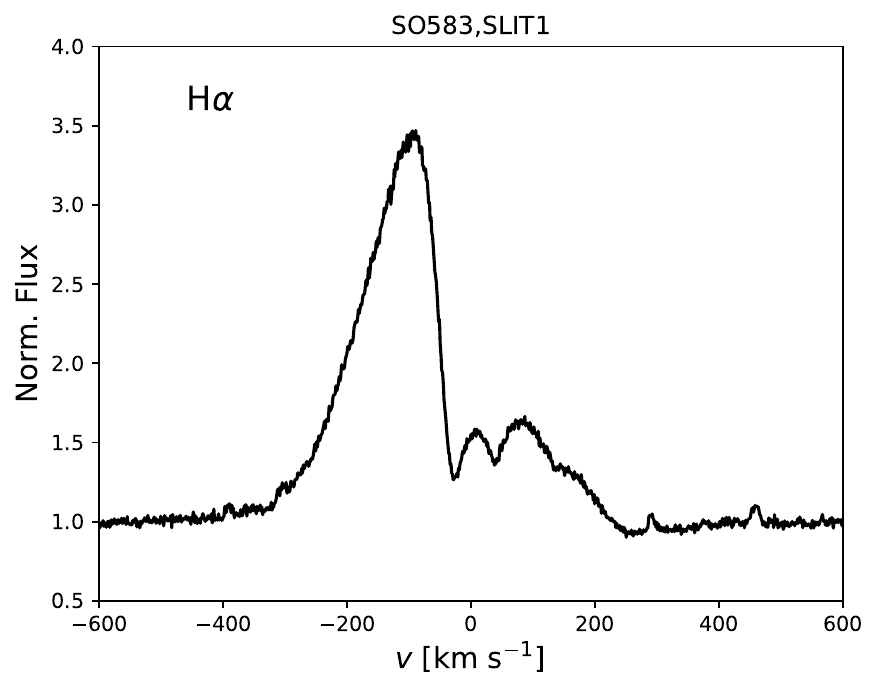}}
\hfill
\subfloat{\includegraphics[trim=0 0 0 0, clip, width=0.3 \textwidth]{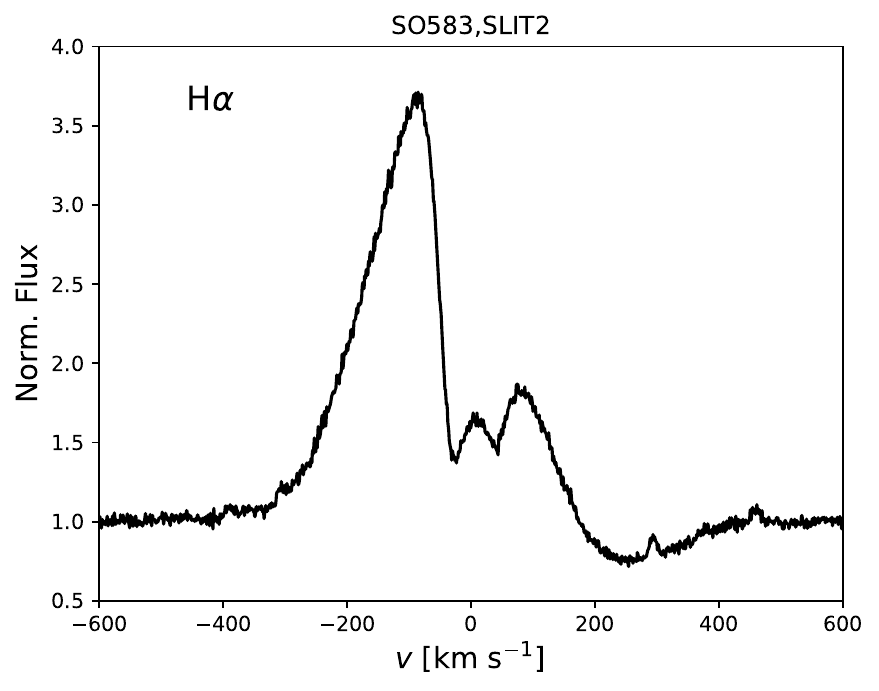}}
\hfill
\subfloat{\includegraphics[trim=0 0 0 0, clip, width=0.3 \textwidth]{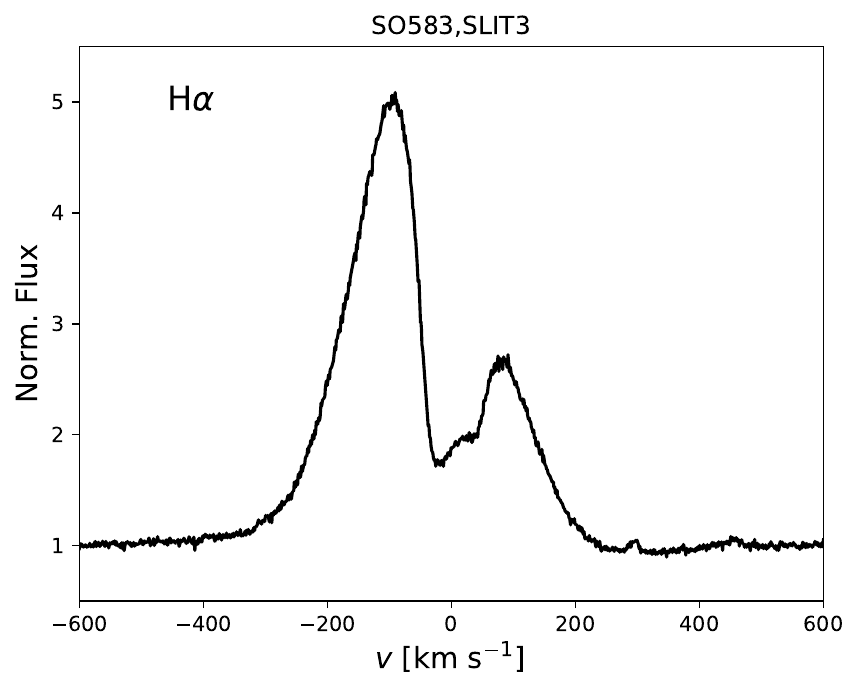}}
\hfill  
\subfloat{\includegraphics[trim=0 0 0 0, clip, width=0.3 \textwidth]{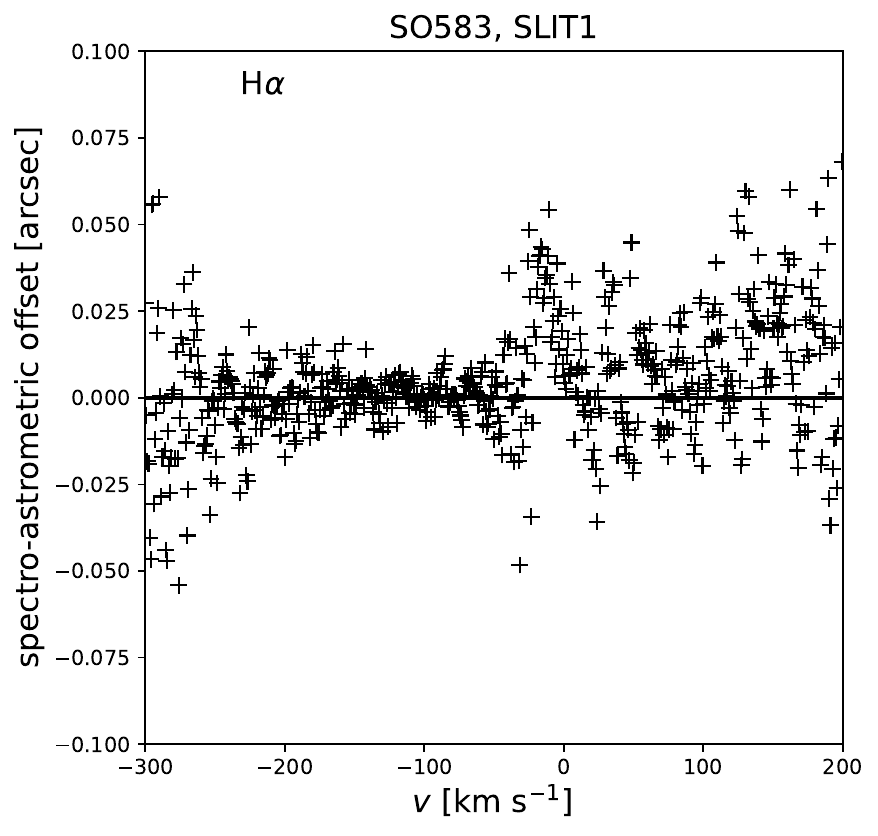}}
\hfill
\subfloat{\includegraphics[trim=0 0 0 0, clip, width=0.3 \textwidth]{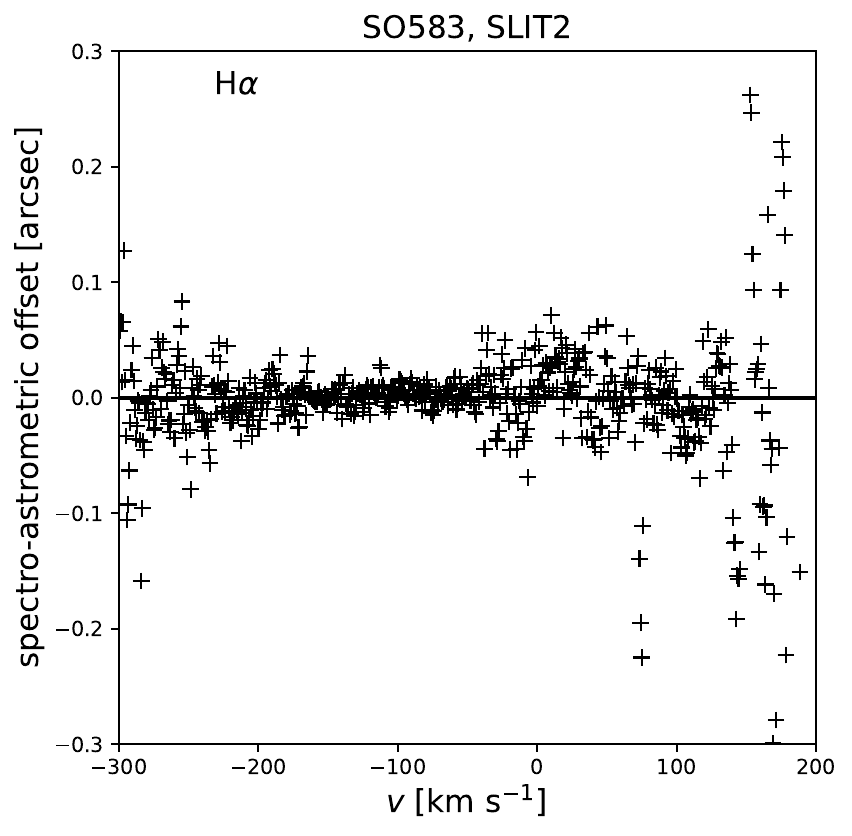}}
\hfill
\subfloat{\includegraphics[trim=0 0 0 0, clip, width=0.3 \textwidth]{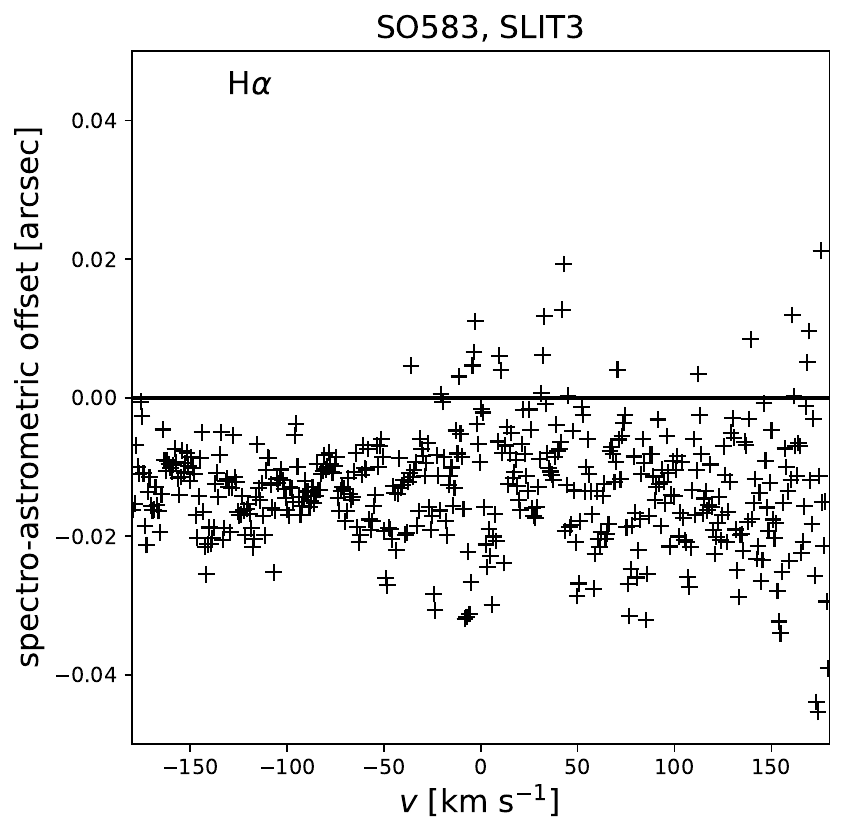}} 
\hfill
\subfloat{\includegraphics[trim=0 0 0 0, clip, width=0.3 \textwidth]{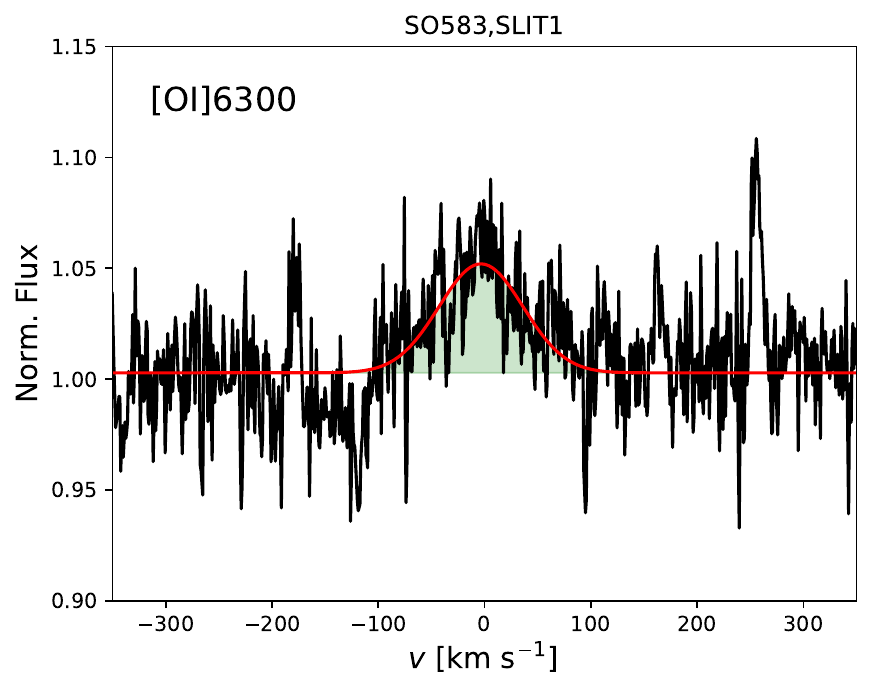}}
\hfill
\subfloat{\includegraphics[trim=0 0 0 0, clip, width=0.3 \textwidth]{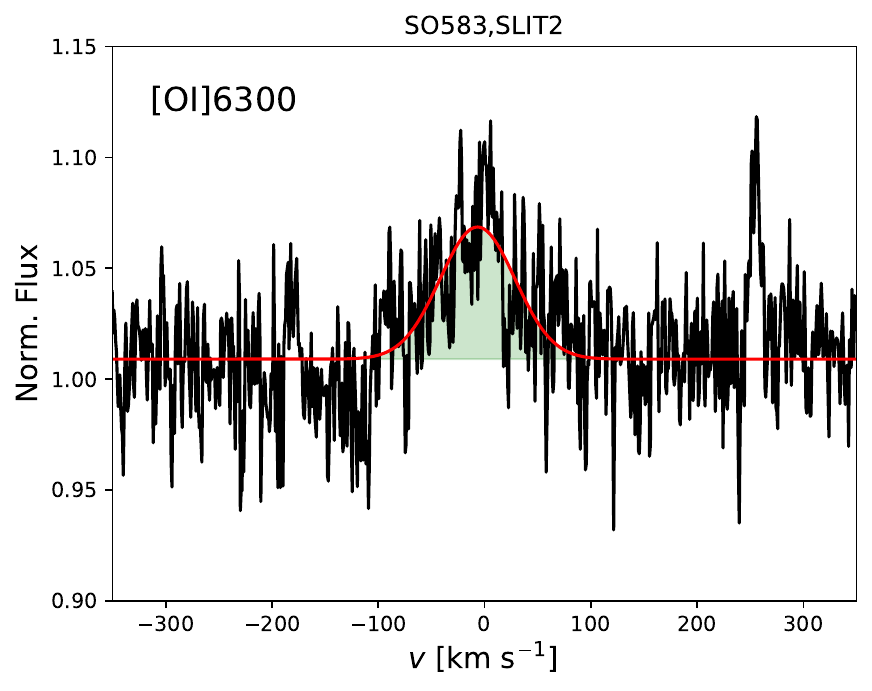}}
\hfill
\subfloat{\includegraphics[trim=0 0 0 0, clip, width=0.3 \textwidth]{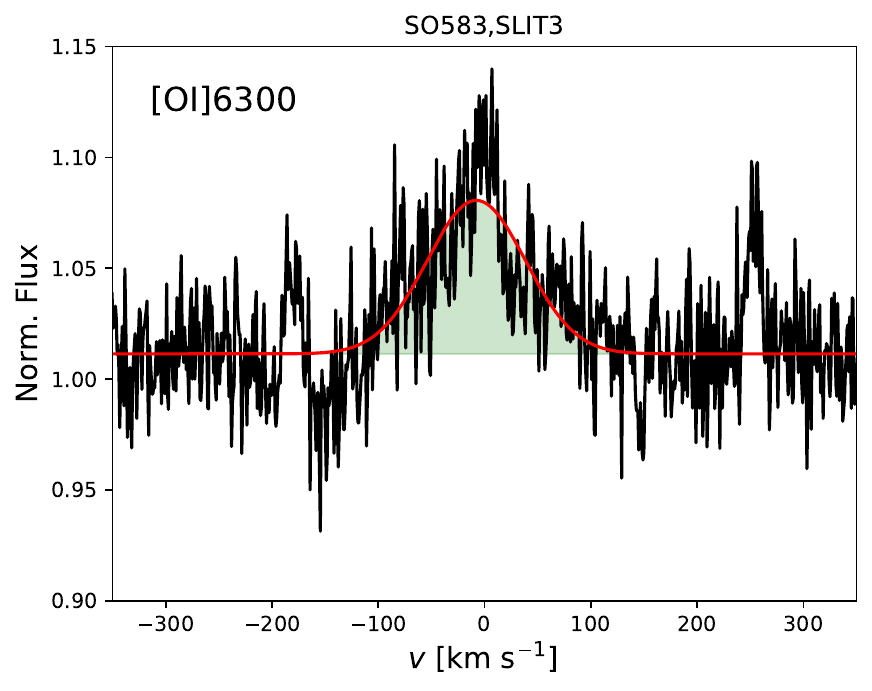}} 
\hfill   
\subfloat{\includegraphics[trim=0 0 0 0, clip, width=0.3 \textwidth]{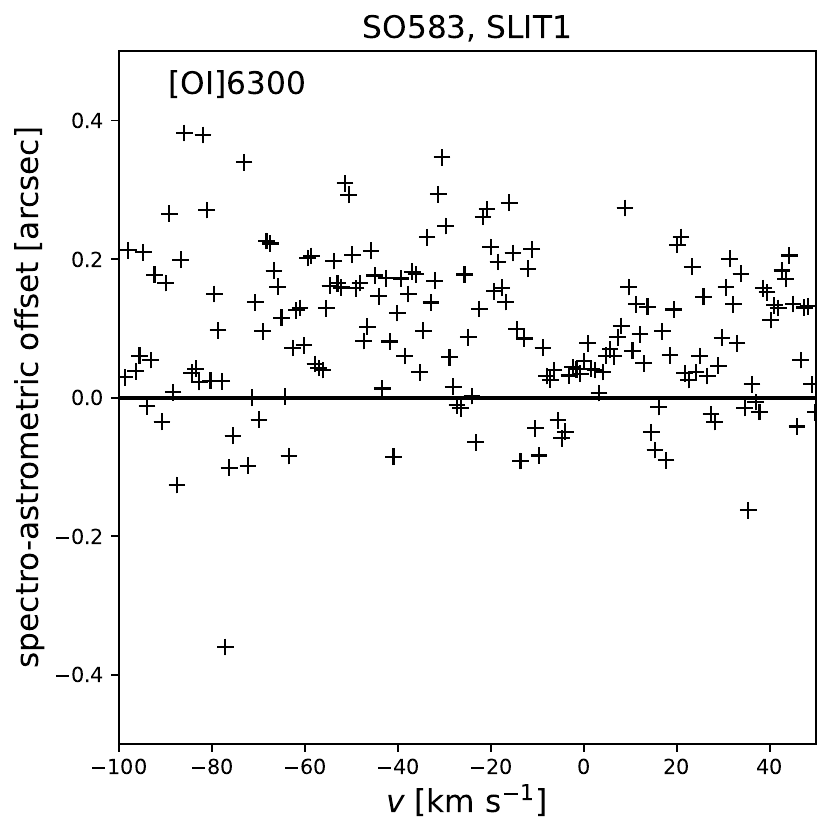}}
\hfill
\subfloat{\includegraphics[trim=0 0 0 0, clip, width=0.3 \textwidth]{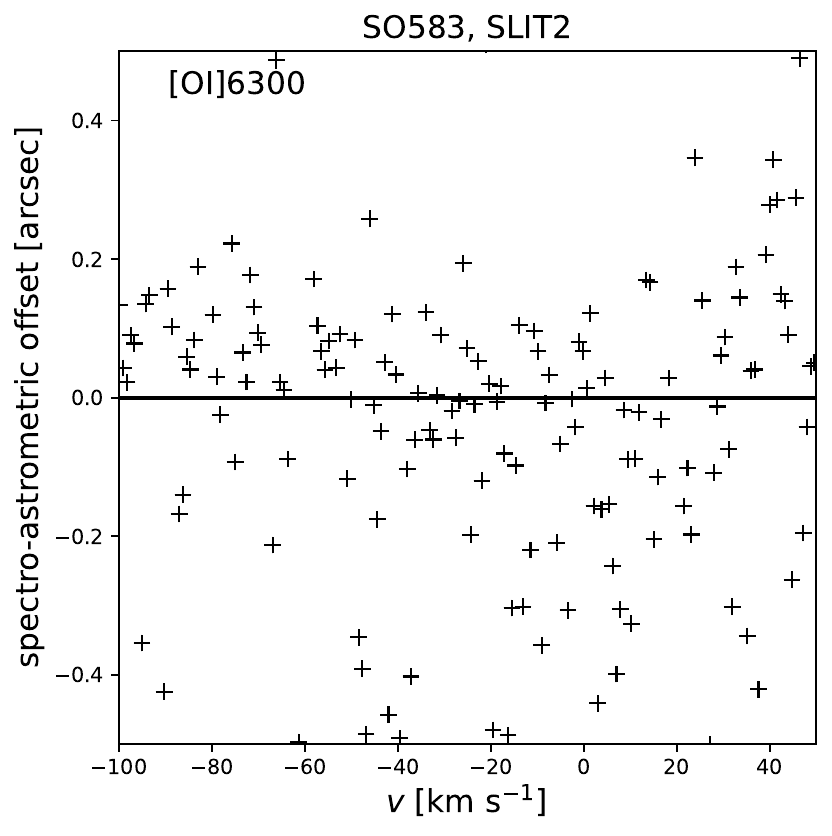}}
\hfill
\subfloat{\includegraphics[trim=0 0 0 0, clip, width=0.3 \textwidth]{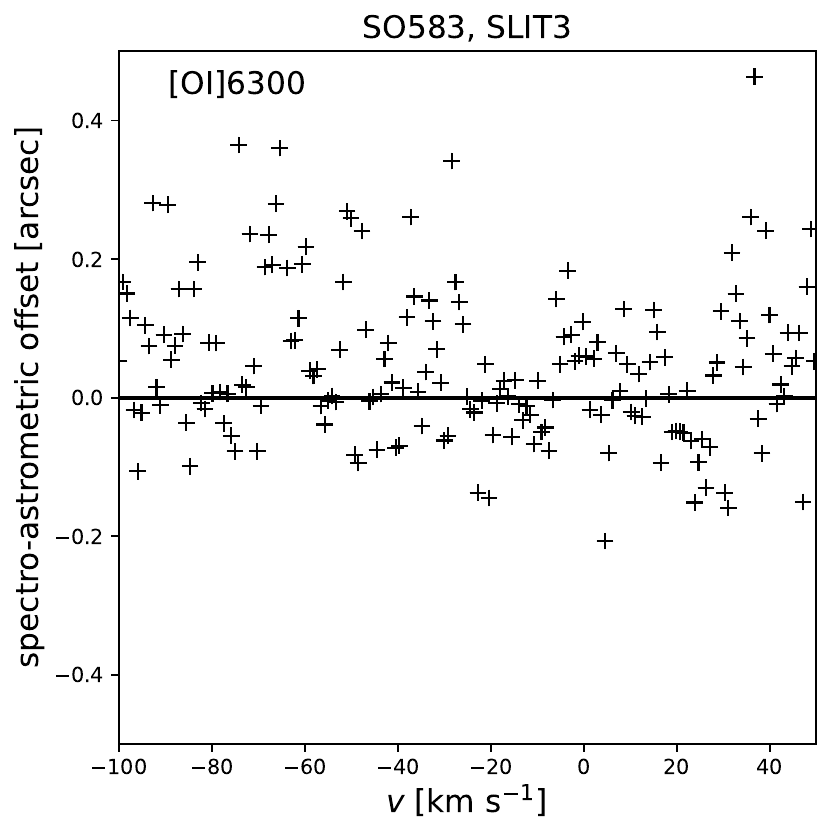}} 
\hfill 
\caption{\small{Line profiles of H$\alpha$ and [OI]$\lambda$6300 for all slit positions of SO583.}}\label{fig:all_minispectra_SO583}
\end{figure*} 

\begin{figure*} 
\centering
\subfloat{\includegraphics[trim=0 0 0 0, clip, width=0.3 \textwidth]{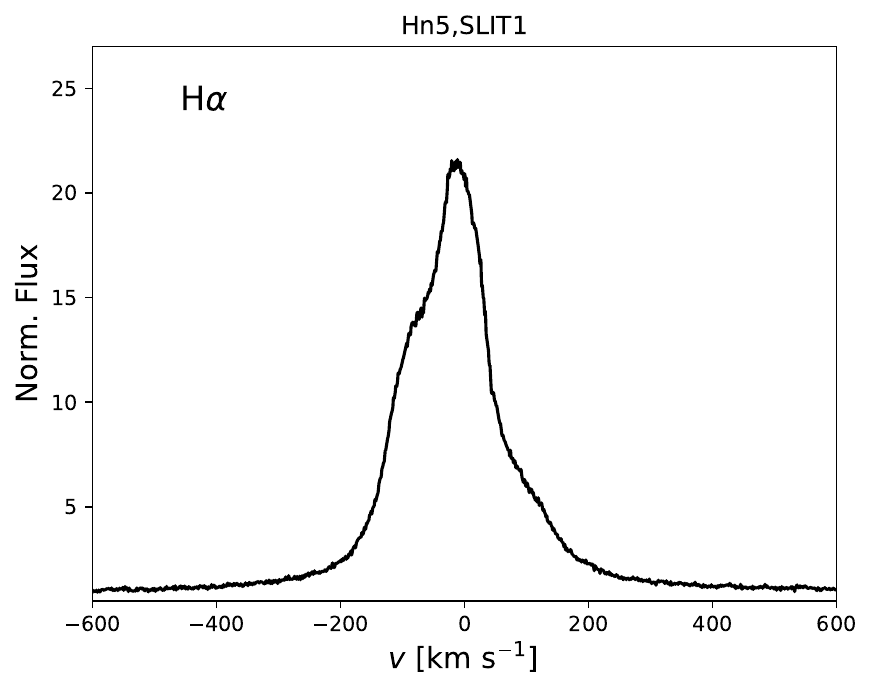}}
\hfill
\subfloat{\includegraphics[trim=0 0 0 0, clip, width=0.3 \textwidth]{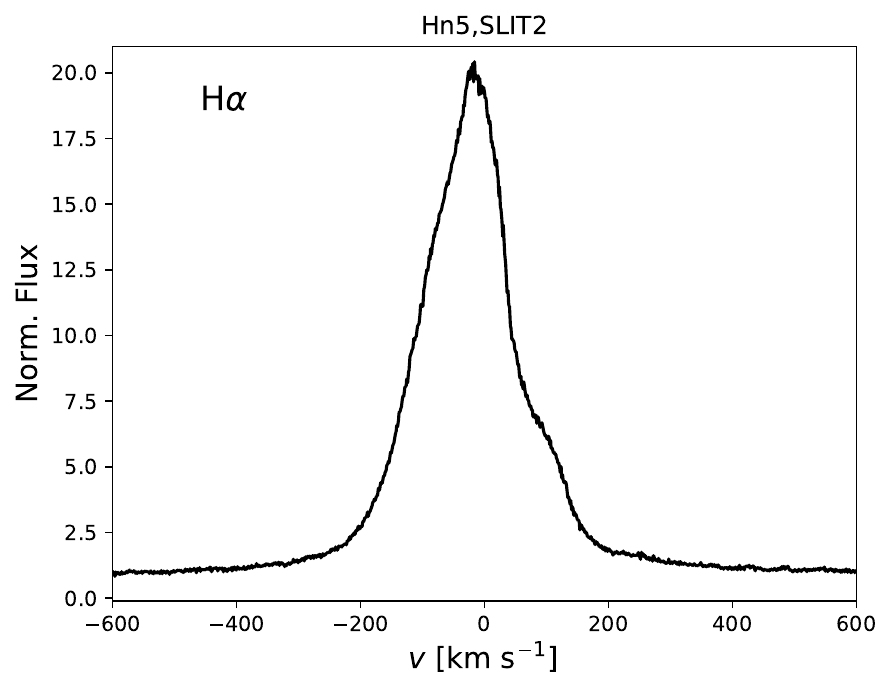}}
\hfill
\subfloat{\includegraphics[trim=0 0 0 0, clip, width=0.3 \textwidth]{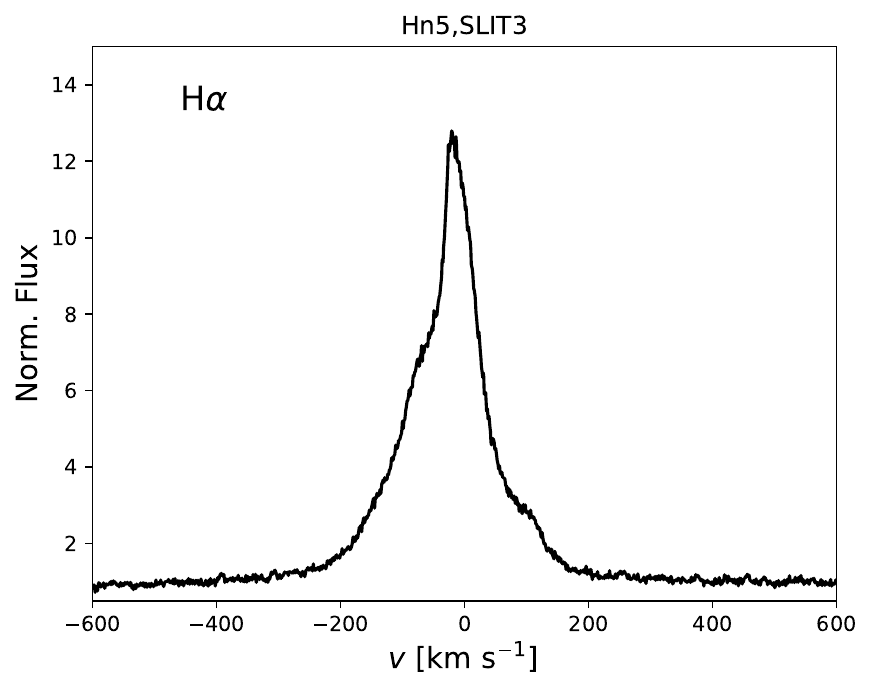}}
\hfill  
\subfloat{\includegraphics[trim=0 0 0 0, clip, width=0.3 \textwidth]{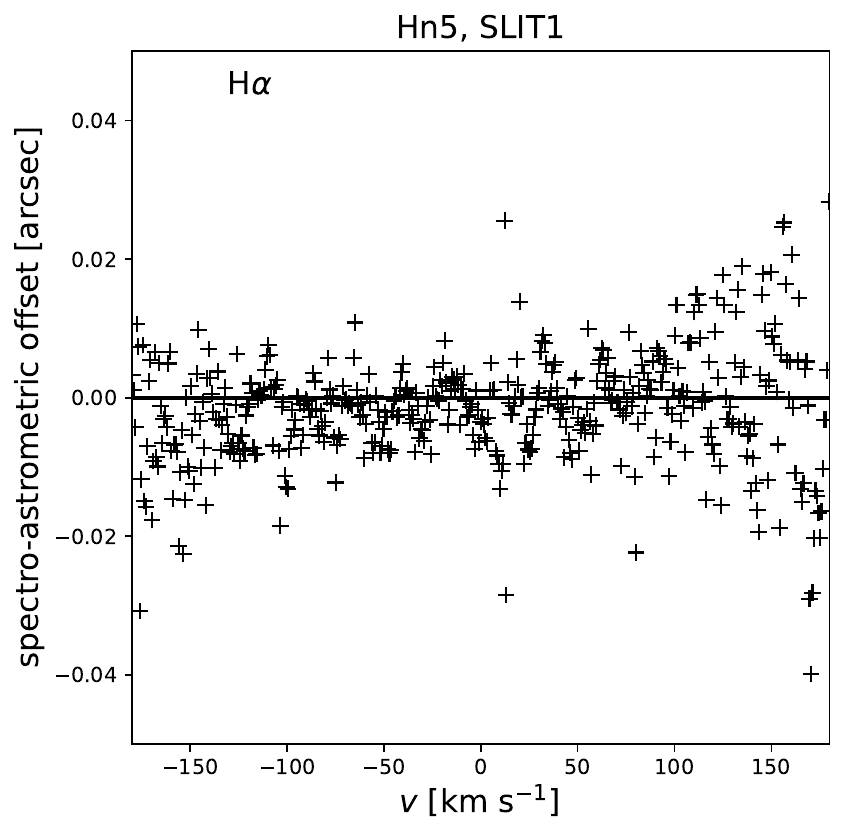}}
\hfill
\subfloat{\includegraphics[trim=0 0 0 0, clip, width=0.3 \textwidth]{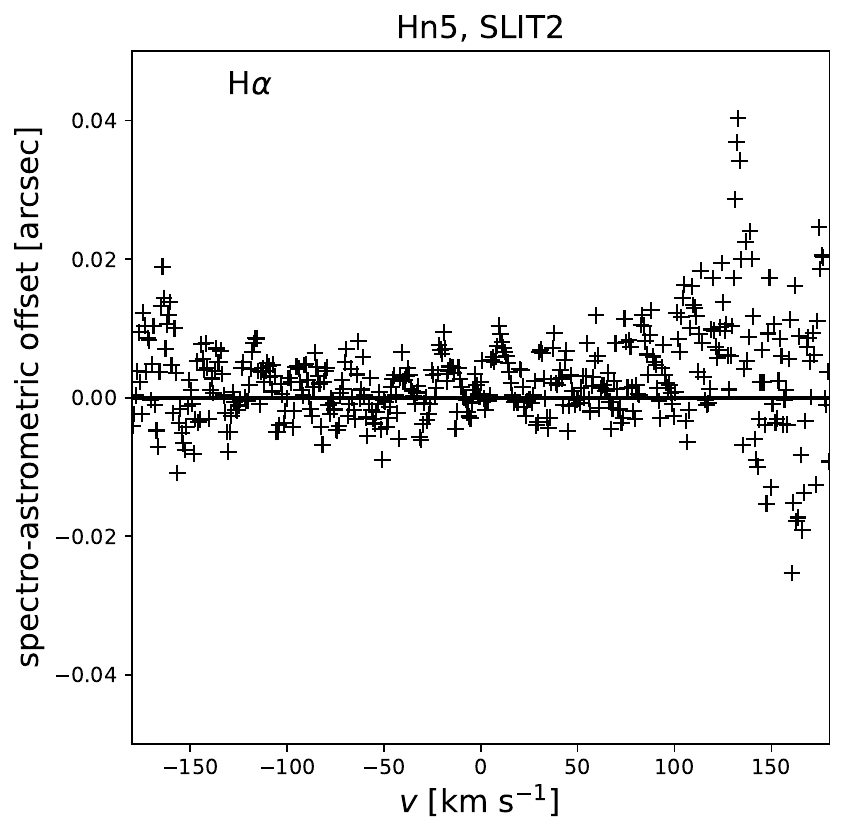}}
\hfill
\subfloat{\includegraphics[trim=0 0 0 0, clip, width=0.3 \textwidth]{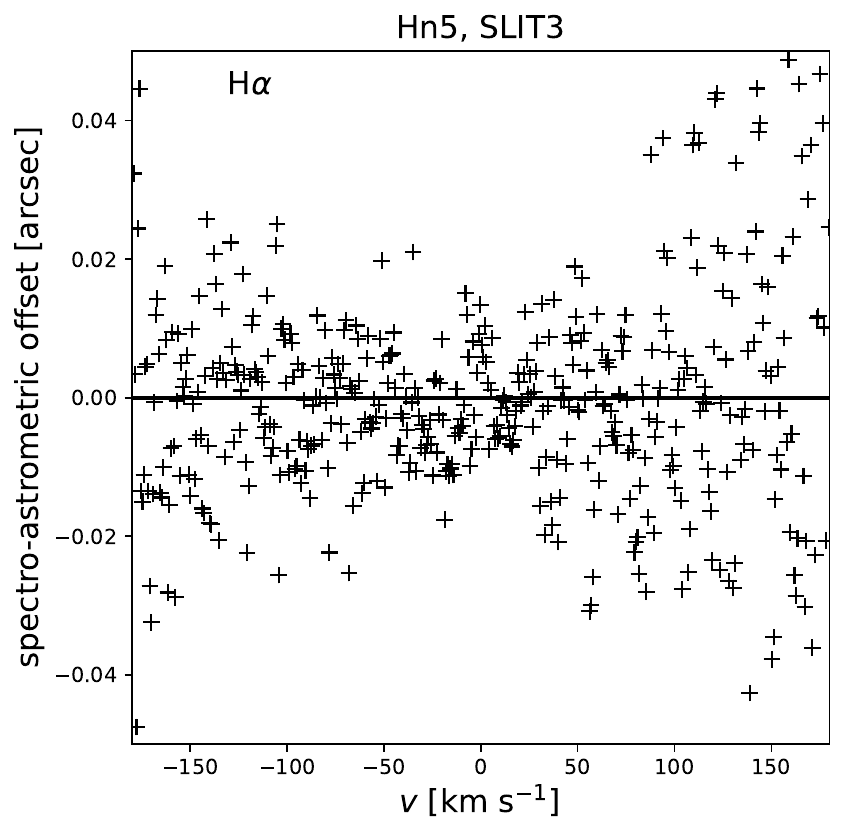}} 
\hfill
\subfloat{\includegraphics[trim=0 0 0 0, clip, width=0.3 \textwidth]{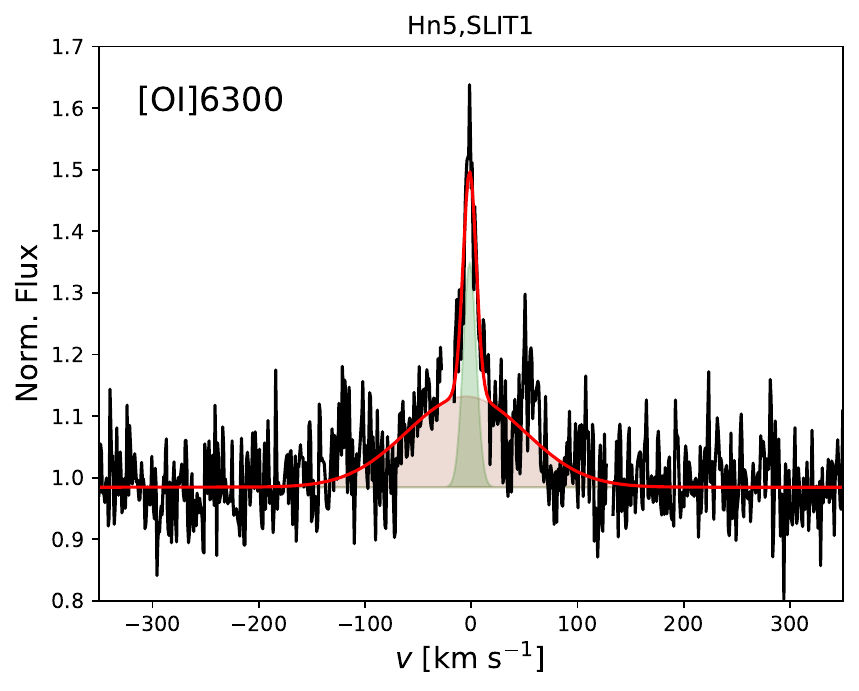}}
\hfill
\subfloat{\includegraphics[trim=0 0 0 0, clip, width=0.3 \textwidth]{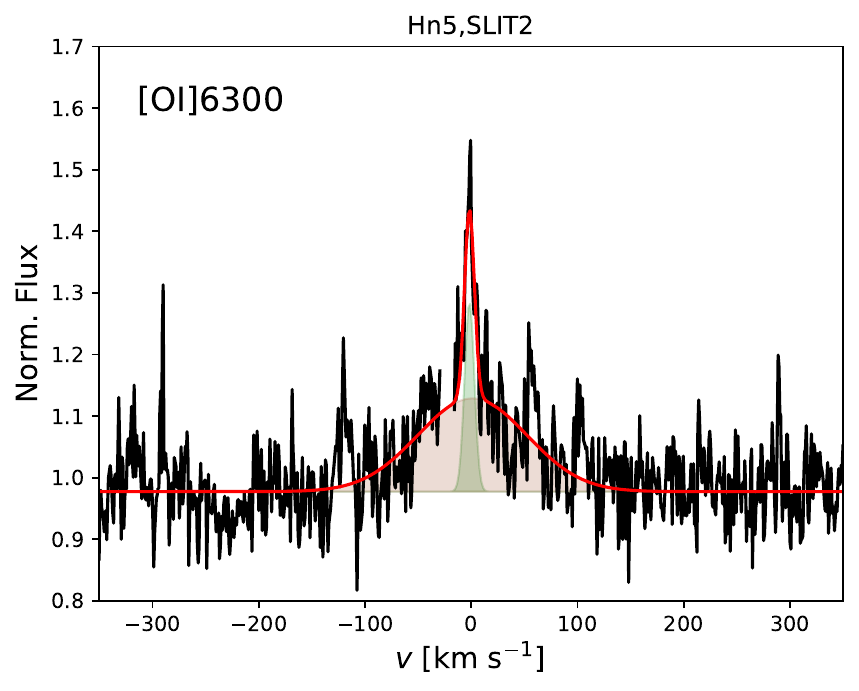}}
\hfill
\subfloat{\includegraphics[trim=0 0 0 0, clip, width=0.3 \textwidth]{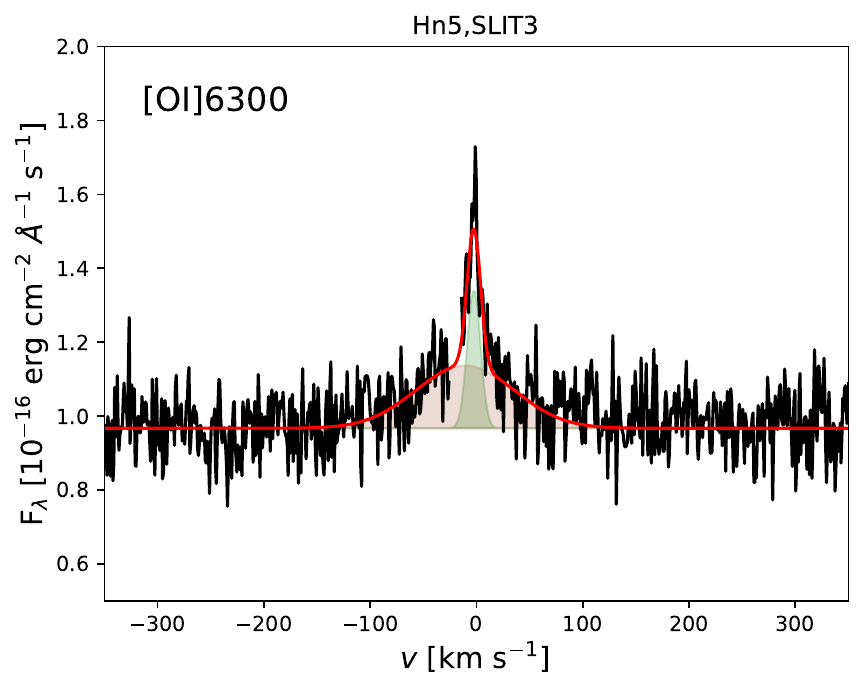}} 
\hfill   
\subfloat{\includegraphics[trim=0 0 0 0, clip, width=0.3 \textwidth]{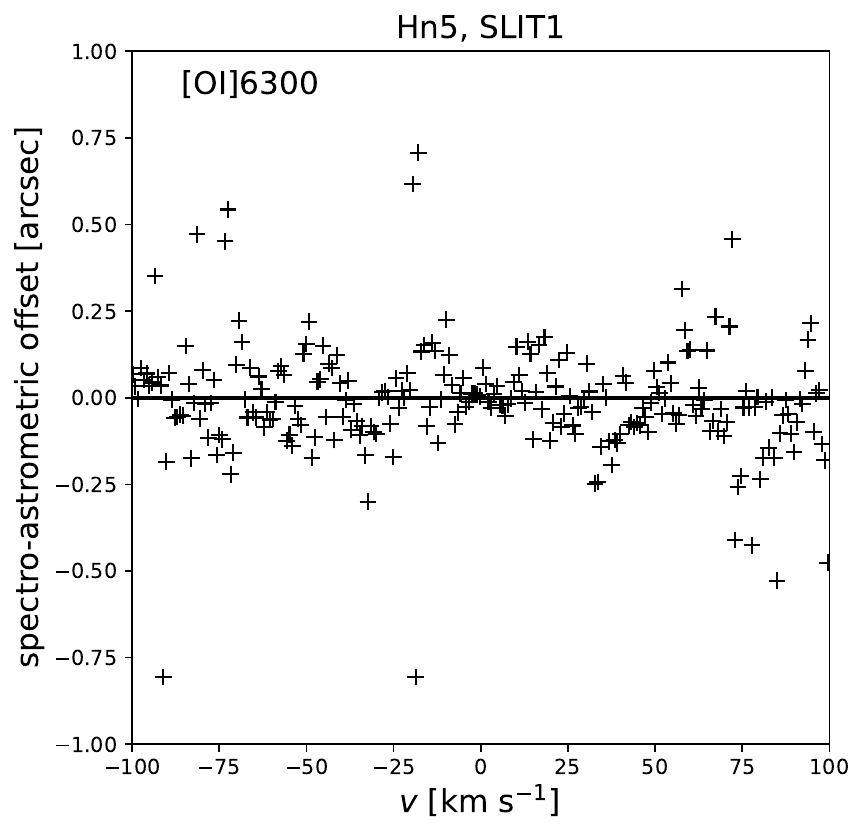}}
\hfill
\subfloat{\includegraphics[trim=0 0 0 0, clip, width=0.3 \textwidth]{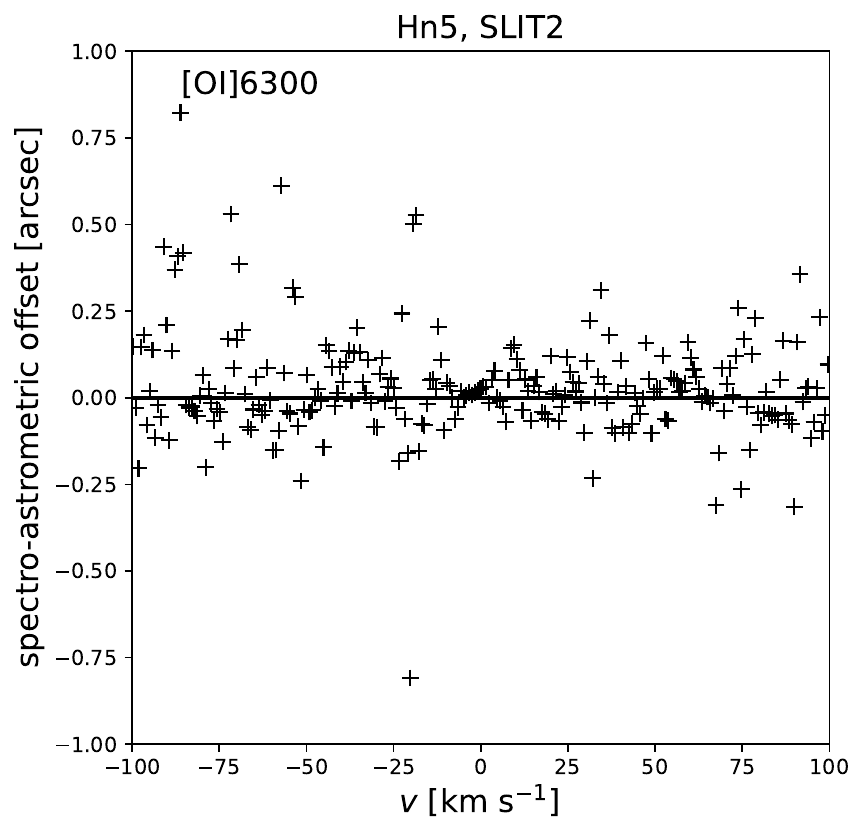}}
\hfill
\subfloat{\includegraphics[trim=0 0 0 0, clip, width=0.3 \textwidth]{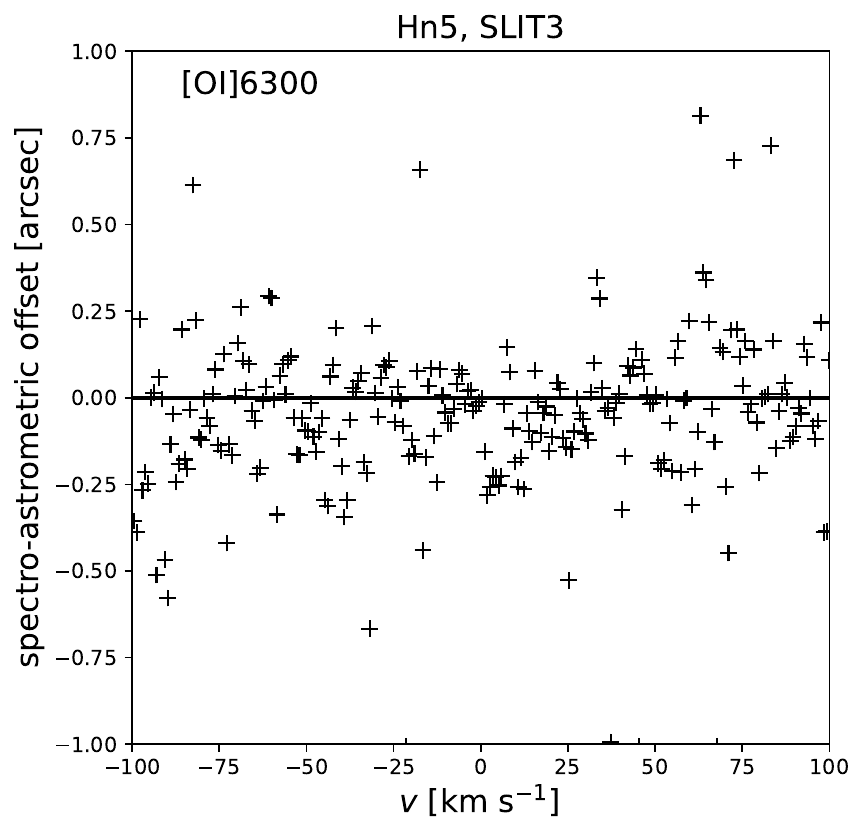}} 
\hfill 
\caption{\small{Line profiles of H$\alpha$ and [OI]$\lambda$6300 for all slit positions of  Hn\,5.}}\label{fig:all_minispectra_Hn5}
\end{figure*} 

\begin{figure*} 
\centering
\subfloat{\includegraphics[trim=0 0 0 0, clip, width=0.3 \textwidth]{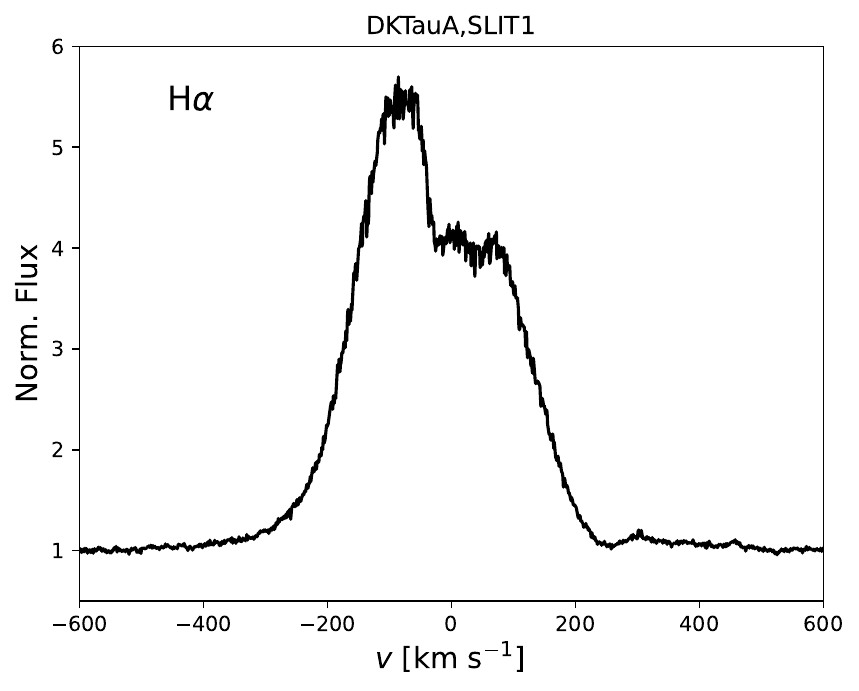}}
\hfill
\subfloat{\includegraphics[trim=0 0 0 0, clip, width=0.3 \textwidth]{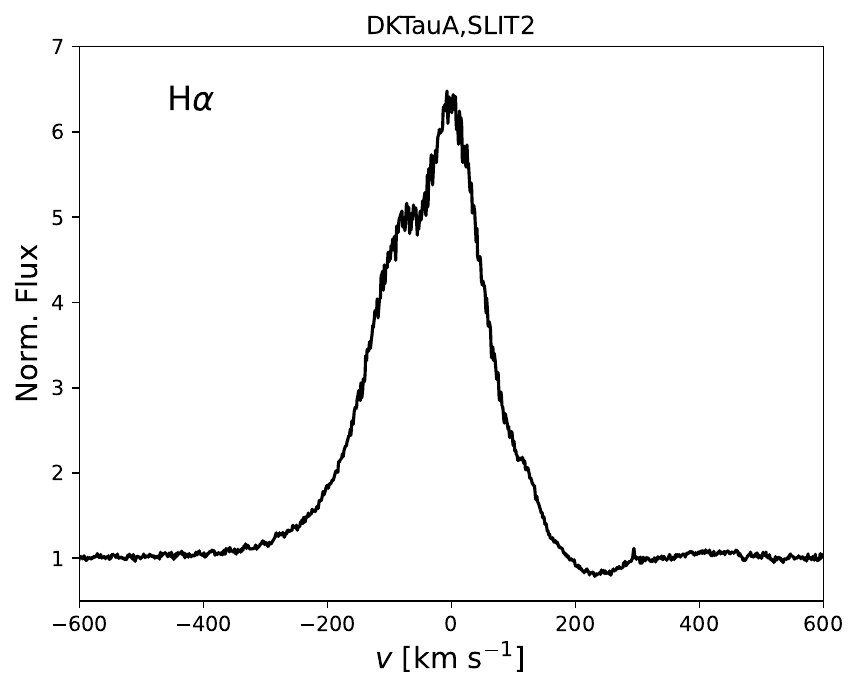}}
\hfill
\subfloat{\includegraphics[trim=0 0 0 0, clip, width=0.3 \textwidth]{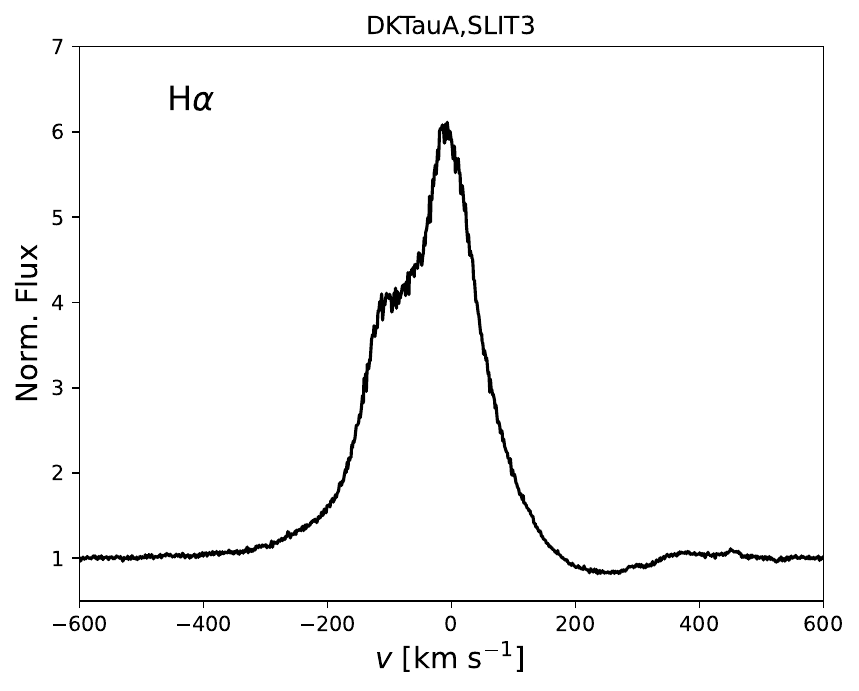}}
\hfill  
\subfloat{\includegraphics[trim=0 0 0 0, clip, width=0.3 \textwidth]{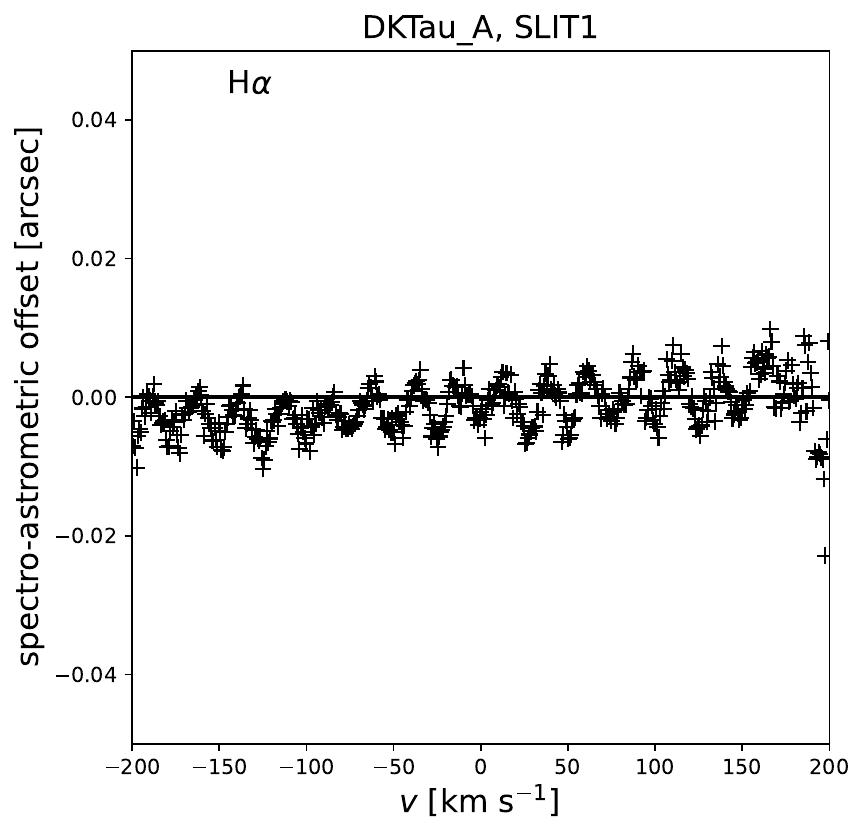}}
\hfill
\subfloat{\includegraphics[trim=0 0 0 0, clip, width=0.3 \textwidth]{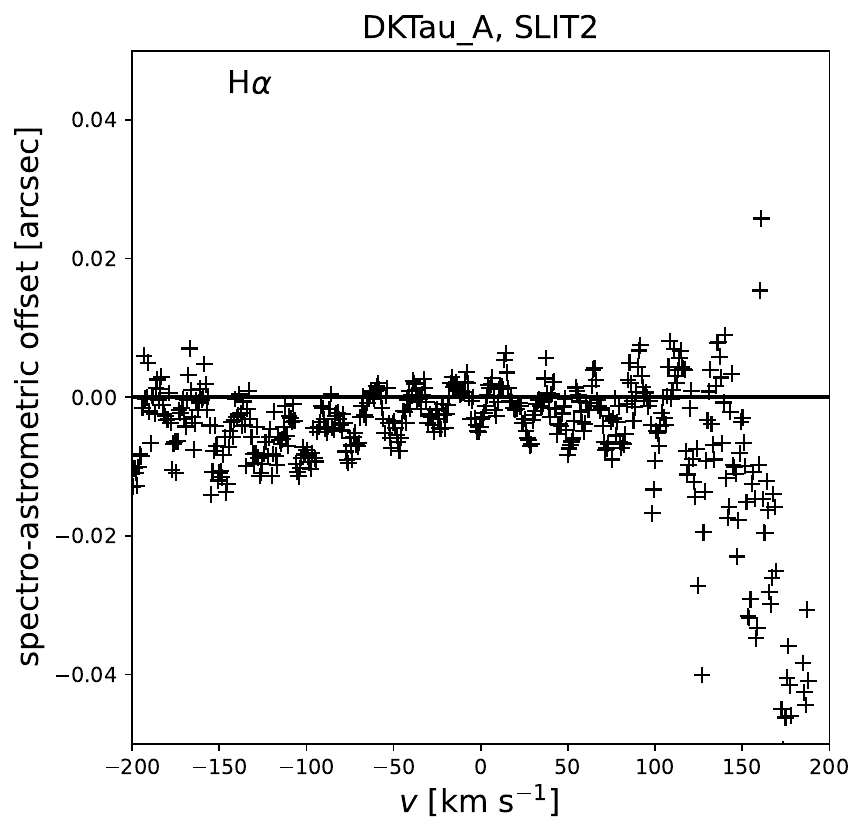}}
\hfill
\subfloat{\includegraphics[trim=0 0 0 0, clip, width=0.3 \textwidth]{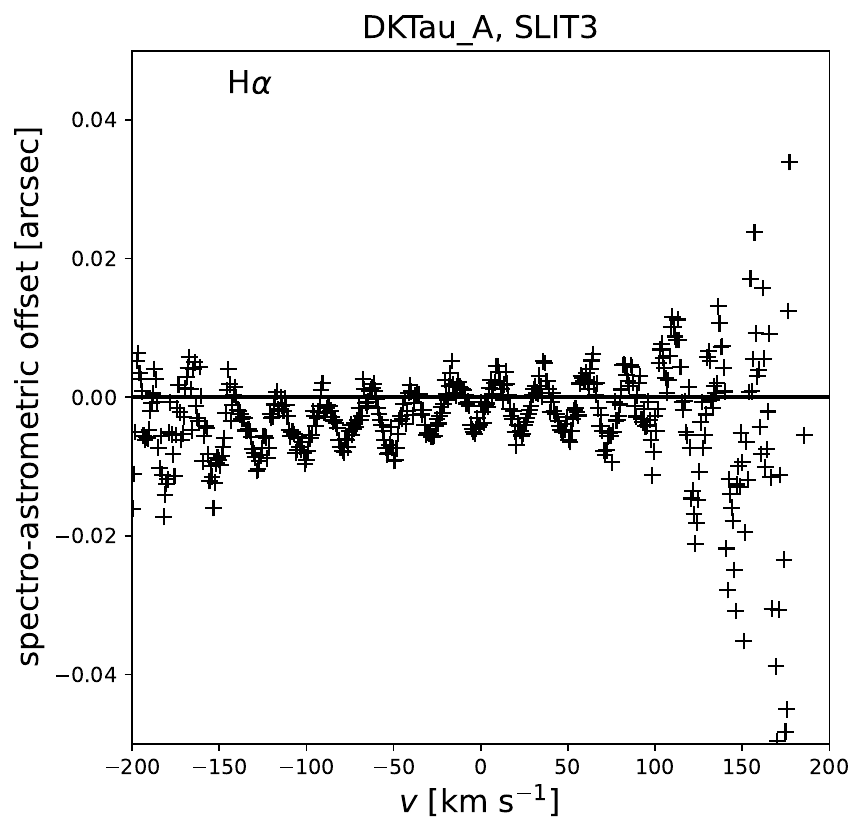}} 
\hfill
\subfloat{\includegraphics[trim=0 0 0 0, clip, width=0.3 \textwidth]{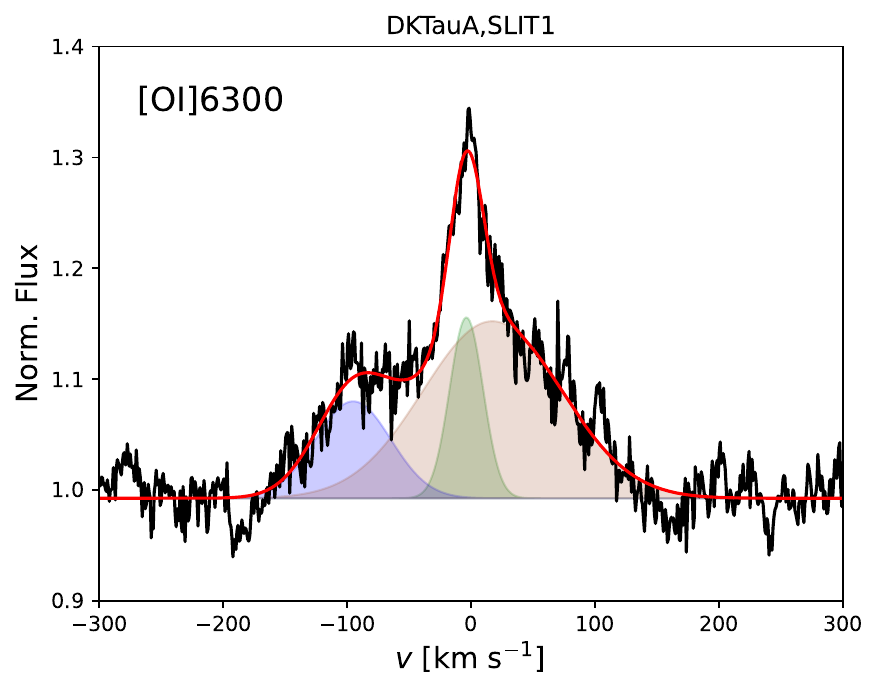}}
\hfill
\subfloat{\includegraphics[trim=0 0 0 0, clip, width=0.3 \textwidth]{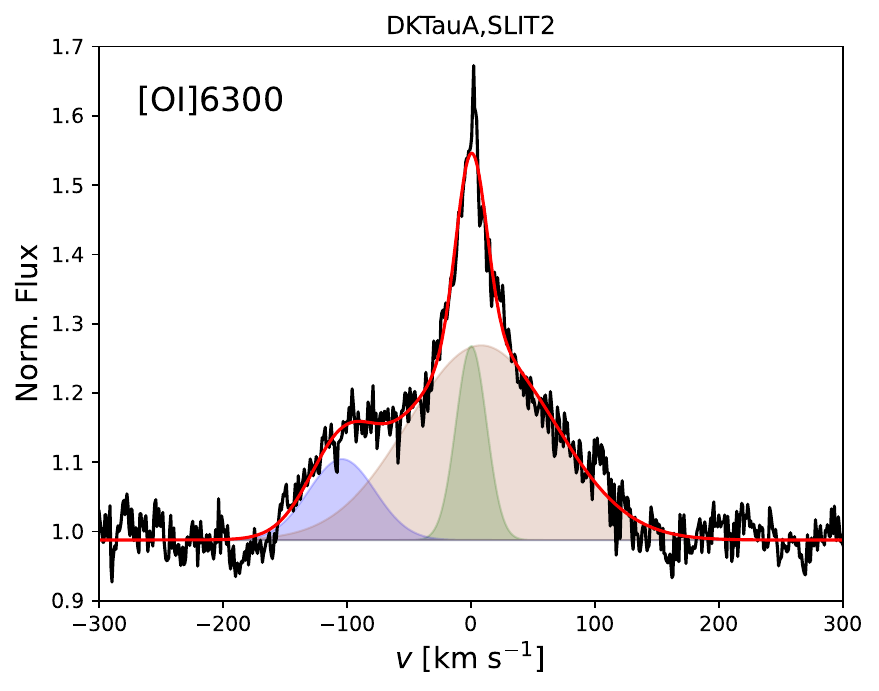}}
\hfill
\subfloat{\includegraphics[trim=0 0 0 0, clip, width=0.3 \textwidth]{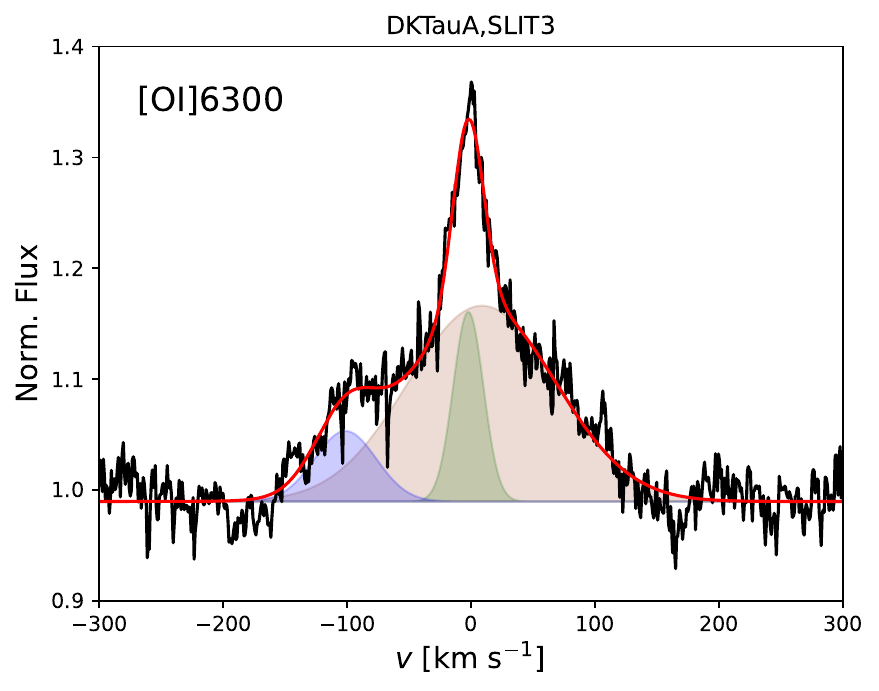}} 
\hfill   
\subfloat{\includegraphics[trim=0 0 0 0, clip, width=0.3 \textwidth]{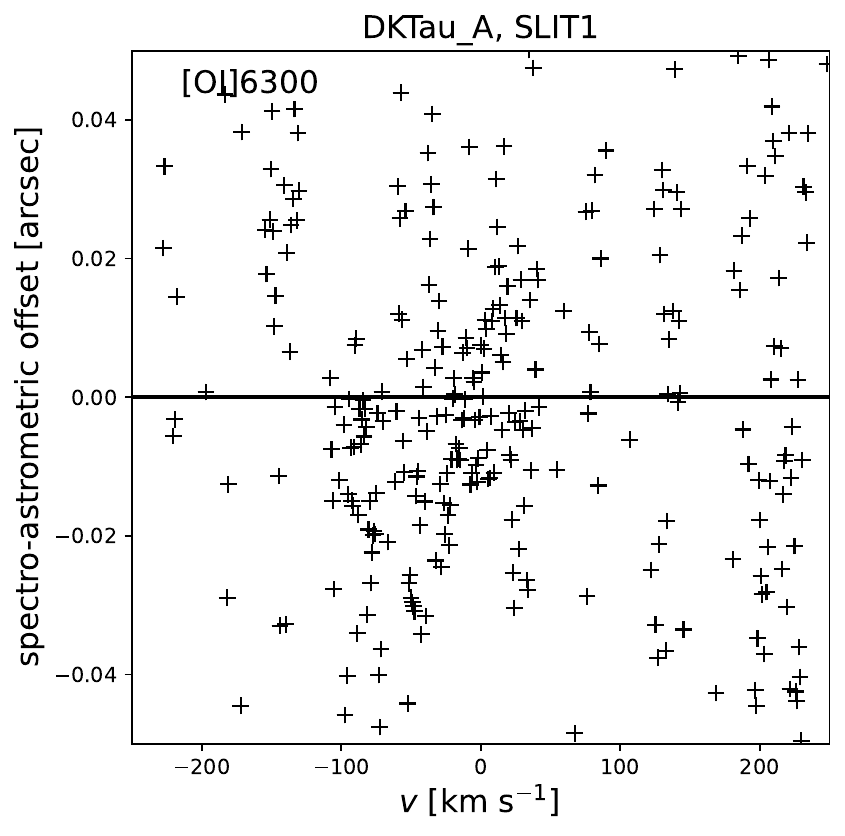}}
\hfill
\subfloat{\includegraphics[trim=0 0 0 0, clip, width=0.3 \textwidth]{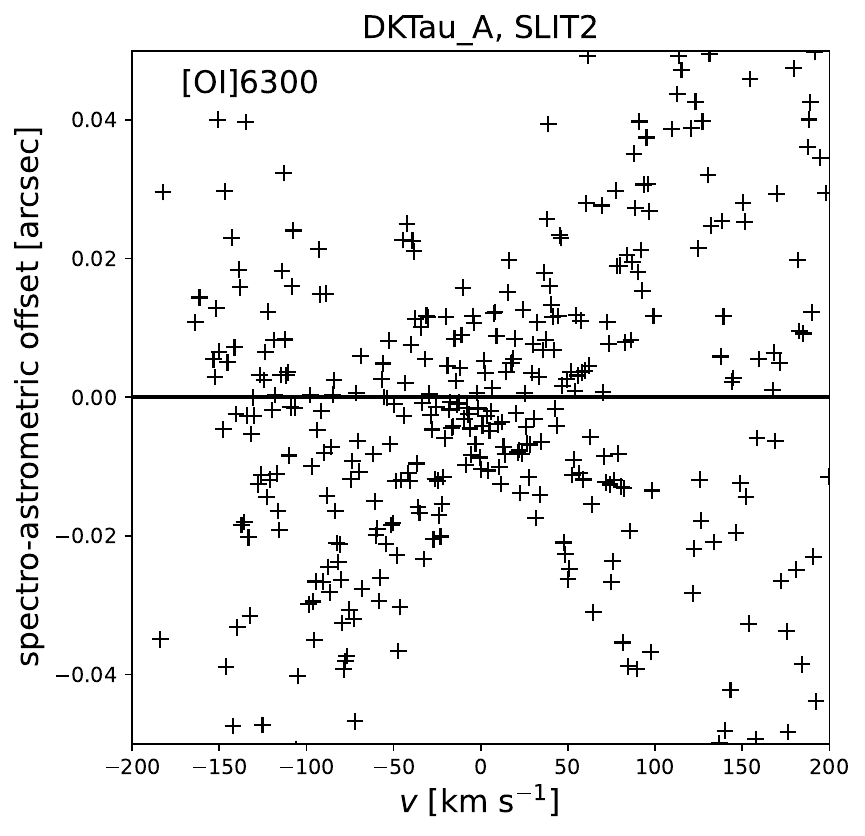}}
\hfill
\subfloat{\includegraphics[trim=0 0 0 0, clip, width=0.3 \textwidth]{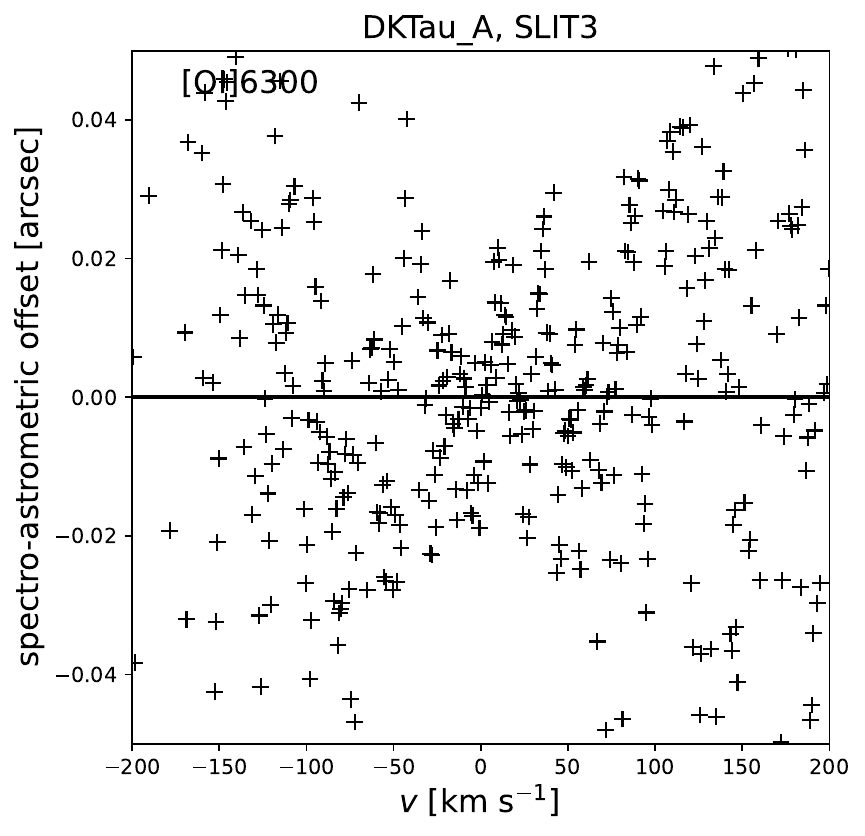}} 
\hfill
\caption{\small{Line profiles of H$\alpha$ and [OI]$\lambda$6300 for all slit positions of DK\,Tau\,A.}}\label{fig:all_minispectra_DKTau_A}
\end{figure*} 

\begin{figure*} 
\centering
\subfloat{\includegraphics[trim=0 0 0 0, clip, width=0.3 \textwidth]{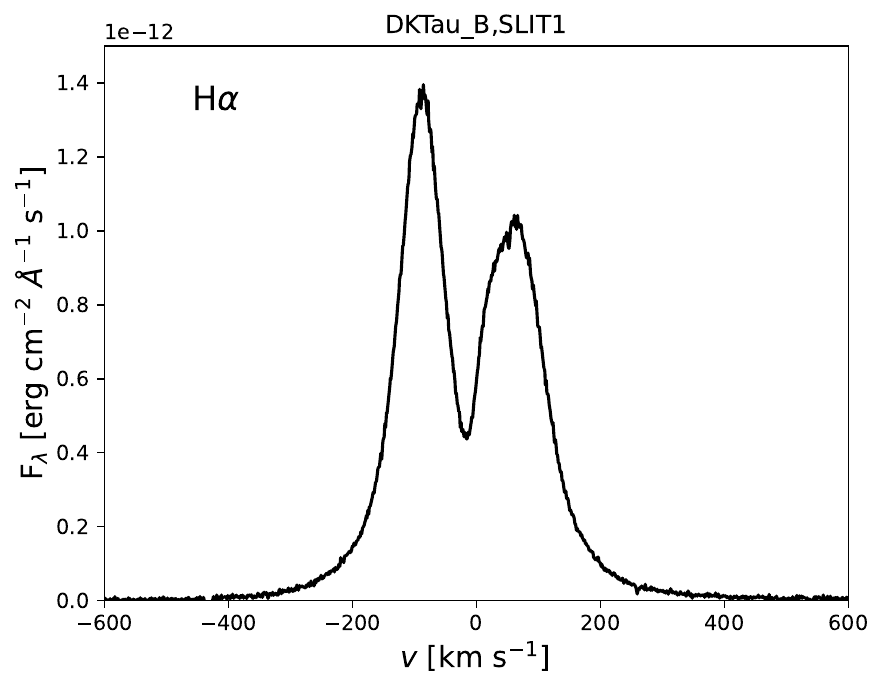}}
\hfill
\subfloat{\includegraphics[trim=0 0 0 0, clip, width=0.3 \textwidth]{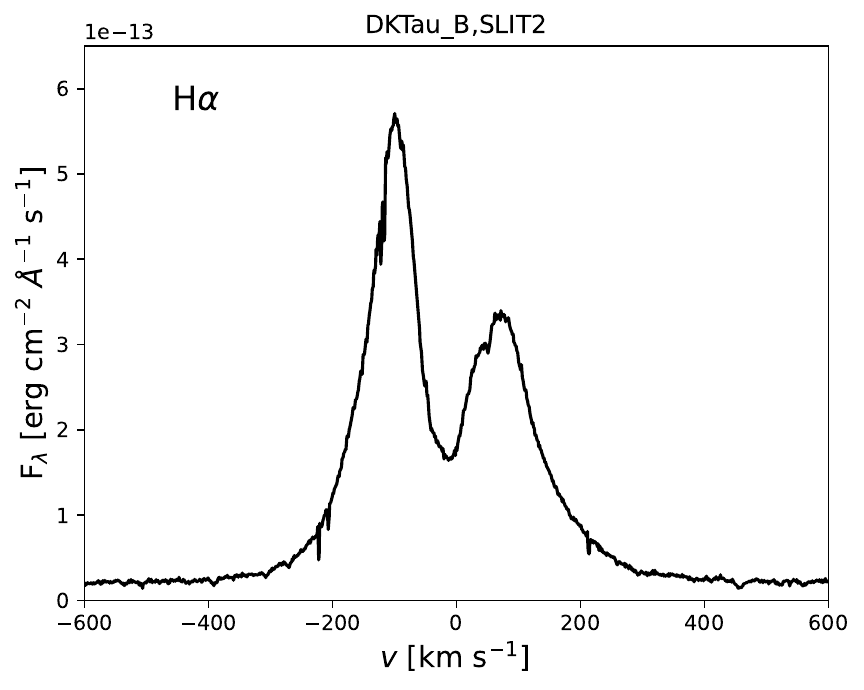}}
\hfill
\subfloat{\includegraphics[trim=0 0 0 0, clip, width=0.3 \textwidth]{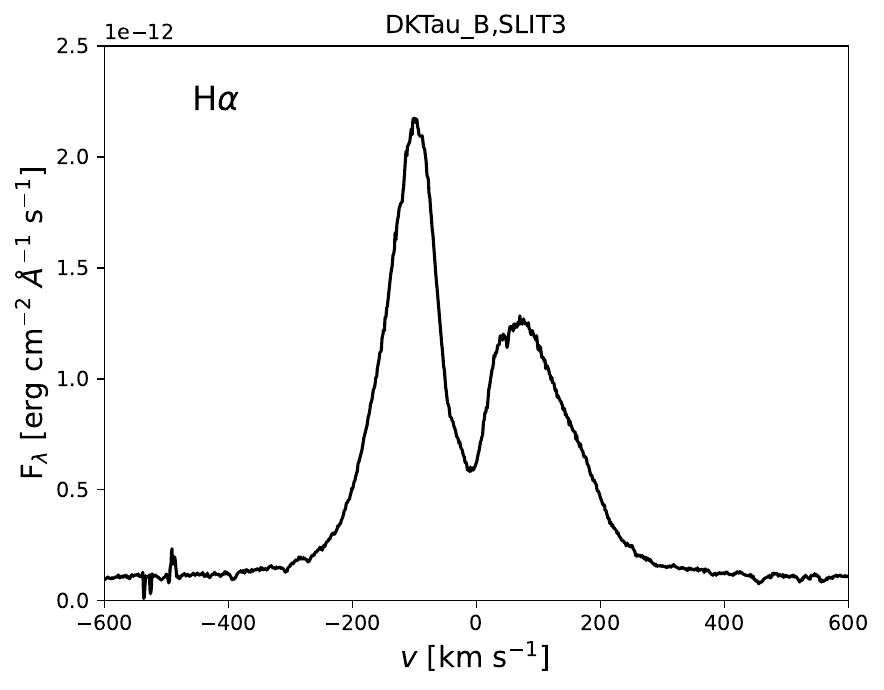}}
\hfill  
\subfloat{\includegraphics[trim=0 0 0 0, clip, width=0.3 \textwidth]{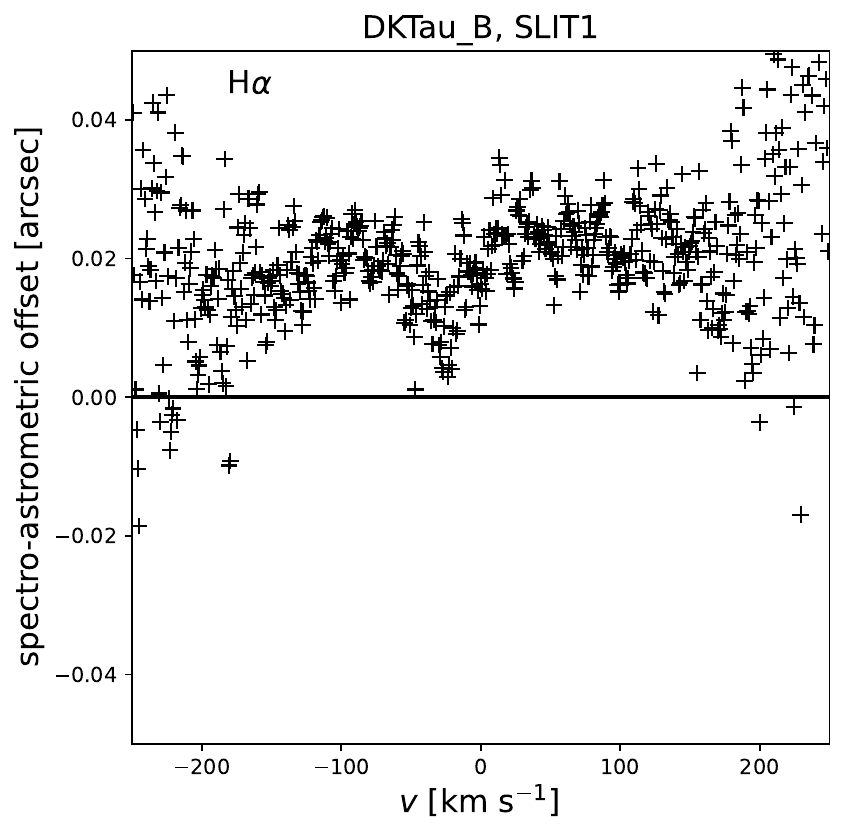}}
\hfill
\subfloat{\includegraphics[trim=0 0 0 0, clip, width=0.3 \textwidth]{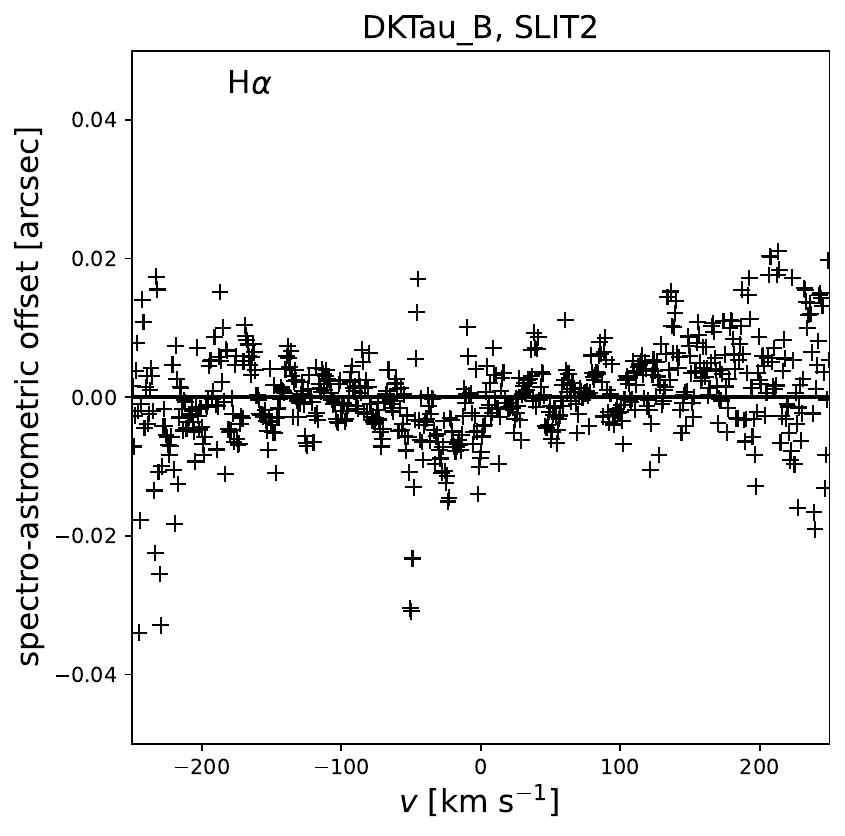}}
\hfill
\subfloat{\includegraphics[trim=0 0 0 0, clip, width=0.3 \textwidth]{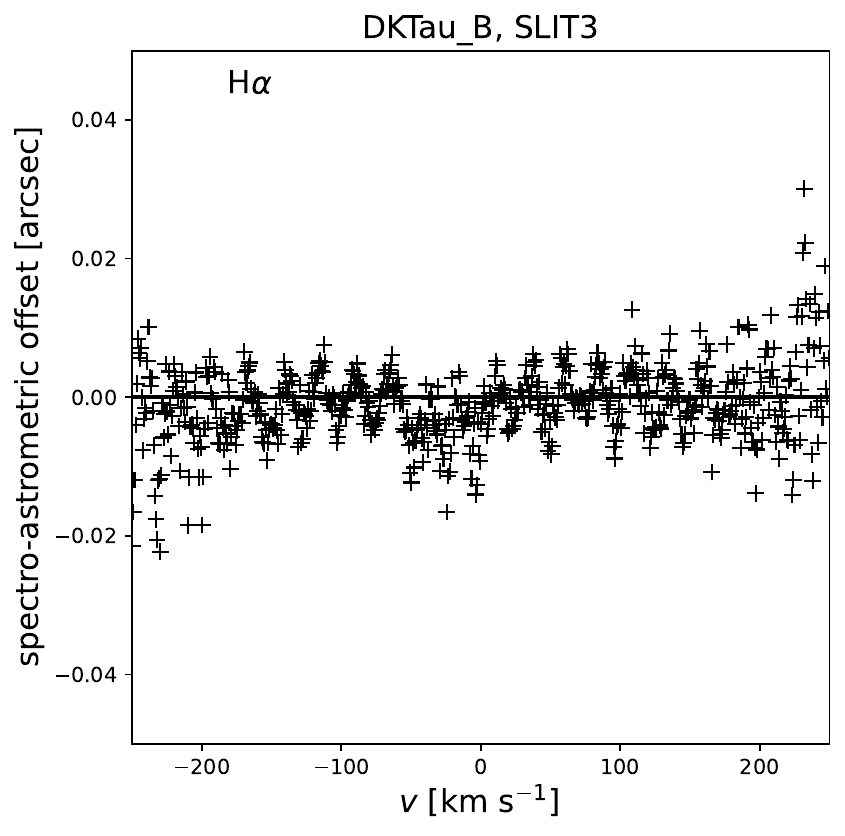}} 
\hfill
\subfloat{\includegraphics[trim=0 0 0 0, clip, width=0.3 \textwidth]{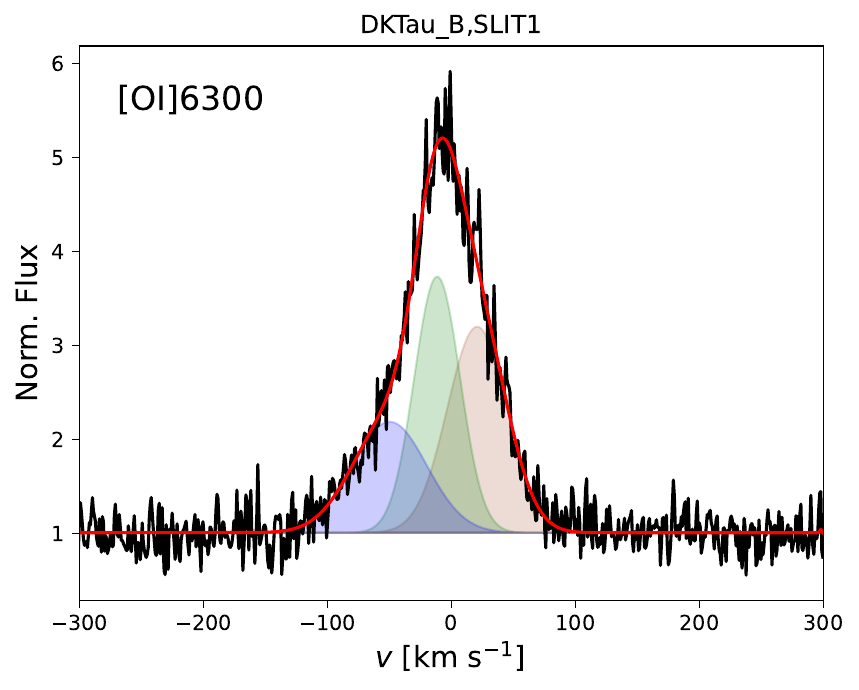}}
\hfill
\subfloat{\includegraphics[trim=0 0 0 0, clip, width=0.3 \textwidth]{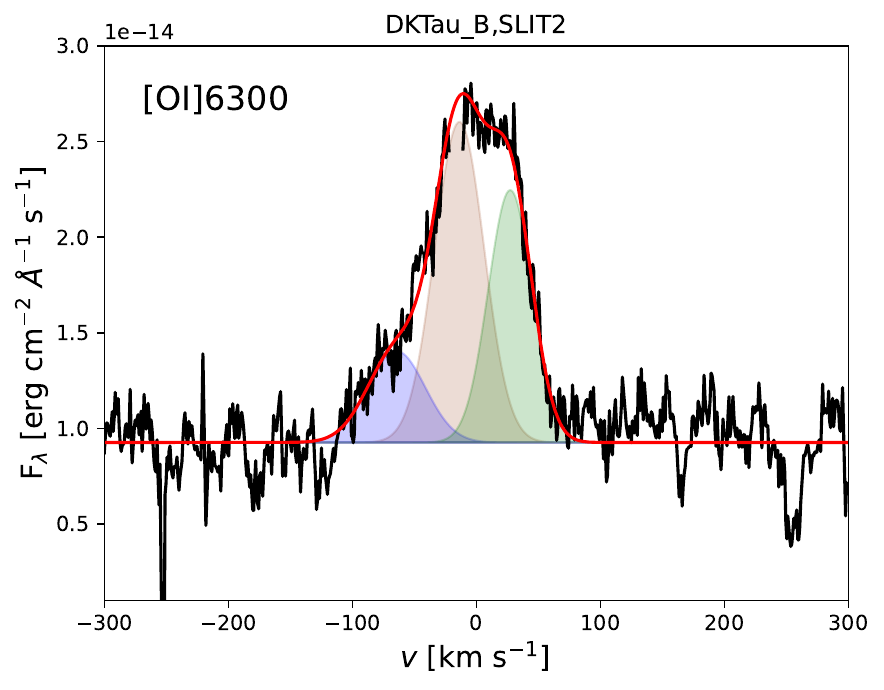}}
\hfill
\subfloat{\includegraphics[trim=0 0 0 0, clip, width=0.3 \textwidth]{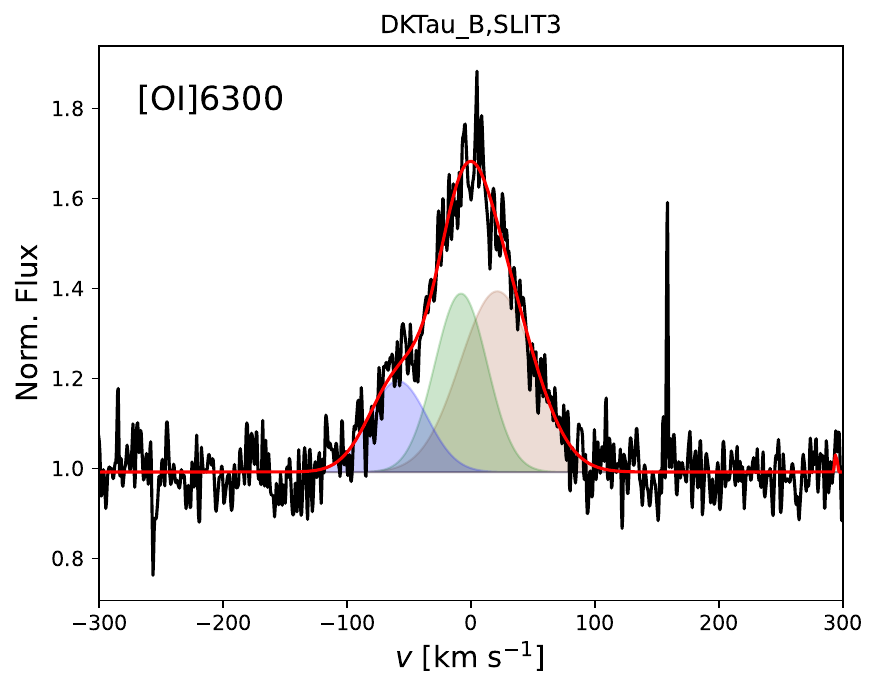}} 
\hfill   
\subfloat{\includegraphics[trim=0 0 0 0, clip, width=0.3 \textwidth]{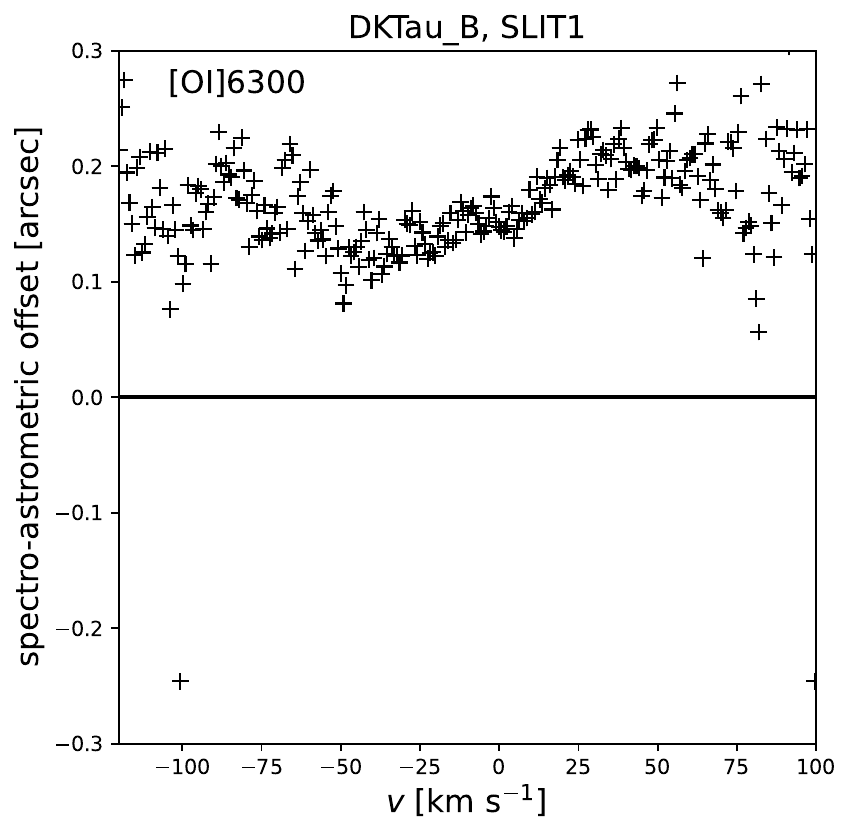}}
\hfill
\subfloat{\includegraphics[trim=0 0 0 0, clip, width=0.3 \textwidth]{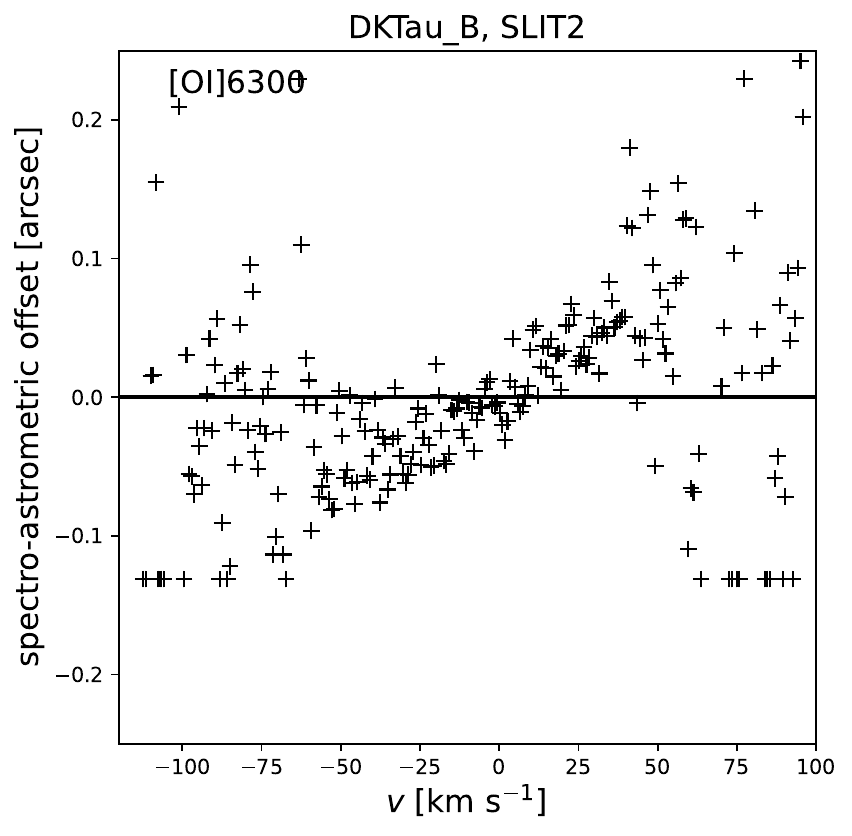}}
\hfill
\subfloat{\includegraphics[trim=0 0 0 0, clip, width=0.3 \textwidth]{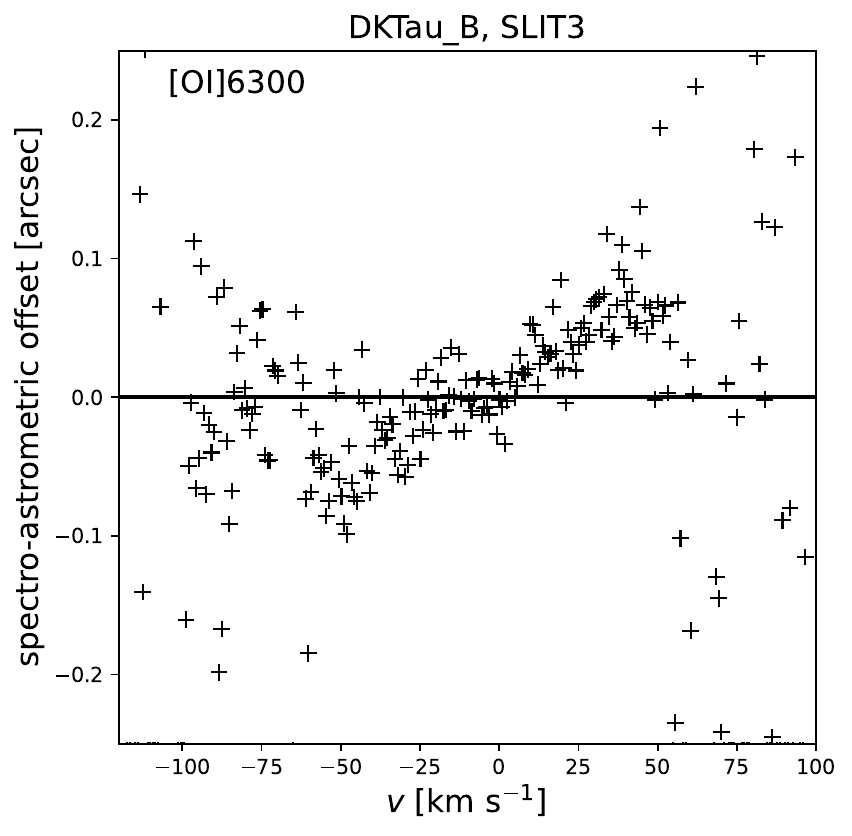}} 
\hfill
\caption{\small{Line profiles of H$\alpha$ and [OI]$\lambda$6300 for all slit positions of DK\,Tau\,B.}}\label{fig:all_minispectra_DKTau_B}
\end{figure*} 

\begin{figure*} 
\centering
\subfloat{\includegraphics[trim=0 0 0 0, clip, width=0.3 \textwidth]{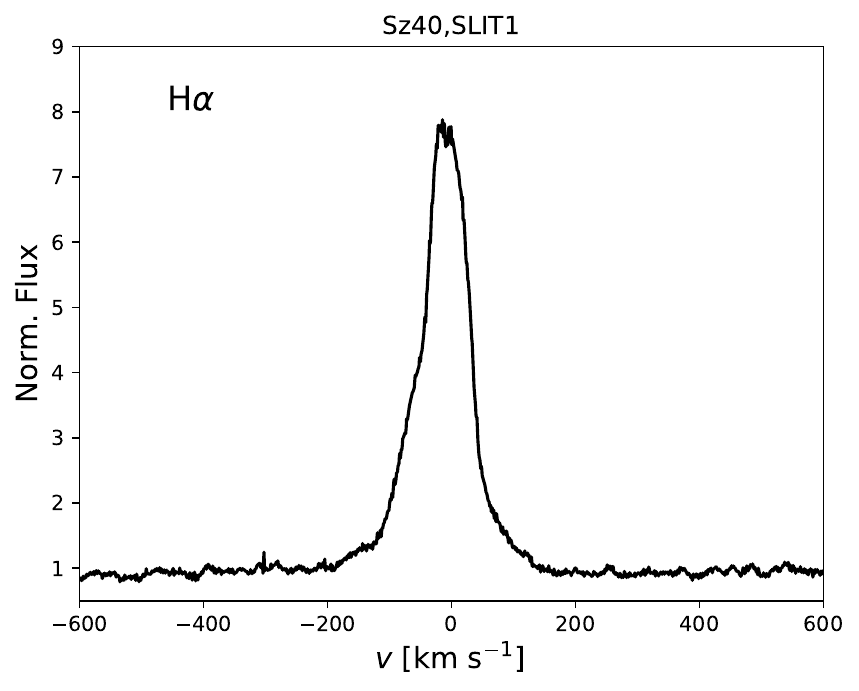}}
\hfill
\subfloat{\includegraphics[trim=0 0 0 0, clip, width=0.3 \textwidth]{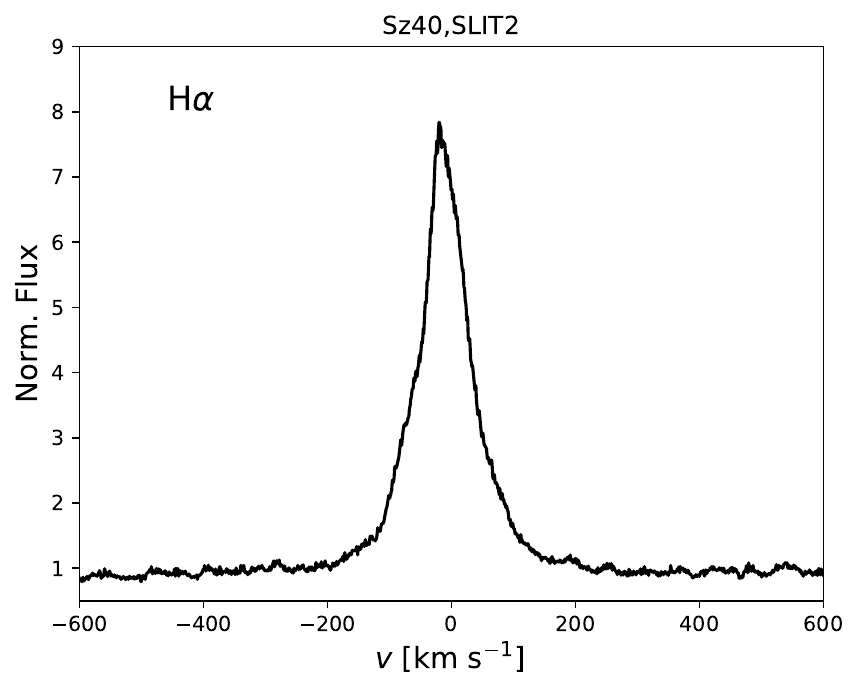}}
\hfill \\
\subfloat{\includegraphics[trim=0 0 0 0, clip, width=0.3 \textwidth]{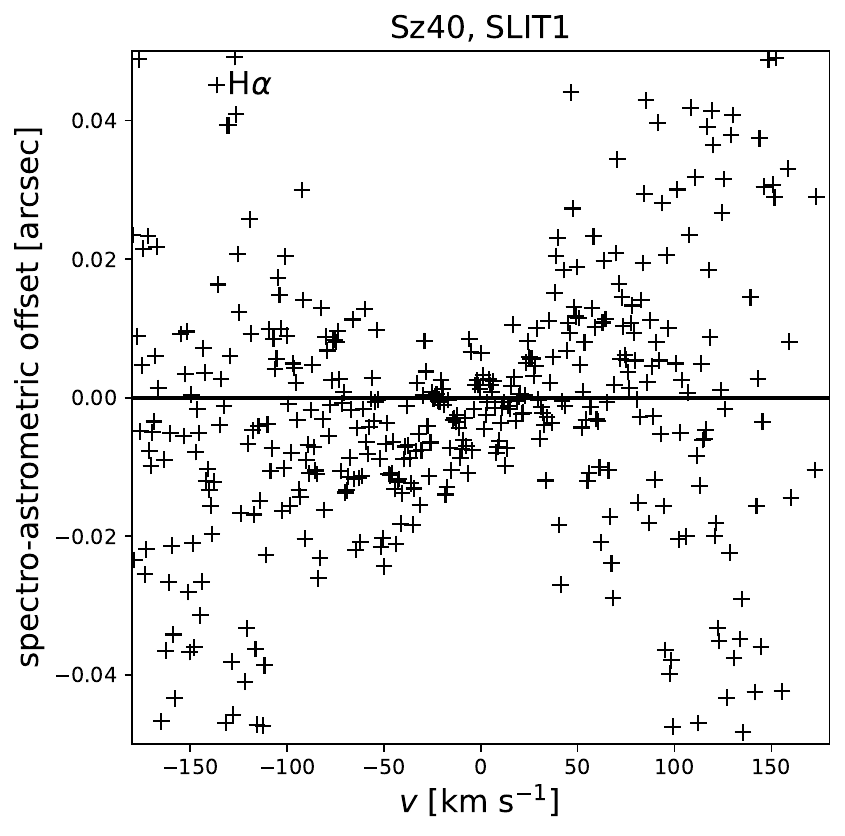}}
\hfill
\subfloat{\includegraphics[trim=0 0 0 0, clip, width=0.3 \textwidth]{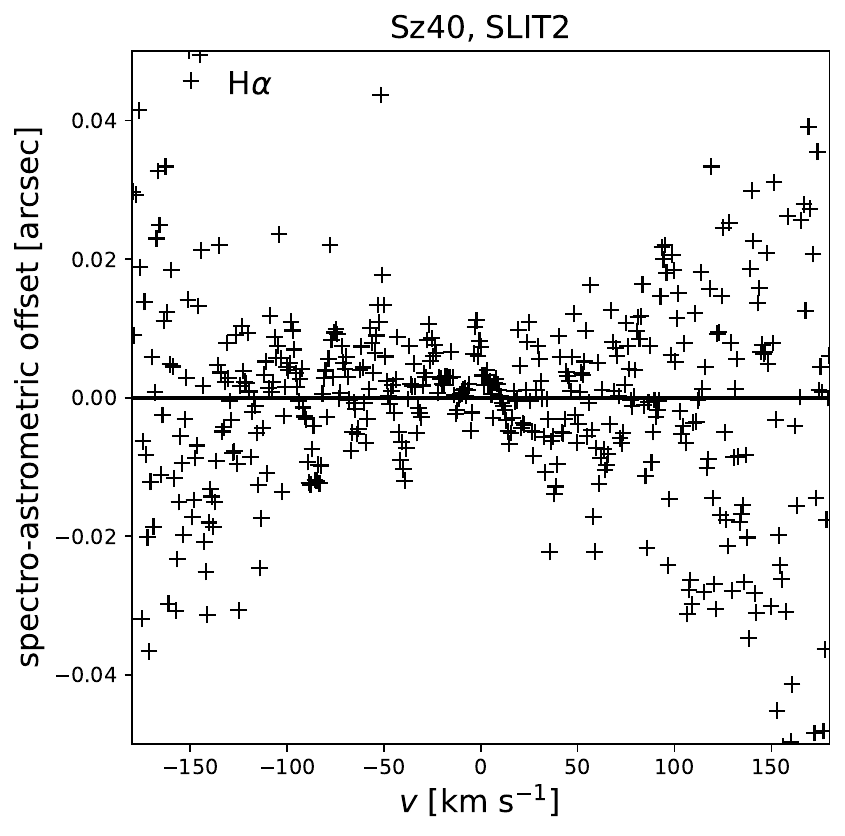}}
\hfill \\
\subfloat{\includegraphics[trim=0 0 0 0, clip, width=0.3 \textwidth]{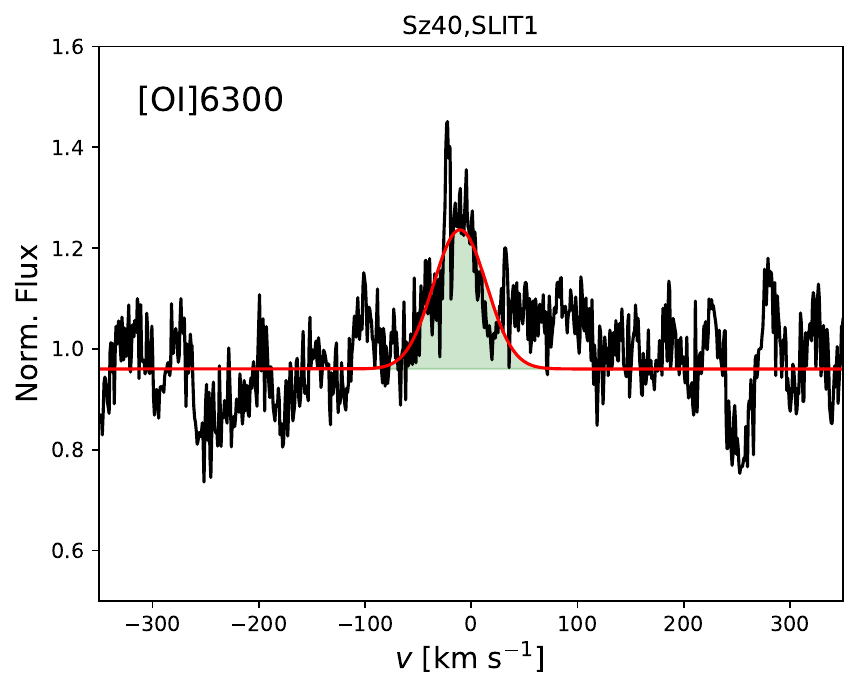}}
\hfill
\subfloat{\includegraphics[trim=0 0 0 0, clip, width=0.3 \textwidth]{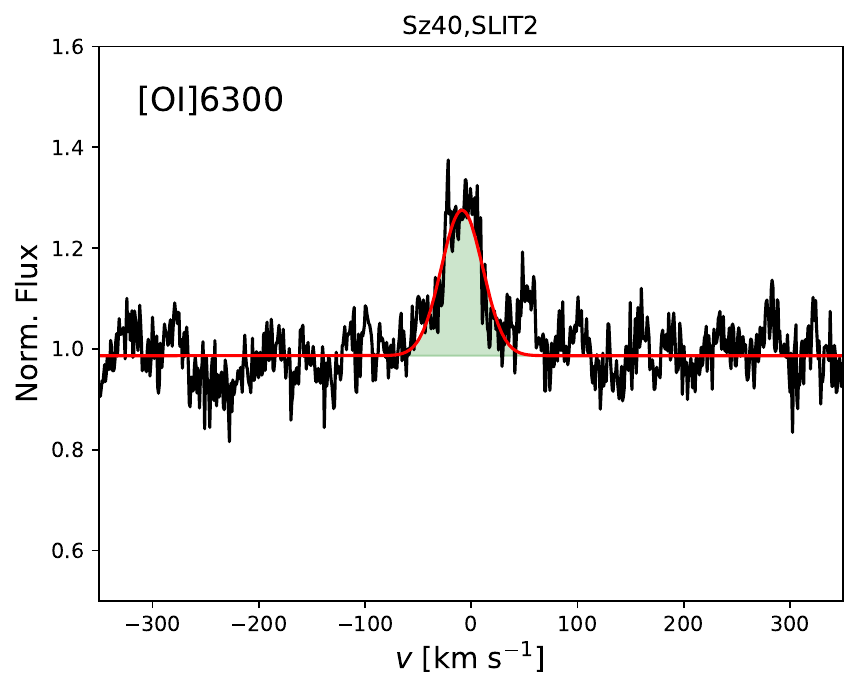}}
\hfill \\
\subfloat{\includegraphics[trim=0 0 0 0, clip, width=0.3 \textwidth]{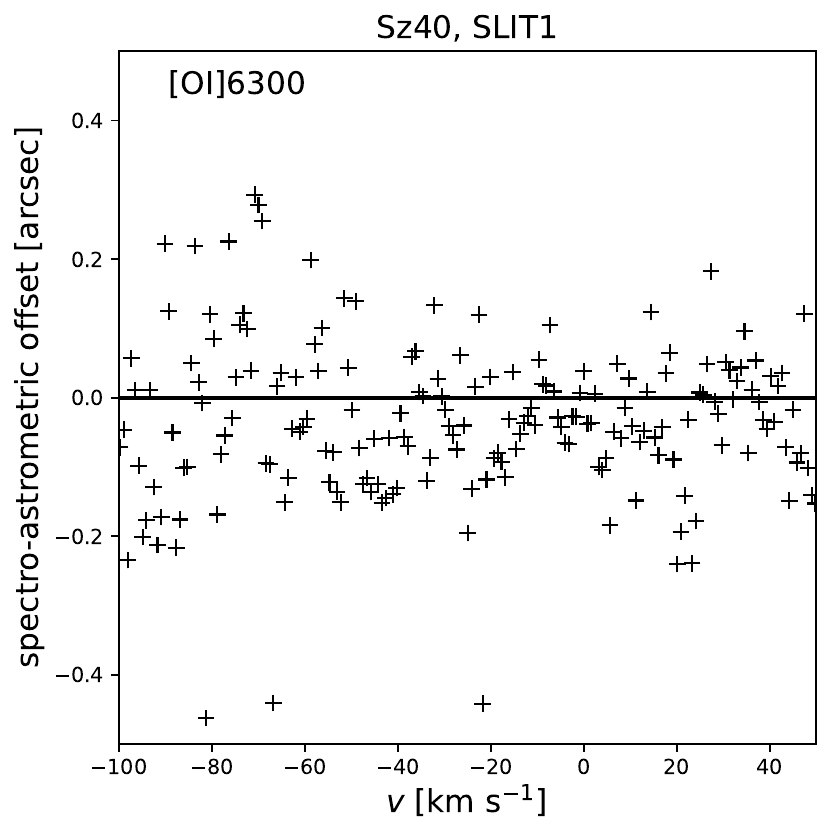}}
\hfill
\subfloat{\includegraphics[trim=0 0 0 0, clip, width=0.3 \textwidth]{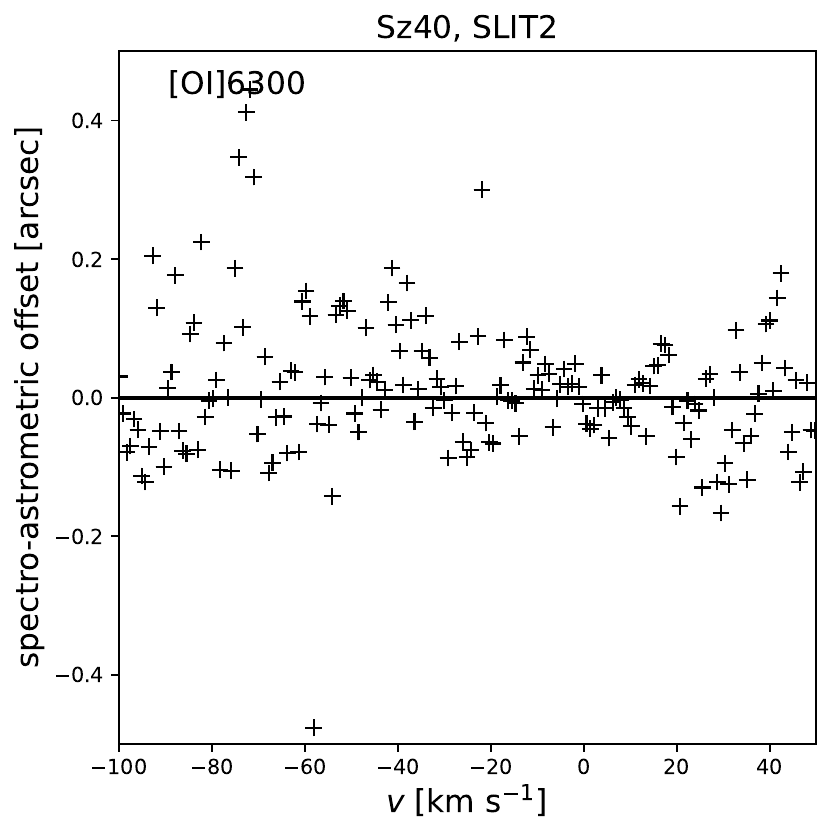}}
\hfill
\caption{\small{Line profiles of H$\alpha$ and [OI]$\lambda$6300 for all slit positions of Sz\,40.}}\label{fig:all_minispectra_Sz40}
\end{figure*} 

\begin{figure*} 
\centering
\subfloat{\includegraphics[trim=0 0 0 0, clip, width=0.3 \textwidth]{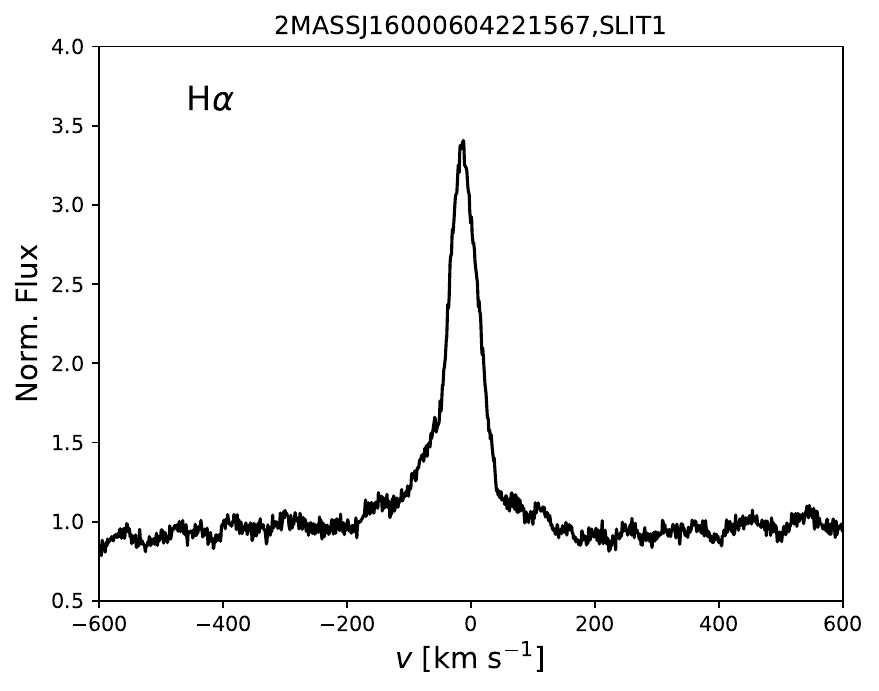}}
\hfill
\subfloat{\includegraphics[trim=0 0 0 0, clip, width=0.3 \textwidth]{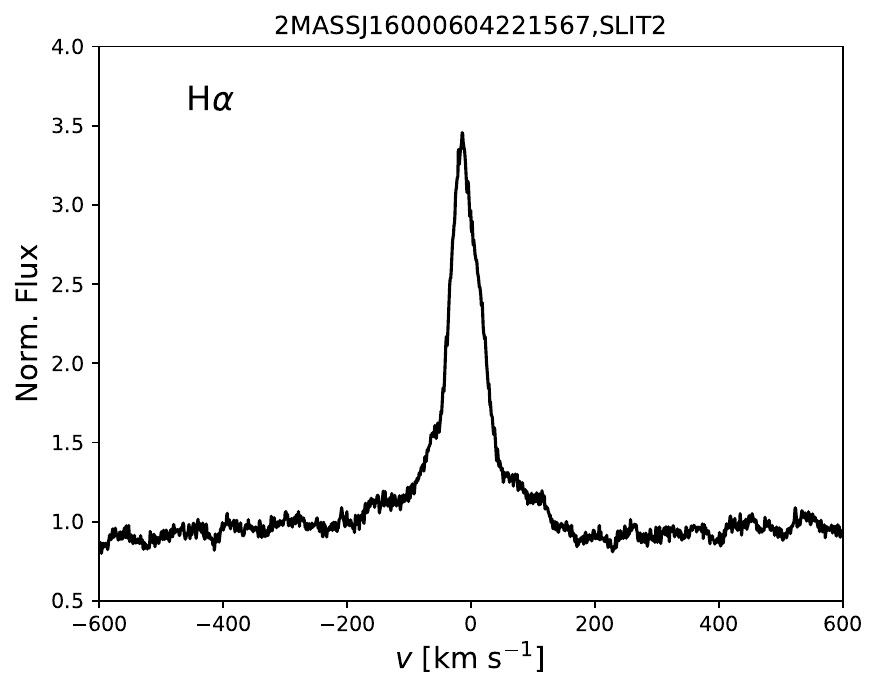}}
\hfill \\
\subfloat{\includegraphics[trim=0 0 0 0, clip, width=0.3 \textwidth]{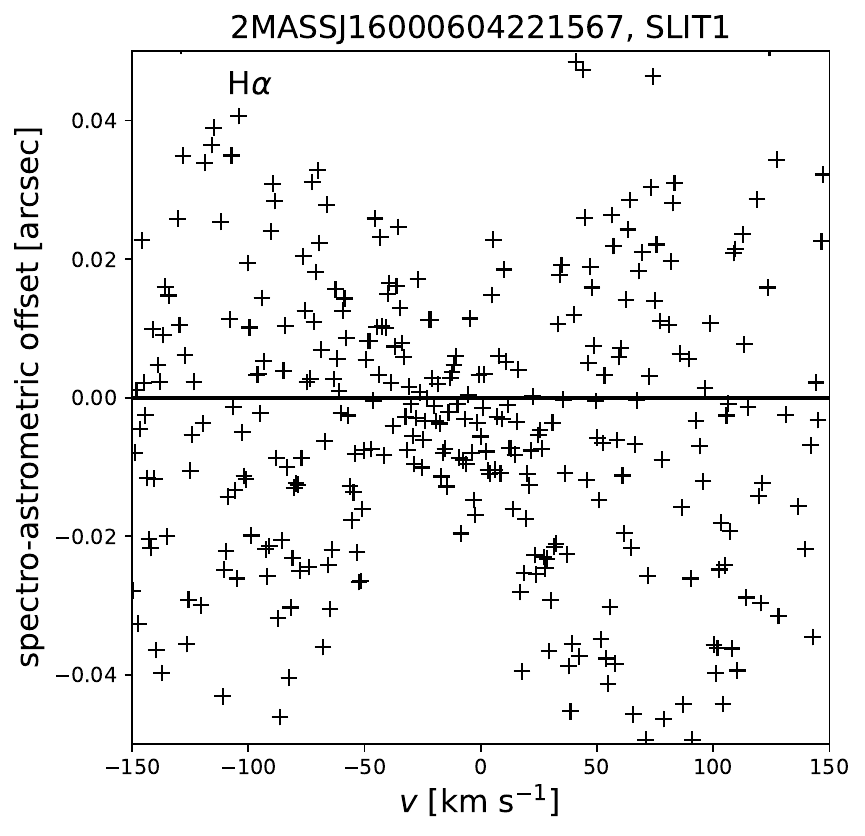}}
\hfill
\subfloat{\includegraphics[trim=0 0 0 0, clip, width=0.3 \textwidth]{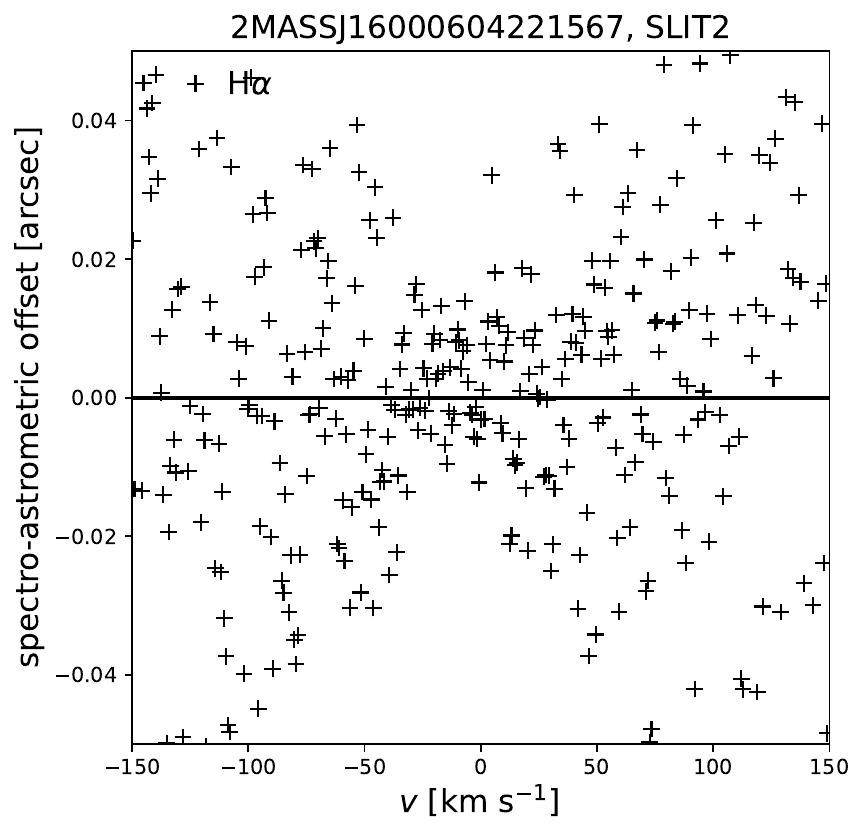}}
\hfill \\
\subfloat{\includegraphics[trim=0 0 0 0, clip, width=0.3 \textwidth]{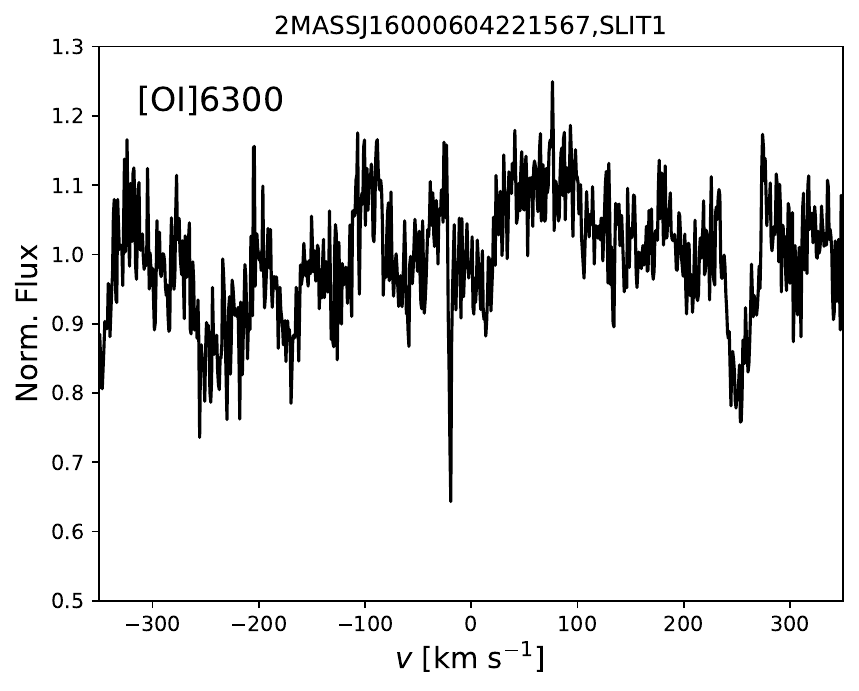}}
\hfill
\subfloat{\includegraphics[trim=0 0 0 0, clip, width=0.3 \textwidth]{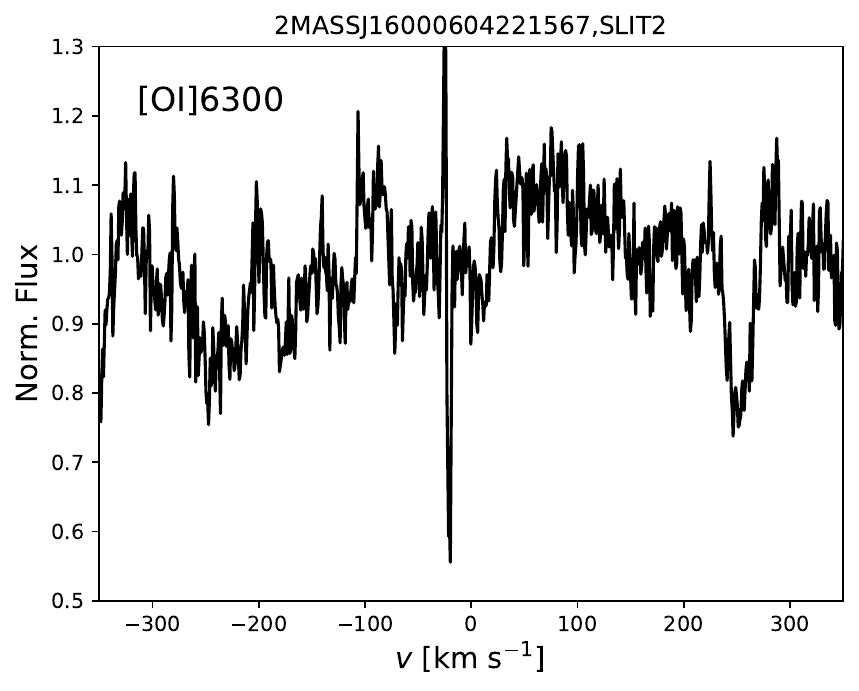}}
\hfill 
\caption{\small{Line profiles of H$\alpha$ and [OI]$\lambda$6300 for all slit positions of 2MASSJ16000604221567.}}\label{fig:all_minispectra_2MASSJ16000604221567}
\end{figure*} 

\begin{figure*} 
\centering
\subfloat{\includegraphics[trim=0 0 0 0, clip, width=0.24 \textwidth]{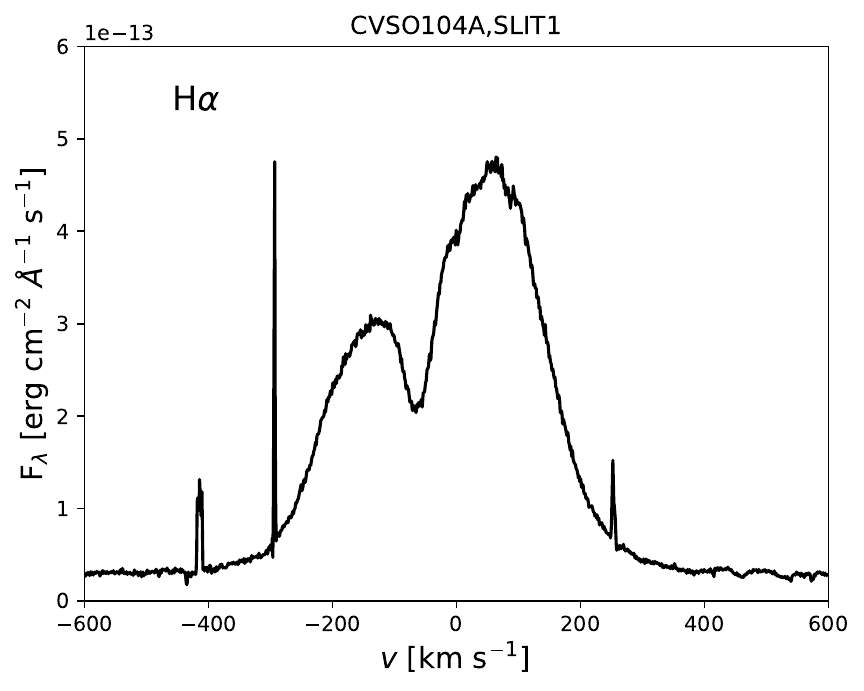}}
\hfill
\subfloat{\includegraphics[trim=0 0 0 0, clip, width=0.24 \textwidth]{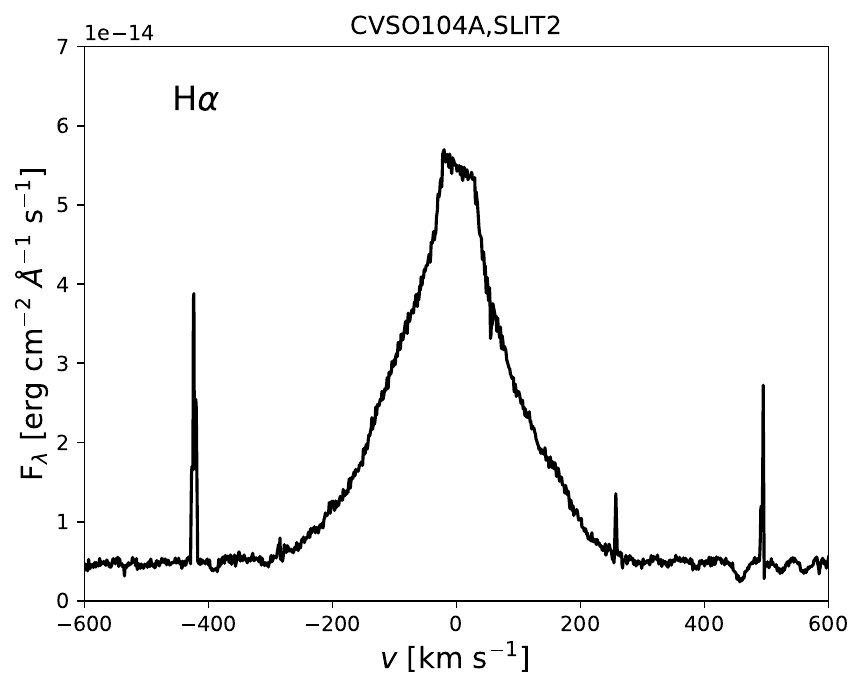}}
\hfill
\subfloat{\includegraphics[trim=0 0 0 0, clip, width=0.24 \textwidth]{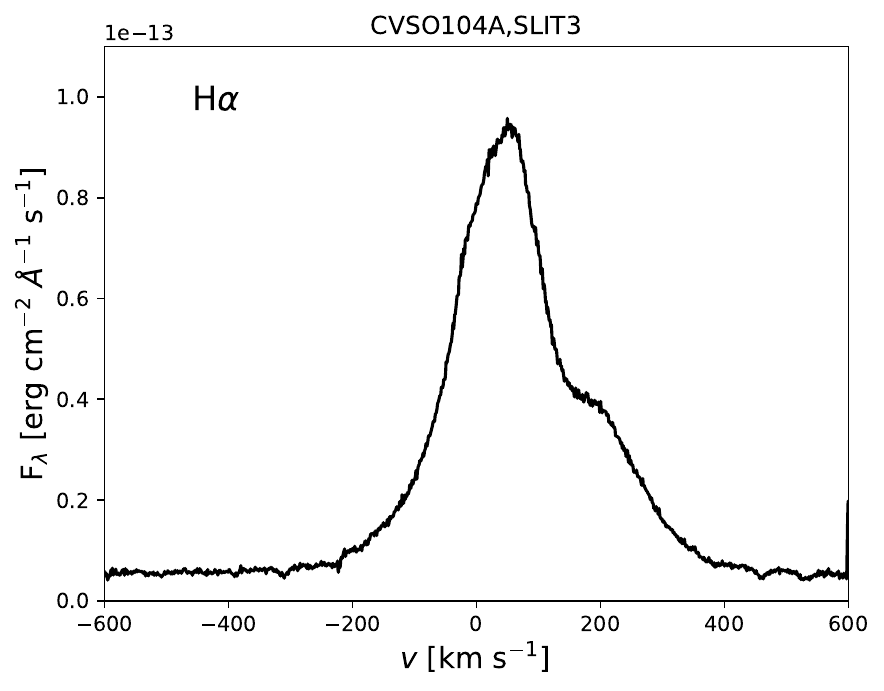}}
\hfill  
\subfloat{\includegraphics[trim=0 0 0 0, clip, width=0.24 \textwidth]{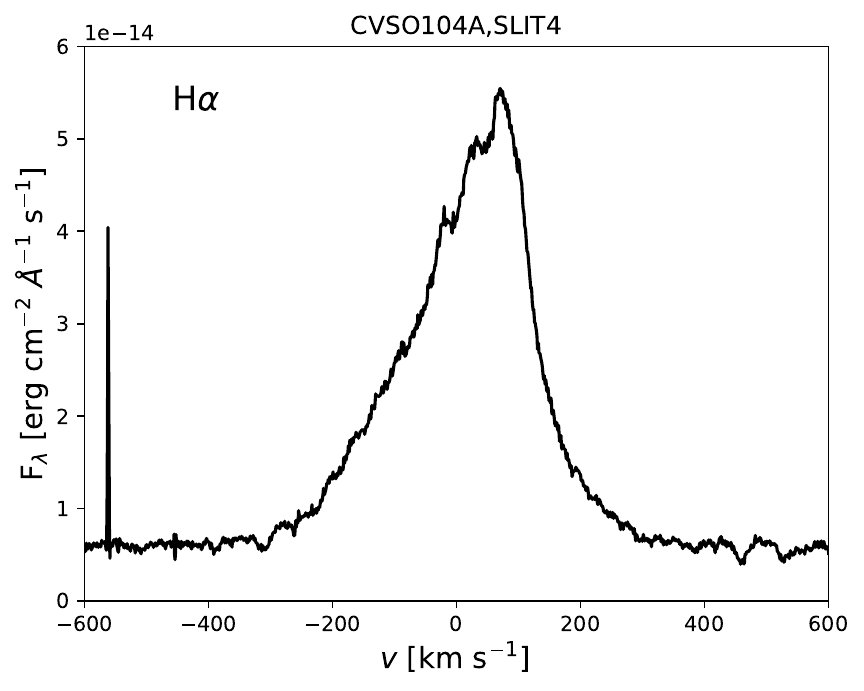}}
\hfill
\subfloat{\includegraphics[trim=0 0 0 0, clip, width=0.24 \textwidth]{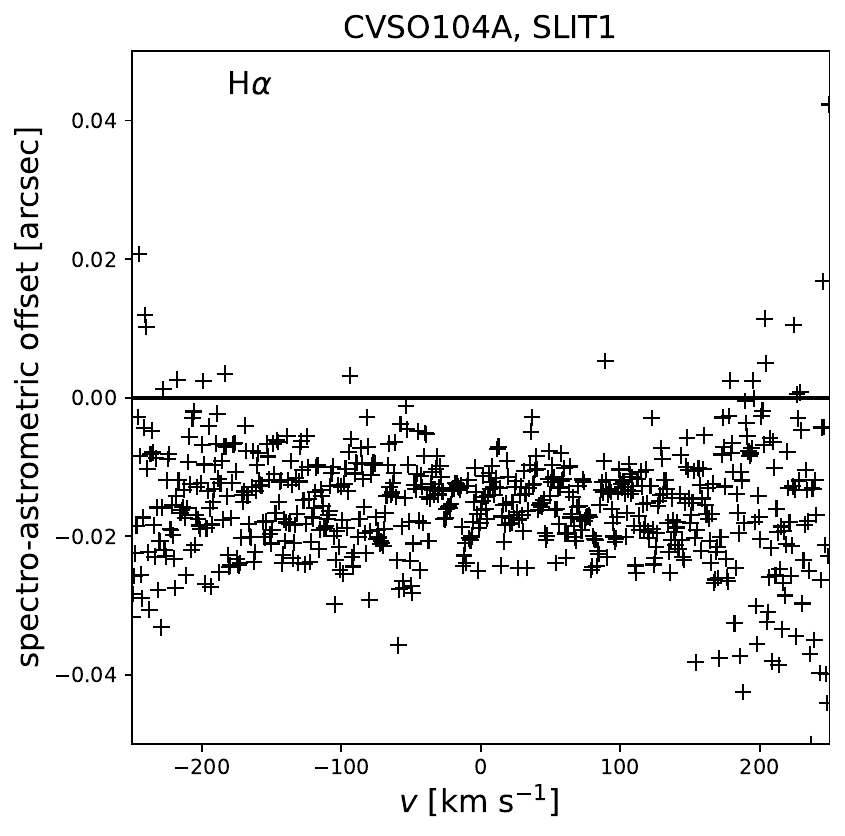}}
\hfill
\subfloat{\includegraphics[trim=0 0 0 0, clip, width=0.24 \textwidth]{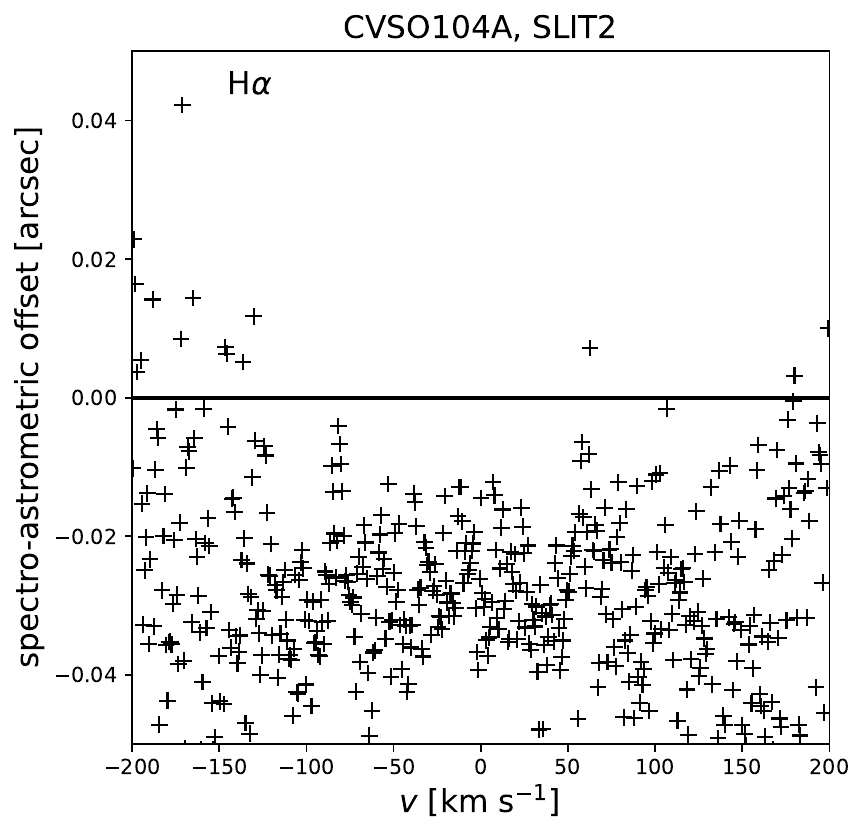}}
\hfill
\subfloat{\includegraphics[trim=0 0 0 0, clip, width=0.24 \textwidth]{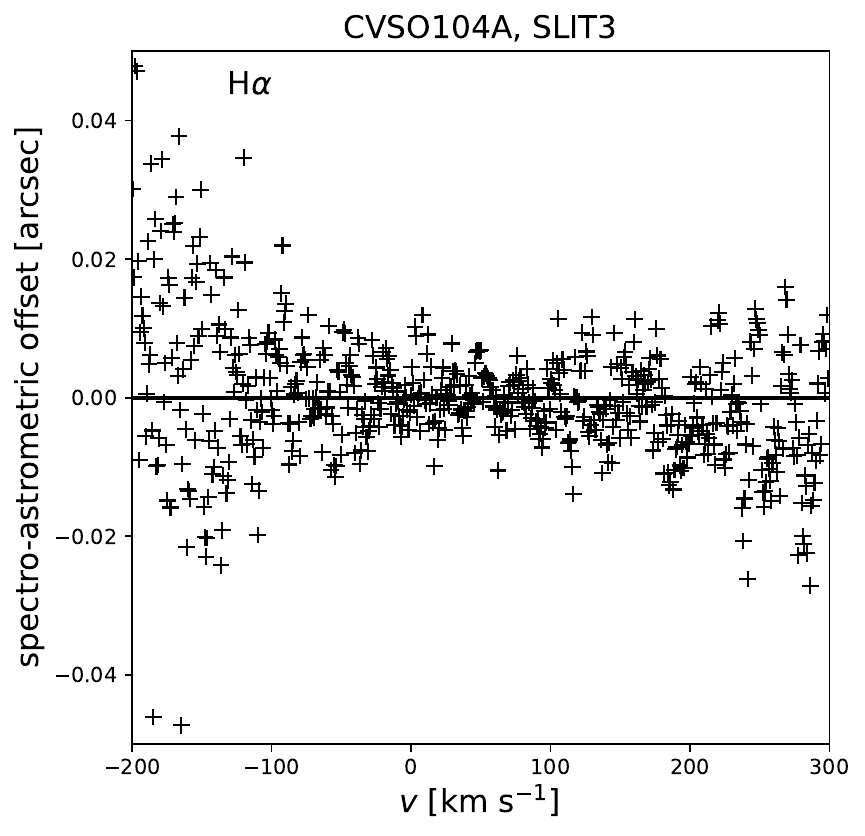}} 
\hfill
\subfloat{\includegraphics[trim=0 0 0 0, clip, width=0.24 \textwidth]{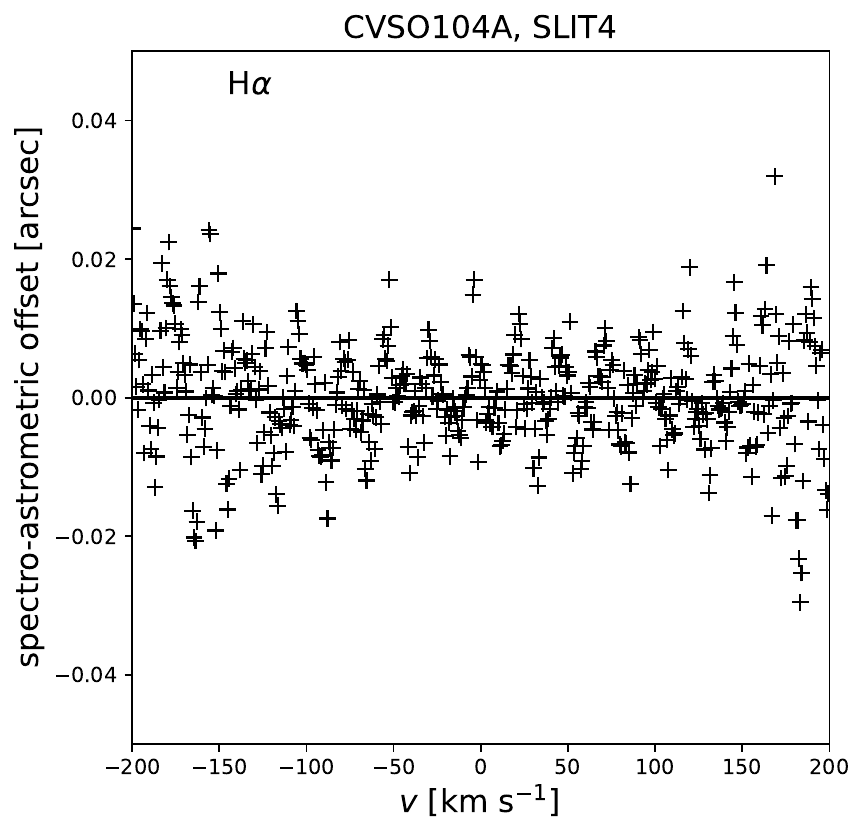}} 
\hfill
\subfloat{\includegraphics[trim=0 0 0 0, clip, width=0.24 \textwidth]{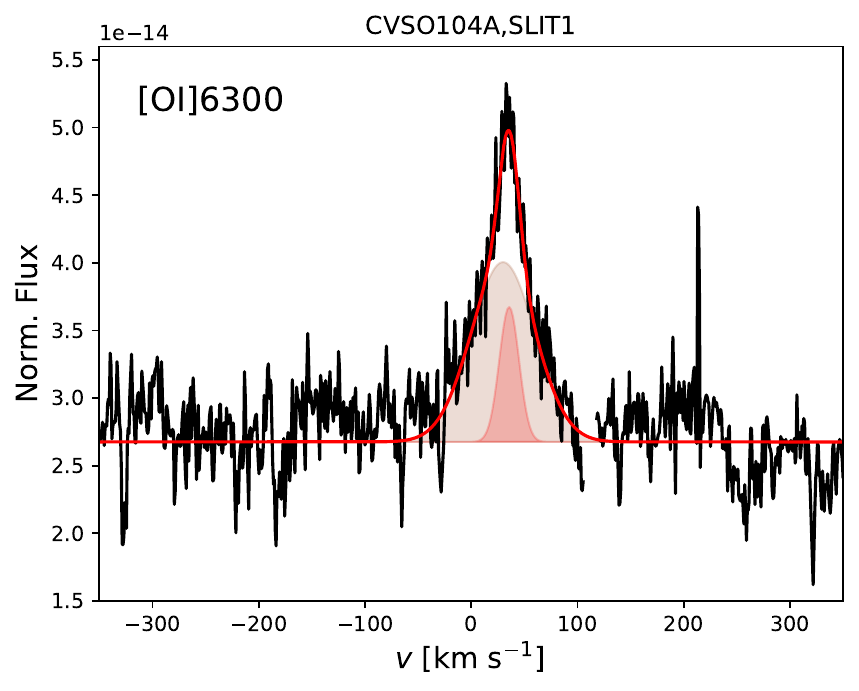}}
\hfill
\subfloat{\includegraphics[trim=0 0 0 0, clip, width=0.24 \textwidth]{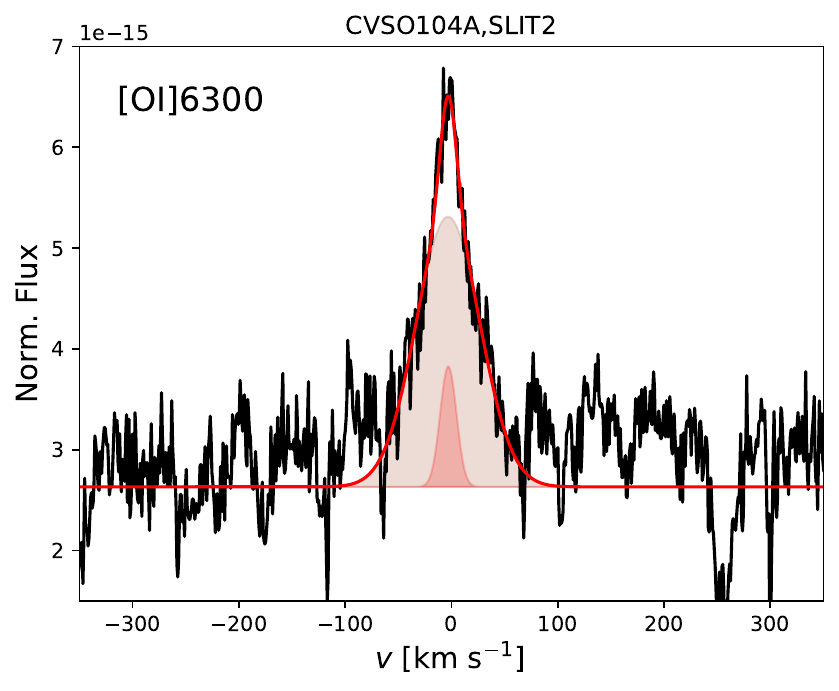}}
\hfill
\subfloat{\includegraphics[trim=0 0 0 0, clip, width=0.24 \textwidth]{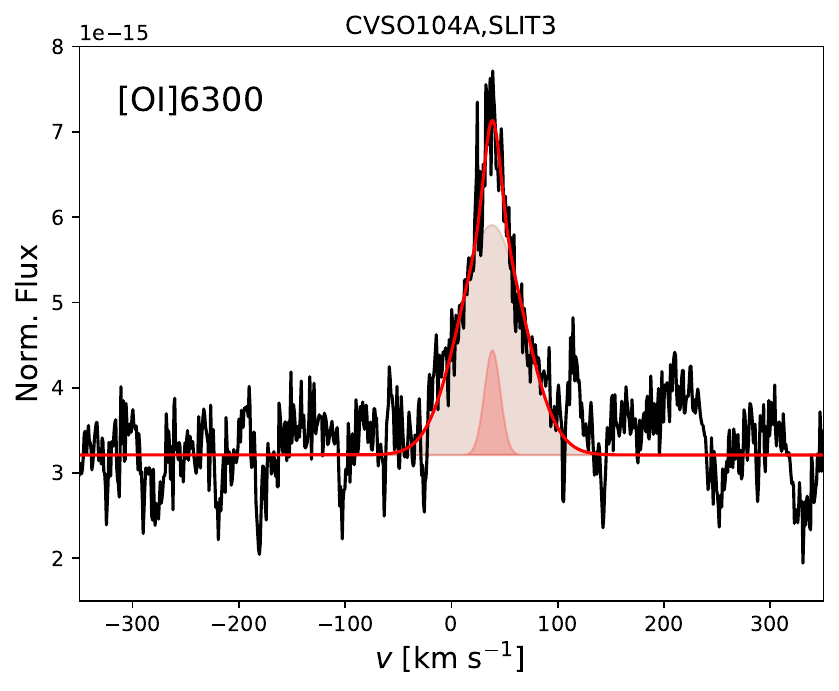}} 
\hfill   
\subfloat{\includegraphics[trim=0 0 0 0, clip, width=0.24 \textwidth]{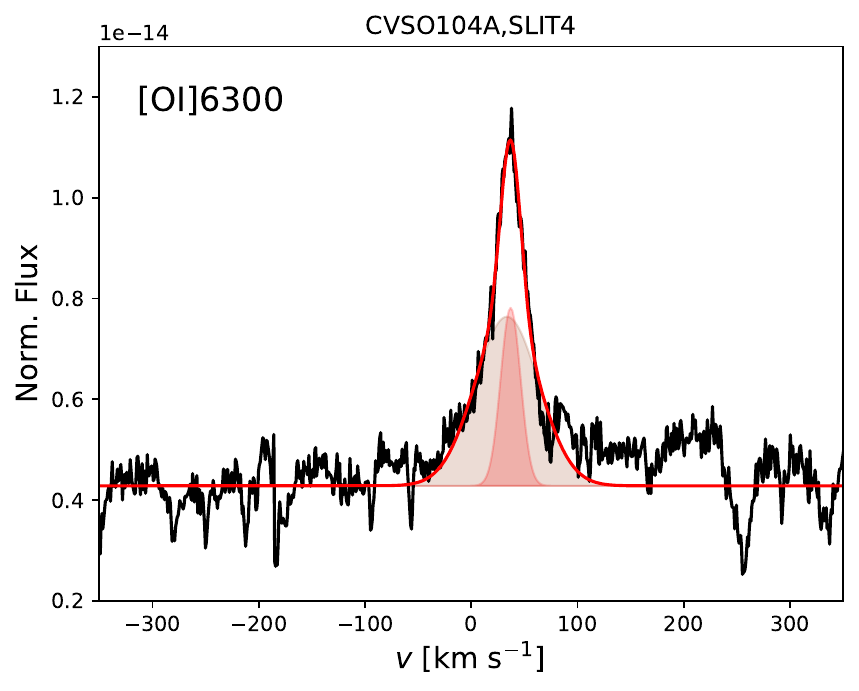}} 
\hfill   
\subfloat{\includegraphics[trim=0 0 0 0, clip, width=0.24 \textwidth]{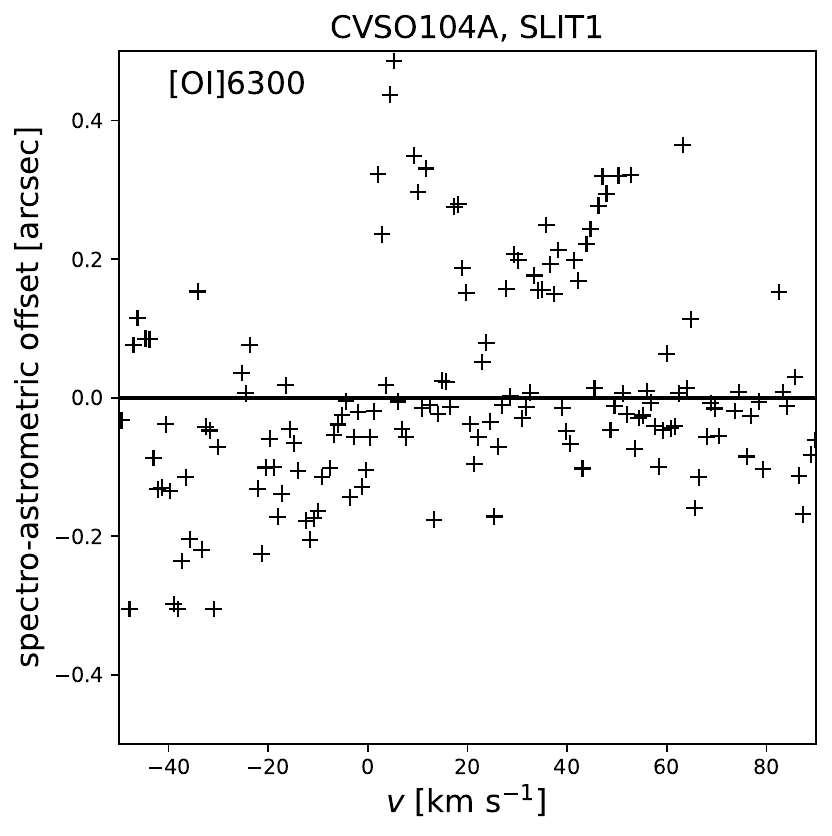}}
\hfill
\subfloat{\includegraphics[trim=0 0 0 0, clip, width=0.24 \textwidth]{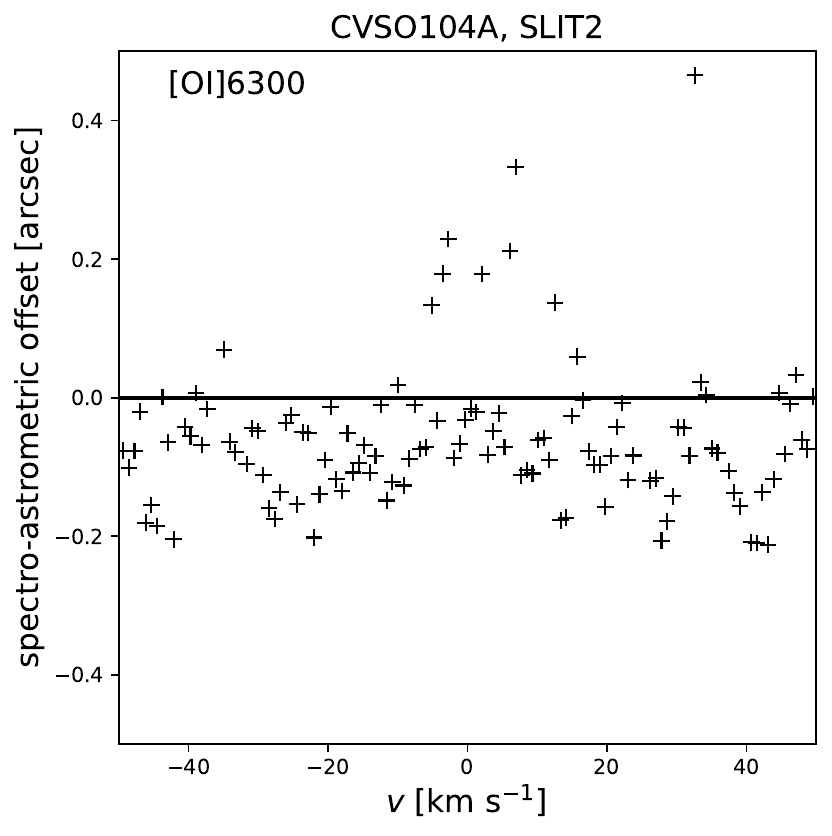}}
\hfill
\subfloat{\includegraphics[trim=0 0 0 0, clip, width=0.24 \textwidth]{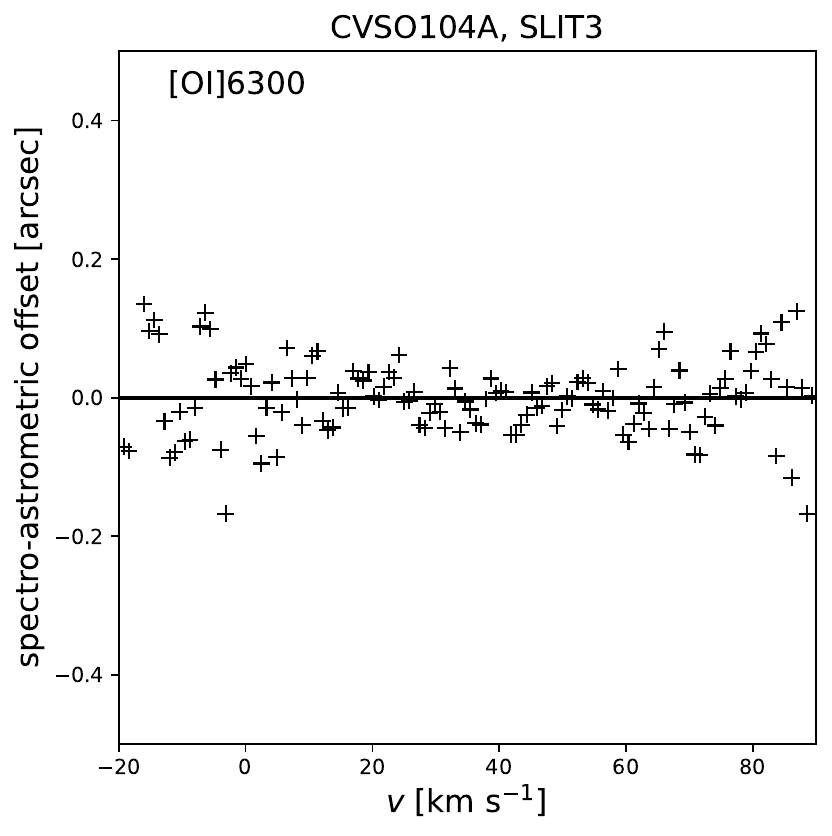}} 
\hfill
\subfloat{\includegraphics[trim=0 0 0 0, clip, width=0.24 \textwidth]{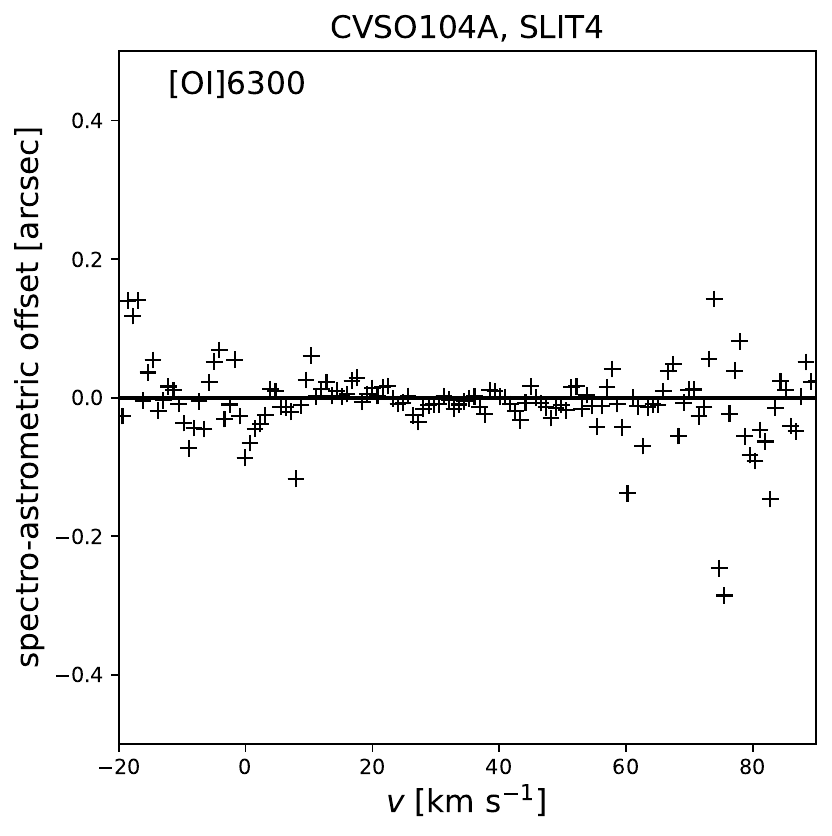}} 
\hfill
\caption{\small{Line profiles of H$\alpha$ and [OI]$\lambda$6300 for all slit positions of CVSO104\,A.}}\label{fig:all_minispectra_CVSO104A}
\end{figure*} 

\setcounter{table}{0}
\renewcommand{\thetable}{E.\arabic{table}}
  
 \setcounter{figure}{0}
\renewcommand\thefigure{\thesection E.\arabic{figure}}  

\setcounter{equation}{0}
\renewcommand{\theequation}{E.\arabic{equation}}  

\begin{figure*} 
\centering
\subfloat{\includegraphics[trim=0 0 0 0, clip, width=0.25 \textwidth]{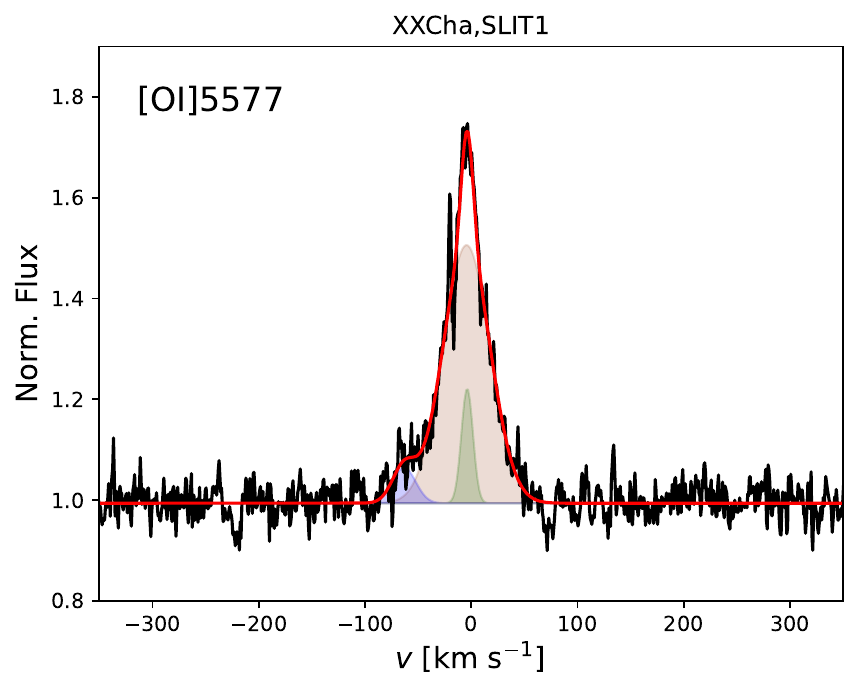}}
\hfill
\subfloat{\includegraphics[trim=0 0 0 0, clip, width=0.25 \textwidth]{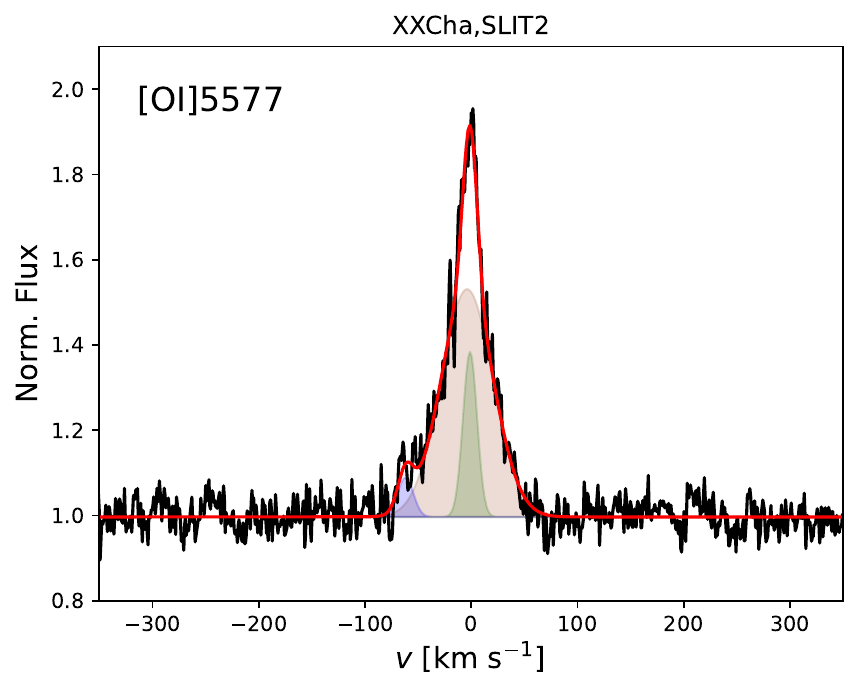}}
\hfill
\subfloat{\includegraphics[trim=0 0 0 0, clip, width=0.25 \textwidth]{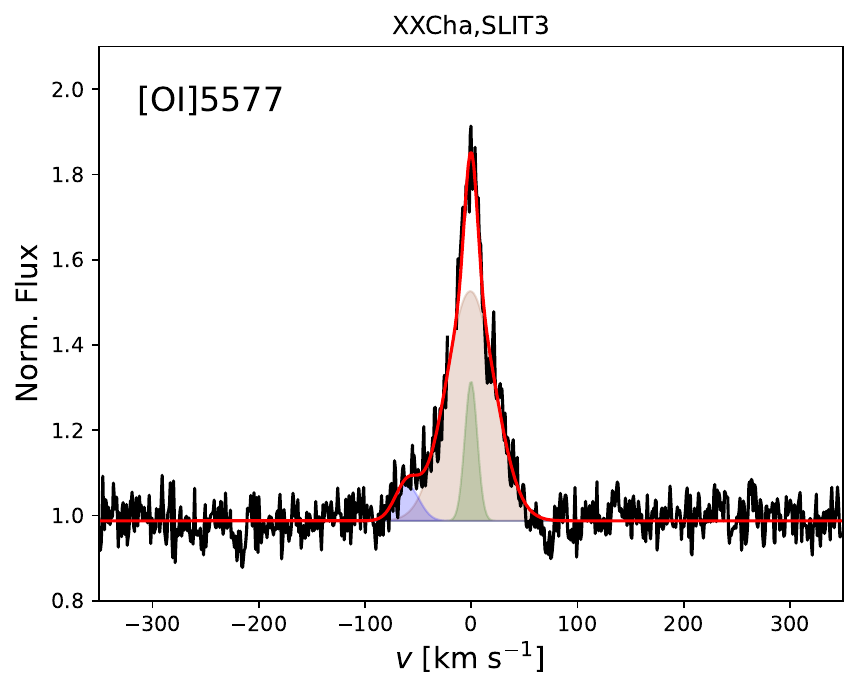}}
\hfill  \\
\subfloat{\includegraphics[trim=0 0 0 0, clip, width=0.25 \textwidth]{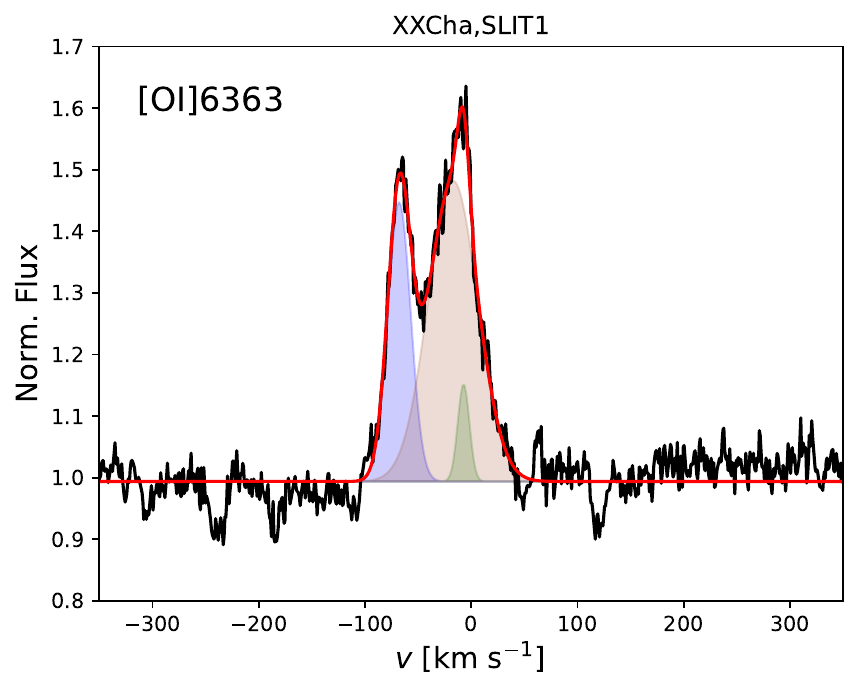}}
\hfill
\subfloat{\includegraphics[trim=0 0 0 0, clip, width=0.25 \textwidth]{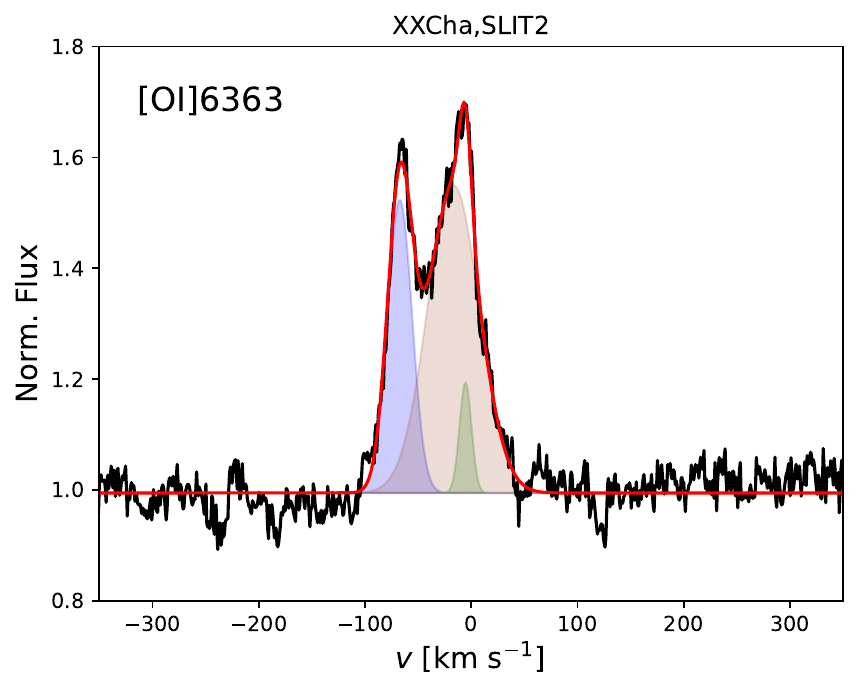}}
\hfill
\subfloat{\includegraphics[trim=0 0 0 0, clip, width=0.25 \textwidth]{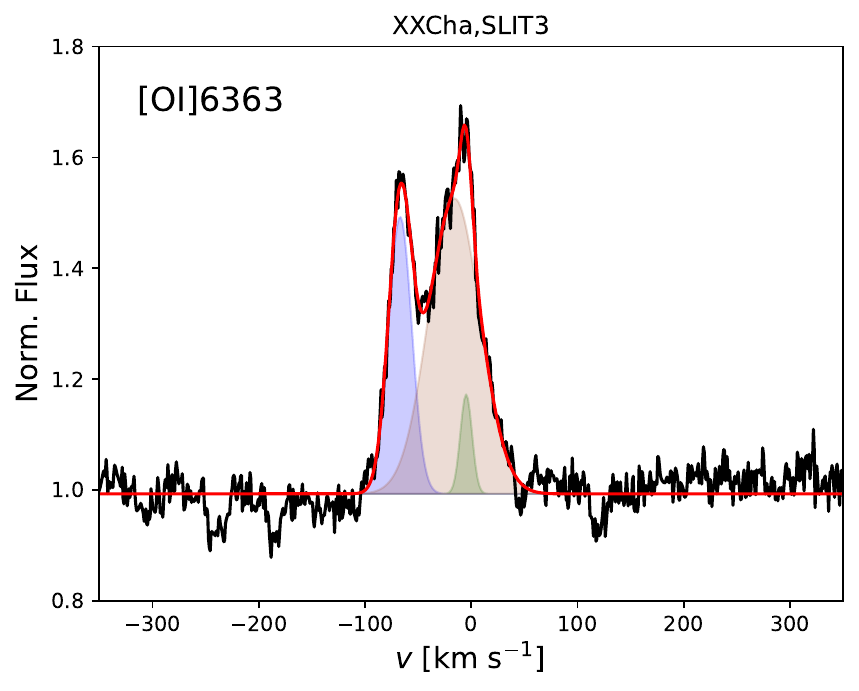}} 
\hfill \\
\subfloat{\includegraphics[trim=0 0 0 0, clip, width=0.25 \textwidth]{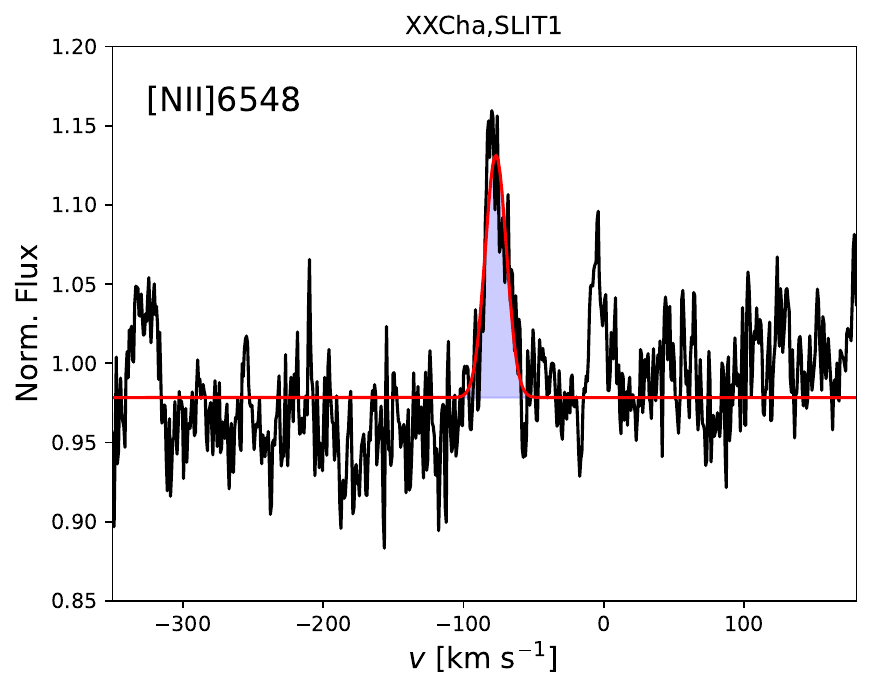}}
\hfill
\subfloat{\includegraphics[trim=0 0 0 0, clip, width=0.25 \textwidth]{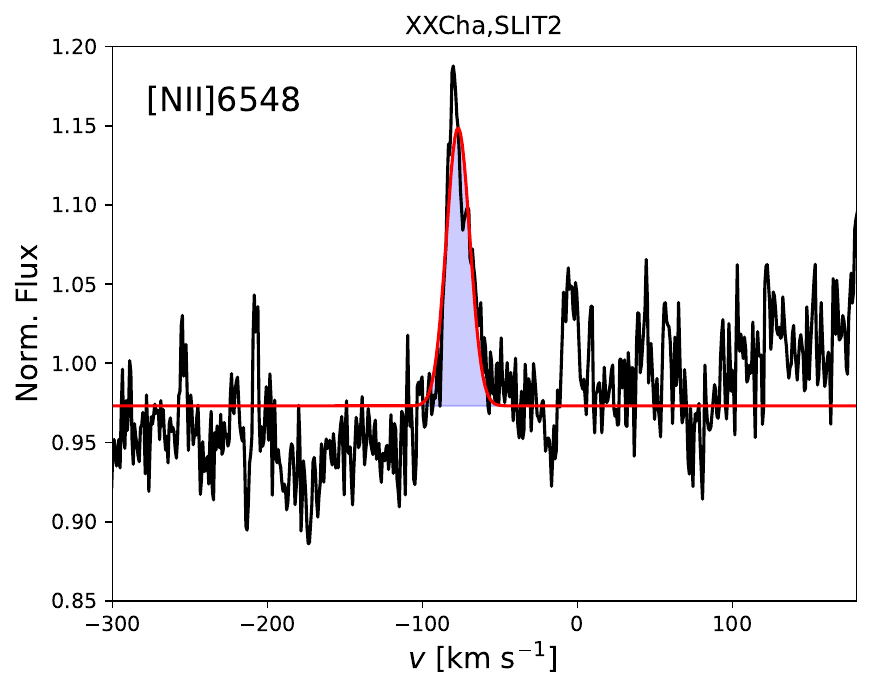}}
\hfill
\subfloat{\includegraphics[trim=0 0 0 0, clip, width=0.25 \textwidth]{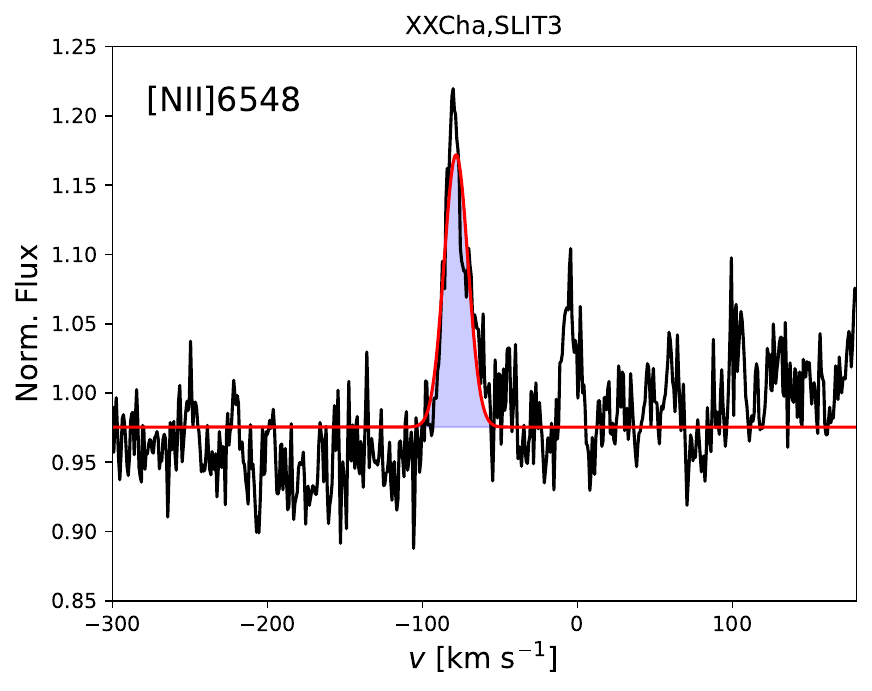}} 
\hfill  \\  
\subfloat{\includegraphics[trim=0 0 0 0, clip, width=0.25 \textwidth]{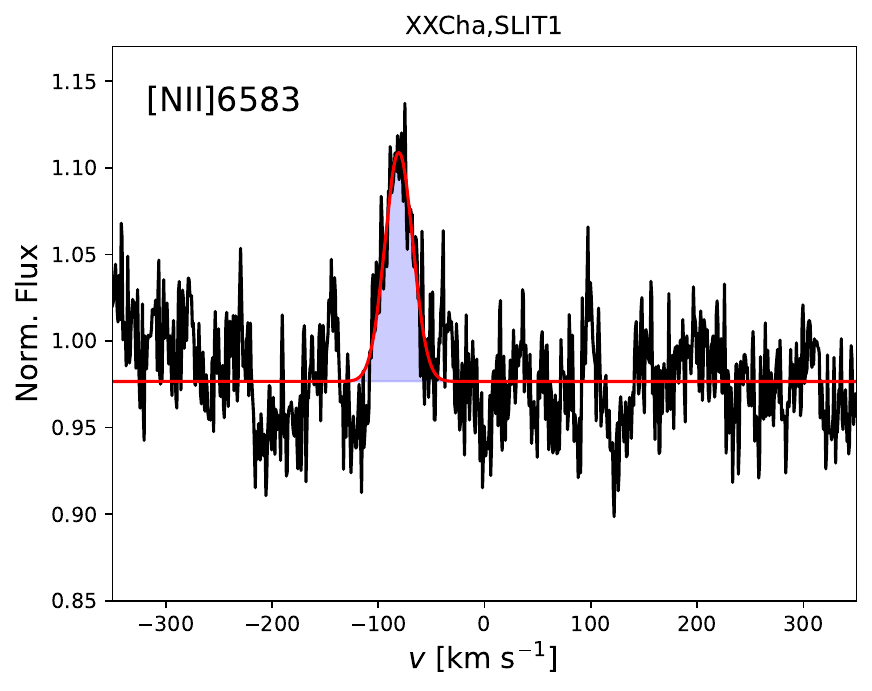}}
\hfill
\subfloat{\includegraphics[trim=0 0 0 0, clip, width=0.25 \textwidth]{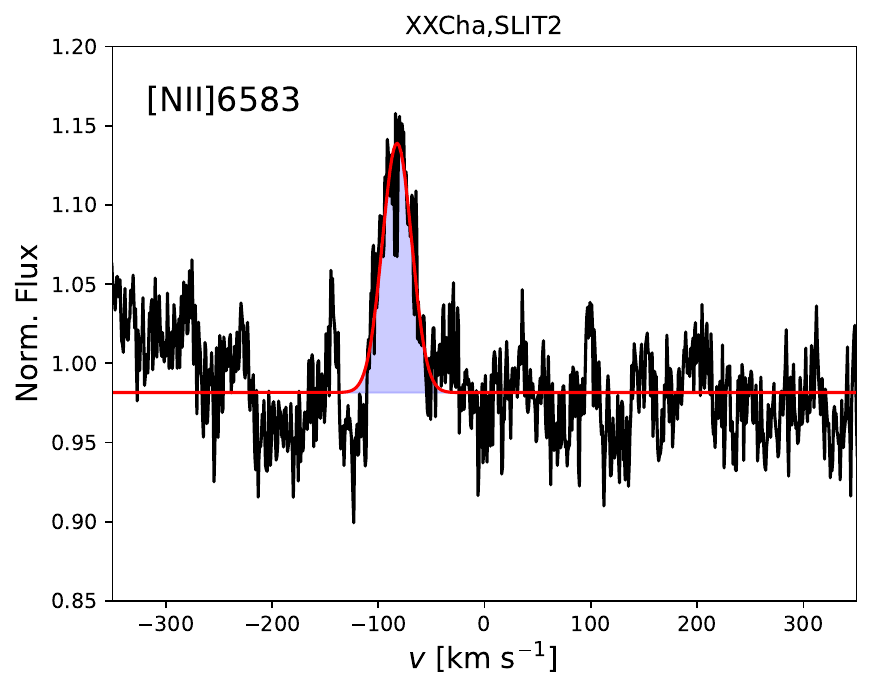}}
\hfill
\subfloat{\includegraphics[trim=0 0 0 0, clip, width=0.25 \textwidth]{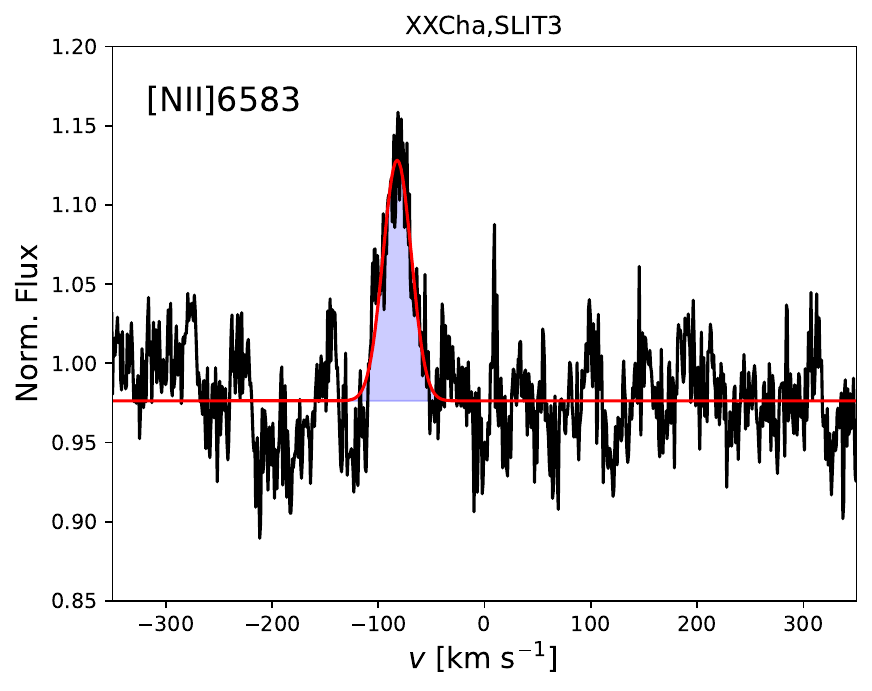}} 
\hfill \\
\subfloat{\includegraphics[trim=0 0 0 0, clip, width=0.25 \textwidth]{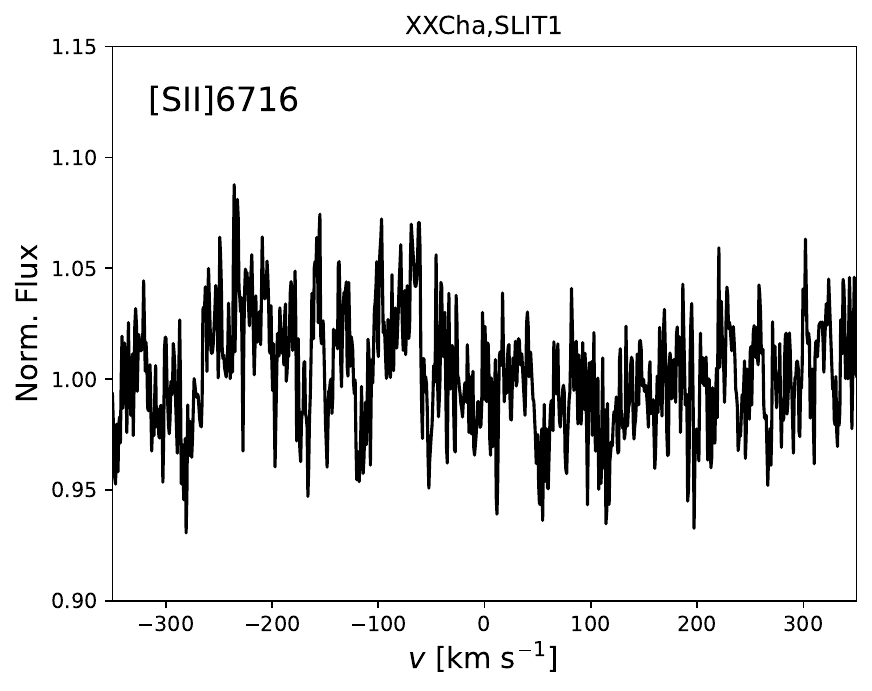}}
\hfill
\subfloat{\includegraphics[trim=0 0 0 0, clip, width=0.25 \textwidth]{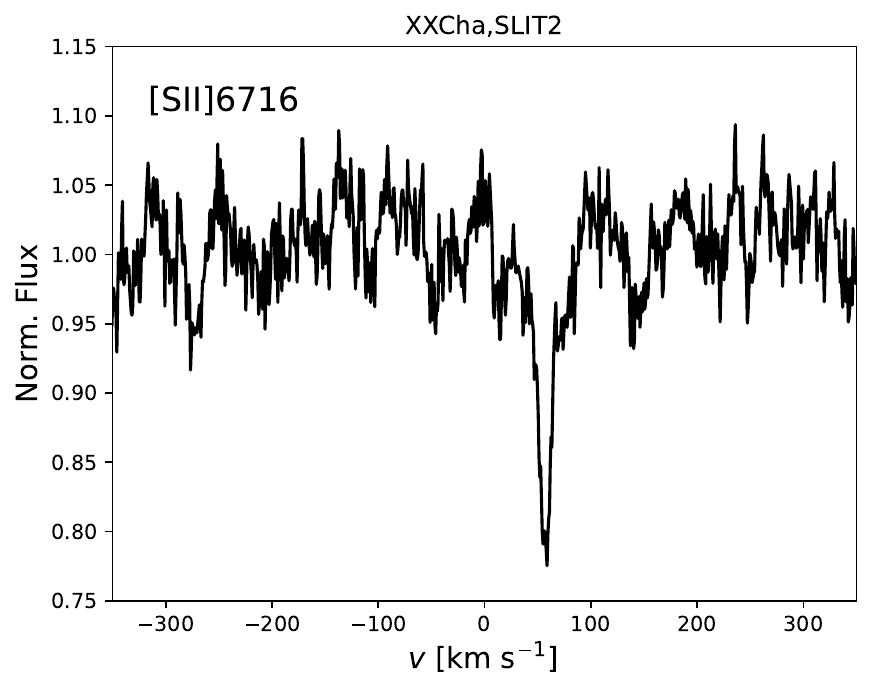}}
\hfill
\subfloat{\includegraphics[trim=0 0 0 0, clip, width=0.25 \textwidth]{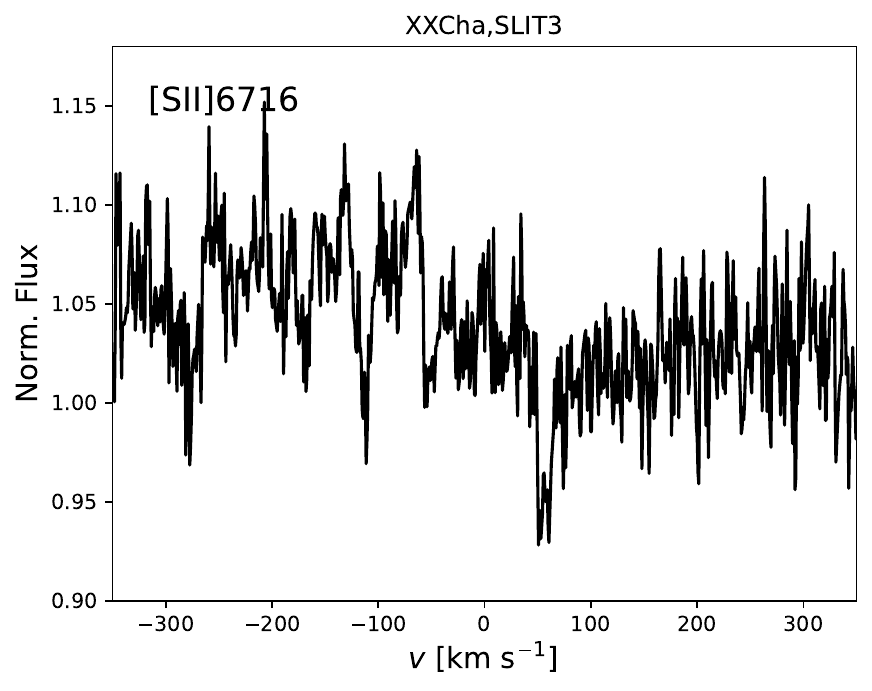}} 
\hfill \\
\subfloat{\includegraphics[trim=0 0 0 0, clip, width=0.25 \textwidth]{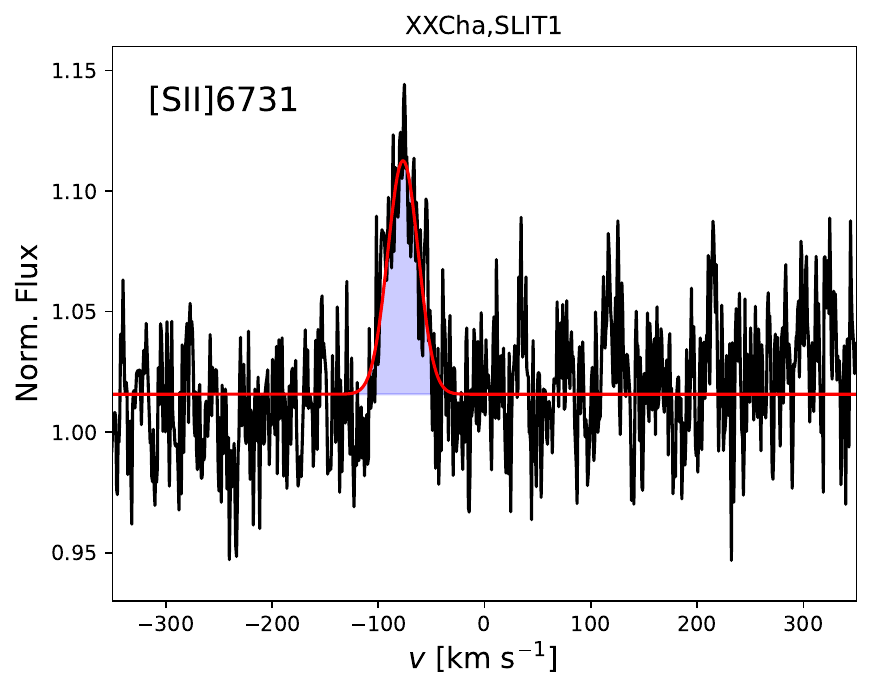}}
\hfill
\subfloat{\includegraphics[trim=0 0 0 0, clip, width=0.25 \textwidth]{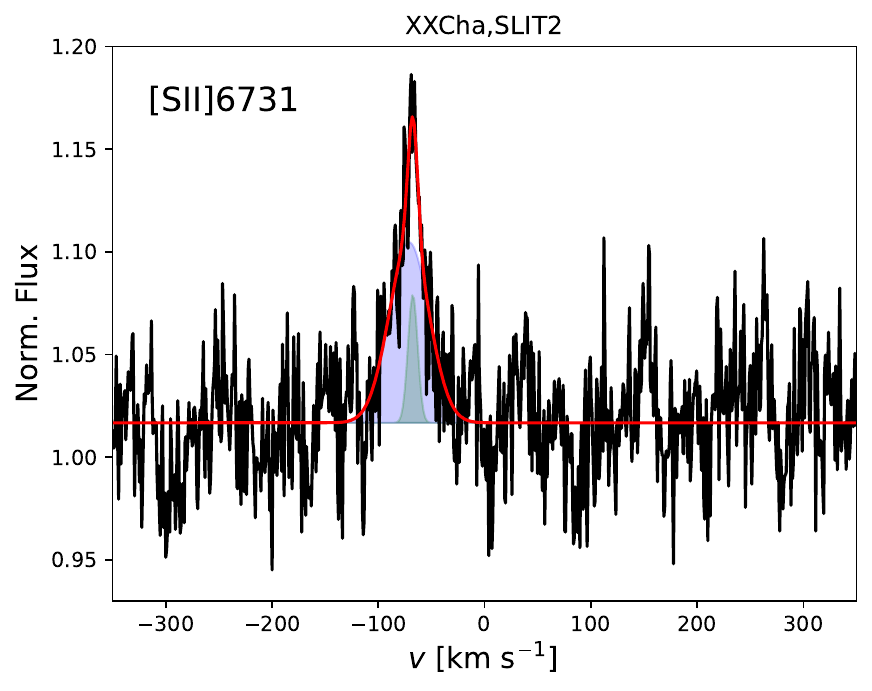}}
\hfill
\subfloat{\includegraphics[trim=0 0 0 0, clip, width=0.25 \textwidth]{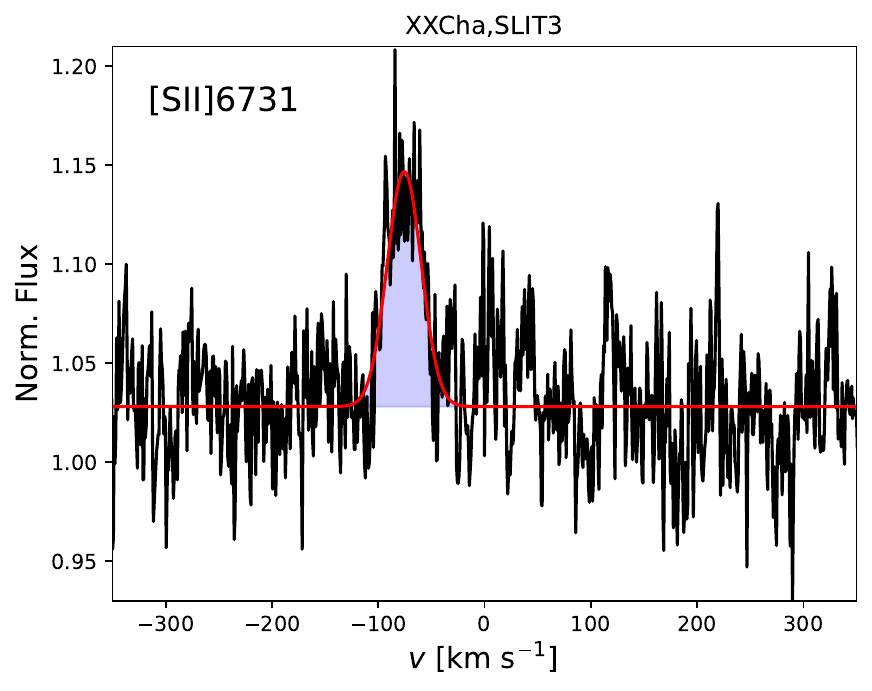}} 
\hfill
\caption{\small{Line profiles XX\,Cha.}}\label{fig:XXCha}
\end{figure*} 

\begin{figure*} 
\centering
\subfloat{\includegraphics[trim=0 0 0 0, clip, width=0.25 \textwidth]{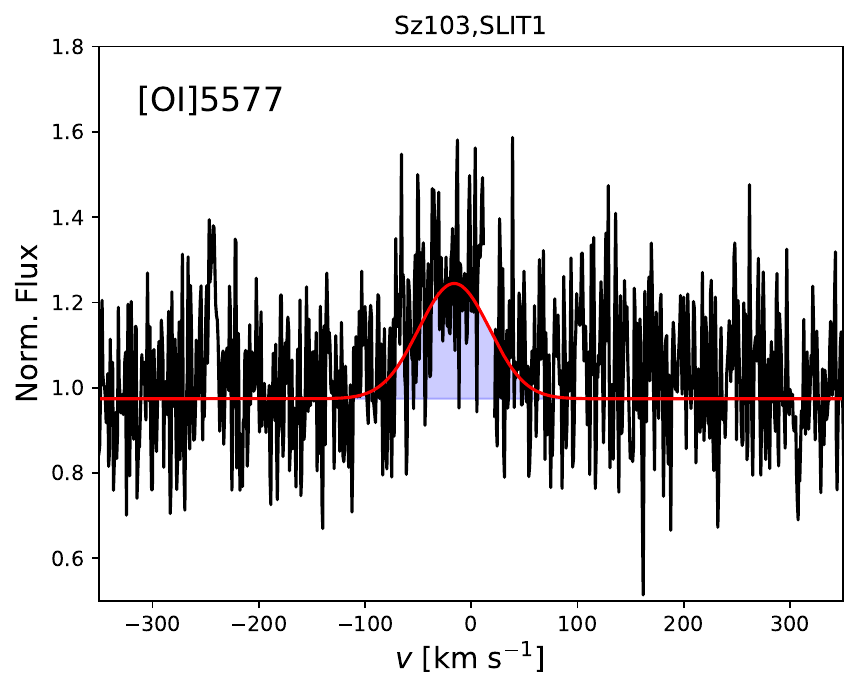}}
\hfill
\subfloat{\includegraphics[trim=0 0 0 0, clip, width=0.25 \textwidth]{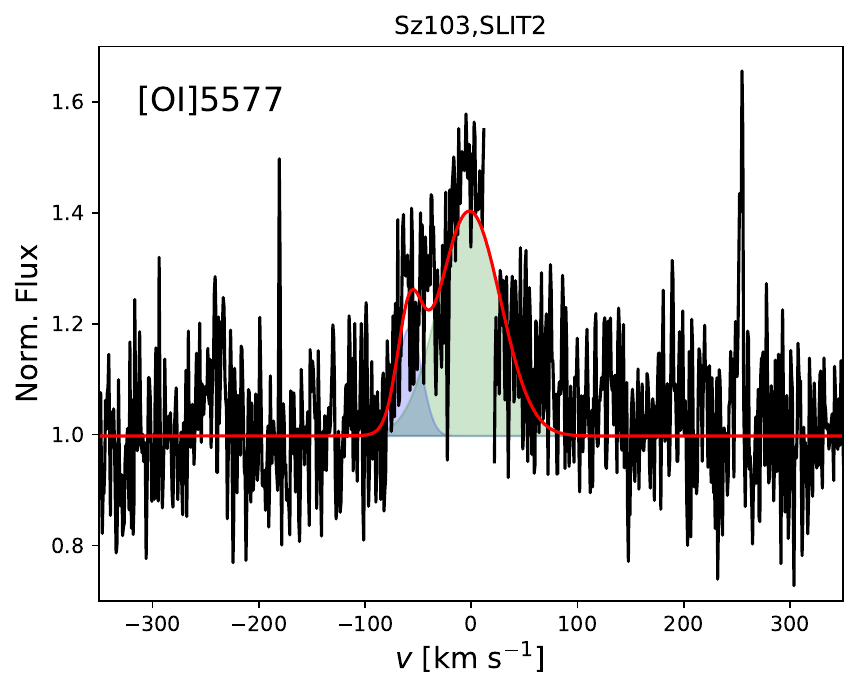}}
\hfill
\subfloat{\includegraphics[trim=0 0 0 0, clip, width=0.25 \textwidth]{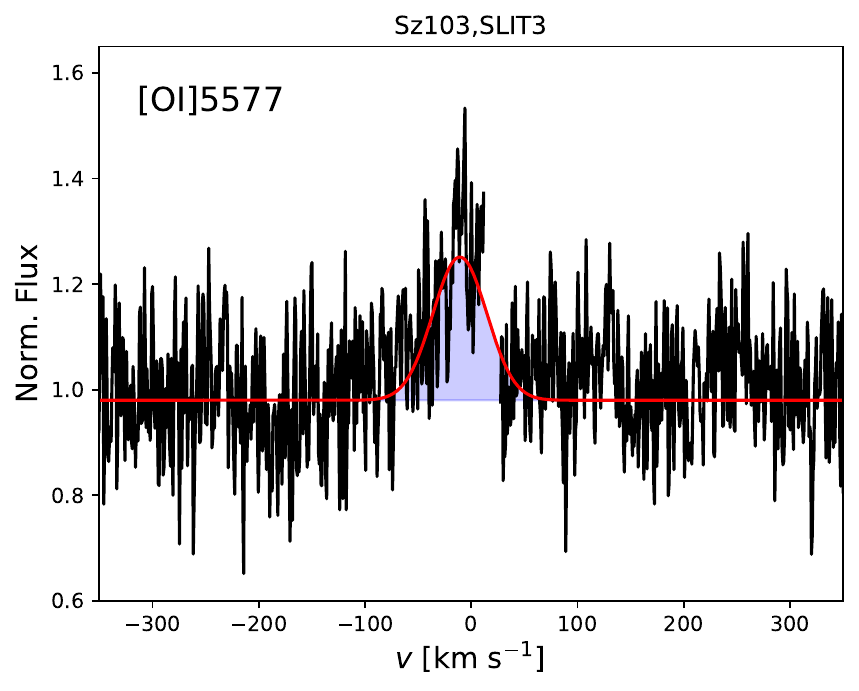}}
\hfill   \\
\subfloat{\includegraphics[trim=0 0 0 0, clip, width=0.25 \textwidth]{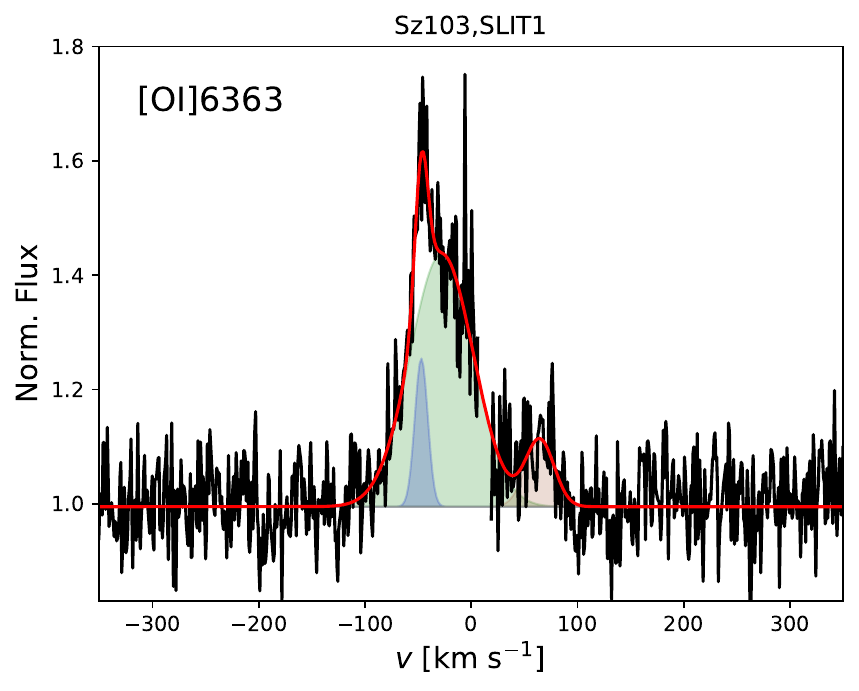}}
\hfill
\subfloat{\includegraphics[trim=0 0 0 0, clip, width=0.25 \textwidth]{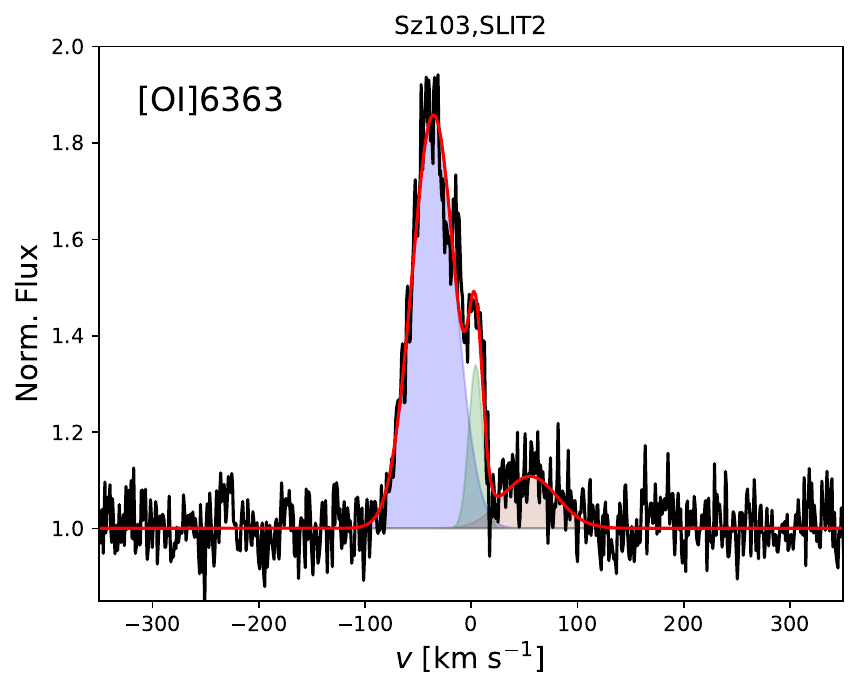}}
\hfill
\subfloat{\includegraphics[trim=0 0 0 0, clip, width=0.25 \textwidth]{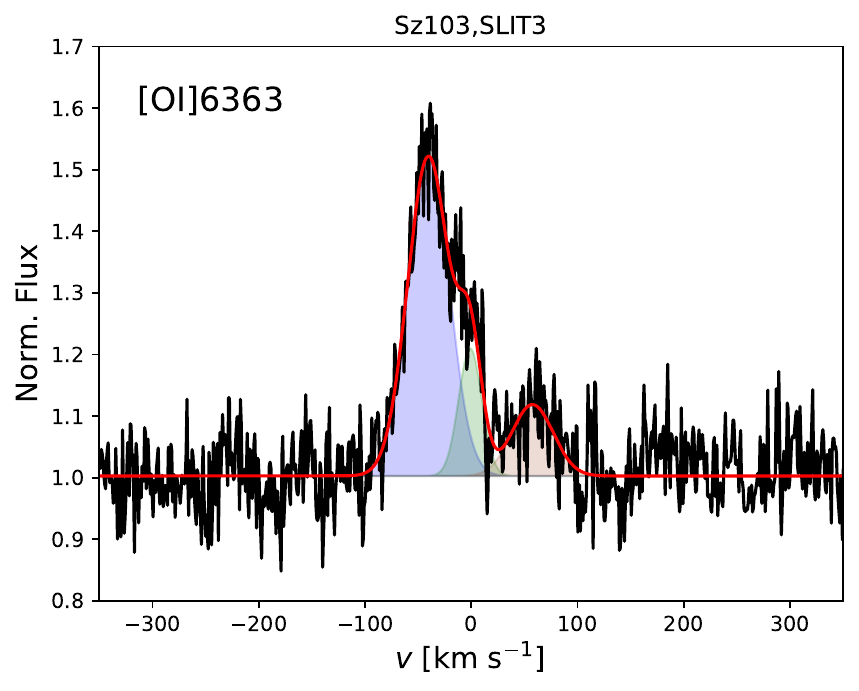}} 
\hfill \\
\subfloat{\includegraphics[trim=0 0 0 0, clip, width=0.25 \textwidth]{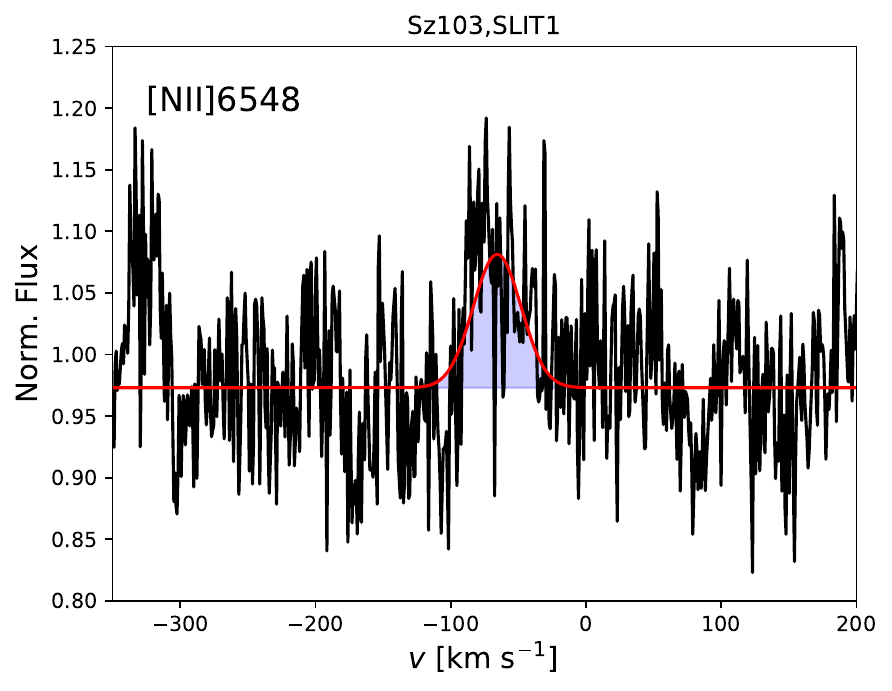}}
\hfill
\subfloat{\includegraphics[trim=0 0 0 0, clip, width=0.25 \textwidth]{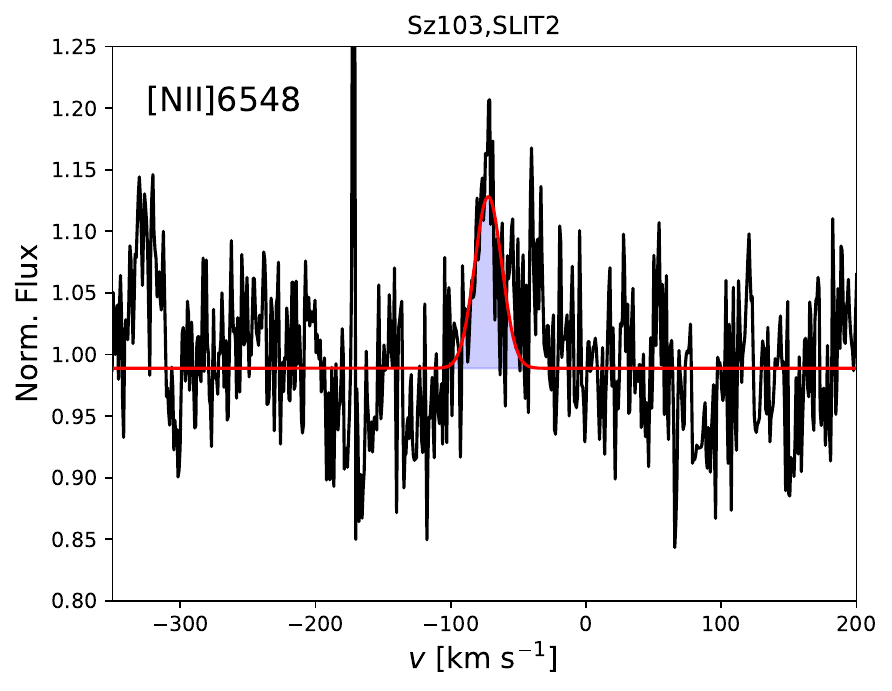}}
\hfill
\subfloat{\includegraphics[trim=0 0 0 0, clip, width=0.25 \textwidth]{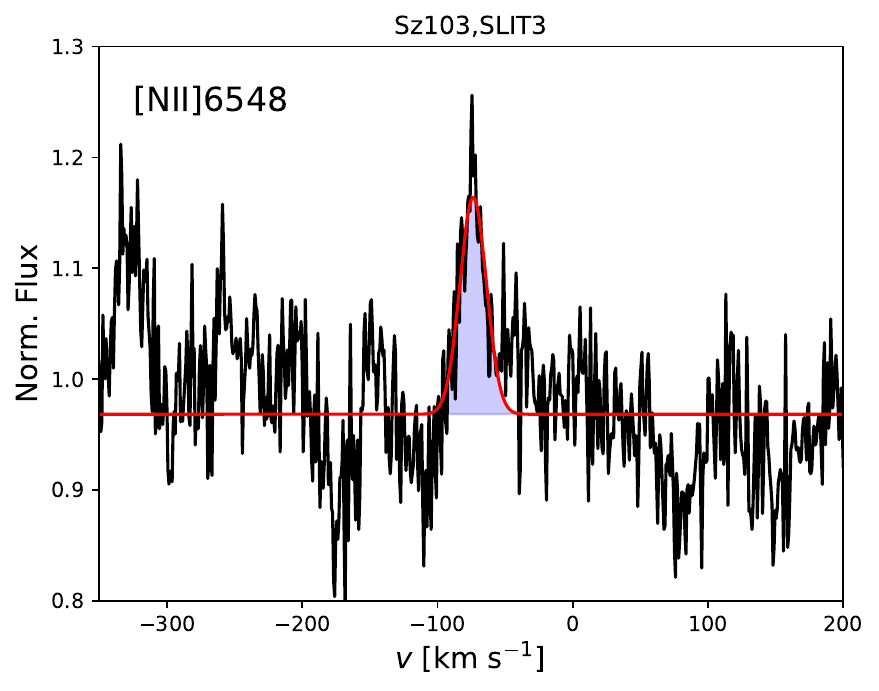}} 
\hfill    \\
\subfloat{\includegraphics[trim=0 0 0 0, clip, width=0.25 \textwidth]{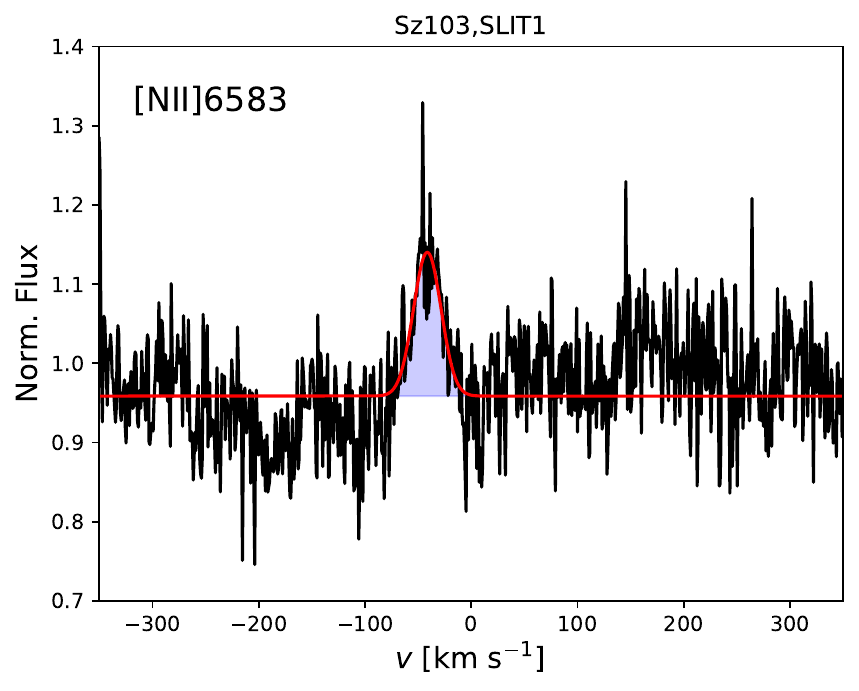}}
\hfill
\subfloat{\includegraphics[trim=0 0 0 0, clip, width=0.25 \textwidth]{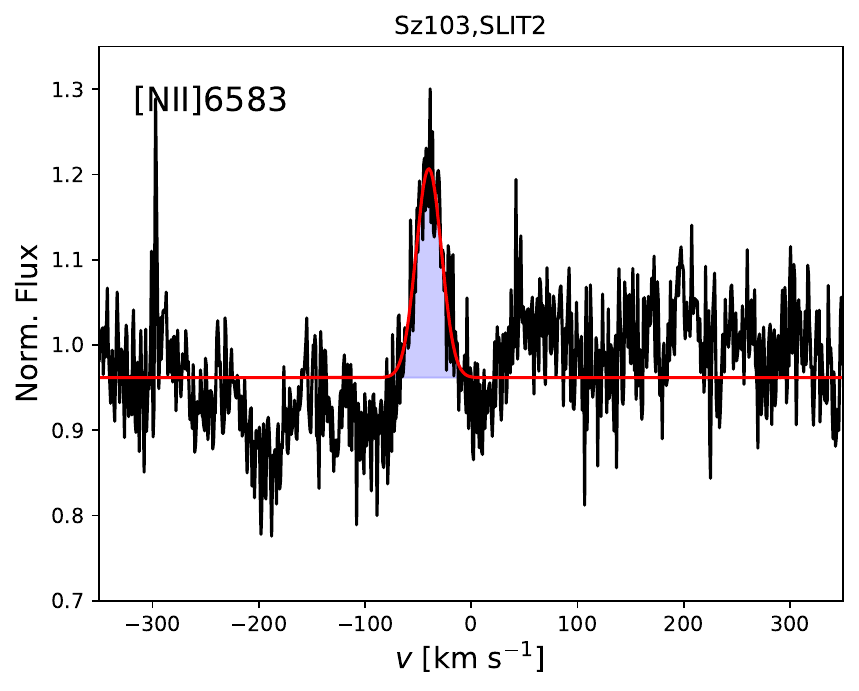}}
\hfill
\subfloat{\includegraphics[trim=0 0 0 0, clip, width=0.25 \textwidth]{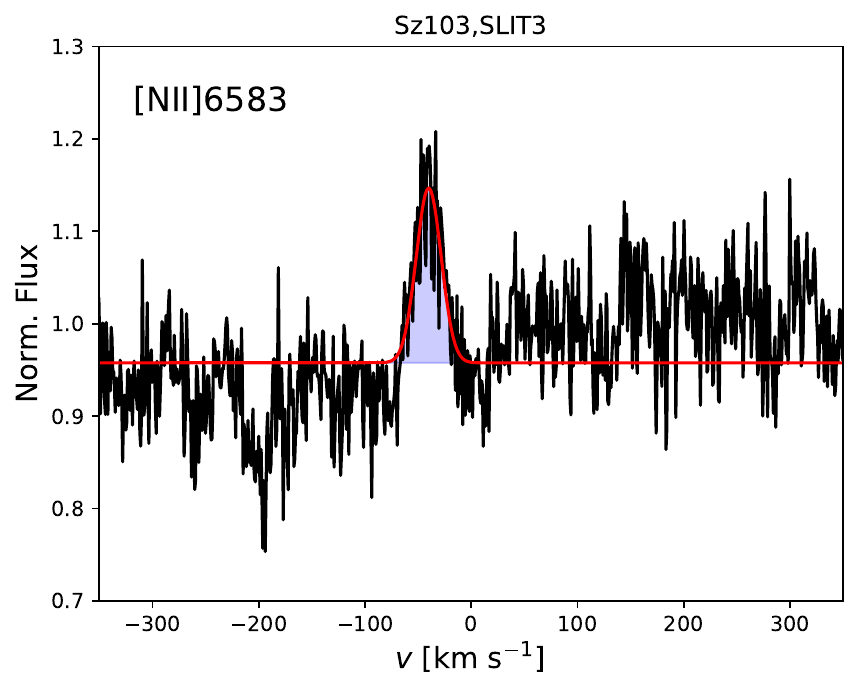}} 
\hfill \\ 
\subfloat{\includegraphics[trim=0 0 0 0, clip, width=0.25 \textwidth]{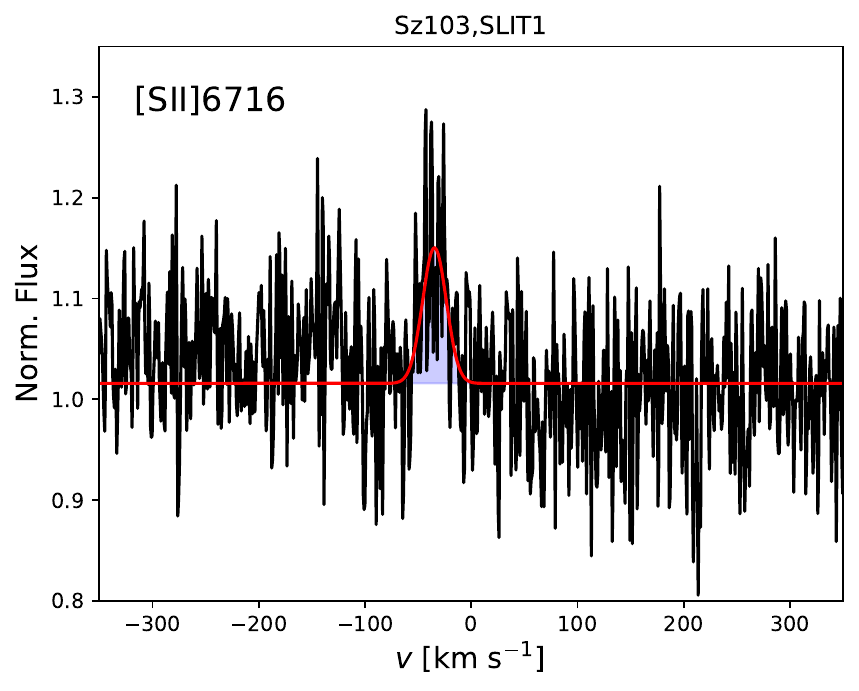}}
\hfill
\subfloat{\includegraphics[trim=0 0 0 0, clip, width=0.25 \textwidth]{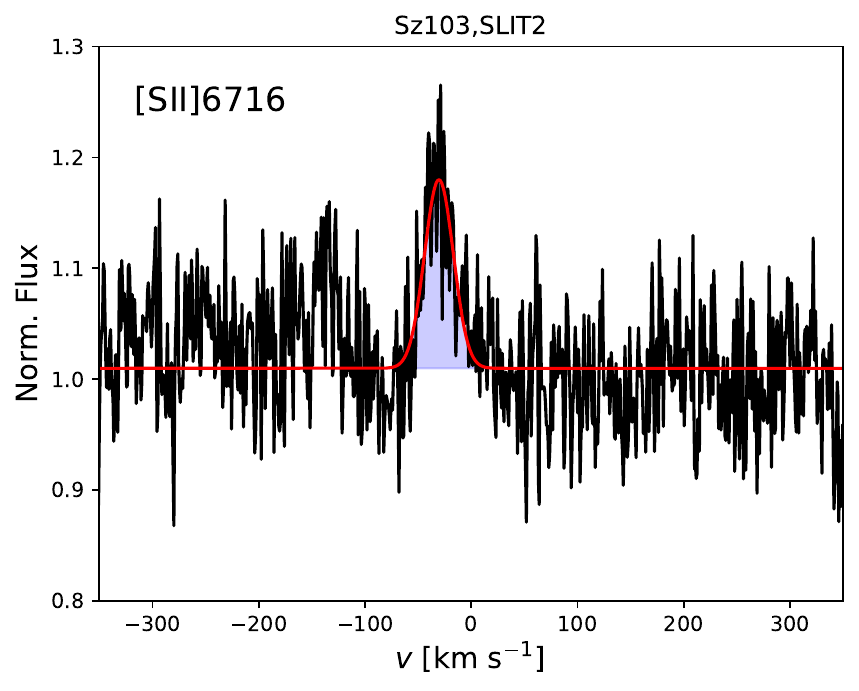}}
\hfill
\subfloat{\includegraphics[trim=0 0 0 0, clip, width=0.25 \textwidth]{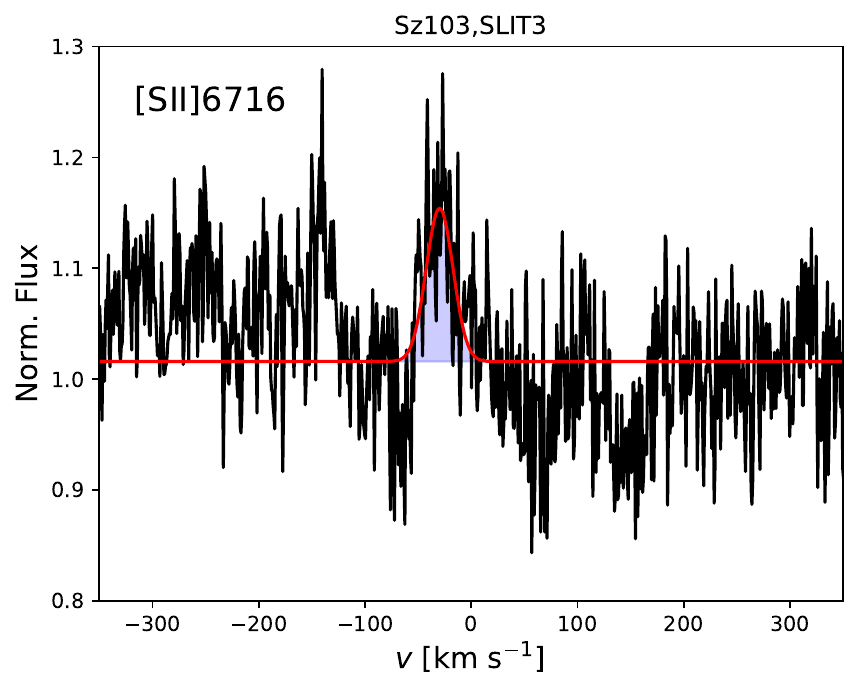}} 
\hfill \\
\subfloat{\includegraphics[trim=0 0 0 0, clip, width=0.25 \textwidth]{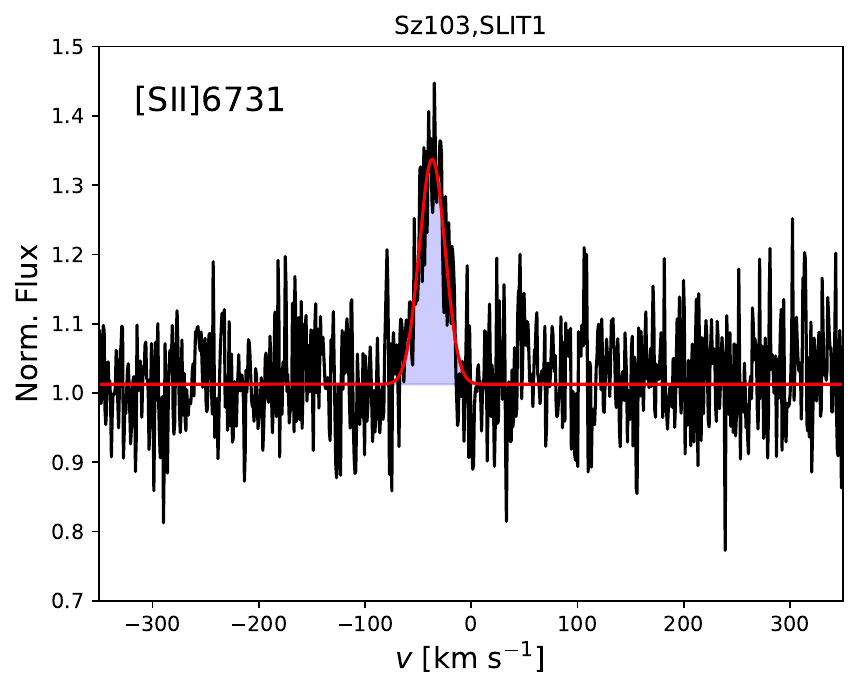}}
\hfill
\subfloat{\includegraphics[trim=0 0 0 0, clip, width=0.25 \textwidth]{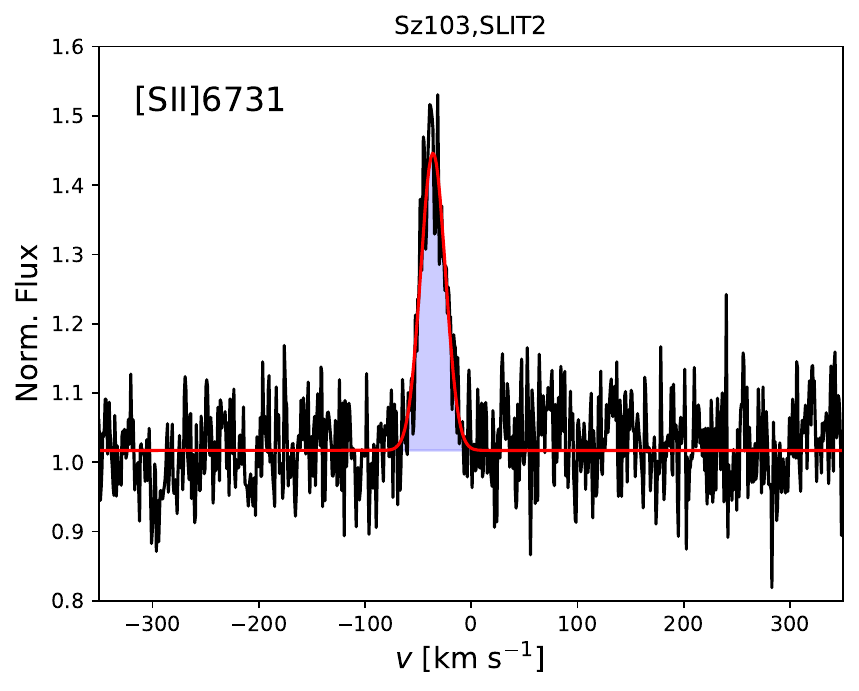}}
\hfill
\subfloat{\includegraphics[trim=0 0 0 0, clip, width=0.25 \textwidth]{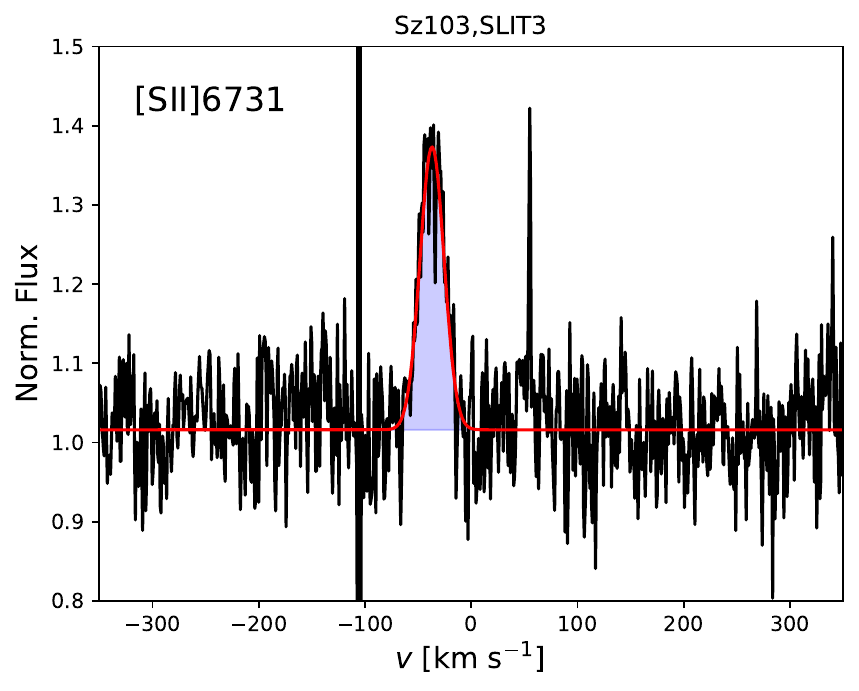}} 
\hfill
\caption{\small{Line profiles Sz\,103.}}\label{fig:Sz103}
\end{figure*} 

\begin{figure*} 
\centering
\subfloat{\includegraphics[trim=0 0 0 0, clip, width=0.25 \textwidth]{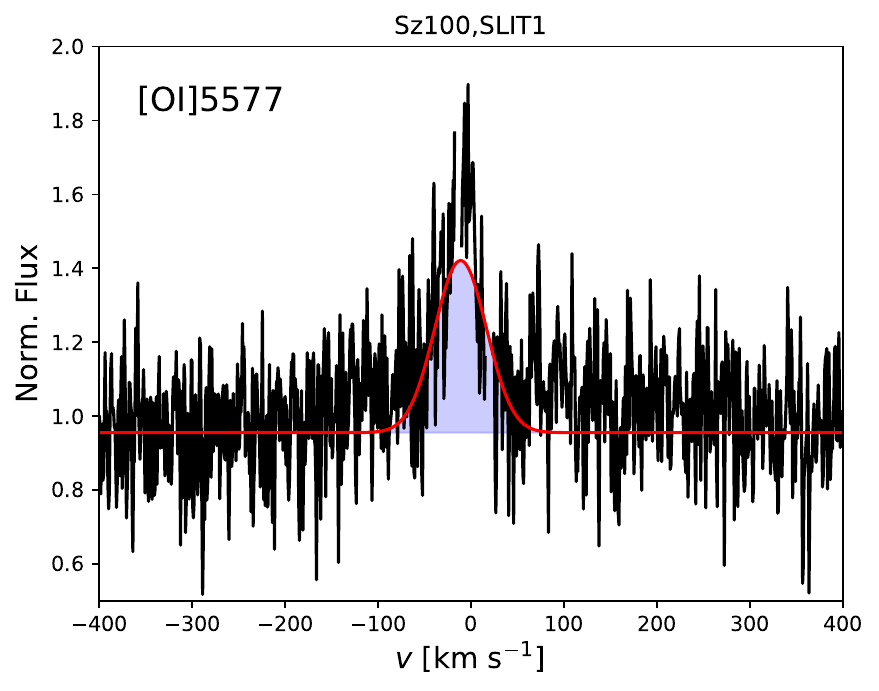}}
\hfill
\subfloat{\includegraphics[trim=0 0 0 0, clip, width=0.25 \textwidth]{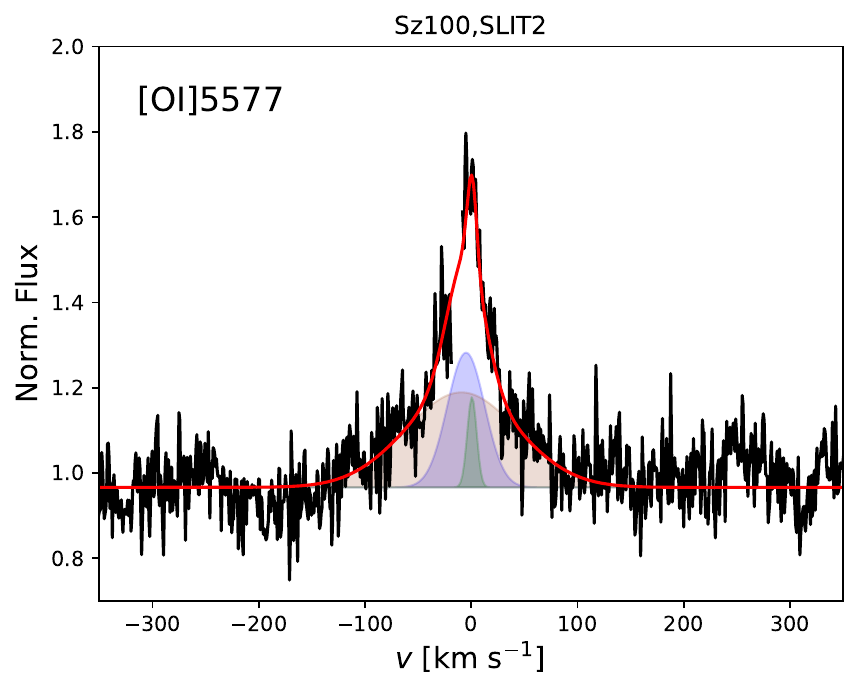}}
\hfill
\subfloat{\includegraphics[trim=0 0 0 0, clip, width=0.25 \textwidth]{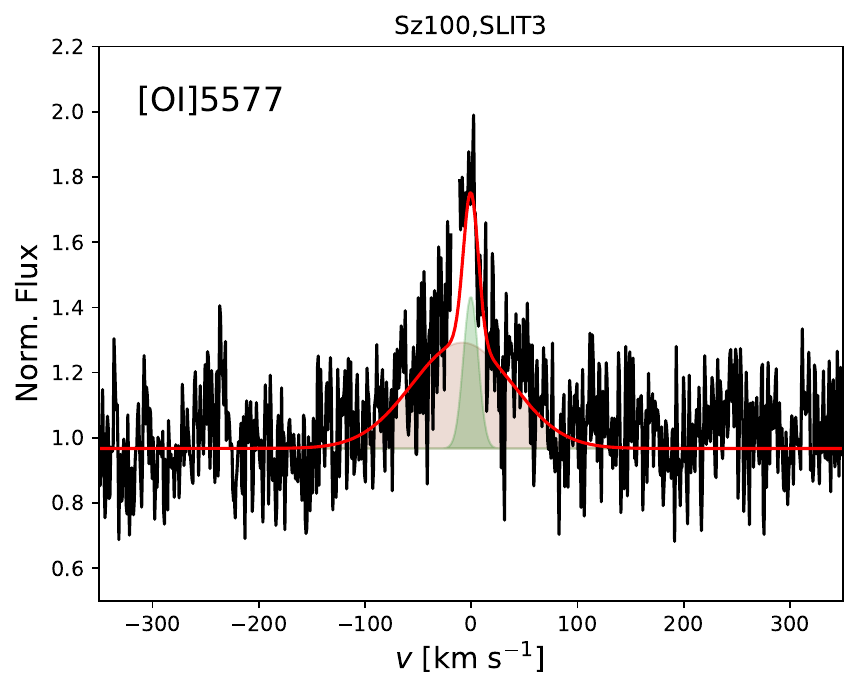}}
\hfill  \\
\subfloat{\includegraphics[trim=0 0 0 0, clip, width=0.25 \textwidth]{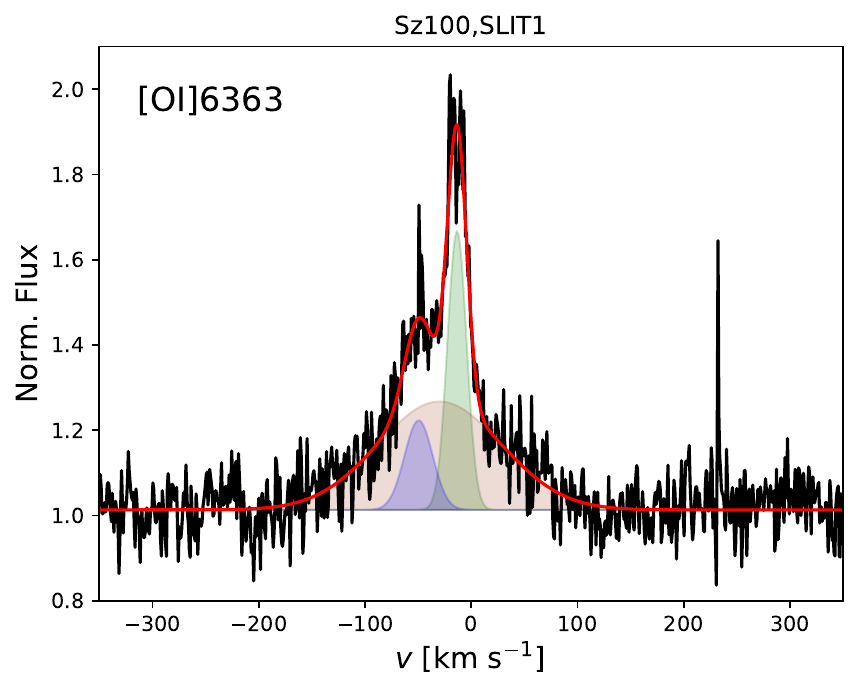}}
\hfill
\subfloat{\includegraphics[trim=0 0 0 0, clip, width=0.25 \textwidth]{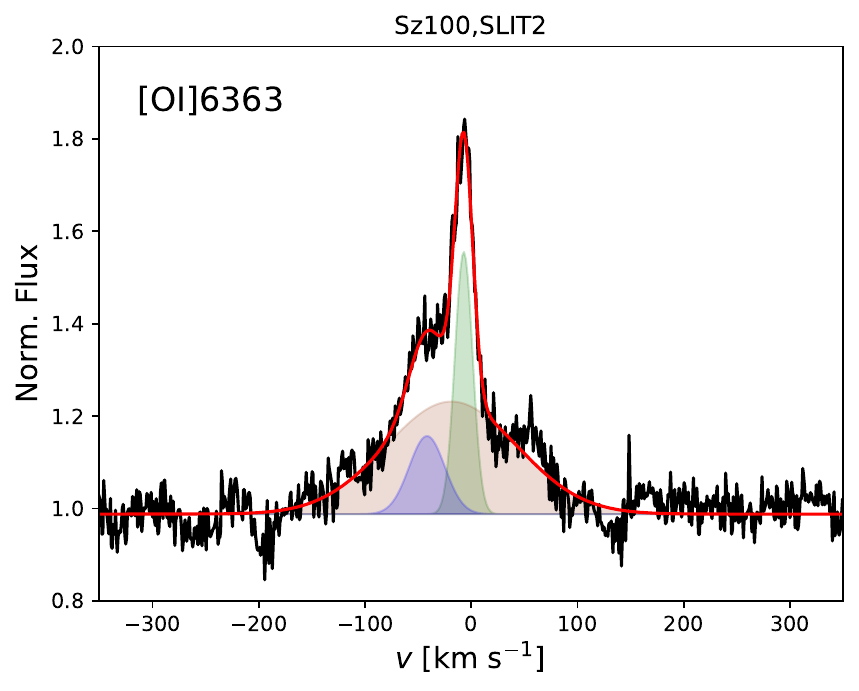}}
\hfill
\subfloat{\includegraphics[trim=0 0 0 0, clip, width=0.25 \textwidth]{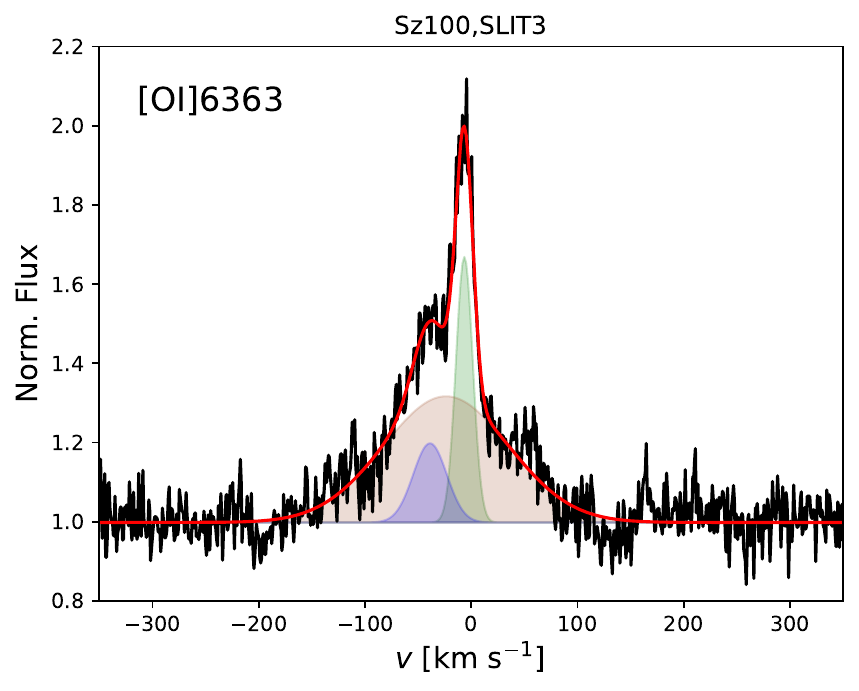}} 
\hfill \\
\subfloat{\includegraphics[trim=0 0 0 0, clip, width=0.25 \textwidth]{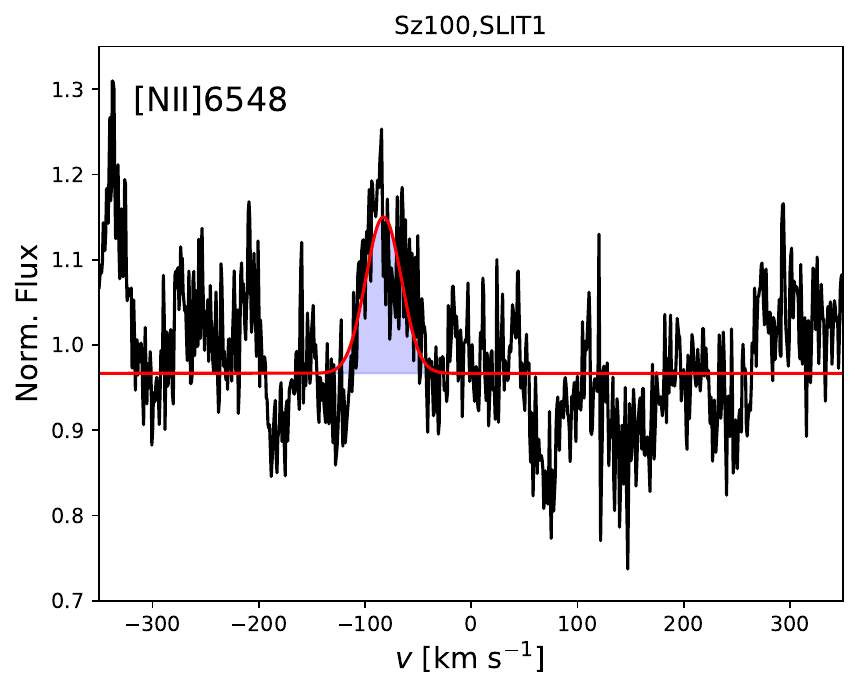}}
\hfill
\subfloat{\includegraphics[trim=0 0 0 0, clip, width=0.25 \textwidth]{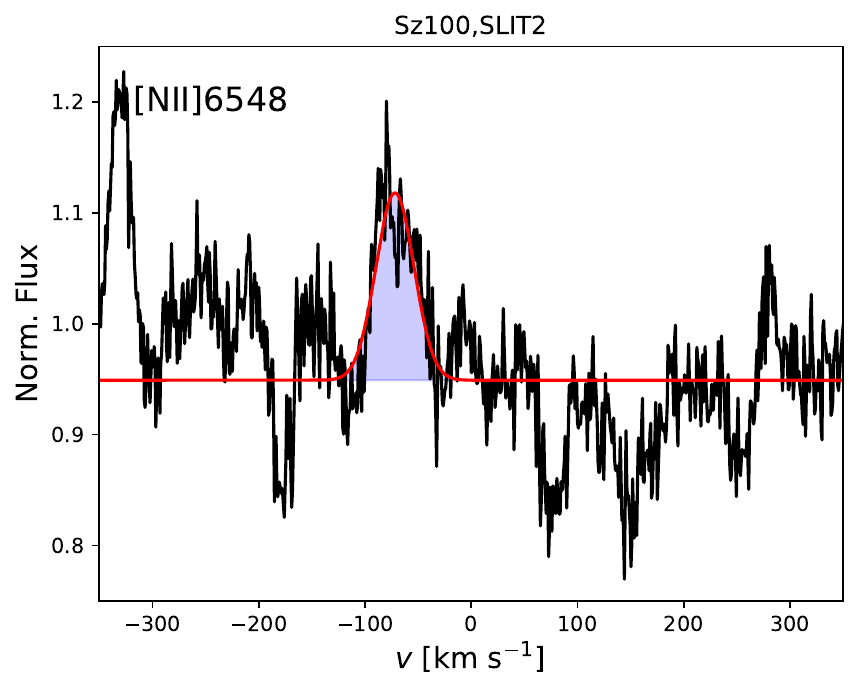}}
\hfill
\subfloat{\includegraphics[trim=0 0 0 0, clip, width=0.25 \textwidth]{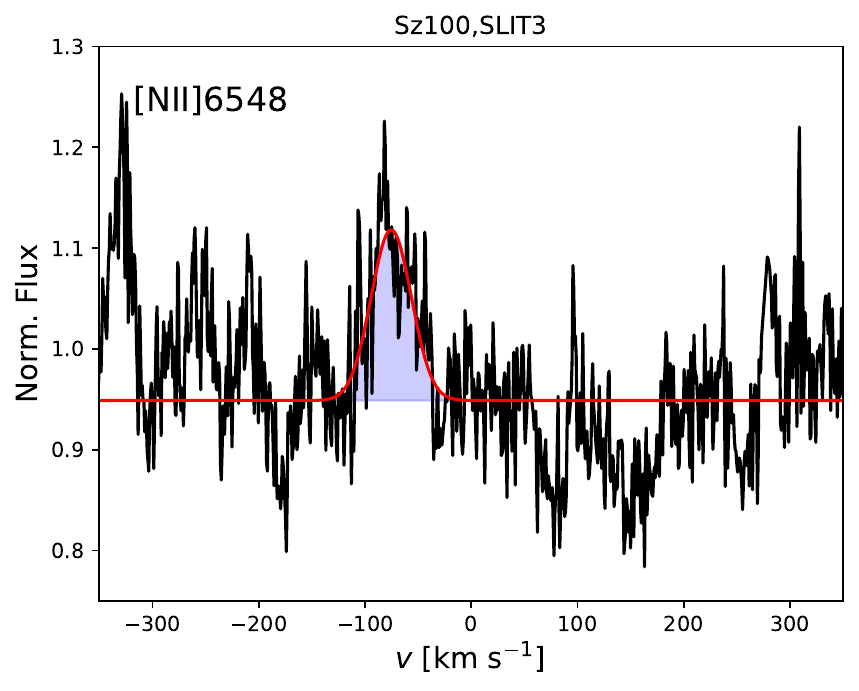}} 
\hfill    \\
\subfloat{\includegraphics[trim=0 0 0 0, clip, width=0.25 \textwidth]{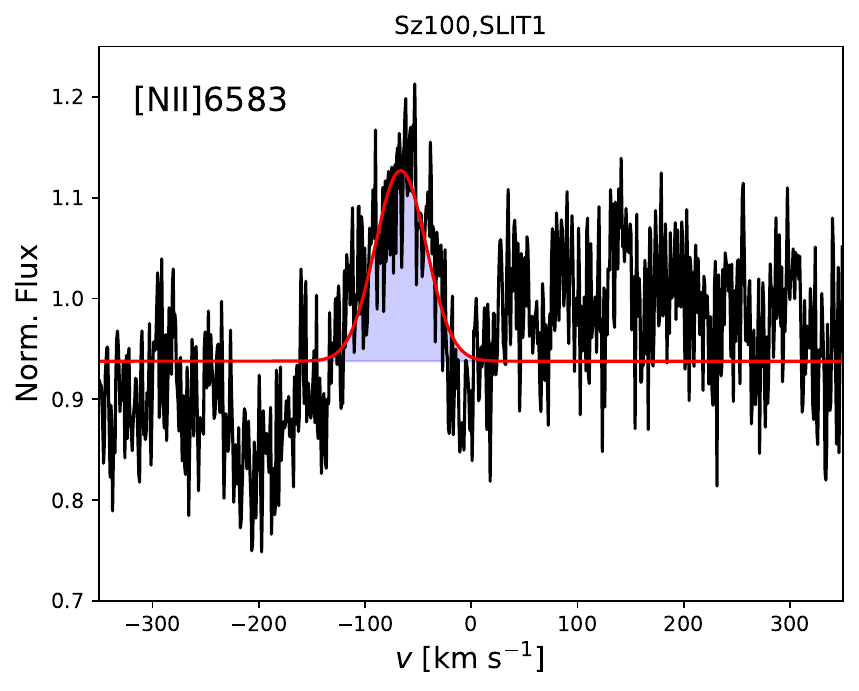}}
\hfill
\subfloat{\includegraphics[trim=0 0 0 0, clip, width=0.25 \textwidth]{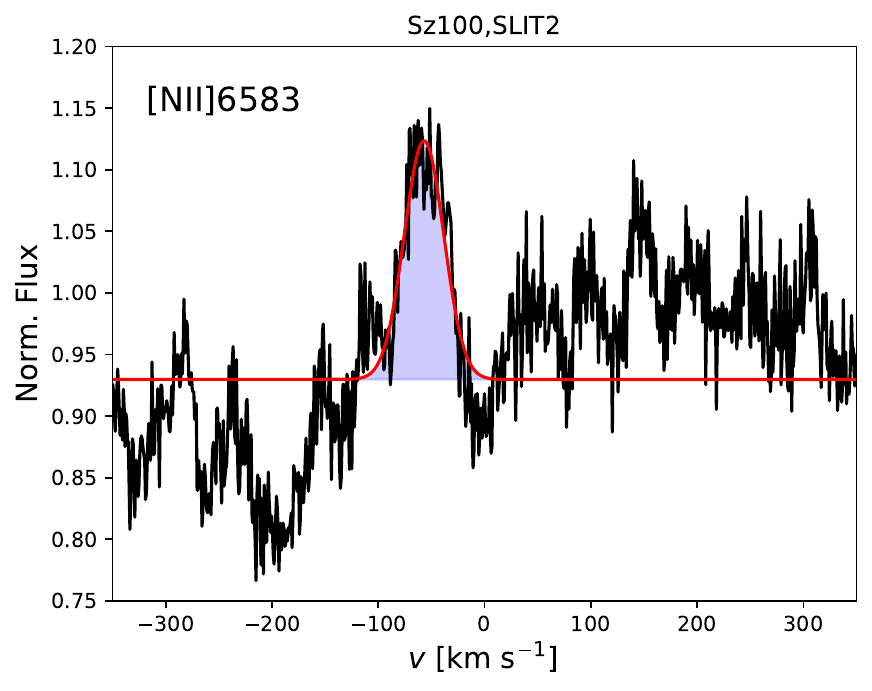}}
\hfill
\subfloat{\includegraphics[trim=0 0 0 0, clip, width=0.25 \textwidth]{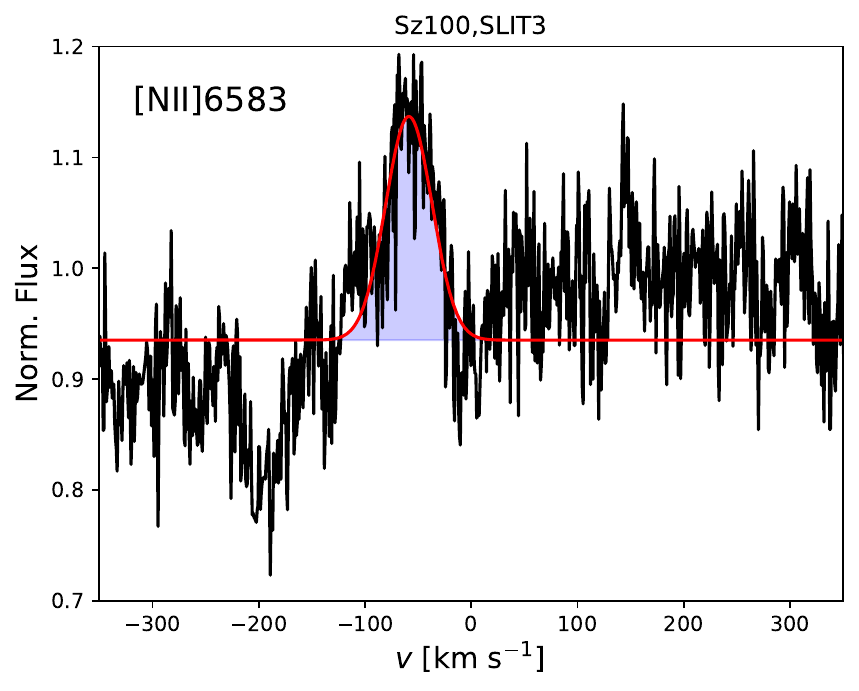}} 
\hfill \\
\subfloat{\includegraphics[trim=0 0 0 0, clip, width=0.25 \textwidth]{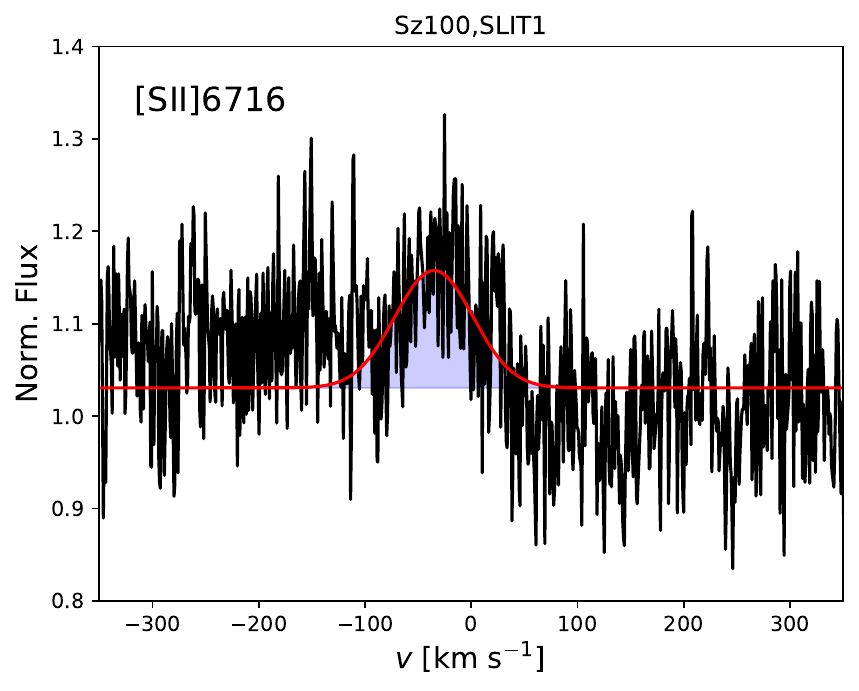}}
\hfill
\subfloat{\includegraphics[trim=0 0 0 0, clip, width=0.25 \textwidth]{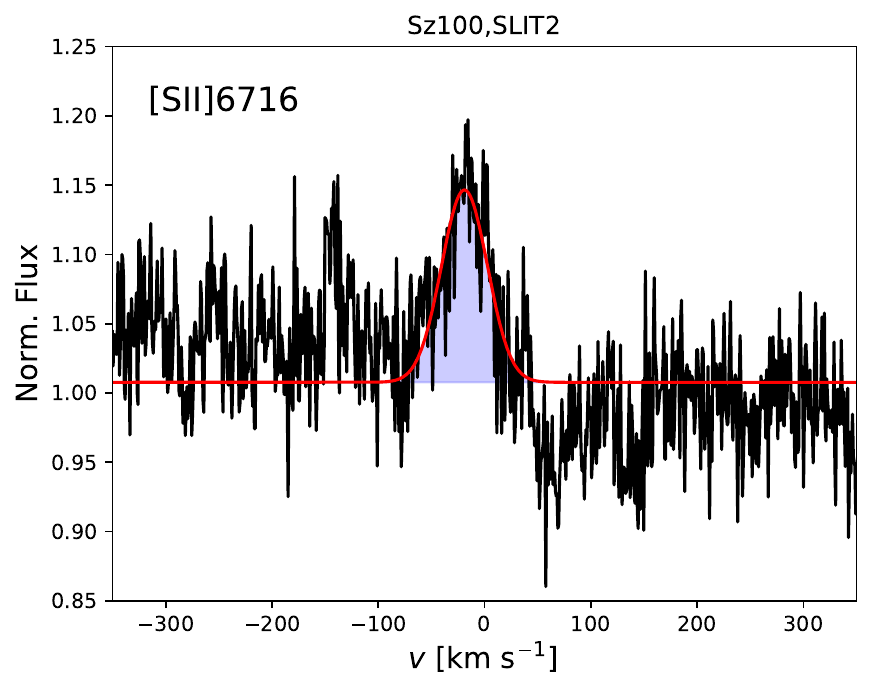}}
\hfill
\subfloat{\includegraphics[trim=0 0 0 0, clip, width=0.25 \textwidth]{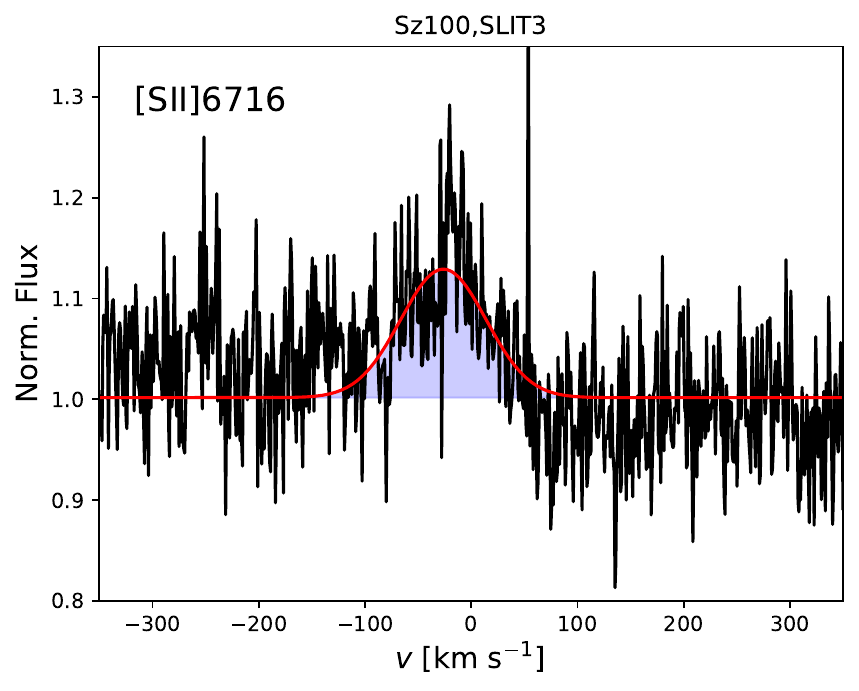}} 
\hfill \\
\subfloat{\includegraphics[trim=0 0 0 0, clip, width=0.25 \textwidth]{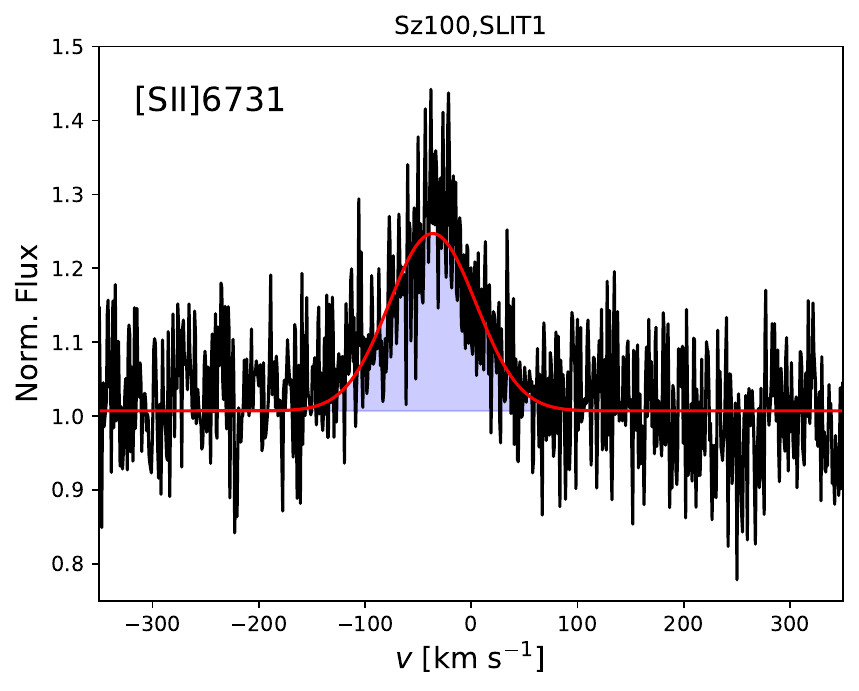}}
\hfill
\subfloat{\includegraphics[trim=0 0 0 0, clip, width=0.25 \textwidth]{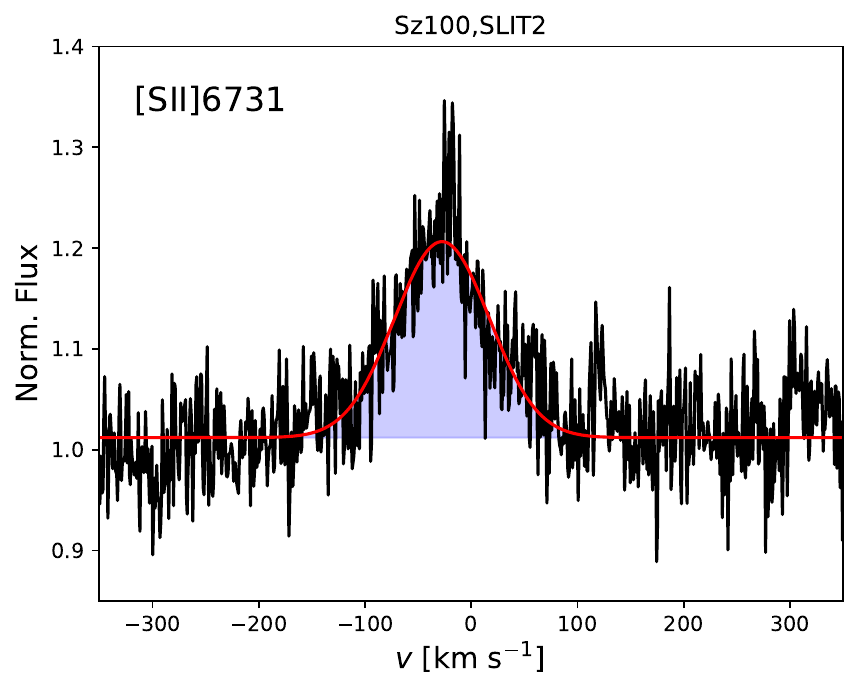}}
\hfill
\subfloat{\includegraphics[trim=0 0 0 0, clip, width=0.25 \textwidth]{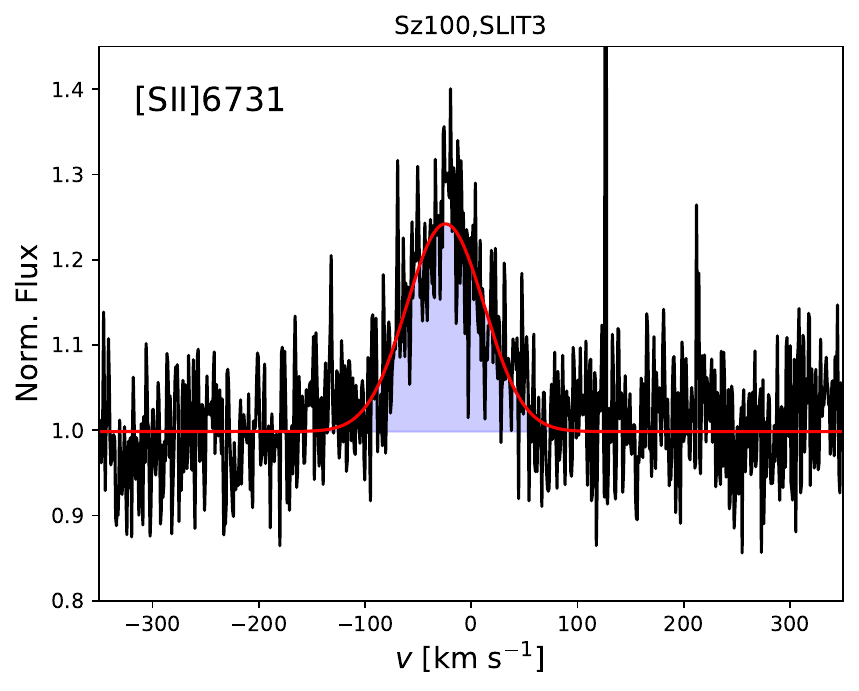}} 
\hfill
\caption{\small{Line profiles Sz\,100.}}\label{fig:Sz100}
\end{figure*} 

\begin{figure*} 
\centering
\subfloat{\includegraphics[trim=0 0 0 0, clip, width=0.25 \textwidth]{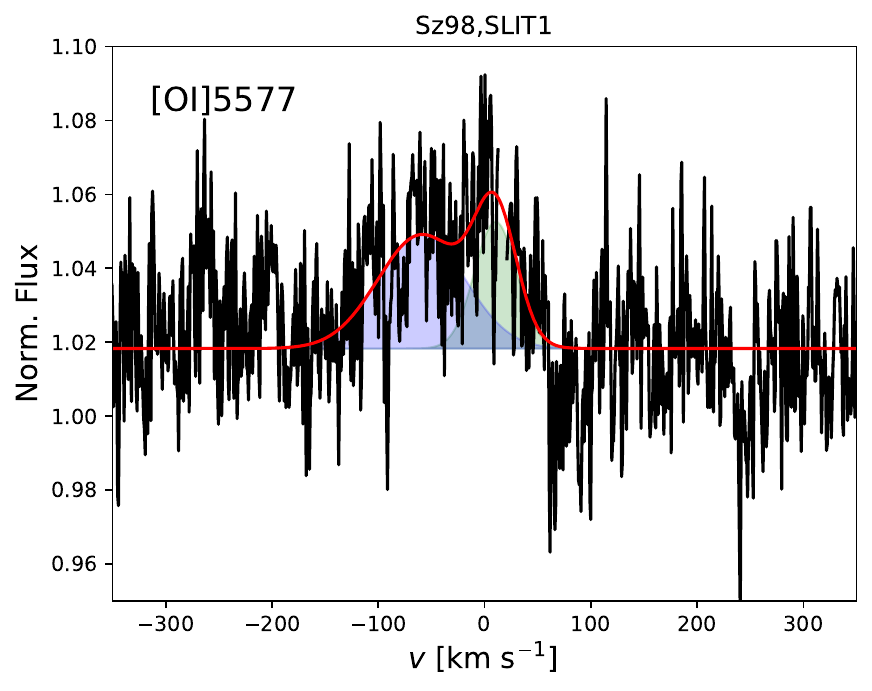}}
\hfill
\subfloat{\includegraphics[trim=0 0 0 0, clip, width=0.25 \textwidth]{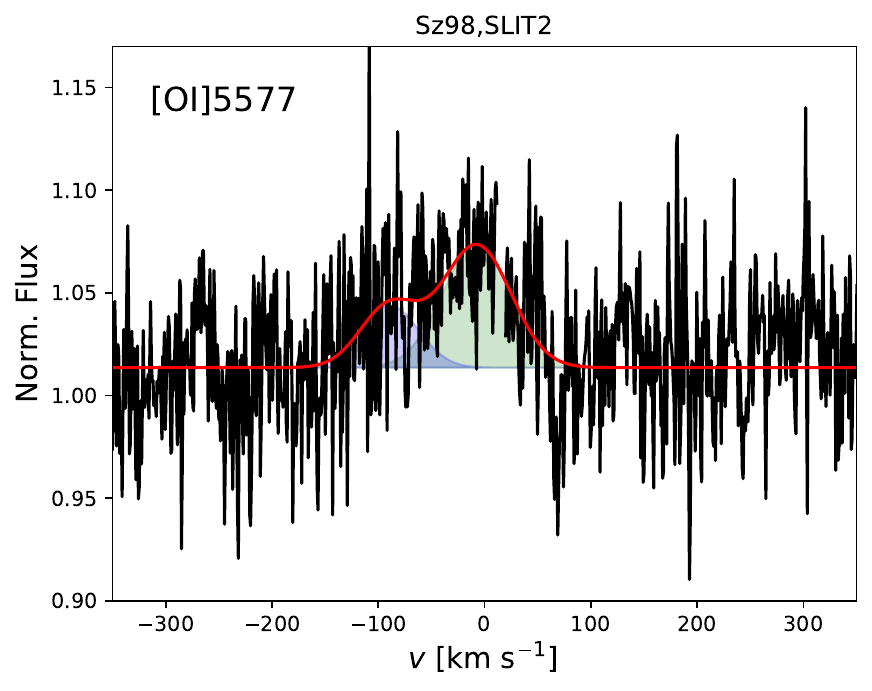}}
\hfill
\subfloat{\includegraphics[trim=0 0 0 0, clip, width=0.25 \textwidth]{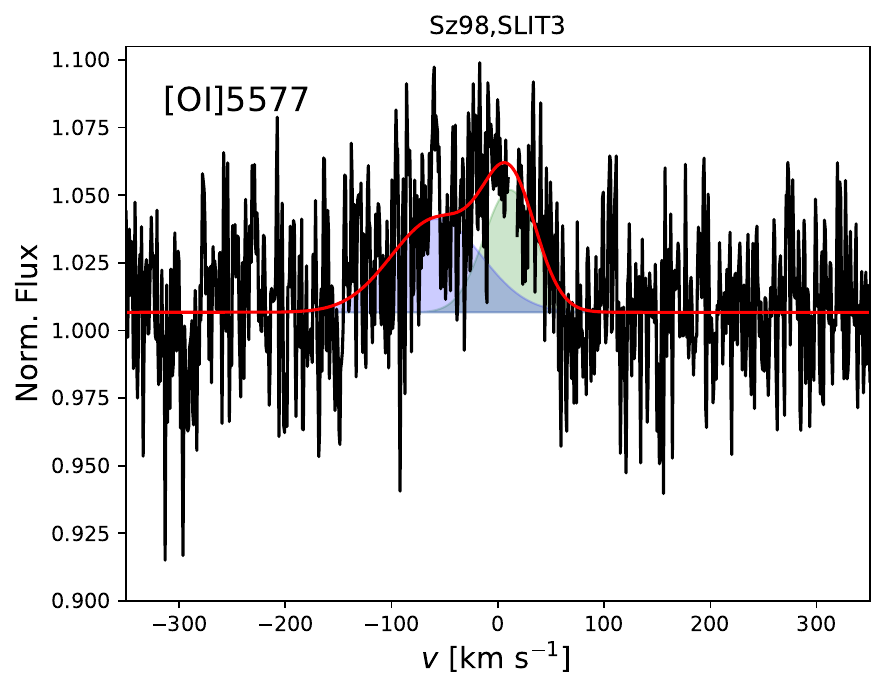}}
\hfill  \\
\subfloat{\includegraphics[trim=0 0 0 0, clip, width=0.25 \textwidth]{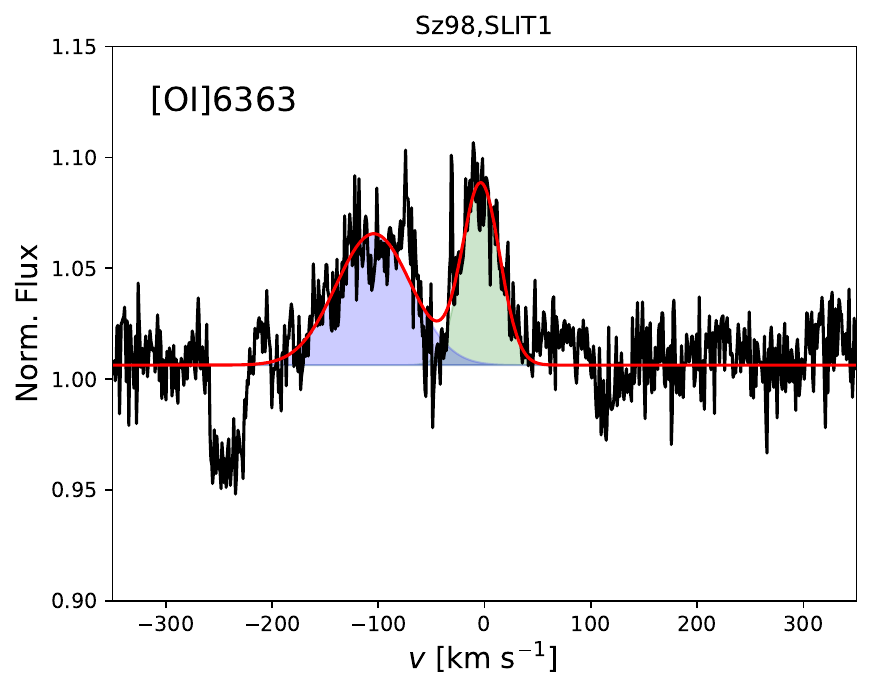}}
\hfill
\subfloat{\includegraphics[trim=0 0 0 0, clip, width=0.25 \textwidth]{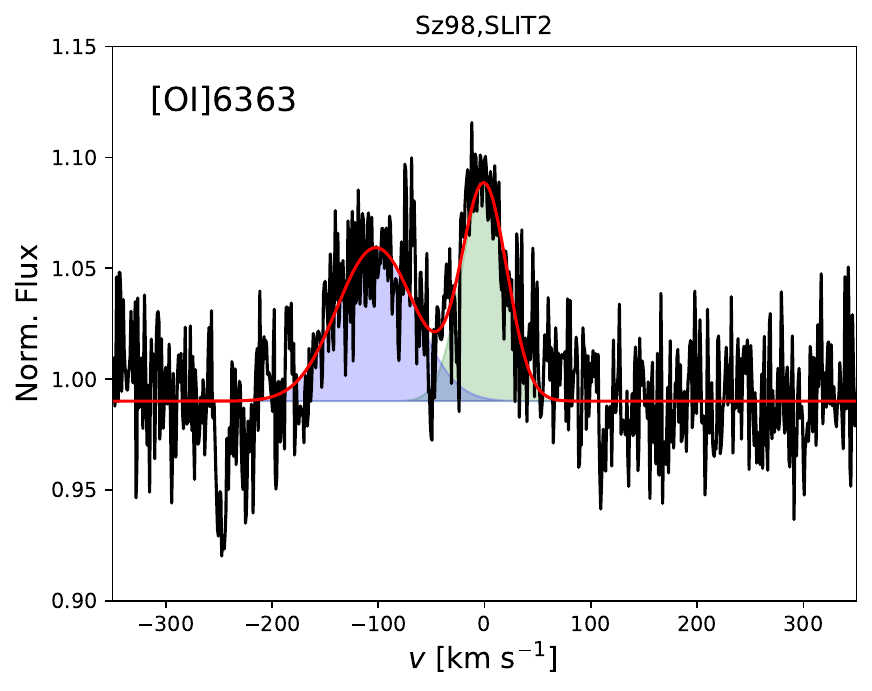}}
\hfill
\subfloat{\includegraphics[trim=0 0 0 0, clip, width=0.25 \textwidth]{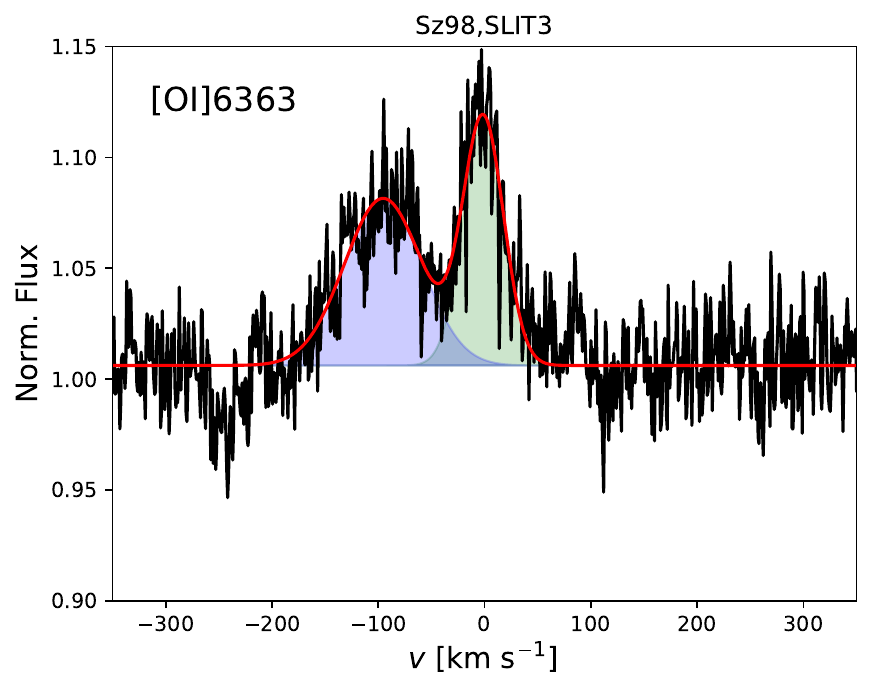}} 
\hfill \\
\subfloat{\includegraphics[trim=0 0 0 0, clip, width=0.25 \textwidth]{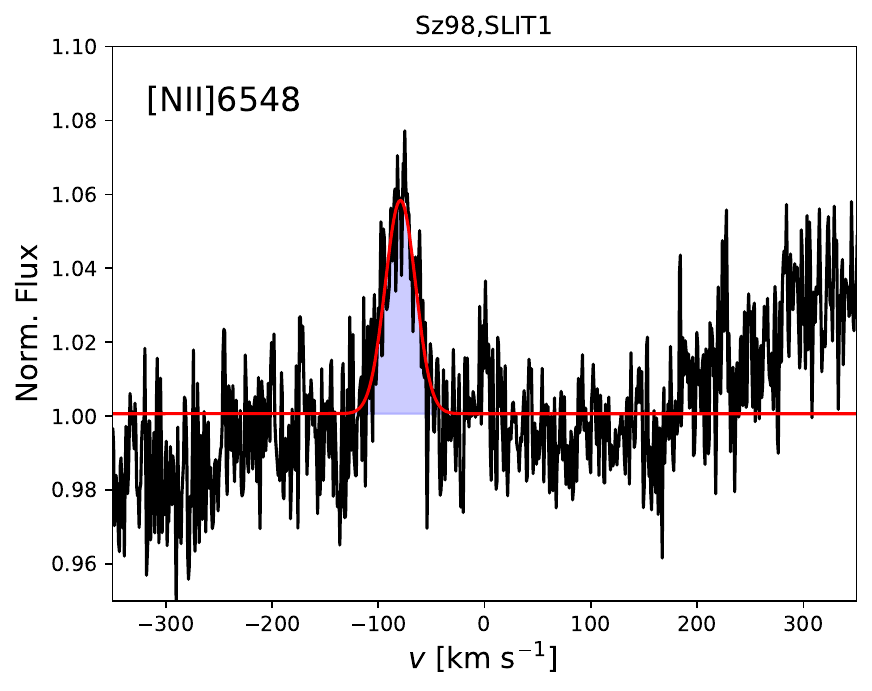}}
\hfill
\subfloat{\includegraphics[trim=0 0 0 0, clip, width=0.25 \textwidth]{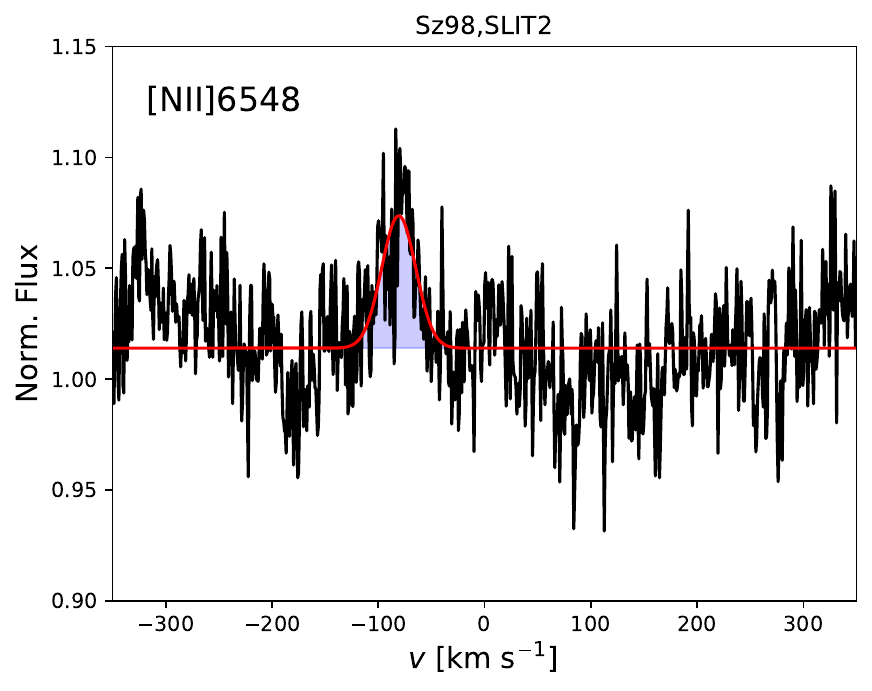}}
\hfill
\subfloat{\includegraphics[trim=0 0 0 0, clip, width=0.25 \textwidth]{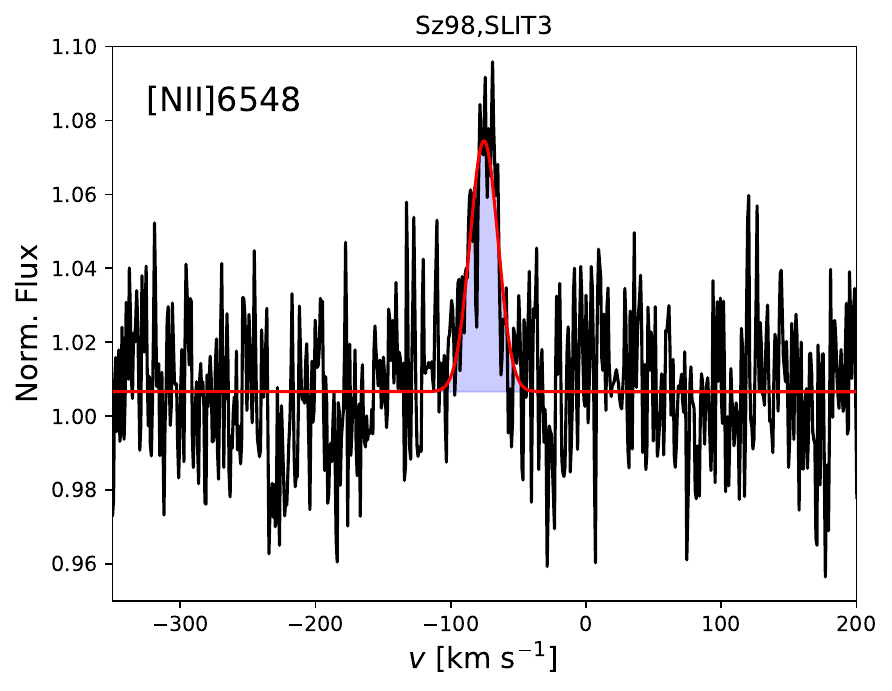}} 
\hfill    \\
\subfloat{\includegraphics[trim=0 0 0 0, clip, width=0.25 \textwidth]{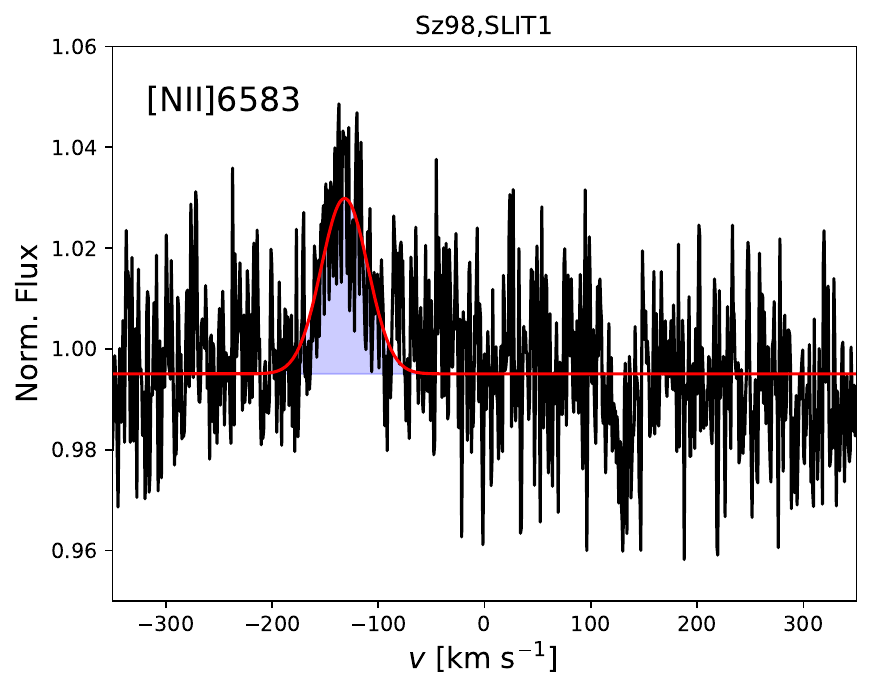}}
\hfill
\subfloat{\includegraphics[trim=0 0 0 0, clip, width=0.25 \textwidth]{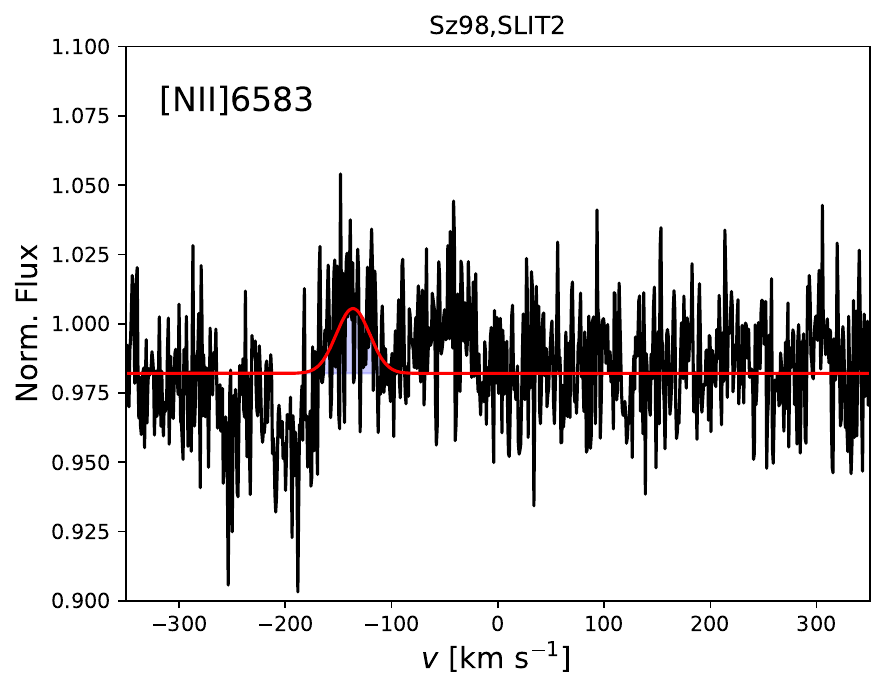}}
\hfill
\subfloat{\includegraphics[trim=0 0 0 0, clip, width=0.25 \textwidth]{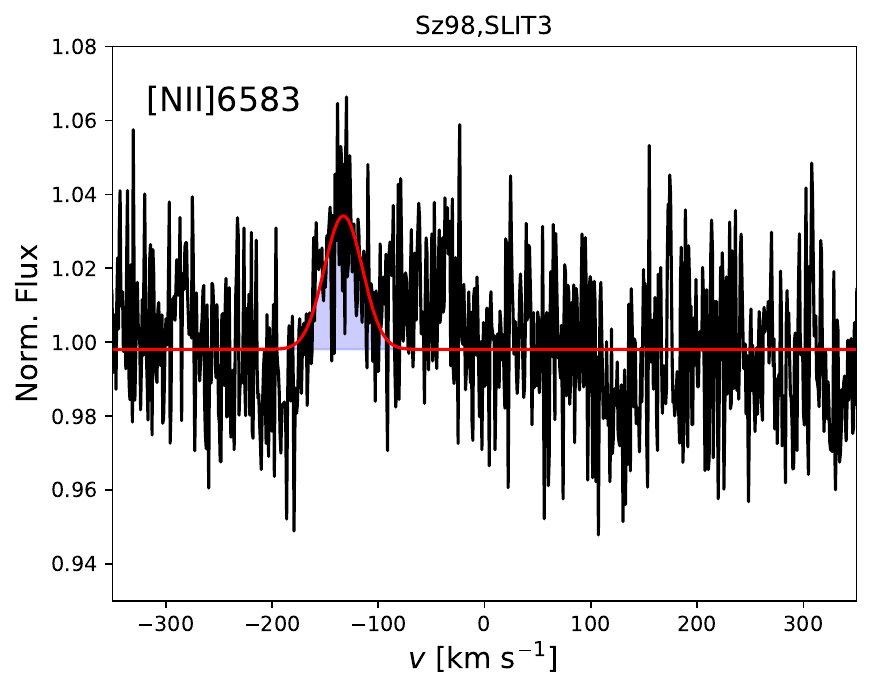}} 
\hfill  \\
\subfloat{\includegraphics[trim=0 0 0 0, clip, width=0.25 \textwidth]{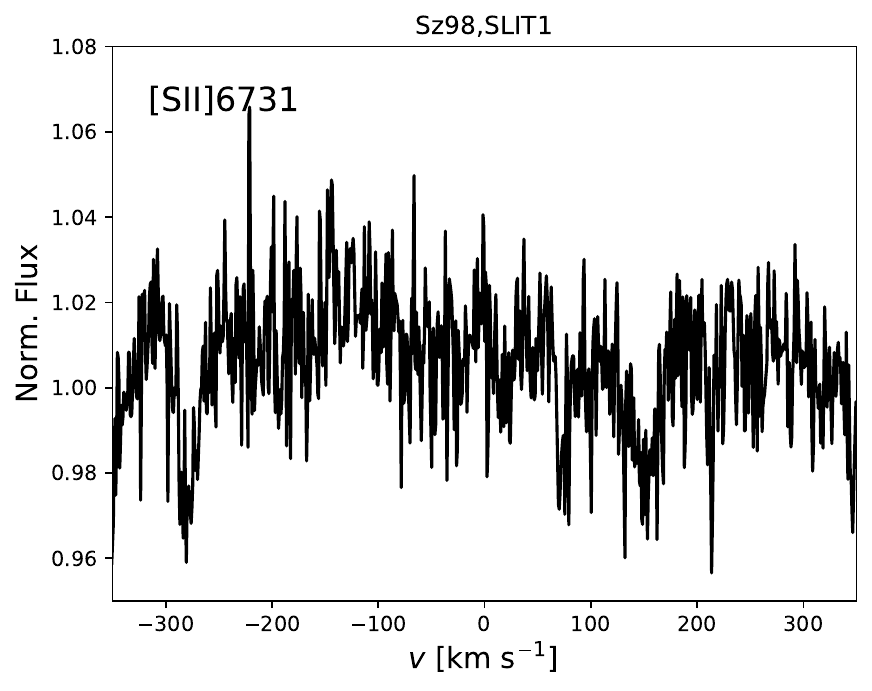}}
\hfill
\subfloat{\includegraphics[trim=0 0 0 0, clip, width=0.25 \textwidth]{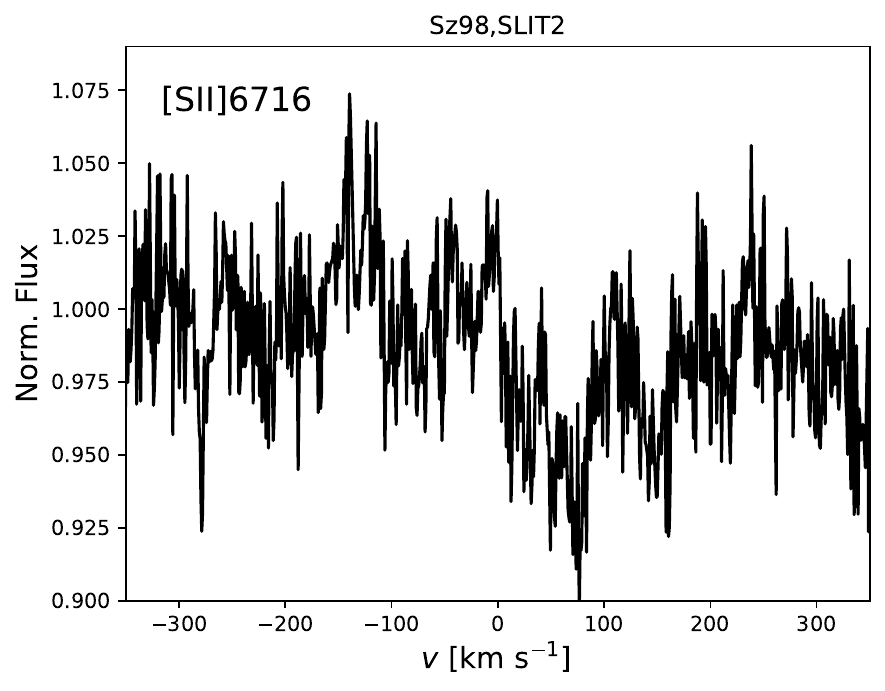}}
\hfill
\subfloat{\includegraphics[trim=0 0 0 0, clip, width=0.25 \textwidth]{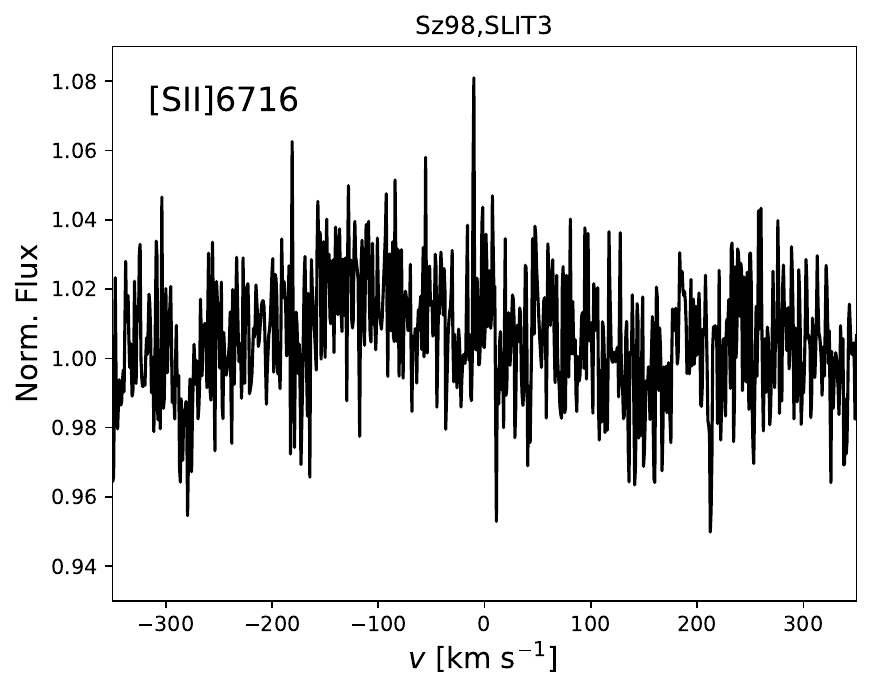}} 
\hfill \\
\subfloat{\includegraphics[trim=0 0 0 0, clip, width=0.25 \textwidth]{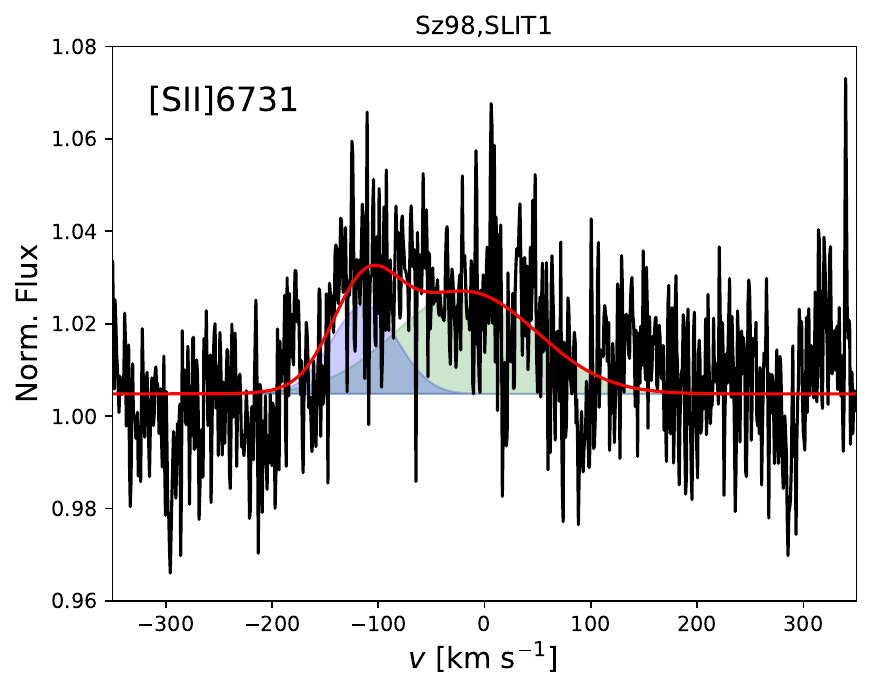}}
\hfill
\subfloat{\includegraphics[trim=0 0 0 0, clip, width=0.25 \textwidth]{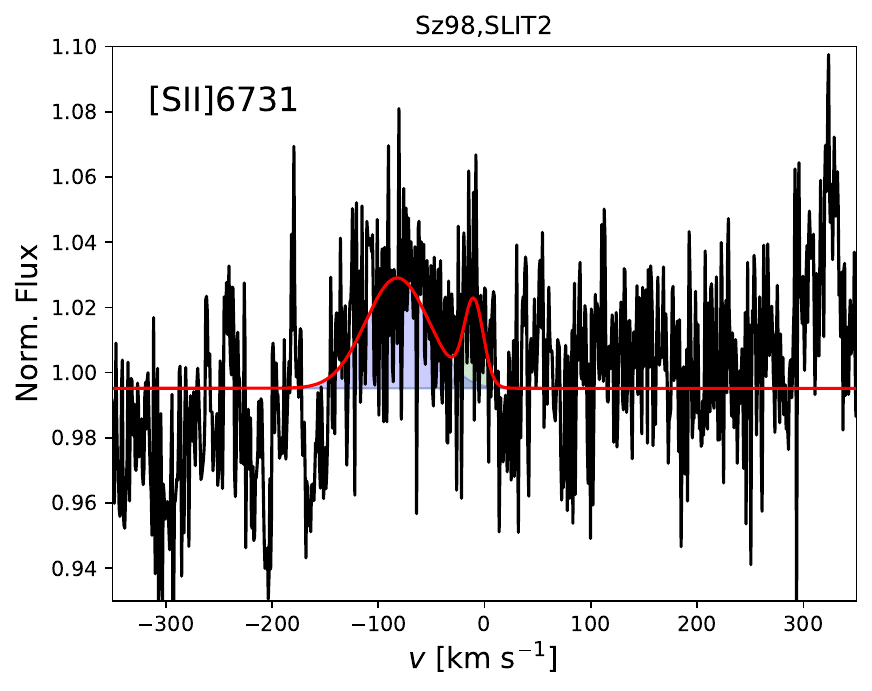}}
\hfill
\subfloat{\includegraphics[trim=0 0 0 0, clip, width=0.25 \textwidth]{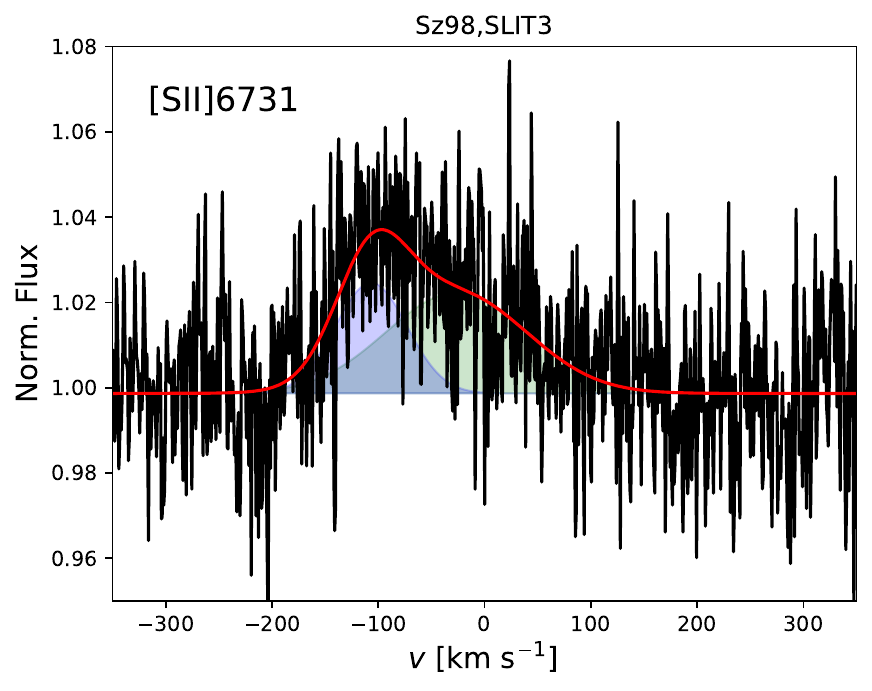}} 
\hfill
\caption{\small{Line profiles Sz\,98.}}\label{fig:Sz98}
\end{figure*} 

\begin{figure*} 
\centering
\subfloat{\includegraphics[trim=0 0 0 0, clip, width=0.25 \textwidth]{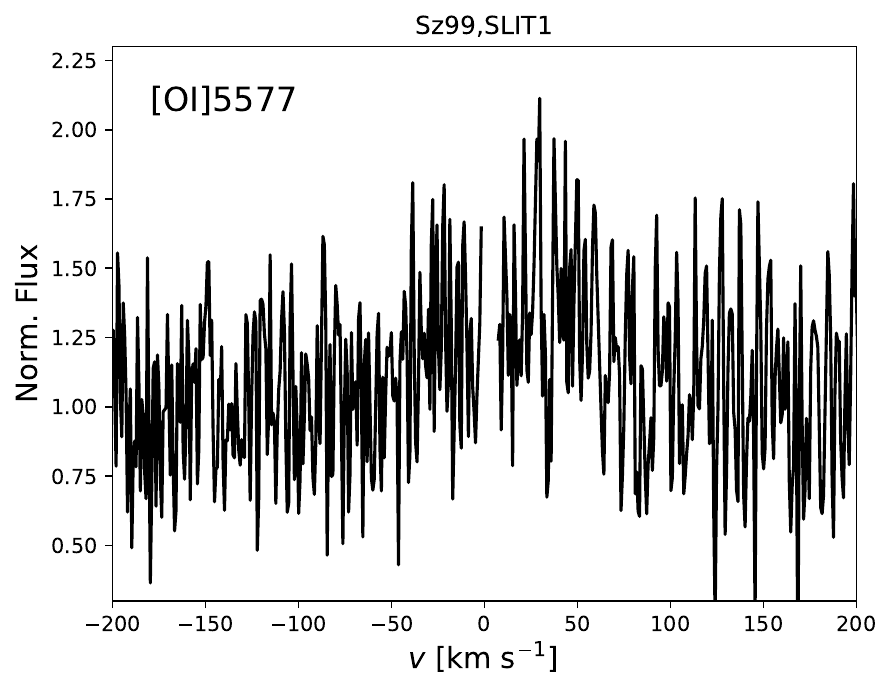}}
\hfill
\subfloat{\includegraphics[trim=0 0 0 0, clip, width=0.25 \textwidth]{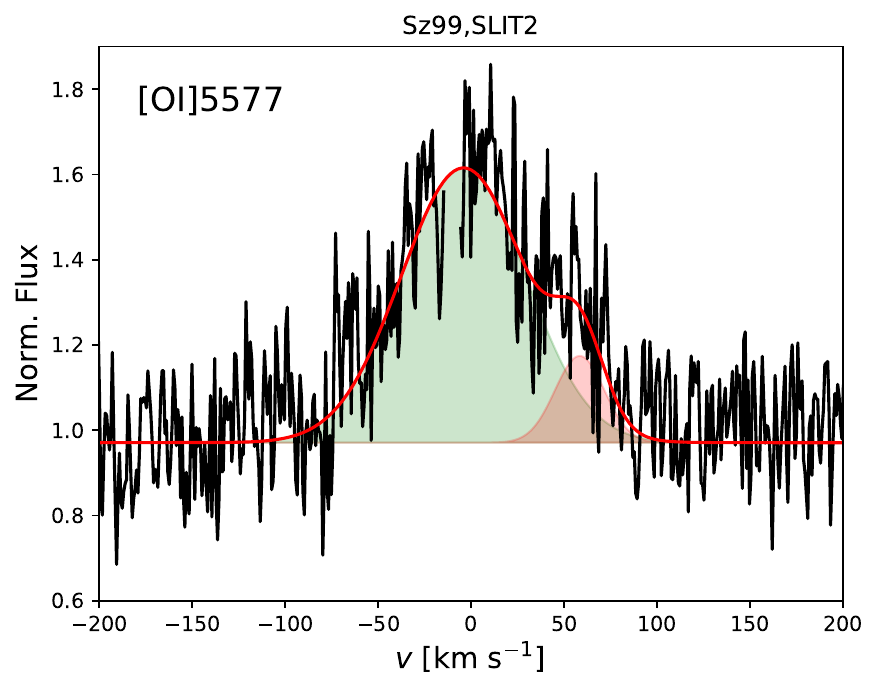}}
\hfill
\subfloat{\includegraphics[trim=0 0 0 0, clip, width=0.25 \textwidth]{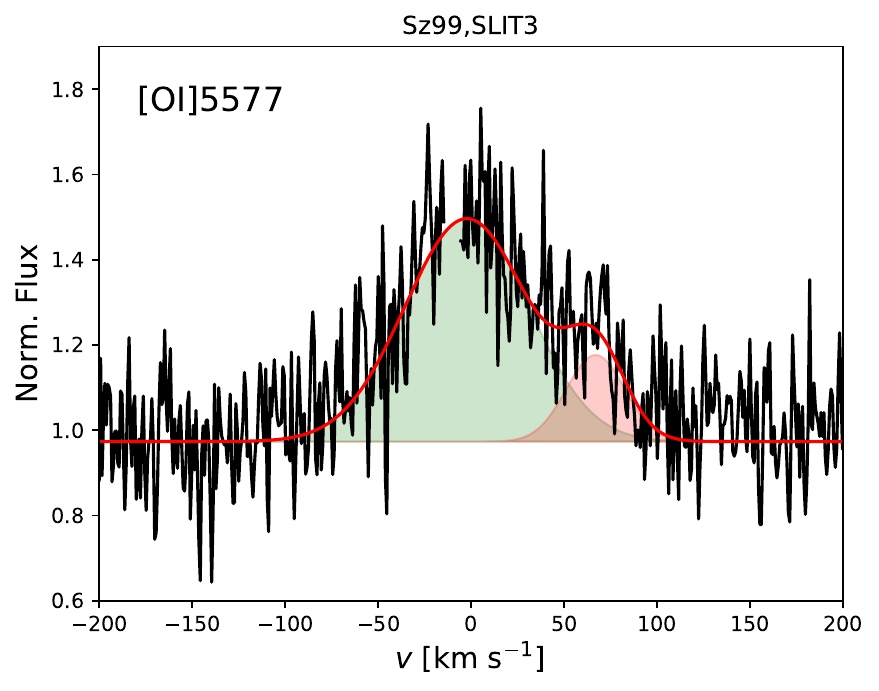}}
\hfill   \\
\subfloat{\includegraphics[trim=0 0 0 0, clip, width=0.25 \textwidth]{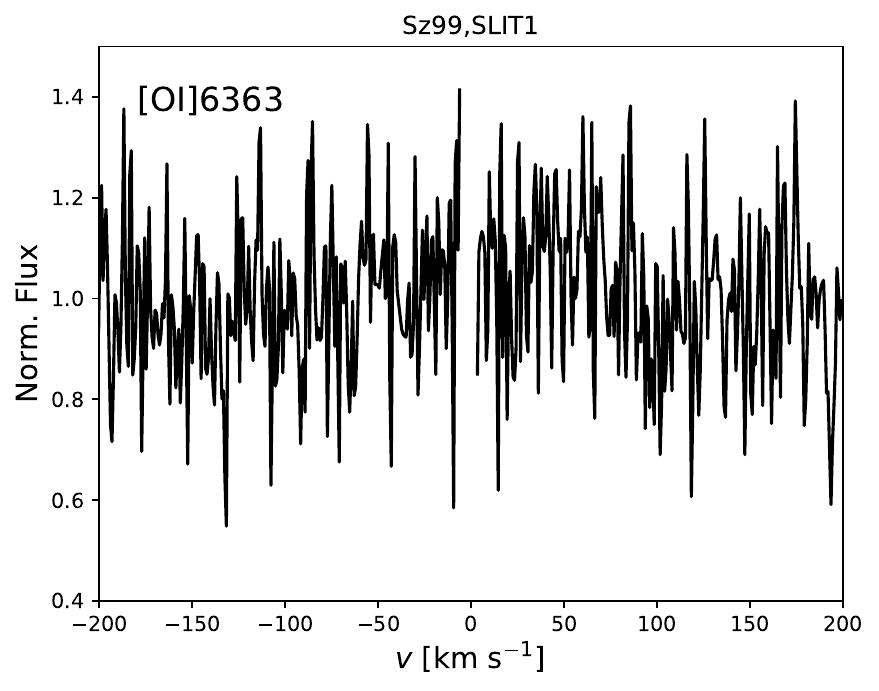}}
\hfill
\subfloat{\includegraphics[trim=0 0 0 0, clip, width=0.25 \textwidth]{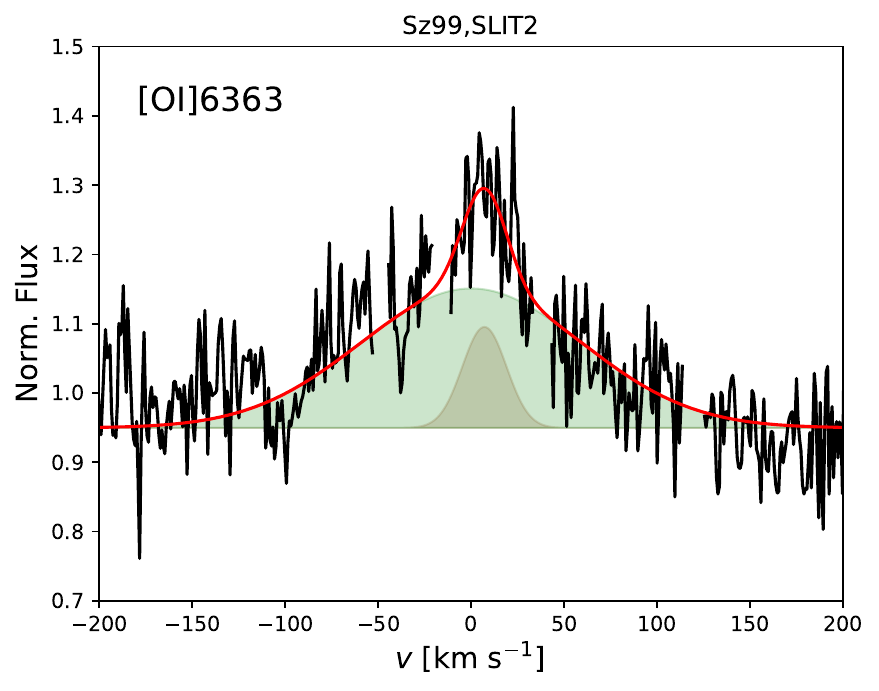}}
\hfill
\subfloat{\includegraphics[trim=0 0 0 0, clip, width=0.25 \textwidth]{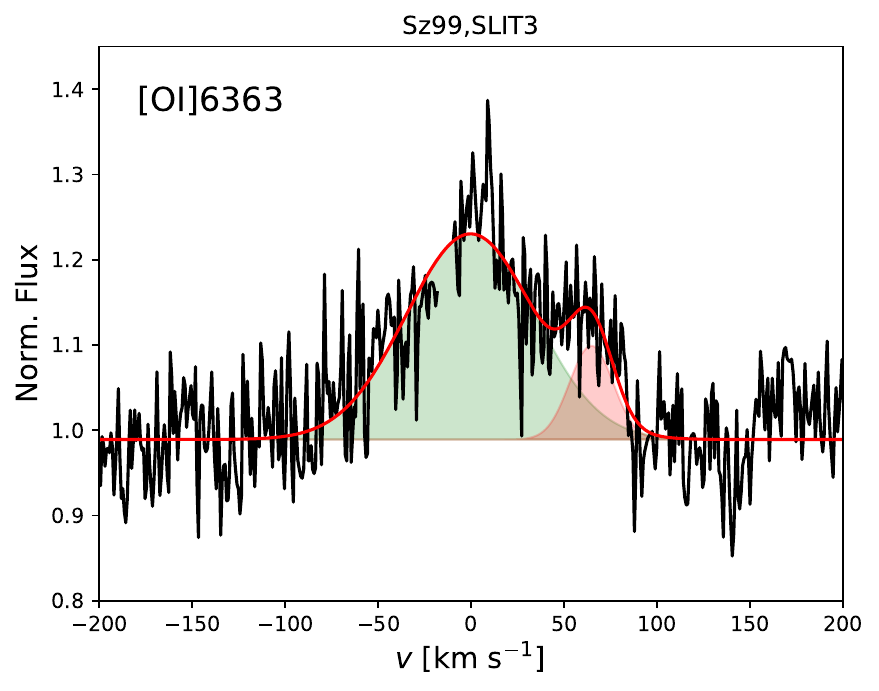}} 
\hfill \\
\subfloat{\includegraphics[trim=0 0 0 0, clip, width=0.25 \textwidth]{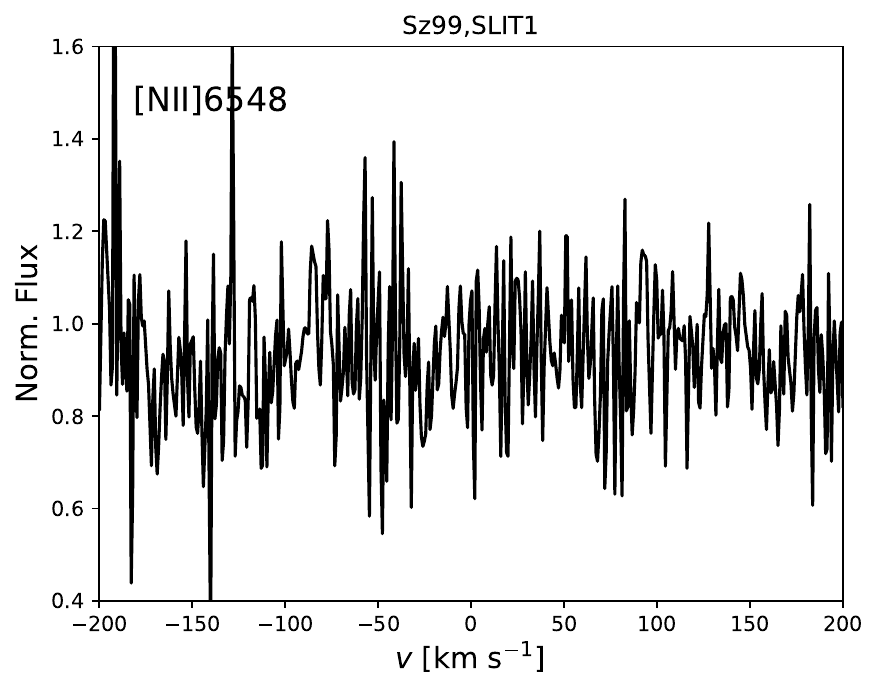}}
\hfill
\subfloat{\includegraphics[trim=0 0 0 0, clip, width=0.25 \textwidth]{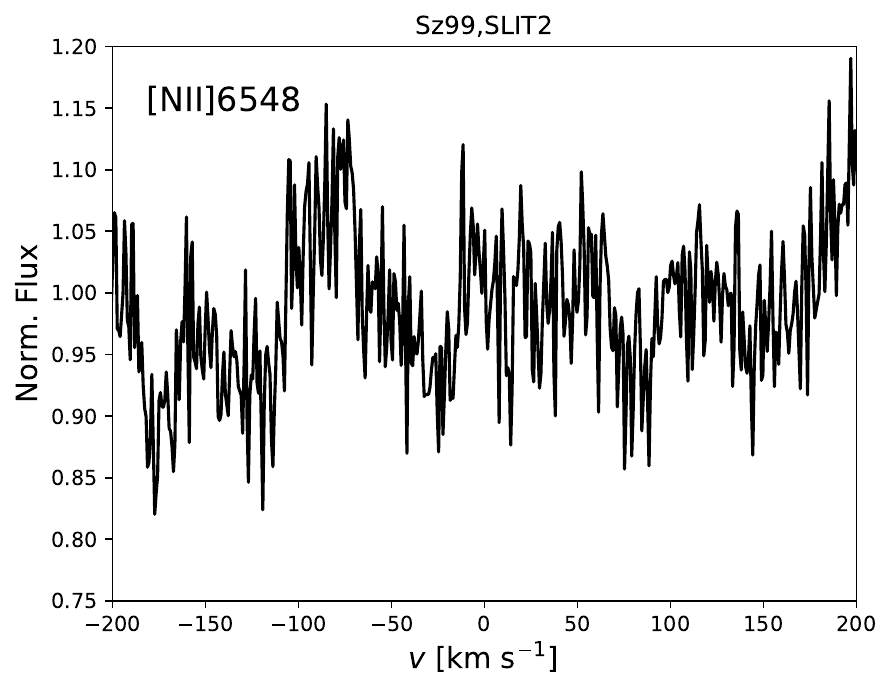}}
\hfill
\subfloat{\includegraphics[trim=0 0 0 0, clip, width=0.25 \textwidth]{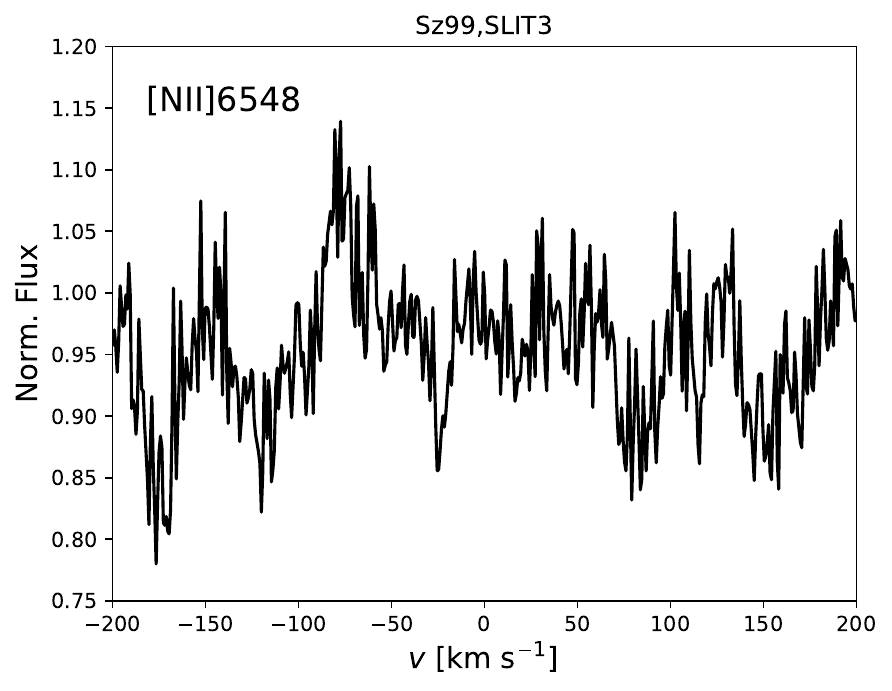}} 
\hfill    \\
\subfloat{\includegraphics[trim=0 0 0 0, clip, width=0.25 \textwidth]{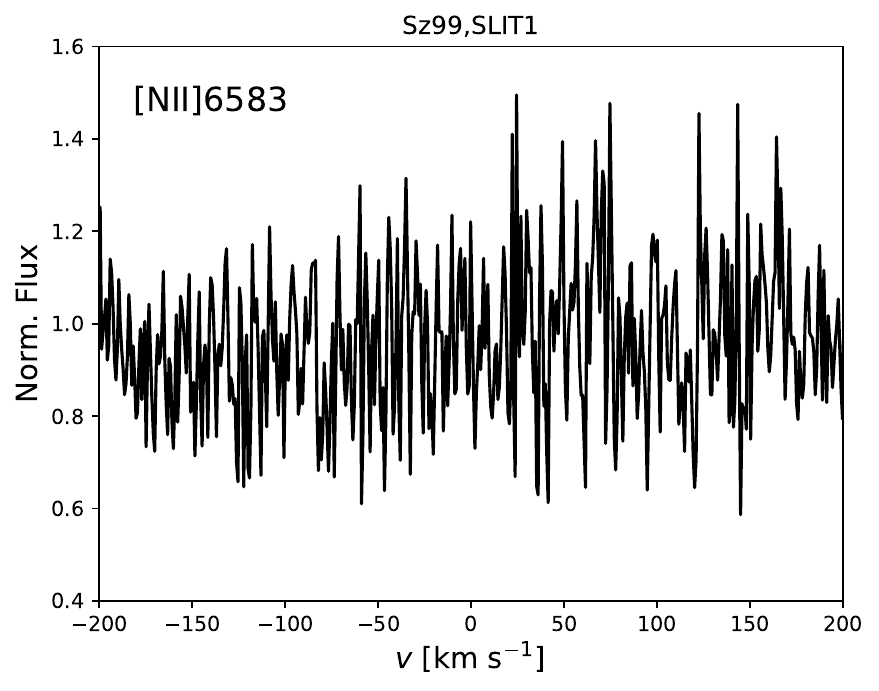}}
\hfill
\subfloat{\includegraphics[trim=0 0 0 0, clip, width=0.25 \textwidth]{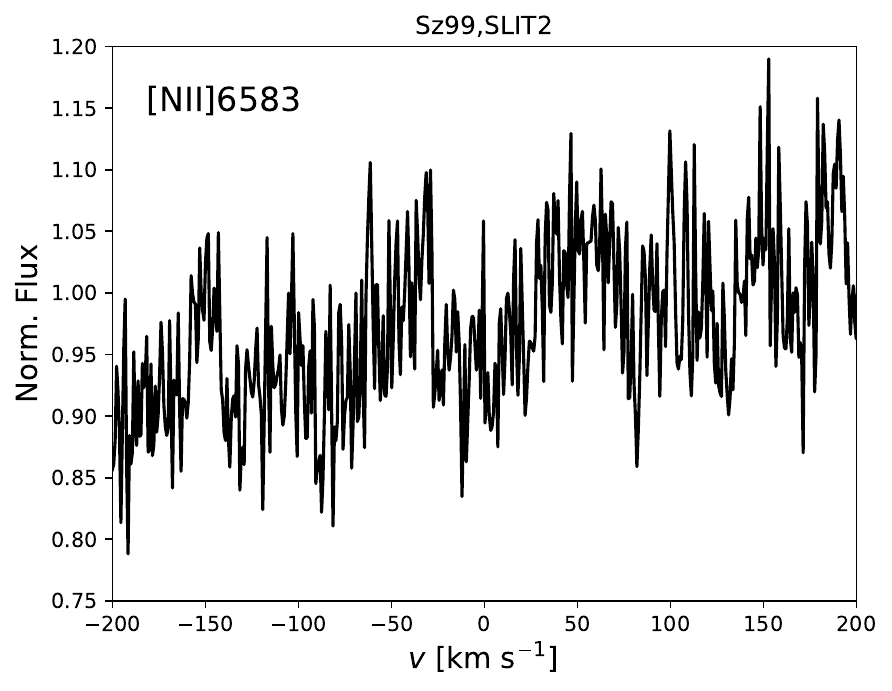}}
\hfill
\subfloat{\includegraphics[trim=0 0 0 0, clip, width=0.25 \textwidth]{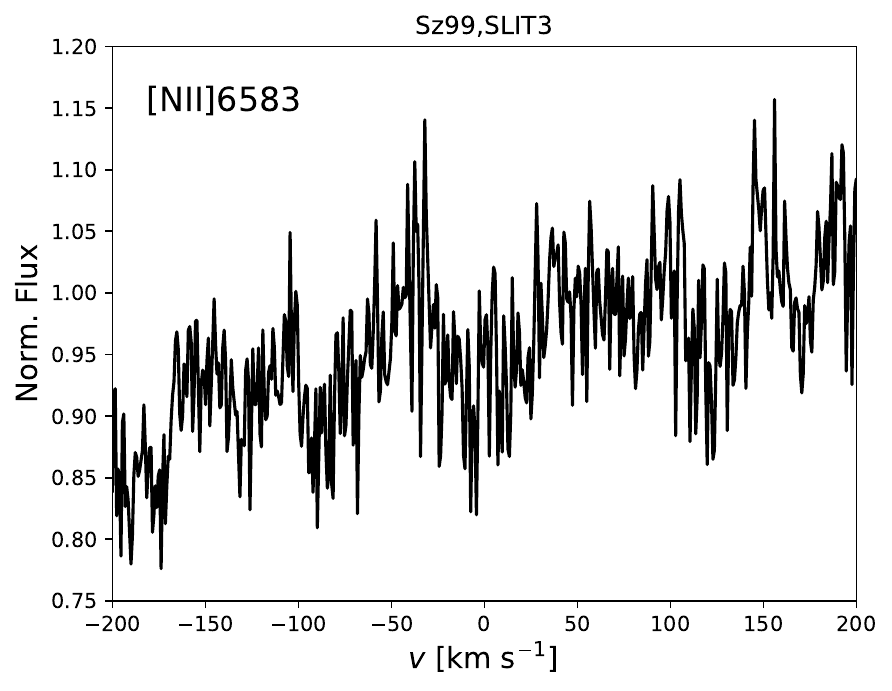}} 
\hfill \\
\subfloat{\includegraphics[trim=0 0 0 0, clip, width=0.25 \textwidth]{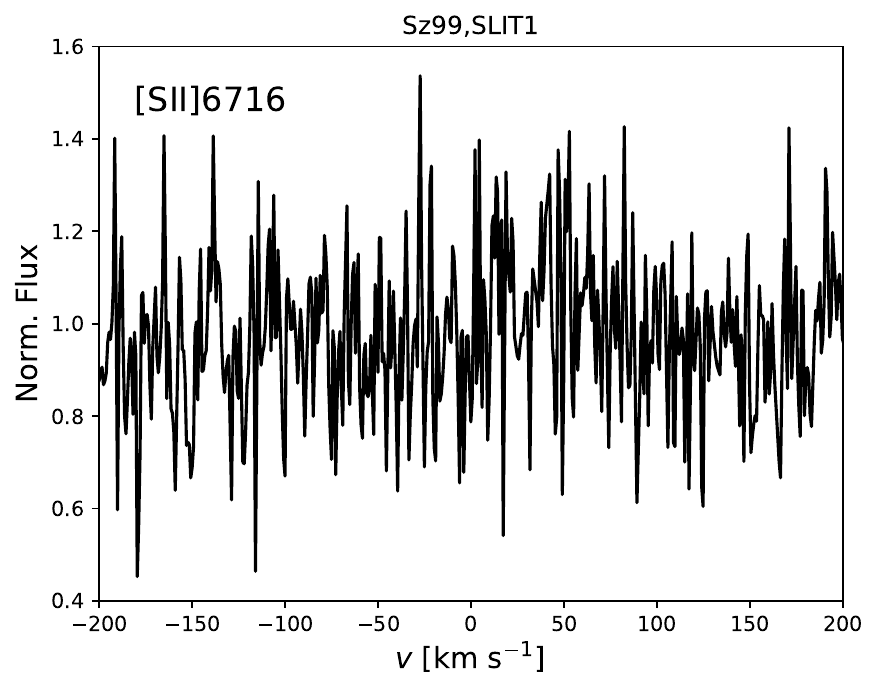}}
\hfill
\subfloat{\includegraphics[trim=0 0 0 0, clip, width=0.25 \textwidth]{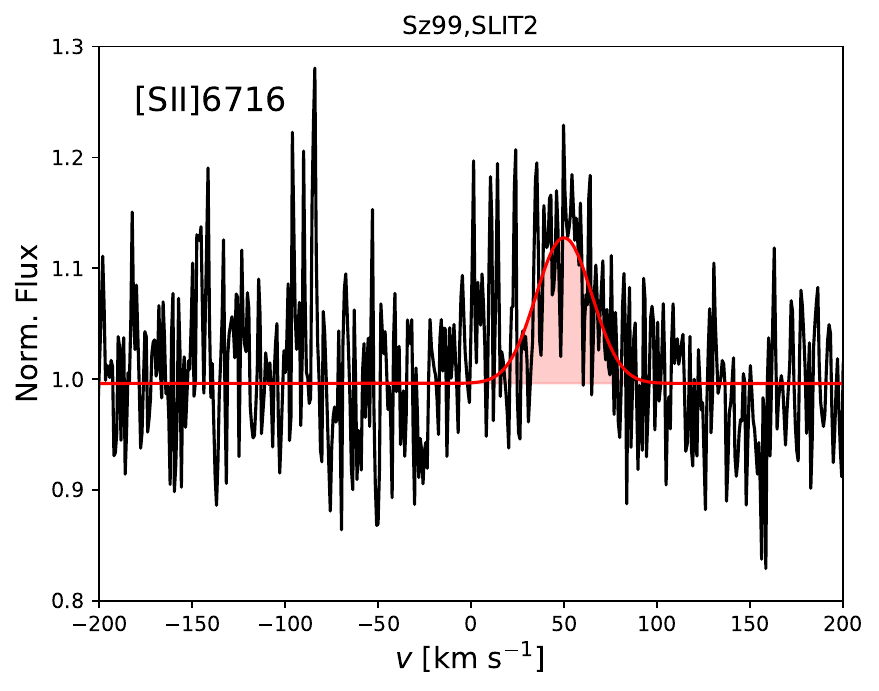}}
\hfill
\subfloat{\includegraphics[trim=0 0 0 0, clip, width=0.25 \textwidth]{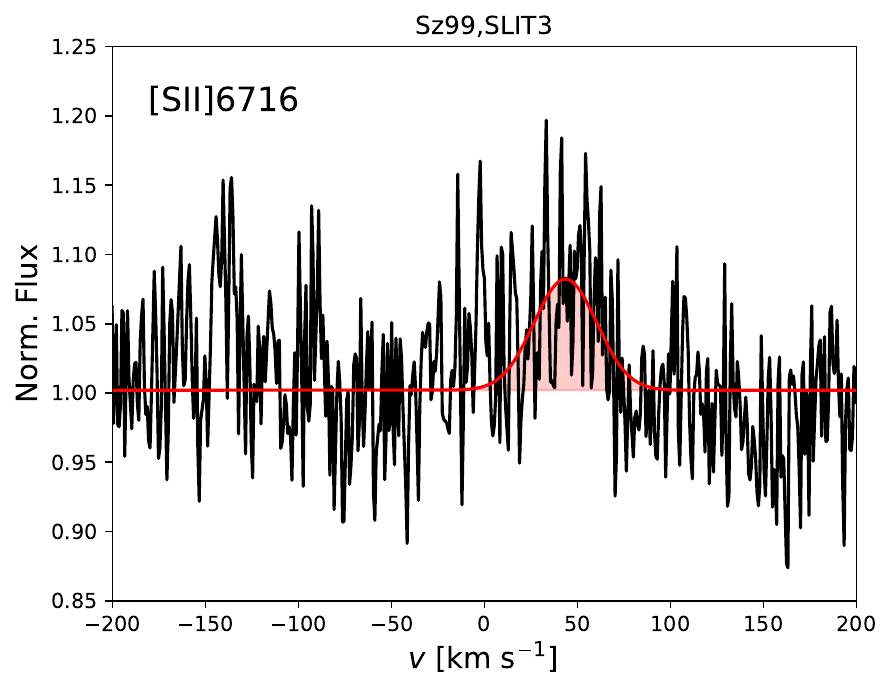}} 
\hfill \\
\subfloat{\includegraphics[trim=0 0 0 0, clip, width=0.25 \textwidth]{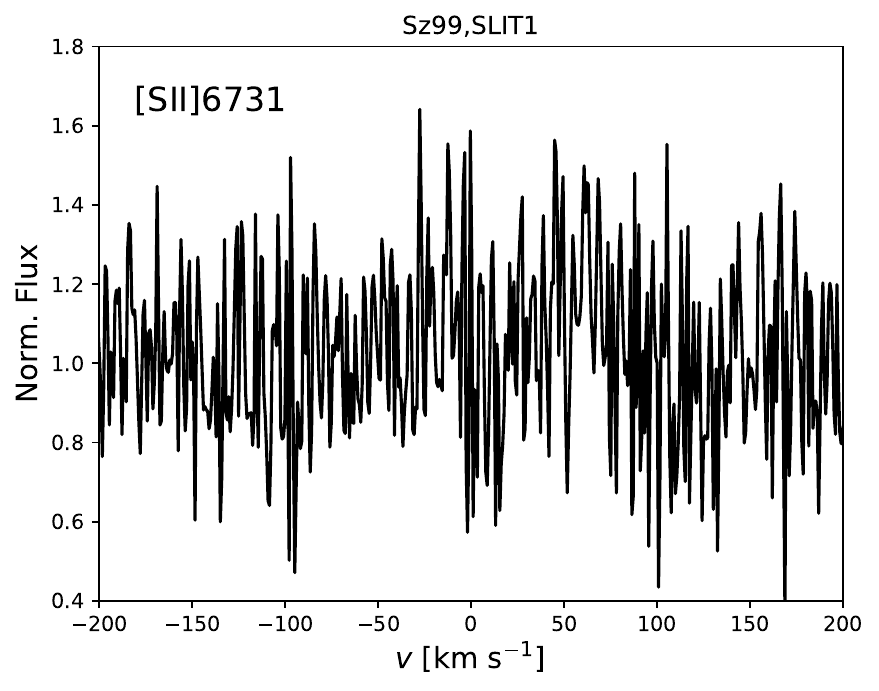}}
\hfill
\subfloat{\includegraphics[trim=0 0 0 0, clip, width=0.25 \textwidth]{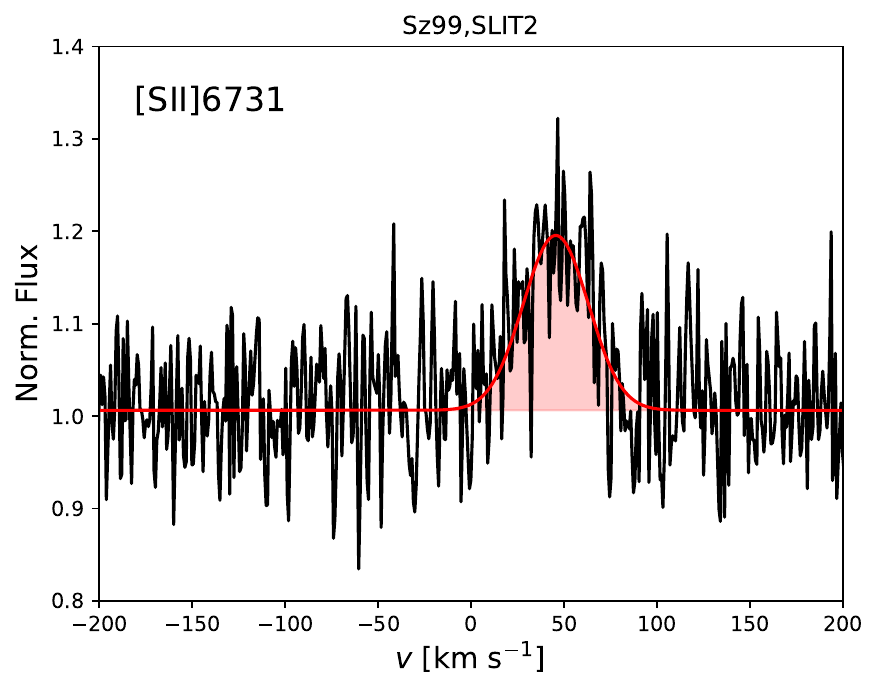}}
\hfill
\subfloat{\includegraphics[trim=0 0 0 0, clip, width=0.25 \textwidth]{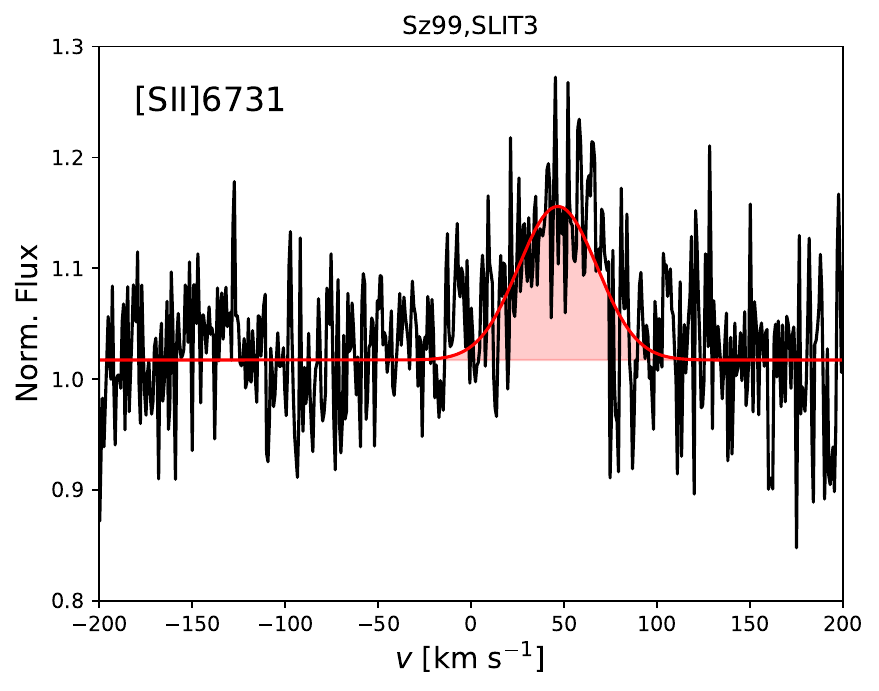}} 
\hfill
\caption{\small{Line profiles Sz\,99.}}\label{fig:Sz99}
\end{figure*} 

\begin{figure*} 
\centering
\subfloat{\includegraphics[trim=0 0 0 0, clip, width=0.25 \textwidth]{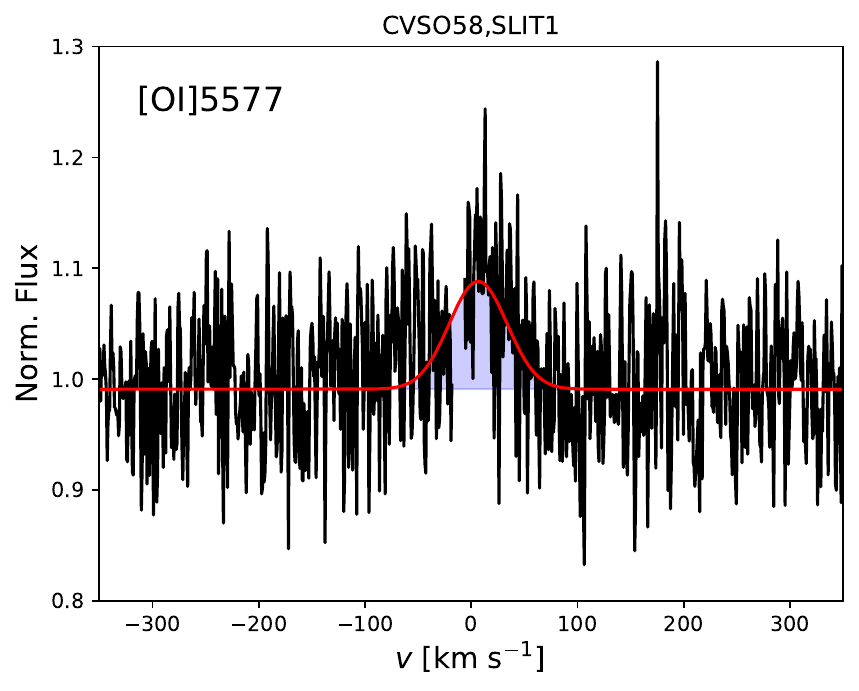}}
\hfill
\subfloat{\includegraphics[trim=0 0 0 0, clip, width=0.25 \textwidth]{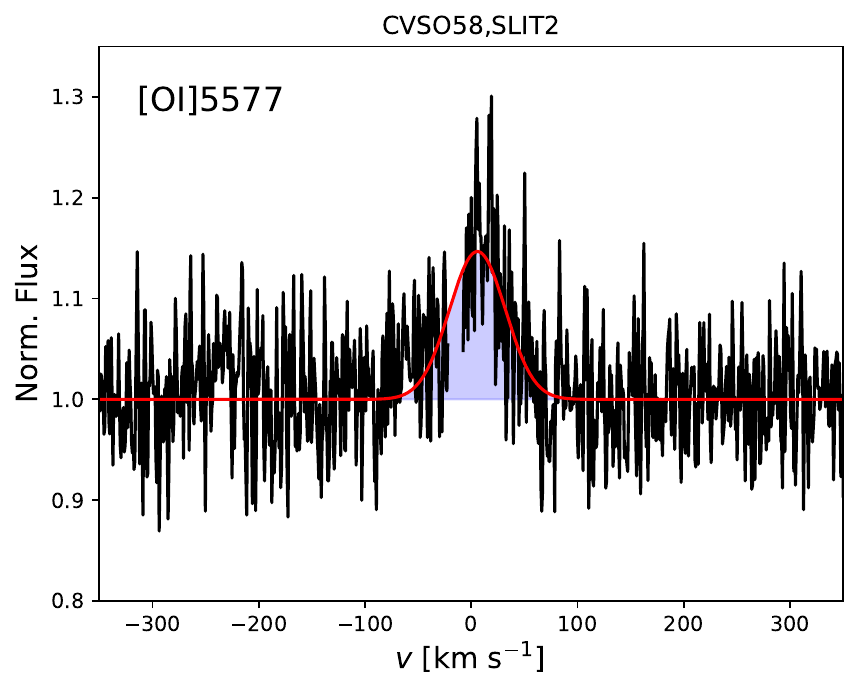}}
\hfill
\subfloat{\includegraphics[trim=0 0 0 0, clip, width=0.25 \textwidth]{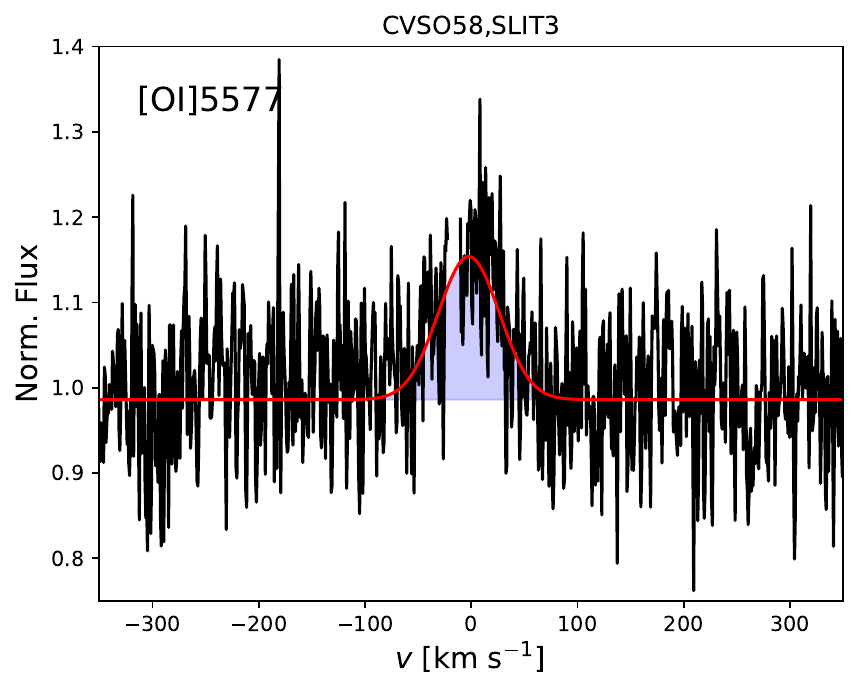}}
\hfill  \\
\subfloat{\includegraphics[trim=0 0 0 0, clip, width=0.25 \textwidth]{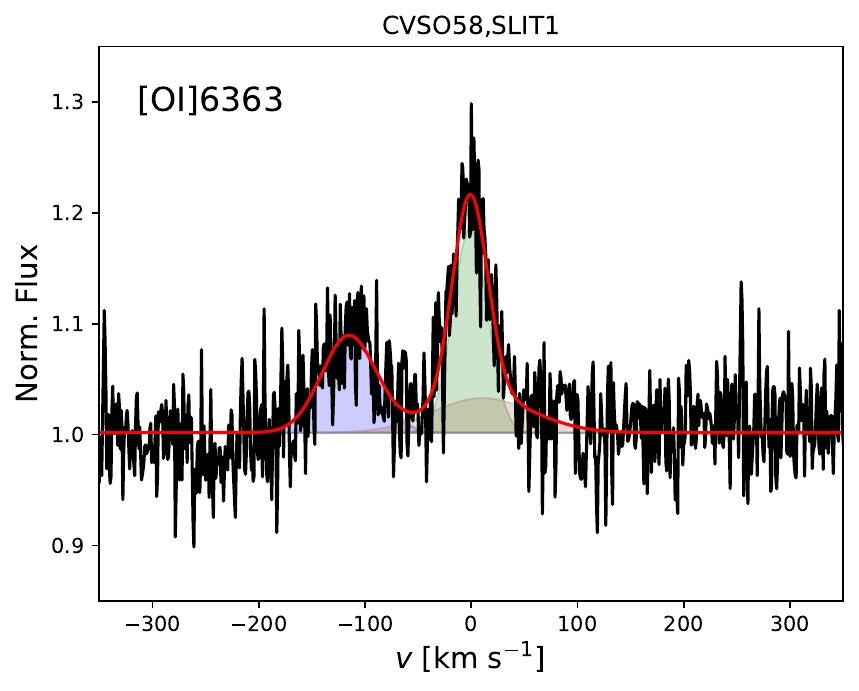}}
\hfill
\subfloat{\includegraphics[trim=0 0 0 0, clip, width=0.25 \textwidth]{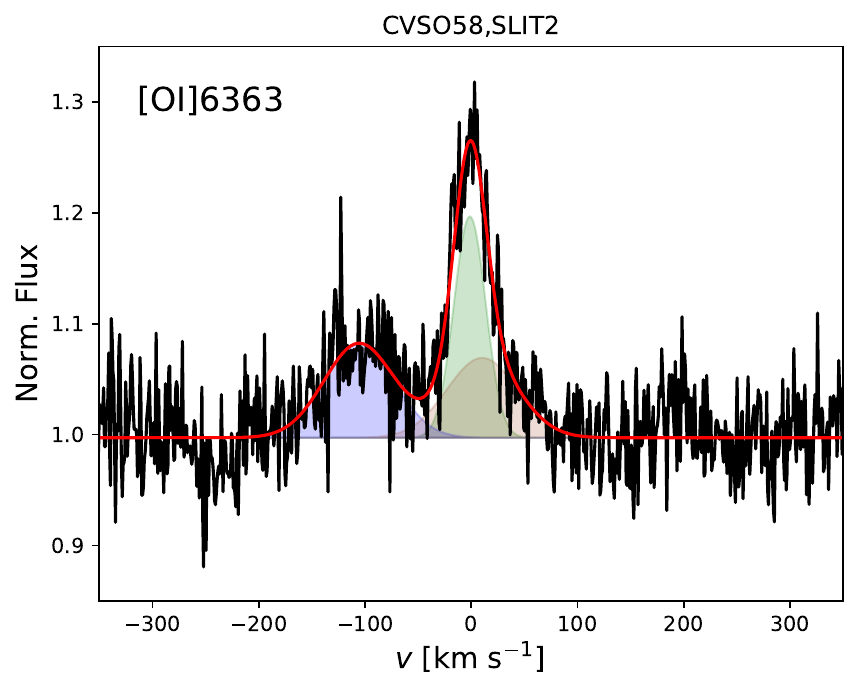}}
\hfill
\subfloat{\includegraphics[trim=0 0 0 0, clip, width=0.25 \textwidth]{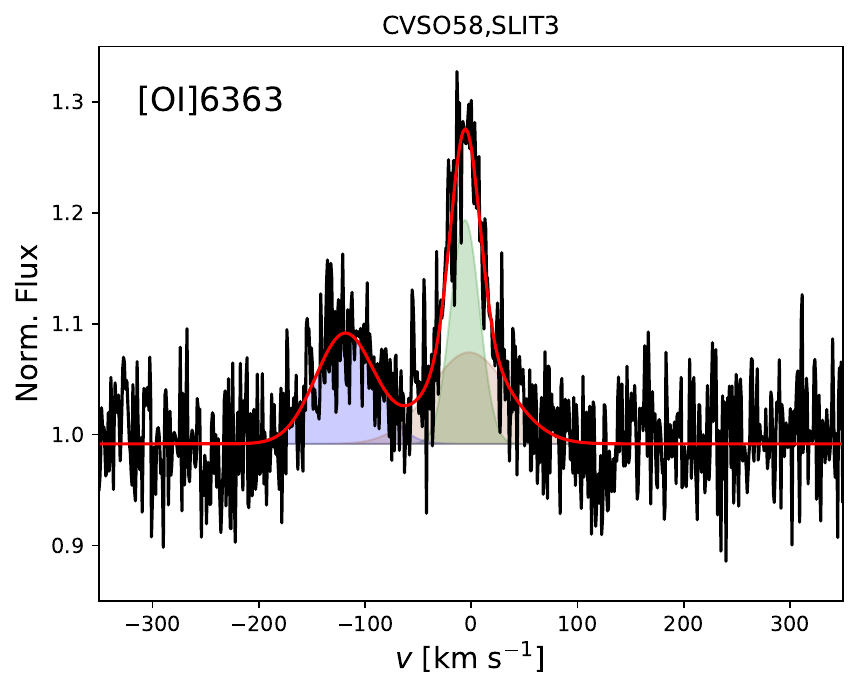}} 
\hfill \\
\subfloat{\includegraphics[trim=0 0 0 0, clip, width=0.25 \textwidth]{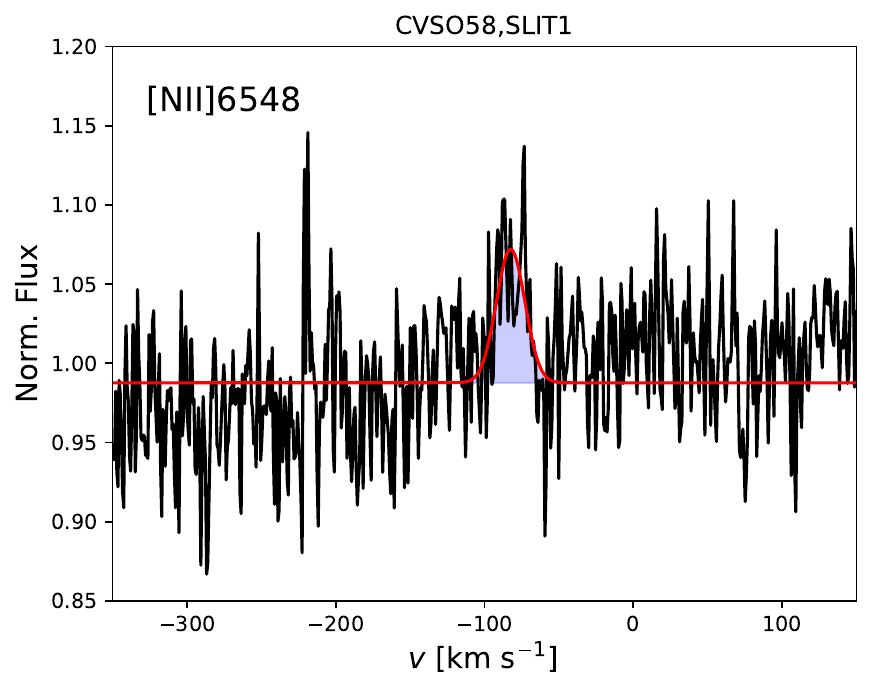}}
\hfill
\subfloat{\includegraphics[trim=0 0 0 0, clip, width=0.25 \textwidth]{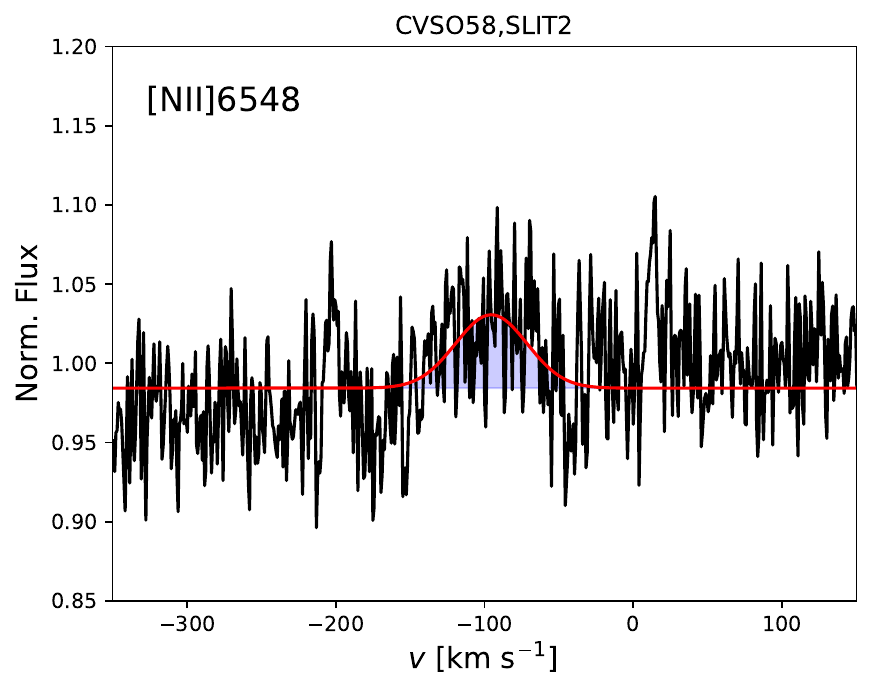}}
\hfill
\subfloat{\includegraphics[trim=0 0 0 0, clip, width=0.25 \textwidth]{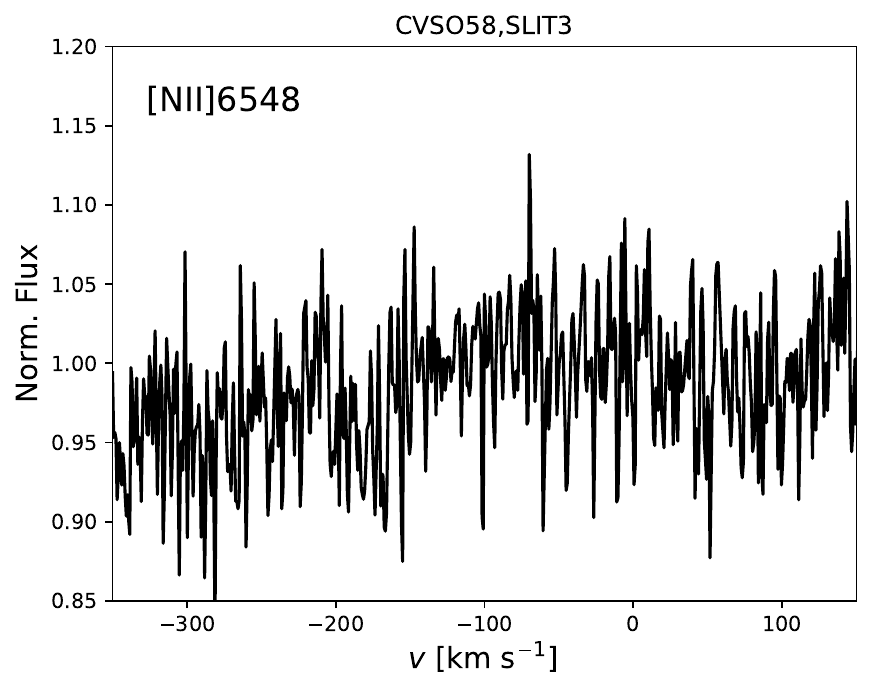}} 
\hfill    \\
\subfloat{\includegraphics[trim=0 0 0 0, clip, width=0.25 \textwidth]{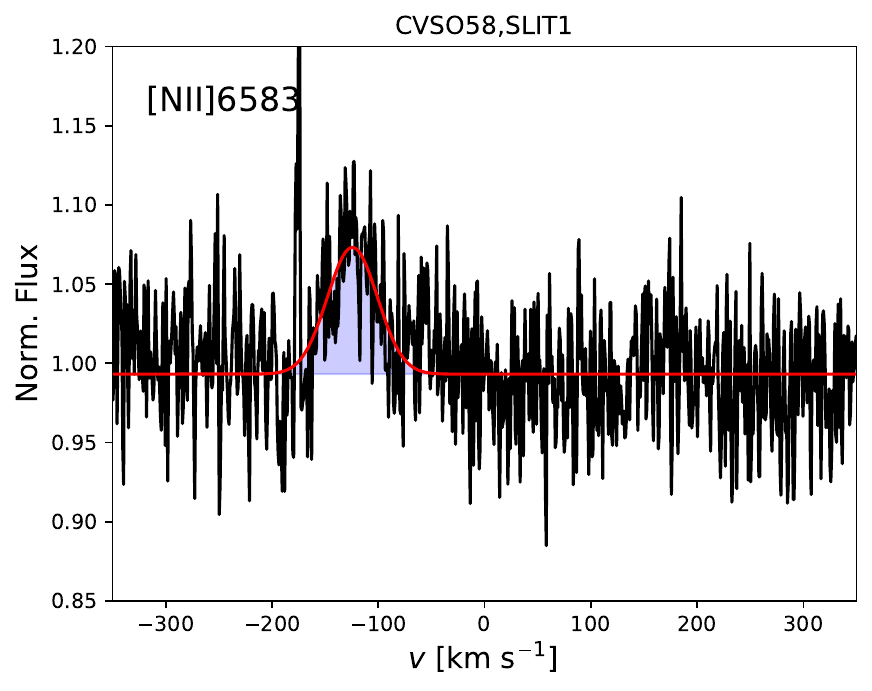}}
\hfill
\subfloat{\includegraphics[trim=0 0 0 0, clip, width=0.25 \textwidth]{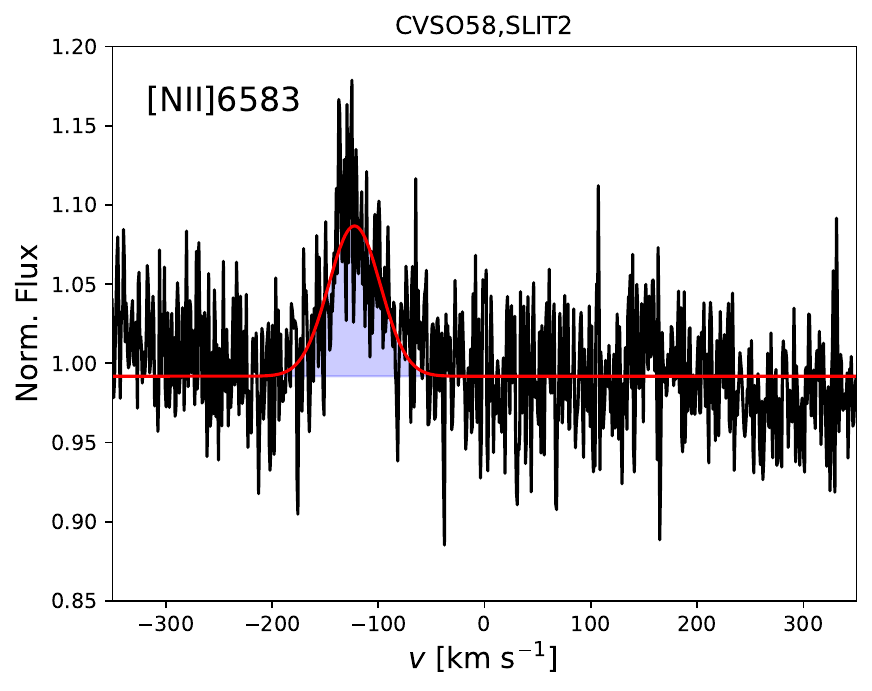}}
\hfill
\subfloat{\includegraphics[trim=0 0 0 0, clip, width=0.25 \textwidth]{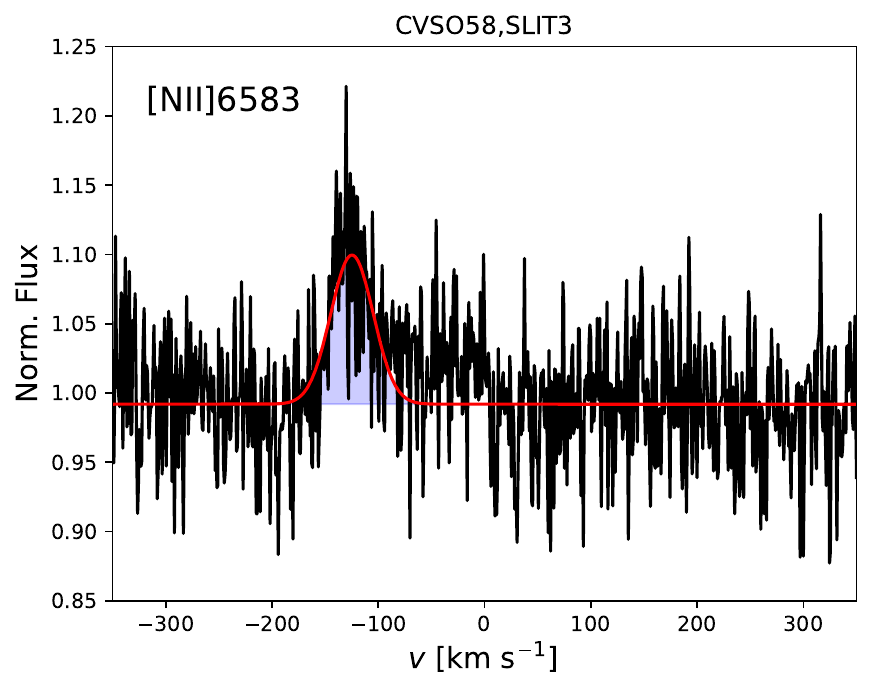}} 
\hfill \\
\subfloat{\includegraphics[trim=0 0 0 0, clip, width=0.25 \textwidth]{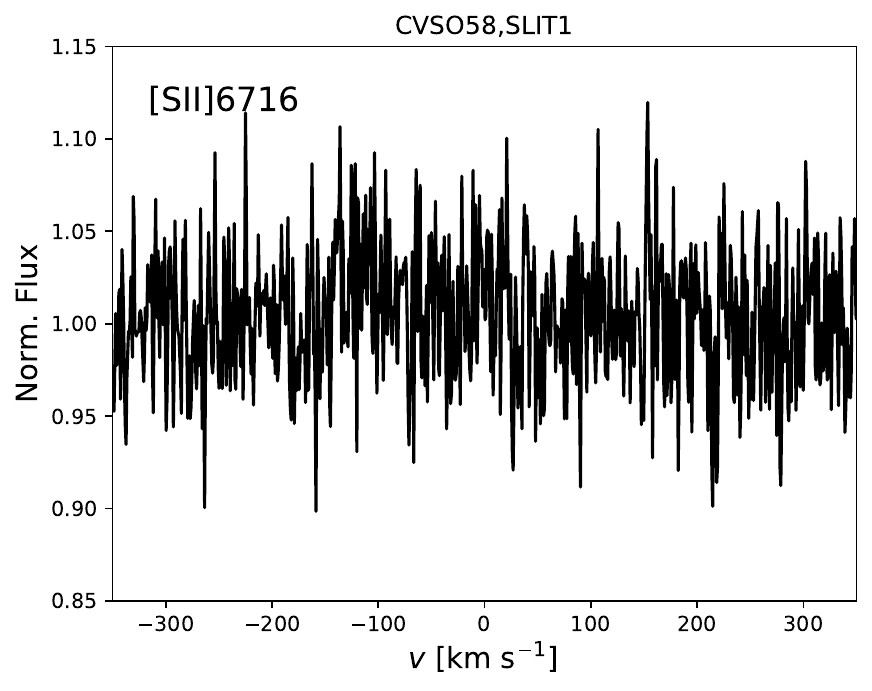}}
\hfill
\subfloat{\includegraphics[trim=0 0 0 0, clip, width=0.25 \textwidth]{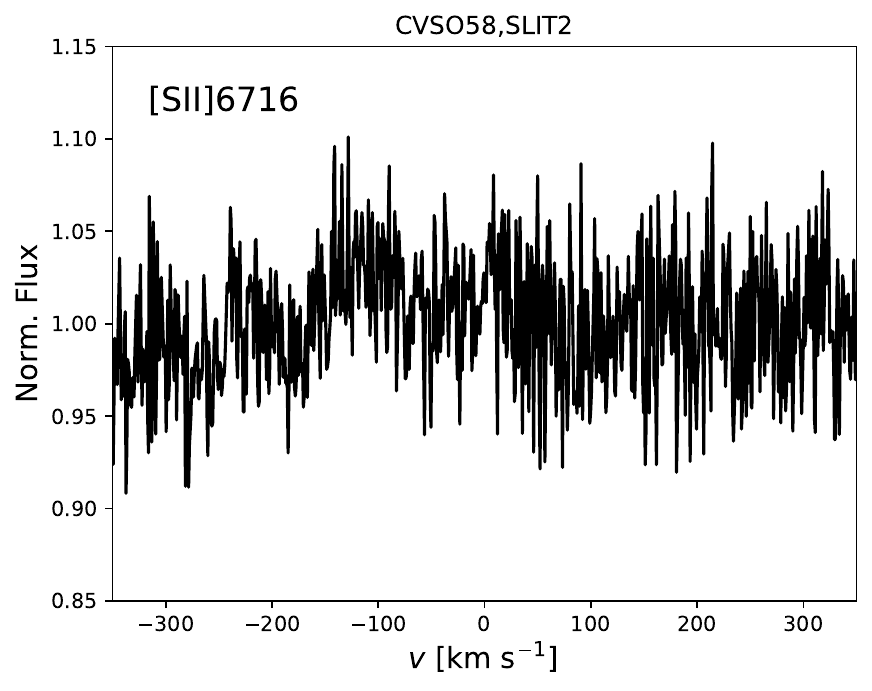}}
\hfill
\subfloat{\includegraphics[trim=0 0 0 0, clip, width=0.25 \textwidth]{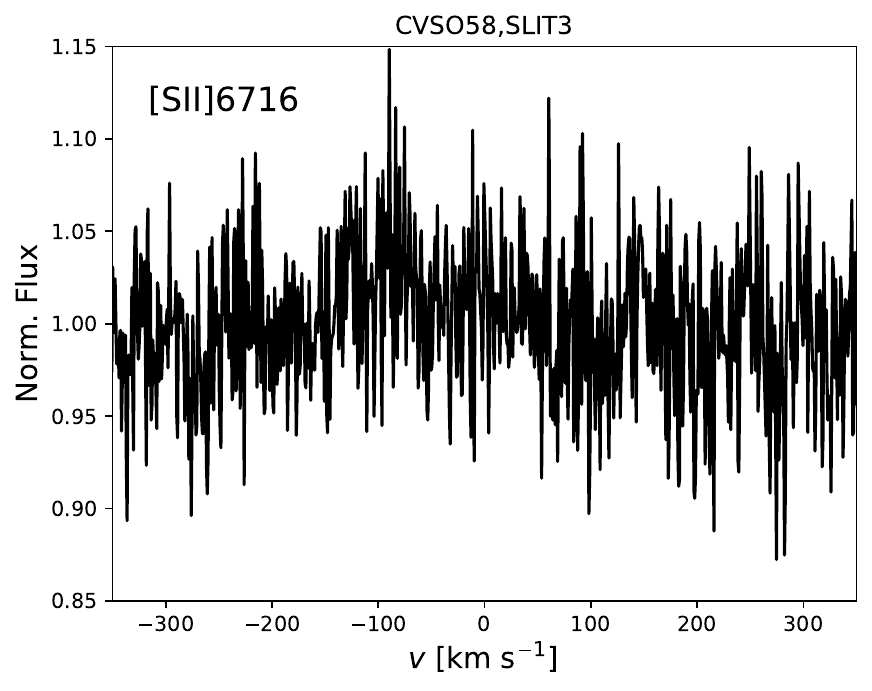}} 
\hfill \\
\subfloat{\includegraphics[trim=0 0 0 0, clip, width=0.25 \textwidth]{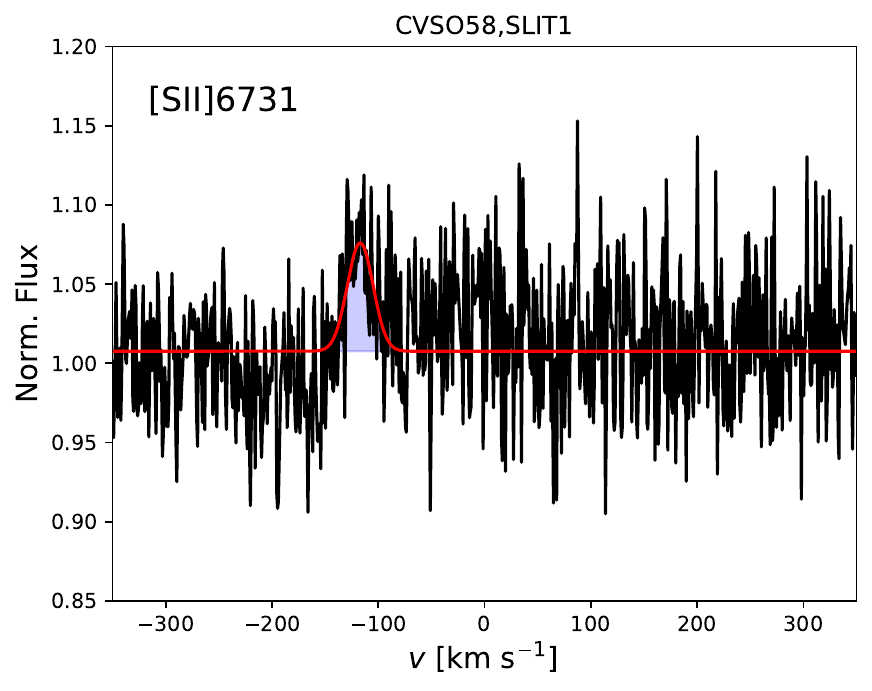}}
\hfill
\subfloat{\includegraphics[trim=0 0 0 0, clip, width=0.25 \textwidth]{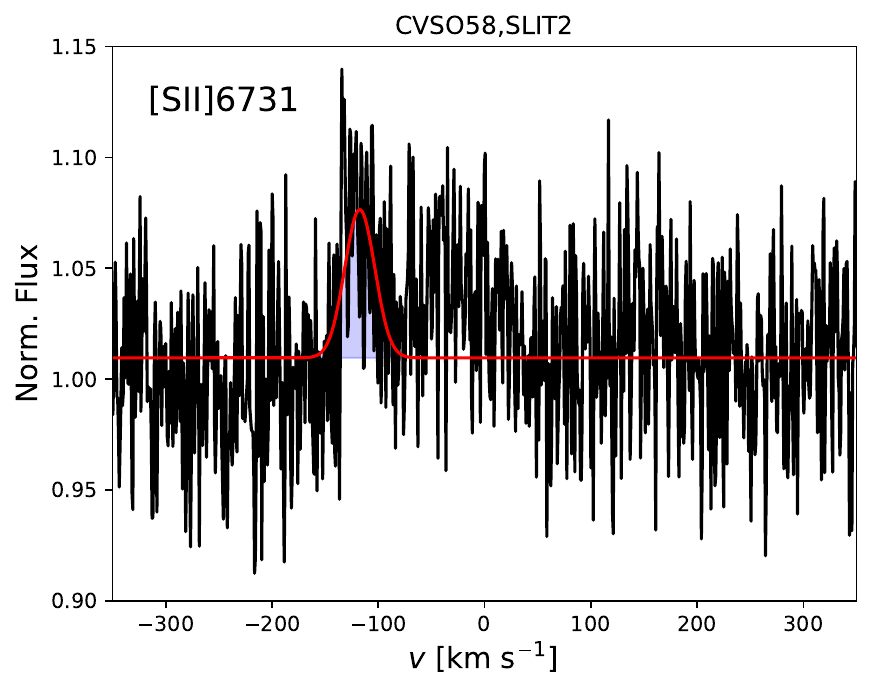}}
\hfill
\subfloat{\includegraphics[trim=0 0 0 0, clip, width=0.25 \textwidth]{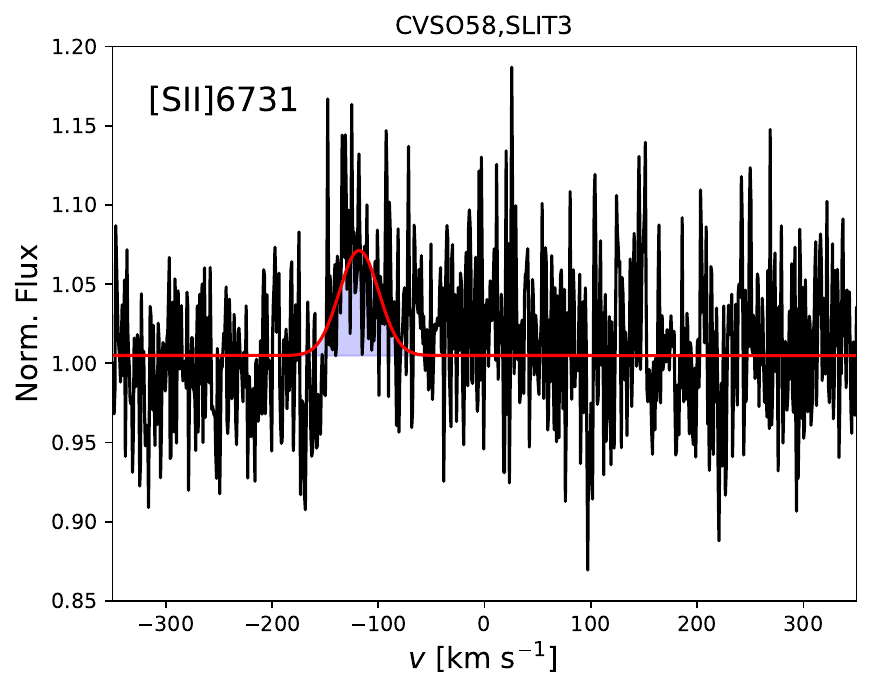}} 
\hfill
\caption{\small{Line profiles CVSO\,58.}}\label{fig:CVSO58}
\end{figure*} 

\begin{figure*} 
\centering
\subfloat{\includegraphics[trim=0 0 0 0, clip, width=0.25 \textwidth]{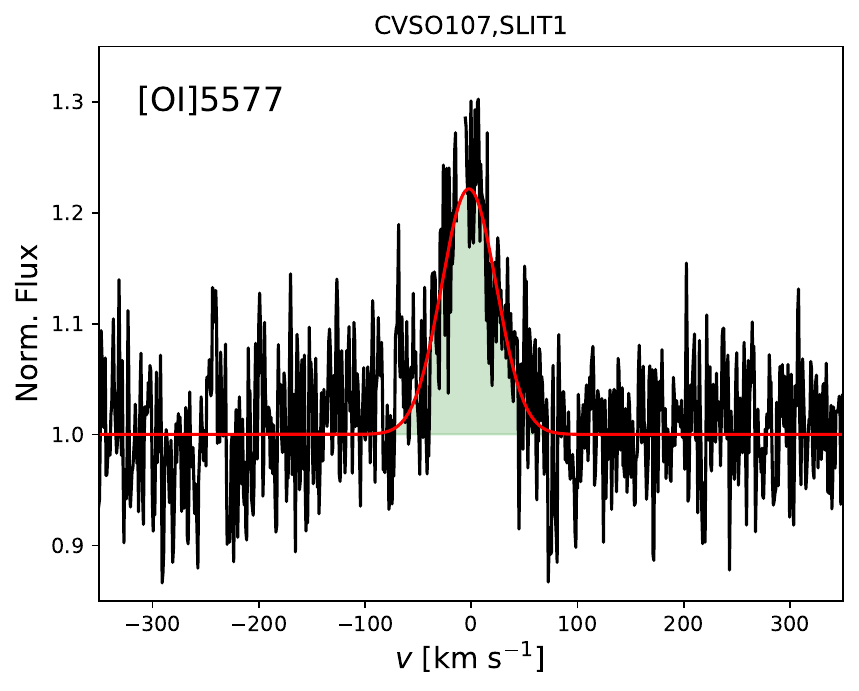}}
\hfill
\subfloat{\includegraphics[trim=0 0 0 0, clip, width=0.25 \textwidth]{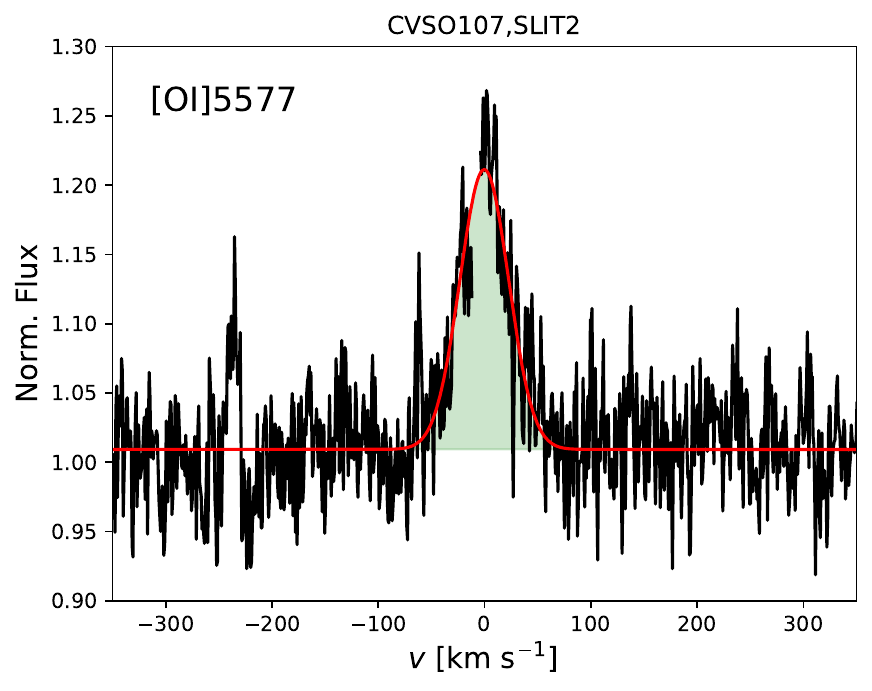}}
\hfill
\subfloat{\includegraphics[trim=0 0 0 0, clip, width=0.25 \textwidth]{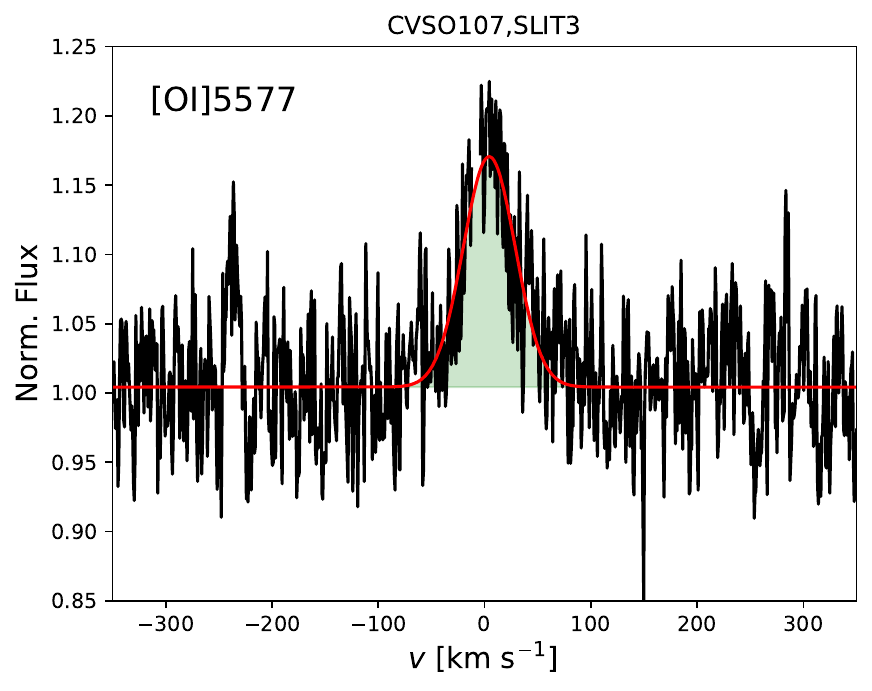}}
\hfill   \\
\subfloat{\includegraphics[trim=0 0 0 0, clip, width=0.25 \textwidth]{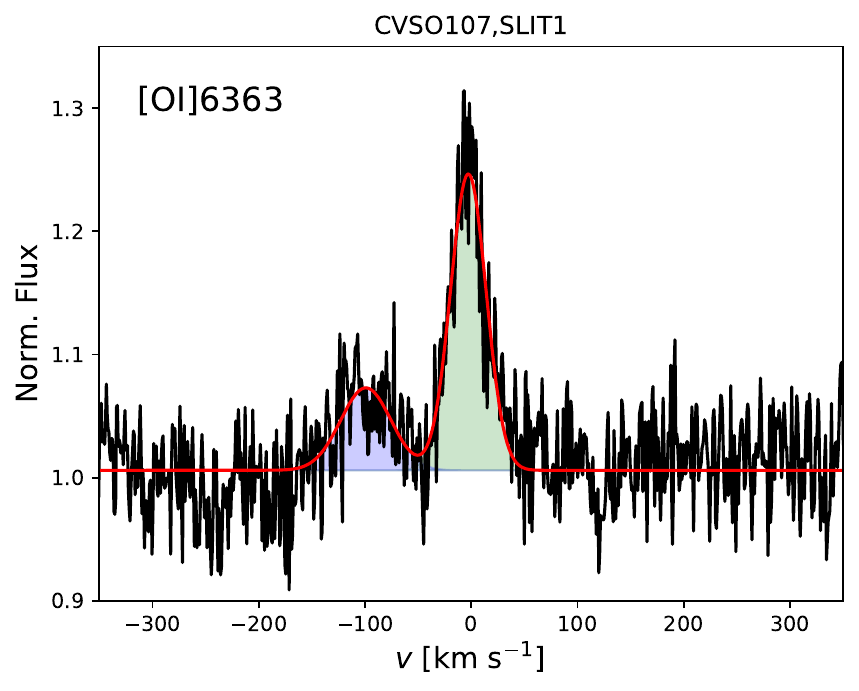}}
\hfill
\subfloat{\includegraphics[trim=0 0 0 0, clip, width=0.25 \textwidth]{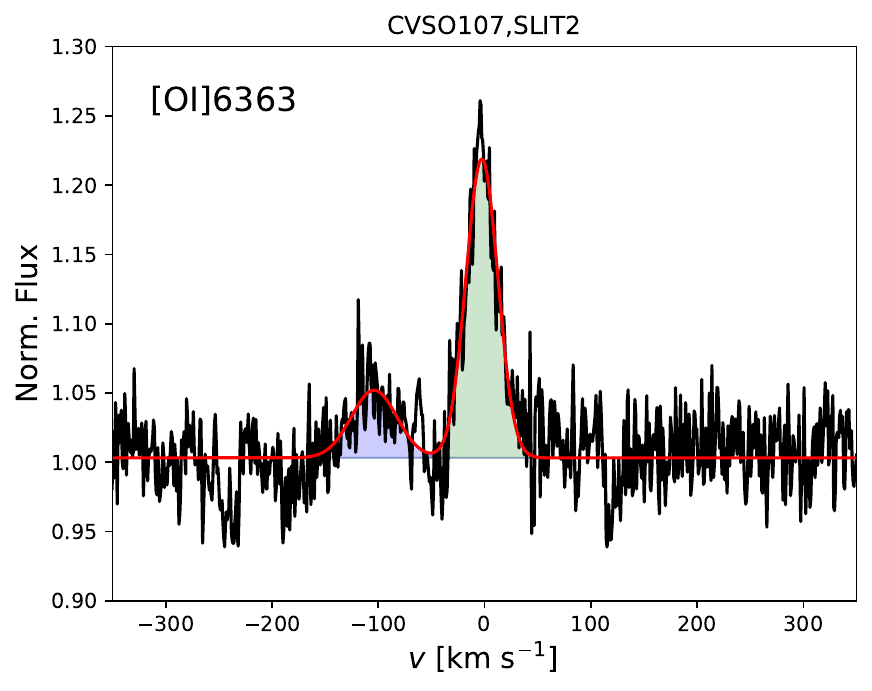}}
\hfill
\subfloat{\includegraphics[trim=0 0 0 0, clip, width=0.25 \textwidth]{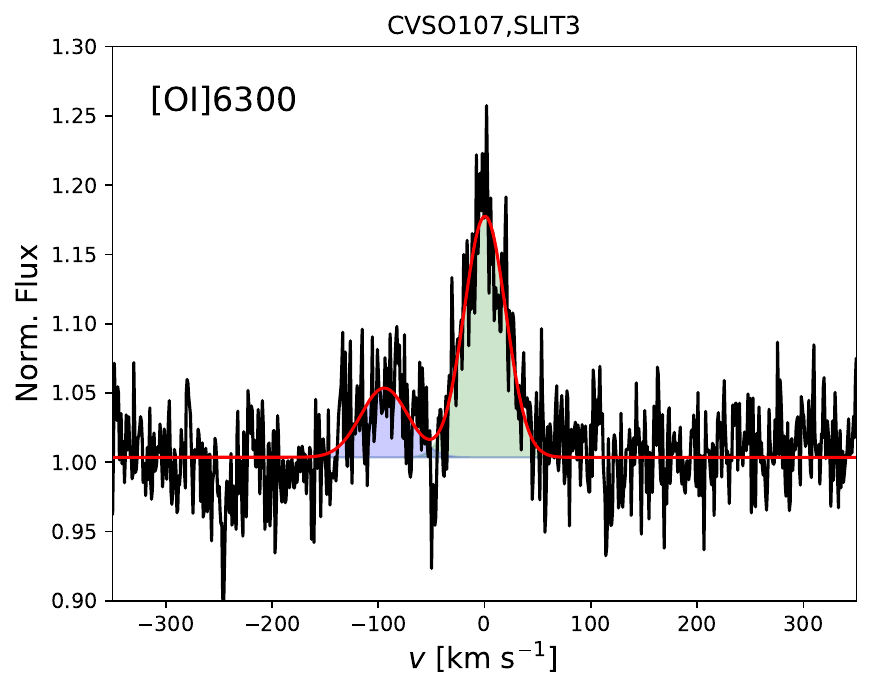}} 
\hfill \\
\subfloat{\includegraphics[trim=0 0 0 0, clip, width=0.25 \textwidth]{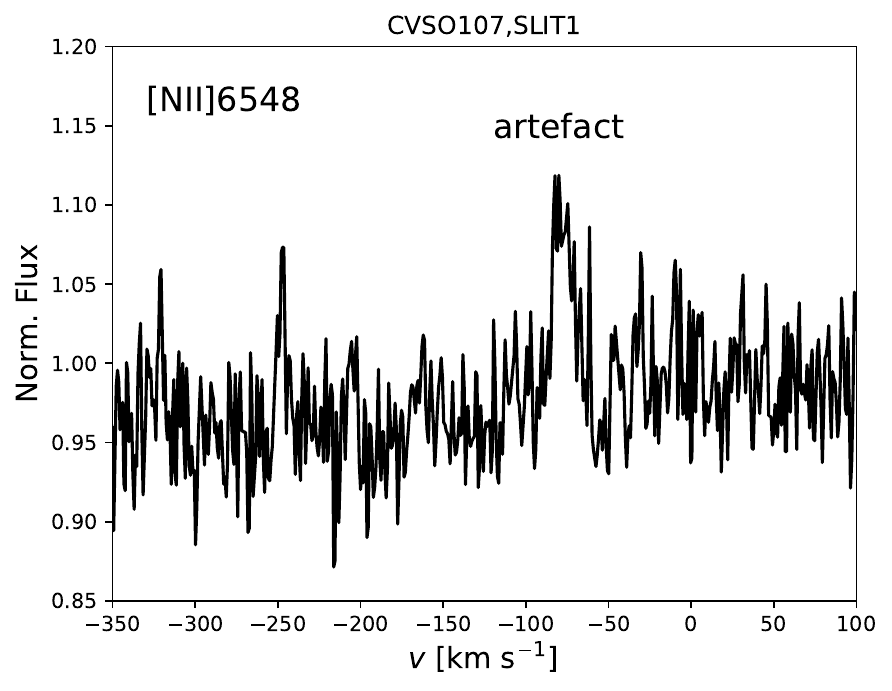}}
\hfill
\subfloat{\includegraphics[trim=0 0 0 0, clip, width=0.25 \textwidth]{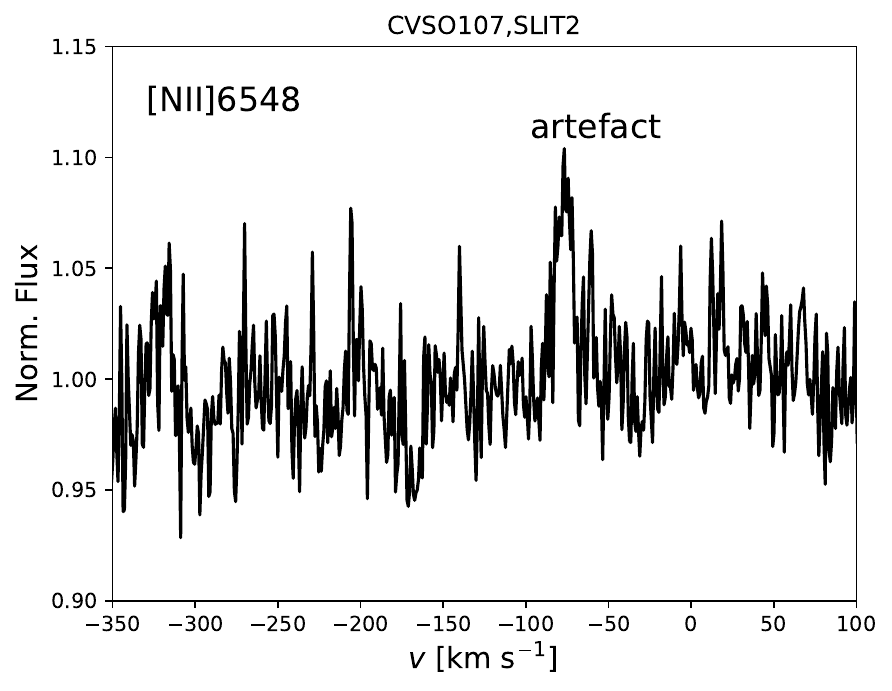}}
\hfill
\subfloat{\includegraphics[trim=0 0 0 0, clip, width=0.25 \textwidth]{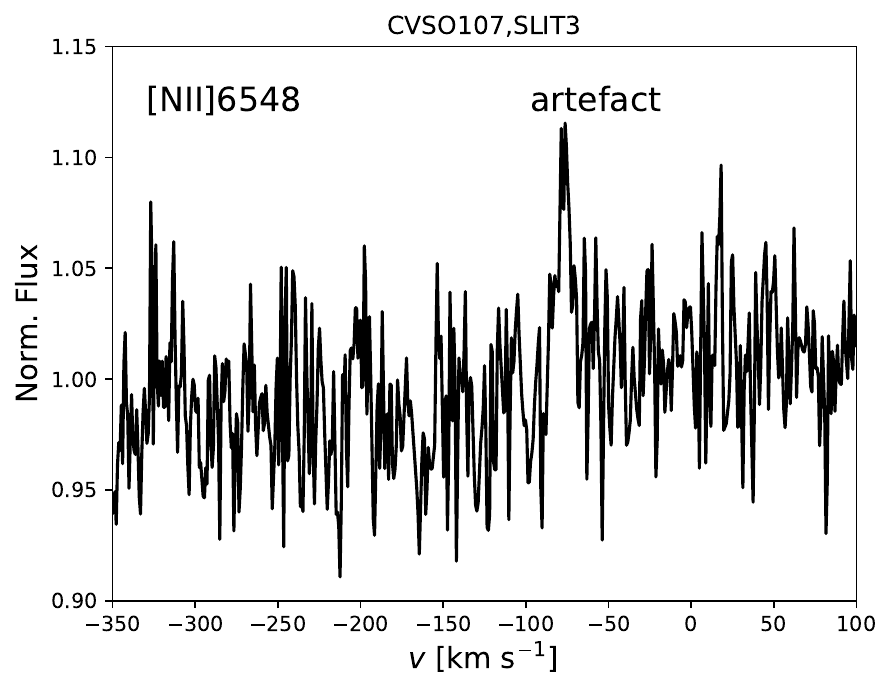}} 
\hfill    \\
\subfloat{\includegraphics[trim=0 0 0 0, clip, width=0.25 \textwidth]{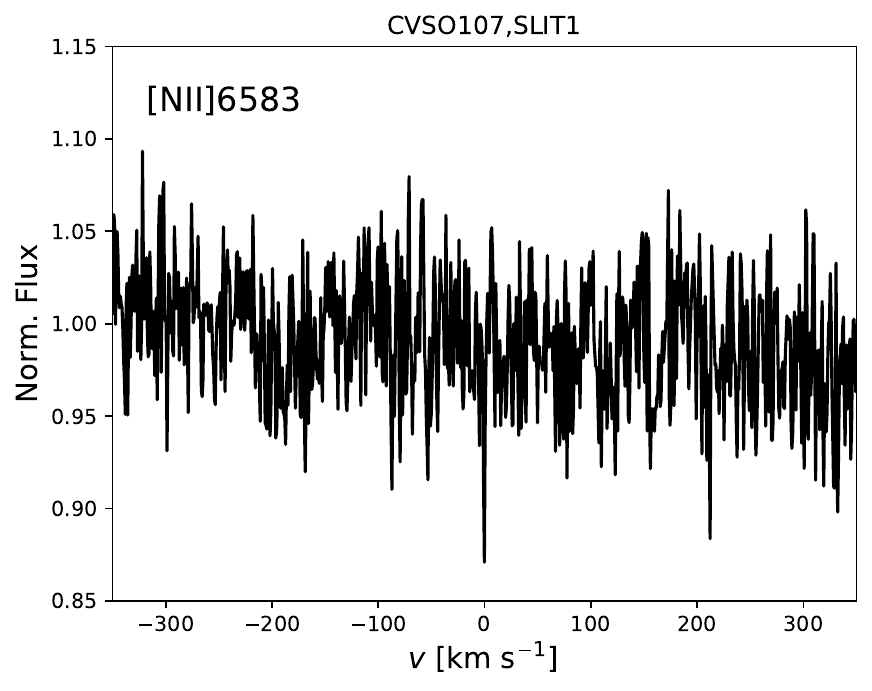}}
\hfill
\subfloat{\includegraphics[trim=0 0 0 0, clip, width=0.25 \textwidth]{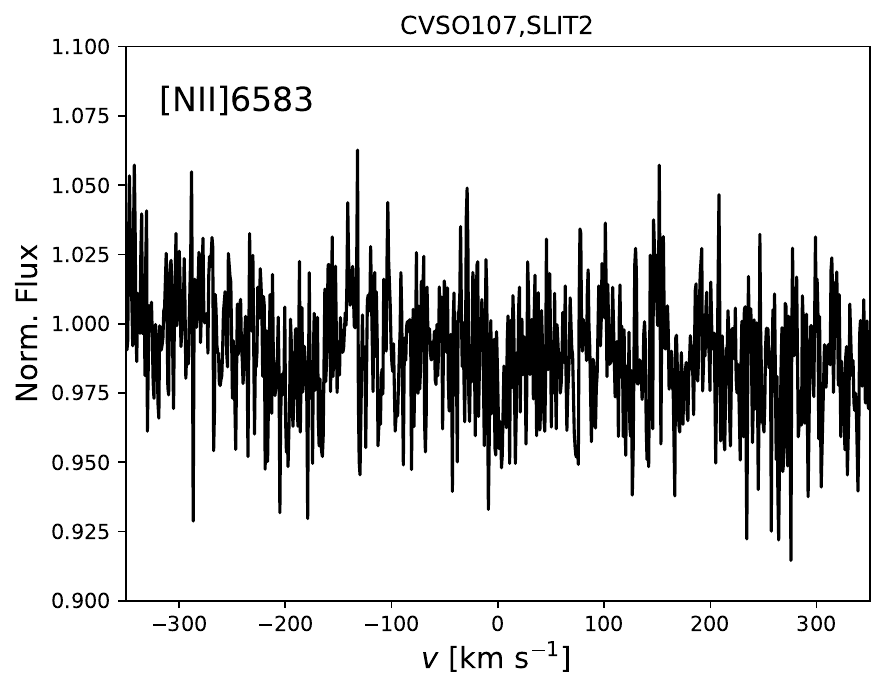}}
\hfill
\subfloat{\includegraphics[trim=0 0 0 0, clip, width=0.25 \textwidth]{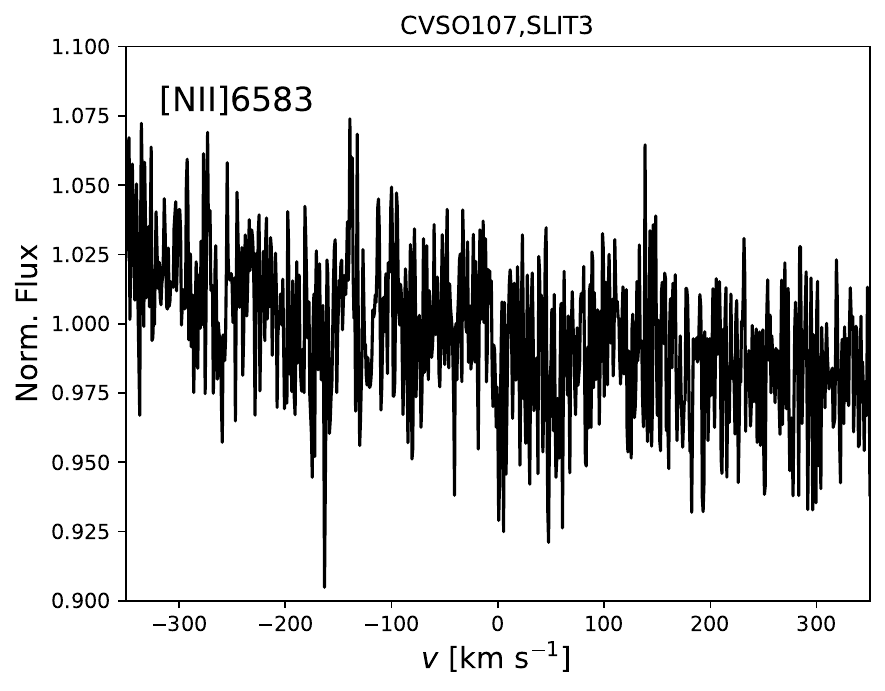}} 
\hfill \\
\subfloat{\includegraphics[trim=0 0 0 0, clip, width=0.25 \textwidth]{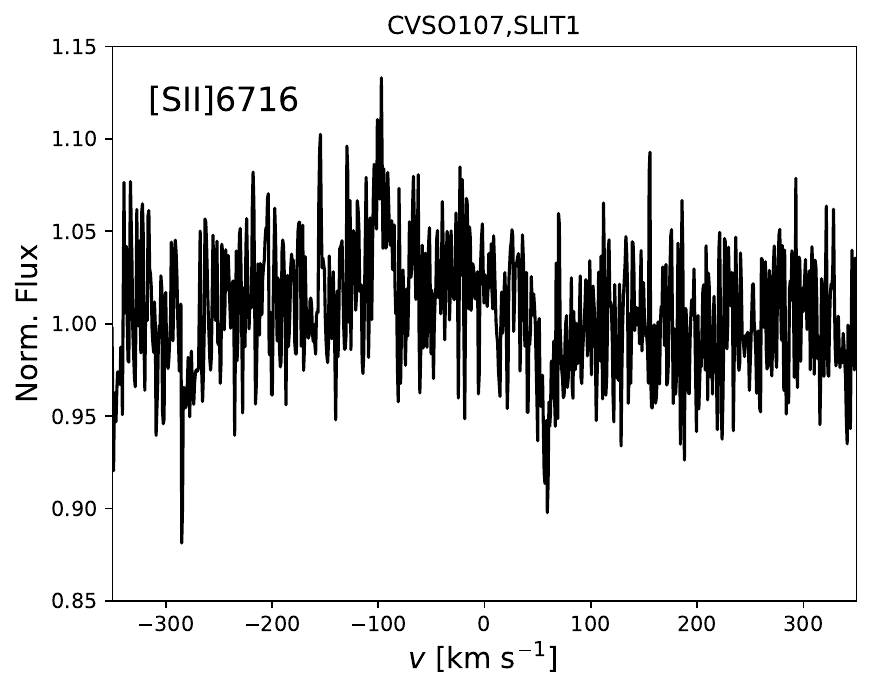}}
\hfill
\subfloat{\includegraphics[trim=0 0 0 0, clip, width=0.25 \textwidth]{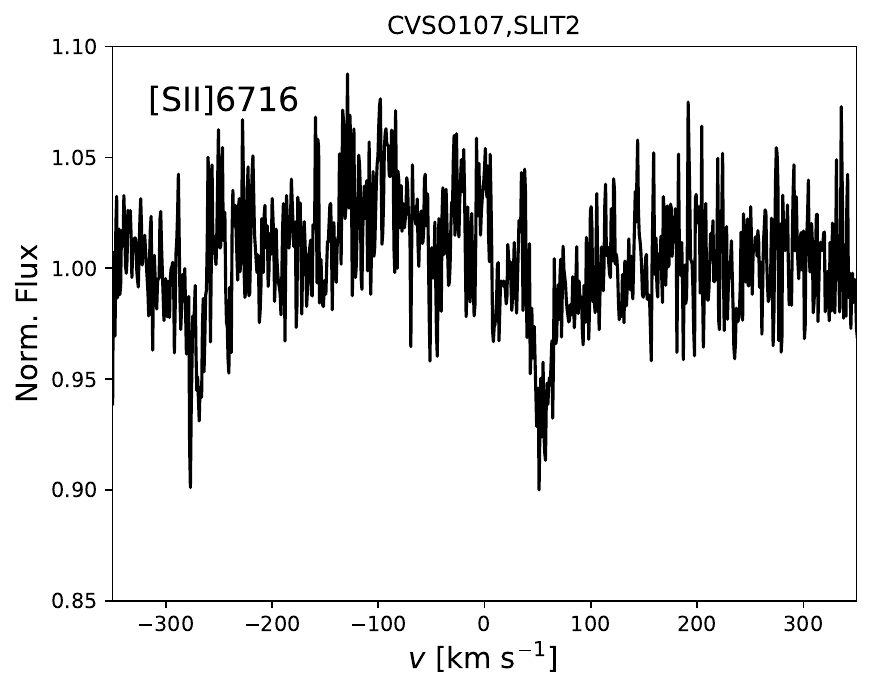}}
\hfill
\subfloat{\includegraphics[trim=0 0 0 0, clip, width=0.25 \textwidth]{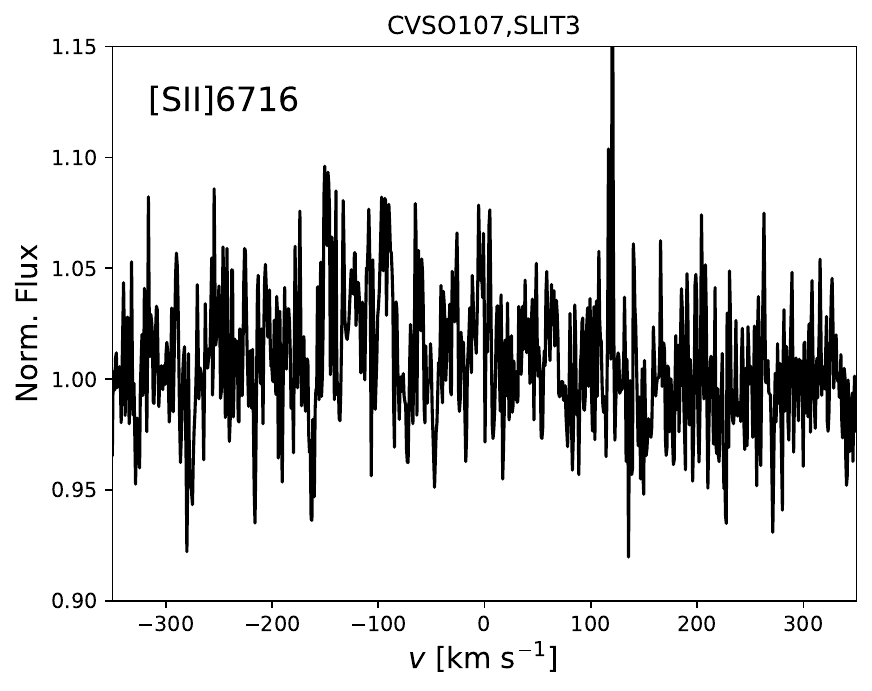}} 
\hfill \\
\subfloat{\includegraphics[trim=0 0 0 0, clip, width=0.25 \textwidth]{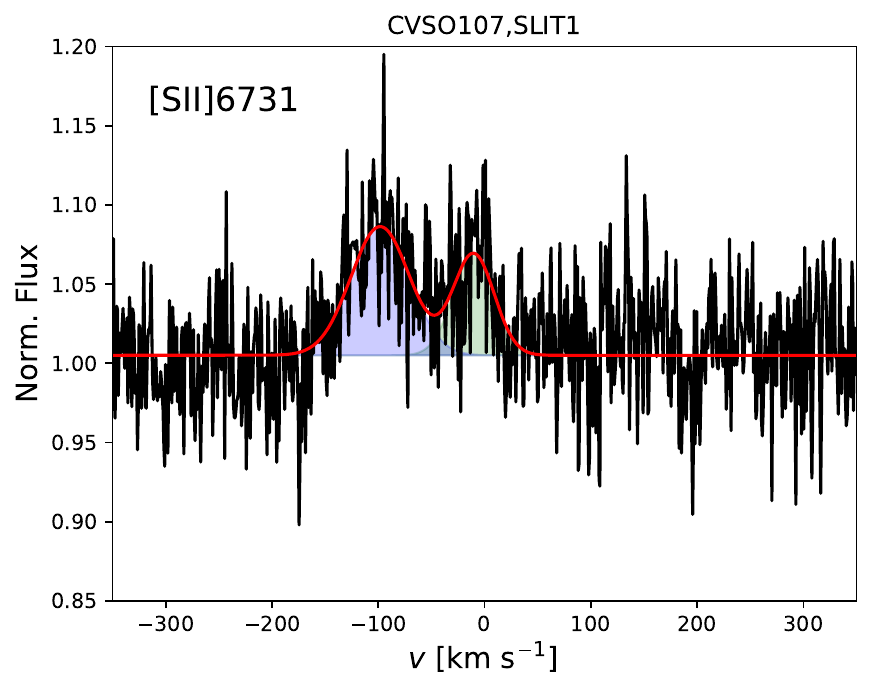}}
\hfill
\subfloat{\includegraphics[trim=0 0 0 0, clip, width=0.25 \textwidth]{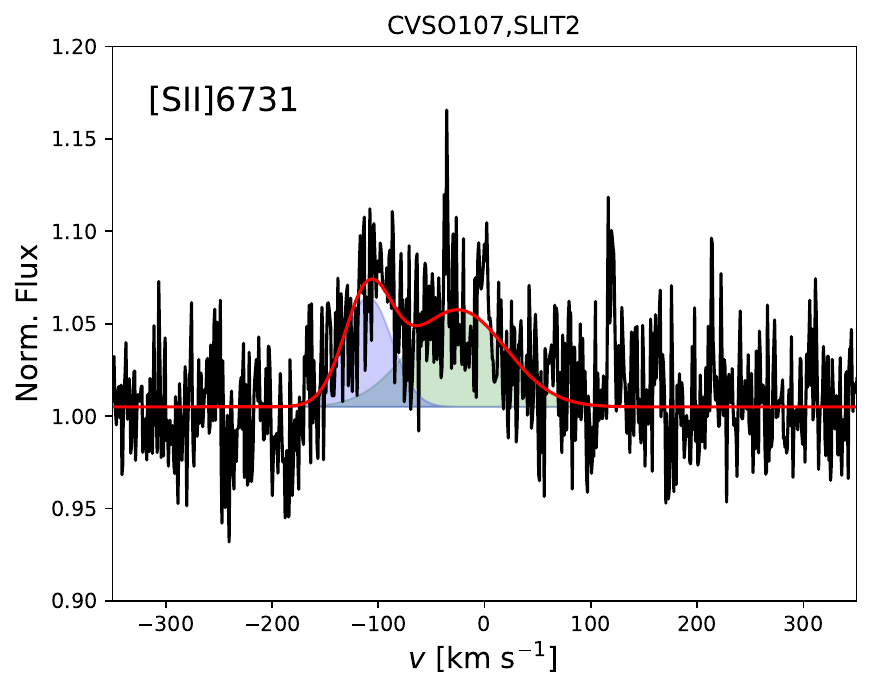}}
\hfill
\subfloat{\includegraphics[trim=0 0 0 0, clip, width=0.25 \textwidth]{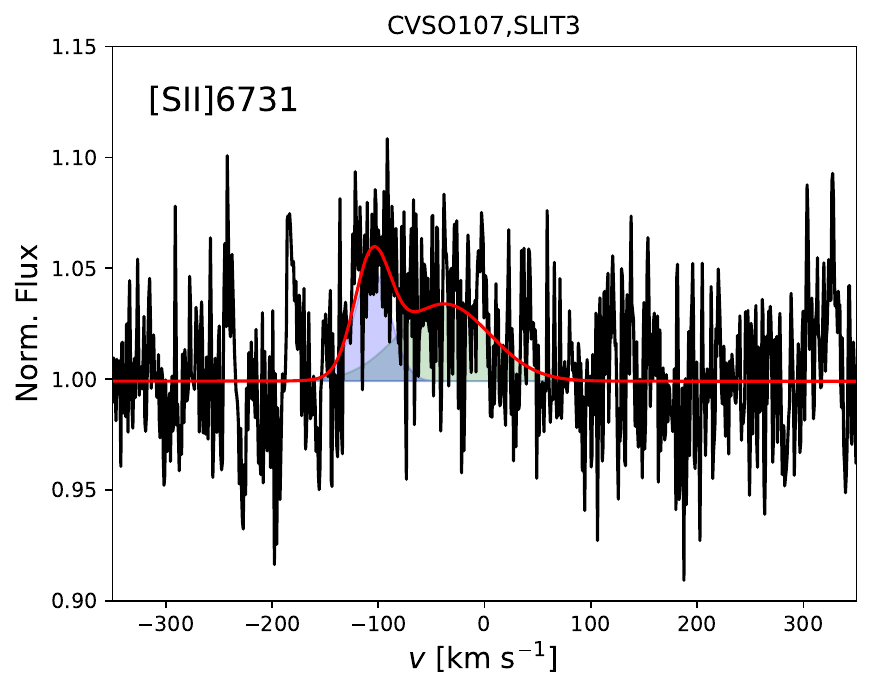}} 
\hfill
\caption{\small{Line profiles CVSO\,107.}}\label{fig:CVSO107}
\end{figure*} 

\begin{figure*} 
\centering
\subfloat{\includegraphics[trim=0 0 0 0, clip, width=0.25 \textwidth]{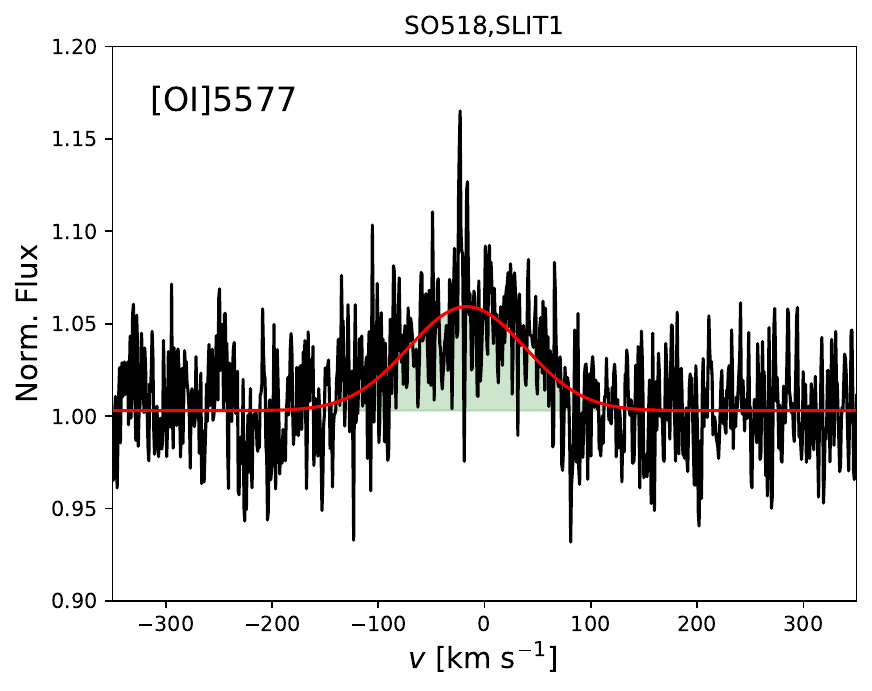}}
\hfill
\subfloat{\includegraphics[trim=0 0 0 0, clip, width=0.25 \textwidth]{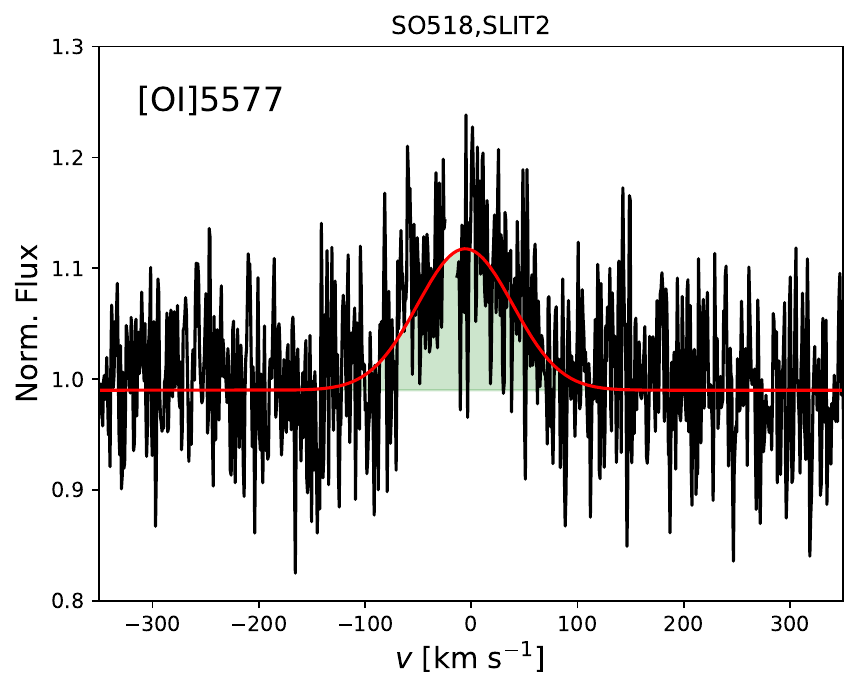}}
\hfill
\subfloat{\includegraphics[trim=0 0 0 0, clip, width=0.25 \textwidth]{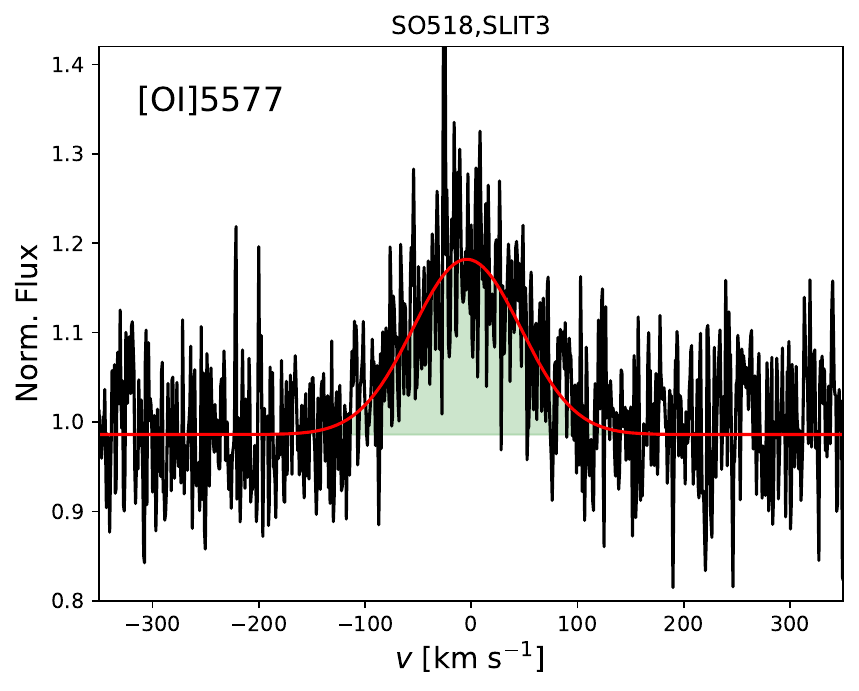}}
\hfill   \\
\subfloat{\includegraphics[trim=0 0 0 0, clip, width=0.25 \textwidth]{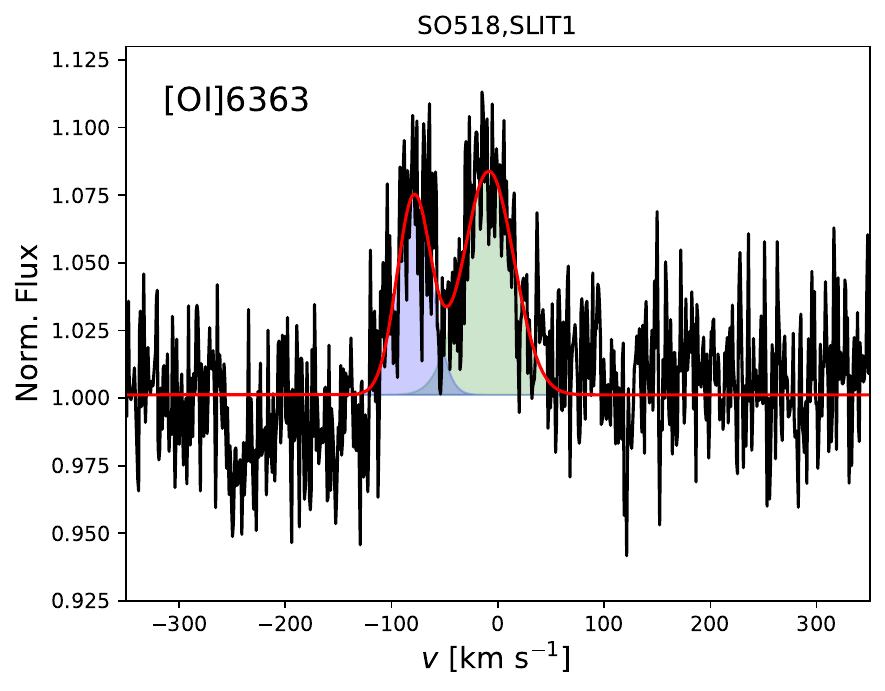}}
\hfill
\subfloat{\includegraphics[trim=0 0 0 0, clip, width=0.25 \textwidth]{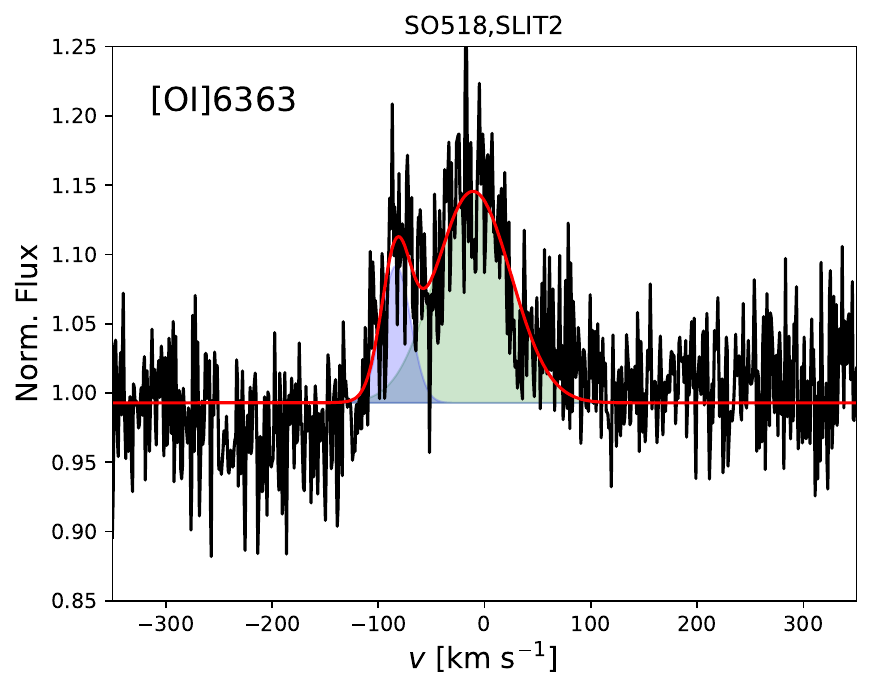}}
\hfill
\subfloat{\includegraphics[trim=0 0 0 0, clip, width=0.25 \textwidth]{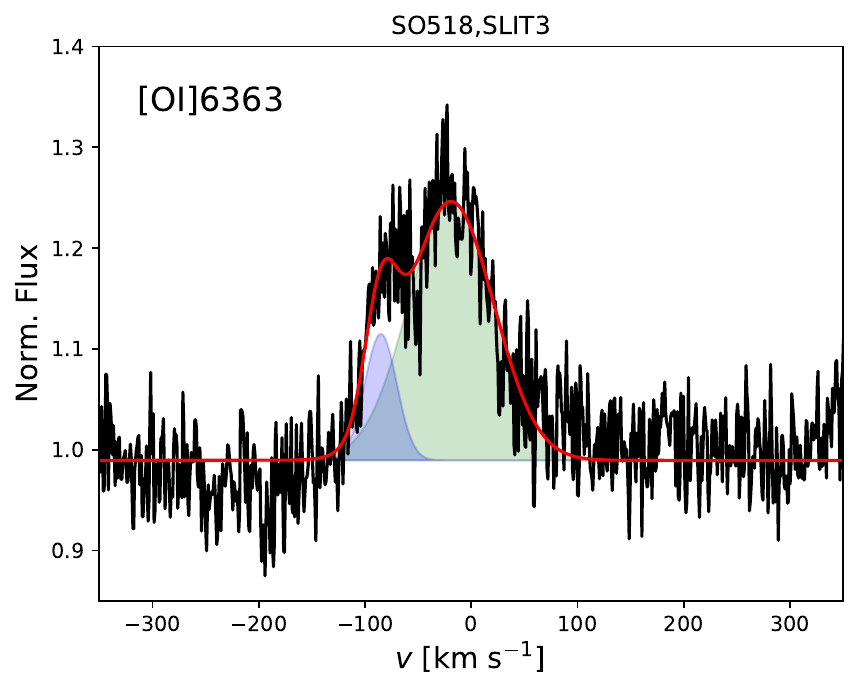}} 
\hfill \\
\subfloat{\includegraphics[trim=0 0 0 0, clip, width=0.25 \textwidth]{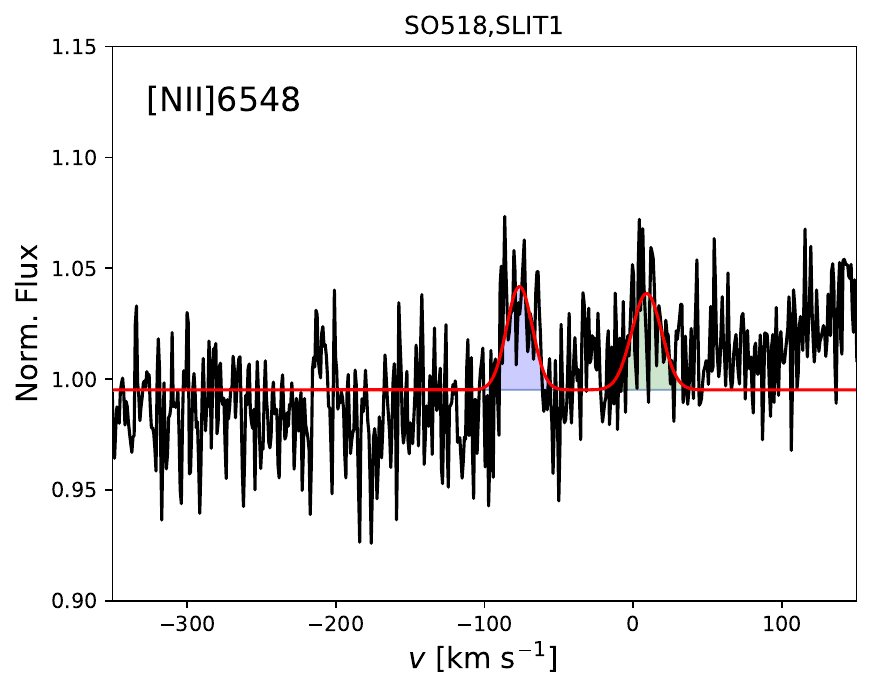}}
\hfill
\subfloat{\includegraphics[trim=0 0 0 0, clip, width=0.25 \textwidth]{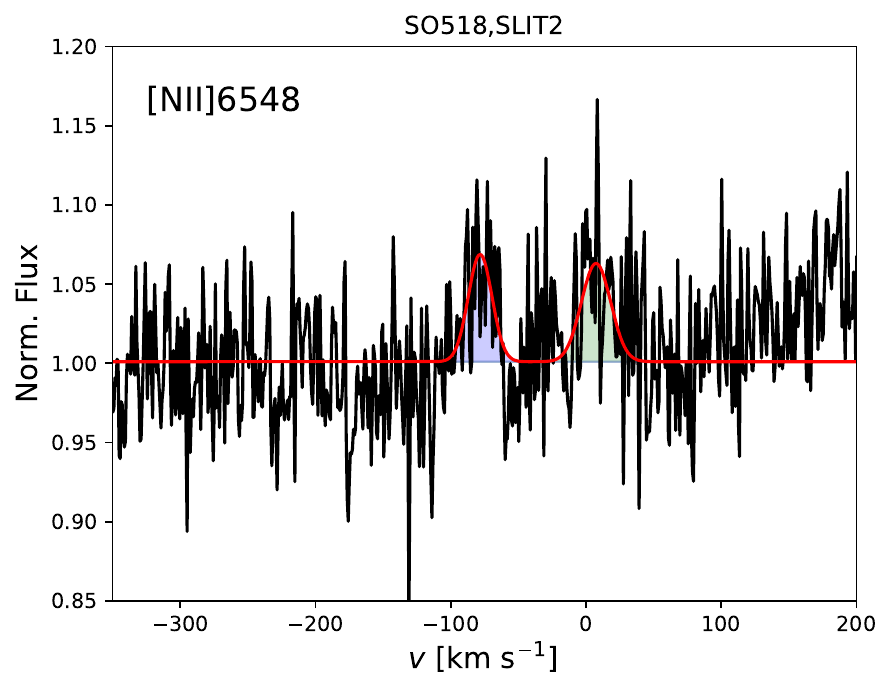}}
\hfill
\subfloat{\includegraphics[trim=0 0 0 0, clip, width=0.25 \textwidth]{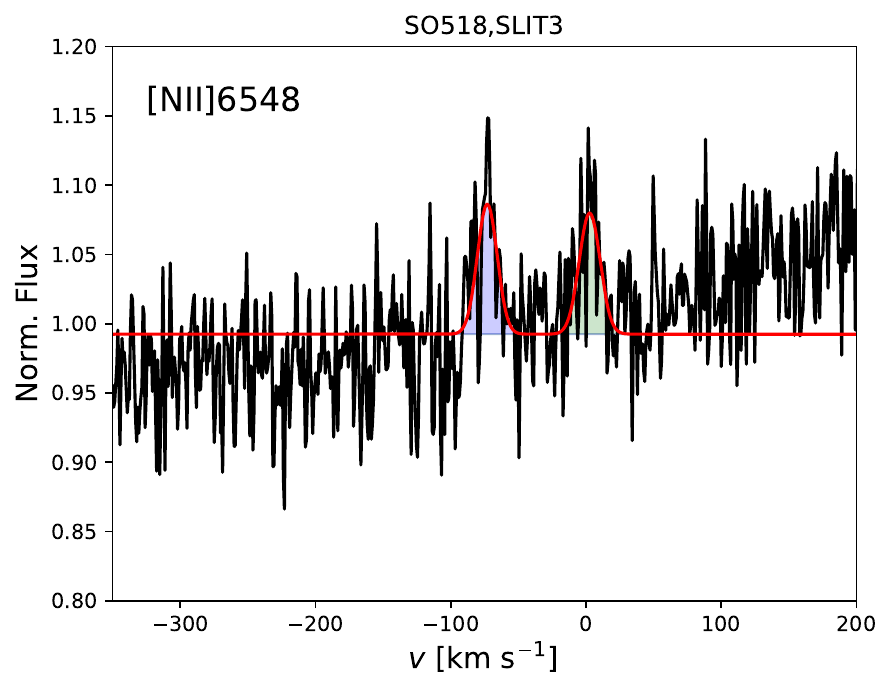}} 
\hfill    \\
\subfloat{\includegraphics[trim=0 0 0 0, clip, width=0.25 \textwidth]{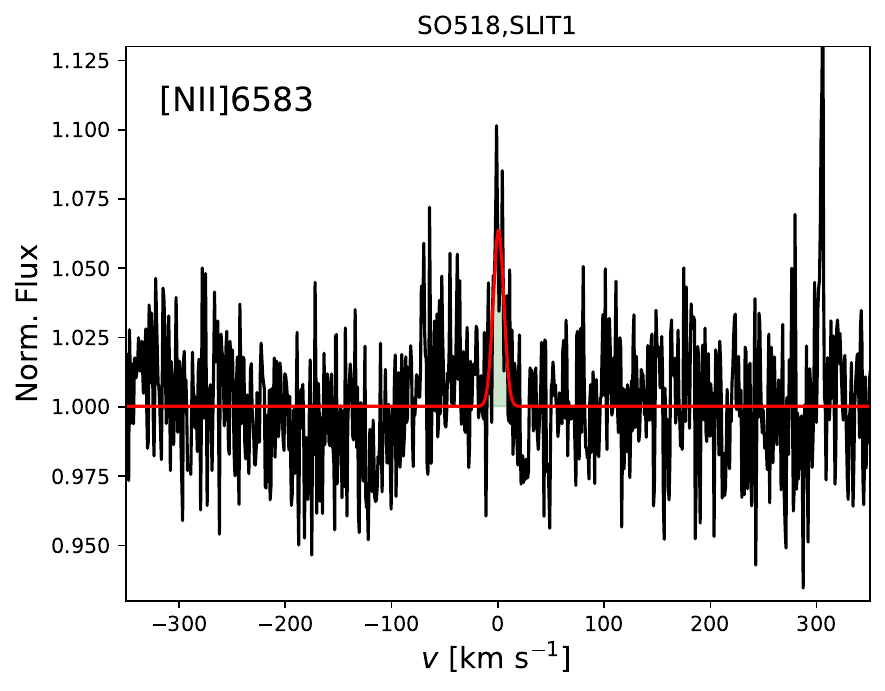}}
\hfill
\subfloat{\includegraphics[trim=0 0 0 0, clip, width=0.25 \textwidth]{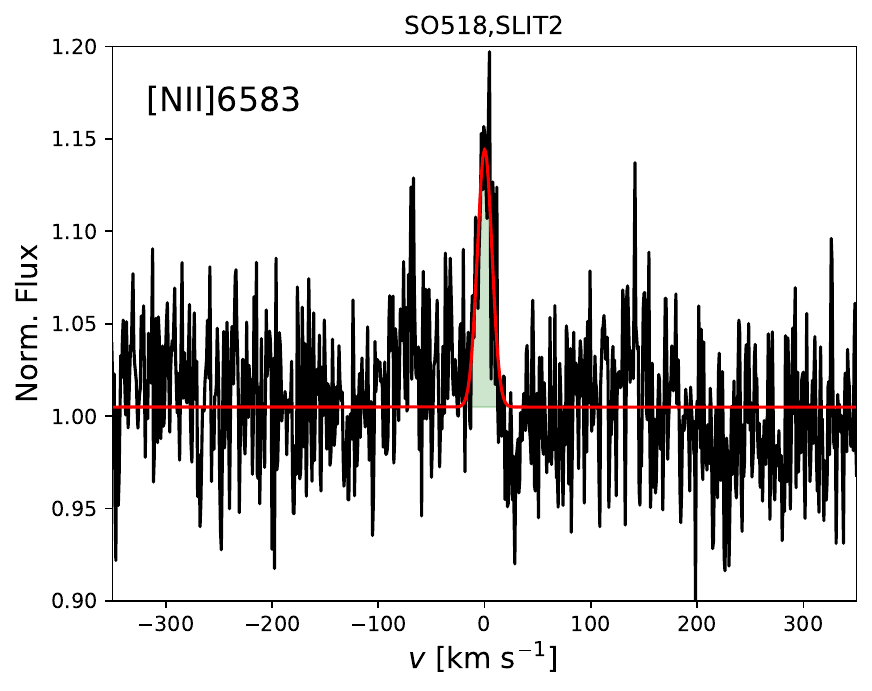}}
\hfill
\subfloat{\includegraphics[trim=0 0 0 0, clip, width=0.25 \textwidth]{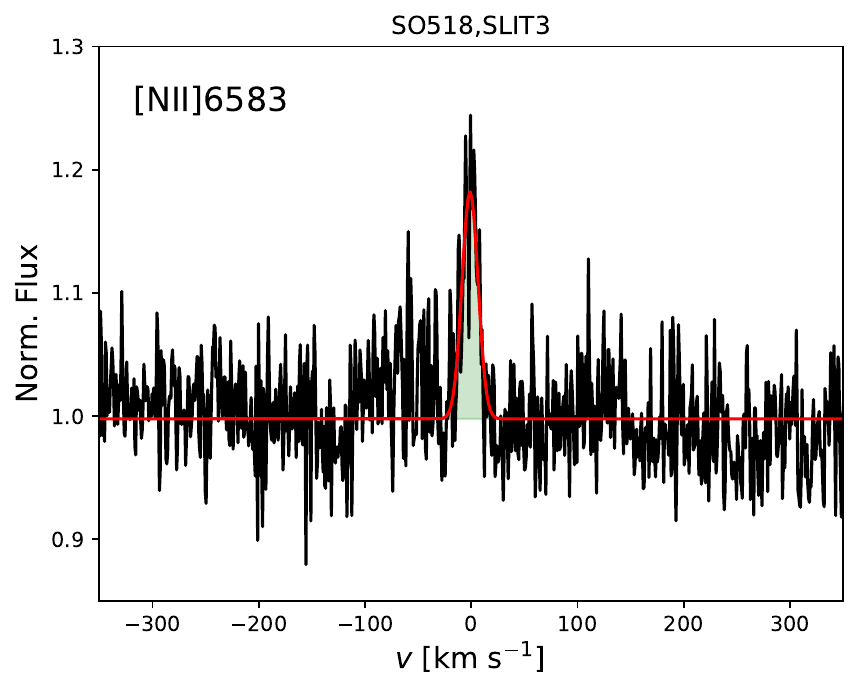}} 
\hfill \\
\subfloat{\includegraphics[trim=0 0 0 0, clip, width=0.25 \textwidth]{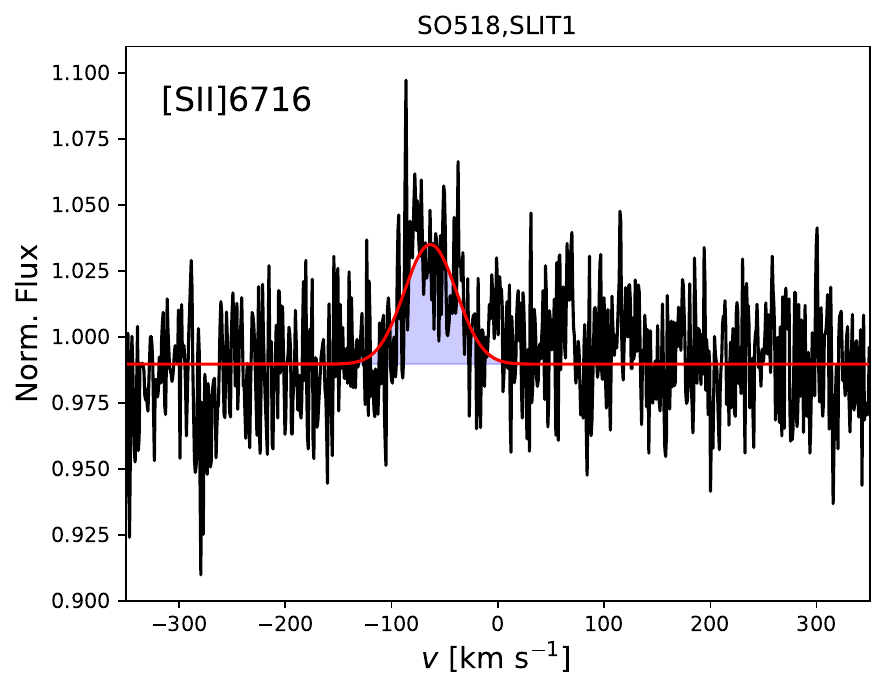}}
\hfill
\subfloat{\includegraphics[trim=0 0 0 0, clip, width=0.25 \textwidth]{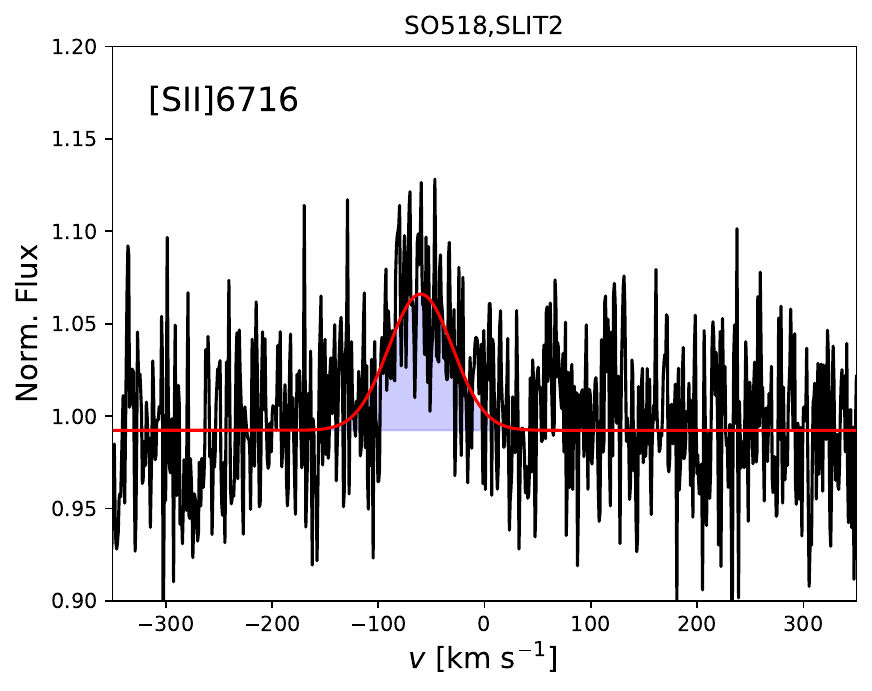}}
\hfill
\subfloat{\includegraphics[trim=0 0 0 0, clip, width=0.25 \textwidth]{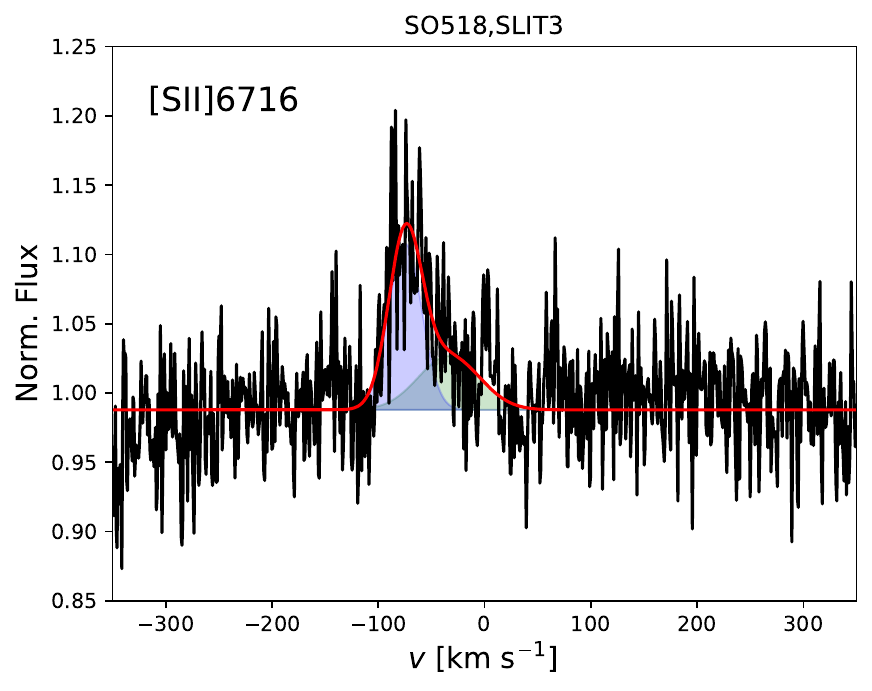}} 
\hfill \\
\subfloat{\includegraphics[trim=0 0 0 0, clip, width=0.25 \textwidth]{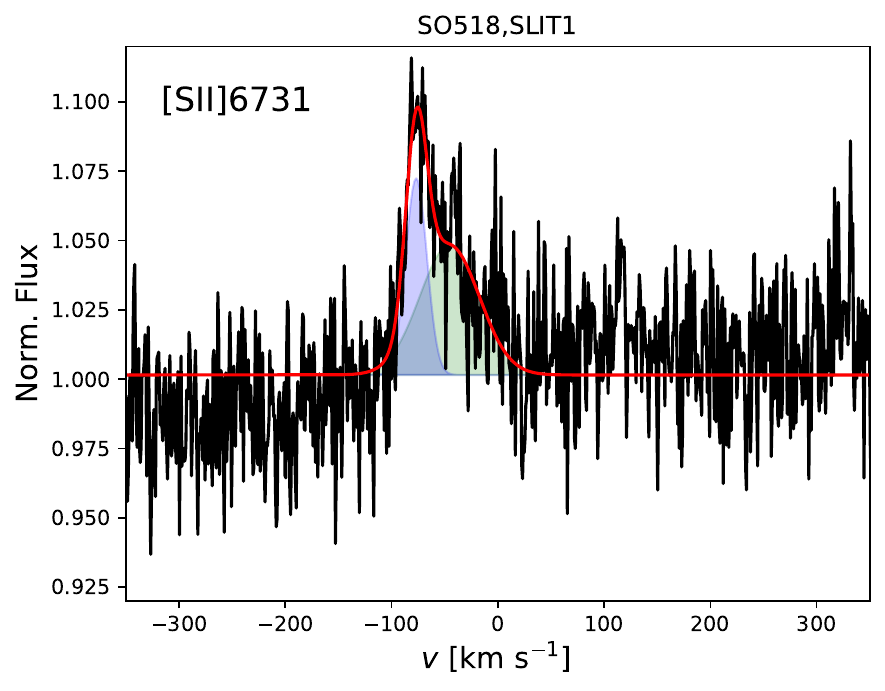}}
\hfill
\subfloat{\includegraphics[trim=0 0 0 0, clip, width=0.25 \textwidth]{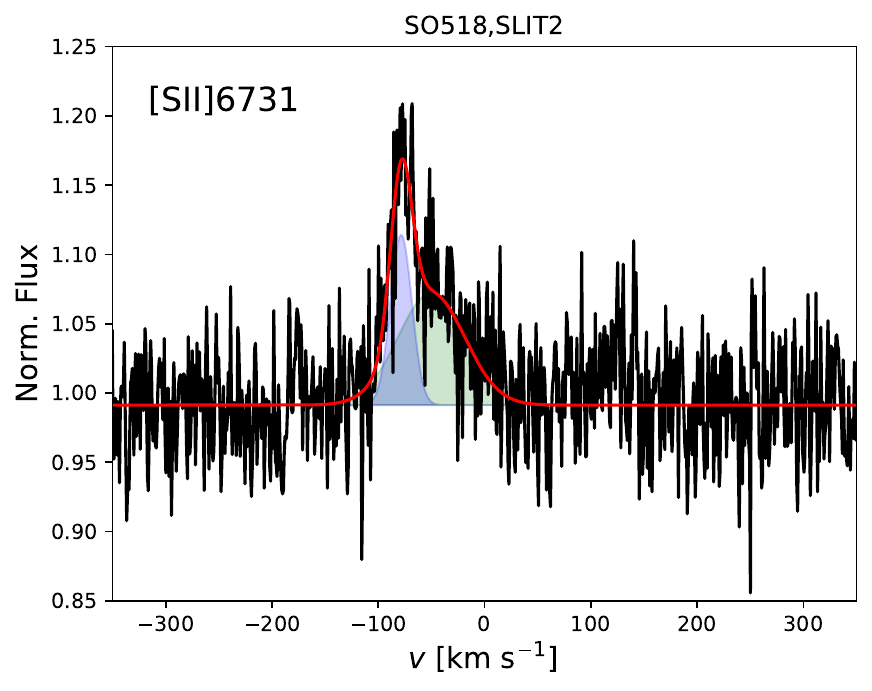}}
\hfill
\subfloat{\includegraphics[trim=0 0 0 0, clip, width=0.25 \textwidth]{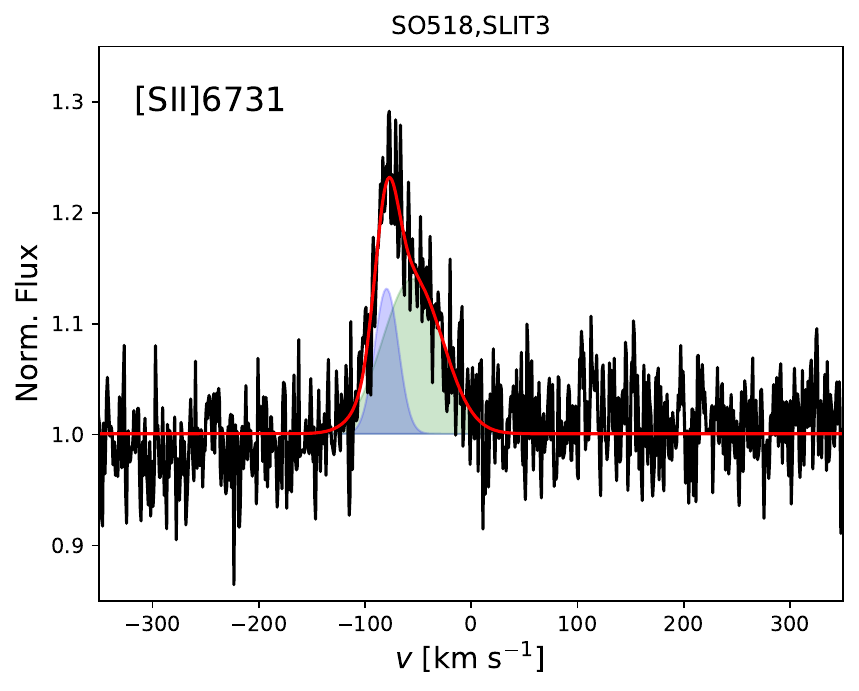}} 
\hfill
\caption{\small{Line profiles SO518.}}\label{fig:SO518}
\end{figure*} 

\pagebreak

\setcounter{table}{0}
\renewcommand{\thetable}{F.\arabic{table}}
  
 \setcounter{figure}{0}
\renewcommand\thefigure{\thesection F.\arabic{figure}}  

\setcounter{equation}{0}
\renewcommand{\theequation}{F.\arabic{equation}}  

\begin{table*}
{\def\arraystretch{2}\tabcolsep=1pt
\tiny
\caption{\small{Properties of the [OI]$\lambda$6300 components.}}\label{table:OI6300components_part_I}
\centering
\begin{tabular}{|c||c|c||c|c||c|c||}
\hline
 \multirow{1}{*}{ \textbf{Target}} &
      \multicolumn{2}{|c|}{\textbf{Comp. 1 (HVC)}}  &
      \multicolumn{2}{|c|}{\textbf{Comp. 2 (NLVC)}} &
      \multicolumn{2}{|c|}{\textbf{Comp. 3 (BLVC)}} \\
\hline\hline
 & $v_p^{1}$ & FWHM$^{2}$ & $v_p^{1}$ & FWHM$^{2}$ & $v_p^{1}$ & FWHM$^{2}$      \\  
 & $\text{km}\,\text{s}^{-1}$ & $\text{km}\,\text{s}^{-1}$ & $\text{km}\,\text{s}^{-1}$ & $\text{km}\,\text{s}^{-1}$  & $\text{km}\,\text{s}^{-1}$ & $\text{km}\,\text{s}^{-1}$  \\ \hline 
 \hline  
 \textbf{Sz\,84}  & - & - & $+0.1\pm 0.4$ & $25.9\pm 0.9$  & - & - \\
ep2            & - & -  & $+0.4\pm 0.5$ & $30.2\pm 1.1$  & - & - \\
ep3            & - & - & $+0.2\pm 0.4$ & $29.7\pm 0.9$   & - & - \\
 \hline 
 \textbf{Sz\,97}  & - & - & $-25.9\pm 3.2$ & $37.4\pm 3.7$ & $+9.7\pm 5.7$ & $127.4\pm 10.0$ \\
ep2             & - & - & $-30.6\pm 1.4$ & $36.6\pm 4.3$ & $27.4\pm 9.5$ & $136.7\pm 16.7$ \\
ep3             & - & - & $-26.8\pm 1.9$ & $37.7\pm 5.4$ & $+8.5\pm 5.5$ & $137.1\pm 10.4$ \\
 \hline 
 \textbf{Sz\,98} & $-107.7\pm 2.4$ & $68.9\pm 7.3$ & $-3.4\pm 0.6$ & $22.9\pm 1.7$ & $-8.7\pm 3.6$ & $158.2\pm 6.7$ \\
 ep2 & $-109.9\pm 2.6$ & $63.0\pm 9.0$ & $-3.3\pm 0.6$ & $19.8\pm 1.5$ & $-19.2\pm 3.2$ & $151.5\pm 6.1$ \\
ep3  & $-113.6\pm 1.4$ & $55.2\pm 5.2$ & $-2.8\pm 0.5$ & $20.9\pm 1.4$ & $-25.2\pm 2.4$ & $145.0\pm 5.0$ \\
 \hline 
\textbf{Sz 99} & $+58.9\pm 5.1$ & $41.5\pm 16.7$ & - & - & $+7.3\pm 15.0$  & $88.8\pm 23.1$ \\
ep2 & $+54.6\pm 1.2$ & $36.5\pm 2.3$ & - &  - & $-4.7\pm 0.9$ & $62.8\pm 1.9$ \\
ep3 & $+60.3\pm 2.4$ & $43.6\pm 4.1$ & - & - & $-0.5\pm 1.3$ & $66.4\pm 2.4$ \\
 \hline 
\textbf{Sz\,100} & $-46.7\pm 1.9$ & $39.8\pm 4.3$ & $-12.2\pm 0.4$ & $20.2\pm 0.8$ &  $-29.9\pm 1.1$ & $152.4\pm 3.3$ \\
ep2 & $-38.2\pm 1.6$ & $37.9\pm 3.3$  & $-6.0\pm 0.3$ & $21.1\pm 0.7$ & $-26.2\pm 0.7$ & $143.0\pm 2.2$ \\
ep3 & $-37.7\pm 1.6$ & $34.7\pm 3.4$  & $-6.3\pm 0.4$ & $20.5\pm 0.8$ & $-24.4\pm 0.8$ & $149.3\pm 2.5$ \\
 \hline 
\textbf{Sz\,103} & $-40.4\pm 0.9$ & $36.1\pm 1.6$ & $-4.3\pm 1.6$ & $29.3\pm 3.2$  & $0.0\pm 4.2$ & $139.7\pm 9.2$ \\
ep2 & $-36.3\pm 0.4$ & $41.7\pm 0.8$ & $+2.6\pm 0.6$  & $23.5\pm 1.3$ &  $5.4\pm 2.3$ & $132.2\pm 4.1$ \\
ep3 & $-39.7\pm 0.7$ & $34.7\pm 1.2$ & $-4.8\pm 1.4$  & $28.8\pm 2.5$ & $1.5\pm 3.0$ & $139.9\pm 6.5$ \\
\hline 
\textbf{Sz 112}    & - & - & $1.6\pm 0.4$ & $38.6\pm 1.3$ & $7.9\pm 3.6$ & $125.6\pm 13.4$ \\
ep2                & - & - &  $1.4\pm 0.6$ & $34.8\pm 2.4$ & $4.6\pm 1.8$ & $78.8\pm 7.5$ \\
ep3                & - & - &   $1.6\pm 0.4$ & $32.0\pm 1.6$ & $3.5\pm 1.3$ & $85.2\pm 5.4$ \\
 \hline 
\textbf{SSTC...503} & - & - & - & - & $-7.7\pm 0.8$  & $94.9\pm 2.0$ \\
ep2                 & - & - & - & - & $-6.0\pm 0.9$  & $96.0\pm 2.1$ \\
ep3                 & - & - & - & - & $-5.7\pm 1.0$ &  $100.4\pm 2.4$\\
 \hline 
\textbf{SSTc...646} & - & - & - & - & $-22.3\pm 1.7$ & $83.9\pm 4.3$ \\
ep2                 & - & - & - & - & $-24.0\pm 2.1$ & $73.5\pm 5.1$ \\
ep3                 & - & - & - & - & $-21.5\pm 2.6$ & $76.2\pm 6.6$ \\
 \hline 
\textbf{Sz\,129}      & - & - & $+1.5\pm 0.2$ & $29.7\pm 0.5$ & $+11.9\pm 1.0$ & $117.7\pm 3.2$ \\
ep2                 & - & - & $+1.3\pm 0.3$ & $29.9\pm 0.8$ & $+10.4\pm 1.4$ & $118.6\pm 4.5$ \\
ep3                 & - & - & $+0.9\pm 0.3$ & $30.8\pm 1.0$ & $+11.4\pm 2.5$ & $129.0\pm 7.8$ \\
\hline\hline
 \end{tabular}
\tablefoot{$^1$ peak velocity, $^2$ Full width half maximum of the emission line deconvolved via $\Delta v = \sqrt{\Delta v_\text{obs}^2 - \Delta v_\text{instr}^2}$.}
 }
 \end{table*}

\begin{table*}     
{\def\arraystretch{2}\tabcolsep=1pt
\tiny
\caption{\small{Properties of the [OI]$\lambda$6300 components.}}\label{table:OI6300components_part_III}
\centering
\begin{tabular}{|c||c|c||c|c||c|c||}
\hline
 \multirow{1}{*}{ \textbf{Target}} &
      \multicolumn{2}{|c|}{\textbf{Comp. 1 (HVC)}}  &
      \multicolumn{2}{|c|}{\textbf{Comp. 2 (NLVC)}} &
      \multicolumn{2}{|c|}{\textbf{Comp. 3 (BLVC)}} \\
\hline\hline
 & $v_p^{1}$ & FWHM$^{2}$ & $v_p^{1}$ & FWHM$^{2}$ & $v_p^{1}$ & FWHM$^{2}$ \\  
 & $\text{km}\,\text{s}^{-1}$ & $\text{km}\,\text{s}^{-1}$ & $\text{km}\,\text{s}^{-1}$ & $\text{km}\,\text{s}^{-1}$  & $\text{km}\,\text{s}^{-1}$ & $\text{km}\,\text{s}^{-1}$ \\ \hline 
 \hline  
\textbf{CVSO\,58} & $-119.7\pm 1.2$ & $49.2\pm 3.9$ & $-2.9\pm 0.4$ & $37.1 \pm 1.4$ & $-17.0\pm 4.6$ & $146.9\pm 12.7$\\
 ep2 & $-119.4\pm 1.0$ & $51.1\pm 3.2$   & $-0.2\pm 0.3$ & $37.3 \pm 1.1$ & $-15.0\pm 3.3$ &    $150.7\pm 9.0$ \\
 ep3 & $-122.2\pm 1.1 $ & $46.6\pm 3.6$  & $-5.1\pm 0.4$ & $46.6\pm 3.6$ & $-27.3\pm 3.9$ & $161.9\pm 9.1$ \\
 \hline 
\textbf{CVSO\,104A} & - & - & $+35.6\pm 1.7$ $^{3}$ & $22.0\pm 5.6$ & $29.9\pm 2.0$ & $70.6\pm 6.5$ \\
ep2                 & - & - & $-2.9\pm 2.8$  $^{3}$& $18.2\pm 8.4$ & $-3.2\pm 1.7$ & $70.6\pm 5.2$ \\
ep3                 & - & - & $+38.6\pm 1.9$  $^{3}$& $17.9\pm 5.6$ & $38.0\pm 1.2$ & $70.6\pm 11.6$ \\
ep4                 & - & - & $+37.0\pm 1.1$  $^{3}$& $22.2\pm 3.5$ & $33.5\pm 1.4$ & $70.6\pm 0.1$ \\
 \hline 
\textbf{CVSO\,107} & $-111.7\pm 1.4$ & $45.3\pm 4.0$ & $-3.4\pm 0.3$ & $26.3\pm 1.0$  & $-12.4\pm 2.2$ & $128.7\pm 6.5$ \\
ep2    & $-107.2\pm 1.1$ & $40.6\pm 2.8$  & $-1.7\pm 0.3$ & $25.9\pm 1.0$   & $-10.0\pm 1.4$ & $109.1\pm 4.7$ \\
ep3    & $-107.8\pm 1.1$ & $33.3\pm 3.2$  & $-1.7\pm 0.3$ & $29.1\pm 1.0$   & $-10.6\pm 2.4$ & $150.0\pm 7.3$ \\
 \hline 
\textbf{CVSO\,109} & - & - & $-0.2\pm 0.4$ & $27.1\pm 1.4$ & $-0.8\pm 2.1$ & $92.2\pm 7.9$ \\
ep2  & - & -  & $-0.5\pm 0.4$ & $30.1\pm 1.1$ & $+6.6\pm 4.3$  & $135.7\pm 14.4$ \\
ep3  & - & -  & $-1.6\pm 0.4$ & $25.7\pm 1.5$ & $-1.2\pm 1.9$  & $82.7\pm 7.3$ \\
 \hline 
\textbf{CVSO176}   & $-100.9\pm 2.5$ & $49.4\pm 5.9$ & - & - & $+3.0\pm 0.8$ & $77.2\pm 2.1$ \\
ep2     & $-99.1\pm 2.4$ & $39.1\pm 5.7$ & - & - & $+1.29\pm 0.8$ & $74.9\pm 2.1$ \\
ep3     & $-97.2\pm 1.7$ & $27.7\pm 4.1$ & - & - & $-0.1\pm 1.1$  & $78.8\pm 2.7$ \\
 \hline 
\textbf{SO518} & $-81.7\pm 1.2$ & $43.6\pm 2.8$ & -& -& $-19.3\pm 1.4$ & $84.2\pm 5.3$ \\
  &  $+84.7\pm 7.5$  &  $79.1\pm 13.9$  & -  & -  & -  & -  \\
ep2 & $-81.9\pm 1.2$ & $43.4\pm 3.2$ & - &  -& $-19.6\pm 1.5$ & $88.2\pm 5.4$ \\
  &  $+90.9\pm 7.9$ & $89.9\pm 14.0$  & -  & -  & -  &  - \\
ep3 & $-84.2\pm 1.2$ & $46.0\pm 2.9$ & - & - & $-21.7\pm 1.3$ & $89.3\pm 5.1$ \\
  & $+81.8\pm 10.8$  & $105.6\pm 15.4$ & -  &  - &  - & -  \\
 \hline 
\textbf{SO583} & - & - & - & - & $-3.3\pm 2.9$ & $93.2\pm 7.4$ \\
ep2            & - & - & - & - & $-6.6\pm 2.5$ & $83.0\pm 6.1$ \\
ep3            & - & - & - & - & $-7.8\pm 2.2$ & $109.2\pm 5.6$ \\
 \hline\hline
 \end{tabular}
\tablefoot{$^1$ peak velocity, $^2$ Full width half maximum of the emission line deconvolved via $\Delta v = \sqrt{\Delta v_\text{obs}^2 - \Delta v_\text{instr}^2}$. $^{3}$}
 }
 \end{table*}

\begin{table*} 
{\def\arraystretch{2}\tabcolsep=1pt
\tiny
\caption{\small{Properties of the [OI]$\lambda$6300 components.}}\label{table:OI6300components_part_IV}
\centering
\begin{tabular}{|c||c|c||c|c||c|c||}
\hline
 \multirow{1}{*}{ \textbf{Target}} &
      \multicolumn{2}{|c|}{\textbf{Comp. 1 (HVC)}}  &
      \multicolumn{2}{|c|}{\textbf{Comp. 2 (NLVC)}} &
      \multicolumn{2}{|c|}{\textbf{Comp. 3 (BLVC)}} \\
\hline\hline
 & $v_p^{1}$ &  FWHM$^{2}$ & $v_p^{1}$ & FWHM$^{2}$  & $v_p^{1}$ & FWHM$^{2}$      \\  
 & $\text{km}\,\text{s}^{-1}$  &  $\text{km}\,\text{s}^{-1}$  &  $\text{km}\,\text{s}^{-1}$  &  $\text{km}\,\text{s}^{-1}$ & $\text{km}\,\text{s}^{-1}$ &   $\text{km}\,\text{s}^{-1}$  \\ \hline 
 \hline 
 \textbf{CS\,Cha}   & - & -  & $-3.2\pm 0.3$ & $16.8\pm 0.8$ &    - & - \\
ep2               & - & -  & $-1.7\pm 0.3$ & $16.5\pm 0.7$ &   - & - \\
ep2               & - & -  & $-2.9\pm 0.3$ & $17.0\pm 0.8$ &   - & - \\
\hline 
\textbf{CV\,Cha}    & $-59.4\pm 6.2$ & $76.1\pm 10.5$ &   $-4.6\pm 1.1$ & $44.6\pm 2.1$  & - & - \\
ep2                & $-56.4\pm 6.7$ & $78.7\pm 11.2$ &   $-3.0\pm 1.0$ & $43.2\pm 2.2$ & - & - \\
ep3                & $-63.1\pm 8.0$ & $92.6\pm 13.6$ &   $-4.8\pm 0.8$ & $45.0\pm 2.5$ & - & - \\
\hline 
\textbf{VW\,Cha}    & - & - & $-4.5\pm 0.3$ & $28.6\pm 1.2$  & $-8.2\pm 1.6$ & $86.5\pm 5.8$ \\
ep2               & - & - & $-3.0\pm 0.4$ & $29.0\pm 1.4$  & $-5.1\pm 1.2$ & $78.7\pm 4.9$ \\
ep3               & - & - & $-3.9\pm 0.3$ & $27.4\pm 1.1$ & $-8.1\pm 1.3$ & $77.0\pm 4.7$ \\
\hline 
\textbf{VZ\,Cha}   & - & - & $1.0\pm 0.4$ & $19.2\pm 1.0$ & $6.4\pm 3.7$ & $136.3\pm 11.2$ \\
ep2             & - & - & $1.8\pm 0.3$ & $21.8\pm 0.9$ & $14.2\pm 4.4$ & $155.4\pm 13.1$ \\
ep3            & - & - & $1.9\pm 0.3$ & $22.7\pm 0.9$ & $-1.5\pm 4.1$ & $150.2\pm 13.1$ \\
\hline \hline 
 \textbf{Hn\,5}  & - & -  & $-1.5\pm 0.5$ & $13.8\pm 1.2$ & $-5.1\pm 2.8$ & $131.4\pm 8.5$ \\
ep2             & - & - & $-1.7\pm 0.5$ & $10.8\pm 1.3$ & $1.0\pm 2.9$ & $120.9\pm 7.9$ \\
ep3            & - & -  & $-2.4\pm 0.7$ & $14.2\pm 1.8$  & $-10.7\pm 3.1$ & $108.6\pm 9.2$ \\
 \hline 
 \textbf{XX\,Cha}   & $-67.2\pm 0.3$ & $21.1\pm 0.7$   & $-6.7\pm 0.3$ & $18.2\pm 0.8$   & $-27.4\pm 0.4$ & $79.8\pm 0.7$ \\
ep2                & $-66.1\pm 0.2$ & $21.2\pm 0.6$   & $-6.2\pm 0.3$ & $19.0\pm 0.8$   & $-27.5\pm 0.4$ & $81.7\pm 0.6$ \\
ep3                & $-66.0\pm 0.3$ & $22.1\pm 0.8$   & $-5.5\pm 0.4$ & $18.6\pm 0.9$   & $-25.6\pm 0.5$ & $80.7\pm 0.8$ \\
 \hline 
 \textbf{Sz\,40}     & - & -     & - & -    & $-10.7\pm 1.6$ & $58.7\pm 1.3$ \\
 ep2                & - & -     & - & -    & $-8.7\pm 0.9$  & $45.5\pm 2.0$ \\
\hline \hline
\textbf{TWHya} & - & - & $0.0\pm 0.1$ & $9.5\pm 0.3$ & $-1.3\pm 0.2$ & $22.8\pm 0.8$   \\
ep2             & - & - & $-0.2\pm 0.1$ &$9.7\pm 0.2$ & $-1.3\pm 0.2$ & $24.0\pm 0.9$ \\
\hline 
\textbf{DK\,Tau\,A}  & $-95.2\pm 3.4$ & $71.7\pm \pm 6.3$ & $-3.9\pm 0.6$ & $31.8\pm 2.1$ & $+17.0\pm 4.2$ & $132.1\pm 5.8$ \\
ep2             &  $-104.4\pm 1.7$ & $63.3\pm 4.0$ & $0.0\pm 0.4$ & $29.0\pm 1.2$ & $+8.0\pm 1.5$ & $138.5\pm 3.5$ \\
ep3             &  $-101.5\pm 2.1$ & $58.8\pm 5.3$ & $-2.4\pm 0.5$ & $28.8\pm 1.4$ & $+8.6\pm 1.9$ & $140.6\pm 4.1$ \\
\hline
\textbf{DK\,Tau\,B}  & $-71.1\pm 5.1$ & $66.6\pm 7.4$  & $-9.0\pm 1.2$ & $34.8\pm 3.9$  &  $-0.7\pm 1.4$ & $77.2\pm 10.2$  \\
ep2             & $-64.5\pm 5.4$ & $54.6\pm 6.8$  & $27.3\pm 2.9$ & $41.8\pm 5.0$  & $-13.7\pm 1.8$ & $47.0\pm 6.3$  \\
ep3             & $-59.8\pm 4.7$ & $73.3\pm 8.1$  & $-0.6 \pm   1.3$ & $27.6 \pm 3.2$  &  $-0.5\pm 1.9$ & $90.2\pm 8.4$  \\
 \hline \hline
 \end{tabular}
\tablefoot{$^1$ peak velocity, $^2$ Full width half maximum of the emission line, $^3$ deconvolved line width calculated via $\Delta v = \sqrt{\Delta v_\text{obs}^2 - \Delta v_\text{instr}^2}$.}
 }
 \end{table*}

\setcounter{table}{0}
\renewcommand{\thetable}{G.\arabic{table}}
  
 \setcounter{figure}{0}
\renewcommand\thefigure{\thesection G.\arabic{figure}}  

\setcounter{equation}{0}
\renewcommand{\theequation}{G.\arabic{equation}}

%
\begin{figure*} 
\centering
\subfloat{\includegraphics[trim=0 0 0 0, clip, width=0.25 \textwidth]{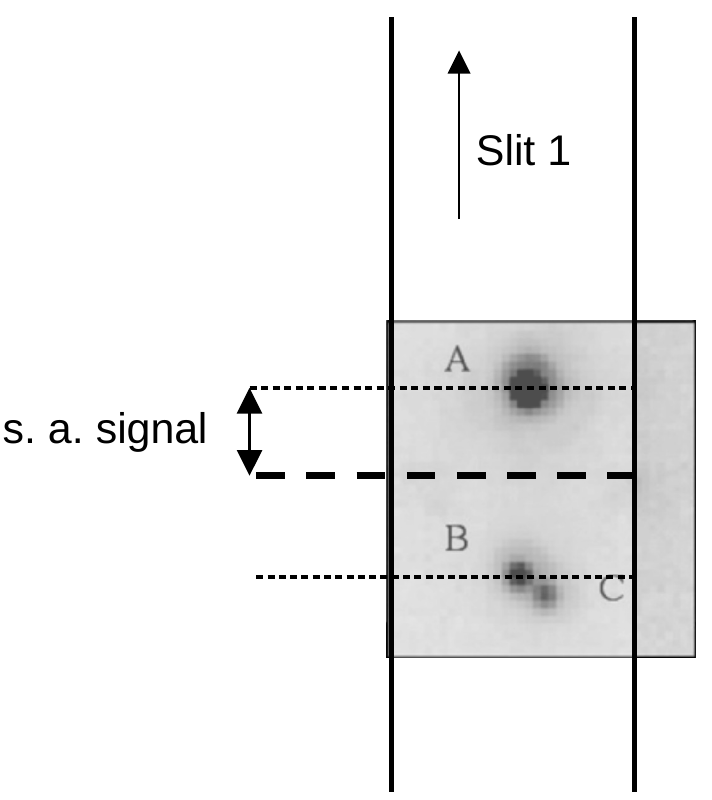}}
\hfill
\subfloat{\includegraphics[trim=0 0 0 0, clip, width=0.3 \textwidth]{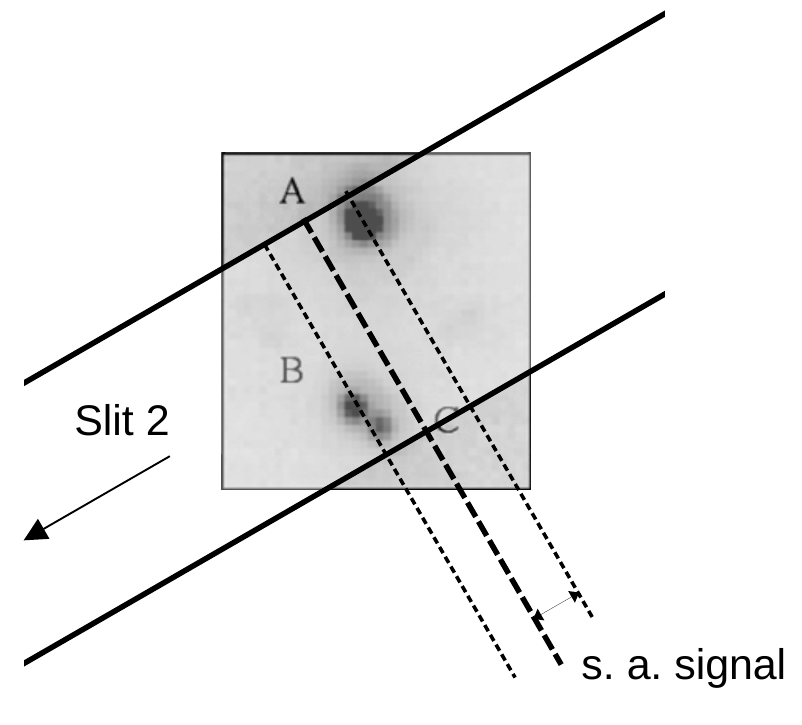}}
\hfill
\subfloat{\includegraphics[trim=0 0 0 0, clip, width=0.25 \textwidth]{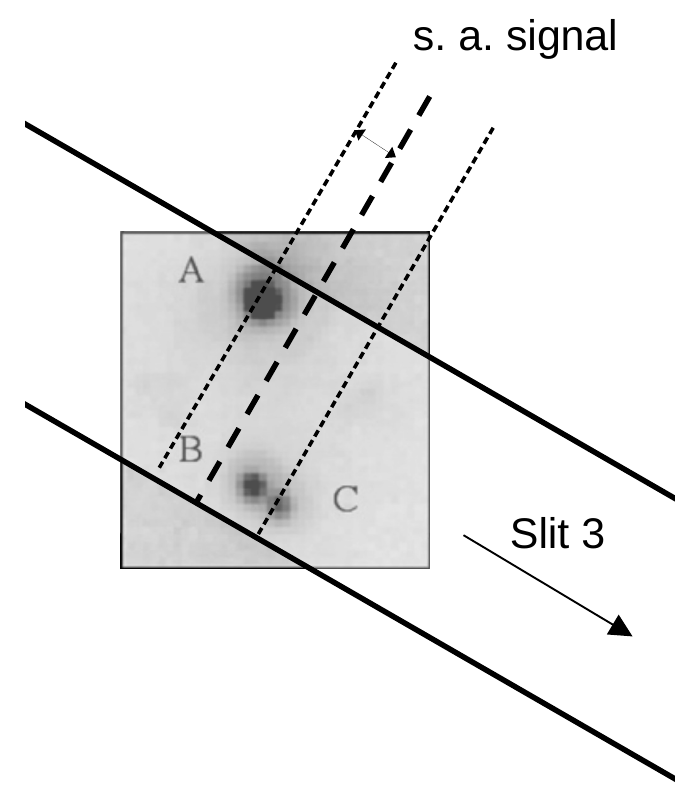}}
\hfill   
\caption{\small{Explanation of the spectro-astrometric signals detected towards VW\,Cha. VW\,Cha forms a triple system \citep[Fig.\,2 in][]{correia_2006}. The constituents A and B are separated by $0\farcs 65$ at P.A. 177$^\text{o}$, that is, aligned with slit position 1 of this study. The constituents B and C are separated by $0\farcs 11$ at P.A. 233$^\text{o}$. In the three  observations the $0\farcs 6$ wide UVES slit (thick solid lines) covers all three components of VW\,Cha. The photocentre of VW\,Cha (marked as thick dashed line) lies in between A and B+C depending on the slit orientation and the brightnesses of A, B, and C. The detected different spectro-astrometric offsets with respect to the photocentre are therefore an effect of the slit orientation.}}\label{fig:vwcha_slits}
\end{figure*} 

\end{document}